# THE INTERNATIONAL LINEAR COLLIDER

Technical Design Report | Volume 3.II: Accelerator Baseline Design

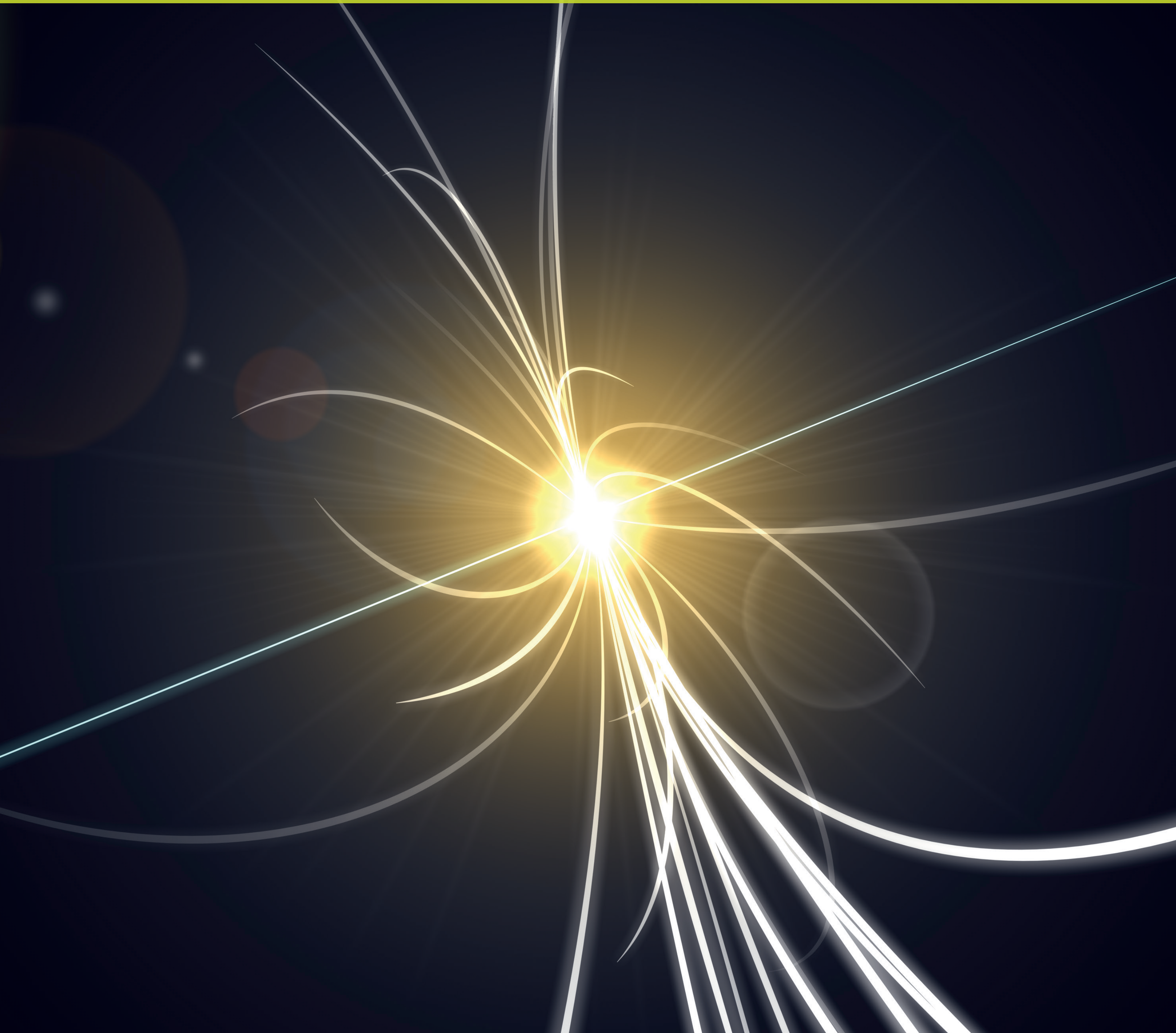

# The International Linear Collider

# Technical Design Report

2013

international linear collider

Volume 1: Executive Summary

Volume 2: Physics

Volume 3: Accelerator

Part I: R&D in the Technical Design Phase

Volume 3: Accelerator

Part II: Baseline Design

Volume 4: Detectors

# Volume 3

# Accelerator

## Part II

## Baseline Design

Editors

Chris Adolphsen, Maura Barone, Barry Barish, Karsten Buesser, Phil Burrows, John Carwardine (Editorial Board Chair), Jeffrey Clark, Hélène Mainaud Durand, Gerry Dugan, Eckhard Elsen, Atsuto Enomoto, Brian Foster, Shigeki Fukuda, Wei Gai, Martin Gastal, Rongli Geng, Camille Ginsburg, Susanna Guiducci, Mike Harrison, Hitoshi Hayano, Keith Kershaw, Kiyoshi Kubo, Vic Kuchler, Benno List, Wanming Liu, Shinichiro Michizono, Christopher Nantista, John Osborne, Mark Palmer, James McEwan Paterson, Thomas Peterson, Nan Phinney, Paolo Pierini, Marc Ross, David Rubin, Andrei Seryi, John Sheppard, Nikolay Solyak, Steinar Stapnes, Toshiaki Tauchi, Nobu Toge, Nicholas Walker, Akira Yamamoto, Kaoru Yokoya

# Acknowledgements

We acknowledge the support of BMWF, Austria; MinObr, Belarus; FNRS and FWO, Belgium; NSERC, Canada; NSFC, China; MPO CR and VSC CR, Czech Republic; Commission of the European Communities; HIP, Finland; IN2P3/CNRS, CEA-DSM/IRFU, France; BMBF, DFG, Helmholtz Association, MPG and AvH Foundation, Germany; DAE and DST, India; ISF, Israel; INFN, Italy; MEXT and JSPS, Japan; CRI(MST) and MOST/KOSEF, Korea; FOM and NWO, The Netherlands; NFR, Norway; MNSW, Poland; ANCS, Romania; MES of Russia and ROSATOM, Russian Federation; MON, Serbia and Montenegro; MSSR, Slovakia; MICINN-MINECO and CPAN, Spain; SRC, Sweden; STFC, United Kingdom; DOE and NSF, United States of America.

# Contents

# Common preamble to Parts I and II

The International Linear Collider (ILC) is a linear electron-positron collider based on 1.3 GHz superconducting radio-frequency (SCRF) accelerating technology. It is designed to reach 200-500 GeV (extendable to 1 TeV) centre-of-mass energy with high luminosity. The design is the result of over twenty years of linear collider R&D, beginning in earnest with the construction and operation of the SLC at SLAC. This was followed by extensive development work on warm X-band solutions (NLC/JLC) and the pioneering work by the TESLA collaboration in the 1990s on superconducting L-band RF. In 2004, the International Technology Review Panel, set up by the International Committee for Future Accelerators, ICFA, selected superconducting technology for ILC construction. The Global Design Effort (GDE) was set up by ICFA in 2005 to coordinate the development of this technology as a worldwide international collaboration. Drawing on the resources of over 300 national laboratories, universities and institutes worldwide, the GDE produced the ILC *Reference Design Report* (RDR) [1–4] in August 2007. The report describes a conceptual design for the ILC and gives an estimated cost and the required personnel from collaborating institutions.

The work done by the GDE during the RDR phase identified many high-risk challenges that required R&D, which have subsequently been the focus of the worldwide activity during the Technical Design Phase. This phase has achieved a significant increase in the achievable gradient of SCRF cavities through a much better understanding of the factors that affect it. This improved understanding has permitted the industrialisation of the superconducting RF technology to more than one company in all three regions, achieving the TDP goal of 90 % of industrially produced cavities reaching an accelerating gradient of 31.5 MV/m. A further consequence is an improved costing and construction schedule than was possible in the RDR. Other important R&D milestones have included the detailed understanding of the effects of, and effective mitigation strategies for, the "electron-cloud" effects that tend to deteriorate the quality of the positron beam, particularly in the ILC damping rings. The achievement of the R&D goals of the TDR has culminated in the publication of this report, which represents the completion of the GDE's mandate; as such, it forms a detailed solution to the technical implementation of the ILC, requiring only engineering design related to a site-specific solution to allow the start of construction.

Volume 3 (Accelerator) of the *Technical Design Report* is divided into two separate parts reflecting the GDE's primary goals during the Technical Design Phase period (2007–2012):

**Part I: R&D in the Technical Design Phase** summarises the programmes and primary results of the risk-mitigating worldwide R&D including industrialisation activities.

**Part II: Baseline Design** provides a comprehensive summary of the reference layout, parameters and technical design of the accelerator, including an updated cost and construction schedule estimate.

The R&D results and studies of cost-effective solutions for the collider presented in Part I directly support the design presented in Part II, which is structured as a technical reference.

# Chapter 1
# Introduction

This reference report contains the technical specifications and design for an International Linear Collider that is based on mature technology and is relatively low risk. The heart of the accelerator consists of two approximately 11-km long SCRF main linacs, based on the technology developed by the TESLA collaboration and proposed in 2001 for the TESLA linear collider [5]. The updated design reflects the significant worldwide developments in this technology, with the establishment of R&D infrastructure as well as a significant industrial base in the Americas, Asia and Europe. The global high-gradient SCRF R&D driven by the GDE has succeeded in routinely establishing the required 35 MV/m average performance, with every indication that this could be exceeded in future years. Integrated systems tests at the TTF2/FLASH accelerator in DESY, Hamburg have demonstrated many of the design and performance parameters for the ILC, and this currently unique facility will soon be joined by similar test accelerators in both KEK, Japan, and Fermilab, USA.

The design evolution since the original RDR reflects the results of this R&D, a re-evaluation of cost-performance trade-offs, and a more detailed considerations of site-specific cost-optimum design options. Beyond the fundamental R&D, the on-going industrialisation of the technology has enabled the GDE to provide realistic industrial studies for globally mass-producing the approximately 18,000 SCRF nine-cell cavities required and assembling them into 1750 cryomodules. These studies have resulted in a relatively robust and defensible cost estimate, as well as clear concepts as to how the machine could be constructed as an international project based predominantly on in-kind contributions, complete with a realistic construction and installation schedule. The system designs and associated cost estimates reported here are considered sufficiently complete to form a sound basis for a "Proposal to Construct" soon after an International ILC Organisation has been formalised and a specific site has been selected.

Extensive studies of the physics potential of the ILC have taken place over many years [2, 6]. They have explored the complementarity of the ILC with the Large Hadron Collider (LHC) as well as the unique discovery potential of the ILC. The identification of a Higgs boson at the LHC [7–10] validates these studies of Standard Model physics at the ILC, not only with regard to the Higgs but also top physics and precisions studies. The lack to date of any signal for physics beyond the Standard Model gives no explicit motivation to go to energies higher than the 500 GeV of the first stage of the ILC while placing a premium on the flexibility of the ILC to be upgraded to energies up to and beyond 1 TeV. In addition, the precision studies possible at the ILC may well give indications of new physics at much higher energies.

The ILC design detailed in this volume can achieve the performance during the first years of operation that fullfil the physics potential of the ILC as detailed by the above process. These design criteria are:

- A continuous centre-of-mass energy range between 200 GeV and 500 GeV
- A peak luminosity of approximately $2 \times 10^{34}\ \mathrm{cm^{-2}\,s^{-1}}$ at 500 GeV centre-of-mass

- 80 % electron polarisation at the Interaction Point (IP)
- A relative energy stability and precision of ≤0.1%
- An option for 50 to 60 % positron polarisation

In addition the machine must be upgradeable to a centre-of-mass energy of 1 TeV, which at a minimum implies a site that can be expanded from 30 to 50 km in length. The ILC design documented here guarantees a rich, varied and flexible physics program to complement that of the Large Hadron Collider (LHC).

The parameters, the basic layout of the machine and the design of most of the technical subsystems represent a single generic solution, independent of considerations of possible site constraints. In particular, the design of the main linac technical systems (cavities, cryomodules, klystrons and modulators etc.) are independent of the final location of the machine. In addition, the design of the other accelerator systems (damping rings, electron and positron sources, beam delivery system) are also 'generic'. However, the optimum systems design and the dependence of choice of site – in particular the solutions for the main SCRF linac RF-power distribution and the Conventional Facilities and Siting (CFS) – have been addressed during the technical design phase. As a result, two variants of Civil and Technical design are elaborated in this report:

**Flat topography** refers to a site-specific design where relatively flat surface areas are readily available for equipment and service buildings, with access being provided to the underground accelerator tunnels via vertical shafts. The LHC is an example of such a topography, and both the European and Americas regional sample sites (CERN and FNAL respectively) are based on this design variant;

**Mountainous topography** refers to a site-specific design more suited to a steeply sloping surface environment where available space for 'surface buildings' is at a premium. In this case, the majority of equipment is housed underground, and access is provided by horizontal (or gently sloping) access tunnels. The Asian sample sites in Japan are based on this design variant.

In both cases, a significant difference in underground geology and local experience has also strongly influenced the choice of underground tunnel solutions. The two site-dependent variants are further differentiated by the approach adopted in supplying the RF power for the superconducting linacs: one predominantly surface based (Klystron Cluster Scheme, or KCS), and thus more suited to the flat-topography variant; and one a more traditional Distributed Klystron Scheme (DKS) suitable for underground implementation in the mountainous-region design. Each approach has significant differences in the criteria and requirements for the conventional facilities and civil engineering. They provide mature solutions which provide the flexibility to allow the ILC to be adapted relatively quickly to any emerging potential host site.

Despite the major differences of the two site variants, the core requirements, accelerator layout and technologies remain the same. Figure 1.1 describes the structure of the design work presented in the remainder of this report.

The ILC has from the outset been set up as a large science project with international governance from design through construction to operation. There have been several regional studies of the pros and cons of different approaches to the governance of large international science projects and the ILC GDE independently studied the issues and produced a Project Implementation Planning document [11]. This is summarised in Chapter 13. The issues discussed include funding models with both common funds and in-kind contributions from international partners, and the unique and extensive responsibilities of the host region or nation.

The remainder of this report – TDR Part II: ILC Baseline Design – provides a comprehensive description of the complete current baseline technical design and cost of the ILC, including the site dependent variants. An overview of the design is given in Section 2.1. In brief, the 500 GeV centre-of-mass energy ILC has a footprint of ~31 km. At its heart are two 11 km long linacs based on superconducting RF niobium cavities running at 1.3 GHz, operating at an average gradient of 31.5 MV/m and with a pulse length of 1.6 ms. The linacs are designed to accelerate beams of electrons and positrons to energies up to 500 GeV and collide them at energies up to 1 TeV. The electron and positron beams themselves are produced in different ways: the electron beams are obtained from a polarised source; positrons are produced via pair-conversion of high-energy photons produced in an undulator, which means that polarised positrons are more difficult to produce and are thus a design option. The high luminosity required to fulfil the ILC's ambitious physics programme can only be obtained if both electron and positron beams are "cooled" significantly, compressing their phase space at 5 GeV via damping rings with a circumference of 3.2 km. This low emittance is maintained by a beam-transport system followed by a two-stage compressor which produces bunch trains consisting of 1312 bunches in a train of length ~500 ns at a repetition rate of 5 Hz on entry to the accelerating linacs. After acceleration, the beams are brought into collision by 2.25 km long beam-delivery systems, which bring the two beams into collision with a 14 mrad crossing angle and with the optimum parameters to maximise the produced luminosity.

**Figure 1.1**
Approach to the site-specific design variants for the ILC.

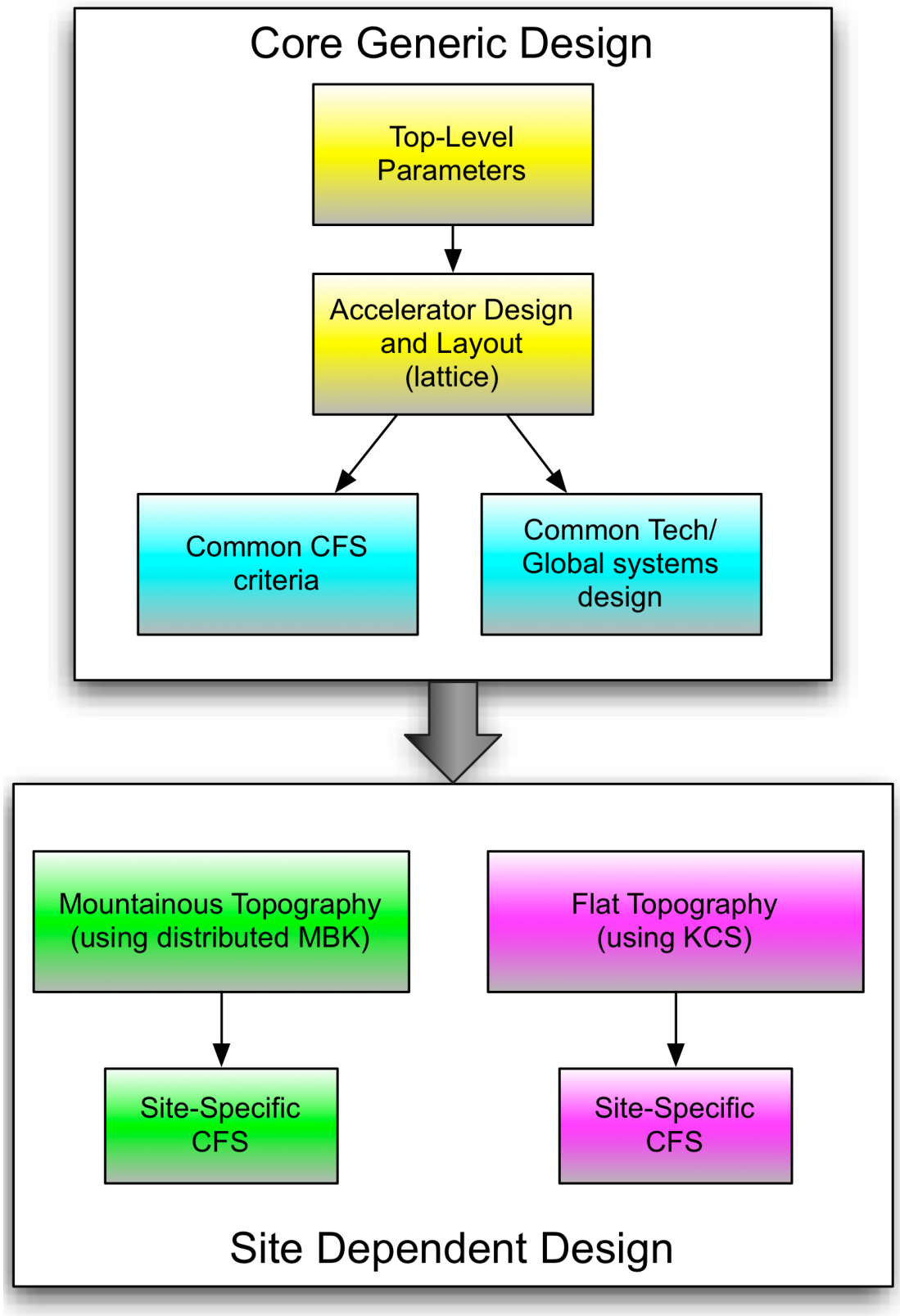

Further detailed technical documentation is available in the ILC Technical Design Documentation EDMS system (http://ilc-edms.desy.de) or directly from http://www.linearcollider.org/ILC/GDE/technical-design-documentation, and are referenced in this report where appropriate. For details of the Technical Design Phase R&D programme, see TDR Part I.

# Chapter 2
# General Parameters, Layout and Systems Overview

## 2.1 Introduction

This chapter is intended to provide an introductory overview of the ILC machine design, its top-level parameters and sub-system functionality, in preparation for the more detailed descriptions in the remaining chapters of the report. Figure 2.1 shows a schematic view of the overall layout of the ILC, indicating the location of the major sub-systems:

- a polarised electron source based on a photocathode DC gun;
- an undulator-based positron source, driven by a the high-energy main electron beam;
- 5 GeV electron and positron damping rings (DR) with a circumference of 3.2 km, housed in a common tunnel at the centre of the ILC complex;
- beam transport from the damping rings to the main linacs, including acceleration to 15 GeV followed by a two-stage bunch compressor system prior to injection into the main linac;
- two 11 km long main linacs, utilising 1.3 GHz SCRF cavities, operating at an average gradient of 31.5 MV/m, with a pulse length of 1.6 ms;
- a $2{\times}2.25$ km-long beam-delivery system, which brings the two beams into collision with a 14 mrad crossing angle, at a single interaction point which can be shared by two detectors (push-pull).

The total footprint is ~31 km. The electron source, positron source (including a low-powered auxiliary source), and the electron and positron damping rings are centrally located around the interaction region (IR) in the *central region*. The damping ring complex is laterally displaced by a sufficient distance so as not to interfere with the detector hall, and is connected to the main accelerator housing via transfer tunnels. The electron and positron sources themselves are housed in the same (main accelerator) tunnels as the beam-delivery systems, to reduce the overall cost and scope of the underground construction of the central region.

In the remainder of this chapter, Sections 2.2 and 2.3 provide an overview of the top-level parameters and the common accelerator description. Section 2.4 provides an introduction to the two site-dependent solutions, mostly pertaining to conventional facilities and siting. Finally, Section 2.5 briefly introduces the scope of the optional luminosity and energy upgrades.

**Figure 2.1**
Schematic layout of the ILC complex for 500 GeV CM.

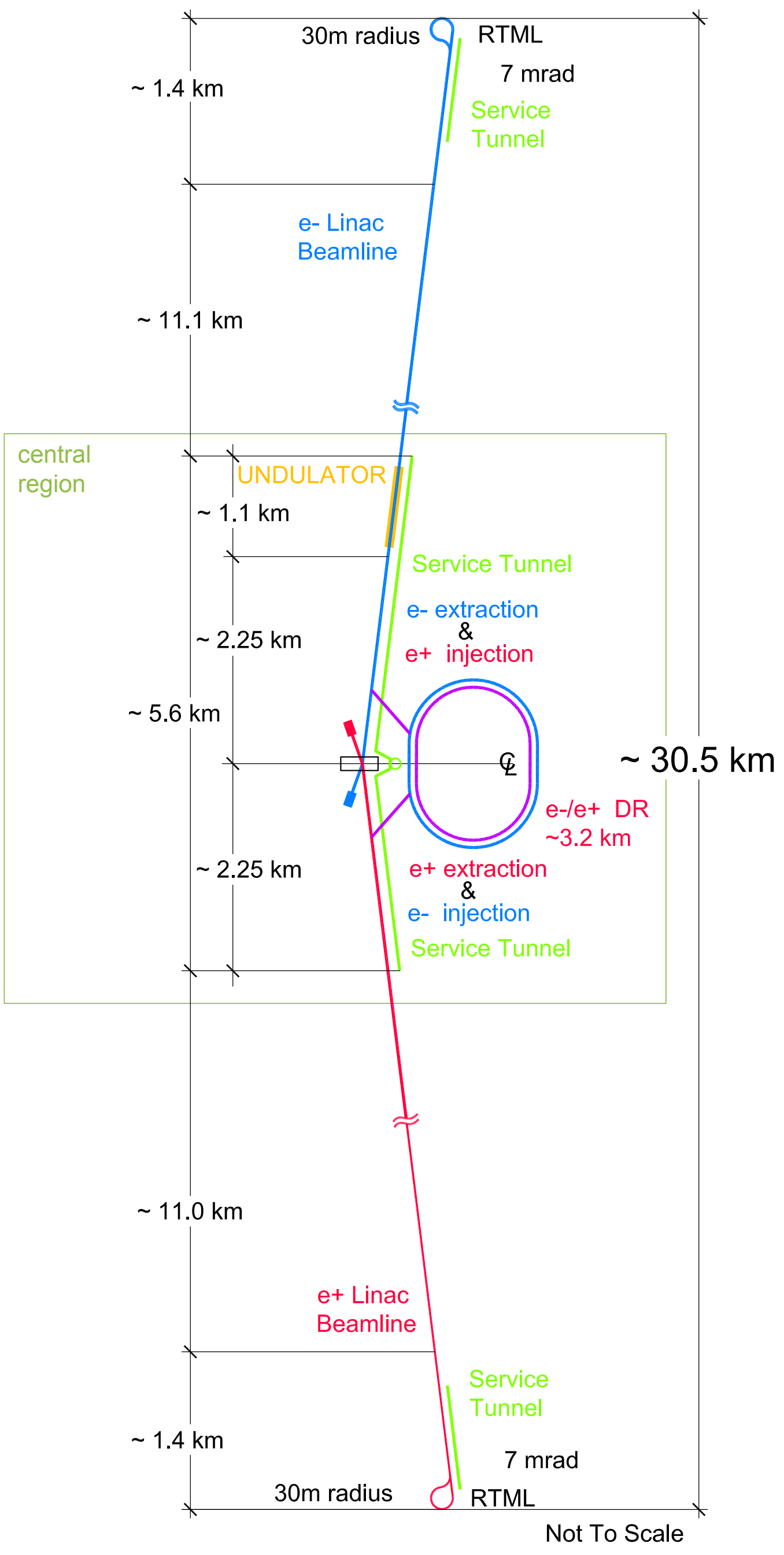

## 2.2 Top-Level Parameters

### 2.2.1 Physics related machine parameters for 200–500 GeV centre-of-mass running

The top-level parameters for the baseline operational range of centre-of-mass energies from 200 to 500 GeV have been optimised to provide the maximum attainable physics performance with a relatively low risk and minimum cost. Table 2.1 shows the primary parameters for 200, 250, 350 and 500 GeV centre-of-mass operation.

The choice of parameters represent trade-offs between the constraints imposed by the various accelerator sub-systems:

- For the damping rings, bunch charge, bunch spacing and the total number of bunches are limited by various instability thresholds. The most important is the electron cloud in the positron ring; other factors include realistically achievable injection and extraction kicker pulse rise-times and the desire to minimise the circumference of the rings and thereby the cost.

**Table 2.1.** Summary table of the 200–500 GeV baseline parameters for the ILC. The reported luminosity numbers are results of simulation [12]

| Centre-of-mass energy | $E_{CM}$ | GeV | 200 | 230 | 250 | 350 | 500 |
|---|---|---|---|---|---|---|---|
| Luminosity pulse repetition rate | | Hz | 5 | 5 | 5 | 5 | 5 |
| Positron production mode | | | 10 Hz | 10 Hz | 10 Hz | nom. | nom. |
| Estimated AC power | $P_{AC}$ | MW | 114 | 119 | 122 | 121 | 163 |
| Bunch population | $N$ | $\times 10^{10}$ | 2 | 2 | 2 | 2 | 2 |
| Number of bunches | $n_b$ | | 1312 | 1312 | 1312 | 1312 | 1312 |
| Linac bunch interval | $\Delta t_b$ | ns | 554 | 554 | 554 | 554 | 554 |
| RMS bunch length | $\sigma_z$ | μm | 300 | 300 | 300 | 300 | 300 |
| Normalized horizontal emittance at IP | $\gamma\epsilon_x$ | μm | 10 | 10 | 10 | 10 | 10 |
| Normalized vertical emittance at IP | $\gamma\epsilon_y$ | nm | 35 | 35 | 35 | 35 | 35 |
| Horizontal beta function at IP | $\beta_x^*$ | mm | 16 | 14 | 13 | 16 | 11 |
| Vertical beta function at IP | $\beta_y^*$ | mm | 0.34 | 0.38 | 0.41 | 0.34 | 0.48 |
| RMS horizontal beam size at IP | $\sigma_x^*$ | nm | 904 | 789 | 729 | 684 | 474 |
| RMS vertical beam size at IP | $\sigma_y^*$ | nm | 7.8 | 7.7 | 7.7 | 5.9 | 5.9 |
| Vertical disruption parameter | $D_y$ | | 24.3 | 24.5 | 24.5 | 24.3 | 24.6 |
| Fractional RMS energy loss to beamstrahlung | $\delta_{BS}$ | % | 0.65 | 0.83 | 0.97 | 1.9 | 4.5 |
| Luminosity | $L$ | $\times 10^{34}$ cm$^{-2}$ s$^{-1}$ | 0.56 | 0.67 | 0.75 | 1.0 | 1.8 |
| Fraction of $L$ in top 1% $E_{CM}$ | $L_{0.01}$ | % | 91 | 89 | 87 | 77 | 58 |
| Electron polarisation | $P_-$ | % | 80 | 80 | 80 | 80 | 80 |
| Positron polarisation | $P_+$ | % | 30 | 30 | 30 | 30 | 30 |
| Electron relative energy spread at IP | $\Delta p/p$ | % | 0.20 | 0.19 | 0.19 | 0.16 | 0.13 |
| Positron relative energy spread at IP | $\Delta p/p$ | % | 0.19 | 0.17 | 0.15 | 0.10 | 0.07 |

- The maximum length of the beam pulse is constrained by the decision to limit the length of the Main Linac RF pulse to the $\sim$ 1.6 ms now routinely achieved in the available 1.3 GHz 10 MW multi-beam klystrons and modulators. The beam current is further constrained by the need to minimise the required number of klystrons (peak power), as well as from consideration of high-order modes (cryogenic load and beam dynamics). Dynamic cryogenic load (refrigeration) is also a cost driver, and limits the repetition rate of the machine.
- Both the electron and positron sources set constraints on the achievable beam-pulse parameters. For the laser-driven photo-cathode polarised electron source, the limits are set by the laser; for the undulator-based positron source, the limits are set by consideration of power deposition in the photon target. The beam-pulse length is further constrained by the achievable performance of the warm RF capture sections (both sources).
- At the interaction points, single bunch parameters are limited by the strong beam-beam effects and a desire to control both beam-beam backgrounds and the kink instability.

Finally, a careful reevaluation of the cost-performance balance has resulted in a choice of parameters which are considered relatively low risk and cost effective. All the primary cost related parameters have been either directly demonstrated, or represent justifiable extrapolations from the current state of the art.

### 2.2.2 Special considerations for running at low centre-of-mass energy

While the maximum energy performance requirement dictates many of the key parameters and the overall geometry and cost of the machine, attention needs to be given to providing sufficient luminosity at the lower centre-of-mass-energy range, and in particular <300 GeV. Two issues limit the possible performance at these lower energies:

- positron production from the undulator-based source is significantly degraded for electron beam energy below 150 GeV;
- the beam divergence at the interaction point is nominally constrained by the collimation depth, which results in a $\gamma^2$ scaling of the luminosity, rather than the traditionally assumed $\gamma$-scaling.

The solution adopted for the current baseline for the positron source is to have an additional electron pulse at 150 GeV energy to make positrons. This additional pulse would be interleaved with the nominal 5 Hz luminosity production pulse. This so-called 10 Hz operation mode leads to several design criteria for the baseline:

- Both electron and positron damping rings must now damp the beam in 100 ms instead of the nominal 200 ms. This requires additional wigglers and RF in the ring.
- The positron damping ring is 'empty' for 100 ms, after which the current is ramped up in $\sim$1 ms (and similarly ramped down during extraction). Dealing with transient beam loading requires an additional RF power overhead (approximately 15%).
- All the linacs for the electron machine (capture RF, 5 GeV booster linac, bunch compressors and main linac) must run at 10 Hz.
- The positron production pulse (150 GeV beam) must be safely extracted after the source undulator and dumped, requiring an additional pulsed magnet and extraction beamline system.
- A pulsed-magnet steering system is required upstream of the source undulator (downstream of the main electron linac) to compensate for the difference in trajectory between the 150 GeV positron production and <150 GeV luminosity pulses.

The 10 Hz mode is made cost effective by the fact the total RF power and cryoload for the main (electron) linac does not exceed the 500 GeV case when the beam energy (and therefore the main linac gradient) is reduced below 150 GeV. For the electron bunch compressor and source linacs, the AC power requirement effectively doubles for the 10 Hz operation mode. The 10 Hz mode also drives the design criteria and power requirements for the damping rings.

To mitigate the beam-divergence constraint at the IP, a shorter FD arrangement is used for $E_{\mathrm{cm}} \leq 300\,\mathrm{GeV}$, which increases the collimation depth and hence the IP beam divergence (by up to 30 % in the horizontal plane). The FD will be implemented in a modular design to accommodate both high- and low-energy running configurations, thus avoiding the need to exchange the magnet cryostat.

There are no issues with running the main SCRF linacs at reduced gradient in order to produce lower centre-of-mass energy. The lower average gradient results in a shorter fill time and overall higher RF-to-beam power efficiency. Simulations of the beam dynamics have indicated no significant additional degradation of vertical emittance.

## 2.3 Accelerator Layout and Design

### 2.3.1 Superconducting RF Main Linacs

The ILC Main Linacs accelerate the beams from 15 GeV (after acceleration in the upstream bunch compressors) to a maximum energy of 250 GeV. Beam acceleration in each linac is provided by approximately 7,400 $\sim$1 m-long superconducting niobium cavities consisting of nine elliptical cells (see Fig. 2.2) operating at 2 K, assembled into $\sim$850 cryomodules. The average gradient of the cavities is 31.5 MV/m (for 500 GeV centre-of-mass beam energy), with a corresponding $Q_0 \geq 10^{10}$. A random cavity-to-cavity gradient spread of $\pm$20% is assumed to accommodate expected mass-production variations in the maximum achievable gradient.

For an average of 31.5 MV/m operation with the nominal beam current of 5.8 mA, the optimal matched $Q_L \approx 5.4 \times 10^6$. This corresponds to a cavity fill time of 925 µs, which, together with the nominal beam pulse of 727 µs, requires a total RF pulse length of 1.65 ms.

As well as the adjustable high-power coupler, the cavity package includes the cavity mechanical tuner, which is integrated into the titanium helium vessel of the cavity. In addition to a slow mechanical tuner (used for initial tuning and slow drift compensation), a fast piezo-driven tuner is also included to dynamically compensate Lorentz-force detuning during the RF pulse.

**Figure 2.2**
A 1.3 GHz superconducting nine-cell niobium cavity.

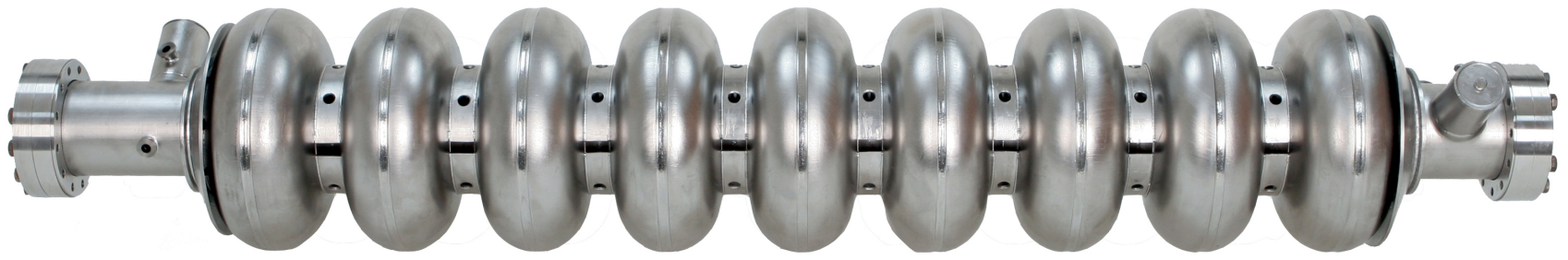

The Main Linacs are constructed almost entirely from the two standard variants of ILC cryomodule, both 12.65 m long: a Type A module with nine 1.3 GHz nine-cell cavities; and Type B with eight nine-cell cavities and one superconducting quadrupole package located at the centre of the module. The Main Linac has a FODO lattice structure, with a quadrupole (Type B module) every third cryomodule.

The cryomodule design is a modification of the Type-3 version (Fig. 2.3) developed and used at DESY in the TTF2/FLASH accelerator, and also being used for the 100 cryomodules currently being produced by industry for the European X-Ray FEL (XFEL), also based at DESY. Within the cryomodules, a 300 mm-diameter helium-gas return pipe serves as a strongback to support the nine cavities and other beam-line components in the case of the Type-A module. For the Type-B module, the central cavity package is replaced by a superconducting quadrupole package that includes the quadrupole itself, a cavity BPM, and superconducting horizontal and vertical corrector dipole magnets. The quadrupoles establish the main-linac magnetic lattice, which is a weak-focusing FODO optics with an average beta function of ∼80 m. Every cryomodule also contains a 300 mm-long higher-order-mode beam-absorber assembly that removes energy through the 40-80 K cooling system from beam-induced higher-order modes above the cavity-cutoff frequency.

**Figure 2.3**
SCRF cryomodule. Left: a type-III module being installed at DESY's FLASH facility. Right: the ILC type-IV module.

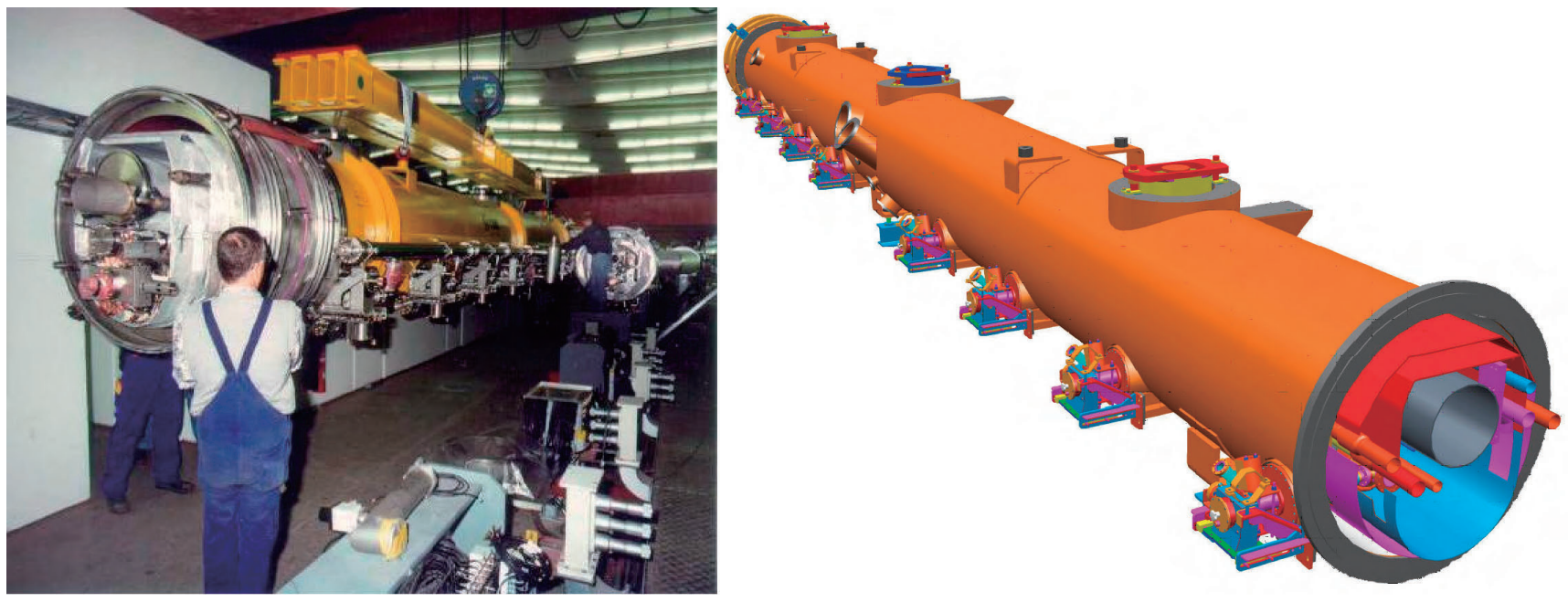

To operate the cavities at 2 K, they are immersed in a saturated He II bath, and helium gas-cooled shields intercept thermal radiation and thermal conduction at 5–8 K and at 40–80 K. The estimated static and dynamic cryogenic heat loads per cryomodule at 2 K are approximately 1.7 W and 9.8 W, respectively. Liquid helium for the main linacs and the bunch-compressor RF is supplied from a total of 10–12 large cryogenic plants, each of which has an installed equivalent cooling power of ∼20 kW at 4.5 K. The plants are located in pairs approximately every 5 km along the linacs, with each plant cooling ∼2.5 km of contiguous linac. The main linacs follow the Earth's average curvature to simplify the liquid-helium transport.

The RF power is provided by 10 MW multibeam klystrons (MBK), each driven by a 120 kV Marx modulator. The 10 MW MBK is now a well established technology having achieved the ILC specifications and has several vendors worldwide (Fig. 2.4). The 120 kV Marx-modulator prototypes (Fig. 2.5) have achieved the required specifications and are now undergoing design for manufacture and cost.

**Figure 2.4**
Examples of industry produced 10 MW Multibeam Klystrons.

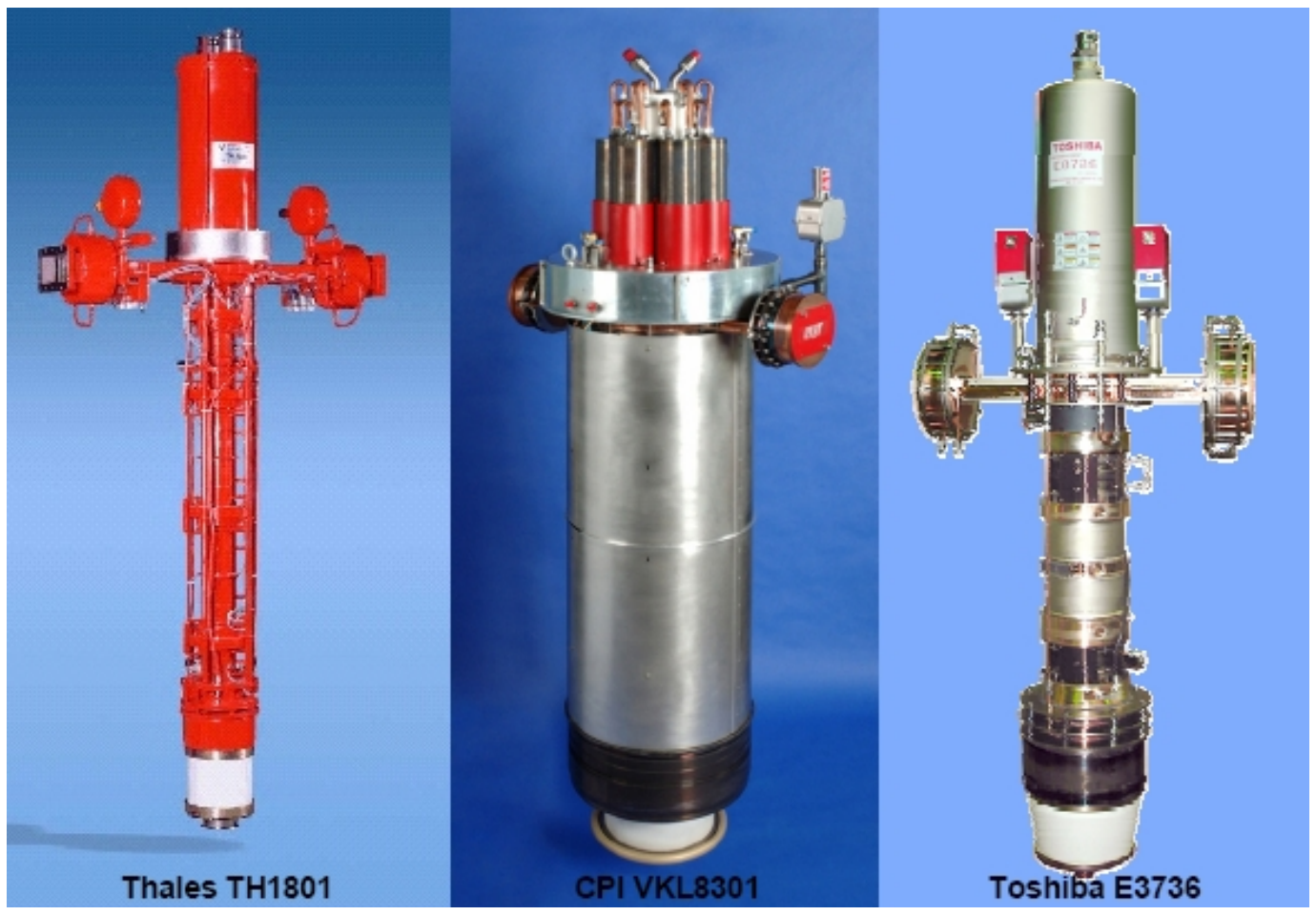

While the RF power source remains fundamentally the same, two cost-effective design variants for transporting the RF microwave power to the accelerator are considered in the baseline:

**Figure 2.5**
Prototype 120 kV Marx modulator.

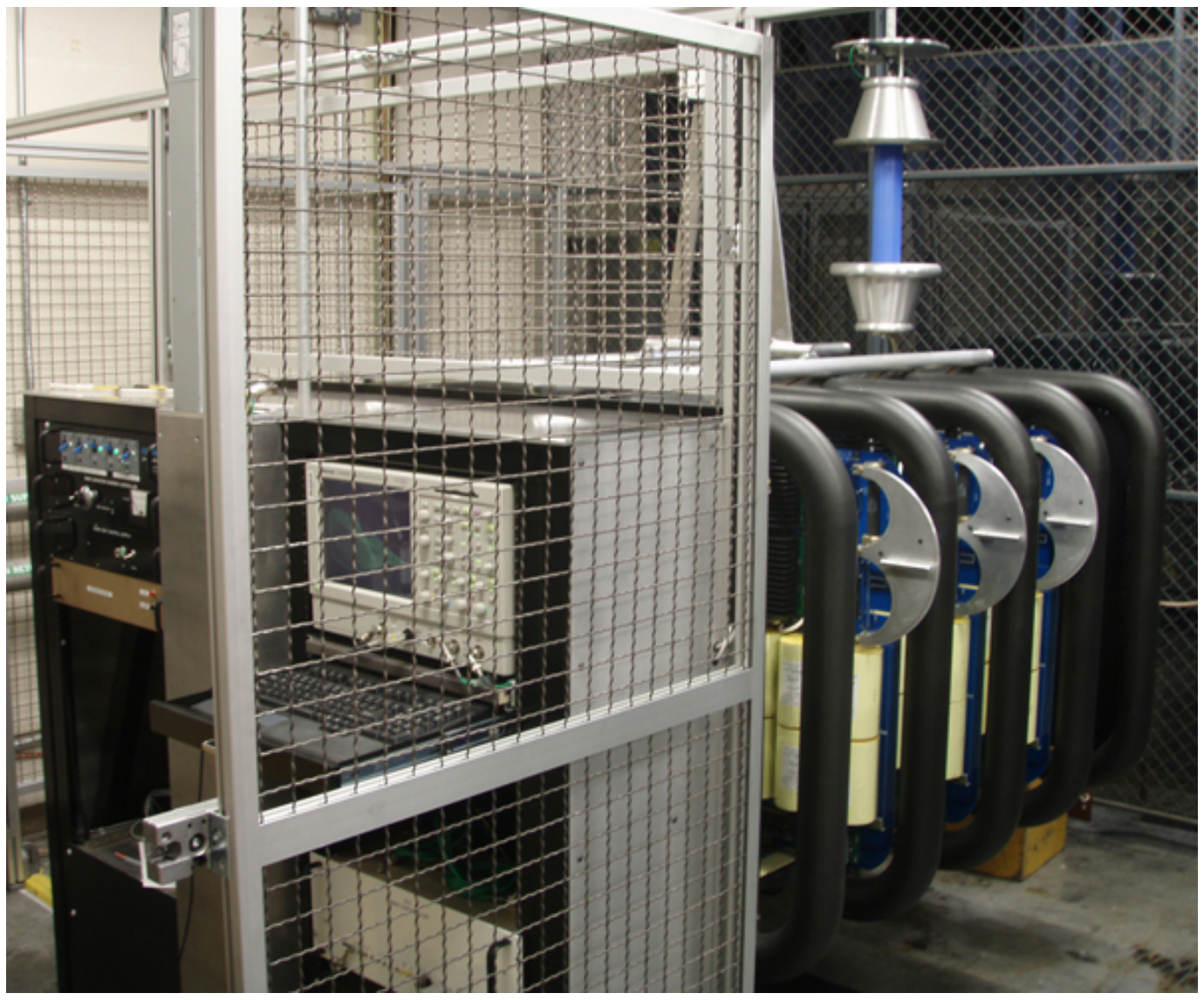

- A more traditional Distributed Klystron Scheme (DKS), where a klystron is used to drive 39 cavities. The klystrons and modulators are distributed along the entire length of the SCRF linacs, in the same tunnel but shielded from the accelerator itself;
- A novel Klystron Cluster Scheme (KCS – see Fig. 2.6), where all the klystrons are located in 'clusters' in surface buildings located periodically along the linacs. The power from a single cluster of 19 klystrons (~190 MW) is combined into an over-moded waveguide, which then transports the power down into the tunnel and along an approximately 1 km section of linac. For every three cryomodules, a Coaxial Tap Off (CTO) extracts ~6.7 MW of power to a local power-distribution system feeding 26 cavities.

The advantages of KCS are primarily in transferring a large fraction of the heat load to the surface where it can be more cost-effectively removed, at the same time as reducing the required underground volume. The disadvantages are the need for additional surface buildings and shafts (one every 2 km of linac), and additional losses in the long waveguide distribution systems. In addition, significant R&D is still required compared to the mature and tested distributed system. Nonetheless, the estimated cost savings associated with KCS make it an attractive solution which has been adopted for the Main Linacs in the flat-topography design variant.

The need for the extensive surface infrastructure does not make KCS a cost-effective solution for the mountainous topography, for which DKS has been adopted.

**Figure 2.6**
The Klystron Cluster Scheme (KCS).

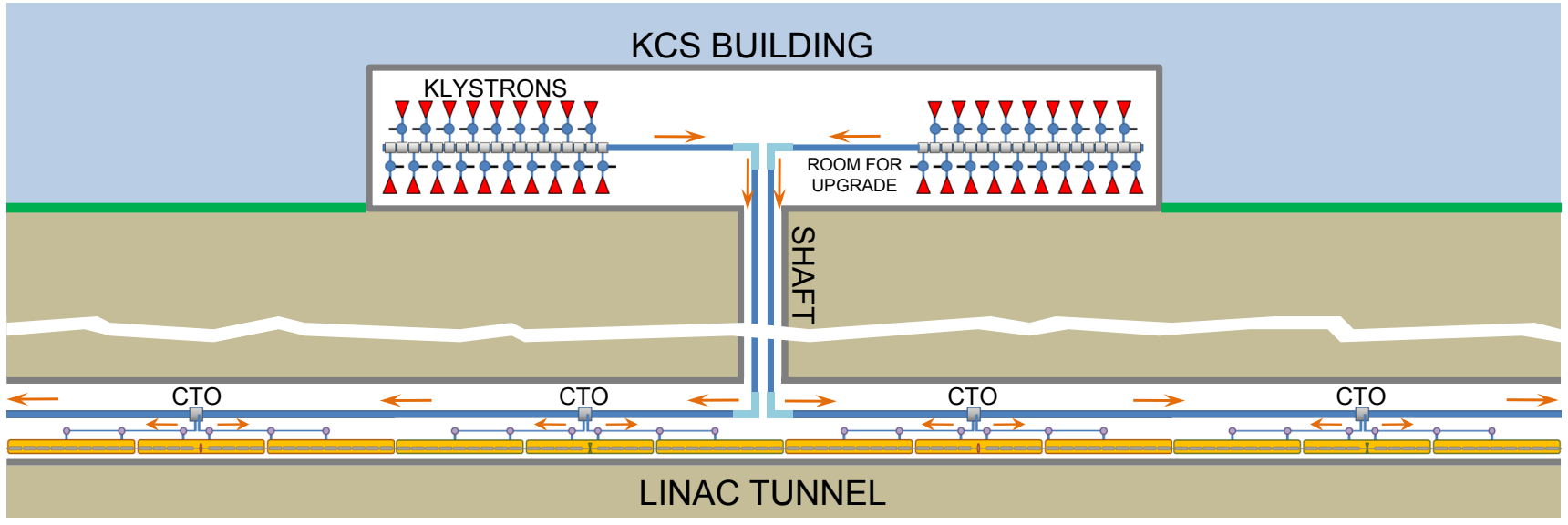

For both KCS and DKS, the local power-distribution systems are essentially identical, other than the number of cavities being driven. A key requirement is the ability to tune remotely both the phase and forward power to each individual cavity, thereby supporting the ±20% gradient spread in the accelerator, and thereby maximising the average available gradient.

## 2.3.2 Electron Source

The polarised electron source is located in the central-region accelerator tunnel together with the positron Beam Delivery System. The beam is produced by a laser illuminating a strained GaAs photocathode in a DC gun, providing the bunch train with 90 % polarisation. Two independent laser and gun systems provide redundancy. Normal-conducting structures are used for bunching and pre-acceleration to 76 MeV, after which the beam is accelerated to 5 GeV in a superconducting linac using 21 standard ILC cryomodules. Before injection into the damping ring, superconducting solenoids rotate the spin vector into the vertical, and a separate superconducting RF structure is used for energy compression. The layout of the polarised electron source is shown in Fig. 2.7.

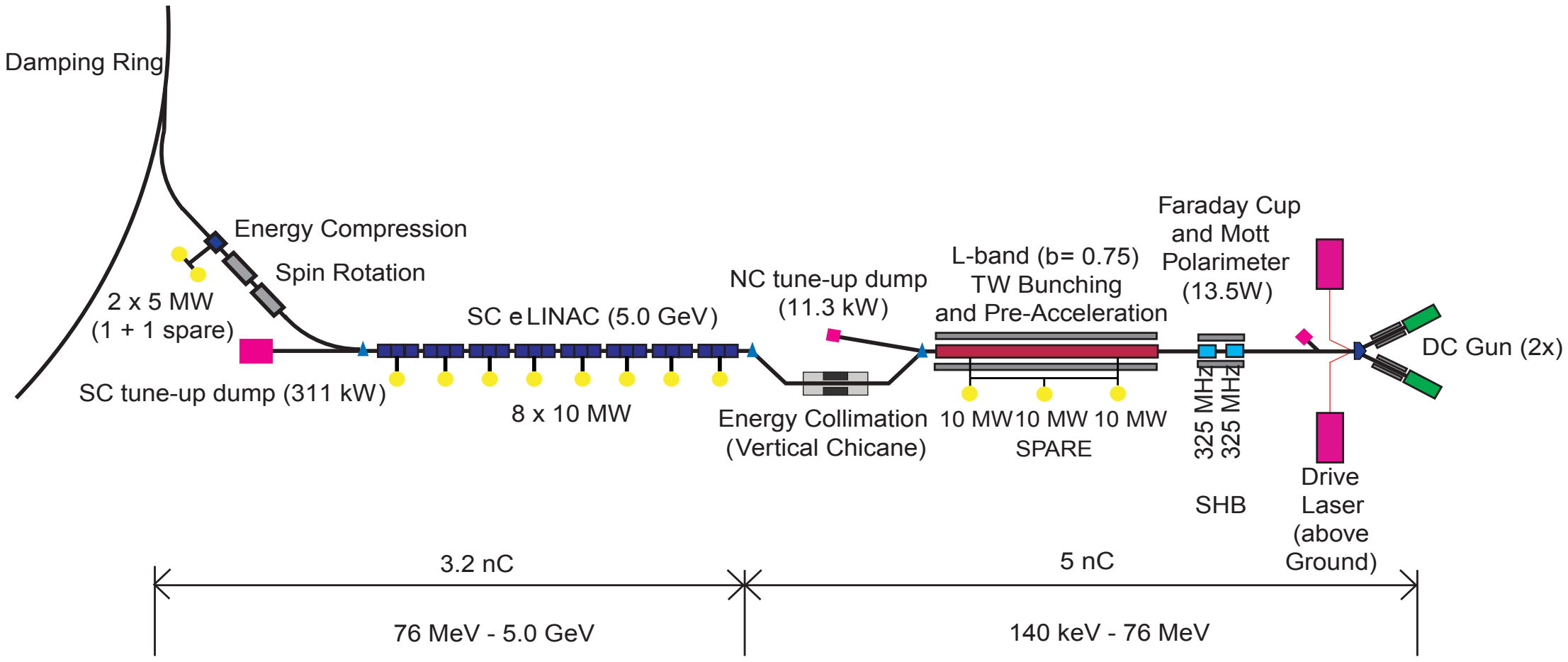

**Figure 2.7.** Schematic View of the Polarised Electron Source.

### 2.3.3 Positron Source

The major elements of the ILC positron source are shown in Fig. 2.8. The source uses photoproduction to generate positrons. After acceleration in the main linac, the primary electron beam is transported through a 147 m superconducting helical undulator which generates photons with energies from ~10 MeV up to ~ 30 MeV depending on the electron beam energy. The electron beam is then separated from the photon beam and displaced horizontally by approximately 2 m using a low-emittance chicane. The photons from the undulator are directed onto a rotating 0.4 radiation-length Ti-alloy target ~ 500 meters downstream, producing a beam of electron and positron pairs. This beam is then matched using an optical-matching device (a pulsed flux concentrator) into a normal conducting (NC) L-band RF and solenoidal-focusing capture system and accelerated to 125 MeV. The electrons and remaining photons are separated from the positrons and dumped. The positrons are accelerated to 400 MeV in a NC L-band linac with solenoidal focusing. The beam is then accelerated to 5 GeV using superconducting L-band RF. Before injection into the damping ring, superconducting solenoids rotate the spin vector into the vertical, and a separate superconducting RF structure is used for energy compression.

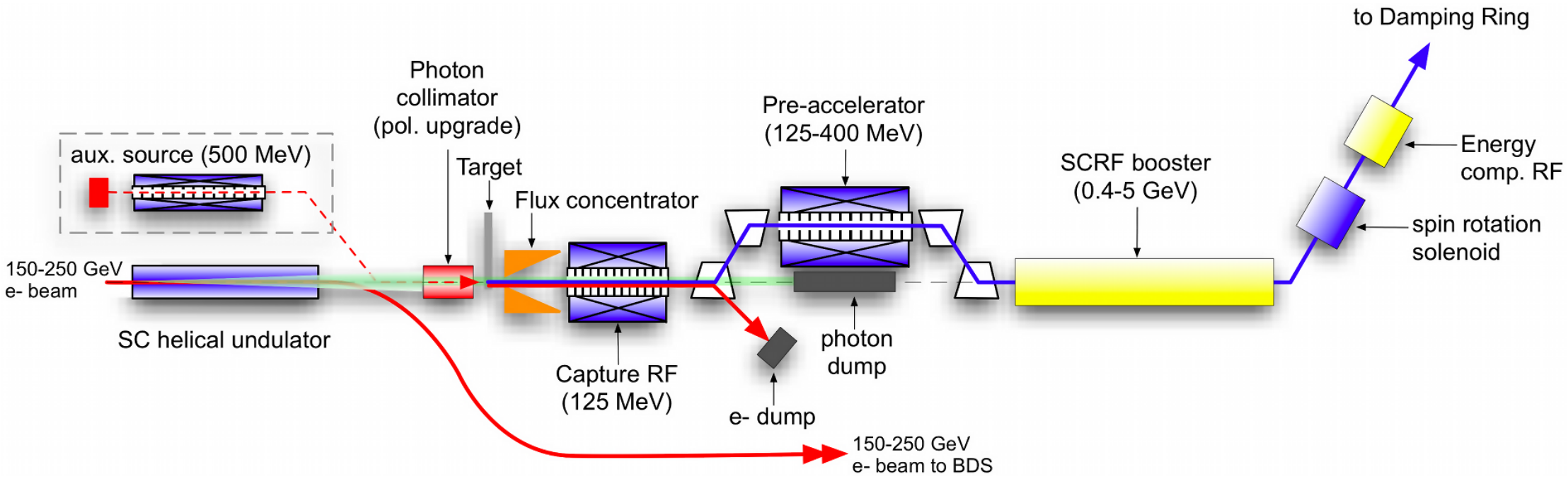

**Figure 2.8.** Overall Layout of the Positron Source.

The baseline design provides a polarisation of 30 %. Space for a ~ 220 m undulator has been reserved for an eventual upgrade to 60 % polarisation, which would also require a photon collimator upstream of the target.

To allow commissioning and tuning of the positron and downstream systems when the high-energy electron beam is not available, a low-intensity auxiliary positron source is provided. This is effectively a conventional positron source, which uses a 500 MeV warm linac to provide an electron beam which is directed onto the photon target, providing a few percent of the nominal positron current.

To accommodate 10 Hz operation, a separate pulsed extraction line is required immediately after the undulator, to transport the 150 GeV electron-beam positron-production pulse to the high-power tune-up dump, located downstream in the Beam Delivery System.

The target and capture sections are high-radiation areas which require appropriate shielding and remote-handling facilities.

### 2.3.4 Damping Rings

The damping rings must accept $e^-$ and $e^+$ beams with large transverse and longitudinal emittances and damp them (by five orders of magnitude for the positron vertical emittance) to the low-emittance beam required for luminosity production, within the 200 ms between machine pulses (100 ms for 10 Hz mode). In addition, they must compress on injection and decompress on extraction the ~1 ms beam pulse by roughly a factor of 90 to fit into the ring circumference of 3.2 km.

The baseline design consists of one electron and one positron ring operating at a beam energy of 5 GeV. Both rings are housed in a single tunnel with one ring positioned directly above the other.

Space is foreseen in the tunnel for a third ring (second positron ring) as a possible upgrade. The damping ring complex is located in the central region, horizontally offset from the interaction region by approximately 100 m to avoid the detector hall. Two transfer tunnels connect the damping ring tunnel to the electron and positron main accelerator tunnels respectively (see Fig. 2.9).

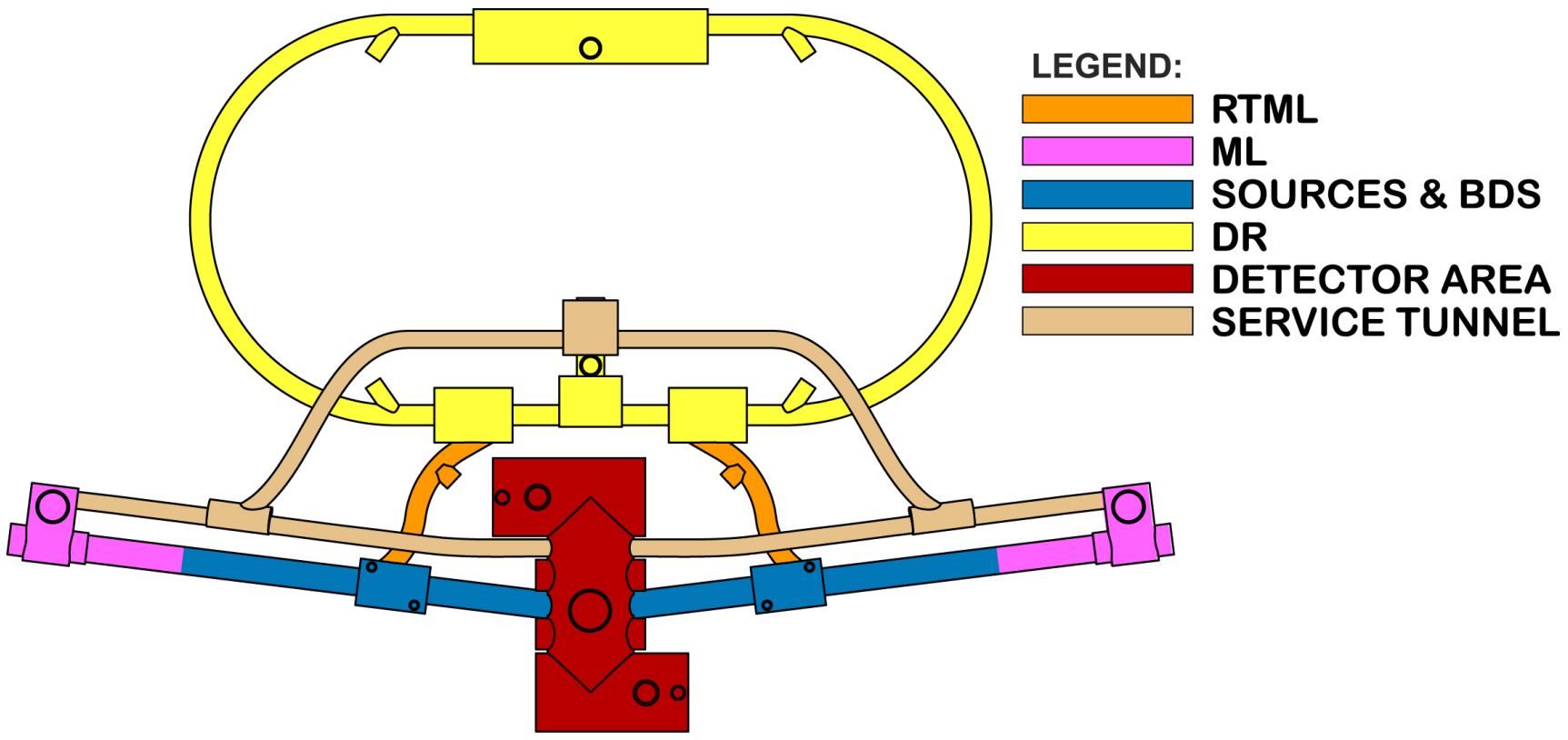

**Figure 2.9**
Damping ring location in the central region

The damping-ring lattice follows the race-track design shown schematically in Fig. 2.10. The two arc sections are constructed from 75 Theoretical Minimum Emittance (TME) cells. One of the two 712 m-long straight sections accommodates the RF cavities, damping wigglers, and a variable path length to accommodate changes in phase (phase trombone), while the other contains the injection and extraction systems, and a circumference-adjustment chicane.

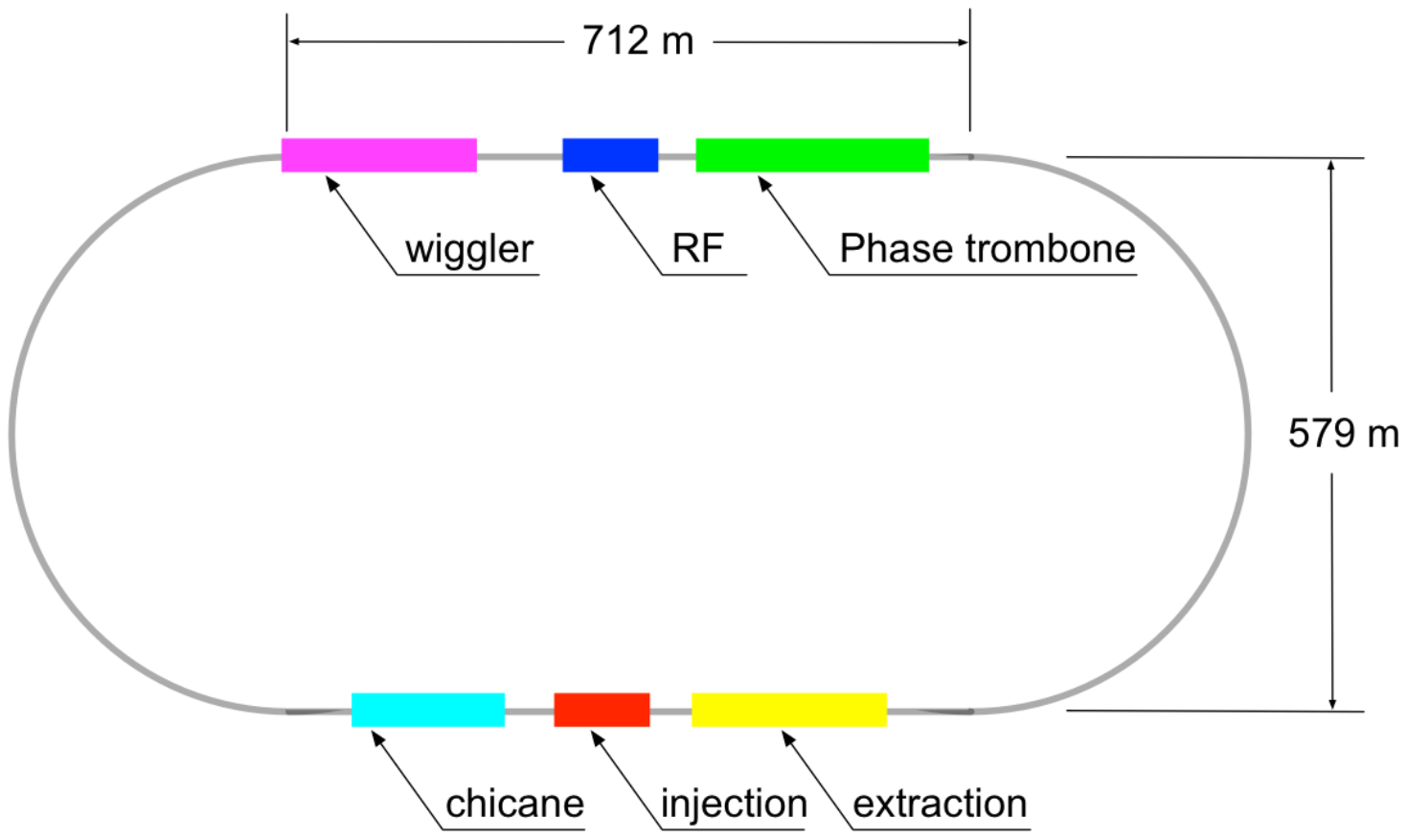

**Figure 2.10**
Schematic of the damping-ring layout.

Approximately 113 m of superferric wigglers (54 units $\times$2.1 m) are used in each damping ring. The wigglers operate at 4.5 K, with a peak field requirement of 2.16 T (positron ring 10 Hz mode).

The superconducting RF system is operated in continuous wave (CW) mode at 650 MHz, and provides a maximum of 20 MV for each ring, again specified for the positron ring in 10 Hz mode (nominal 5 Hz operation requires 14 MV for both electron and positron). The frequency is chosen to be half the linac RF frequency to maximise the flexibility for different bunch patterns. The single-cell cavities operate at 4.5 K and are housed in twelve 3.5 m-long cryomodules. The RF section of the lattice can accommodated up to 16 cavities, of which 12 are assumed to be installed for the baseline.

The momentum compaction of the lattice is relatively large, which helps to maintain single bunch stability, but requires a relatively high RF voltage to achieve the design RMS bunch length (6 mm). The dynamic aperture of the lattice is sufficient to allow the large-emittance injected beam to be captured with minimal loss.

The electron-cloud effect in the positron damping ring, which can cause instability, tune spread, and emittance growth, has been seen in a number of other rings and is relatively well understood. Extensive R&D and simulations (Part I Section 3.5) indicate that it can be controlled by proper surface treatment and design of the vacuum chamber to suppress secondary emission, and by the use of solenoids and clearing electrodes to suppress the buildup of the cloud. A full specification of mitigation techniques based on the R&D results is included in the baseline and cost estimate.

Mitigation of the fast ion instability in the electron damping ring is achieved by limiting the pressure in the ring to below 1 nTorr, and by the use of short gaps in the ring fill pattern.

For the baseline parameters, the bunch spacing within trains is approximately 8 ns which sets the limit for the rise and fall time for the injection and extraction kicker systems. (For the luminosity upgrade this number reduces to $\sim$4 ns.) Short stripline kicker structures can achieve this, and extensive R&D on the pulser has demonstrated several technologies that can meet the specifications (Part I Section 4.4).

## 2.3.5 Ring to Main Linac

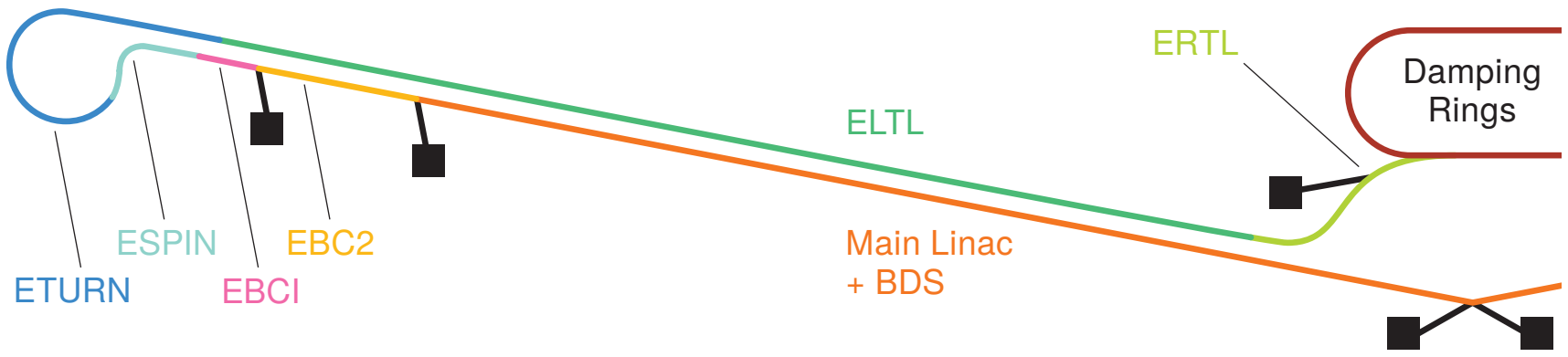

**Figure 2.11** Schematic of the electron RTML (the positron system is a mirror image, with labels prefixed with 'P'). See text for explanation of the subsystems.

The electron and positron Ring to Main Linac (RTML) systems are the longest contiguous beamlines in the ILC. The layout of the RTML is identical for both electrons and positrons, and is shown in Fig. 2.11. The RTML consists of the following subsystems, representing the various functions that it must perform:

- a $\sim$ 15 km long 5 GeV transport line (ELTL);
- betatron- and energy-collimation systems (in ERTL);
- a $180^\circ$ turn-around, which enables feed-forward beam stabilisation (ETURN);
- spin rotators to orient the beam polarisation to the desired direction (ESPIN);
- a two-stage bunch compressor to compress the beam bunch length from several millimetres to a few hundred microns, as required at the IP (EBC1 and EBC2).

The two-stage bunch compressor includes acceleration from 5 GeV to 15 GeV in order to limit the increase in fractional energy spread associated with bunch compression. The acceleration is provided by sections of SCRF main-linac technology. A primary challenge for the RTML systems is the preservation of the emittance extracted from the damping rings; the combination of the long uncompressed bunch from the damping ring and large energy spread (after compression) make the tuning and tolerances particular demanding. However, tuning techniques developed from detailed simulations have demonstrated acceptable emittance growth.

In addition to the beam-dynamics challenges, an RMS phase jitter of $\leq$0.24° between the electron and positron bunch-compressor RF systems is specified to limit bunch arrival-time jitter at the interaction point to an acceptable level. Beam-based feedback systems integrated into the bunch-compressor low-level RF system should be able to limit the phase jitter to this level.

## 2.3.6 Beam-Delivery System

The ILC Beam-Delivery System (BDS) is responsible for transporting the $e^+e^-$ beams from the exit of the high-energy linacs, focusing them to the sizes required to meet the ILC luminosity goals, bringing them into collision, and then transporting the spent beams to the main beam dumps. In addition, the BDS must perform several other critical functions:

- characterise the incoming (transverse) beam phase space and match it into the final focus;

- remove any large-amplitude particles (beam-halo) from the linac to minimize background in the detectors;
- measure and monitor the key physics parameters such as energy and polarisation before and after the collisions.

The layout of the beam-delivery system is shown in Fig. 2.12. There is a single collision point with a 14 mrad total crossing angle. The 14 mrad geometry provides space for separate extraction lines but requires crab cavities to rotate the bunches in the horizontal plane for effective head-on collisions. There are two detectors in a common interaction region (IR) hall in a so-called "push-pull" configuration.

The geometry of the BDS has been designed to accommodate the 1 TeV centre-of-mass-energy upgrade, in particular to minimise the emittance growth due to synchrotron radiation to a few percent at these beam energies. The baseline lattice uses fewer magnets (predominantly dipoles) for the lower-energy operation.

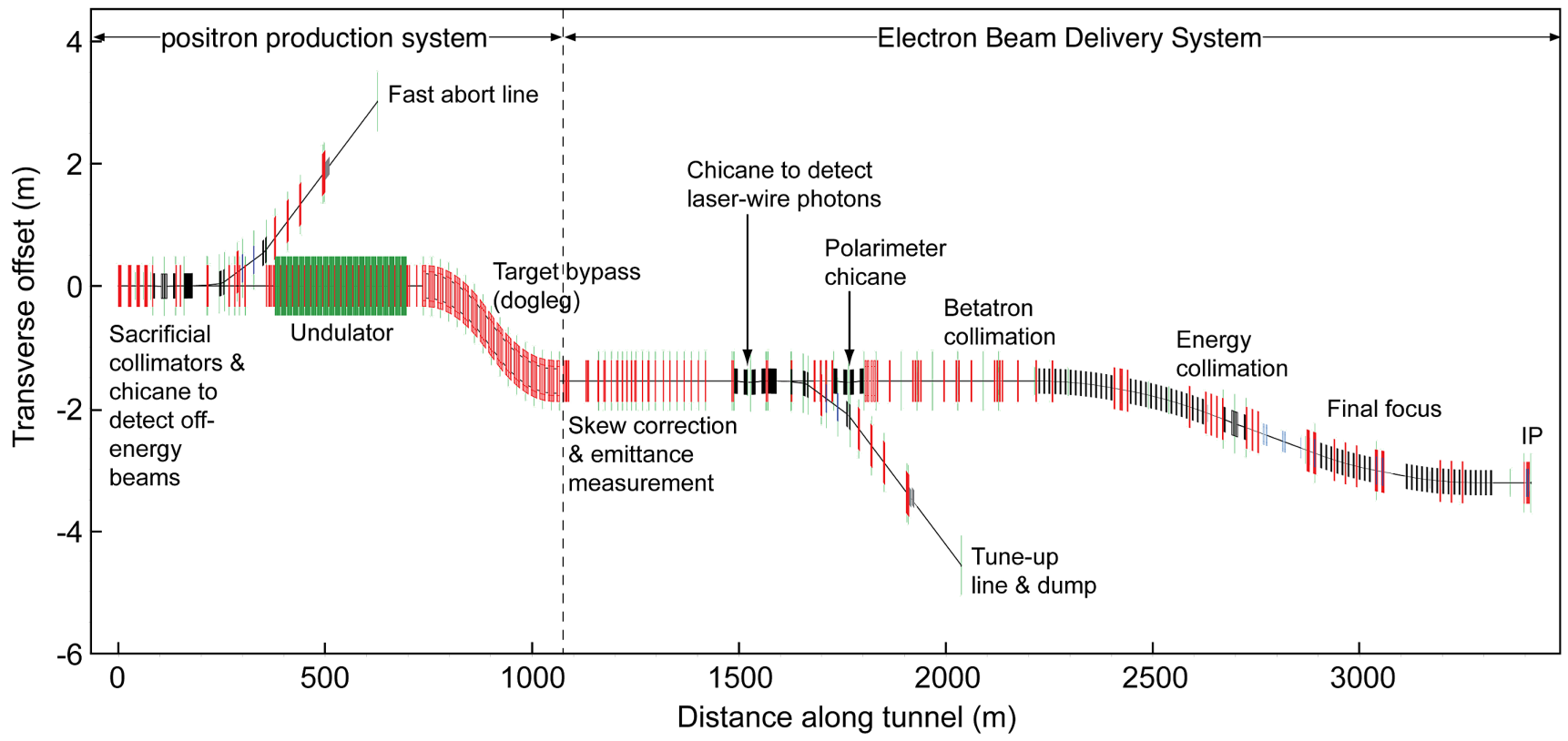

**Figure 2.12**
BDS lattice layout, showing the major sub-systems. Shown is the electron BDS, which starts at the vertical dotted line. (Also shown is the positron system upstream of the electron BDS). The positron BDS is a mirror image.

The main subsystems of the BDS are (beam direction):

- a section containing emittance measurement and matching (correction) sections, trajectory feedback, polarimetry and energy diagnostics;
- a collimation section which removes beam-halo particles that would otherwise generate unacceptable background in the detector, and also contains magnetised iron shielding to deflect and/or absorb muons generated in the collimation process;
- the final focus (FF), which uses strong compact superconducting quadrupoles to focus the beam at the IP, with sextupoles providing local chromaticity correction;
- the interaction region, containing the experimental detectors. The final-focus quadrupoles closest to the IP are integrated into the detector to facilitate detector "push-pull";
- the extraction line, which has a large enough bandwidth to transport the heavily disrupted beam cleanly to a high-powered water-cooled dump. The extraction line also contains important polarisation and energy diagnostics.

The beam-delivery optics provides demagnification factors of typically several hundreds in the beam size, resulting in very large beta functions (several thousand kilometres) at critical locations, leading to the tightest alignment tolerances in the entire machine. In addition, careful correction of the strong chromaticity and geometric aberrations requires a delicate balance of higher-order optical terms. The tight tolerances on magnet motion (down to tens of nanometres), makes continuous trajectory correction and the use of fast beam-based feedback systems mandatory. Furthermore, several critical components (e.g. the final focusing doublet) may well require mechanical stabilisation. Beam-based alignment and beam phase-space tuning algorithms are necessary to adjust and tune the optical aberrations that would otherwise significantly degrade the luminosity. The ability to tune the beams to the required levels relies extensively on remote precision mechanical adjustment of the magnets, as well as precision diagnostics. Many of the techniques and required instrumentation are being successfully developed in the ATF2 programme (Part I Section 3.6).

The tight tolerance on the relative uncorrelated phase jitter between the electron and positron superconducting crab-cavity systems requires timing precisions at the level of tens of femtoseconds. Although this tolerance is tight, it is comparable to that achieved at modern linac-driven FELs.

Control of machine-generated backgrounds is performed by careful optics control and tuning of an extensive collimation system, as well as by the use of non-linear elements ("tail-folding" octupoles). The design of the collimation system carefully considers wakefield effects at small apertures; this requires careful electromagnetic design of the mechanical collimators themselves, as well as precision control of the beam motion using fast trajectory-correction (feedback).

The main beam dumps, which use a high-pressure high-velocity water design, represent a major installation. Since the dumps will be significantly activated during operation, they are designed and rated for the full average beam power at 1 TeV of 14 MW, in order to avoid having to replace them for the energy upgrade.

## 2.4 Site Dependent Designs

Conventional Facilities and Siting (CFS) is the designation for all aspects of the design relating to civil engineering, power distribution, water cooling and air conditioning systems. The CFS and the main-linac SCRF technology represent the two largest elements of the total project cost. The CFS design and costs can be broken down into three main areas:

1. Civil construction, including underground and surface structures, shafts and access tunnels;
2. Electrical systems (AC power distribution etc.);
3. Mechanical systems (water cooling and air handling etc.).

The CFS solutions and associated cost are developed based on the requirements defined by the accelerator layout and parameters briefly discussed in the previous sections. In order to minimise (optimise) the total CFS cost it is necessary to understand how it depends on the accelerator design, and if necessary re-evaluate and iterate the design approach. Reduction of the scope of the underground civil construction (for example) was considered a primary design goal during the Technical Design Phase, which resulted in significant modifications to the accelerator design and parameter space. In addition, the criteria for the electrical and mechanical systems, as well as the rationale and approach to access shafts and tunnels, have been scrutinised to reduce costs.

While the accelerator-systems layout and requirements are the primary driver for the CFS design, the solutions are heavily influenced by regional considerations of site topography and geology, as well as local legislation (such as safety requirements). Geology will determine the most cost-effective approach to tunnelling method, while topography can influence the surface structures and lengths and depths of access tunnels or shafts. All of these factors can shift the balance of the cost-optimisation and influence the accelerator design. As a result, the final machine construction will be strongly influenced by the choice of site.

In the absence of a definitive site for the ILC, it was necessary to evaluate, as far as possible, different sites with different characteristics. To this end, several sample sites have been developed:

- The Americas sample site lies in Northern Illinois near Fermilab. The site provides a range of locations to position the ILC in a north-south orientation. The site chosen has approximately one-quarter of the machine on the Fermilab site. The surface is primarily flat. The long tunnels are bored in a contiguous Dolomite rock stratum ("Galena Platteville"), at a typical depth of 30–100 m below the surface;

- For the Asian sites, two possible ILC candidate sites have been identified: Kitakami in the Tohoku district in northern Japan; and Sefuri in the Kyushu district in the south. Both potential sites provide a uniform terrain located along a mountain range, with a tunnel depth ranging from 40–600 m. The chosen geology is uniform granite highly suited to modern tunneling methods (e.g. New Austrian Tunneling Method (NATM));
- The European site is located at CERN, Geneva, Switzerland, and runs parallel to the Jura mountain range, close to the CERN site. The majority of the machine is located in the 'Molasse' (a local impermeable sedimentary rock), at a typical depth of 100–150 m.

The Americas and European sample sites are relatively similar and are examples of 'flat topography' sites. Both these sites use the Klystron Cluster Scheme for the RF power distribution. The geology lends itself well to the use of Tunnel Boring Machines (TBM) which provide a round tunnel cross section.

The Japanese sites are examples of mountainous topology, where available space for surface infrastructure is severely limited, requiring most of the accelerator infrastructure to be housed underground. The most cost-effective solution in this geology and topology is NATM. The Japanese sites use the Distributed Klystron Scheme.

In the following two sections, the site-dependent designs will be briefly described, with a focus on the main-linac accelerator tunnel, the central region and detector hall.

### 2.4.1 Flat-topography site-dependent design (Americas and European sample sites)

Figure 2.13 shows an artist's rendition of the civil construction layout for the flat-topography site. The shafts leading to surface installations (not shown) are clearly indicated. The KCS RF system requires an additional 3 shafts (per side).

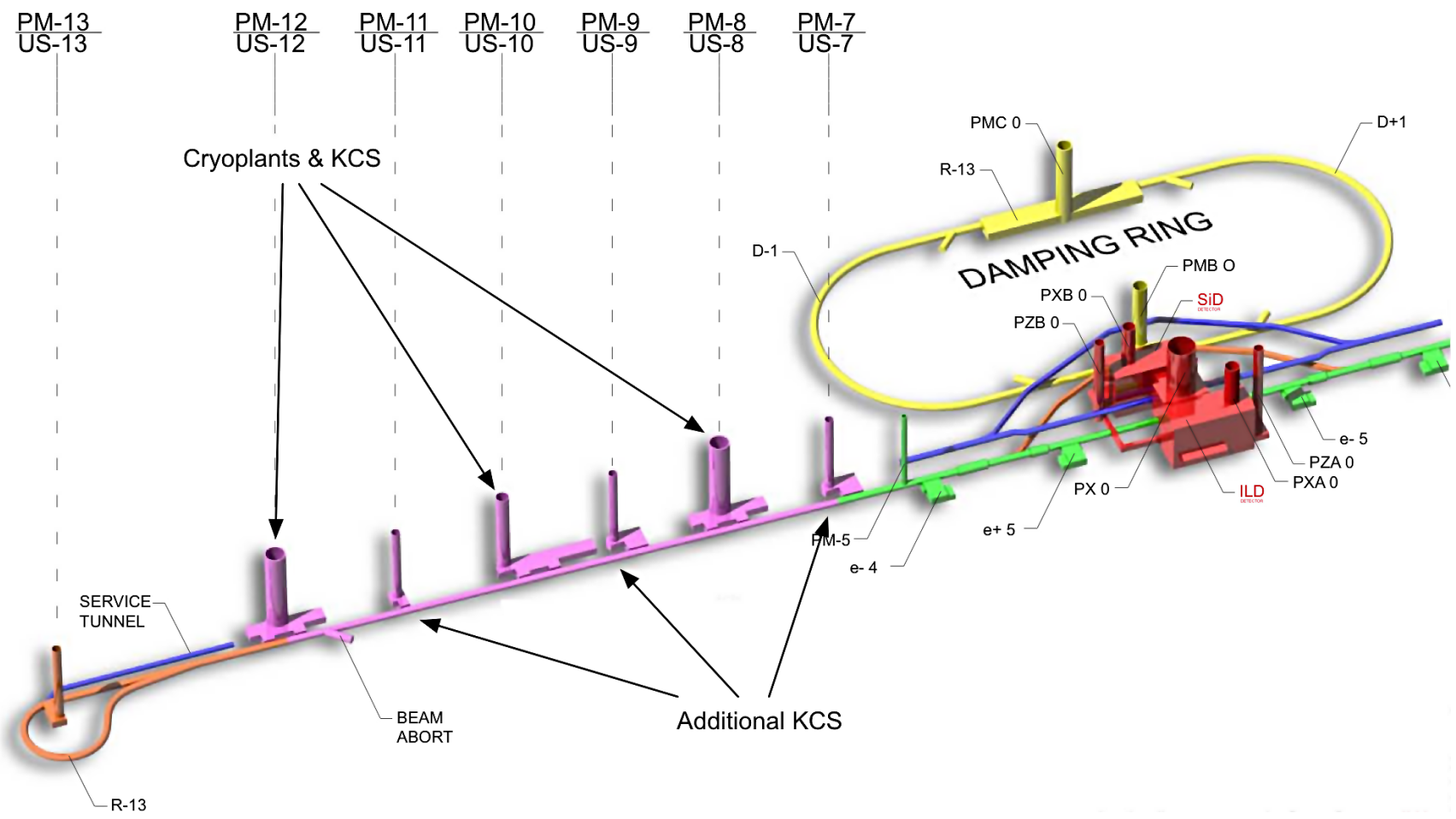

**Figure 2.13** Artist's impression of the International Linear Collider (not to scale). Shown is the electron side of the machine and the central region. The layout shows the CFS solution for the flat topography, which utilises vertical shafts for access, and includes additional shafts for the KCS RF power distribution.

As the klystrons and modulators are housed on the surface, the single main-linac tunnel (main accelerator tunnel) can have a relatively small diameter. Figure 2.14 shows the cross-section of the tunnel.

For the bunch-compressor RF (RTML) and the central region (containing the source linacs as well as several parallel beam lines), a separate underground service tunnel is provided, which is connected to the main accelerator tunnel via penetrations. (The RF in these regions uses DKS rather than KCS, and the service tunnel is used to house the klystrons and modulators.)

The design of the detector hall (Fig. 2.15) accommodates the two detectors in a push-pull detector arrangement. The requirements for the hall and access shafts are primarily driven by the concepts of on-surface construction (similar to CMS at LHC), and the need to have sufficient access to the detectors while in the parked (i.e. off-beam) position. The large 18 m-diameter shaft located directly over the interaction point serves as the primary access for lowering large parts of both detectors into the underground hall.

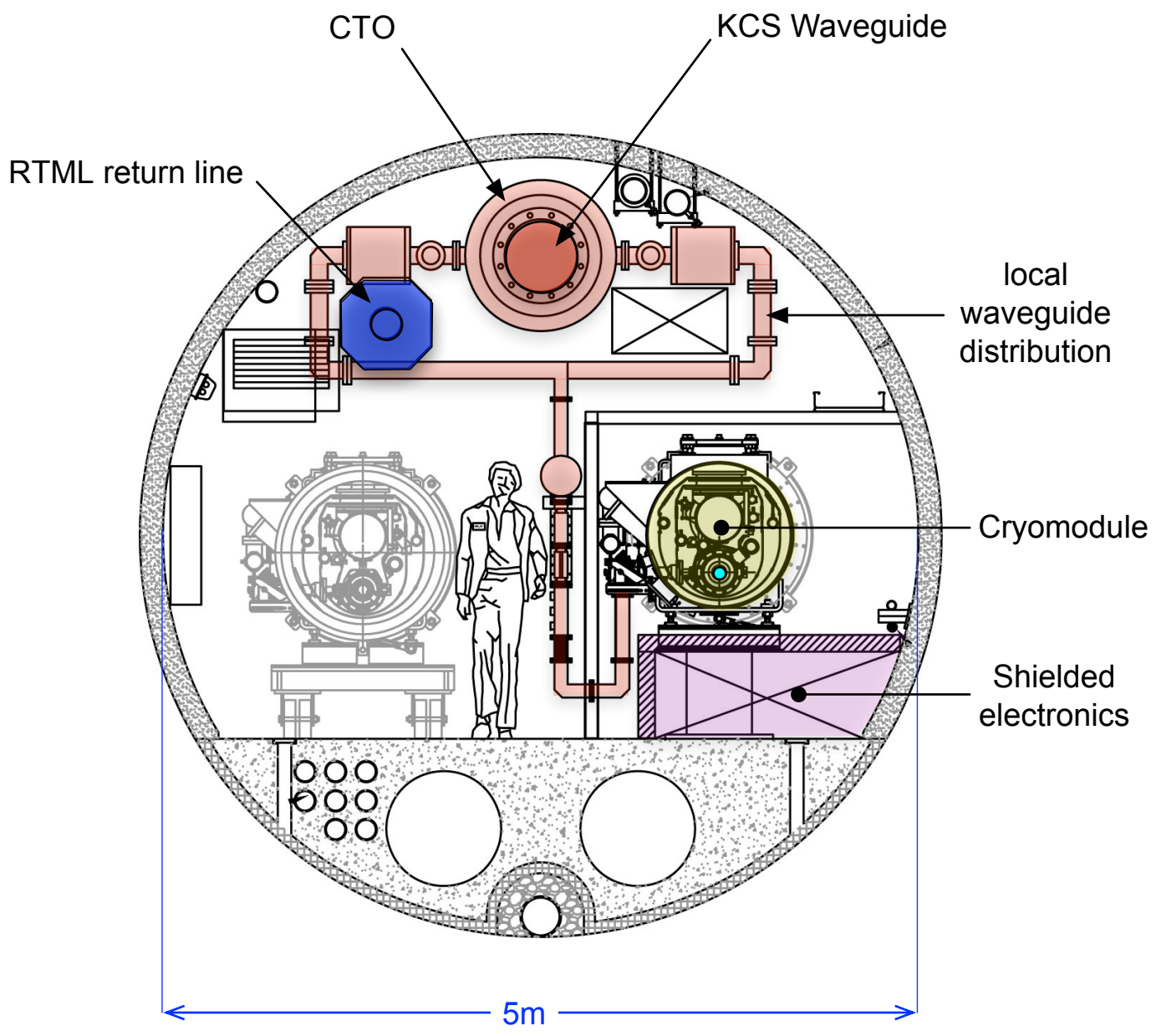

**Figure 2.14**
Cross section of the main linac tunnel for the flat-topography variant, using KCS.

The overall power loads for the entire machine, including mechanical (water cooling), is strongly influenced by the main-linac configuration and in particular the possibility of having a large fraction of the heat load from the RF power source and cryogenics on the surface, allowing more cost-effective solutions.

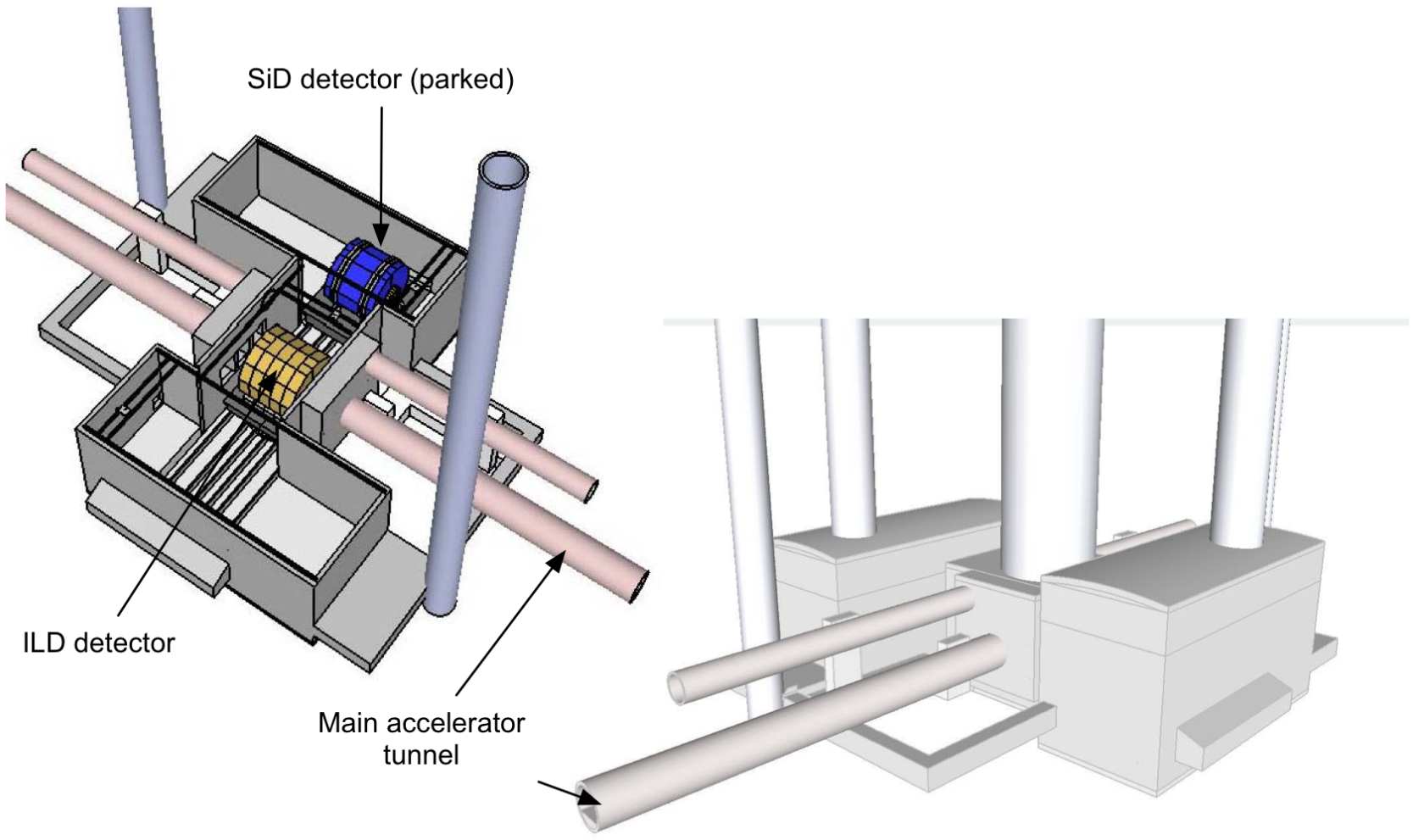

**Figure 2.15**
Two views of the design for the detector hall for the flat-topography site variant.

## 2.4.2 Mountainous-topography site-dependent design (Asian sample sites)

For the mountainous topography proposed for the two Japanese sample sites, the accelerator infrastructure must be predominantly housed in underground caverns. In addition, access is provided via gently sloping horizontal access tunnels. For the uniform hard granite geology, a single wide-tunnel solution constructed with NATM is the most cost-effect solution. In order to house both the accelerator (cryomodules) and the distributed RF power sources and associated electronics, an 11-m wide dome-shaped tunnel will be excavated. The tunnel is wide enough to accommodate a concrete shielding wall between the accelerator itself and the RF power systems, effectively providing a cost-effective twin-tunnel solution. Figure 2.16 shows a perspective view of a section of tunnel.

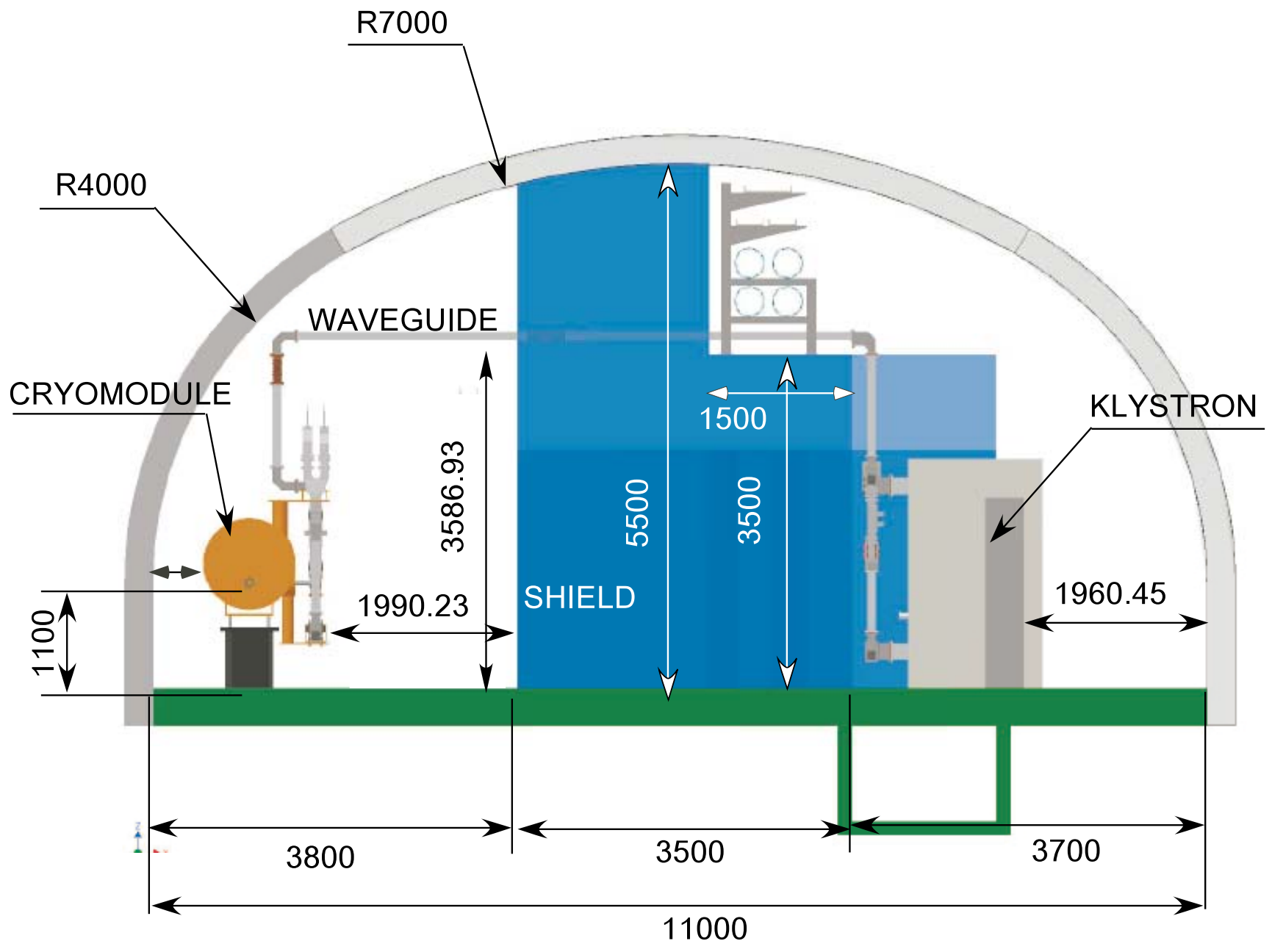

**Figure 2.16** Cross section of the mountainous-topography tunnel (so-called 'Kamaboko').

The cryoplants and associated cooling and power systems are housed in caverns adjacent to the main accelerator tunnel (Fig. 2.17). The single wide-tunnel structure runs the entire length of the accelerator, and therefore removes the need for a separate service tunnel in the RTML (bunch compressor) and centra

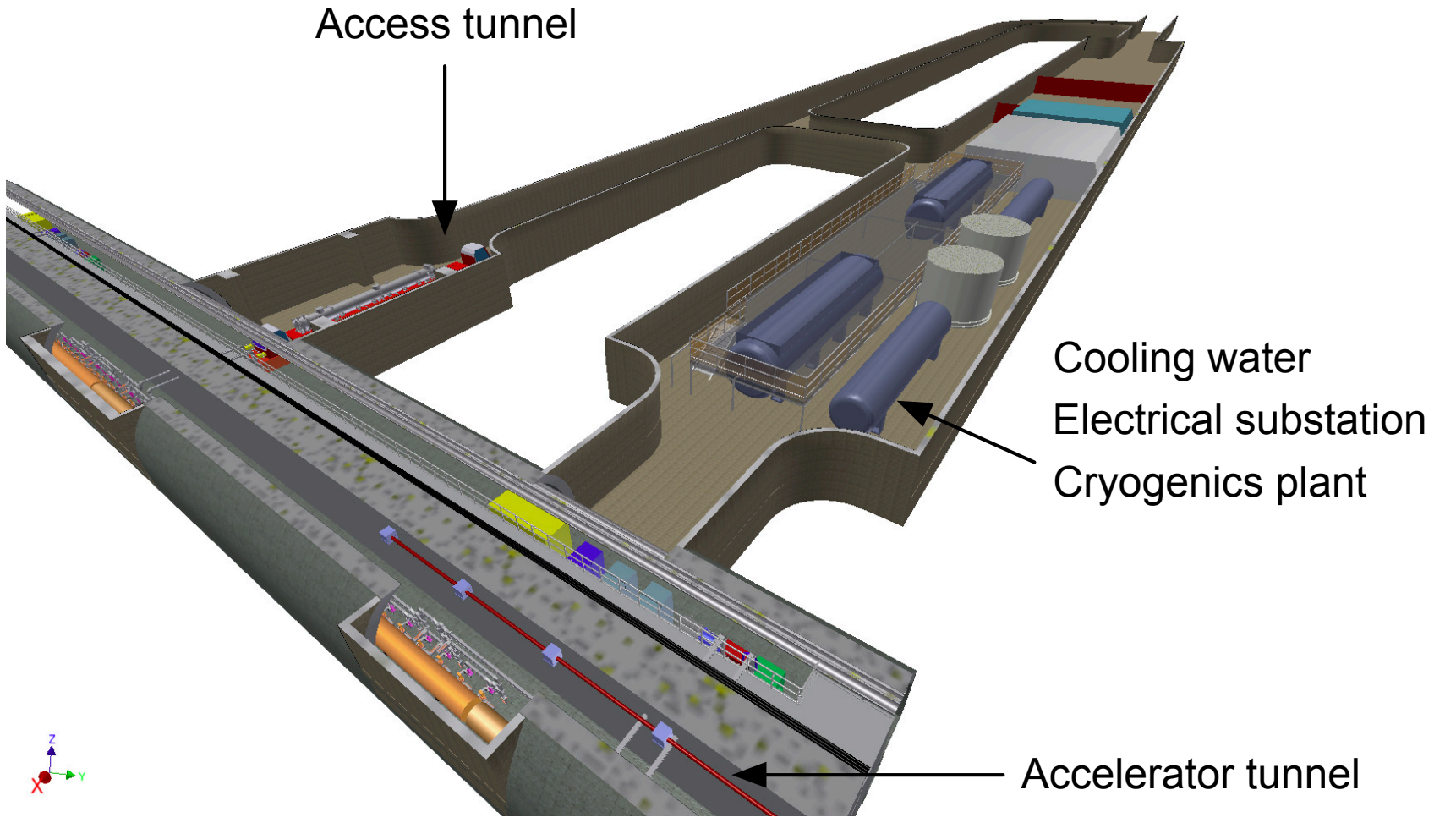

**Figure 2.17** Perspective view of the underground cavern arrangement for the cryogenic plants, power and cooling systems.

Unlike the flat-topography site designs which utilise the KCS system, the mountainous-topography solution has all the primary heat loads located underground, and those associated with the RF power sources are distributed along the entire length of the linacs. This influences the design approach to mechanical and electrical systems, resulting in a different optimised solution from the flat-topography sites.

The need for horizontal access in the mountainous topography also strongly influences the design of the detector hall (Fig. 2.18). A CMS-like surface assembly is not considered cost-effective in this situation, and instead the underground hall is designed to accommodate underground *in situ* assembly. The single wide-access tunnel also serves the damping-ring installation.

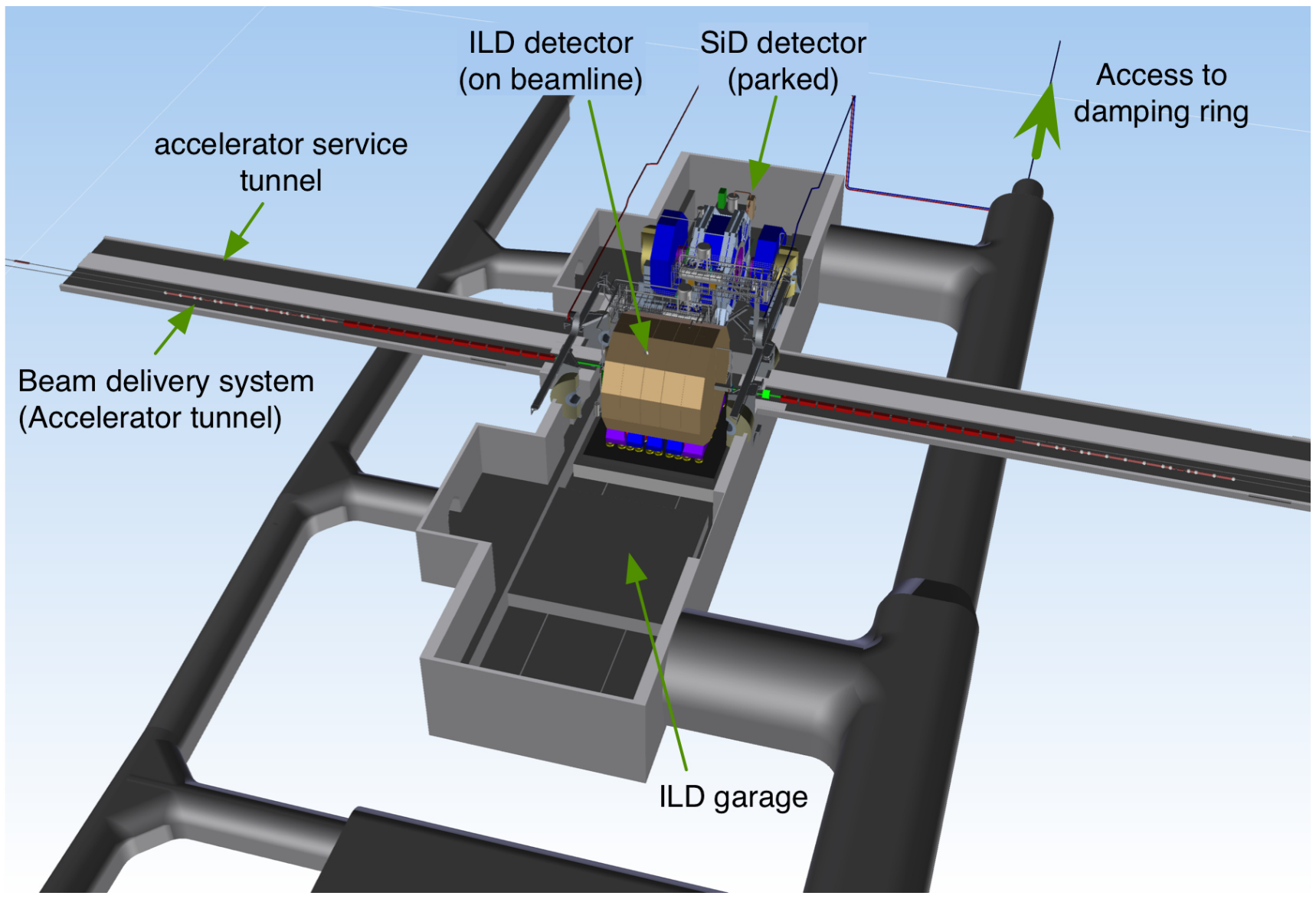

**Figure 2.18**
Perspective view of the underground detector hall for the mountainous topography, showing the two detectors and service tunnels.

## 2.5 Luminosity and Energy Upgrade Options

The technical design, cost estimate and construction schedule reported in this report have been optimised for the 500 GeV baseline scenario. Although considered in significantly less detail, two upgrade scenarios are foreseen:

- A luminosity upgrade (up to a factor of two), which is accomplished by doubling the number of bunches per beam pulse (doubling the beam power). This requires increasing the number of klystrons and modulators by approximately 50 %. The baseline design also foresees the possibility of constructing a second positron damping ring in the same tunnel, should electron-cloud effects require. All other accelerator systems are already rated for the higher beam power. Basic (minimum) provisions for the required conventional facilities are included in the baseline design to support the luminosity upgrade, although upgrades to the cooling systems will be required;

- An increase in the centre-of-mass energy up to 1 TeV by increasing the length of the SCRF main linacs. This requires a relocation of the bunch compressors and 180-degree turn-around, as well as an extension of the long return line from the damping rings to the turnaround and the extension of the main linacs. The latter upgrade is assumed to be based on a forward-looking SCRF technology compatible with the existing installation (RF pulse length etc.), but likely to have higher-performance specifications both in gradient and quality factor. The overall site-footprint requirement for the 1 TeV machine is approximately 50 km, with a site power requirement of approximately 300 MW. In order to minimise the impact on the existing machine during the upgrade construction, the baseline BDS geometry and high-power beam dumps are already to be compatible with 1 TeV operation.

Chapter 12 deals with both the luminosity and centre-of-mass energy upgrades in more detail.

# Chapter 3
# Main Linac and SCRF Technology

## 3.1 Overview of the ILC Main Linacs

### 3.1.1 Introduction

The two main linacs accelerate the electron and positron beams from their injected energy of 15 GeV to their final collision energy of between 100 GeV and 250 GeV, over a combined length of 22 km. The linacs utilise superconducting technology, consisting of approximately 16,000 L-band (1.3 GHz) nine-cell standing-wave niobium cavities operating at an average gradient of 31.5 MV/m in a 2 K superfluid-helium bath, integrated into ~1,700 12 m-long cryomodules. The choice of operating frequency is a balance between the higher cost of larger, lower-frequency cavities and the increased cost at higher frequency associated with the lower sustainable gradient from the increased surface resistivity. The optimum frequency is in the region of ~1.5 GHz, but during the early R&D on the technology, 1.3 GHz was chosen due to the commercial availability of high-power klystrons at that frequency [13].

The choice of accelerating gradient is the largest single cost-driver; it defines the required number of cavities and tunnel lengths of the Main Linacs. An average accelerating gradient of 31.5 MV/m is required for 500 GeV centre-of-mass-energy operation. However, the main linac systems — and in particular the RF power systems — are specified to accommodate up to a ± 20% spread in individual cavity performance. The gradient achieved in the low-power vertical test (mass production acceptance test) is specified ~ 10% higher (35 MV/m) to allow for operational gradient overhead for low-level RF (LLRF) controls, as well as some degradation during cryomodule installation (few MV/m).

The TESLA elliptical cavity has been chosen for the ILC baseline design due to its maturity and the experience accumulated over the past decade and a half. In particular, approximately 800 TESLA cavities are currently under production for the European XFEL.

The design average acceleration gradient (31.5 MV/m) and qualify factor ($Q_0 = 10^{10}$ — see Section 3.2) has been achieved and exceeded in many cavities, several of which accelerate beam in the TTF/FLASH facility at DESY, Hamburg (see Part I Section 3.2). Mass production of high-performance cavities by industry has progressed significantly in recent years, giving confidence that the required parameters can be achieved (see Part I Section 2.3).

The cryomodule is similar in design to that developed by the TESLA collaboration, of which over ten examples have been constructed, six of which are operational at TTF/FLASH. For the ILC, two types of modules are foreseen, one integrating nine cavities (Type-A), and one integrating eight cavities, with a superconducting quadrupole package located at the centre of the string (Type-B). Both modules are designed to have the identical length of 12.652 m.

### 3.1.2 Linac layout

Table 3.1 provides an overview of the parameters and component counts for the ILC main linacs. The linacs are constructed from a near-contiguous string of cryomodules, interrupted only by the segmentation of the cryogenic strings (see below). The linacs are housed in underground tunnels which generally follow the curvature of the earth, primarily to simplify the flow of the two-phase helium at 2 K. The electron linac has an additional nine cryomodules to provide the $\sim$ 2.6 GeV needed to compensate the energy loss in the undulator-based positron source (Chapter 5). In addition both electron and positron linacs have $\sim$ 1.5 % energy-overhead to increase availability.

**Table 3.1**
Summary of key numbers for the SCRF Main Linacs for 500 GeV centre-of-mass-energy operation. Where parameters for positron and electron linacs differ, the electron parameters are given in parenthesis.

| | | |
|---|---|---|
| *Cavity (nine-cell TESLA elliptical shape)* | | |
| Average accelerating gradient | 31.5 | MV/m |
| Quality factor $Q_0$ | $10^{10}$ | |
| Effective length | 1.038 | m |
| R/Q | 1036 | Ω |
| Accepted operational gradient spread | ±20% | |
| | | |
| *Cryomodule* | | |
| Total slot length | 12.652 | m |
| Type A | 9 cavities | |
| Type B | 8 cavities | 1 SC quad package |
| | | |
| *ML unit (half FODO cell)* | 282 (285) | units |
| (Type A - Type B - Type A) | | |
| | | |
| *Total component counts* | | |
| Cryomodule Type A | 564 (570) | |
| Cryomodule Type B | 282 (285) | |
| Nine-cell cavities | 7332 (7410) | |
| SC quadrupole package | 282 (285) | |
| | | |
| Total linac length – flat top. | 11027 (11141) | m |
| Total linac length – mountain top. | 11072 (11188) | m |
| Effective average accelerating gradient | 21.3 | MV/m |
| | | |
| *RF requirements (for average gradient)* | | |
| Beam current | 5.8 | mA |
| beam (peak) power per cavity | 190 | kW |
| Matched loaded $Q$ ($Q_L$) | $5.4 \times 10^6$ | |
| Cavity fill time | 924 | μs |
| Beam pulse length | 727 | μs |
| Total RF pulse length | 1650 | μs |
| RF–beam power efficiency | 44% | |

Either 26 or 39 adjacent cavities are effectively driven by a common RF power source as indicated in Fig. 3.1. The local power distribution system provides flexibility in adjusting the forward power to each cavity, necessary in dealing with the expected spread in individual cavity gradient performance (see Section 3.6.4). The RF power is provided by 10-MW multi-beam klystrons driven by a solid-state Marx modulator (see Section 3.6.3 and Section 3.6.2 respectively).

Two possible schemes (flat and mountain topography respectively) have been developed during the Technical Design Phase for the layout of the tunnels, and in particular the approach to delivering RF power to the local distribution system and ultimately the cavities.

For a *mountainous topography*, such as the candidate sites in Japan, the more standard Distributed Klystron Scheme (DKS) would be used in a 11 m-wide, "kamaboko-shaped" tunnel whose interior is divided into two corridors by a thick (2.0 m to 3.5 m) concrete wall. The cryomodules occupy one side of the tunnel while the RF systems including modulators, klystrons, power supplies, and instrumentation racks, are located on the other side. This arrangement permits access to the RF equipment for maintenance, repair, or replacement during beam operation, and limits radiation exposure to most of the electronics (except for equipment placed in or near the cryomodules). For a *flat topography*, a novel Klystron Cluster Scheme (KCS) is envisioned, with all RF-generating equipment located in surface facilities where the klystron power is combined (up to 300 MW) and transferred through overmoded circular waveguides to the linac tunnel below. The tunnel, which is about 5 m in diameter, mainly contains the cryomodules and waveguides with some electronics (e.g. quadrupole-magnet power supplies, LLRF monitoring and control electronics) that is housed in radiation-shielded, 2 m-wide racks under the cryomodules.

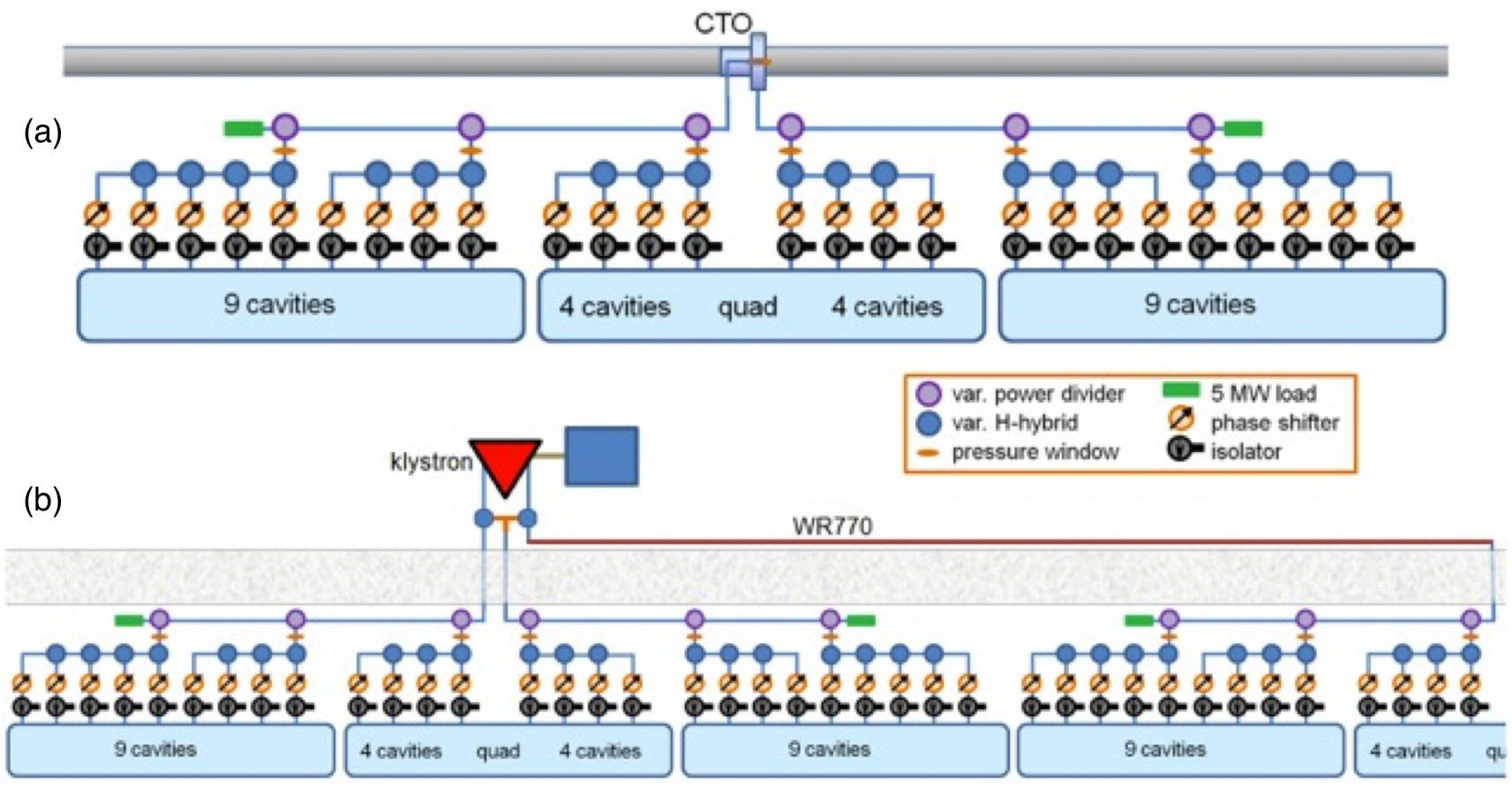

**Figure 3.1** Schematic of the Local Power Distribution System (LPDS) which delivers RF power to 13 accelerating cavities in the main linacs; (a) and (b) show the KCS and DKS options respectively.

The tunnels are assumed to be deep underground (∼100 m) and connected to the surface through vertical shafts (flat topography) or sloped access routes (mountainous topography). The number, location and size of these shafts or access-ways is determined by the maximum length of a cryogenic unit (and maximum available size of a cryoplant, see below), and, in the case of KCS, the maximum distance over which the RF power can be realistically transported via the large overmoded circular waveguide (Section 3.9.3).

The cryogenic segmentation of the main linacs is organised as:

- an **ML unit** which consists of three cryomodules in a Type A – Type B – Type A arrangement (26 cavities and 1 quadrupole package);
- a **cryo string**, which consists of 4 ML units (long string with 12 cryomodules) or 3 ML units (short string with 9 cryomodules), followed by a 2.5 m cold-box;
- A **cryo unit** comprising of between 10 to 16 cryo strings, with the final cold-box being replaced by an 2.5 m service box.

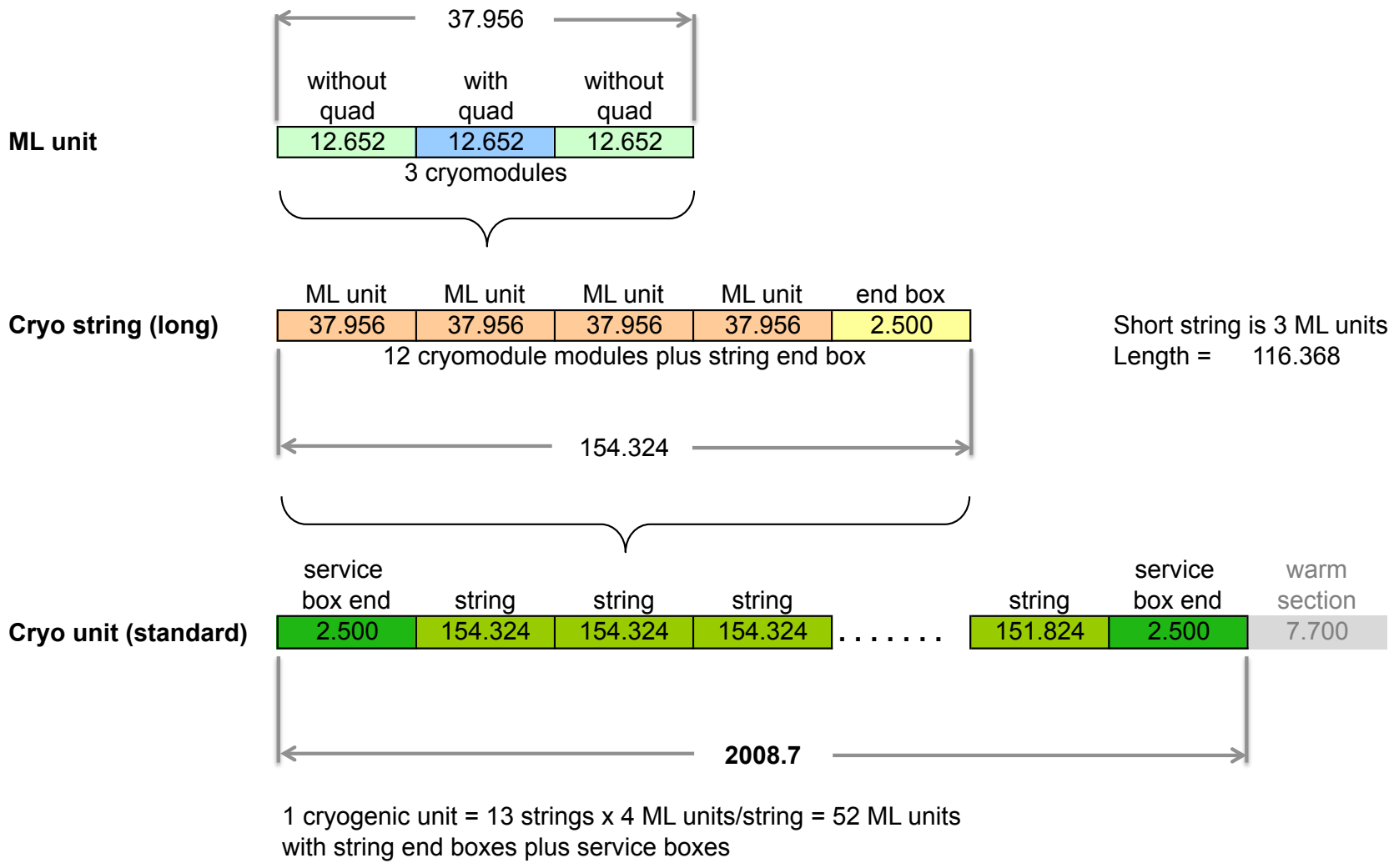

**Figure 3.2** Basic cryogenic segmentation in the main linacs. Note that the length of the cryo units varies depending on the number of strings. (All lengths given in metres.)

Figure 3.2 shows an example configuration for a single cryo unit based on 13 long cryo strings. The maximum length for a cryo unit is approximately 2.5 km, and is set by consideration of the largest practical cryoplant size (approximately 4 MW, comparable to those running at LHC). This includes a 40 % overcapacity to account for pre-cooling from room temperature, variation in cooling water temperature, and operational overhead. (About half of the AC power consumed in the cryoplants is used to remove the RF energy dissipated in the 2 K cavities.) Two cryoplants are typically located together, with one plant feeding an upstream cryo unit and the other a downstream cryo unit. This results in a typical spacing of cryo plant installations (vertical shafts or access ways) of approximately 5 km. The most upstream cryoplant also provides cooling for the accelerator sections for the bunch compressors in the RTMLs (Chapter 7). The exact linac segmentation and number of cryoplants differs for the two site-dependent variants considered, although the number of cryomodules in the linacs are the same. In particular, five cryoplants are envisaged for the mountainous topography, while for the flat topography the total load is distributed over six plants. These differences are driven by the approach to the RF power distribution for each site variant. Section 3.8 and Section 3.9 provide details of the main linac segmentation for the flat and mountainous topographies respectively.

Each cryo unit is separated by a short ~7.7 m warm section that includes vacuum-system components and a 'laser wire' to measure beam size.

At the exit of main linacs there is a section of warm beamline which acts as the matching interface to the downstream systems. This section of beamline provides:

- matching and machine-protection collimation for the transition between the relatively large apertures in the main linac, to the smaller ones in the downstream (warm) systems (most notably the positron-source undulator located at the exit of the electron linac);
- beam-trajectory correction using a fast intra-train feedback/feedforward system, which should reduce pulse-to-pulse jitter to approximately 10 percent of the vertical beam height. The fast kickers will also be used to correct repetitive bunch-to-bunch variation possibly arising from long-range wake fields. On the electron linac side, in addition to the fast feedback correction, a 10 Hz pulsed magnet system is required to adjust the 150 GeV positron production beam during 10 Hz operation at low energy operation (see Section 3.1.4).

Table 3.2 summarises the combined power consumption of the Main Linacs. Of this power, 9.9 MW goes into the beams and the corresponding wall-plug to beam power efficiency is 9.6 %.

**Table 3.2**
Main Linac AC power consumption for both site-dependent variants. Details can be found in Chapter 11

| System | Flat Topography AC power (MW) | Mountain Topography AC power (MW) |
|---|---|---|
| Modulators | 58.1 | 52.1 |
| Other RF system and controls | 5.8 | 5.5 |
| Conventional facilities | 13.3 | 16.4 |
| Cryogenics | 32.0 | 32.0 |
| Total | 109.2 | 106.1 |

### 3.1.3 Accelerator Physics

Table 3.3 lists the basic beam parameters for the main linacs. The main-linac lattice uses FODO optics, with a quad spacing of 37.96 m, corresponding to one quad per three cryomodules (ML unit). Each quadrupole magnet is accompanied by horizontal and vertical dipole correctors and a cavity BPM which operates at 1.3 GHz. The lattice functions are not perfectly regular due to the interruptions imposed by the cryogenic system, but do not change systematically along the linac so the focusing strength is independent of beam energy. Figure 3.3 shows the lattice for the last cryo-unit of the main linac. The average lattice beta function is approximately 80 m and 90 m in the horizontal and vertical planes respectively. The mean phase advance per cell is 75° in the horizontal plane and 60° in the vertical plane. The small vertical bending required to follow the Earth's curvature is provided by vertical correctors near the quadrupole locations, and gives rise to $\sim 1\,\mathrm{mm}$ of vertical dispersion

**Table 3.3**
Nominal Linac Beam Parameters for 500 GeV CMS operation.

| Parameter | Value | Unit |
|---|---|---|
| Initial beam energy | 15 | GeV |
| Final (max.) beam energy | 250 | GeV |
| Particles per bunch | $2.0 \times 10^{10}$ | |
| Beam current | 5.8 | mA |
| Bunch spacing | 554 | ns |
| Bunch train length | 727 | µs |
| Number of bunches | 1312 | |
| Pulse repetition rate | 5 | Hz |
| Initial $\gamma\epsilon_x$ | 8.4 | µm |
| Final $\gamma\epsilon_x{}^\dagger$ | 9.4 | µm |
| Initial $\gamma\epsilon_y$ | 24 | nm |
| Final $\gamma\epsilon_y{}^\dagger$ | 30 | nm |
| $\sigma_z$ | 0.3 | mm |
| Final $\sigma_E/E^\dagger$ | 0.07 | % |
| Bunch phase relative to RF crest | 5 | degrees ahead |

$^\dagger$ at exit of linac

(peak). Dispersion matching and suppression at the beginning and end of the linac are achieved by supplying additional excitation to small numbers of correctors in "dispersion-bump" configurations.

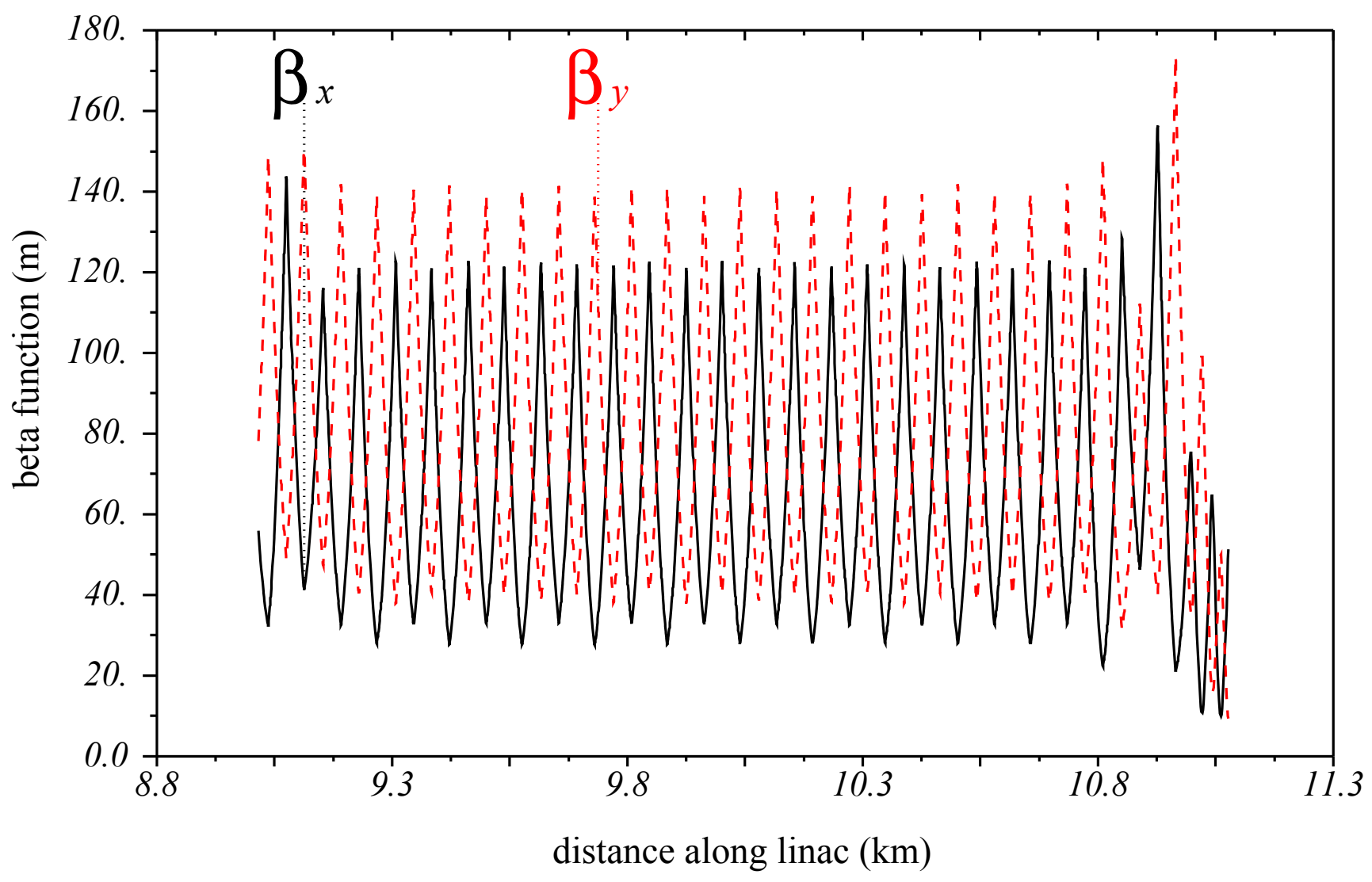

**Figure 3.3**
Example lattice functions for the main linac. The plot shows the beta functions for the last cryo-unit of the linac. (The warm post-linac collimation system is also included.)

The beam emittance at the damping ring extraction is $\gamma\epsilon_x = 8\,\mu\text{m}$ and $\gamma\epsilon_y = 20\,\text{nm}$. The biggest challenge is in keeping the small vertical emittance from being degraded during the beam transport to the beam interaction point (IP). This cannot be done perfectly, and the ILC parameters specify a target emittance at the IP of $\gamma\epsilon_x = 10\,\mu\text{m}$ and $\gamma\epsilon_y = 35\,\text{nm}$. An emittance growth budget has been set at $\Delta\epsilon_y \leq 10\,\text{nm}$ for the total of the RTML and the main linac. The goal for the alignment and tuning procedures is to ensure that the emittance growth is within the budget.

To limit the emittance dilution, the position and orientation of the beamline elements are set fairly precisely during installation (Table 3.4), and beam-based alignment methods are then used to adjust the corrector magnet strengths to establish an orbit that minimises the beam emittance growth. The task is made easier by the fact that the long-range (bunch-to-bunch) wakefields are weak and the initial bunch trajectories are very similar, so minimising the emittance growth of one bunch will do it for all. As discussed above, any slow variation of the relative trajectories of the bunches along the trains will be removed after the linacs using a fast-kicker-based feedback system. The relatively weak single-bunch longitudinal wakefield can be compensated a small off-crest phase (5 degree at 31.5 MV/m).

To suppress any resonate buildup of the wakefields and their effect on the beam, a higher-order-mode (HOM) damping system has been carefully designed into the cavities and the HOM frequencies are effectively detuned cavity-to-cavity at the $10^{-3}$ level as a result of geometric differences within the fabrication tolerances. Such fabrication variations can also lead to diagonal polarisation of the dipole modes instead of horizontal and vertical polarisation. The difference in horizontal and vertical betatron phase advance noted above prevents such $x$-$y$ coupling from causing orbit jitter to couple between the horizontal and vertical planes. The high-impedance HOMs for TESLA cavities have been calculated and experimentally verified at TTF/FLASH at DESY, Hamburg [14]. A table of the HOMS and their $r/Q$ values can be found in [13].

**Table 3.4**
Installation alignment errors (rms) of the linac beam-line elements. BPM specifications are also included.

| Error | with respect to | value | |
|---|---|---|---|
| Cavity offset | module | 300 | µm |
| Cavity tilt | module | 300 | µrad |
| BPM offset | module | 300 | µm |
| BPM resolution | | 5 | µm |
| BPM calibration | | ≤10 | % |
| Quadrupole offset | module | 300 | µm |
| Quadrupole roll | module | 300 | µrad |
| Module offset | beamline reference | 200 | µm |
| Module tilt | beamline reference | 20 | µrad |

The assumed installation errors are listed in Table 3.4. Cavities and Quadrupole magnets are inaccessible once installed into the cryomodules, and need to be mounted and carefully aligned during assembly that allows for thermal contractions of the support system during the cryostat cool down. Results of stretched-wire measurements in cryomodules (see Part I Section 2.6) have demonstrated that the specifications can be reproducibly met over several thermal cycles [15]. After installation in the tunnel, the offsets of the quadrupole and BPM are ultimately established by beam-based techniques at the micron level (i.e. the quadrupole centres are shifted with corrector magnets and the BPM offsets are determined with a quadrupole-shunting technique). The bunch-emittance dilution is dominated by chromatic (dispersive) effects and wakefield kicks arising from misaligned quadrupole magnets and cavities respectively. Emittance growth from these perturbations is mainly corrected through local or quasi-local steering algorithms such as Ballistic Alignment (BA), Kick Minimisation (KM), or Dispersion Free Steering (DFS), with additional correction achieved through local orbit distortions, which produce offsetting amounts of dispersion at a given phase (‘dispersion bumps’). A more complete description of the emittance-dilution mechanisms and the steering algorithms be can found in Part I Section 4.6.

A BPM with horizontal and vertical readout and micron-level single-bunch resolution is located adjacent to each quadrupole magnet. For beam-size monitoring, a single laser wire is located in each of the warm sections between main linac cryogenics units (about every 2.5 km). Upstream quadrupole magnets are varied to make local measurements of the beam emittances at these points.

### 3.1.4 Operation

Fast trajectory control is implemented in the warm regions upstream and downstream of the Main Linacs but not within the linacs themselves, as the trajectory jitter generated by magnetic and RF field variations is expected to be small (see Part I Section 4.6). Likewise, beam energy and energy spread are only measured upstream and downstream of the Main Linacs, and there are no beam-abort systems or energy collimation chicanes along the linacs. The Machine Protection System will only allow beam into the RTML if the trajectory is within a defined phase space, and if the RF phases in the RTML and Main Linac cavities are within a prescribed range prior to the beam extraction from the damping rings (during the approximately 800 µs fill time for the RF). The limiting apertures in the cryomodules are the 70 mm diameter cavity irises.

The linac length (number of cryomodules) includes a 1.5% energy overhead for 250 GeV operation. This overhead can compensate for failed cavities or RF systems. (See Section 10.2 for more details.)

The required beam energy is first 'coarsely' adjusted by setting the required RF power and cavity $Q_{ext}$, and 'finely' adjusted by cross-phasing RF units near the end of the linacs. For operation at low beam energy (low gradient), the modulator voltage and RF pulse length would be reduced to save energy.

As noted in Section 2.2.2 for operation below 250 GeV beam energy, the electron linac will be operated at 10 Hz to provide alternatively a 150 GeV beam for positron production, followed by a $\leq 125\,\mathrm{GeV}$ beam for luminosity production. (The positron linac only runs at the nominal 5 Hz.) For this reason, all linac RF devices are specified at 10 Hz, although at reduce peak power requirements (approximately one-half of that required at 250 GeV beam-energy operation). Furthermore, at the reduced gradient there is already sufficient RF and cryogenic AC power installed to run the linac at the higher repetition rate. (This is helped by the fact that the RF fill time is reduced by approximately one-half, and thus shortens the RF pulse; this has the effect of increasing both the RF-to-beam power efficiency, as well as reducing still further the dynamic cryogenic load.) Transport of the two different beam energies in 10 Hz mode has been simulated (see Part I Section 4.6). The main linac will be tuned to preserve the low emittance of the luminosity production pulse (lower energy); the emittance of the higher-energy positron-production pulse is not critical, and the simulations have shown that the trajectory offset at the exit of the linac is typically a few millimetres — well within the aperture of the linac. However, this offset will require adjustment to bring the beam on-axis of the source undulator, which requires a 10 Hz pulsed magnet system in the warm section immediately downstream of the electron linac.

### 3.1.5 Linac Systems

The remaining sections of this chapter describe in detail the main components of the Main Linacs, starting from the SC cavities and working outward through the cryomodule, high power RF systems and finally to the low-level RF (LLRF) controls:

**Section 3.2 Cavity performance and production specifications** covers the cavity design, performance specifications, and baseline industrial production process, including the required surface preparation to achieve the required high-performance.

**Section 3.3 Cavity integration** discusses the complete cavity package and how it is assembled, including the high-power RF coupler, HOM couplers, helium tank and mechanical frequency tuner.

**Section 3.4 Cryomodule design including quadrupole and cryogenic systems** describes the mechanical design of the 12.7 m long cryomodules, which comprise the vacuum vessel and the items within, including cavities, thermal shielding, cryogenic feed and return lines, beamline absorber and a quad 'package,' consisting of a quadrupole magnet, horizontal and vertical corrector magnets and an RF BPM. Estimates of the cryogenic heat loads are also presented.

**Section 3.5 Cryogenic cooling system** describes the layout of the cryogenic plants and required plant capacities.

**Section 3.6 RF power source** presents the common components of the RF system for the KCS and DKS systems, i.e. the 120 kV Marx Modulators and the 10 MW Multiple Beam Klystrons (MBKs) that they power. Also, the local RF distribution system that divides up the feed power to the cavities in the tunnel is described.

**Section 3.7 Low-level RF-control concept** covers the design and operational aspects of the low-level RF system (LLRF) that is required to stabilise the vector sum of cavity voltages to within less than 1 % across the beam pulse. This includes the more 'global' control via closed-loop feedback on the klystrons, as well as local (per cavity) compensation of Lorentz-force frequency detuning using piezo-electric controllers.

**Section 3.8 Main linac layout for a mountain topography** discusses those design features specific to the mountain topography site-specific design, and in particular the linac layout and DKS.

**Section 3.9 Main linac layout for a flat topography** discusses those design features specific to the flat topography site-specific design, and in particular the linac layout and KCS.

## 3.2 Cavity production specifications

### 3.2.1 Cavity Design

Figure 3.4 shows schematics of a baseline 9-cell superconducting cavity and the assembly with liquid-helium (LHe) tank. Table 3.5 summarises the main design parameters of the cavity.

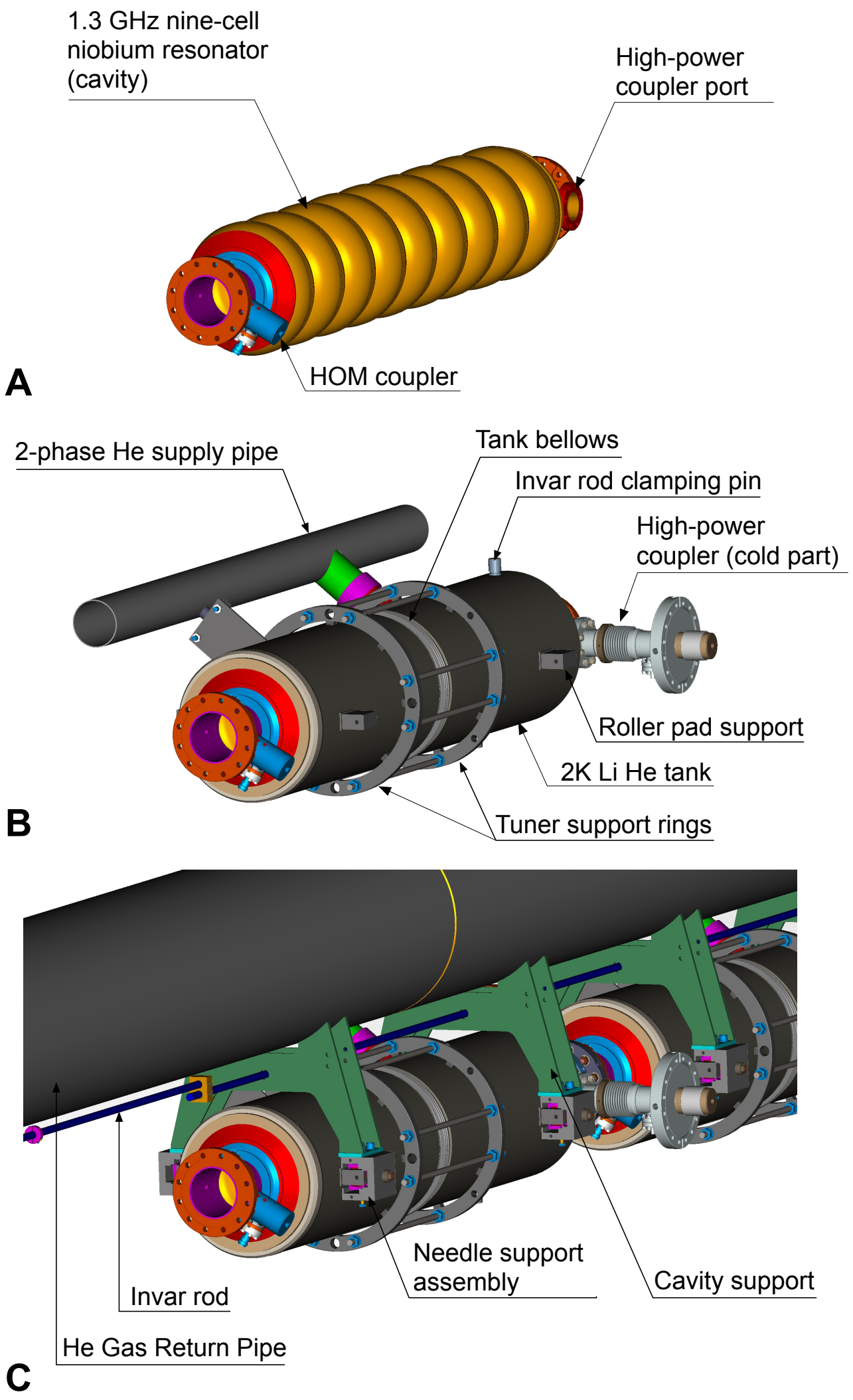

**Figure 3.4**
The baseline cavity package and string assembly: (A) the nine cell cavity (resonator); (B) the "dressed" cavity, showing the helium tank, 2-phase helium supply, high-power coupler (cold part) and the mount for the cavity tuner; (C) cavity package mounted into the cavity string and cryomodule. (Note the "blade" cavity tuner is not shown.)

**Table 3.5**
Cavity parameters for the SCRF cavities.

| Parameter | Value |
|---|---|
| Type of accelerating structure | Standing wave |
| Accelerating mode | $TM_{010}, \pi$ mode |
| Type of cavity-cell shape | Tesla (or Tesla-like) |
| Fundamental frequency | 1.300 GHz |
| Operation: | |
| – Average gradient (range allowed) | 31.5 MV/m (±20% ) |
| – Quality factor (at 31.5 MV/m) | $\geq 1 \times 10^{10}$) |
| Qualification: | |
| – Average gradient (range allowed) | 35.0 MV/m (±20% ) |
| – Quality factor (at 35 MV/m) | $\geq 0.8 \times 10^{10}$ |
| – Acceptable radiation (at 35 MV/m) | $\leq 10^{-2}\mathrm{mGy/min}^{\dagger}$ |
| Active length | 1038.5 mm |
| Total length (beam flanges, face-to-face) | 1247.4 mm |
| Input-coupler pitch distance, including inter-connection | 1326.7 mm |
| Number of cells | 9 |
| Cell-to-cell coupling | 1.87% |
| Iris aperture diameter (inner/end cell) | 70/78 mm |
| Equator inner diameter | ~210 mm |
| $R/Q$ | 1036 Ω |
| $E_{peak}/E_{acc}$ | 2.0 |
| $B_{peak}/E_{acc}$ | 4.26 mT/(MV/m) |
| Tunable range | ±300 kHz |
| $\Delta f/\Delta L$ | 315 kHz/mm |
| Number of HOM couplers | 2 |
| $Q_{ext}$ for high-impedance HOM | $< 1.0 \times 10^5$ |
| Nb material for cavity (incl. HOM coupler and beam pipe): | |
| - RRR | ≥300 |
| - Mechanical yield strength (annealed) | $\geq 39\,\mathrm{MPa}$ |
| Material for helium tank | Nb-Ti Alloy |
| Max design pressure (high-pressure safety code) | 0.2 MPa |
| Max hydraulic-test pressure | 0.3 MPa |

† Example number taken from [16] — see text for more details

## 3.2.2 Cavity fabrication and surface processing

The fabrication process of ILC superconducting cavities and their surface treatment have substantially matured during the Technical Design Phase. The R&D leading to these procedures — as well as a more detailed discussion of the steps involved — can be found in Part I Section 2.3. The procedure is summarised in Table 3.6.

There are two key issues concerning the mechanical fabrication of the cavities for the ILC. The first is the quality assurance of the niobium materials. The second is the process quality control of electron-beam welding. The sheet and bulk niobium which are supplied by vendors must be scanned for detecting and avoiding materials with defects[1]; once accepted, they have to be the protected from mechanical damages and dust throughout the manufacturing process. Defective materials can become limit the performance of completed cavies. Impurities introduced into the welds and in the heat-affected zones next to welds will also limit the gradient performance. Weld joints must have smooth beads without surface irregularities and without sharp edges on locations where the weld puddle meets the bulk material. A defect on the equator weld will in general result in a local enhancement of the magnetic field. A single such defect can cause a quench, leading to a degradation of gradient performance. Current production experience suggests some 10–20% of cavities produced could suffer from this problem, and therefore procedures for repairing the cavity surface has been developed (see Part I Section 2.2.7). For ILC mass-production rates, however, it is expected that — after some initial ramp-up period — the electron-beam welding process will be improved to significantly reduce the number of such defects, to a level of <10%.

[1] During ILC mass production, it is conceivable that such scanning will be at a reduced rate for QC only, once the production has been established.

**Table 3.6.** Summary of steps required to fabricate a nine-cell cavity.

| Steps | Reference parameters | Notes: |
|---|---|---|
| *9-cell cavity fabrication:* | | |
| - Raw material preparation | Nb sheet: $t = 2.8\,\mathrm{mm}$, $RRR \geq 300$ | Acceptance with sheet inspection including visual and non-destructive defect inspection. |
| - Component fabrication | | Using press, machining, and electron-beam welding (EBW). |
| - Assembly of 9-cell cavity | | Using EBW |
| - Inner-surface inspection | | Using optical inspection method†. |
| *Inner-surface treatment:* | | |
| - Light etching with BCP | 5–20 µm | (Optional: EP, 5–20 µm) |
| - Heavy EP | 100–120 µm | ~24 µm/hour at $30\,°\mathrm{C} \leq T \leq 35\,°\mathrm{C}$ |
| - Post-heavy-EP cleaning | | |
| - Out-gasing | 800 °C, ≥2 hours | |
| - RF tuning | | Using tuning machine and non-contacting bead-pull method*. |
| - Light EP | 20–30 µm | ~12 µm/hour at $20\,°\mathrm{C} \leq T \leq 30\,°\mathrm{C}$. |
| - Ethanol or detergent rising | ~1 hour | |
| - First HPR rinsing | 6 hours, 3 passes | Purity level of water: Resistivity 18 MΩ cm. |
| - First clean-room assembly | | |
| - Final HPR rinsing | 6 hours, 3 passes | |
| - Final assembly | | In class 10 clean-room. |
| - Leak check | | Sensitivity: $\leq 2 \times 10^{-10}\,\mathrm{Pa\,m^3\,s^{-1}}$. |
| - In-situ baking | ~48 hours at ~120 °C | |
| *Assembly of LHe tank:* | | |
| - Pre-assembly & check-out | | Check-out of tank components. Validation of hermeticity and mechanical sturdiness under over-pressurised conditions from safety standpoint. |
| - Assembly of LHe tank | | Assembly with 9-cell cavity part, using EB or TIG welding |
| - HPC inspection | 1.5 (or 1.25)×2 bar | Differential. Depending on HPC in region. |
| - Leak-check | | Sensitivity: $\leq 1 \times 10^{-9}\,\mathrm{Pa\,m^3\,s^{-1}}$. |
| - General inspection | | Dimensions etc. |
| *Cavity RF performance test:* | | |
| - Cool-down | | Cool-down time: several hours |
| - $Q_0$ vs. gradient | | $\pi$ and pass-band mode, including radiation monitoring. |
| *Post-performance test assembly and check:* | | |
| - coupler and HOM assembly | | Including leak-check. |
| - tuner assembly | | Including functioning test. |
| - General inspection | | As an acceptance test for cavity-string assembly. |

Terms:
RRR: residual resistance ratio
EBW: electron-beam welding
BCP: buffered chemical polishing
EP: electro-polishing
HPR: high-pressure (pure water) rinsing
TIG: tungsten inert-gas welding

† Dedicated tooling/facility provided by laboratories.

The surface-preparation steps have developed over many years into the established recipe outlined above. The details can be found in Part I Section 2.3. In summary, the process steps are designed to:

1. remove material damage incurred during the fabrication process or handling by using chemical procedures;
2. remove the chemical residues left over from the material removal steps;
3. remove hydrogen in the bulk niobium which has been captured during the chemical procedures in step 1;
4. remove any particulate contamination which entered during the cleaning and assembly steps; and
5. close up the cavity to form a hermetically sealed structure.

Figure 3.5 provides an overview of the cavity production process, and in particular the approach to testing. A key issue for mass production is achieving the required performance yield ($> 90\%$) in a cost-effective manner. The current approach — based on existing R&D experience discussed in Part I Section 2.3 — is to allow specific steps to be iterated in the production process. The first test in this respect is an optical inspection (Part I Section 2.2.2) of the cavity directly after fabrication, but before any surface treatment. This inspection is intended to identify candidate surface defects as described above, which may limit the cavity performance to below 20 MV/m. These cavities (an estimated 10% or less after the production process has matured) would be removed from the production line and mechanically repaired using the techniques described in Part I Section 2.2.7. A second optical inspection is made after initial surface treatment (bulk electro-polishing, 800° heat treatment, followed by mechanical RF tuning), to identify weld defects that may have been uncovered by the removal of 150 µm niobium during bulk electro-polishing (an estimated few percent of the total production). The cavities then undergo the final surface preparation steps and the high-Q RF antenna, two HOM couplers and the helium tank are mounted. The final performance (acceptance) test is a low-powered RF test in a vertical cryostat at 2 K (so-called vertical test), where the cavity ultimate performance is measured (maximum acceptable gradient, quality factor, field emission etc.)[2]. Based on the current status of the R&D, it is expected that some fraction of cavities (≤20%) will be limited to <28 MV/m and will require some remedial action, the exact nature of which depends on the mode of failure. If the cavity gradient is limited by field emission (excessive X-rays), then it is highly likely that only an additional high-pressure rinse (6 hours) will suffice, a process which is relatively straightforward. A limitation due to a breakdown (i.e. quench) will require an additional light electro-polishing step ($\sim 25\,\mu\mathrm{m}$), which is more process intensive, and requires the removal (at least) of HOM couplers and high-Q RF antenna, which must then be re-mounted after the surface treatment. It is anticipated that the removal of the helium tank can be avoided for this second-pass treatment. Irrespective of which procedure is followed, the cavity must then undergo a second vertical test. Although ultimately cavities with a gradient performance ≥28 MV/m will be accepted, this represents only the lower limit of the assumed gradient spread ($28\,\mathrm{MV/m} \leq G \leq 42\,\mathrm{MV/m}$). Therefore all cavities that fail to make 35 MV/m on the first-pass test undergo a second cycle (either HPR or light EP). On the second-pass test, cavities achieving the minimum required 28 MV/m are accepted for cryomodule assembly, as indicated in Fig. 3.5. Although a third-pass is feasible, it is not considered necessary and is not included in the cost

The acceptance criteria for gradient and quality factor have been well established and standardised during the technical design phase. The measurement and acceptable levels of X-rays generated in the vertical test (an indication of field emission) remains the least well-defined quantity. Methods and standards that can be unambiguously applied to test infrastructures around the world requires additional R&D. Table 3.5 quotes a radiation value for qualification of $\leq 10^{-2}\mathrm{mGy/min}$ at 35 MV/m, a number based in the current European XFEL experience [16]. However this number is particular to the DESY vertical test set-up, and cannot be universally adopted by other test infrastructures. Definition of such practical standards is especially important when the anticipated globally-distributed nature of the cavity manufacturing and testing is considered.

It is expected that the mechanical fabrication and surface preparation will be performed by industry. However, the final RF test will be performed by a collaborating laboratory or institute, which will host the required cryogenic and RF test infrastructure and personnel. The second-pass process steps could alternatively be dealt with locally by the lab hosting the test facility (as is the case for the European XFEL), or returned to industry.

[2]The decision to assemble the tank and HOM couplers before the vertical test is driven by mass production considerations, and follows the same approach taken for the cavity production for the European XFEL project.

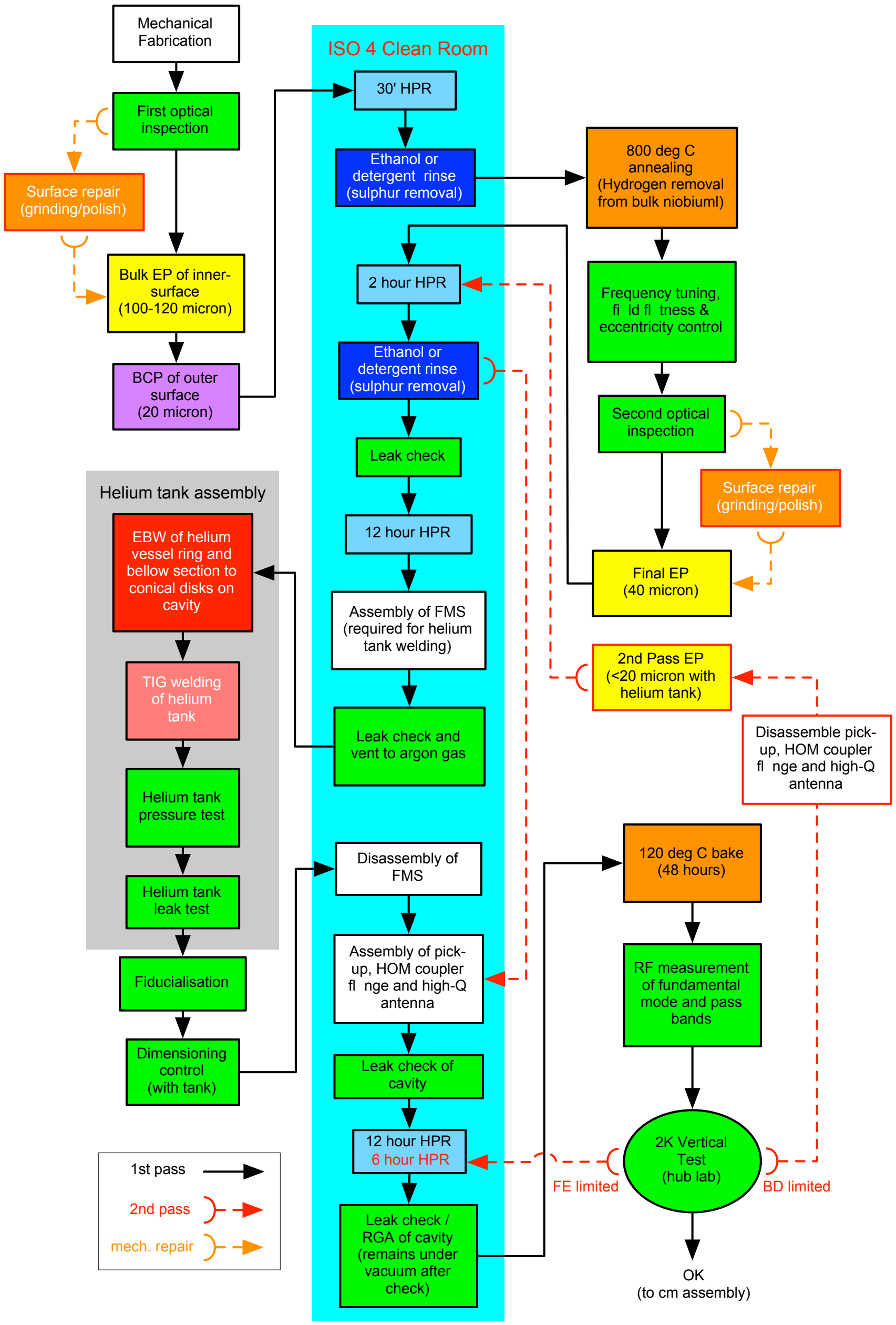

**Figure 3.5** Flow chart showing the various steps involved in the surface treatment of high-gradient cavities. The 'first pass' (black arrows) is based on the process currently being used by XFEL and is slightly different from the steps presented in Table 3.6. Red arrows indicate possible 'second pass' procedures for a cavity which fails either optical inspection or the final 2 K vertical performance test. Note that the second-pass EP is assumed to be made with the helium tank still attached.

## 3.3 Cavity integration

The most important part of the cryomodule is the cavity package, which is an integrated system consisting of a 9-cell cavity, contained within a titanium-alloy helium tank connected to a helium supply pipe, a fundamental-mode power coupler, a frequency tuner, and a magnetic shield (see Fig. 3.4 A and B). The design of the ILC cavity system is based on the original TESLA design used in TTF/FLASH and currently being produced in industry for the European XFEL project. The ILC cavity package consists of the following:

- a nine-cell niobium resonator (cavity), complete with two HOM couplers and RF antenna, flanges etc.;
- a titanium-alloy helium tank (cryostat), split with a bellows to support the mechanical tuner;
- the mechanical tuner itself (so-called blade tuner), mounted on the two halves of the helium tank;
- a high-power fundamental-mode RF coupler;

- a magnetic shield which surrounds the cavity and is installed inside of the helium tank.

The cavity and its manufacture are discussed in the previous section (Section 3.2). The remainder of this section will describe the baseline high-power coupler, frequency tuner, helium tank and HOM couplers.

## 3.3.1 Fundamental-mode power coupler

The 'TTF-III' input coupler was originally developed for TESLA [17], [18], and has since been modified by a collaboration of LAL and DESY for use in the European XFEL [19]. Due to the maturity of the design and extensive experience with this coupler, it has been adopted as the baseline design of the fundamental power coupler for the ILC. The main specifications of this input coupler are listed in Table 3.7.

**Table 3.7**
Main specifications of the input coupler. The parameters represent the approximate maximum expected values during operation, including possible upgrades.

| Parameter | Specifications |
|---|---|
| Frequency | 1.3 GHz |
| Operation pulse width | 1.65 ms |
| Operation Repetition rate | 5 Hz / 10 Hz |
| Maximum beam current | 8.8 mA |
| Accelerating gradient of cavity | 31.5 MV/m ± 20% |
| Required RF power in operation | ∼ 400 kW |
| Range of external $Q$ value | (1.0 ∼ 10.0) $\times 10^6$ (tunable) |
| RF process in cryomodule | > 1200 kW for ≤ 400 μs pulse width |
| | > 500 kW for > 400 μs pulse width |
| RF process with reflection mode in test stand. | > 600 kW for 1.6 ms pulse width |
| RF process time | < 50 hours in warm state |
| | < 20 hours in cold state |
| Approximate heat loads | < 0.01 mW (2K static) |
| | 0.07 W (5K static) |
| | 0.6 W (40K static) |
| | < 0.02 W (2K dynamic) |
| | 0.12 W (5K dynamic) |
| | 1.6 W (40K dynamic) |
| Number of windows | 2 |
| Bias voltage capability | Required |

The coupler is a complex device assembled from roughly 130 parts. As with the cavities, the couplers must be assembled in very clean environments.

### 3.3.1.1 Mechanical design

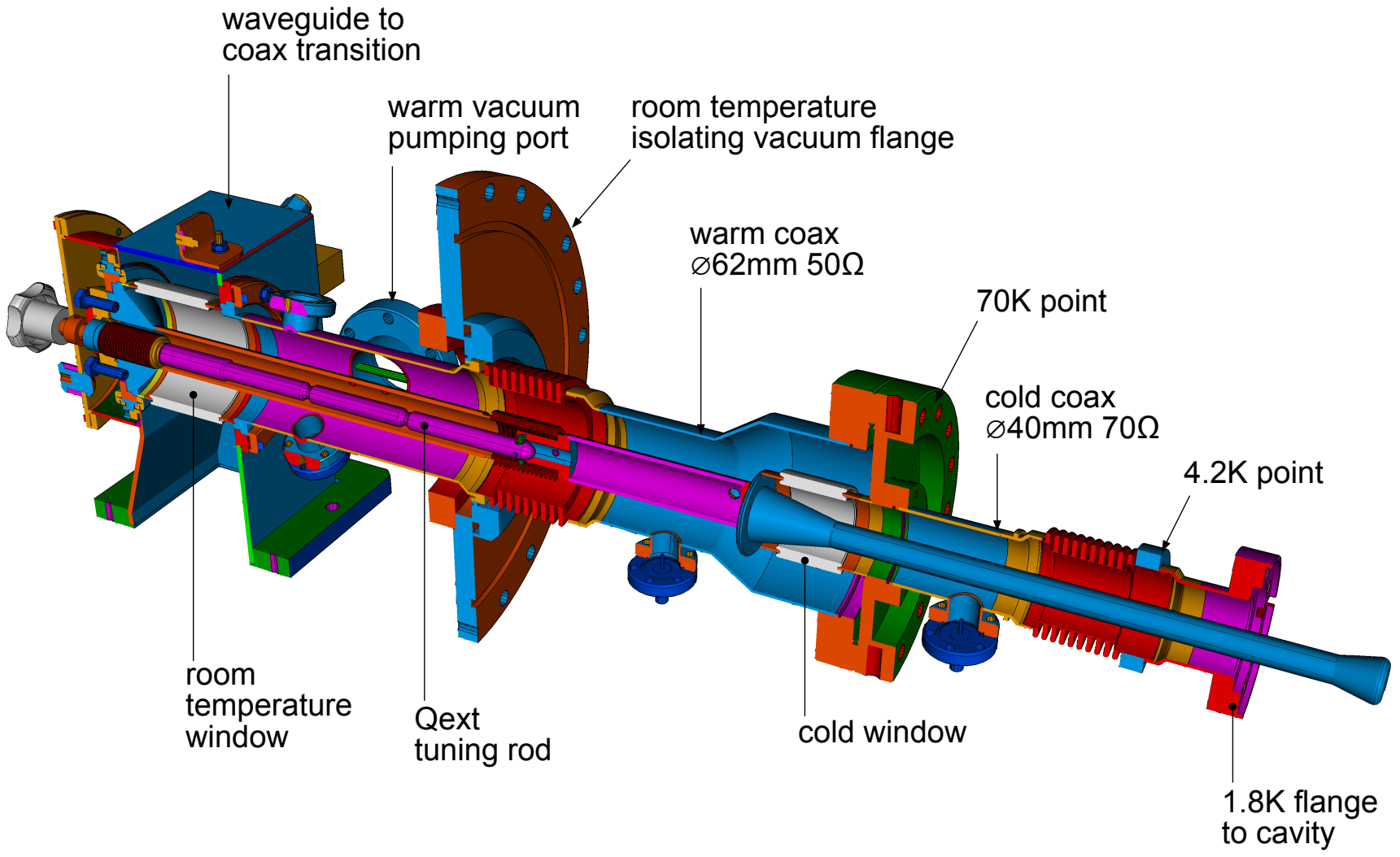

**Figure 3.6**
Schematic drawing of TTF-III (XFEL) input coupler.

A 3-dimensional sectioned schematic of the coupler assembly is shown in Fig. 3.6. The coupler is separated into a warm and cold part as shown, the latter of which is mounted into the cavity at 2 K. RF power is brought in via a rectangular waveguide (WR650) into the "door-knob" mode converter at the warm transitionapproximate max . The RF power then propagates through a coaxial transmission line into the cavity beam pipe via the antenna (at 2 K). Both the warm and cold parts have a ceramic RF window, which protects and separates the cavity vacuum and the vacuum inside the warm coupler part. (The cold part shares the same vacuum with the cavity.) Both RF windows are cylindrical ceramic pieces made of $Al_2O_3$, the vacuum surfaces of which are coated with a few nanometers thickness of titanium-nitride in order to prevent multipacting. The two bellows (warm and cold) in the outer conductor allow a $\pm$ 10 mm adjustment of the antenna penetration into the cavity beam pipe to change the coupling to the cavity, providing a range of $Q_{ext}$ of 1–10 $\times$ $10^6$. The antenna position ($Q_{ext}$) is adjusted via a tuning rod housed in the central conductor, and driven by a remote actuator at the end of the warm transition.

The outer conductor is made of thin stainless steel whose inner surface has a 10 µm thick copper plating. The required thickness of the copper plaiting is a trade-off balance between providing enough electrical skin depth to prevent penetration of the RF into the stainless-steel outer conductor, thus minimising ohmic losses, and achieving a thermal balance between heat conduction from the warm end of the coupler (static load) while providing cooling for the RF losses (dynamic load).

Each input coupler is equipped with three electron-current pick-up probes for monitoring discharges inside the coupler. Provision is made to DC-bias the inner conductor to suppress the onset of multipacting. The warm coupler vacuum is maintained by a separate vacuum pumping system at a pressure of $< 10^{-8}$ mbar.

#### 3.3.1.2 Initial coupler processing (acceptance testing) and final assembly

After receipt from industry and before assembly into the cavity, the input couplers undergo warm RF conditioning, which also forms part of the coupler acceptance testing. Coupler test facilities will likely be located at collaborating institutes, such as the one for the European XFEL at LAL, Orsay. The coupler test facility requires clean room facilities for the handling and cleaning the coupler parts, pumping and baking systems, and high-power RF systems for processing of the couplers.

The approach to the coupling processing, including clean-room assembly of the parts and subsequent cleaning and in-situ bake-out, is the result of extensive R&D at LAL for the European XFEL [20], which has resulted in a significant reduction in the time required to condition the couplers (now approximately 20 hours). Figure 3.7 shows the steps in preparing the coupler for the warm RF processing. First, the interior of coaxial parts and window ceramics of both the warm and cold parts of the couplers are inspected and cleaned in a clean-room environment, after which they are assembled together. A pair of couplers are then installed in a special rectangular waveguide system for RF processing. The typical conditioning procedure is to raise the RF power and pulse width in steps from near zero to predetermined maximums, avoiding out-gassing in excess of a prescribed vacuum trip level ($\sim 2 \times 10^{-7}$ mbar). The RF pulse width starts from 20 µs, and is then increased to 50, 100, 200, 400, 800, 1300, and 1500 µs. The entire procedure is automated.

Once successfully processed, the couplers are disassembled in a clean room to avoid any contamination of their interior surfaces, and then sealed and transported to a cryomodule assembly facility. Here, the cold and warm parts are separated and the cold part of the coupler mounted into the cavity in a class-10 clean room during the cavity string assembly. The warm part of the coupler is installed only after the complete string and cold mass have been installed into the cryomodule vacuum vessel. Final assembly of the warm couple part is made in a clean environment provided by mobile clean-room cabins.

After installation of the complete coupler into the cryomodule, further light conditioning is required at both room temperature and 2 K. This processing is performed either as part of the cryomodule tests, or in-situ after installation in the accelerator tunnel (see Section 3.4.3).

**Figure 3.7**
Flow chart of the input coupler process and test.

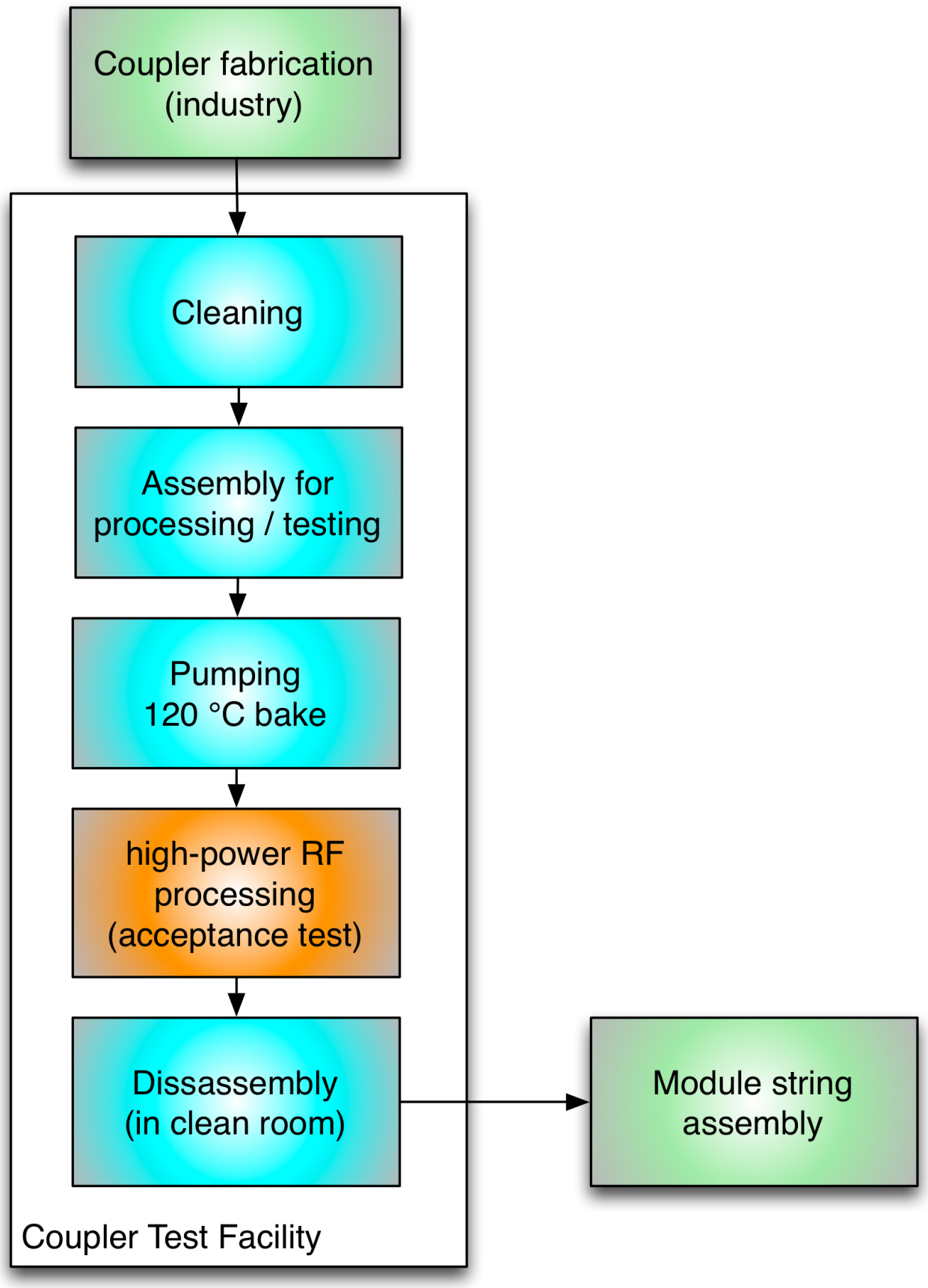

### 3.3.2 Frequency tuner

The mechanical cavity tuner is required to provide two functions:

- a slow mechanically adjustment of the frequency of the cavity and bring it on resonance (static tuning);
- a fast 'pulsed' adjustment using a piezo system to dynamically compensate Lorentz-force detuning during the RF pulse.

Specifications for the frequency tuner system is summarised in Table 3.8. The "Blade Tuner" design [21–24], which has been developed by INFN Milano-LASA as a coaxial and light tuning solution for TESLA-type cavities, has been adopted for the ILC baseline.

**Figure 3.8**
Schematic of the blade tuner.

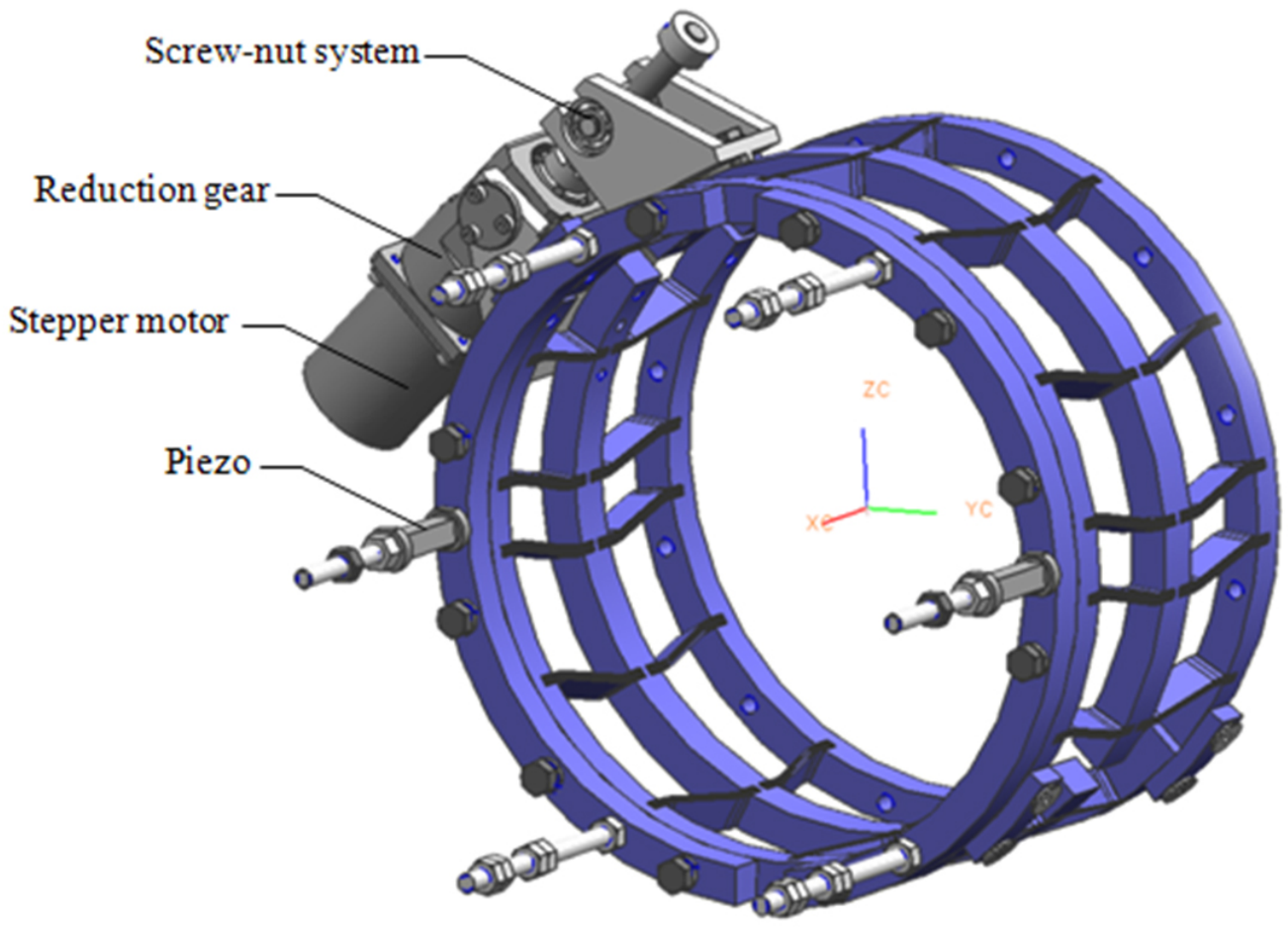

**Table 3.8**
Main specifications of the frequency tuner.

| Tuner | Parameter | Specifications |
|---|---|---|
| Slow tuner | | |
| | Tuning range | $>$ 600 kHz |
| | Hysteresis | $<$ 10 μm |
| | Motor characteristics | Step motor, power-off holding, magnetically shielded |
| | Motor location | Inside 5K shield, accessible from outside |
| | Magnetic shield | $<$ 20mG |
| | Heat load by motor | $<$ 50 mW at 2 K |
| | Motor lifetime | $> 20 \times 10^6$ steps |
| Fast tuner | | |
| | Tuning range | $>$1KHz at 2 K |
| | LFD residuals | $<$ 50 Hz at 31.5 MV/m flat-top |
| | Actuator | Piezo actuator, located inside 5K shield, Two actuators for redundancy |
| | Heat load by actuator | $<$ 50 mW at 2 K |
| | Magnetic shield | $<$ 20mG |
| | Actuator lifetime | $> 10^{10}$ pulses |

Figure 3.8 shows the tuner. The azimuthal movement of the central ring is converted into the required longitudinal cavity strain without backlash via the elastic blades. The tuner mechanics as well as the blades are made of titanium, which provides both mechanical strength and a small thermal expansion coefficient. The slow tuning action is generated by a stepper motor operating at 5 K, coupled via a mechanical reduction gear, to rotate a threaded shaft which moves the central ring azimuthally. A CuBe threaded shaft is used as a screw-nut system. Fast tuning action is driven by two piezoelectric ceramic actuators mounted symmetrically on either side of the cavity as shown in Fig. 3.8, which efficiently allows the transfer of their stroke to the helium tank, in series with the slow mechanical tuner. The coaxial tuner is installed on a mid location of the helium tank that is split in two halves by a bellow. This arrangement allows for simplification of end-cone regions of the cavity which need to accommodate fundamental mode and HOM couplers.

The blade-tuner and in particular the piezo actuators need to be under compression to operate. This is achieved by applying an initial pre-load using a calibrated cavity tensioning, which provides an initial frequency de-tuning and the correct amount of compression for the tuner.

The tuner mechanics, motor, gearbox and piezo actuators must be designed for high reliability, since a failure of the tuner mechanism will seriously hinder the optimal operation of that cavity, and in general these devices are not easily accessible once installed into the cryomodule. Possible solutions which could allow limited access to (for example) the motor and gearbox are being considered, but require much more detailed investigation of the impact on the cryomodule design. It should be noted that the tuners for the European XFEL can only accessed by removing and disassembling the module; the concept adopted here is the use of pre-testing for the components at cryogenic temperatures and careful design of the mechanical systems, such that a lifetime of $>$20 years can be expected. Once operational, the European XFEL will provide important experience on the reliability of such imbedded tuner systems. Part I Section 2.4 provides more detailed discussion on tuner designs and options.

### 3.3.3 HOM couplers

The higher-frequency eigenmodes in the cavity excited by the intense beam bunches must be damped to avoid multibunch instabilities and beam breakup. This is accomplished by extracting the stored energy via higher-order mode (HOM) couplers mounted on the both sides of the beam pipe of the 9-cell cavity [13, p. II-42]. The design of the HOM coupler is shown in Fig. 3.9. The superconducting pickup antenna is well cooled and insensitive to $\gamma$ radiation and electron bombardment. A tuneable 1.3 GHz notch filter suppresses power extraction from the accelerating mode. The $Q_{\mathrm{ext}}$ for the high-impedance modes should be reduced to $< 10^5$ [13, p. II-55].

A TE121 ‘trapped mode’ which is concentrated in the centre cells and has a low amplitude in the end cells is damped by an asymmetric shaping of the end half cells. By using this asymmetric end half cells, one can enhance the field amplitude of the TE121 mode in one end cell, while preserving the field homogeneity of the fundamental mode and the good HOM coupling to the untrapped modes TE111, TM110 and TM011.

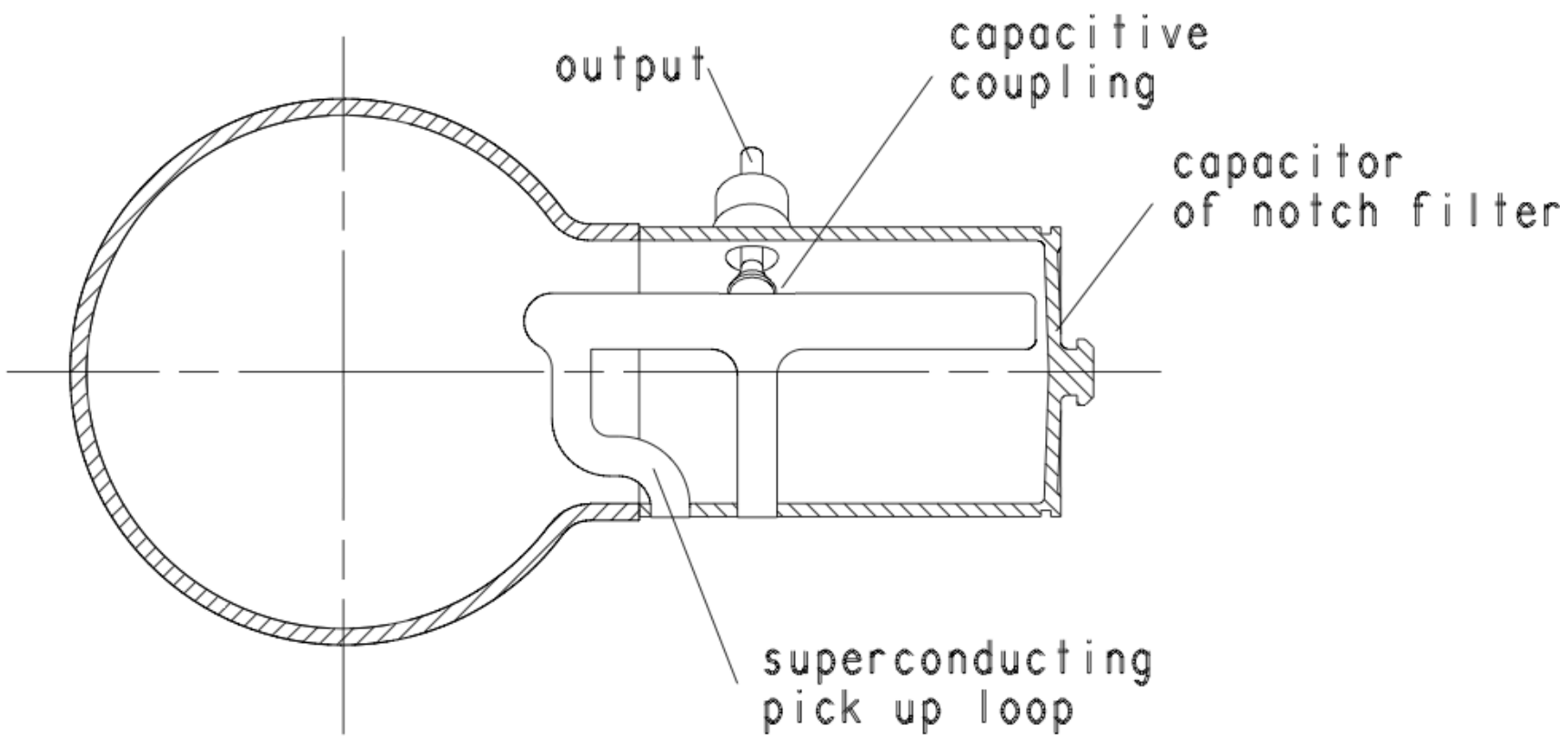

**Figure 3.9**
Cross section of the higher-order mode (HOM) coupler.

The two polarisation states of dipole modes in principle require two orthogonal HOM couplers at each side of the cavity. The optimum angle between the two couplers is 110°, with both couplers being installed at 35° to the horizontal, as shown in Fig. 3.10. Each HOM antenna plane is rotated 30° in its cylinder axis from the perpendicular plane of the beam pipe. The parts of the HOM antenna loops which project into beam pipe are positioned close to the end cells. Beam dynamics studies of the asymmetry arrangement of HOM coupler antenna together with RF field asymmetry of the input coupler antenna on the beam axis [25] have shown that the resulting effects on both the beam emittance and centroid are negligible.

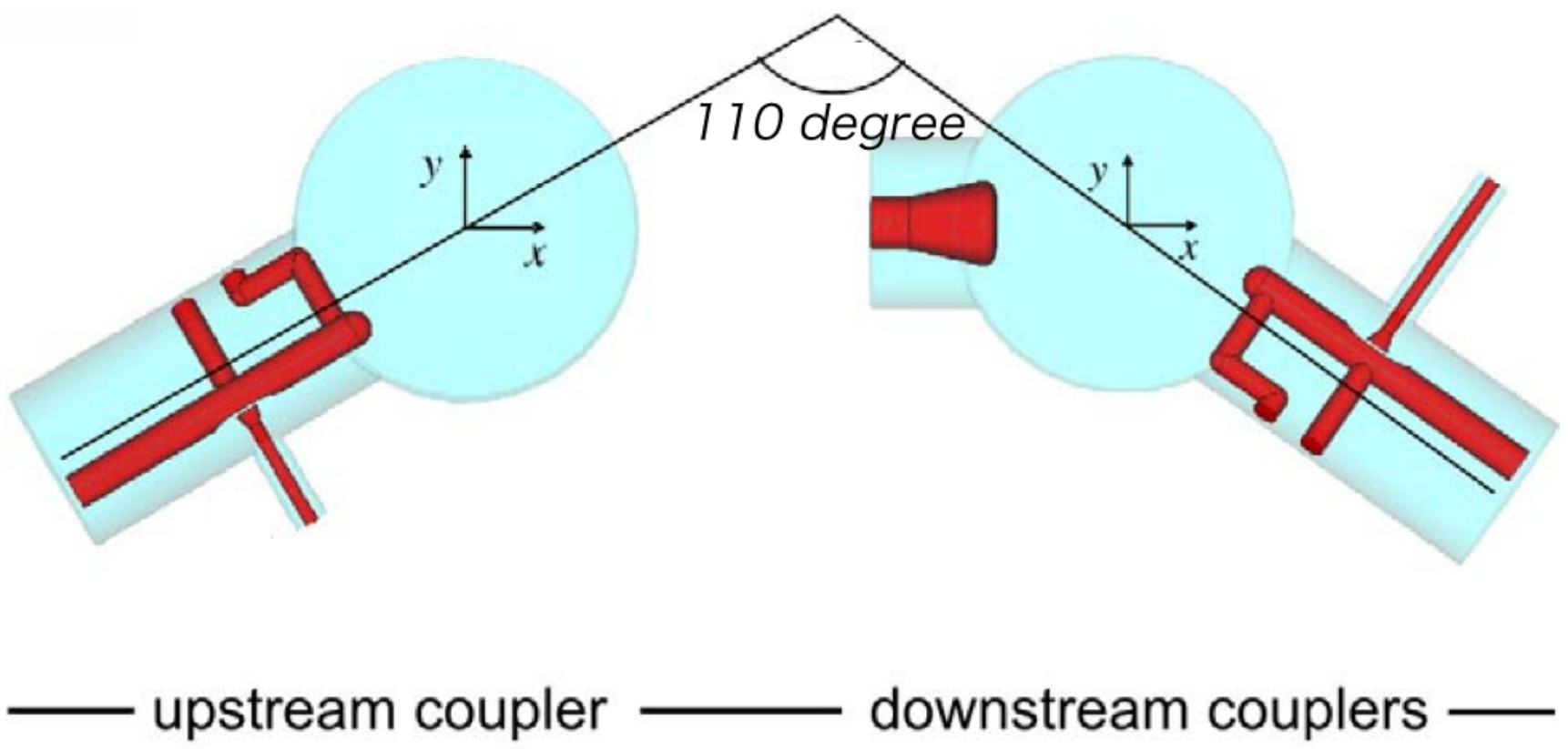

**Figure 3.10**
Orientations of the upstream HOM coupler (left) and the downstream HOM and main power couplers (right).

## 3.3.4 Helium tank and its interface

### 3.3.4.1 Helium tank

The design of the heliumtank for the ILC cavity package consists of a cylinder connected to a 2-phase-helium supply pipe, both of which are constructed from titanium (see Fig. 3.4 B).

Each helium tank has two pairs of "roller pad supports" made of titanium alloy, and welded at the horizontal mid-plane of the tank. The supports are used to hang the cavity from the cavity-support arms which extend underneath the gas return pipe (Fig. 3.4 C). The tank also has a clamping pin to connect it to the invar rod that runs the entire length of the cryomodule. The clamp (and the invar rod) prevents the cavities from moving longitudinally during cool down and warm up, keeping the locations of the high-power couplers fixed with respect to the outside of the cryostat. The roller-pad supports are mounted in slide bearings and adjuster bolts (needle support assembly), which allows contraction and expansion of the helium-gas return pipe, to which the cavity-support arms are fixed.

The helium tank has to accommodate the tuner system as discussed in Section 3.3.2. For this purpose the tank has a thin titanium cylindrical bellows located at its centre. On both sides of the bellows, two flanges are welded for installation of the blade tuner.

The 2-phase-helium pipe has a short branch made of another pipe with the same diameter which is welded to an adaptor hole provided on the helium tank. The 2-phase-helium pipes of neighbouring cavities have their lengths chosen such that they can be readily welded to each other via bellows when they are assembled into a cryomodule.

#### 3.3.4.2 Magnetic shield

The superconducting cavities have to be shielded from external magnetic fields to achieve their maximum performance. The cryoperm shield must cover the entire cavity including the end groups (down to the cavity beam pipe).The current philosophy for ILC is that the cryoperm shield will be placed inside the helium jacket, which is expected to simplify the design of the shield itself. The European XFEL design has the magnetic shield external to the tank, and is installed during the cryomodule assembly. It is expected that the cost of installing an internal shield (during cavity production) is less than the cost of assembling the external shield during module assembly. Furthermore, this simplifies some of the issues concerning the design of an external shield which is compatible with a mid-mounted tuner. There is currently no final design for a suitable shield which can be inserted inside the helium tank. Some experience at KEK with an internal cylindrical shield have demonstrated feasibility. The current TESLA design requires re-design to accommodate such a concept. Once a suitable mechanical solution is found, the effectiveness of the shield will need to be determined by measurement of the cavity quality factor at 2 K. Although this requires further engineering and prototyping, it is expected that cost-effective solutions can be found.

#### 3.3.4.3 Flanges and seals

All the cavity flanges are made of Nb-Ti alloy and use a hexagonal ring seal made of aluminium alloy (Al-Mg-Si) for vacuum sealing. Flanges are required for the two beam pipes, input coupler port, fundamental power pick-up port, and two HOM pick-up ports. The surfaces of the flanges which meet the seal should be machined and polished to a very smooth surface finish. The edges of the hexagonal seal which meet the flange surface need to be sharp and firm.

### 3.3.5 Plug-compatible design

In order to allow various designs for sub-components to work together in the same cryomodule, a set of interface definitions have been internationally agreed upon [26]. To date, the interface definitions cover: the cavity (Section 3.2); fundamental-mode coupler (Section 3.3.1); mechanical tuner and helium tank (Section 3.3.2 and Section 3.3.4, respectively).

#### 3.3.5.1 Cavity resonator

The boundary and the interfaces for the cavity are defined as shown in Fig. 3.11. The length of a cavity as measured from the surface of the two beam-port flanges is 1247 mm. The beam-port flanges are DN78, and a DN40 flange is used for the input coupler.

#### 3.3.5.2 Helium tank

The interfaces of the helium tank are defined by the four roller-pad supports and two ends of the 2-phase-helium supply pipe as shown in Fig. 3.12. The end finish of the 2-phase-helium supply pipe has to have a weld-ready finish for connection of the titanium bellows. The roller-pad supports have to have smooth surfaces, compatible with the needle-support assembly.

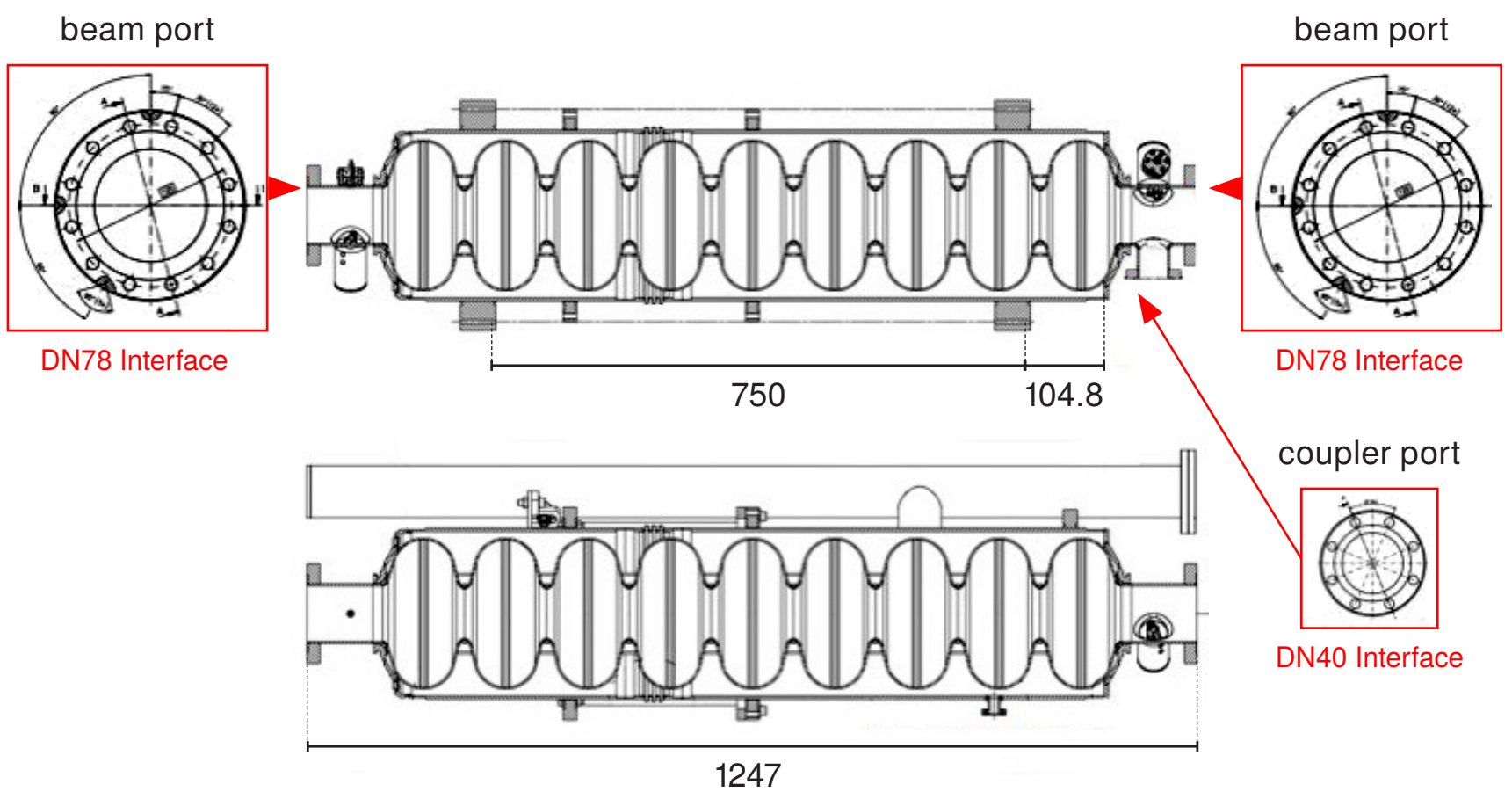

**Figure 3.11** Interface definition of the cavity.

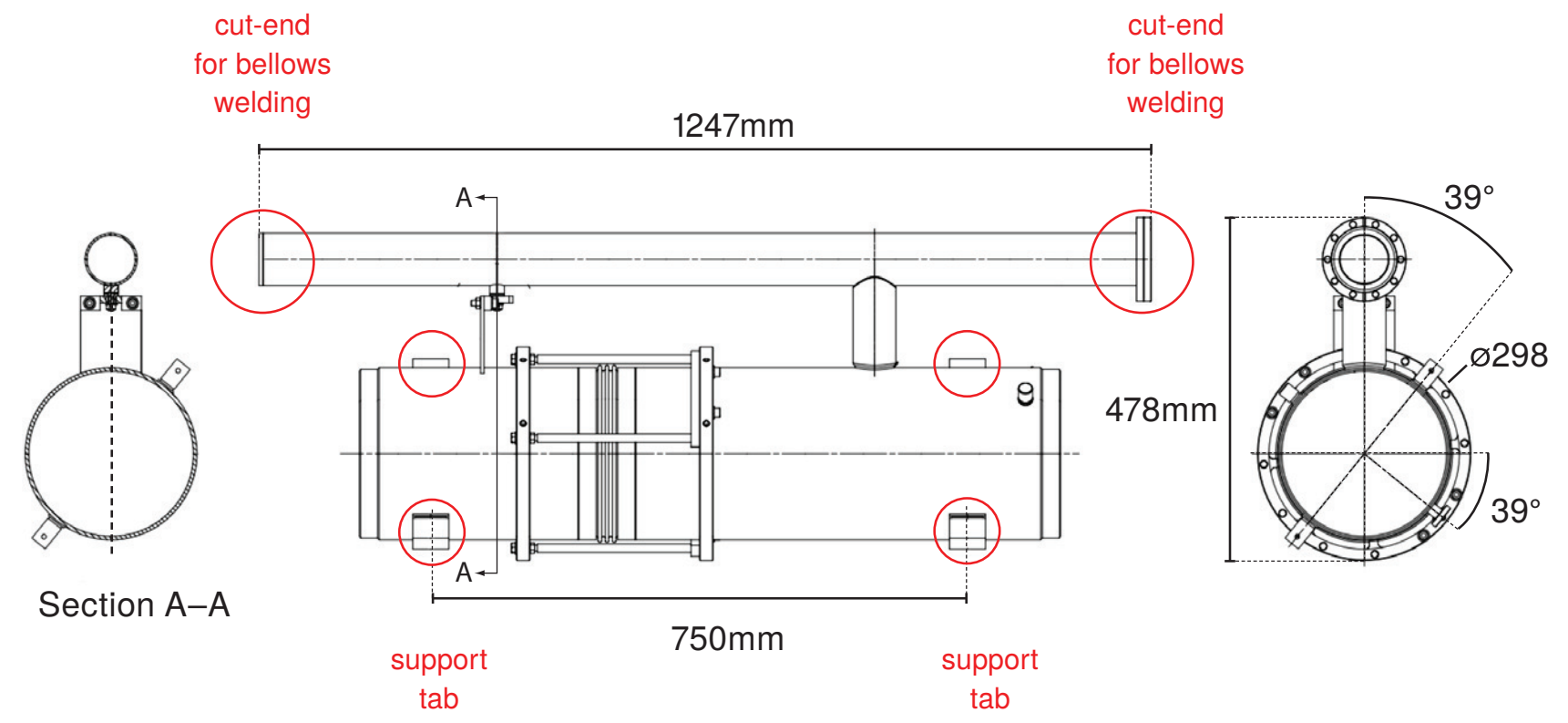

**Figure 3.12** Interface definition of the helium jacket.

#### 3.3.5.3 Fundamental-mode input coupler

The interfaces of the input coupler are defined by the cavity coupler port, cryomodule coupler port and the rectangular waveguide port, as shown in Fig. 3.13. The cavity coupler port is a DN40 interface flange which uses aluminium-made hexagonal sealing. The cryomodule coupler port is a flange with an outer diameter of 260 mm which uses a DN200 O-ring seal.

## 3.4 Cryomodule design including quadrupoles

### 3.4.1 Overview

Cryomodules are the modular building blocks of the ILC superconducting main linacs, and need to fulfil the following main functions:

1. provide mechanical support for beamline elements such as cavities and focussing elements;
2. facilitate achievement of the necessary alignment tolerance and stability according to beam dynamics specifications;
3. create and maintain in an efficient way the cold environment needed for the cavity and magnet operation.

The cryomodules represent the major heat loads at LHe temperatures, and therefore play an important role in the overall cryogenic system optimisation (See Section 3.5).

The highly-integrated design concept for the cryogenic systems leading to a high filling factor and reduced overall cost has been introduced in Section 3.1. In particular the concept of the use of single large cryoplants to cool kilometre-long *cryo-units* (similar to the LHC). Shorter *cryo-strings* are required to achieve segmentation of the insulating vacuum and of the two-phase-helium line.

Each of the 12.652 m-long cryomodules contains either nine cavities (Type A), or eight cavities and one superconducting quadrupole package (including horizontal and vertical dipole correctors and a BPM) located at the centre of the cryomodule (Type B). The cavities and quadrupole package are integrated into the cryomodules along with their supporting structures, thermal shields and insulation, and all of the associated cryogenic piping required for the coolant flow distribution along a cryogenic unit without the need for additional external cryogenic distribution lines.

**Figure 3.13**
Interface definition of the input coupler.

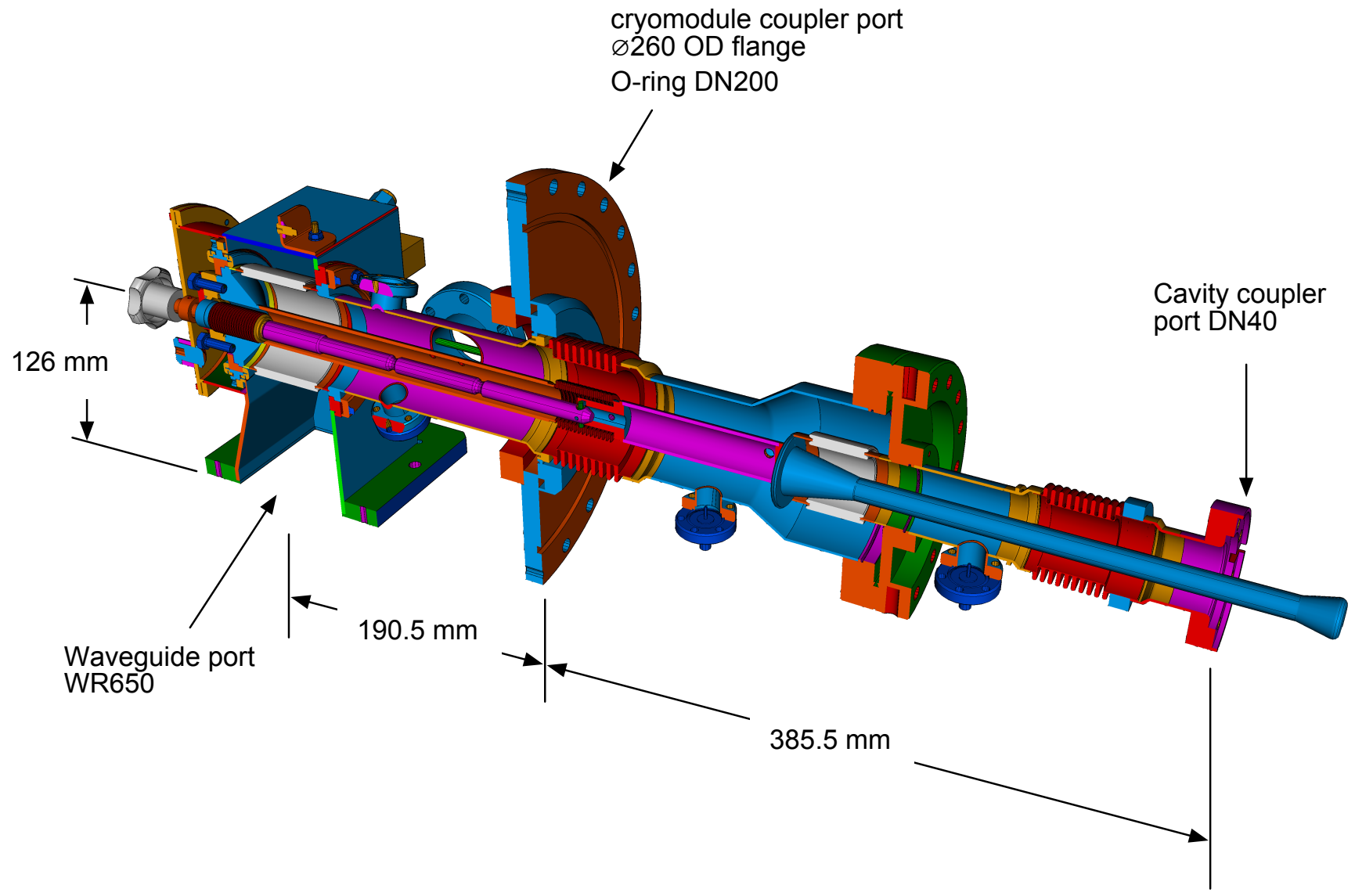

All the 14,742 1.3 GHz cavities in the ILC main linacs are grouped into 1,701 cryomodules (1,134 Type A, and 567 Type B). Another 152 cryomodules are located in the $e^+$ and $e^-$ sources and RTML bunch compressors. Most of these are either the standard Type A or Type B cryomodules, although the sources contain a few with special configurations of cavities and quadrupoles.

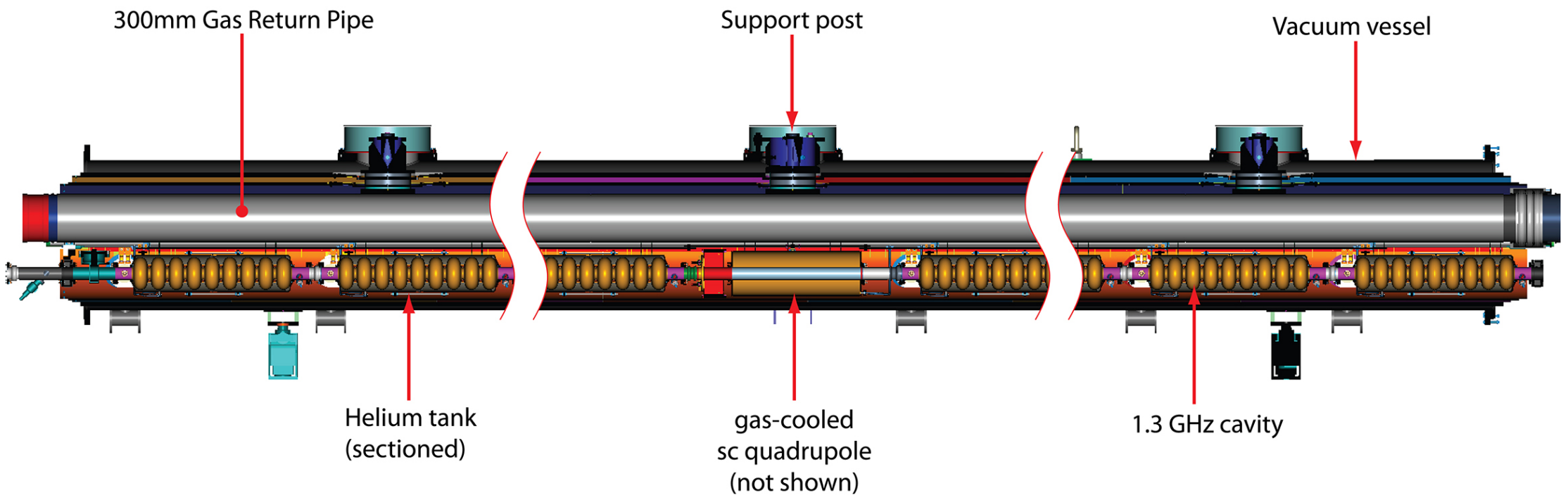

**Figure 3.14.** Longitudinal View of a Type IV Cryomodule (Type B), with eight cavities and a central quadrupole.

### 3.4.2 Cryomodule technical description

Figure 3.14 shows a longitudinal sectioned view of the Type-IV cryomodule (Type B). The design is a modification of the type developed and used in the TESLA Test Facility (TTF) at DESY, with three separate vacuum envelopes (beam vacuum, isolation vacuum and power-coupler vacuum) [27]. The cavity spacing within the cryomodule is $(6 - 1/4)\lambda_0 = 1.327$ m.

Copper-coated flanged bellows are located between beamline components to allow differential thermal contractions. Fundamental-mode couplers (described in Section 3.3.1) provide the RF power to the cavities, and are connected to ports on the vacuum vessel on one side and to the cavity coupler ports on the opposite side. RF cables bring the signals from the field pickup and the HOM antennas (See Section 3.3.3) to the LLRF control system outside the cryomodule for the control of the cavity field amplitude and phase and to extract HOM power from the 2 K level. Manually operated valves required by the clean-room assembly terminate the beam pipe at both module ends. The valves are fitted with simple RF shields.

The decision to place the quadrupole package in the middle of the cryomodule (as in the Type IV design) allows the definition of a standard interconnection interface for all main-linac cryomodules, irrespective of their sub-type, simplifying the tunnel assembly procedures for module connections.

#### 3.4.2.1 The cryomodule cross section

Figure 3.15 shows a cross section of the Type IV ILC Cryomodule derived from the TTF-III design [13, 28]. The largest component of the transverse cross section is the 300 mm-diameter helium-gas return pipe (GRP) which acts as the structural backbone for supporting the string of beamline elements and allows recovery of the mass flow of He vapours at a negligible pressure drop along the cryo-strings, to preserve temperature stability.

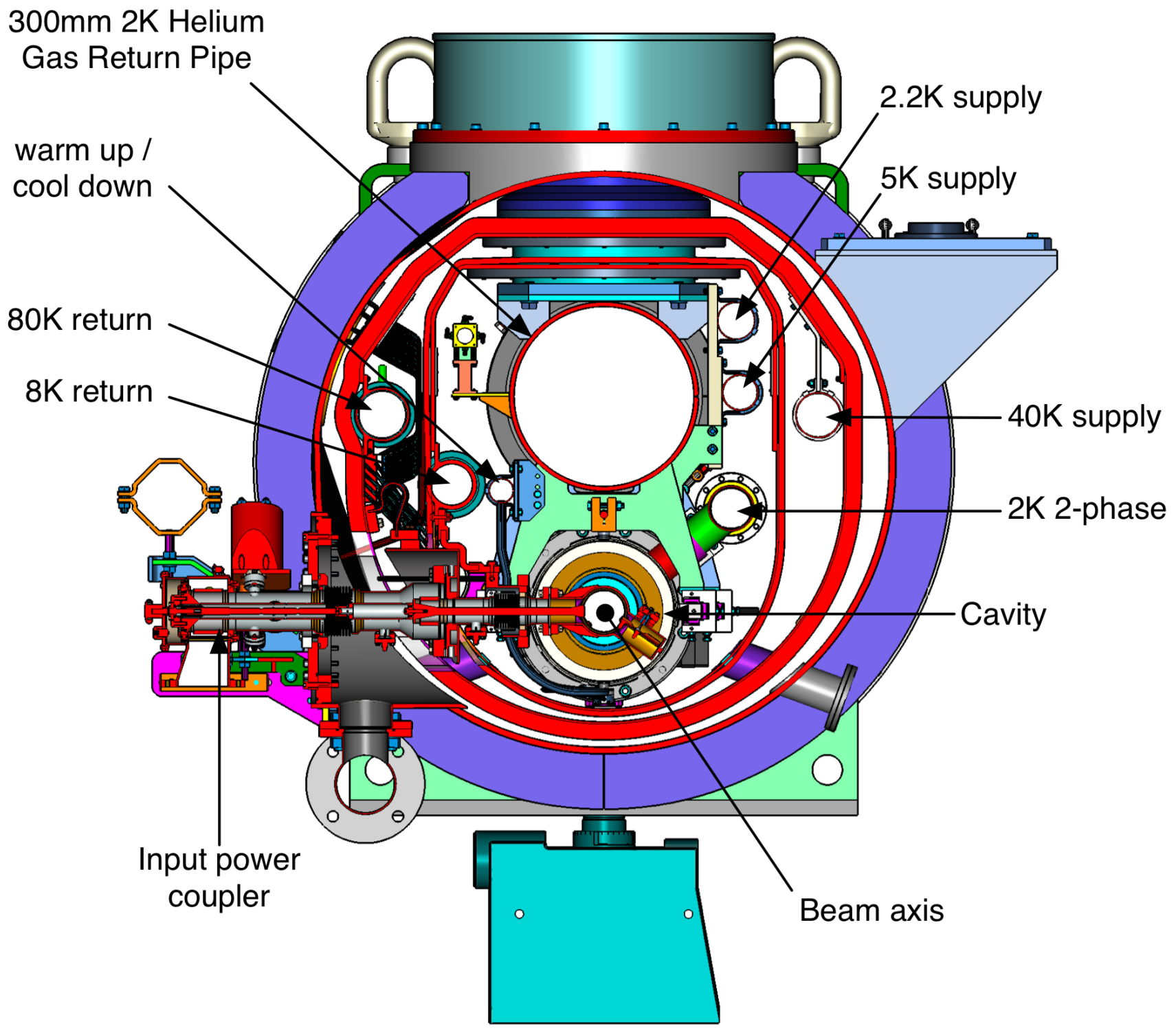

**Figure 3.15** Representative cryomodule cross section

The GRP is supported from the top by three composite posts with small thermal conduction from the room-temperature environment. The posts are connected to adjustable suspension brackets resting on large flanges placed on the upper part of the vacuum vessel. This suspension scheme allows the correct alignment of the axis of the cavities and quadrupole magnets independently of the flange position, without requiring expensive precision machining of these vacuum-vessel components. The centre post is fixed to the vacuum vessel, while the two remaining posts are laterally adjustable and can slide on the flanges to allow the GRP longitudinal contraction/expansion with respect to the vacuum vessel during thermal cycling. Each post consists of a fibreglass pipe terminated by two shrink-fit stainless-steel flanges. Two additional shrink-fit aluminium flanges are provided to allow intermediate heat flow intercept connections to the 5–8 K and 40–80 K thermal shields; the exact location of these flanges has been optimised to minimise heat leakage [29].

#### 3.4.2.2 Transverse / longitudinal cavity positioning and alignment

The cavities and magnet are supported from the GRP by means of stainless-steel brackets holding the four titanium pads on the helium tanks via a longitudinal sliding mechanism, which also provides adjusting screws and pushers for alignment in the transverse (vertical-horizontal) planes. During the module assembly, while the GRP is suspended on the assembly jig before insertion into the vacuum vessel, the beamline components are aligned and the alignment information is first transferred to references points on the GRP, and later transferred to reference points of the vessel for the cryomodule alignment, in order to achieve the installation alignment errors given in Table 3.4 (Section 3.1.3). All TTF cryomodules have been equipped with stretched-wire sensors to monitor the cold-mass displacement and positional reproducibility between cool downs [15] to qualify the alignment procedure, which will be used for the European XFEL.

A mechanical, coaxial (blade) and a piezo-electric tuner are mounted on the cavity vessels.

During cool down, the two ends of the ∼12 m-long GRP move by up to 18 mm toward the centre of the module. The cavity sliding support allows the cavity position to completely decouple from the large GRP contraction induced by the cool down, and avoids large stresses acting on the cavities due to differential shrinkage. To maintain the longitudinal position of the cavity-coupler flange within 1 mm from the coupler port on the warm vacuum vessel—in order to limit large coupler movements occurring with differential contraction—each cavity is clamped to a long invar rod, which is in turn longitudinally anchored at the neutral fixed point of the GRP at the centre post.

The beam-pipe interconnection between the cryomodules consists of a 0.38 m-long section between the end valves that incorporates a HOM absorber (similar to the XFEL design [30]), a bellows, and a vacuum pumping port; the latter is connected to a flange in the vacuum vessel every ninth cryomodule.

#### 3.4.2.3 Thermal radiation shields

The cryostat includes two aluminium thermal-radiation shields operating in the temperature range of 5–8 K and 40–80 K, respectively [31]. The use of a double thermal-radiation shielding reduces the radiative thermal load at 2 K to a negligible amount. Each shield is constructed from a stiff upper part, and multiple lower sections (according to the number of the cold active components, e.g. cavities, magnets). The upper part is supported by the intermediate flanges on the fibreglass posts, constrained at the centre post but slides on the two lateral posts to which they are still thermally connected. The 'finger-welding' technique [31] is used both to connect each thermal shield to its properly shaped aluminium cooling pipe, and the lower-shield parts to the upper ones, by providing good thermal conduction without inducing high stresses on the structure.

Blankets of multi-layer insulation (MLI) are placed on the outside of the 5–8 K and the 40–80 K shields. The 5–8 K shield blanket is made of 10 layers of doubly aluminised mylar separated by insulating spacers while the 40–80 K blanket contains 30 layers. In addition, helium jackets for cavity and magnet packages, gas return pipe and 5–8 K pipes are wrapped with 5 layers of MLI as a mitigating provision to reduce heat transfer in the event of a vacuum failure.

#### 3.4.2.4 The vacuum vessel

The cryostat outer vacuum vessel is constructed from carbon steel and has a standard outer diameter of 38″. Adjacent vacuum vessels are connected to each other by means of a flanged cylindrical sleeve with a bellows. Adjacent vessels have a flange-to-flange distance of 0.85 m, allowing sufficient space to perform the cryogenic connections between modules by means of automated orbital welders. In the event of accidental spills of liquid helium from the cavity vessels, a relief valve on the main-vessel body together with venting holes on the shields prevent excessive pressure build-up in the vacuum vessel. Wires and cables from each module are extracted from the module using metallic sealed flanges with vacuum-tight connectors. The insulating-vacuum system is pumped during normal operation by permanent pump stations located at appropriate intervals. Additional pumping ports are available for movable pump stations, which are used for initial pump down, and in the event of a helium leak. The RF-power coupler needs an additional vacuum system on its room temperature side; this is provided by a common pump line for all couplers in a module, which is equipped with an ion getter and a titanium sublimation pump.

#### 3.4.2.5 Cryogenic lines in the module

The following helium lines [32] are integrated into the cryomodules, as shown in Fig. 3.15.

- The 2 K supply line transfers pressurised single-phase helium through the cryomodule to the end of the cryogenic unit.
- The titanium 2 K two-phase supply line is connected to the cavity helium vessels. It supplies the cavities with liquid helium and returns cold gas to the 300 mm GRP at each module interconnection.
- The 2 K GRP returns the cold gas pumped off the saturated He II baths to the refrigeration plant. It is also a key structural component of the cryomodule
- The 5–8 K supply and return lines. The 5 K supply line is used to transfer the He gas to the end of the cryogenic unit. The 5–8 K return line directly cools the 5–8 K radiation shield and, through the shield, provides the heat-flow intercept for the main coupler and diagnostic cables, and the HOM absorber located in the module interconnection region.
- The 40–80 K supply and return lines. The 40 K supply line is used to transfer He gas to the cryogenic unit end and cools the high-temperature superconductor (HTS) current leads for the quadrupole and correction magnets. The 40–80 K return line directly cools the 40–80 K radiation shield and the HOM absorber and, through the shield, provides an additional heat-flow intercept for the main coupler and diagnostic cables.
- The warm-up/cool-down line connects to the bottom of each cavity helium vessel.

The helium vessels surrounding the cavities, the two-phase supply line and the GRP operate at low-pressure conditions (30 mbar, corresponding to 2 K) while all other cryogenic lines operate at a maximum pressure of 20 bar.

To provide sufficient cooling speed during cool down, the low-pressure lines around the cavities need to sustain a Maximum-Allowable Working Pressure (MAWP) of 2 bar differential, at room temperature. All components in the cryogenic system sustaining pressure conditions need to be assessed for pressure-code conformance [33], as discussed in Section 3.5.

The helium lines of adjacent modules are welded at the module interconnection regions. Only the vacuum flange incorporates a mechanical seal at the cryomodule interconnect. Thermal-shield lines are extruded aluminium with transition joints to stainless steel (similar to those used in the HERA magnets and in TTF and XFEL) at each interconnect, allowing the use of stainless-steel bellows. Similarly, the titanium two-phase line has transition joints to stainless steel in the interconnection region. The cryostat maintains the cavities and magnets at their operating temperature of 2 K.

#### 3.4.2.6 Thermal design and module-heat-loss estimations

A low static heat load is an essential feature required of the cryostat design; the total heat load is dominated by the RF losses, and is thus principally determined by the cavity performance (and its spread). Table 3.9 lists the heat load assumed per cryomodule. The table reports the average values corresponding to one Main Linac unit (ML unit), i.e. three modules in a Type A – Type B – Type A configuration. The values reported here are based on the heat load of a 12-cavity cryomodule which has been calculated for the TESLA TDR [13], and refinement made on the basis of further assessments and static-load measurements obtained from S1-Global (Part I Section 2.7.1) and for the European XFEL prototypes. To scale to the ILC parameters, it is assumed that the gradient is 31.5 MV/m, the cavity $Q_0$ is $1 \times 10^{10}$, and the beam and RF parameters are those listed in Table 3.1 in Section 3.1. These values are used to define cryogenic heat loads and cryoplant parameters for the two variants of the cryogenic systems for the flat and mountainous topography respectively [34].

**Table 3.9**
Average heat loads per module in a ML unit, for the baseline parameter in Table 3.1. All values are in watts [27].

| | 2 K | | 5–8 K | | 40–80 K | |
|---|---|---|---|---|---|---|
| | Static | Dynamic | Static | Dynamic | Static | Dynamic |
| RF Load | | 8.02 | | | | |
| Radiation Load | | | 1.41 | | 32.49 | |
| Supports | 0.60 | | 2.40 | | 18.0 | |
| Input coupler | 0.17 | 0.41 | 1.73 | 3.06 | 16.47 | 41.78 |
| HOM coupler (cables) | 0.01 | 0.12 | 0.29 | 1.17 | 1.84 | 5.8 |
| HOM absorber | 0.14 | 0.01 | 3.13 | 0.36 | -3.27 | 7.09 |
| Beam tube bellows | | 0.39 | | | | |
| Current leads | 0.28 | 0.28 | 0.47 | 0.47 | 4.13 | 4.13 |
| HOM to structure | | 0.56 | | | | |
| Coax cable (4) | 0.05 | | | | | |
| Instrumentation taps | 0.07 | | | | | |
| Diagnostic cable | | | 1.39 | | 5.38 | |
| Sum | 1.32 | 9.79 | 10.82 | 5.05 | 75.04 | 58.80 |
| Total | 11.11 | | 15.87 | | 133.84 | |

Frequencies above the 1.3 GHz fundamental-mode operating frequency and below the beam-pipe cutoff are extracted by input- and HOM-couplers (in order to avoid additional power deposition at cold temperatures), but higher-frequency fields will propagate along the structure and be reflected at normal and superconducting surfaces. In order to reduce the losses at normal conducting surfaces at 2 K and 4 K, the cryomodule includes a special HOM absorber that operates at 70 K, where the cooling efficiency is much higher. The absorber basically consists of a pipe of absorbing material mounted in a cavity-like shielding, and integrated into the connection between two modules. As the inner surface area of this absorber (about 280 cm$^2$) is small compared to that of all the normal conductors in one cryomodule, the absorber has to absorb a significant part of all the RF power incident upon it. In field propagation studies, which assume a gas-like behaviour for photons, it has been shown that an absorber with a reflectivity below 50 % is sufficient [35, 36]. Theoretical and experimental studies indicate that the required absorption can be obtained with ceramics like MACOR or with artificial dielectrics. Figure 3.16 shows the design for the design implemented for the European XFEL, which has been successfully tested at FLASH [37]. The results show very good agreement with the theoretical predictions.

It is worth noting here that a substantial effort has been performed during the Technical Design Phase for the S1-Global module and for the European XFEL Project in the consolidation and benchmarking of the static heat-load assessments, as reported in Part I Section 2.7. The S1-Global measurements show a very good consistency with heat-load estimations when all conduction paths and heat-transfer mechanisms are taken properly into account in the budget, indicating that the module design is well understood and proven [38]. Values for the static loads in Table 3.9 are consistent with the experience gained during the Technical Design phase; the low estimates for static losses reflect the assumption of the reduced diagnostic instrumentation foreseen for the ILC modules with respect to R&D activities such as the S1-Global module tests.

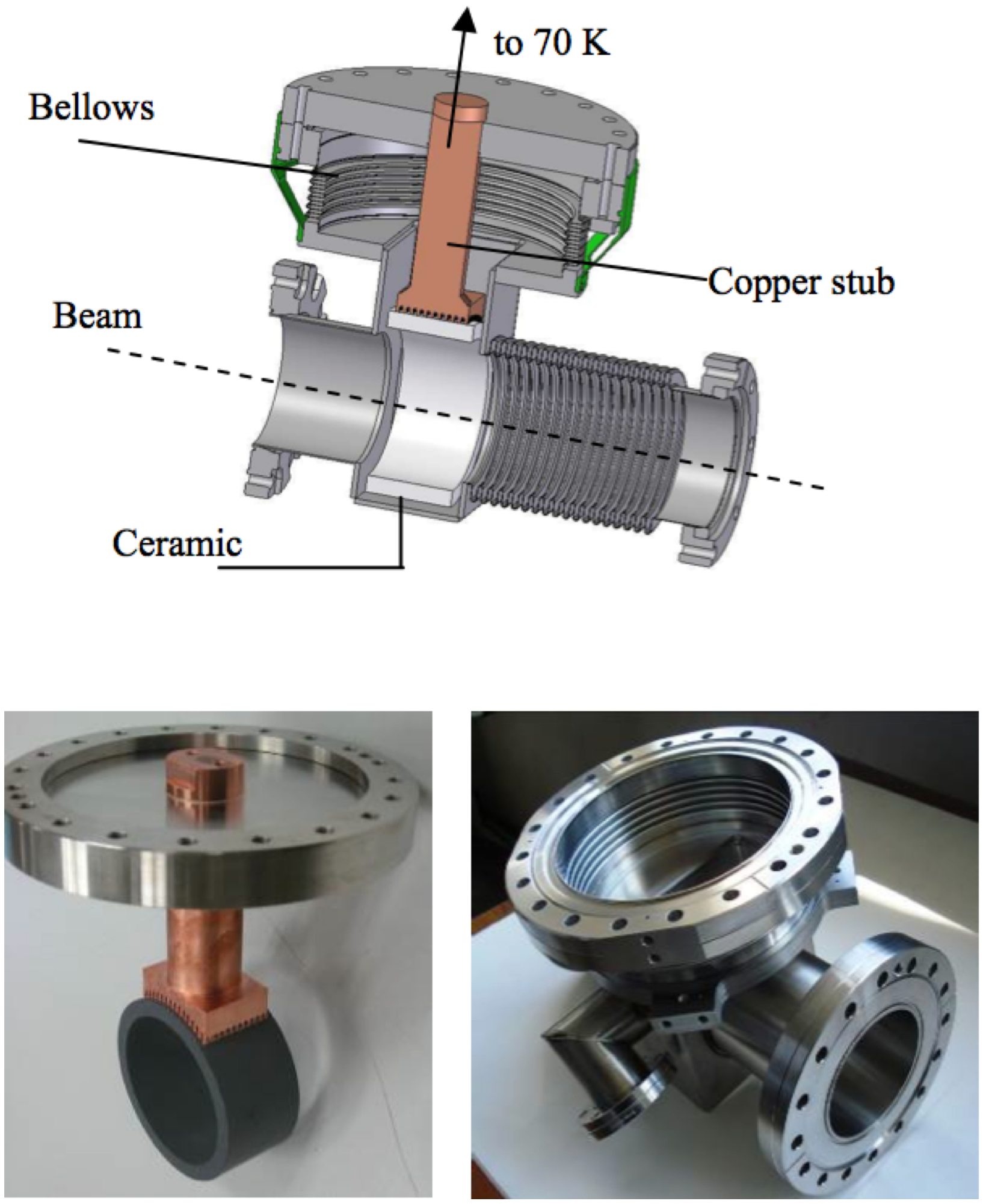

**Figure 3.16**
The design (top) and photographs (bottom left and right) of the HOM absorber for the European XFEL [37].

The experience reported with the European XFEL prototypes has highlighted the importance of the assembly procedures in achieving nominal loads, and "training" effects for the most sensitive 2 K environment [39] (Part I Fig. 2.50).

As a final remark on thermal loads, it must be noted from Table 3.9 that dynamic loads induced by RF are dominant in the 2 K region, and are intrinsically influenced by the spread of cavity performances ($Q_0$ values) and operating point (gradient setting). Much less experience and data is available on dynamic loads, and uncertainty factors need to be taken into account (see Section 3.5).

#### 3.4.2.7 Quadrupole/Corrector/BPM Package

The baseline design for the ILC quadrupole/corrector/BPM package makes use of the conduction-cooled splittable quadrupole [40, 41] developed by FNAL and KEK (Part I Section 2.7). Figure 3.17 shows the magnet assembly; specifications are given in Table 3.10.

A key specification is the magnetic-centre stability of $< 5\,\mu$m for a 20 % change in field strength, which is driven by beam-dynamics requirements (beam-based alignment). Because of superconductor magnetisation effects (cross coupling) between combined quadrupole and dipole coils [42, 43], the quadrupole and dipole correctors are separated.

The split-quadrupole is installed outside of the clean room around a beam pipe, thus decreasing possible contamination of the cavity RF surfaces, and greatly simplifying the string-assembly operation in the clean room.

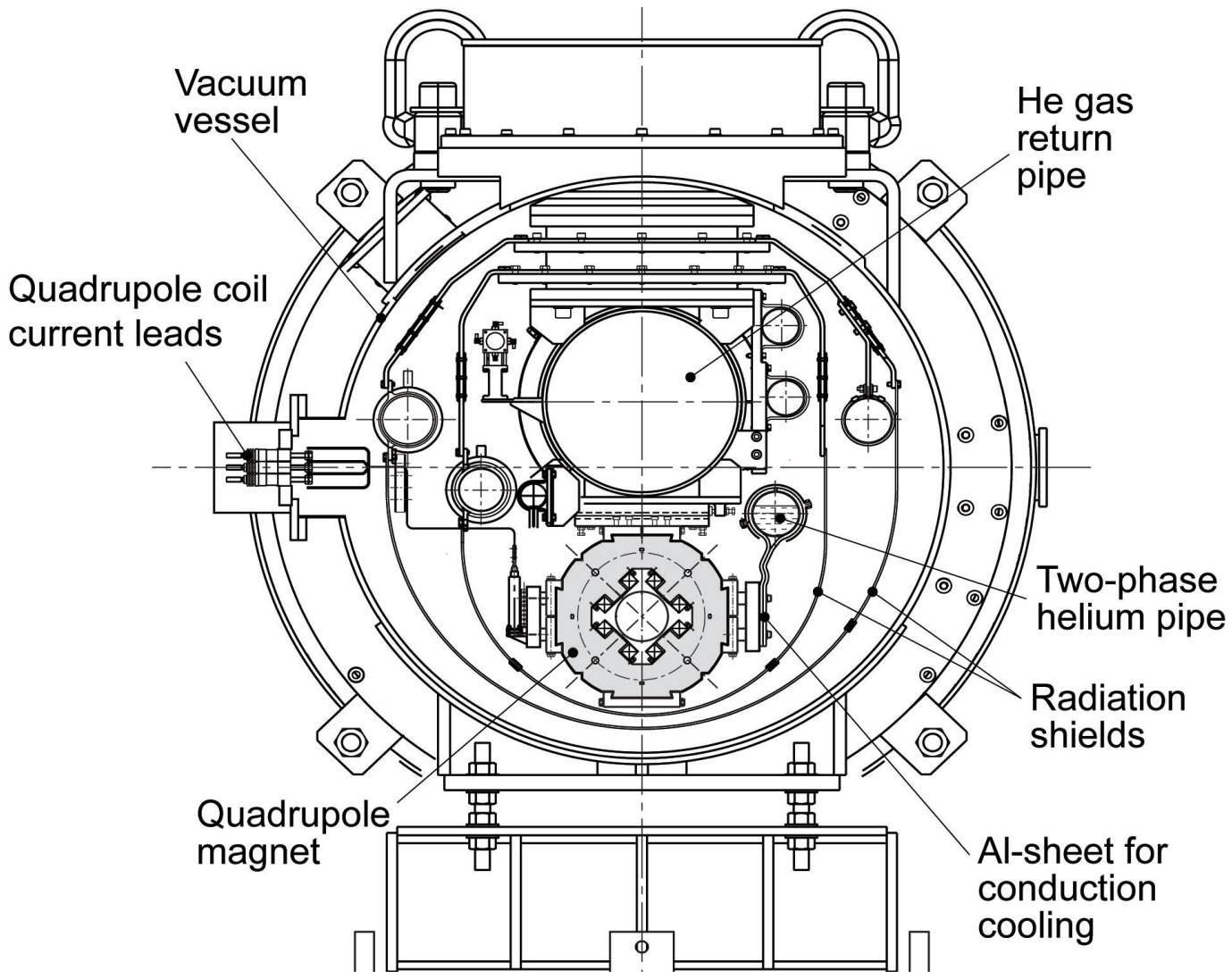

**Figure 3.17**
Cross section of a Type-B cryomodule showing the arrangement of the conduction-cooled split-yoke superconducting quadrupole.

**Table 3.10**
Splittable quadrupole magnet specifications and parameters.

| Parameter | Value | Unit |
|---|---|---|
| Peak gradient | 54 | T/m |
| Peak integrated gradient | 36 | T |
| Field non-linearity at 5 mm radius | 0.05 | % |
| Dipole trim coils integrated strength | 0.075 | Tm |
| Aperture | 78 | mm |
| Pole-to-pole distance | 90 | mm |
| Magnetic stability (20 % field change) | $< 5$ | μm |
| Peak operating quadrupole current | 100 | A |
| Magnet total length | 680 | mm |
| SC wire diameter | 0.5 | mm |
| NbTi filament size (vendor value) | 3.7 | μm |
| Cu:SC volume ratio | 1.5 | |
| Superconductor Critical current ( 5 T and 4.2 K) | 200 | A |
| Coil maximum field at 100 A current | 3.3 | T |
| Magnetic field stored energy | 40 | kJ |
| Quadrupole inductance | 3.9 | H |
| Quadrupole coil number of turns/pole | 900 | |
| Yoke outer diameter | 280 | mm |

An important feature that must be addressed with the final engineering design is the package fiducialisation and subsequent transfer of these features to reproducible, external cryomodule fiducials to assure the correct alignment of the package with respect to the cryomodule string.

The accelerator lattice[3] (Section 3.1.3) has approximately constant focusing and FoDo cell length from 5 GeV up to the full beam energy of 250 GeV —a factor of 25 in beam rigidity. Hence it is not possible to use the identical effective magnet length for all magnets along the accelerator, since the lower-energy quadrupoles would then need to be run at fields (currents) which are too low to provide stable and reproducible performance. Two families of quadrupoles are considered sufficient to resolve this problem: above the nominal 25 GeV point, the long (high integrated field) magnet described above will work adequately. Below 25 GeV a shorter version will be used.

The L-band re-entrant BPM for the main linac is designed to effectively pick up the dipole TM110 mode through four symmetrically arranged waveguides. The dipole-mode frequency of 2.04 GHz is selected to avoid the 1.3 GHz and higher harmonics dark-current signals, and to avoid the cavity HOM frequencies. The re-entrant cavity is chosen for its compact size, and its compatibility with the 78 mm inner diameter of beam pipe. Four waveguides with loop pickups give more than 28 dB isolation from the common-mode excitation. The design L-band re-entrant BPM structure are shown in Fig. 3.18. A prototype model (Fig. 3.19) has been fabricated and tested with beam, and demonstrated a position resolution of 0.3 μm [44].

[3] including the RTML bunch compressors and Main Linac.

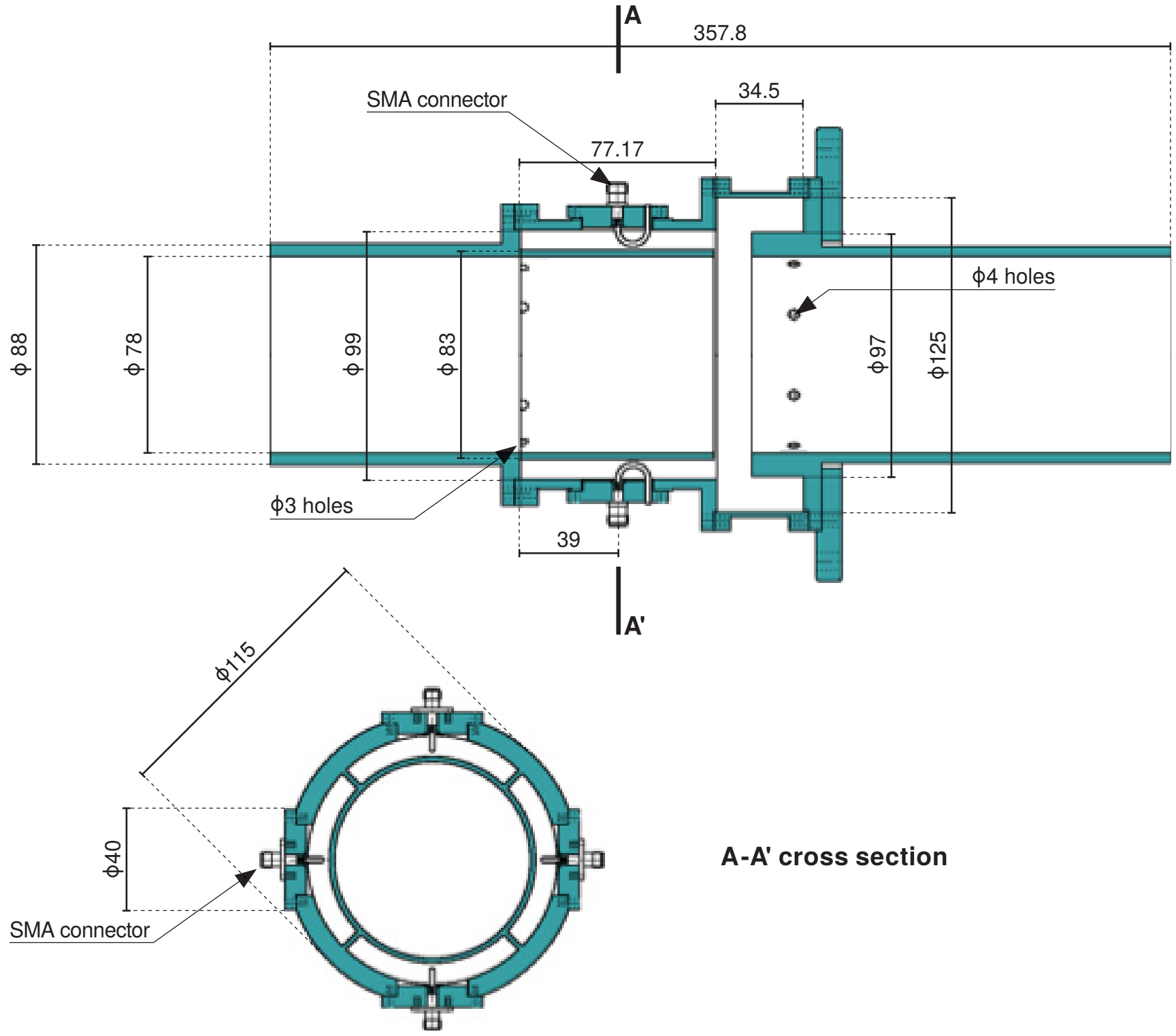

**Figure 3.18**
Schematic of the cold 2.04 GHz re-entrant cavity BPM.

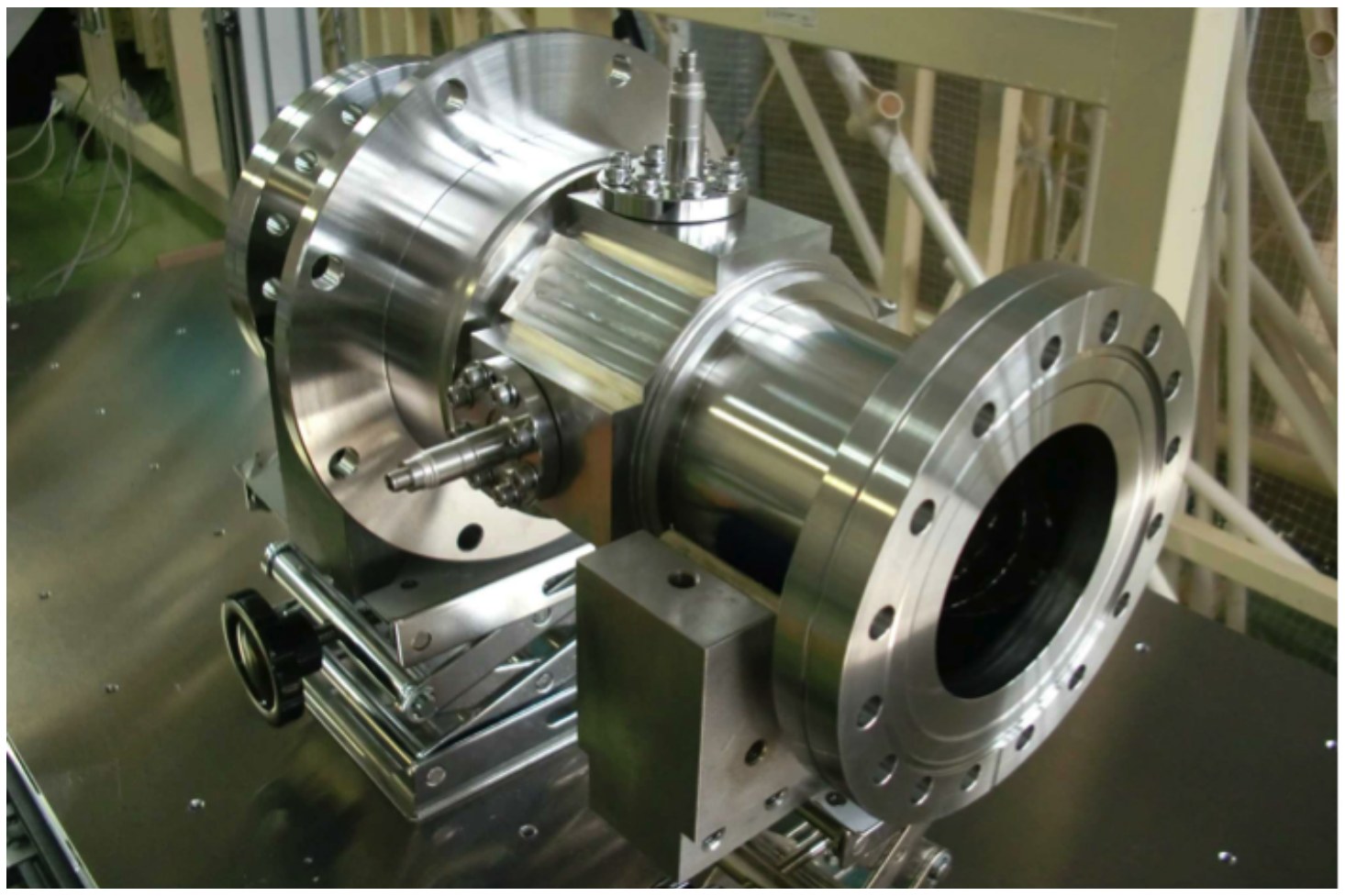

**Figure 3.19**
A photograph of the vacuum-tight prototype for the 2.04 GHz re-entrant cavity BPM.

## 3.4.3 Cryomodule testing

Before installation in the ML tunnel, assembled accelerator cryomodules will be qualified through sampled testing. The qualification includes checking mechanical fit, measuring cryogenic performance, and testing cavity systems at full power up to full voltage capability. A sampled-testing strategy is effective because 100 % of all key individual components, (including cavities, cavity auxiliaries, quadrupoles, instrumentation and cryogenic subsystems), will be fully tested before assembly of the cryomodules, as described in Section 3.3. The random sample cryomodule testing program will be phased in during the ramp-up stage of cryomodule production, during which a greater fraction of cryomodules will be tested, such that a total of one third of the full main-linac cryomodule complement will be tested before installation. All cryomodules will be tested during ramp up at the beginning of production, in order to qualify the production lines; the fraction to be tested will be reduced as the production rate is increased.

The project plan, (see Section 13), calls for an 8 to 10 year construction, installation and commissioning schedule so the peak cryomodule production rate required to produce the total of 1853 cryomodules is 15 per week, roughly 15 times the European XFEL cryomodule production rate (see Part I Section 2.5) During the mass-production cycle, cryomodule-test facilities will require large cryogenic and high-power RF infrastructure. It is expected that these facilities will be hosted and run by collaborating institutes in all three regions, possibly together with a cryomodule-assembly infrastructure. While it is desirable to co-locate the assembly and test facilities, this may prove intractable, and so it is necessary to consider the need to transport the modules (Section 3.4.4). This is the case for the European XFEL, where the cryomodule assembly is hosted at CEA Saclay, Paris, France, while the cavity- and cryomodule-test facilities is located at DESY, Hamburg, Germany.

Figure 3.20 shows the layout of the European cryomodule test facility. Three independent concrete test bunkers, located in the center of the figure, are used to test an average of one cryomodule per week. The test procedure to be followed during the ILC main-linac construction project will largely follow that of the European XFEL and is expected to take fifteen days, including interconnect, warm processing, cool down, high power cold test, warm up, and disconnect. The tests themselves take 8 to 10 days, not including tests and qualification of the local power distribution. At present, cryomodule testing has been done at CERN, DESY, KEK, JLab, and Fermilab and it is reasonable to consider the use of the infrastructures at these collaborating institutes for the purpose of the ILC project, thereby avoiding the delay and cost of building new test facilities.

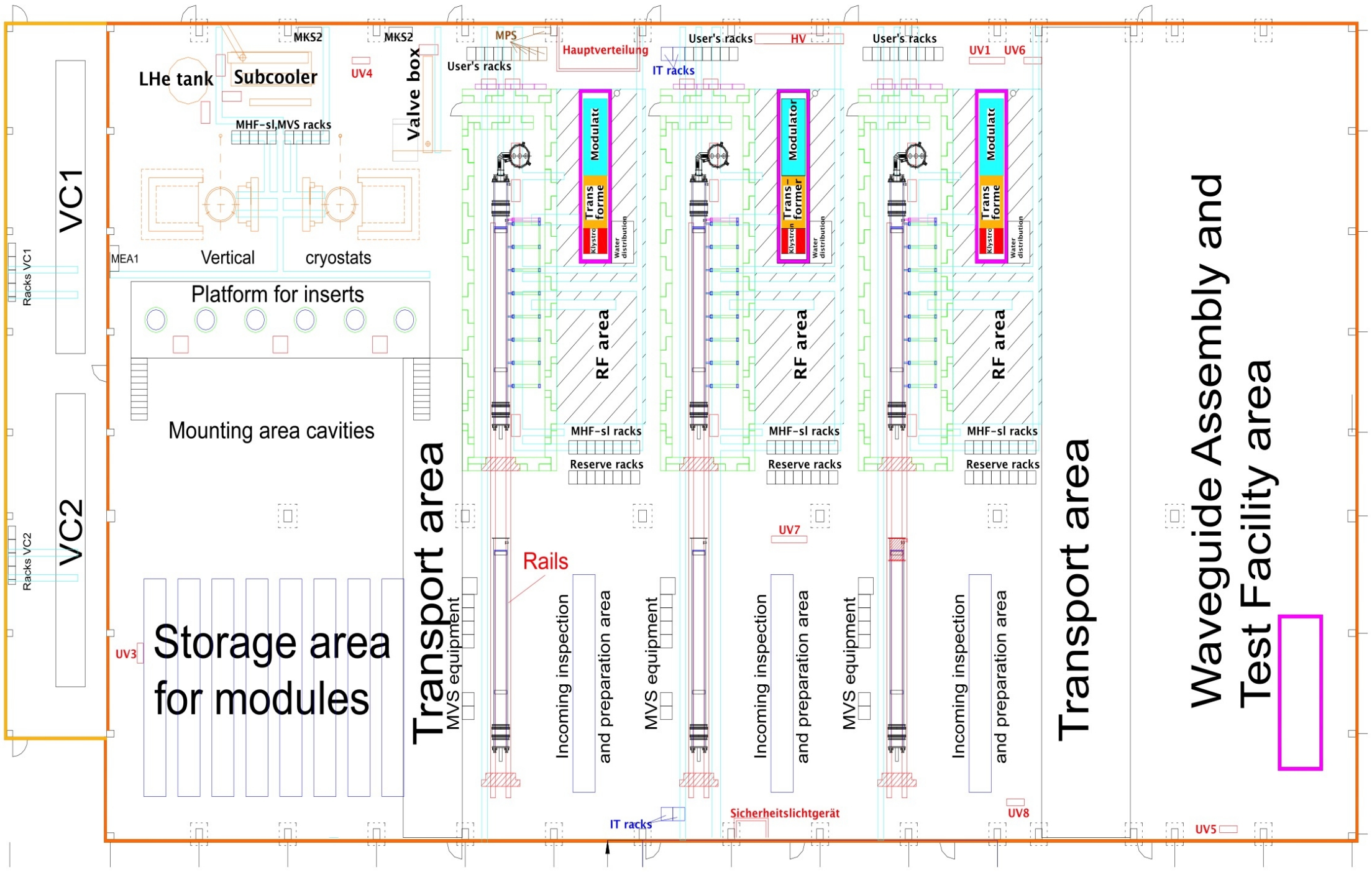

**Figure 3.20.** European XFEL cryomodule test facility Accelerator Module Test Facility, (AMTF), at DESY.

It is expected that a minimum of 15 cryomodule-test stands similar to those in the AMTF must be collectively made available through ILC collaborating institutes to satisfy the aggregate test rate requirement of five per week.

Figure 3.21 illustrates the test steps for cryomodules, starting with equipment hookup. The cryomodule is then placed inside a concrete-shielded test bay and connected to the various vacuum systems (cavity, coupler and insulation vacuum), cryogenic lines and RF wave-guide system. After pumping down the vacuum and performing necessary leak checks, warm RF processing of input couplers is done. The cryomodule is then cooled down to 2 K, and the RF characteristics of each of cavity are measured by feeding low-power RF through a coaxial-to-rectangular waveguide converter. The parameters of the couplers and waveguides are optimised in this condition, before starting high-power RF processing of the couplers, followed by a suite of tests of the cavities and cavity auxiliaries, (including tuner mechanism, piezo tuner, cold coupler and $Q_{\text{ext}}$ tuner), at full gradient.

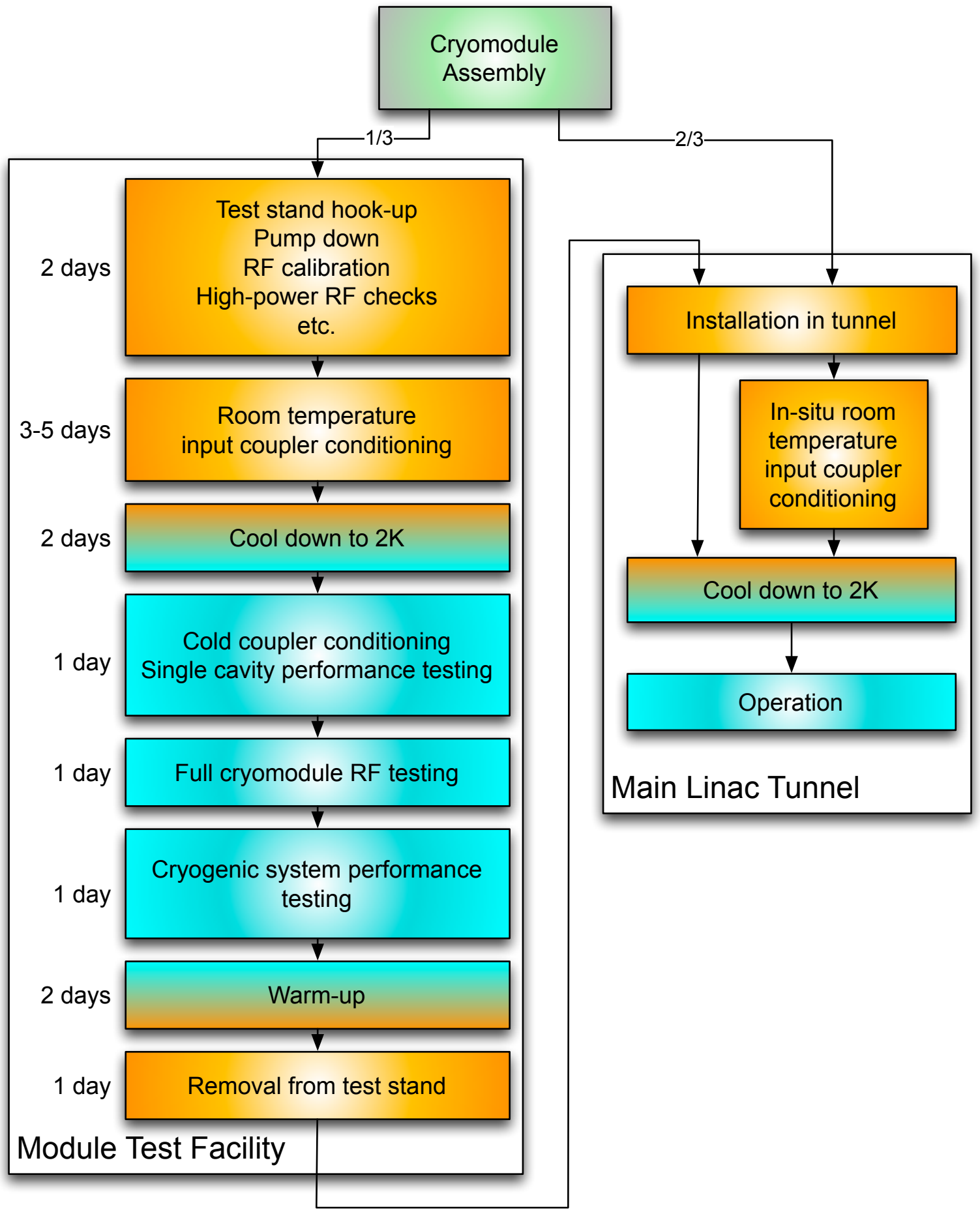

**Figure 3.21**
Flow chart of the cryomodule test. There are two test and processing streams, one third of the cryomodules are tested in the Module Test Facility and the other two thirds will be processed in situ after installation in the tunnel.

Validation of the cryomodule is done by confirming that all the cavities will be able to provide the specified field gradient of 31.5 MV/m $\pm$ 20 %, with acceptable dark current (field emission), after installation in the accelerator. After completing the tests, the cryomodule is warmed up with the tuner tension released.

Figure 3.21 also shows the test and processing procedures to be done following installation in the main-linac tunnel. The remaining two thirds of the cryomodules, those that were not tested in the test facility, are connected to the local power distribution system for coupler conditioning.

### 3.4.4 Shipping of cryomodules between regions

To date, there is limited experience on the shipping of completed cryomodules across the main regions of the ILC collaboration. FNAL shipped the complete ACC39[4] module for FLASH by air transport to DESY (see Fig. 3.22), where the module has been successfully tested up to its specification [45] .

By the end of 2015, the European XFEL will have gained the experience of transporting by road 100 complete modules from the string and module assembly facility at CEA/Saclay to the AMTF testing area at DESY [46], providing a useful statistical sample of data. Figure 3.23 shows an XFEL prototype cryomodule in its transport frame. It is essential that this XFEL experience be incorporated in the development of a reliable method for overseas transport of complete ILC modules.

[4] A special short module derived from the TTF design for the 3.9 GHz cavities

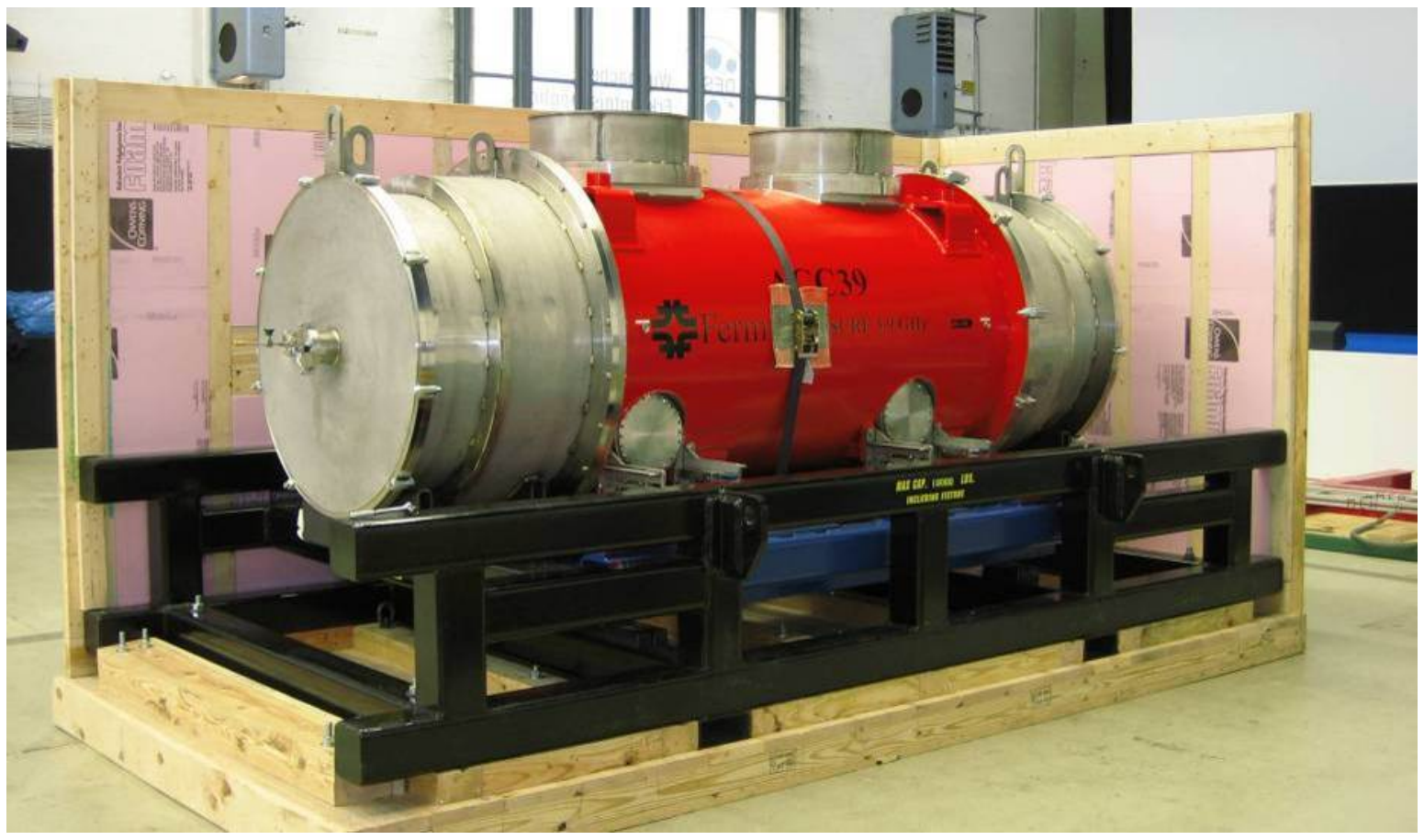

**Figure 3.22**
The ACC39 in its transport box upon arrival in DESY.

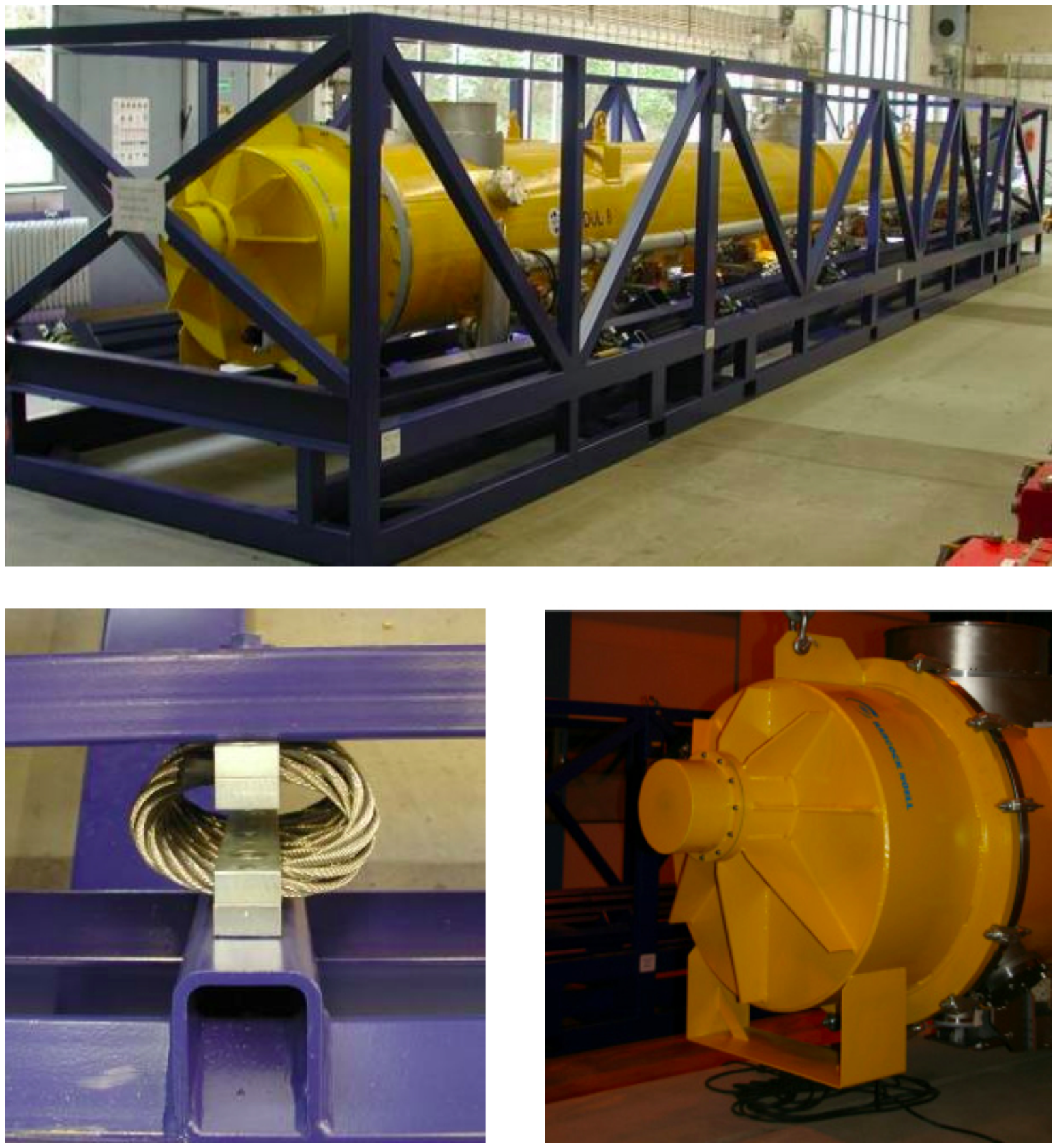

**Figure 3.23**
An European XFEL prototype cryomodule in its transport frame. The module is supported on vibration dampers [46]

## 3.5 Cryogenic cooling scheme

Of the total of 1853 SCRF cryomodules in the ILC, the 1701 Main Linac SCRF cryomodules (92 %) comprise the largest cryogenic cooling load and therefore dominate the design of the cryogenic systems. The 102 cryomodules (6 %) in the bunch compressors (RTML see Chapter 7) are considered extensions to the Main Linacs for the purposes of the cryogenic layout. For this reason, the cryogenic system is described in this chapter.

Figure 3.24 illustrates the cryogenic-system arrangement for ILC, which clearly shows the concept of long 2–2.5 km contiguous cryo-units, cooled by a single large 2 K cryoplant. There are detailed differences in the two site-dependent design variants under consideration, primarily driven by the choice of the RF-power scheme. The most important difference is the choice of a total of 10 cryoplants for the mountainous topography variant (using DKS), and 12 cryoplants for the flat topography (using KCS), as illustrated in Fig. 3.24. The total cryogenic load is however the same, but is distributed

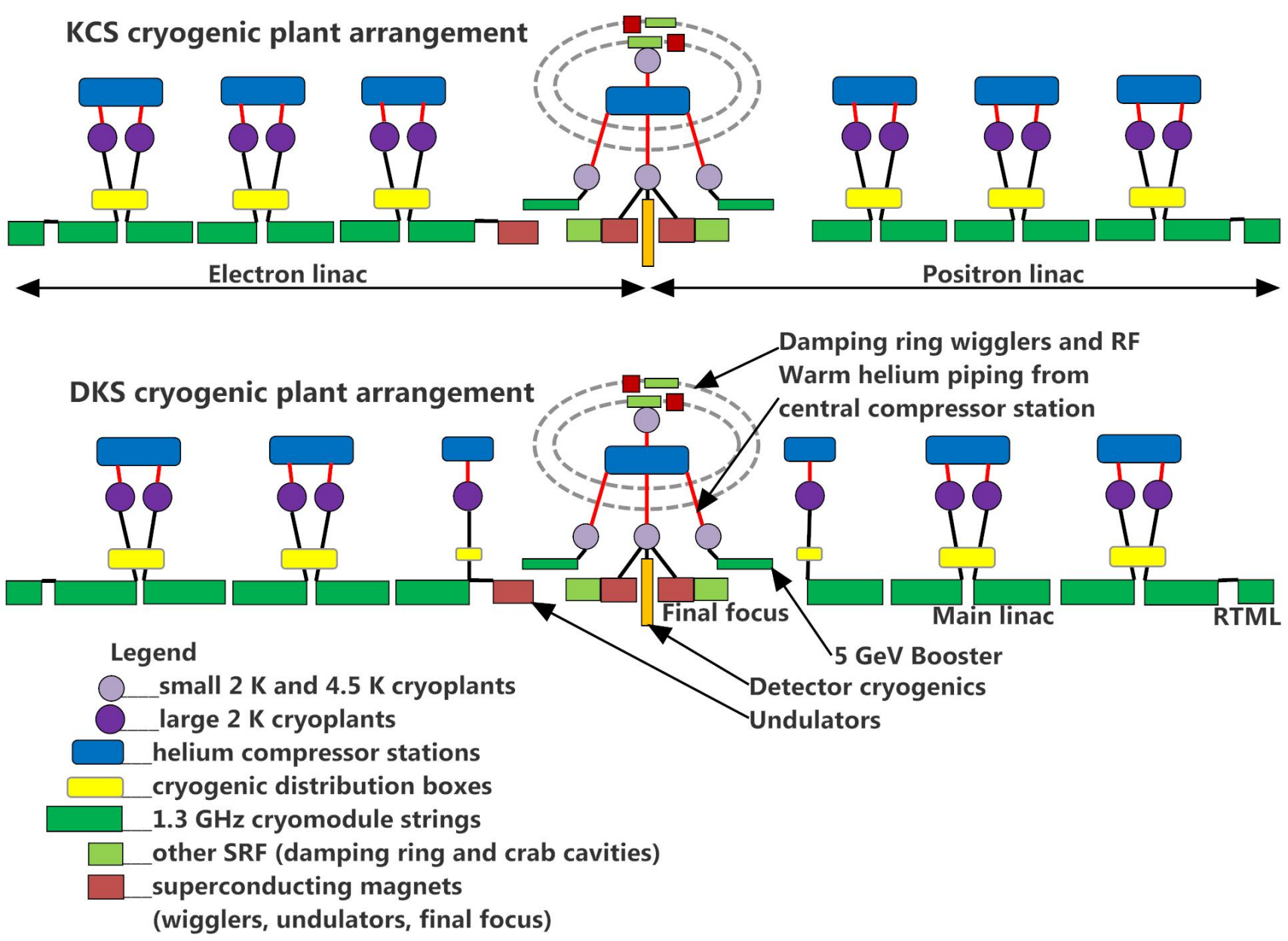

**Figure 3.24** The overall layout concept for the cryogenic systems for both flat (KCS) and mountain (DKS) topography.

differently between the plants for the two schemes.

The most upstream cryoplants in either variant also provide cooling for the 102 cryomodules in the RTML bunch compressors. The remaining loads for the systems are cooled by separate dedicated plants in the central region as shown.

## 3.5.1 Cryogenic cooling scheme for the main linacs

Saturated He II cools RF cavities at 2 K, and helium-gas-cooled shields intercept thermal radiation and thermal conduction at 5–8 K and at 40–80 K. A two-phase line (liquid-helium supply and concurrent vapour return) connects to each helium vessel and connects to the major gas return header once per module. A small diameter warm-up/cool-down line connects the bottoms of the helium vessels. (see Section 3.4 for more details.)

A sub-cooled helium supply line connects to the two-phase line via a Joule-Thomson valve once per cryo-string (9 modules or 12 modules for a short and long string respectively — see Fig. 3.2 in Section 3.1). The 5 K and 40 K heat intercepts and radiation screens are cooled in series through an entire cryogenic unit of up to 2.5 km in length. For the 2 K-cooling of the RF cavities, a parallel architecture is implemented providing parallel cooling of cryo-strings resulting in operational flexibility. Consequently, each cryo-unit is subdivided into about 13 to 21 cryo-strings, each of which corresponds to either 116 m or 154 m-long elementary blocks of the cryogenic refrigeration system, for short and long cryo-strings respectively.

Figure 3.25 shows the cooling scheme of a cryo-string, which contains 12 cryomodules (long string). The cavities are immersed in baths of saturated superfluid helium, gravity filled from a2 K two-phase header. Saturated superfluid helium flows along the two-phase header, which has phase separators located at both ends; the first phase separator is used to stabilise the saturated liquid produced during the final expansion. The second phase separator is used to recover the excess of liquid, which is vaporised by a heater. At the interconnection of each cryomodule, the two-phase header is connected to the pumping return line.

The division of the Main Linac into cryogenic units is driven by various plant size limits and a practical size for the low-pressure return pipe. A cryogenic plant of 25 kW equivalent 4.5 K capacity is a practical limit due to industrial production for heat-exchanger sizes and over-the-road shipping size restrictions. Cryomodule piping pressure drops also start to become rather large with more than 2.5 km distances. Practical plant size and gas-return header pressure drop limits are reached with 189 modules in a 21 short-string cryogenic unit, 2.5 km long.

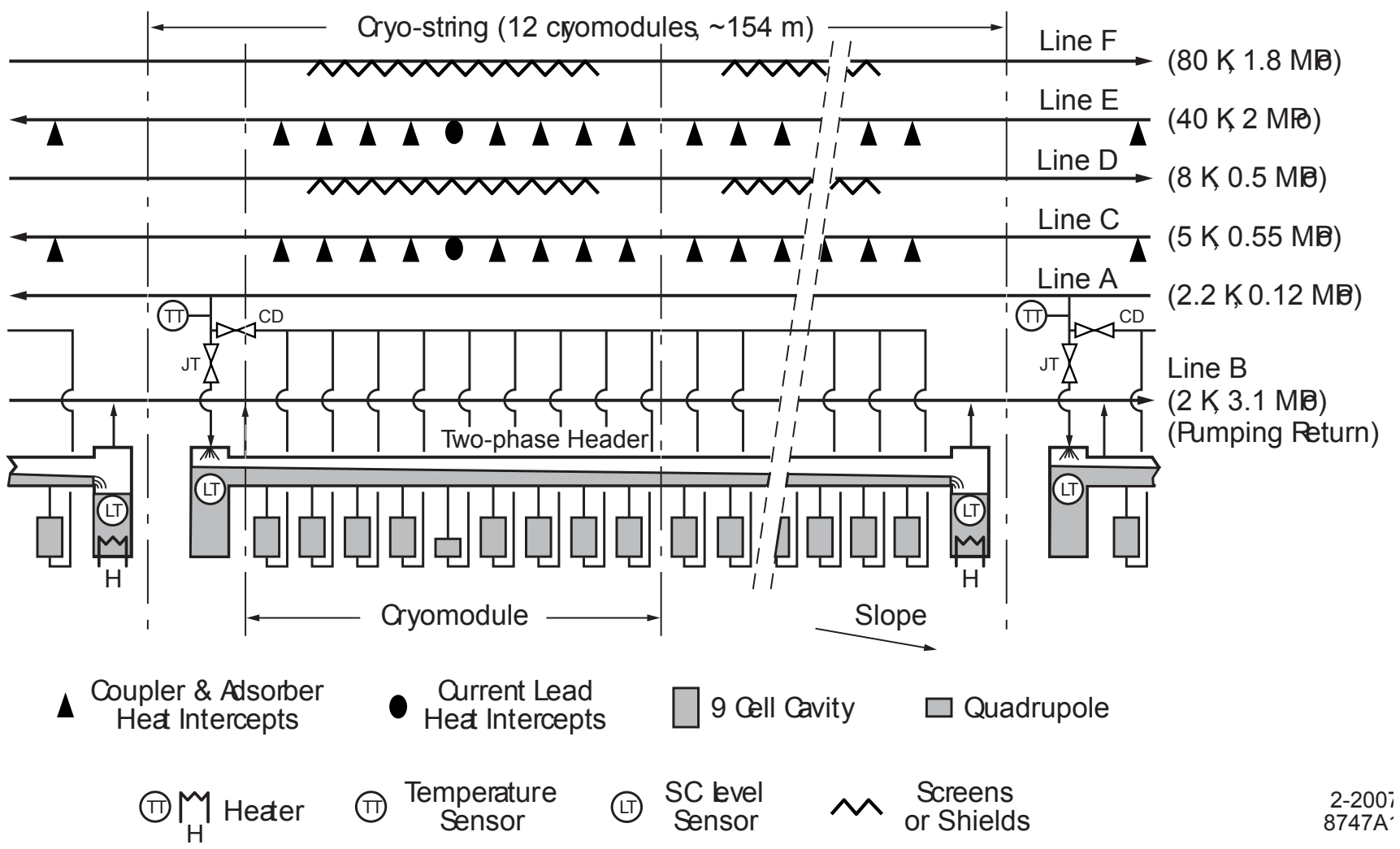

**Figure 3.25** Cooling scheme of a cryo-string.

## 3.5.2 Heat loads and cryogenic-plant power

Table 3.11 shows the predicted heat loads and resulting cryogenic plant sizes for main-linac cryo-units comprised of 13 long cryo-strings for the KCS layout (a total of 156 cryomodules) and for 21 short cryo-strings for the DKS layout (a total of 189 cryomodules). The resulting cryogenic plant capacities are equivalent to 15.4 kW at 4.5 K for the KCS layout and 19.0 kW at 4.5 K for the DKS (mountainous) layout. Both sizes are well within the range of typical large helium cryogenic plant capacities.

**Table 3.11.** Main-linac heat loads and cryogenic plant size [34]. Where there is a site dependence, the values for the flat / mountain topographies are quoted respectively. (The primary difference is in the choice the number of cryo-plants, specifically 6 and 5 plants for flat and mountainous topographies respectively.)

| | | 40–80 K | 5–8 K | 2 K |
|---|---|---|---|---|
| Predicted module static heat load | (W/module) | 75.04 | 10.82 | 1.32 |
| Predicted module dynamic heat load | (W/module) | 58.80 | 5.05 | 9.79 |
| Number of cryomodules per cryogenic unit | | 156 / 189 | 156 / 189 | 156 / 189 |
| Non-module heat load per cryo unit | (kW) | 0.7 / 1.1 | 0.14 / 0.22 | 0.14 / 0.22 |
| Total predicted heat per cryogenic unit | (kW) | 21.58 / 26.40 | 2.61 / 3.22 | 1.87 / 2.32 |
| Efficiency (fraction Carnot) | | 0.28 | 0.24 | 0.22 |
| Efficiency in Watts/Watt | (W/W) | 16.45 | 197.94 | 702.98 |
| Overall net cryogenic capacity multiplier | | 1.54 | 1.54 | 1.54 |
| Heat load per cryogenic unit including multiplier | (kW) | 33.23 / 40.65 | 4.03 / 4.96 | 2.88 / 3.57 |
| Installed power | (kW) | 547/669 | 797/981 | 2028 / 2511 |
| Installed 4.5 K equiv | (kW) | 2.50 / 3.05 | 3.64 / 4.48 | 9.26 / 11.47 |
| Percent of total power at each level | | 0.16 | 0.24 | 0.60 |
| Total operating power for one cryo unit based on predicted heat (MW) | | | 2.63 / 3.24 | |
| Total installed power for one cryo unit (MW) | | | 3.37 / 4.16 | |
| Total installed 4.5 K equivalent power for one cryo unit (kW) | | | 15.40 / 19.01 | |

The table lists an “overall net cryogenic capacity multiplier”, which is a multiplier of the estimated heat loads. This factor accounts for cryogenic plant overcapacity required for control, off-design operation, seasonal temperature variations (which affect compressor operation), and uncertainty in static and dynamic heat loads at the various temperature levels. Note also that cryogenic plant efficiency is assumed to be 28 % at the 40 to 80 K level and 24 % at the 5 to 8 K temperature level. The efficiency at 2 K is only 20%, however, due to the additional inefficiencies associated with producing refrigeration below 4.2 K. All of these efficiencies are in accordance with recent industrial conceptual design estimates.

Table 3.12 summarises the required capacities of the cryogenic plants for the different area systems, including the two configurations under study for the Main Linacs. Total installed power for all the cryogenic systems is about 44 to 46 MW (depending on KCS or DKS configuration), with an expected typical operating power of 34 to 35 MW.

**Table 3.12.** ILC cryogenic plant sizes (also includes sources, damping rings and beam delivery section for completeness) [47].

| Area | # of Plants | Installed Plant Size (each) (MW) | Total Installed Power (MW) | Operating Power (each) (MW) | Total Operating Power (MW) |
|---|---|---|---|---|---|
| Main Linac + RTML flat/mntn | 12 / 10 | 3.37 / 4.16 | 40.44 / 41.60 | 2.63 / 3.24 | 31.56 / 32.40 |
| Positron Source | 1 | 0.65 | 0.65 | 0.35 | 0.35 |
| Electron Source | 1 | 0.70 | 0.70 | 0.48 | 0.48 |
| Damping Rings | 1 | 1.45 | 1.45 | 1.13 | 1.13 |
| BDS | 1 | 0.41 | 0.41 | 0.33 | 0.33 |
| Experiments | 1 | 1.00 | 1.00 | 0.70 | 0.70 |
| Total | 17 / 15 | | 44.65 / 45.81 | | 34.55 / 35.39 |

### 3.5.3 Helium inventory

As illustrated in Fig. 3.26, most of the helium inventory consists of the liquid helium which bathes the RF cavities in the helium vessels. The total helium inventory in ILC will be roughly 63 % of that of the LHC at CERN, about 630 000 liquid litres, or about 82 metric tons (see Table 3.13).

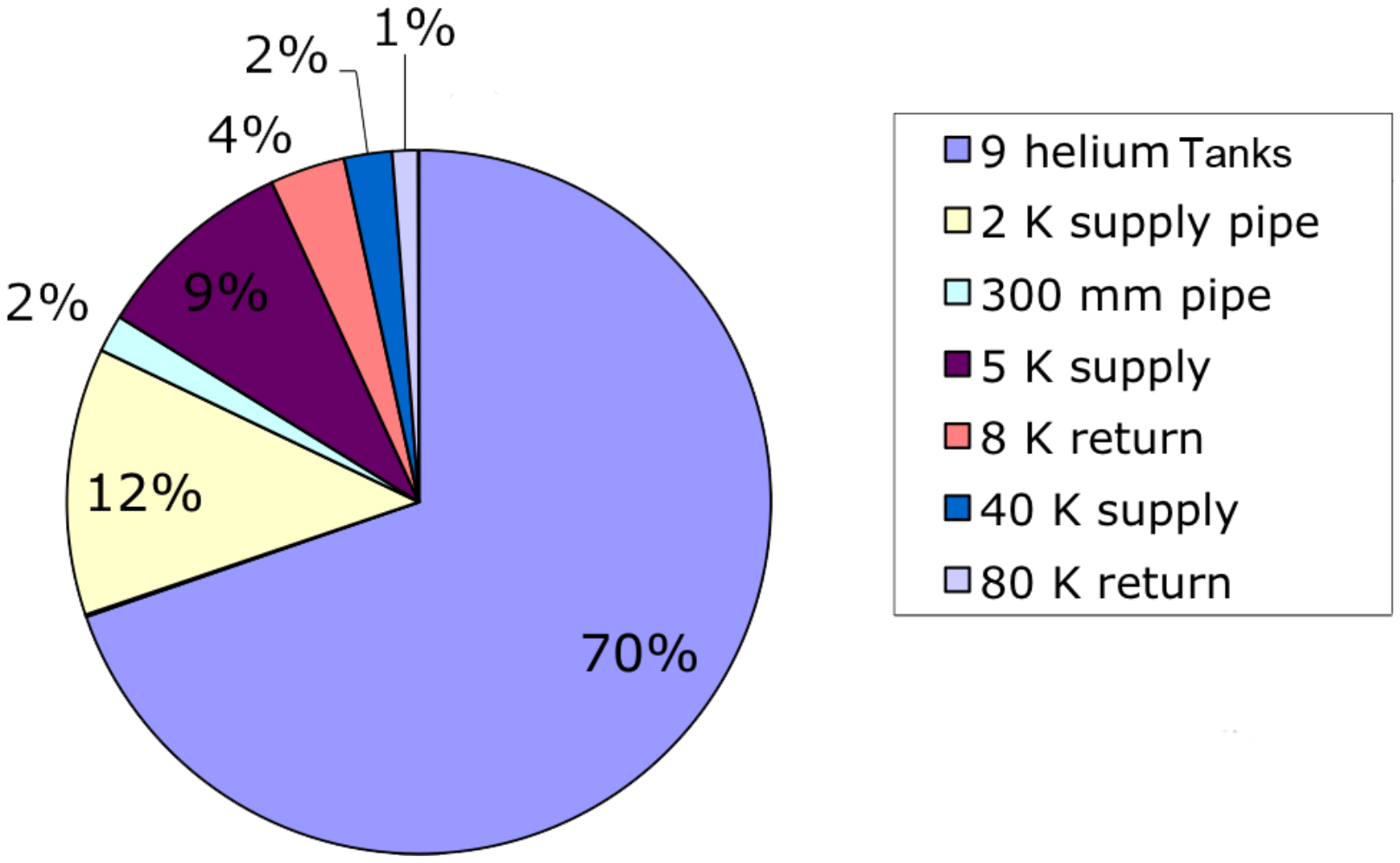

**Figure 3.26**
Helium mass in a module.

**Table 3.13**
Main Linac helium inventory [48].

| Volumes | modules | Helium (liquid liters equivalent) | Tevatron Equiv. | LHC Equiv. |
|---|---|---|---|---|
| One module | 1 | 346 | | |
| String (flat) | 12 | 4153 | 0.07 | |
| String (mountainous) | 9 | 3115 | 0.05 | |
| Cryogenic unit (flat) | 156 | 54 000 | 0.9 | 0.054 |
| Cryogenic unit (mountainous) | 189 | 65 400 | 1.1 | 0.065 |
| ILC Main Linacs | 1825 | 632 000 | 10.5 | 0.63 |

### 3.5.4 Pressure code compliance

The niobium RF cavities limit the maximum allowable pressures at the 2 K level of the cryogenic system. In North America, Europe, and Asia, the titanium helium tanks which surround niobium RF cavities, and part or all of the RF cavity itself, fall under the scope of the local and national pressure vessel rules [33]. Thus, while used for its superconducting properties, niobium must be treated as a material for pressure vessels. Problems with the certification of pressure vessels constructed partially or completely of niobium arise due to the fact that niobium and titanium are not listed as acceptable vessel materials in pressure vessel codes. Considerable effort has been expended in all three regions to gain compliance with pressure vessel codes and permission from authorities to operate ILC-style cryomodules, which contain these exceptional pressure vessels.

Partly due to the constraints of pressure code compliance, and partly to avoid detuning of the RF cavities by high-pressure helium, the cavity helium vessels and associated low-pressure piping (30 mbar corresponding to 2 K), have a Maximum Allowable Working Pressure (MAWP) of 2 bar differential. A higher MAWP for liquid-helium temperature conditions may be established, if necessary, to accommodate pressures during emergency venting with loss of vacuum. Other piping such as the 2 K helium-supply pipe and thermal-shield lines will be rated for 20 bar differential pressure.

Details regarding methods to achieve compliance with pressure codes and permission to operate low-temperature containers made from niobium and titanium will depend on the legal requirements of the regions involved. Documentation and required testing pressures and procedures are not uniform around the world. Testing in one region and operation in another may invoke multiple sets of rules. Laboratories involved in ILC cryomodule development have established methods to satisfy local codes and demonstrate the safety of these systems, sometimes including special arrangements with local authorities for these exceptional vessels. Careful consideration and agreements between all the involved regional authorities will be required for the distributed mass production, testing and finally operation of the cryomodules.

## 3.6 RF power source

### 3.6.1 Overview

The centrepiece of the RF-power system is the 10 MW multibeam klystron (MBK). With the power required by each cavity including a certain overhead for power loss in the waveguides and allowance for tuning, the MBK provides enough peak power in the pulse to drive up to 39 cavities under the nominal beam-loading conditions (see Table 3.14).

**Table 3.14**
Main parameters relevant to the RF power that is required for one 9-cell cavity. The RF-power numbers are intended to give an indication of the power required; they represent the ideal match conditions, and do not include overheads for controls, waveguide losses or the expected spread in operating gradients.

| Parameter | Unit | Value for baseline |
|---:|:---:|:---:|
| RF Frequency | GHz | 1.3 |
| Beam current in the pulse | mA | 5.8 |
| Accelerating gradient | MV/m | 31.5 |
| Cavity length | m | 1.038 |
| $Q_L$ (matched) | | $5.5 \times 10^6$ |
| RF Voltage | MV | 32.7 |
| Beam phase | deg | 5 |
| RF pulse length | ms | 1.65 |
| Beam width | ms | 0.72 |
| Filling time | ms | 0.93 |
| Repetition rate | Hz | 5 |
| RF power into cavity | kW | 188 |
| RF for 26 cavities | MW | 4.9 |
| RF for 39 cavities | MW | 7.4 |

The two site variants (flat and mountainous topography) differ significantly in how the MBK power is supplied to the cavities in the tunnel.

For the mountainous topography, such as the sites considered in Japan, a Distributed Klystron System (DKS) approach is taken where the klystrons are distributed along the main linac tunnel, with each klystron connected directly to 39 cavities (4.5 cryomodules). In this case the tunnel will have a wide flat-bottomed cross section shape referred to as "kamaboko"[5]. The tunnel is divided along its length by a thick concrete radiation shield into two parallel corridors – one for cryomodules and the beamline, the other for klystrons, DC power supplies and control hardware.

For the flat topography, the novel Klystron Cluster Scheme (KCS) is the preferred solution, where all the MBKs, modulators and associated DC power supplies are installed in "clusters" located on the surface. The combined power (∼200 MW) from from the klystrons in a cluster is transported down into and along the accelerator tunnel via a large over-moded circular waveguide.

This section describes those aspects of the components of the RF-power system and Low-level RF (LLRF) control that are common to both flat and mountainous topography site-dependent designs. Details specific to DKS and KCS are given in Section 3.8 and Section 3.9 respectively.

### 3.6.2 Modulator

A Marx-type modulator is used to generate the flat, high-voltage pulses required by the 10 MW klystron. The maximum output-power requirements for the modulator are 120 kV, 140 A, 1.65 ms pulses at a 5 Hz repetition rate. (For 10 Hz mode, the modulators need to provide approximately half the maximum peak power.) Table 3.15 lists the specifications for the modulator, required to drive a klystron producing a peak output power of 10 MW with a microperveance of 3.38 and an efficiency of 65 %.

**Table 3.15**
Parameter specifications for the klystron modulators of the main linacs of ILC.

| Parameter | Unit | Specification |
|---|---|---|
| Output voltage | kV | 120 |
| Output current | A | 140 |
| Pulse width | ms | 1.65 |
| Pulse repetition frequency | Hz | 5 (10) |
| Max. average power | kW | 139 |
| Output pulse flat-top | % | $\pm 0.5$ |
| Pulse-to-pulse voltage fluctuation | % | $\pm 0.5$ |
| Energy deposited into klystron during a gun spark | J | $< 20$ |

The Marx modulator uses solid-state switches to charge capacitors in parallel during the interval between output pulses. During the output pulse, the capacitors are discharged in series to generate a high-voltage output with a magnitude of the charging voltage times the number of stages. With this topology, low-voltage components can be used to produce a high-voltage output without requiring an output transformer. There are several ways to produce a flat output pulse. One method is to integrate a "buck converter" in series with each cell, which uses a closed-loop correction scheme to produce a square output pulse for each cell. A diagram of a Marx modulator and the circuit of one simple Marx cell are shown in Fig. 3.27.

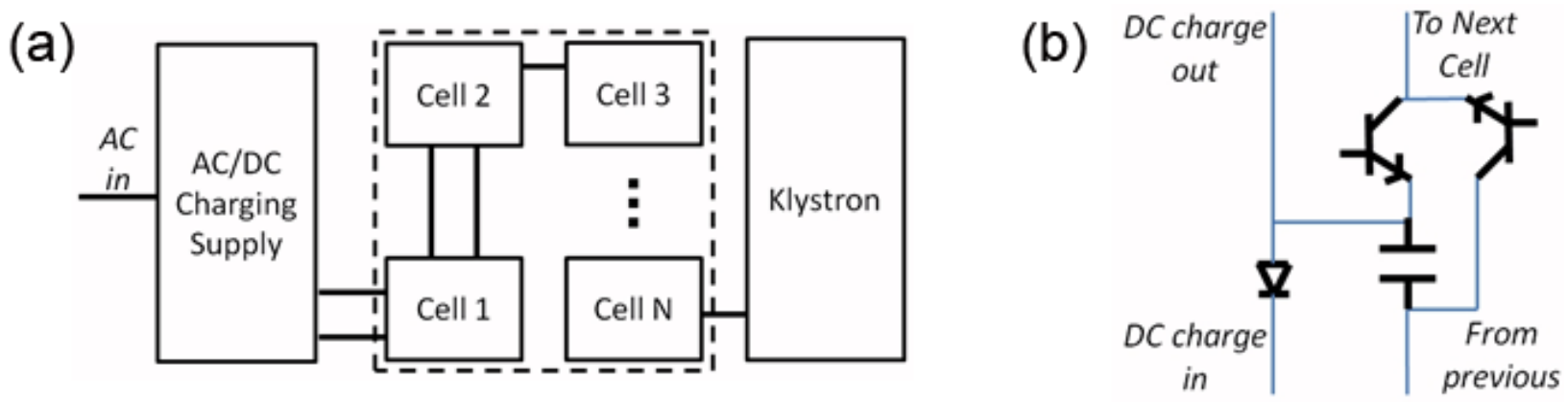

**Figure 3.27**
(a) Simple block diagram of a Marx modulator and (b) simple single cell circuit.

There are several advantageous characteristics of the Marx topology. The modular design simplifies fabrication and allows redundant hardware to be implemented. Solid-state switching is intrinsically long-life, and in conjunction with redundant hardware, a high-availability architecture is possible. Modularity reduces the spares inventory and simplifies maintenance, thereby reducing the mean time to repair. Due to the absence of a high-voltage output transformer, short rise and fall times are possible, further increasing efficiency [54].

[5] A Japanese fish cake which resembles the tunnel cross section.

Multiple R&D programs have been pursued to develop and demonstrate the efficacy of a Marx-topology modulator to drive the klystron. The requirements and many technical advantages are discussed in Part I, Section 2.8. Figure 3.28 shows three prototypes, one built by Diversified technologies and one developed at SLAC, which have demonstrated the technical feasibility of the modulator, as well as providing a cost basis.

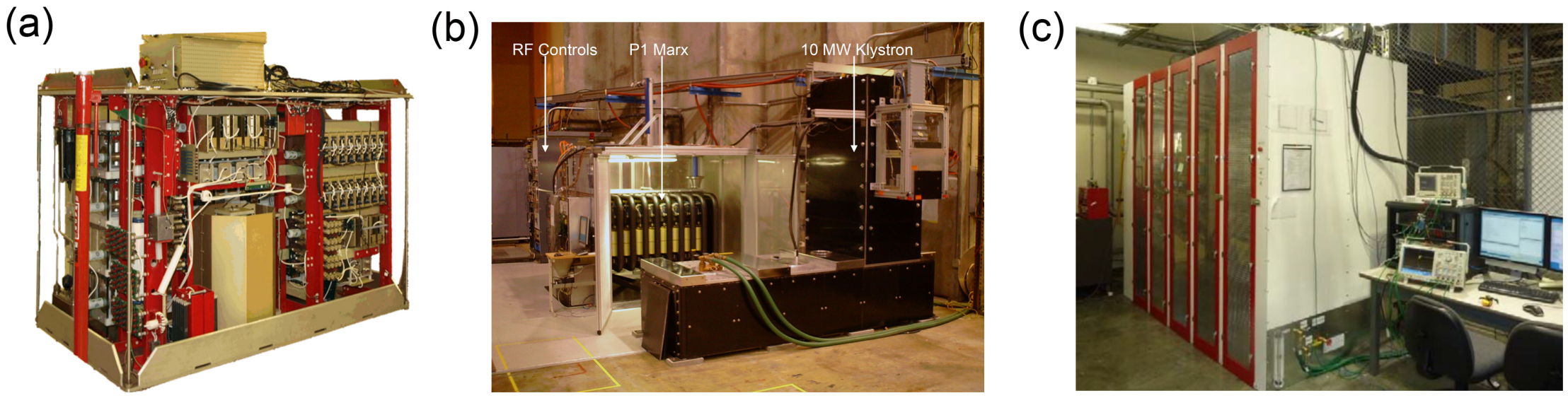

**Figure 3.28.** (a) DTI Marx modulator, (b) SLAC P1 Marx modulator and (c) SLAC P2 Marx.

The SLAC P2 Marx is an embodiment of the Marx concept which has many advanced features. It contains thirty-two identical cells, with $N+2$ redundancy. Pulse-top flatness at the $\pm$ 0.05% level has been demonstrated with operation into a water load. A full-pulse waveform of this Marx and the flatness are shown in Fig. 3.29. The flat pulse is generated using a closed-loop regulation scheme which feeds forward on both the voltages of the individual cells as well as the overall output voltage. In addition, the cells are phase shifted to stagger the ripple of each individual cell with respect to others, which results in a net cancellation, achieving an overall low modulator ripple.

The energy in the rise and fall time of the pulse is dissipated in the klystron collector. Very fast rise and fall times of less than 15 µs were obtained with the water load, approximately 0.5 % of the total energy output from the Marx, corresponding to a very high efficiency.

The AC/DC charging power-supply technology has a low technical risk, but its performance is important in achieving a cost-effective and efficient RF system. A survey of available technologies indicates that a conversion efficiency of 95 % is realisable, which is assumed in the heat loads presented in Table 3.16.

**Table 3.16**
Power efficiencies and heat loads of Marx modulators assuming 10 µs rise and fall times.

| Parameter | Unit | Specification |
|---|---|---|
| SLAC P2 Marx DC to pulse flattop efficiency | % | $95 \pm 1$ |
| Assumed charging supply AC to DC efficiency | % | 95 |
| Usable power delivered to klystron | kW | 138.6 |
| Power delivered to collector during pulse rise and fall | kW | 0.5 |
| Power dissipated to air inside of modulator enclosure | kW | 7.1 |
| Power dissipated in the DC chargers | kW | 7.4 |

Additional characteristics of the SLAC P2 Marx include the use of air insulation and cooling rather than oil. At the marginal expense of compactness, air insulation simplifies maintenance, reduces hazardous-waste containment issues, and simplifies component compatibility. Waste heat is transferred from the modulator via an air-to-water heat exchanger.

The SLAC P2 Marx also utilises an intelligent control system with embedded diagnostic and prognostic systems. These can be used to monitor cell activity in real time and to anticipate the onset of components' end-of-life phase. Twelve 12-bit, 1 MS/s ADC are used within each cell to monitor voltage, current, and temperature values of interest. These are used in the closed-loop regulation

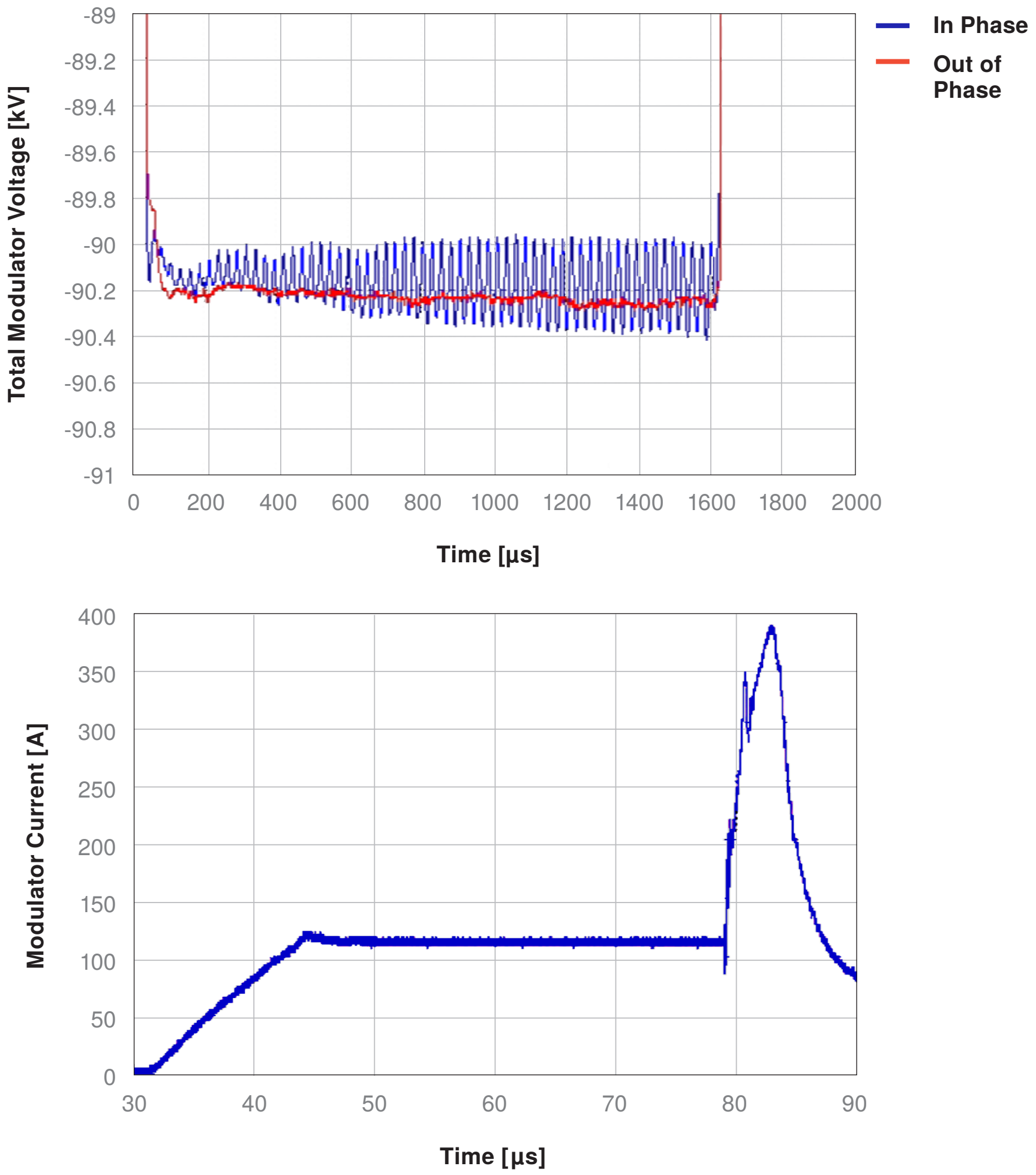

**Figure 3.29** Top: flatness of the Marx modulator pulse flat-top. Bottom: measured modulator current pulse before and after an arc.

scheme and also can be used to troubleshoot the cells in-situ.

Fault susceptibility is another important characteristic. It is necessary not only to prevent damage to the modulator, but also to protect the klystron in the event of a gun arc. The Marx satisfies these requirements. The bottom plot of Fig. 3.29 shows the current waveform from an arc generated in a self-break spark gap that closely simulates a klystron fault. It shows that the IGBT opened with a 0.5 µs delay after sensing the arc, suppressing the energy deposited to less than 10 J, satisfying the requirement for klystron protection. In addition, if a main IGBT fails in the Marx during a gun spark event, the charge IGBT in the cell closes. In this way, the energy within the cell is contained in the cell and is not transferred to the klystron.

The majority of the capabilities of the P2 Marx have been demonstrated. However, to adequately characterise the mean time before failure and the mean time to repair, an extended testing and qualification period is necessary.

### 3.6.3 10 MW Multi-Beam Klystron (MBK)

The RF power to drive the accelerating cavities at the ILC is provided by 10 MW L-band klystrons, whose baseline design is based on a multi-beam scheme. The current baseline multi-beam klystron (MBK) splits the electron current into six beams of low perveance. This arrangement allows a reduction in the beam voltage while weakening the space-charge effect, the net result of which is an improved power efficiency with a lower-voltage modulator. Table 3.17 gives the main parameters for the MBK.

The design effort for the 10 MW-class MBKs began around the time of the TESLA conceptual design and has evolved through the European XFEL project. Vertically mounted prototypes were initially developed by a few electron-tube manufacturers and successfully achieved the 10 MW goal. They were followed by horizontally mounted MBKs, whose construction is compatible with implementation at the European XFEL and ILC. The horizontal MBKs have successfully demonstrated the same RF-power performance as the vertical models. DESY, KEK and SLAC have all procured and operated these MBKs, evaluated their performance and obtained satisfactory results.

**Table 3.17**
Parameters for the 10 MW multi-beam klystron

| Parameter | Specification |
|---|---|
| Frequency | 1.3 GHz |
| Peak power output | 10 MW |
| RF pulse width | 1.65 ms |
| Repetition rate | 5.0 (10) Hz |
| Average power output (5 Hz) | 82.5 kW |
| Efficiency | 65 % |
| Saturated gain | $>$ 47 dB |
| Instantaneous 1 dB BW | $>$ 3 MHz |
| Cathode voltage | $>$ 120 kV |
| Cathode current | $<$ 140 A |
| Filament voltage | 9 V |
| Filament current | 50 A |
| Power asymmetry (between two output windows) | $<$ 1 % |
| Lifetime | $>$ 40,000 hours |

The current MBK designs are now relatively mature. All vendors have provided suitable solutions for both the resonant cavities within the klystron body and the internal beam focusing. Figure 3.30 shows photographs of two L-band MBKs from two different vendors (Thales and Toshiba). Typical performance data is shown in Fig. 3.31.

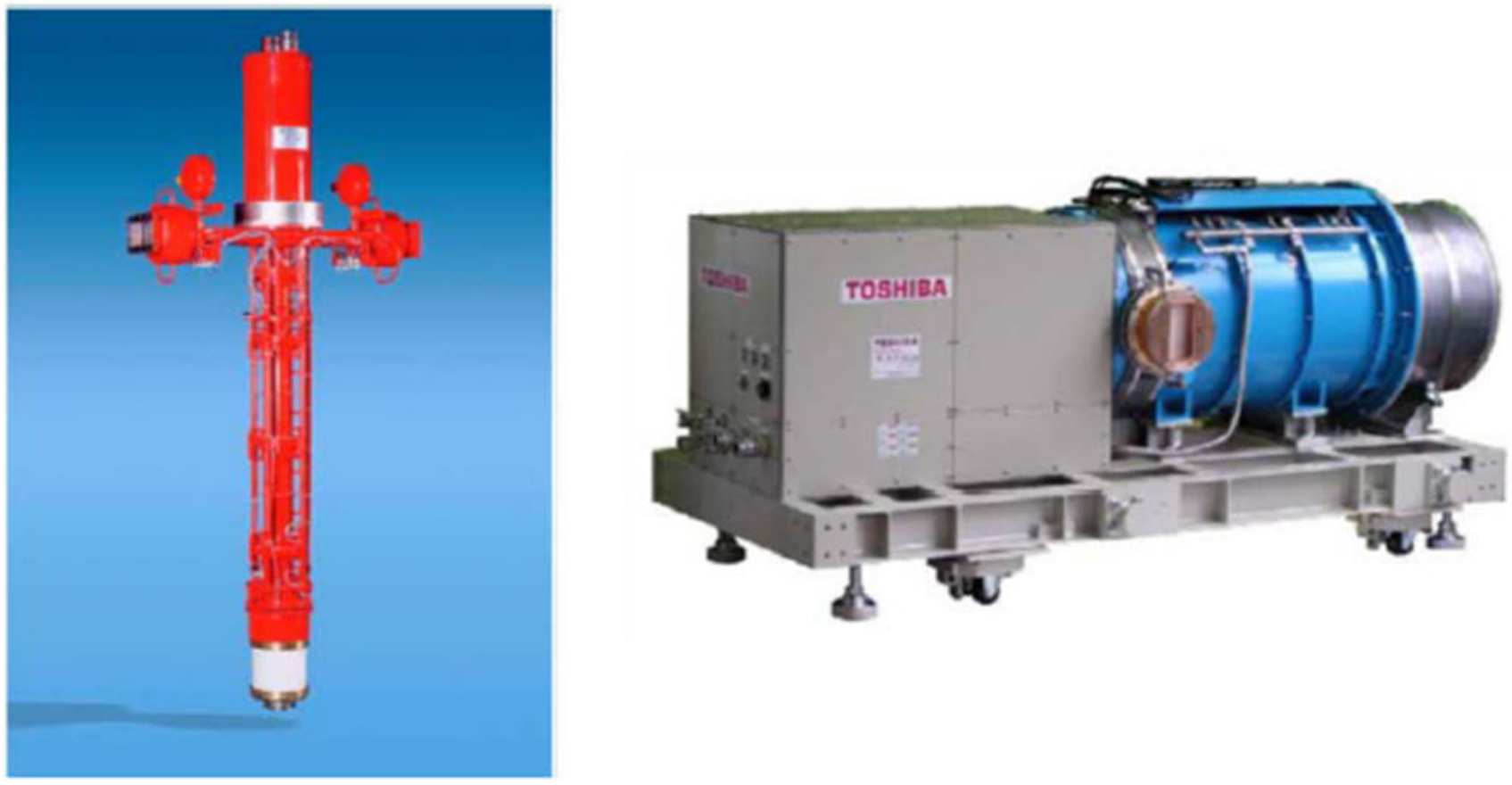

**Figure 3.30**
Thales TH1801 (left) and the horizontally mounted Toshiba E3736.

A crucial aspect for operations of the ILC linacs is the lifetime of the klystrons. The MTBF for the ILC klystrons is specified at $\geq$40,000 hours. The lifetime for linear beam tubes is dominated by the durability of the cathode. With cathode loading as low as 2 A/cm$^2$ achieved by some vendors, the expected (theoretical) lifetime is in excess of 50,000 hours. However, operational experience is required in order to estimate the true lifetime. Lifetime tests are planned, and the $\sim$30 MBKs required for the European XFEL will also provide significant data.

The manufacturability of MBKs is an important issue, since the ILC requires nearly 500 tubes to be prepared within a period of 5 to 7 years. The investment in RF test and processing infrastructure by industry is likely to be cost prohibitive for production at this scale. A more cost-effective model would be for collaborating institutes ("hub laboratories") to host such facilities and to either provide the personnel directly, or make the test infrastructure available to industry (klystron vendors) to use for conditioning and testing.

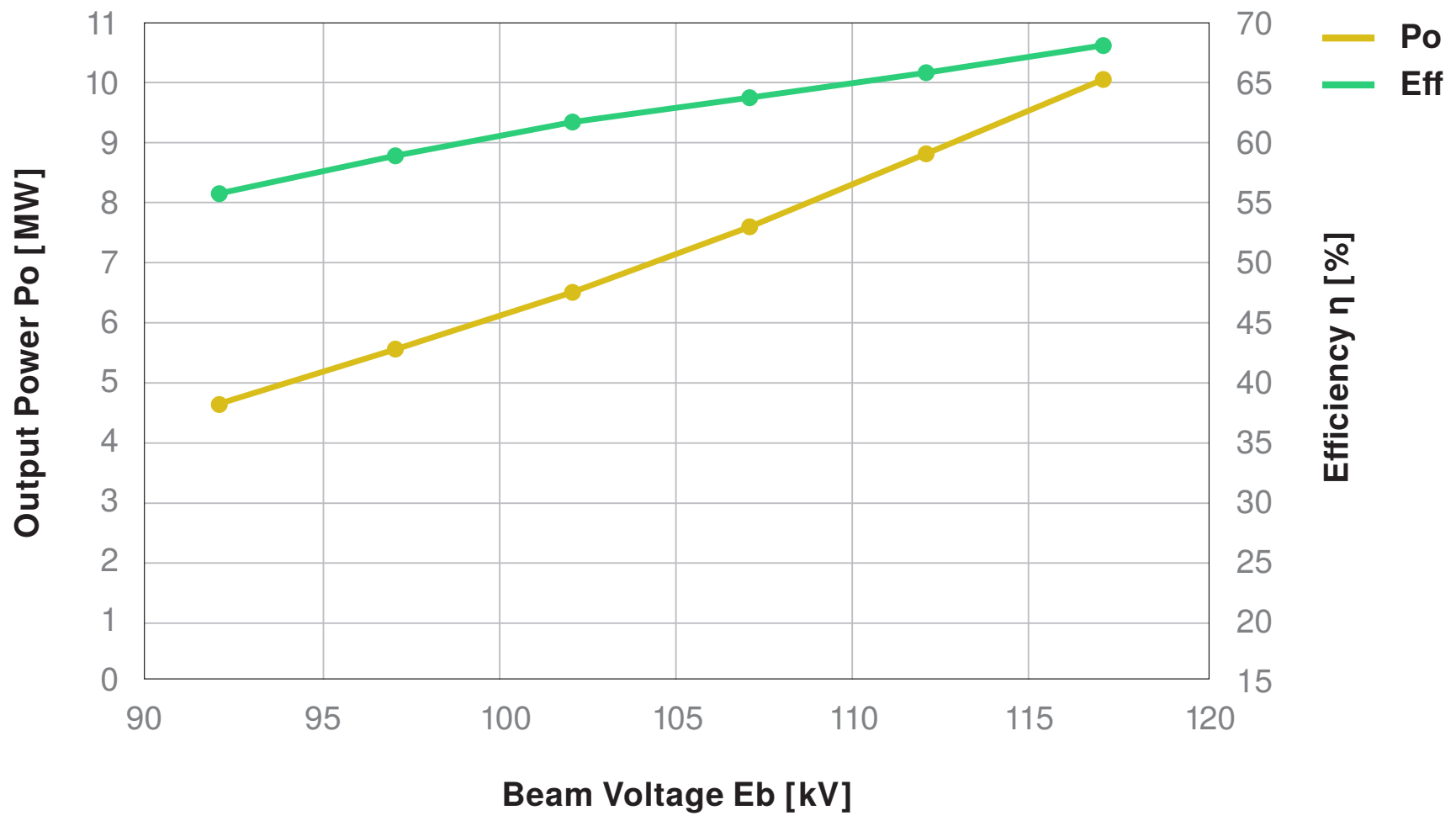

(a)

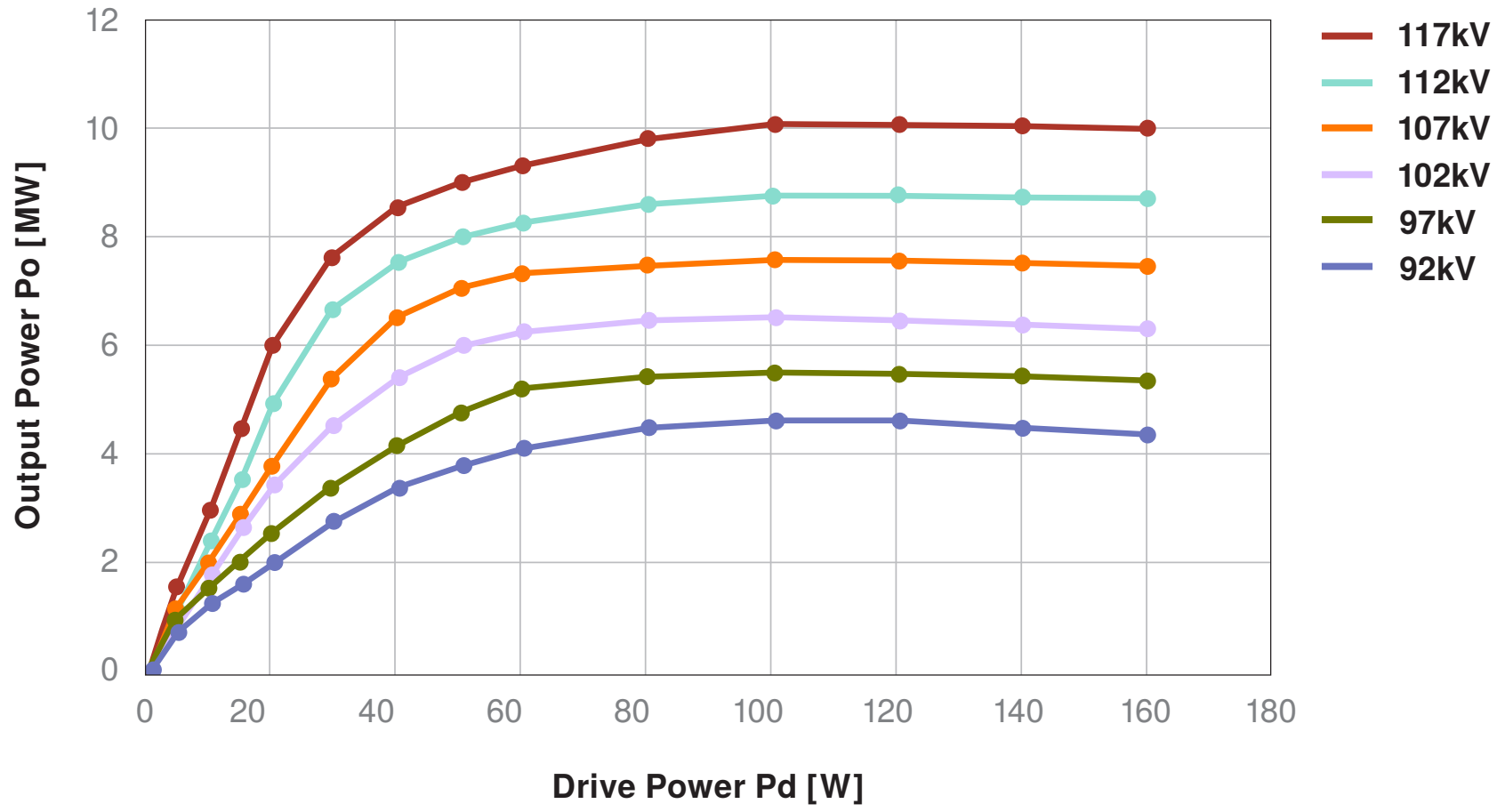

(b)

**Figure 3.31** Measured performance data of Toshiba klystron, showing (top) output power and efficiency as functions of beam voltage and (bottom) gain characteristics.

## 3.6.4 Local power-distribution system

The arrangement and installation of waveguides near the cryomodules is the same in both cases for KCS and DKS, and is commonly referred to as the Local Power-Distribution System (LPDS). The design of the LPDS must provide:

- a cost-effective solution to distributing the RF power to the cavities with minimum RF loss;
- flexibility to remotely and independently adjust the power delivered to each individual cavity to allow for the expected $\pm$ 20% spread in gradient performance.

Furthermore it is desirable to keep as far as possible a common design between DKS and KCS, and in the case of DKS, provide a relatively straightforward reconfiguration to 26 cavities per klystron required for the luminosity upgrade (Section 12.3).

Figure 3.32 shows schematics of distribution of RF power onto cavities in the cryomodules, and a detailed list of components can be found in [49]. Each LPDS drives 13 cavities, and is capable of handling and distributing up to 5 MW of input power. For the KCS (Fig. 3.32(a)), two such LPDS distribute the RF power from one Coaxial Tap-Off (CTO) connected to the high-power overmoded waveguide to 26 cavities (one ML unit). For DKS, three LPDS feeds are used to drive a total of 39 cavities (one and a half ML units) from one single 10 MW klystron (Fig. 3.32(b)). Every third ML unit is thus without a klystron and is fed from the two adjacent ML units.

**Figure 3.32** Schematic of the local power-distribution-system (LPDS) that delivers RF power to accelerator cavities in the main linacs. (a) shows the case of the KCS option, and (b) shows the case of the DKS option. In both cases, cavities are feed in groups of 13. (Reproduced from Fig. 3.1 in Section 3.1.)

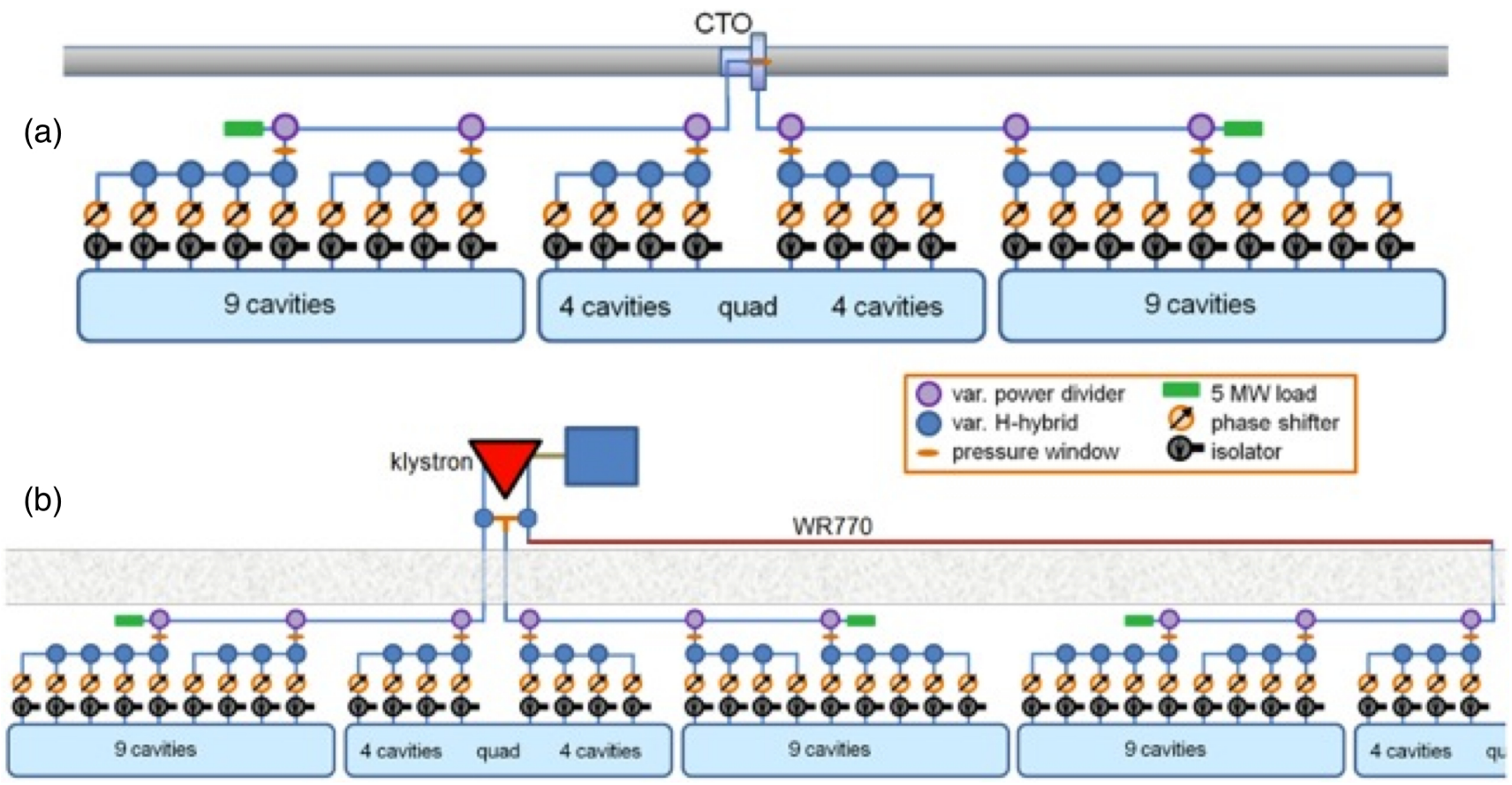

The LPDS distributes the power from a klystron (DKS) or CTO feed (KCS) to the cavities as shown in Fig. 3.32. This is accomplished in 13 cavity groups, one of which is illustrated in Fig. 3.33. The same input power goes to each group of thirteen cavities, with the exception of the third arm for the DKS arrangement (see Section 3.9.3), but the power to each cavity differs due to the variation of up to ± 20% in gradient performance.

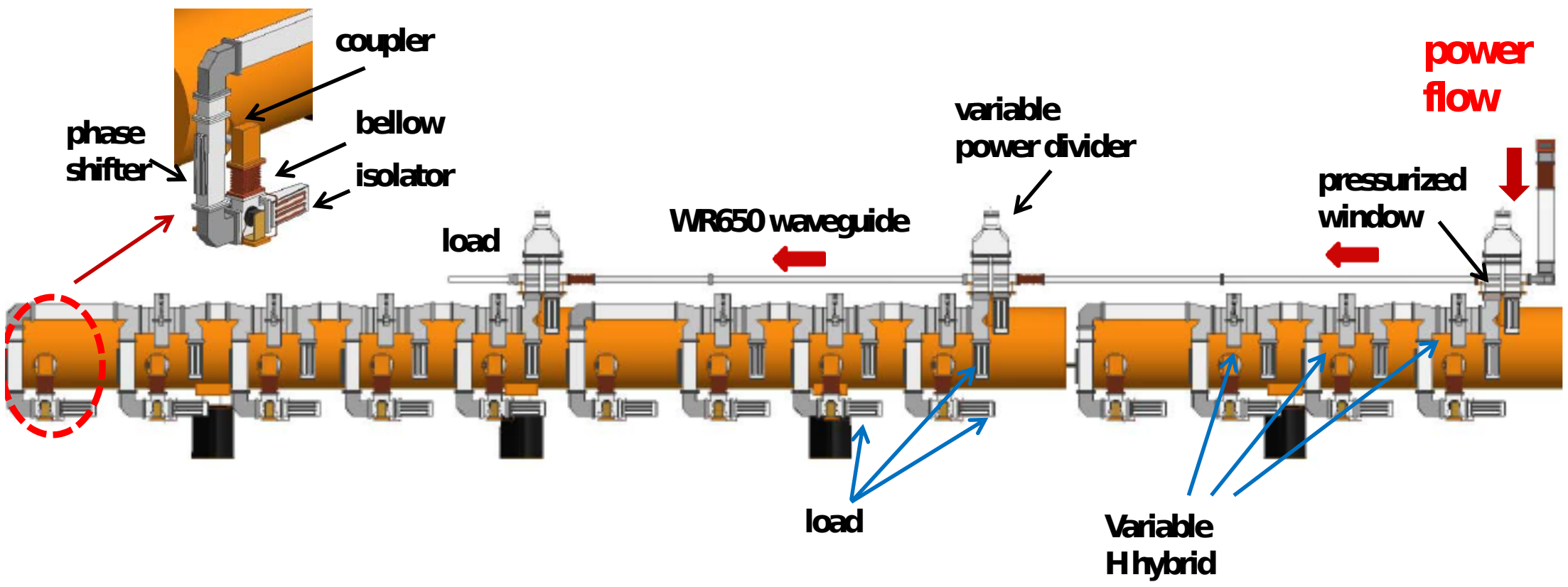

**Figure 3.33.** CAD model of a 13-cavity local power-distribution system (LPDS)

Two types of remotely-controllable variable power-splitters are used to customise the power sent to the cavities. As indicated in Fig. 3.32, three variable power dividers (VPDs) initially split the power into three lines, each feeding either four or five cavities. Any remaining power after the third VPD is dumped in a high-power load. The VPDs are pressurised to one bar $\mathrm{N}_2$, and ceramic RF windows handle the pressure differential to the non-pressurised waveguides. The power division within each set is achieved using variable H-hybrids, and the cavity after the last split uses all the remaining power.

In each cavity feed line is a remotely-controllable phase shifter followed by a ferrite-based isolator (circulator with load). The latter prevents the power that is either reflected or discharged from the cavity from re-entering the waveguide system. RF pickups in the input and load ports of the isolators provide the low-level RF control system with information on the forward and reflected cavity power levels. Finally, a flexible rectangular bellows connects the waveguide to the warm end of the cavity input coupler.

A schematic of the VPD and photograph of the U-bend phase-shifter are shown in Fig. 3.34. The design of the U-bend phase-shifter features an inner waveguide which can be moved like a trombone by an external actuator. The VPD forms a 4-port device, and when one port is loaded it allows full range adjustment of power division between the forward waveguide and the downward extraction waveguide. By moving the phase shifters in opposite directions, the phases of the outputs can be held fixed, allowing a pure amplitude control.

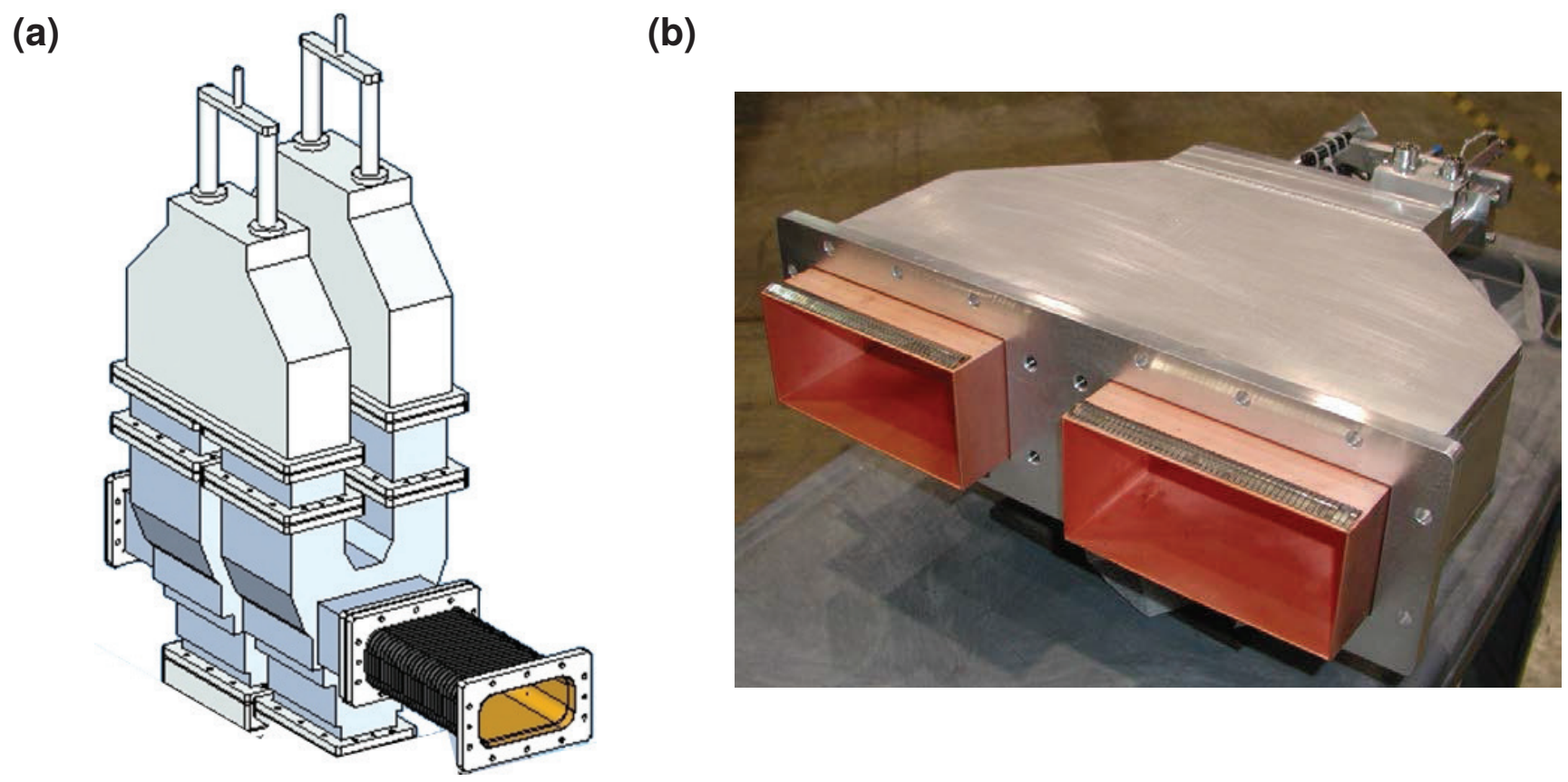

**Figure 3.34**
Schematic of a Variable Power Divider (VPD), comprising two folded magic-T's and two U-bend phase-shifters (left). b) Photograph of a U-bend phase-shifter (Finger-stock provides the contact with the mating waveguide).

Shown in Fig. 3.35 is the original design geometry for the variable H-hybrid, its electromagnetic-field patterns, and an in-line variant that simplifies daisy-chaining for use in the LPDS. This geometry formed the basis of the design for the devices in the 13-cavity LPDS shown previously in Fig. 3.33. The interior of the 4-port hybrid accommodates two electromagnetic-field modes whose relative phase-lengths can be changed by transversally moving the "pontoon-shaped" conductors, resulting in a change in the power division. Any residual phase-change is compensated using the upstream phase-shifter. Although the devices themselves do not allow the power ratios to be adjusted over the full range, the achievable power ratios can deviate significantly from the nominal ratios of 1/5, 1/4, 1/3 and 1/2.

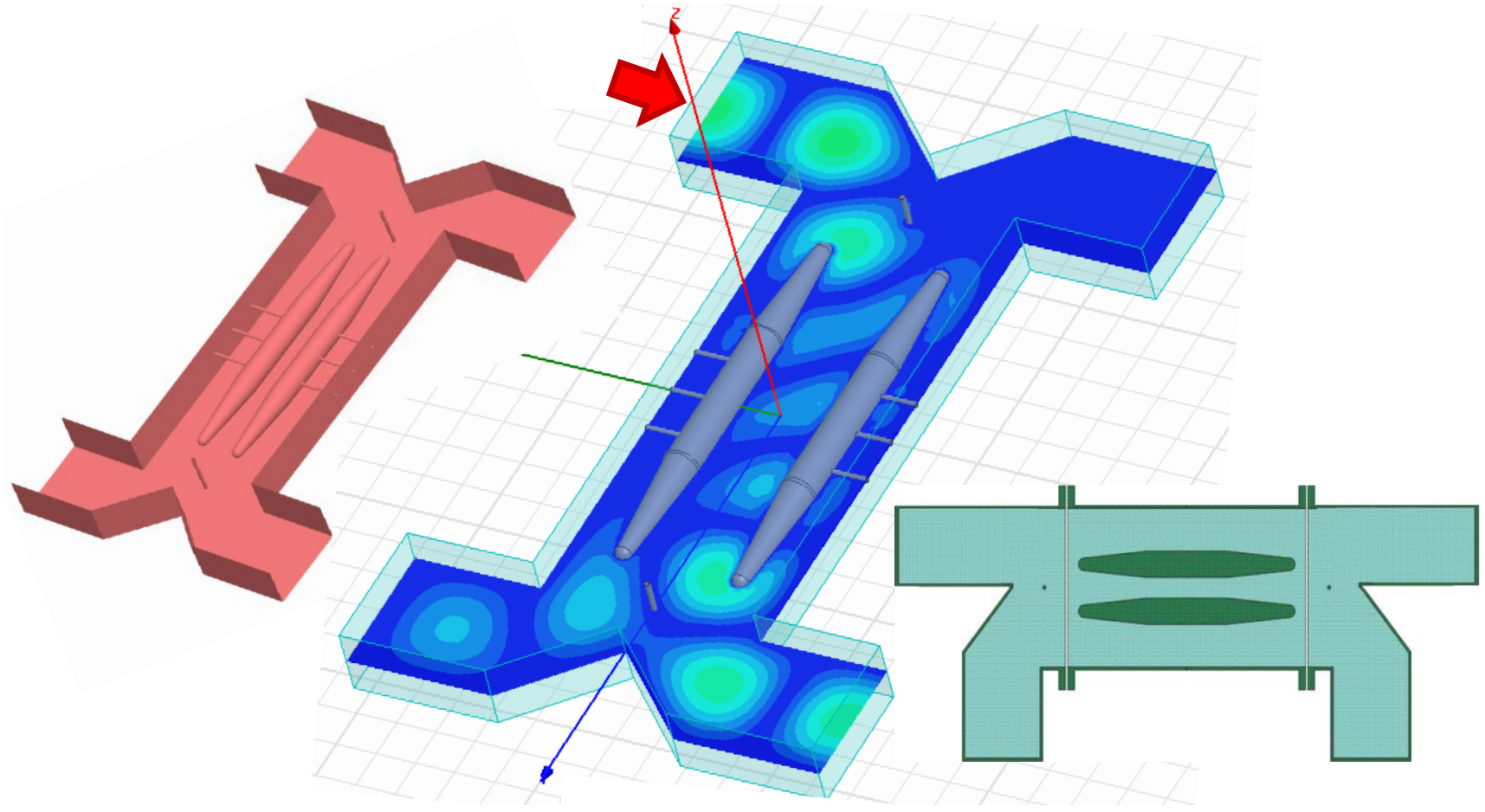

**Figure 3.35**
Illustrations of a variable H-hybrid: original design geometry (left); electromagnetic-field patterns (center); in-line variant used in the LPDS (right).

## 3.6.5 RF power requirements

In order to estimate the total number of klystrons required, it is necessary to take into account all the expected RF losses along the entire power distribution system to the cavity. A further inefficiency can be attributed to the random spread in cavity-gradient performance, since in general it is not possible to perfectly match all the cavities connected to the same power source, since the RF pulse and hence the fill time must be the same for all cavities [51]. This results in reflected power from most cavities. Finally, some small fraction of klystron power is required for LLRF control.

While estimation of the distribution-system losses and control overhead is relatively straightforward, estimation of the impact of the gradient spread can only be described in a statistical sense, since it is currently assumed that sorting the cavities with respect to performance is unlikely during mass production and installation into the cryomodules, resulting in an effectively random cavity distribution in the linac. The approach adopted here is to use Monte Carlo techniques by running many random seeds, and taking the 95 % percentile limit of the power required [53].

Table 3.18 traces the RF-power budget from the cavities back to the klystrons for both DKS and KCS. The random cavity gradients are responsible for the first increase in the table. Some post-installation adjustment of the high-level power division (fine-tuning or changing CTO's and hybrids) is assumed in order to limit this to a few percent. The first number (extra beam power) reflects the statistical fluctuation of the total voltage of the cavities driven by the single source (the source must be able to accommodate the highest voltage, at least at the 95% level). The next effect (reflection) is due to the mismatch of the individual cavities arising from the constant fill time as mentioned above. This represents a real operational power that is dumped in the loads. Folded into this number is about 0.8 % to support the required 5° beam phase with respect to the RF crest. A 7 % overhead (5 % usable) is allotted for LLRF manipulation, and 8 % for average losses in the components of the waveguide circuit that provides local distribution of the power along the cryomodules.

**Table 3.18** RF power budgets for KCS and DKS local power distribution systems

| **Cavity and Local Power Distribution** | | **KCS** (kW) | | **DKS** (kW) |
|---|---|---|---|---|
| *Mean beam power per cavity* | | *189.18* | | *189.18* |
| Extra beam power for ± 20% gradient spread | 2.90 % | 194.67 | 5.30 % | 199.21 |
| s.s. reflection for ± 20% gradient spread | 6.00 % | 206.35 | 6.00 % | 211.16 |
| Required LLRF overhead | 7.00 % | 220.8 | 7.00% | 225.95 |
| Local PDS average losses | 8.00 % | 240 | 8.00 % | 245.59 |
| Multiply by number of cavities fed as a unit | 26 | 6239.9 | 39 | 9578.1 |
| *Required local PDS RF input power* | | *6239.9* | | *9578.1* |
| **Power Combining & Transport (DKS)** | | | | (MW) |
| *RF power to local PDS* | | | | *9.578* |
| Combining/splitting and shielding penetrations | | | 1.10 % | 9.6847 |
| WR770 run loss/3 | | | 1.40 % | 9.8222 |
| *Required power from klystron (DKS)* | | | | *9.822* |
| **Power Combining & Transport (KCS)** | | (MW) | | |
| *RF power to ML Unit* | | *6.2399* | | |
| Multiply by number of ML Units per KCS | 26 (25) | 162.24 (156) | | |
| KCS main waveguide loss | 5.0 % (4.7) | 170.78 (163.69) | | |
| Shaft and bends loss | 1.80 % | 173.91 (166.69) | | |
| CTO string and upgrade WC1375 run loss | 1.50 % | 176.55 (169.23) | | |
| Klystron waveguide into CTO | 5.60 % | 187.03 (186.74) | | |
| Divide by number of klystrons | 19 (18) | 9.8436 (9.9594) | | |
| *Required power from each klystron (KCS)* | | *9.844 (9.959)* | | |

Beyond the LPDS, the accounting diverges for the two options. DKS has additional losses in the WR650 shielding penetrations and dividing/combining components, as well as in the WR770 waveguide run supplying power for half an ML Unit to the vacant klystron position. The higher losses for KCS reflect the much longer waveguide system required to transport the RF power down from the surface cluster. The estimate includes the average loss along the main circular waveguide in the tunnel (assumed to be copper plated) to each CTO, the loss in the bends and shaft waveguide, and the loss along the surface main waveguide and combining CTO string. The latter includes a circular WC1375 waveguide which runs past an area where additional klystrons can be installed for the luminosity upgrade (Section 12). Finally, there is loss budgeted for the waveguide connections from the klystron output ports in the outer region of the CTO. A major contributor here is the 5 MW isolators required to protect the klystron from the reflected power it could see from a failed combining circuit (up to 10 MW). With all these effects taken into account, the final required RF power per klystron is within the specified 10 MW, albeit with relatively little overhead.

## 3.7 Low-level RF (LLRF) control concept

### 3.7.1 Overview of Low-level RF control requirements

The primary function of the main linac and RTML LLRF systems is to control the phase and amplitude of the klystron forward power so that the required cavity fields are reached at the end of the fill time and then remain stable for the duration of the beam pulse. Since many cavities are fed from each individual klystron (or cluster in the case of KCS), the LLRF system regulates the vector sum of all the cavity fields controlled by that klystron (or cluster).

LLRF performance requirements are derived from beam-dynamics considerations of energy stability, luminosity loss, and emittance growth (see Part 1 Section 4.6). RF phase and amplitude jitter tolerances for the bunch compressors and main linacs are given in Tables 3.19 and 3.20, respectively. Beam dynamics considerations of emittance growth also require that voltages in the individual cavities will be corrected to within a few percent over the duration of the pulse.

**Table 3.19**
Bunch Compressor RF dynamic errors, which induce 2 % luminosity loss.

| Error | RMS amplitude | RMS phase |
|---|---|---|
| All klystron correlated change | 0.5 % | 0.32° |
| Klystron to klystron uncorrelated change | 1.6 % | 0.60° |

**Table 3.20**
ML RF dynamic errors, which induce 0.07 % beam energy change.

| Error | RMS amplitude | RMS phase |
|---|---|---|
| All klystron correlated change | 0.07 % | 0.35° |
| Klystron to klystron uncorrelated change | 1.05 % | 5.6° |

### 3.7.2 Vector-sum control of cavity fields

The LLRF system design is based on the digital controller implemented at FLASH and that will be used on the European XFEL [55]. Similar systems have also been implemented at STF at KEK [56] and NML at Fermilab [57]. The main functional elements are illustrated in Fig. 3.36.

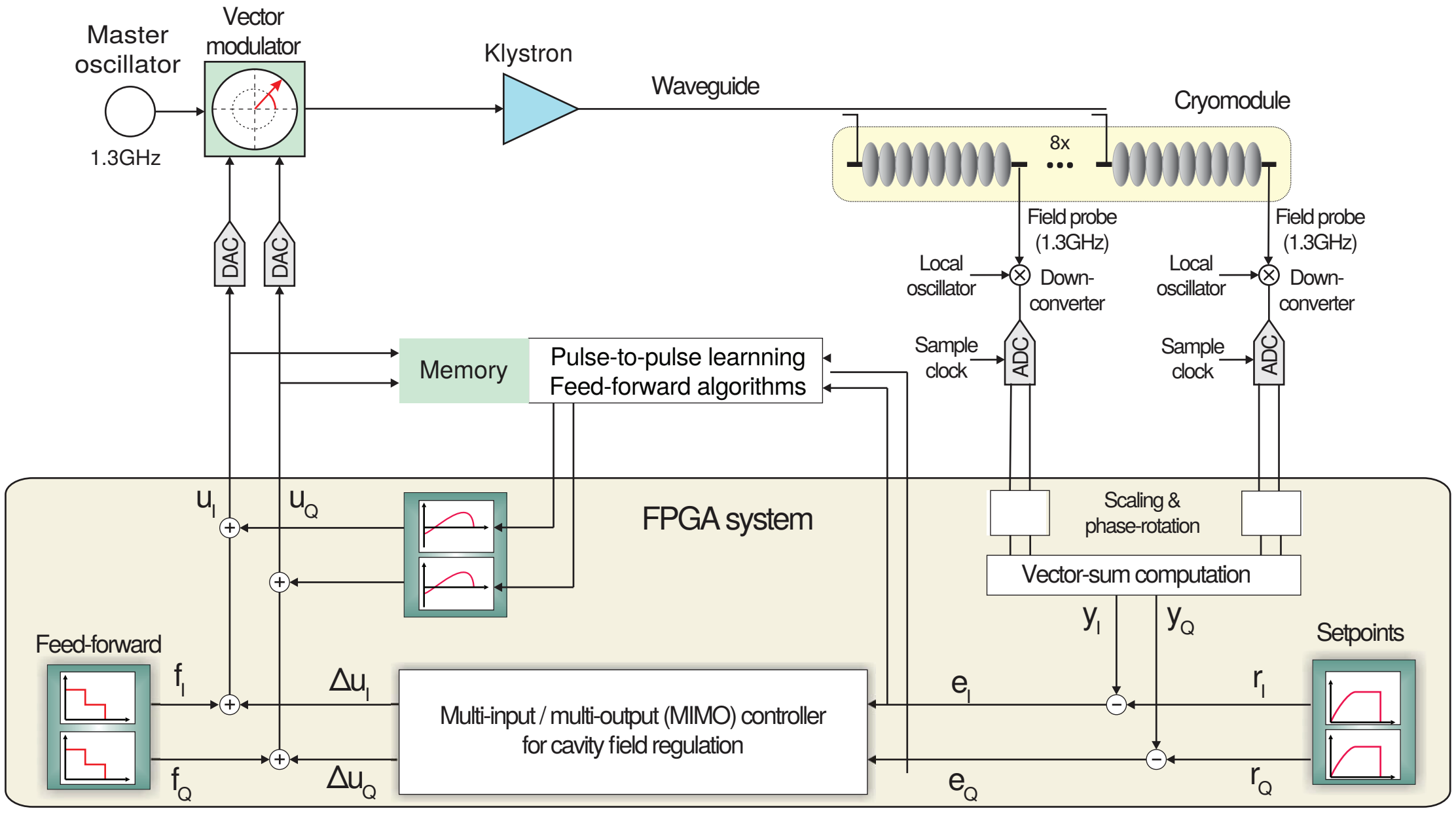

**Figure 3.36.** Functional block diagram of the digital LLRF control system at FLASH (for clarity, only one cryomodule is shown)

The signals from the cavity field probe and forward and reflected RF power arrive at the LLRF electronics as raw 1.3 GHz RF signals, where they are down-converted to an intermediate frequency (IF) and subsequently acquired by fast ADCs. Phase information is preserved by separating the data-stream from the ADCs into in-phase and quadrature terms for processing by the FPGA-based digital LLRF control functions.

The drive signal to the klystron comes from the vector-sum regulator, and comprises two terms: a feed-forward term that is determined a-priori using knowledge of the required cavity-field profile and the expected beam-current profile; and a correction term that is generated dynamically by the feedback regulator based the measured error in cavity-field vector sum that is computed from the partial vector sums from the local LLRF controllers. The resulting controller output is up-converted via a vector modulator that varies the amplitude and phase of the 1.3 GHz master oscillator reference signal that drives the klystron.

#### 3.7.2.1 Learning feed-forward controller

Due to the very low-bandwidth of the superconducting cavities and delays in the closed-loop system, dynamical feedback alone is not sufficient to completely suppress high-frequency distortions or to achieve zero steady-state errors. However, effects that are directly correlated with the 5 Hz pulse structure and that are repeatable from pulse to pulse can be pre-emptively compensated using feed-forward, leaving smaller residual and non-repetitive disturbances to be compensated using the closed-loop feedback regulator. A learning feed-forward system is used to iteratively adjust the shape of the feed-forward waveform in order to compensate for repetitive pulse-to-pulse errors, leaving the intra-pulse feedback system to attenuate pulse-to-pulse jitter and intra-pulse fluctuations. With knowledge about field imperfections in previous pulses, the residual control errors can be minimised. Optimisation of the learning feed-forward system is performed by a model-based learning feed-forward algorithm.

#### 3.7.2.2 Beam loading compensation

The forward power during the fill time is a function of the required cavity fields, while the flat-top power is a function of both the cavity fields and the beam current. With the exception of the ideally matched case (no reflected power), the required forward power during the beam-on period is not the same as that required during the fill time. Since the beam current is assumed to be known a priori from the Damping Ring instrumentation, the LLRF system can pre-emptively step the forward power to the appropriate level immediately before the arrival of the first bunch. This is illustrated in Fig. 3.37, which shows the klystron forward power for operation over a range of beam currents around the nominal design value. Without this feed-forward pre-programming of the RF power, there would be a transient perturbation on the cavity fields as the LLRF feedback system dynamically corrected the forward power.

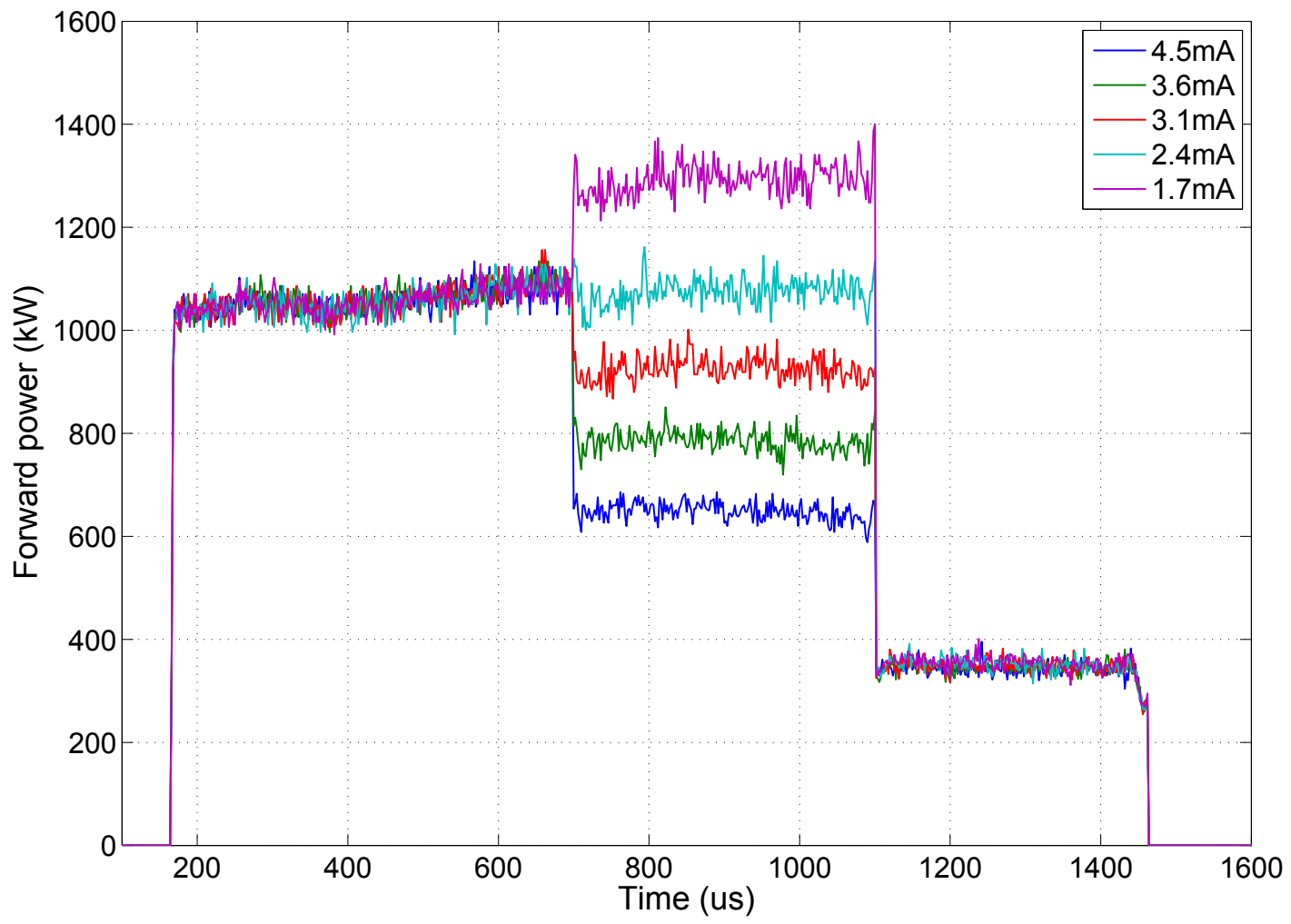

**Figure 3.37**
Klystron forward power for a range of beam currents, measured at FLASH. In this example, the beam-on time extends from 700µs to 1100µs. The power level from 1100µs to the end of the pulse corresponds to the zero beam-loading condition.

#### 3.7.2.3 Intra-pulse dynamical feedback regulator

The feedback regulator operates dynamically within the RF pulse, and its primary function is to attenuate uncorrelated pulse-to-pulse and intra-pulse jitter. The feedback regulator also corrects any residual short-term repetitive errors while the learning feed-forward system adapts to new steady-state conditions or attempts to follow fast-changing pulse-to-pulse drift.

The feedback regulator uses a multi-variable, second-order controller whose coefficients are automatically tuned by model-based controller methods. A more detailed description of the FLASH LLRF system algorithms can be found in [55]. It is worth noting that the FLASH LLRF system makes extensive use of intra-pulse beam-based feedback for additional regulation, specifically bunch arrival time, compression, charge, and energy. Such extensive use of beam-based feedback is not possible in the ILC Main Linac due to the limited availability of intermediate diagnostics, but could be implemented in the bunch compressors, where the required tolerances on regulation are more demanding. Measurement of the final energy of the beam at the exit of the linac will be used to provide a global feedback adjustment.

### 3.7.3 Operation at the limits

Unique and challenging constraints on the main-linac LLRF systems come from the limited operating margins when the linac is operating at maximum operating energy and beam current, specifically

- all cavities must run reliably at up to 95 % of their gradient limits in order to reach the linac design energy of 250 GeV;
- the $\pm$20 % spread in the above limits;
- klystrons will run at up to 95 % of the maximum available forward power. In this region, the klystron gain and phase characteristics are highly non-linear as the klystron output asymptotically approaches saturation.

Studies under these limiting conditions have been performed during the beam tests at the DESY FLASH linac. Details of these studies are described in Part I Section 3.2.

### 3.7.4 Individual cavity control

While the vector-sum controller regulates the net sum (or equivalently the average) of the cavity fields, it does not constrain fields in individual cavities, which may be varying over the beam pulse even if vector-sum is flat. Establishing the optimum setup for individual cavities is achieved through several high-level LLRF applications that have remote control of individual cavity fast and slow tuners and input coupling ($Q_{ext}$), and the fraction of total klystron power to each cavity ($P_k$). The two most important cavity-level high-level functions are compensating Lorentz-force detuning and establishing flat gradients in the presence of beam loading, discussed below.

#### 3.7.4.1 Lorentz-force detuning compensation

When there is RF field in the cavity, the cavity walls experience electrostatic and magnetic forces that act to distort the cavity shape and shift the resonant frequency of the cavity. The magnitude of the detuning ($\Delta f$) due to these Lorentz forces is proportional to the square of the field in the cavity ($E_{acc}$). Since the cavity and its support structure form a mechanical system with mass, resonant frequencies and damping times are slow relative to the 2 ms RF pulse. Resonant frequencies are typically 200-400 Hz with time-constants of tens of milliseconds. The detuning effect over the RF pulse does not occur instantaneously, but instead increases over the duration of the pulse. At 31.5 MV/m, Lorentz forces cause a detuning change of several hundred hertz over the length of the beam-on period.

This detuning must be compensated in order to avoid the significant increase in forward RF power (and the associated higher electric fields at the input coupler) that would otherwise be needed to overcome the reflected power from the detuned cavity. Compensation of the Lorentz-force detuning is accomplished using the fast piezo actuators that are integral to the cavity tuning mechanism (Section 3.3.2). The piezo tuners are driven with a feed-forward waveform to counteract the detuning as it changes over the duration of the pulse. Two different approaches have been successfully demonstrated to determine an appropriate drive waveform for the piezo tuners [58, 59]. Both methods use the piezo tuners to preemptively put the cavity structure into motion before the RF pulse in such a way as to cancel the effect of the Lorentz forces. The methods rely on indirect observations of the cavity detuning inferred from the measured forward- and reflected-power waveforms.

#### 3.7.4.2 Control of cavity gradient flatness

Several strategies have been proposed for optimising the coupling ($Q_{ext}$) and forward power ($P_k$) for each cavity along with the common fill time [51, 52], the goal being to minimise the total klystron power as well as the gradient excursions ("tilts") of the individual cavities. The optimal values of $P_k$ and $Q_{ext}$ for each cavity are dependent on the respective operating gradient and on the beam current. Automated adjustment of the $Q_{ext}$ of the individual cavities to achieve "flat" gradients to well within the required few percent has been successfully demonstrated in FLASH (see Part I, Section 3.2.8).

### 3.7.5 LLRF operations

Besides cavity vector sum control, the LLRF controller performs important control, protection and operations functions for the accelerating cavities.

#### 3.7.5.1 Exception handling

The LLRF system must detect and react to off-normal conditions that could be potentially damaging or could result in machine downtime. Depending on the severity of the condition, the LLRF could either temporarily turn down the klystron output, turn off the RF drive until the next pulse, or turn off the RF drive and wait for operator intervention. All three approaches have been implemented in FLASH and are under study.

#### 3.7.5.2 Automation

It will be essential to automate operation of the main-linac LLRF systems because of the impracticability of manually performing the necessary operational functions on a very large number of technical systems. Examples of automation that are already routinely in operation or are under development at FLASH include: startup and shutdown of the RF systems; cavity resonance control, including compensation of Lorenz forces; vector-sum calibration; quench detection; learning feedforward for the vector sum controller; drift compensation; and loaded-Q optimisation for flat gradients.

### 3.7.6 LLRF system implementation

LLRF system implementation requires high-performance analog RF front-end electronics for conditioning and digitising RF signals and custom real-time algorithms running on high-performance FPGA-based digital processors. Details of the physical implementation (numbers of cavities and signal channels, numbers and locations of electronics racks, cable plant, etc.) are highly dependent on the layout and architecture of the main-linac RF power systems. Implementation of the LLRF systems for KCS and DKS are therefore covered in the site-specific sections of this chapter, in Section 3.9.4 and Section 3.8.4 respectively.

## 3.8 Main-linac layout for a mountainous topography

### 3.8.1 Introduction

This chapter describes the features of the Main Linacs that are unique to the mountainous topography site-dependent design, which utilises the Distributed Klystron Scheme (DKS) for RF distribution. In mountainous regions — such as those sites being discussed in Japan (Section 11.4) — the accelerator is orientated along the side of a valley, and access is provided via near-horizontal access ways. The lack of flat terrain requires that nearly all the equipment including the cryoplants be located in underground caverns. As noted in previous sections, DKS has the RF sources evenly distributed along the linac, and housed in the same tunnel. The modulators and klystrons (Section 3.6.2 and Section 3.6.3 respectively) are separated from the high-radiation environment of the accelerator by a concrete wall up to 3.5 m thick. The RF power from each 10 MW klystron directly feeds 39 cavities (1-1/2 ML units) via the local power distribution system described in Section 3.6.4.

### 3.8.2 Linac layout and cryogenic segmentation

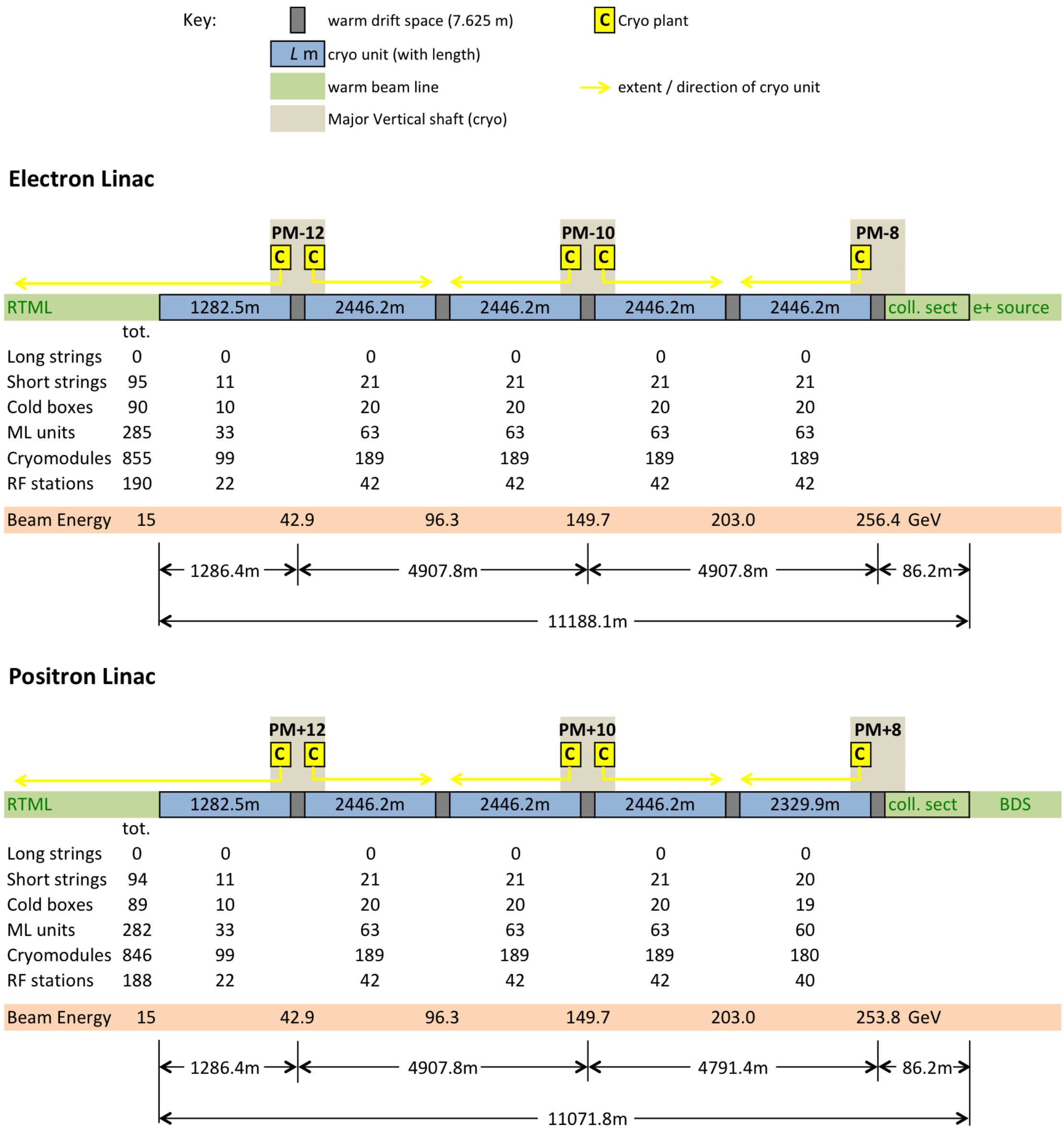

**Figure 3.38.** Schematic layout of the electron (top) and positron (bottom) main linacs for the mountainous topography site-dependent design, using DKS. The primary layout of the shaft arrangements are shown, along with the cryogenic segmentation. Distributions and totals (left-most column) of major linac sub-systems are given.

The main-linac layout is schematically shown in Fig. 3.38. The cryogenic cooling for each linac (and RTML bunch compressors, see Section 3.5) is provided by a total of five large cryoplants space approximately SI5km apart. The cryogenic plants are located underground and are accessible through long horizontal shafts as shown in Fig. 3.39. The cryogenic segmentation is constructed entirely from 'short cryostrings' (9 cryomodules or 2 ML units), as opposed to the more economical long strings (12 cryomodules). This is to accommodate a single design for the local power-distribution system (LPDS, see below). While use of short strings adds cold boxes and a small length increase to the linacs, this is offset by the benefits of having a single LPDS system (both in terms of manufacture and easier installation)

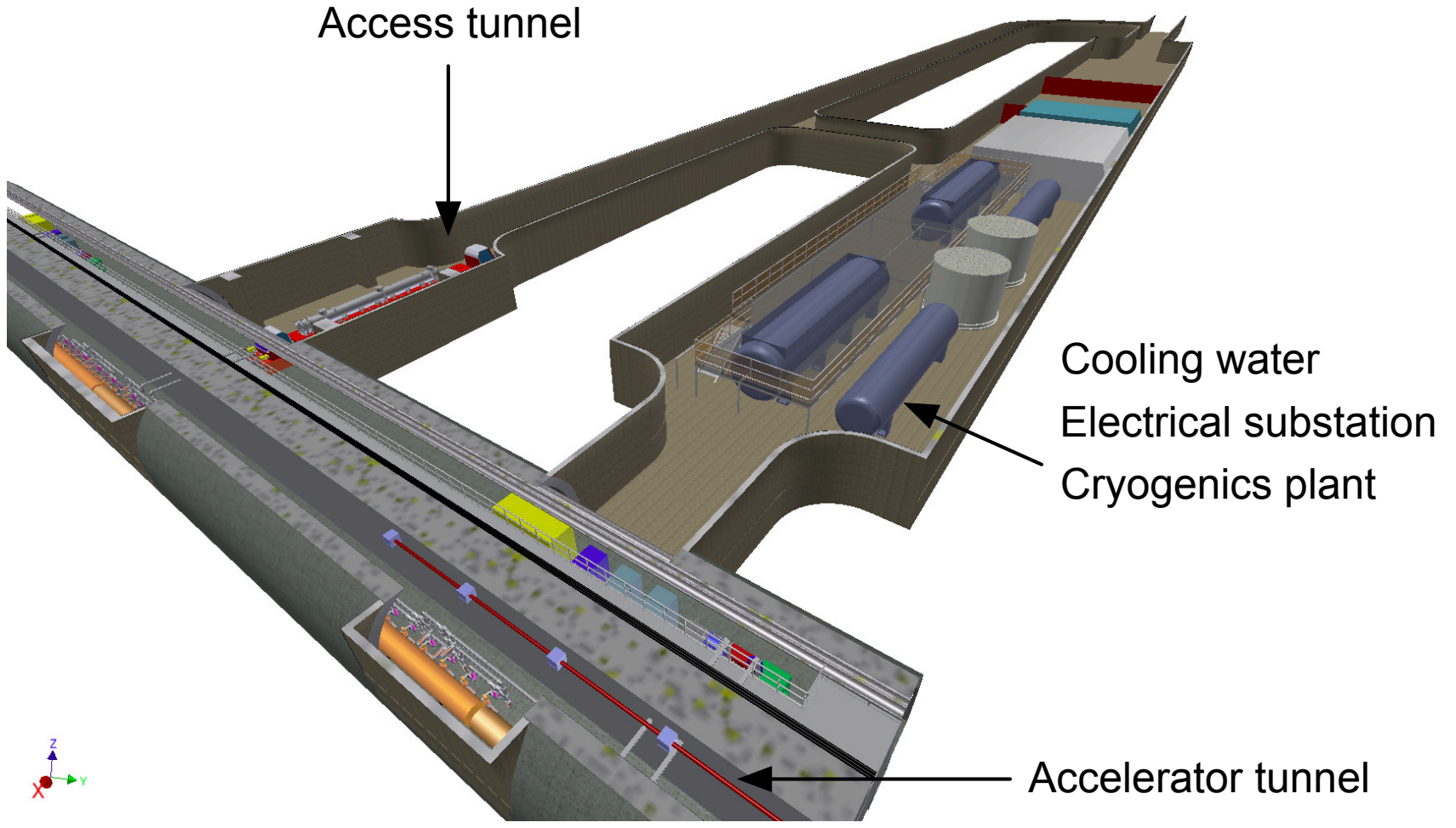

**Figure 3.39**
3D rendering of a cryo cavern and horizontal access way. For more details see Section 11.4.2.

The so-called 'Kamoboko' tunnel cross section is shown in Fig. 3.40. As noted above, this layout provides housing for the klystrons, modulators, electronics and related support infrastructure which is shielded from the radiation environment of the linac. The central-wall shielding is sufficient to permit personnel access to the service area during operations. RF power is brought in waveguides from the klystrons to the linac corridor through penetrations in the wall, which include a jog to prevent line-of-sight radiation. The shape of the tunnel is particularly suited for construction in the mountainous geology found in Japan (for more details see Section 11.4).

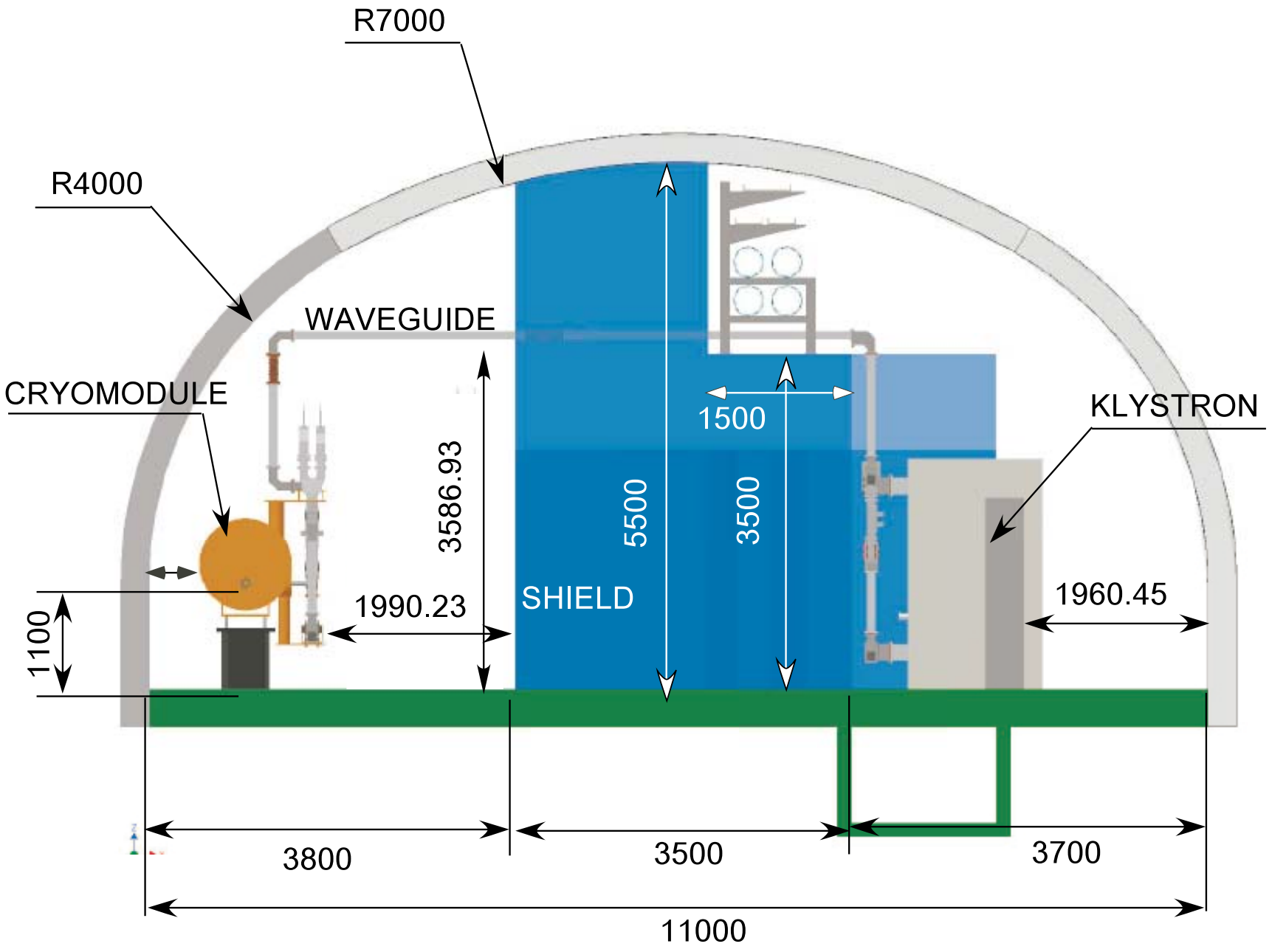

**Figure 3.40**
Cross section of the main-linac tunnel cross section for the mountainous topography site-dependent design.

## 3.8.3 The DKS high-power distribution System

Each unit of the RF system consists of a stand-alone RF source that powers 4 1/2 cryomodules (1-1/2 ML units), containing a total of 39 cavities — the maximum that can be realistically driven by a single 10 MW klystron (see Section 3.6.5).

There are a total of 378 RF sources, each comprising of a high-voltage Marx modulator, a 10 MW klystron, and a power-division waveguide circuit that feeds into the local power-distribution system (LPDS) as described in Section 3.6.

Figure 3.41 shows a schematic of a single DKS unit. The asymmetric layout (which is alternately reflected in each subsequent DKS unit) facilitates a relatively straightforward way to add the additional klystrons required for the luminosity upgrade (Section 12.3).

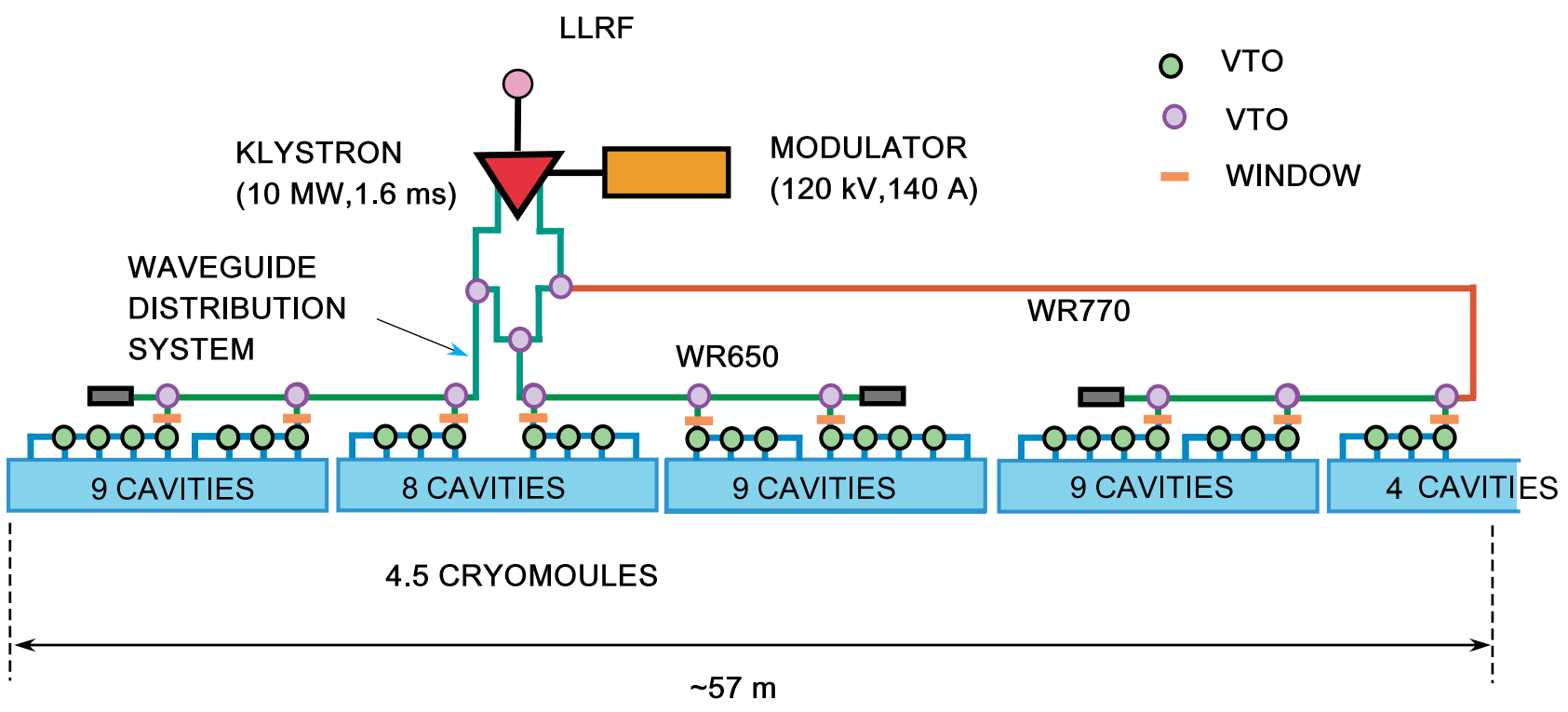

**Figure 3.41** Schematic layout of a DKS RF unit, showing a single klystron driving 39 cavities (1-1/2 ML units).

A unique feature of the DKS approach is the klystron power-division circuit, i.e. the waveguide system connecting the klystron to the local power-distribution system. Three basic LPDS units (13 cavities, see Section 3.6.4) are fed from the two klystron outputs. Power from each output port is first split with a roughly 2:1 ratio through an H-type hybrid. The lower-power outputs from the hybrids are then combined through a T-type hybrid. Two of the resulting three feeds are fed locally through the shield wall; the third runs along the corridor to the location where an upgrade klystron would be situated, and then through one of the shield-wall penetrations prepared for that upgrade klystron. For this 34 m run, the waveguide size is stepped up from WR650 to WR770 to reduce transmission losses. The splitting and combining circuit and the waveguide layout are illustrated in Fig. 3.42.

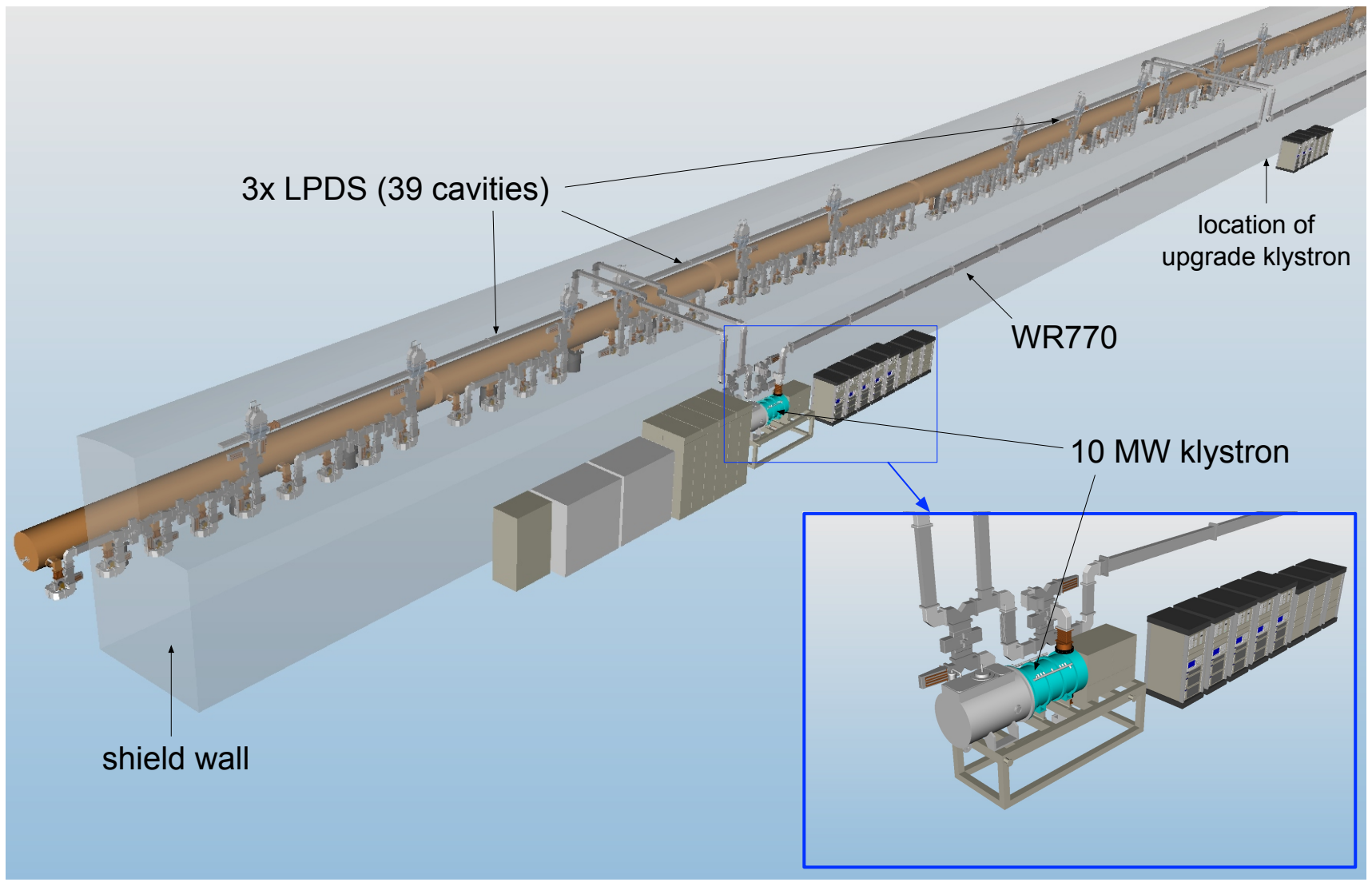

**Figure 3.42** The DKS arrangement in the main-linac tunnel for the mountainous topography. One DKS unit (39 cavities) is shown.

### 3.8.4 LLRF control for DKS

LLRF requirements and design concepts were described in Section 3.7. This section describes the LLRF-system implementation specific for the DKS main-linac configuration.

The DKS layout and scale are very similar that planned for the European XFEL, which uses single 10 MW klystrons to drive groups 32 cavities. Originally proposed for TESLA, this configuration has been the subject of extensive R&D, and in particular there is many years of FEL operations experience from FLASH, as well as the more dedicated studies with ILC-like beams (see Part I Section 3.2).

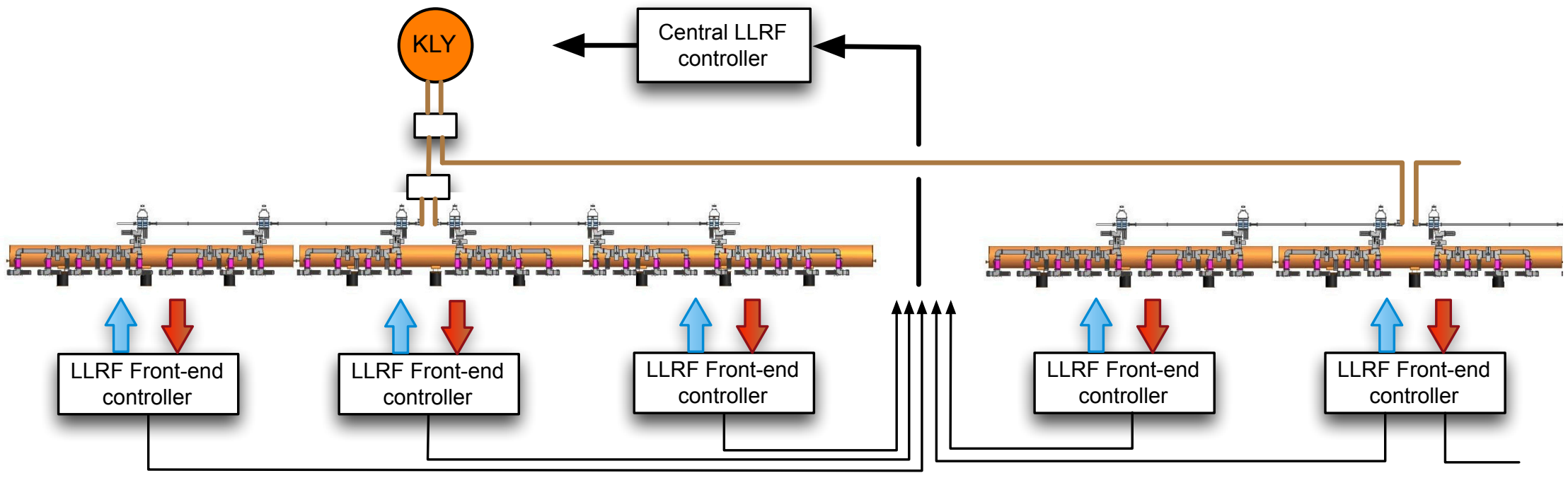

**Figure 3.43.** Implementation block diagram for the DKS LLRF system

An implementation block diagram for the DKS is shown in Fig. 3.43. Front-end LLRF controllers contain the analog interfaces, downconverters and digitisers for the field probe, forward and reflected RF power signals from each cavity. They also provide the control and monitoring interfaces to the cavity mechanical tuners, piezo tuners, cavity input coupler, and RF power dividers on the LPDS Fig. 3.1.

Each front-end controller computes in real-time a partial cavity-field vector sum for its respective cryomodules and sends it to the master LLRF controller over dedicated synchronous data links and fast Ethernet. The master controller, located close to the klystron, performs the vector-sum regulation, exception handling, and overall system coordination. The master controller communicates with the ILC global control system and with high-level applications such as linac energy and energy-profile management.

In addition to computing partial vector sums, the front-end controllers implement the algorithms and control functions associated with individual cavities or cryomodules (see Section 3.7.4).

The relatively short distances (few tens of metres) between cryomodules and klystron mean that cable delays can be kept short and control-loop delays short enough that the LLRF controller can respond almost from bunch to bunch. This allows, for example, fast compensation of jitter at the beginning of the bunch trains.

While Fig. 3.43 shows the front-end controllers located inside the beam enclosure close to each cryomodule, the split-tunnel arrangement of the Kamaboko tunnel offers the possibility of locating the front-end controllers either inside the beam enclosure next to the cryomodule or in the service corridor adjacent to the respective cryomodule. The latter approach has two important benefits: first, sensitive electronics are kept out of the high-radiation environment; second, electronics racks are accessible for service or repair without having to open up the beam enclosure. These benefits must be traded against longer RF-signal cables that must run through the penetrations in the beam enclosure shield wall. A final decision on the locations of electronics crates and cable penetrations will be made during the detailed design phase.

## 3.9 Main-linac layout for a flat topography

### 3.9.1 Introduction

This section describes the primary features of the Main Linacs that are specific to the flat-topography site-dependent design including the use of the Klystron Cluster Scheme (KCS) for the RF power distribution, and the corresponding impact on CFS, cryogenic segmentation and number of cryo plants, as well as the low-level RF control system (LLRF).

As already noted, KCS represents a novel approach to transporting the RF power into the accelerator tunnel. The klystrons and modulators are installed on the surface in 22 groups or 'clusters' (11 per linac). Each cluster contains 18 or 19 klystrons, the combined RF power of which (180–190 MW) is transported via a large 0.48 m-diameter overmoded cylindrical waveguide, first down a vertical shaft into the accelerator tunnel, and then along approximately 1 km of linac, where it drives ∼600 cavities (Fig. 3.44). At every ML unit (26 cavities, or every ∼ 38 m), about 7 MW of power is taped-off from the main KCS waveguide via a specially developed Coaxial Tap-Off (CTO). The CTO is connected directly to the local power-distribution system of the associated ML unit (Section 3.6.4).

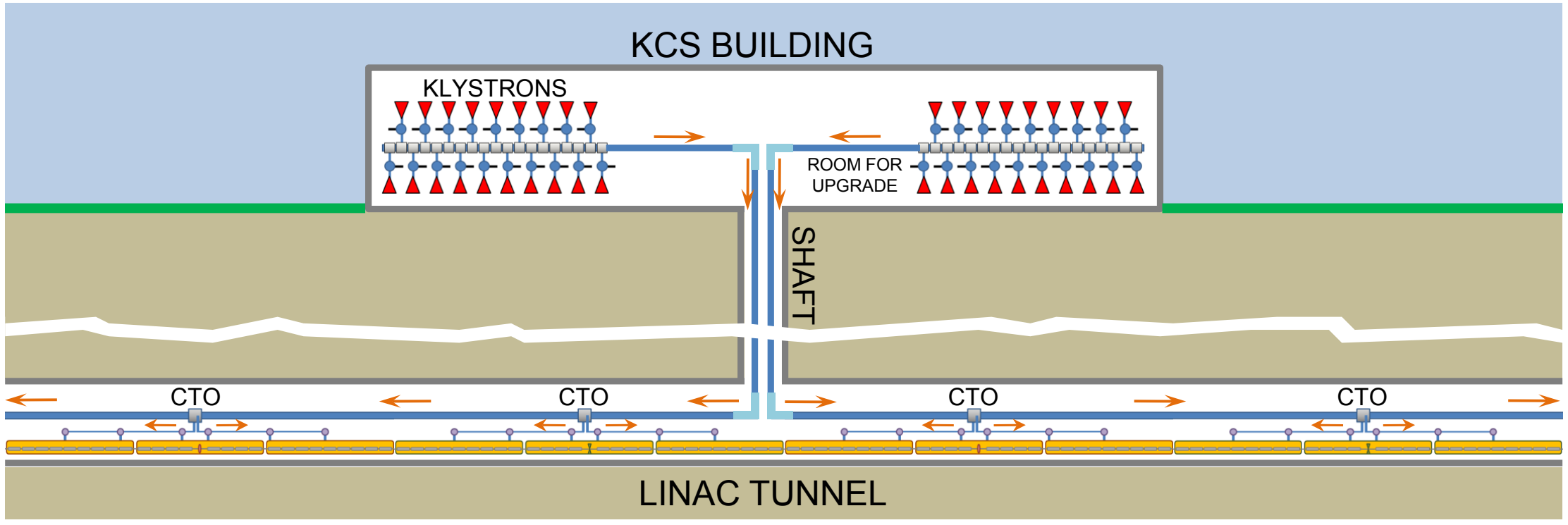

**Figure 3.44.** Schematic illustration (not to scale) of the klystron cluster scheme (KCS) for providing RF power to the main linacs. The upstream and downstream waveguides each extend roughly a kilometre.

The primary advantage of KCS is the removal of the RF power generation and all associated head loads from the underground accelerator tunnel, thus reducing the required tunnel volume, while at the same time easing the requirements on the air and water cooling systems, both of which result in a reduced cost. This must be countered by the need for additional shafts and surface buildings as well as additional waveguide systems and klystron overhead to compensate the additional associated losses. Technical issues with transporting such a high power in a single waveguide and associated components have been the subject of R&D during the Technical Design Phase (See Part I Section 2.8.6), and significant progress has been made in demonstrating individual components. Although further R&D is required, the results so far have proven sufficiently promising to justify adopting the approach for the flat-topography site-dependent baseline.

### 3.9.2 Linac layout and cryogenic segmentation

The most immediate and obvious impact of KCS is on the civil engineering and the need for additional shafts, as shown in Fig. 3.45. Without the need to house the modulators and klystrons underground, a single 5 m-diameter tunnel can be used as shown in Fig. 3.46, suitable for construction with a tunnel boring machine.

Figure 3.47 schematically shows the main-linac configuration, indicating both the KCS configuration and the cryogenic segmentation of the cryomodules and cryoplants. Along each main linac are 6 shafts and KCS buildings containing 11 KCS systems. All but the last building for each linac house two systems, one feeding upstream and one feeding downstream. All but five systems power 26 ML Units ($26 \times 26 = 676$ cavities). The last KCS in the electron linac (11th) and the 7th through 10th in the positron linac power 25 ML Units (650 cavities), as indicated in Fig. 3.47. This non-symmetric situation is due to the three additional ML units in the electron linac, required to provide the additional 2.6 GeV to drive the positron-source undulator (Chapter 5).

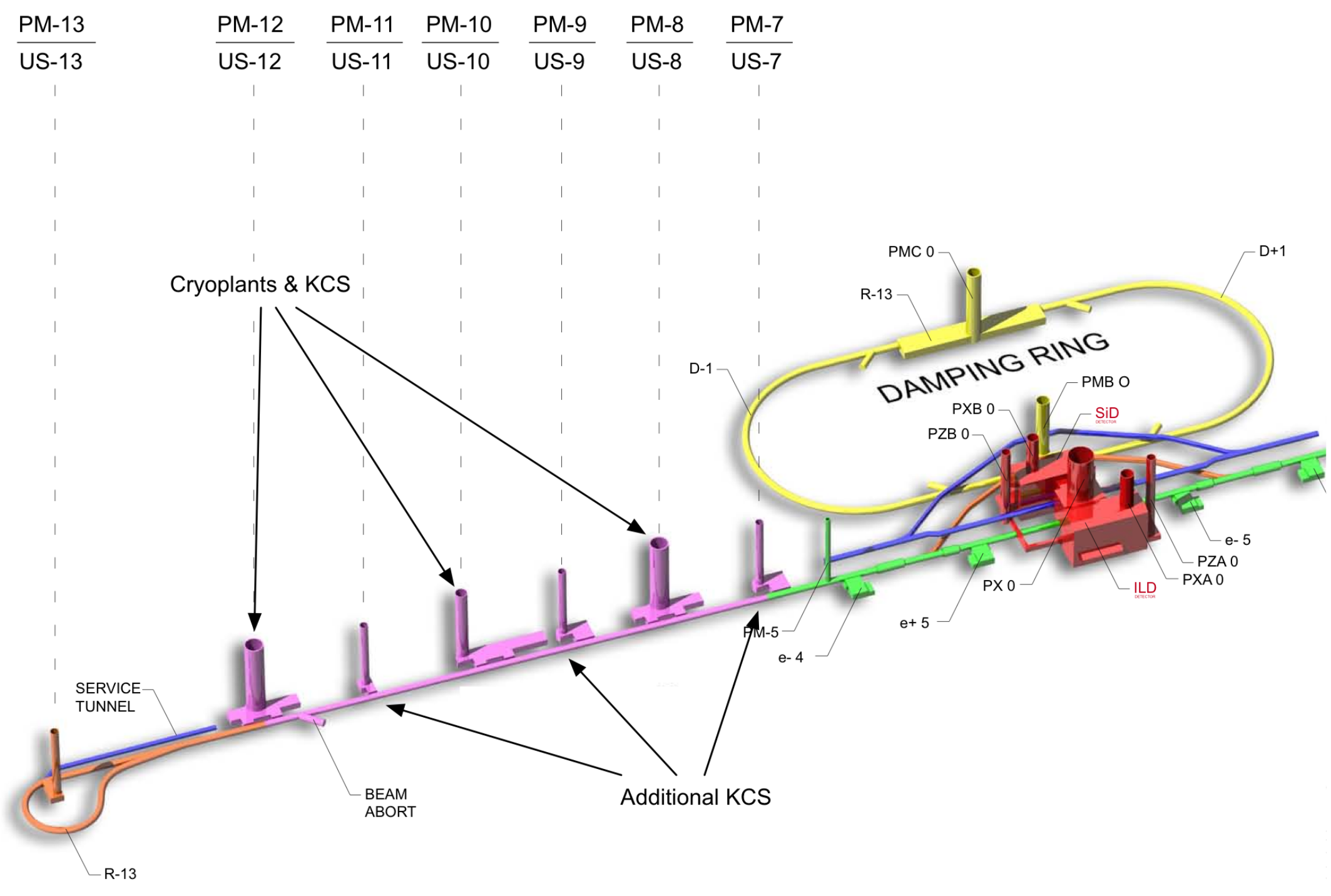

**Figure 3.45.** Schematic illustration (not to scale) of the layout for the flat-topography site-dependent design. The additional shafts required by KCS are indicated.

**Figure 3.46**
Sketch of the cross section of the Main-Linac tunnel for the flat-topography site-dependent design. The KCS cylindrical overmoded waveguide runs along the top of the tunnel as indicated.

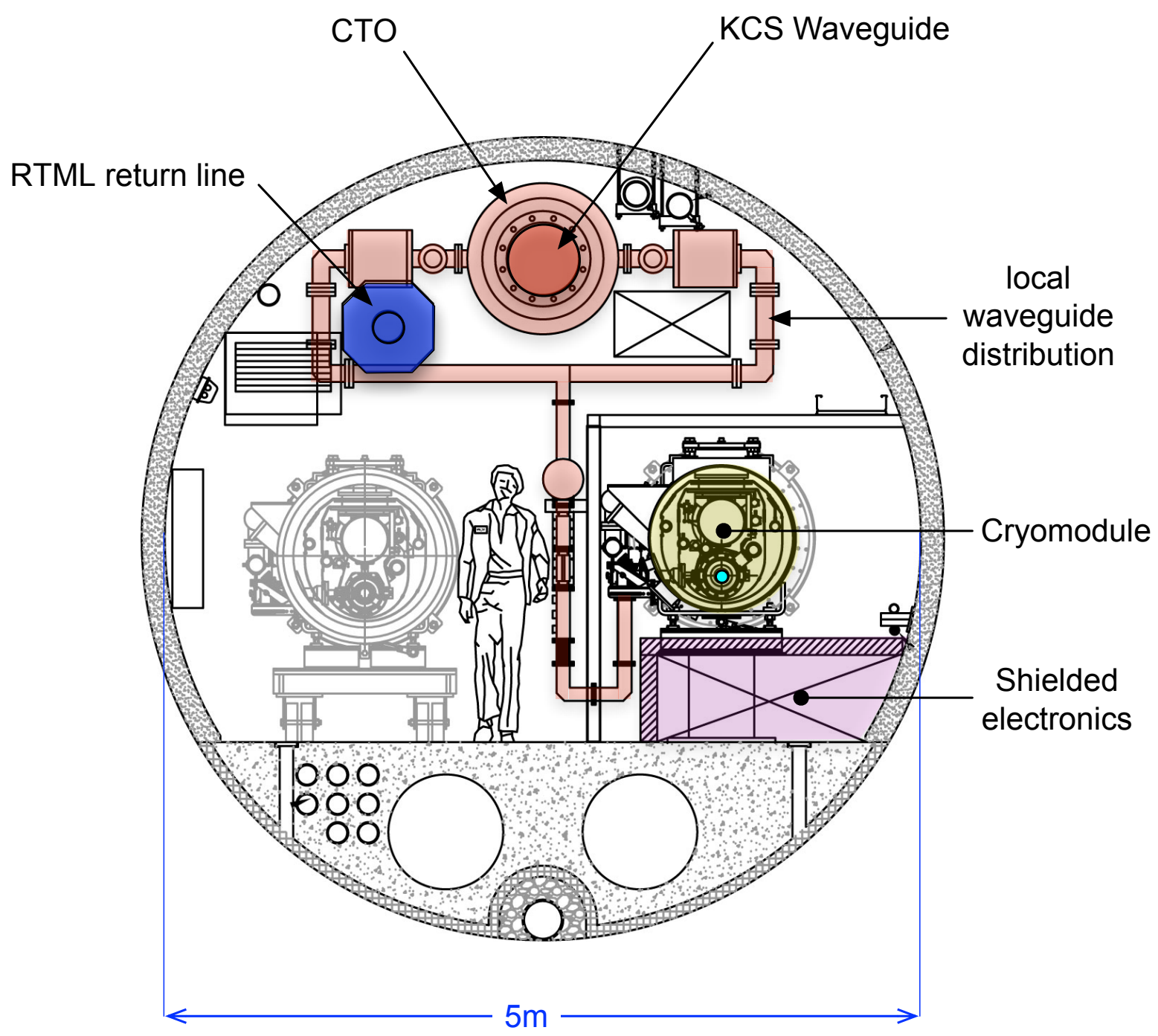

The total cryogenic load of each linac (see Section 3.5) is cooled by six cryoplants located at the major shafts (PM±8, PM±10 and PM±12 in Fig. 3.47), shared with the six klystron clusters at those locations. Although in principle five ~4 MW plants are sufficient, the use of six plants provides an optimum use of the shaft spacing required by KCS (i.e. approximately 2.5 km), constrained by the maximum practical length of the main KCS waveguide.

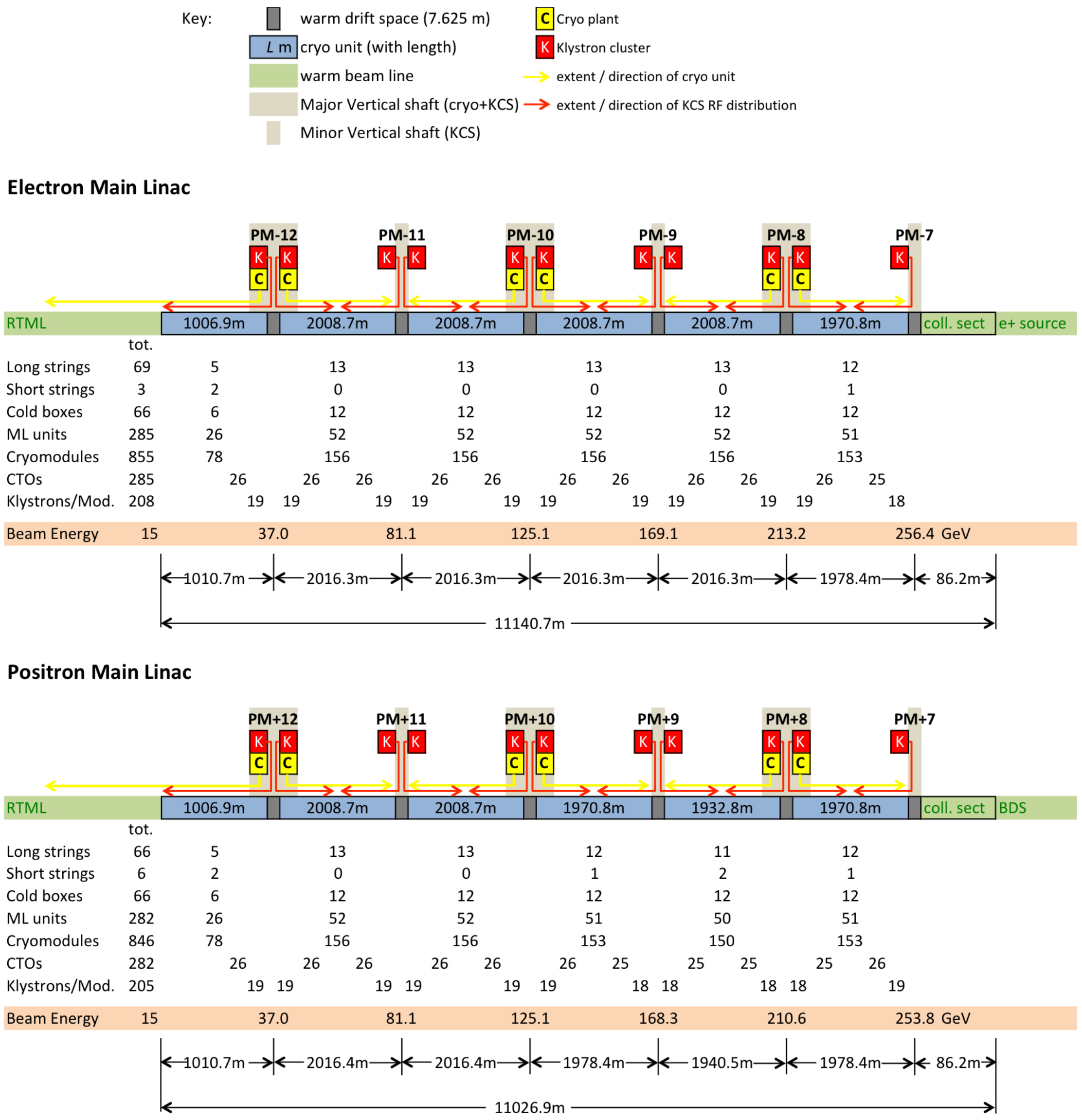

**Figure 3.47.** Schematic layout of the electron (top) and positron (bottom) main linacs for the flat-topography site-dependent design using KCS. The primary layout of the shaft arrangements are shown, along with the cryoplant and KCS segmentation. Distributions and totals (left-most column) of major linac sub-systems are indicated. The choice of six cryo plants is driven in part by compatibility with the KCS RF distribution.

## 3.9.3 The KCS high-power distribution system

Both for high-power handling and for minimal-loss transport over large distances, an overmoded circular waveguide operated in the $TE_{01}$ mode is the preferred RF conduit. It has no surface electric field, and its attenuation constant, at sufficient radius, becomes the lowest. For the main KCS waveguide, a 0.480 m-diameter (WC1890) aluminium pipe is used, pressurised to 2 bar above atmosphere to suppress breakdown. With copper plating for improved surface conductivity, it presents a theoretical attenuation loss of 0.383 dB/km ($\sim$8.44%/km). Because it is overmoded, the tolerances on the circular cross-section of this waveguide and its straightness must be kept fairly tight to avoid buildup of parasitic RF modes. In particular, the circular waveguide is specified to be round to $\pm$0.5 mm (which has been met for the 80 m of bored circular waveguide that has been built). Also, the concentricity of the mating sections for the circular waveguide has a similar tolerance, which is met by using the outer flange surface as a reference. The overall straightness of each KCS waveguide segment should be maintained within a degree or so from beginning to end to limit conversion to the degenerate $TM_{11}$ mode.

The design of the CTO is illustrated in Fig. 3.48. A gap in the inner wall, where the diameter steps above the $TE_{02}$ cutoff, couples a fraction of the power into a surrounding coaxial waveguide. At the termination of the latter, this flowing power is coupled through a set of openings in its outer wall into a wrap-around waveguide and finally through two radial rectangular WR650 ports. As the KCS power is gradually depleted, each successive CTO requires a different coupling. This is achieved via changes in the gap length and matching ridge. With a properly spaced end short, the final CTO becomes a full extractor. Ceramic block windows on the CTO rectangular ports isolate the pressure envelope of the KCS main waveguide. A short, double-stepped taper is used before and after the CTO to match the main KCS 0.48 m circular waveguide to the 0.35 m diameter CTO ports.

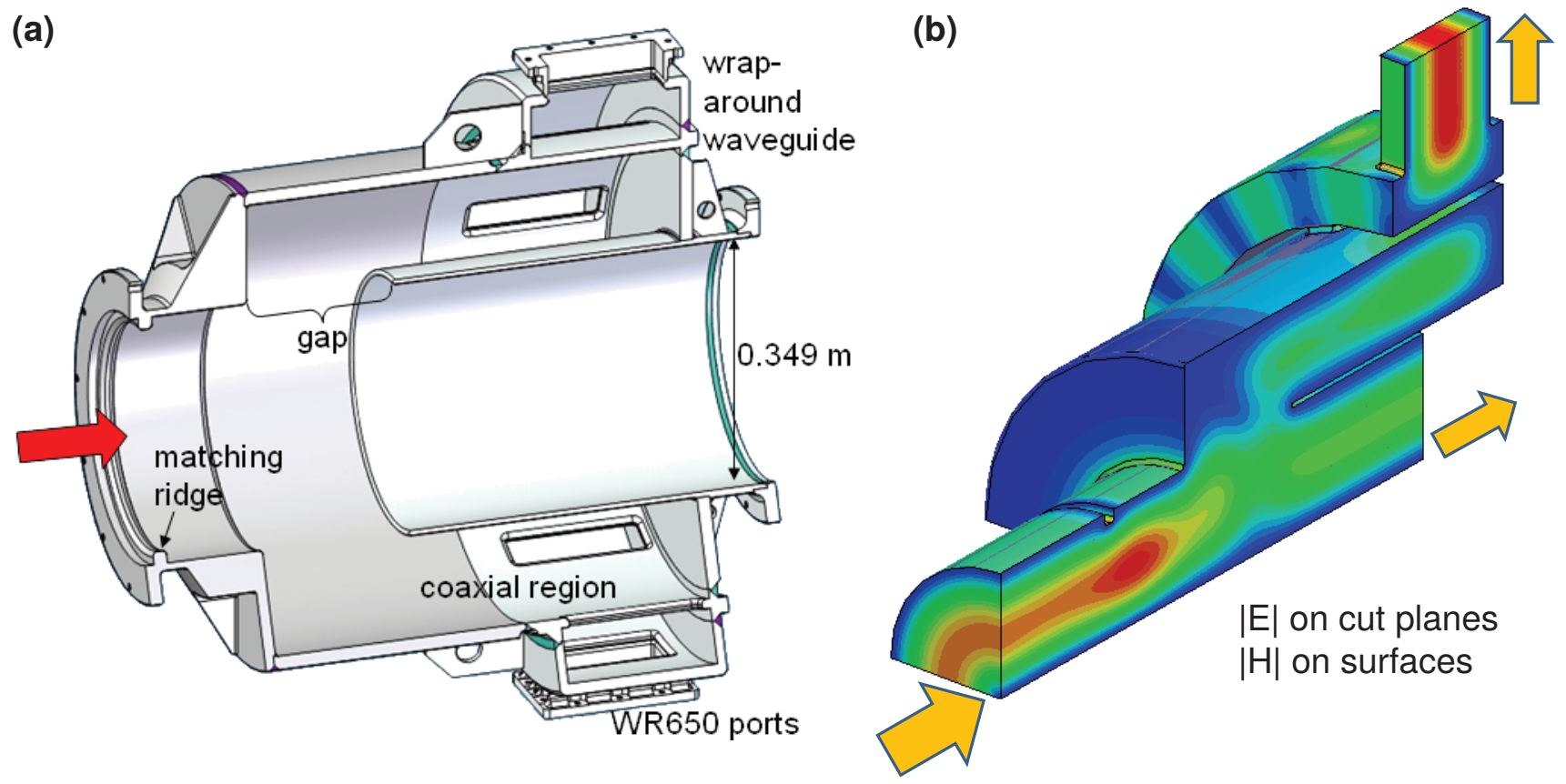

**Figure 3.48** a) cutaway and b) simulated field patterns of a Coaxial Tap-Off (CTO), designed to extract (inject) fractional RF power from (into) a flowing $TE_{01}$ wave in the circular KCS main waveguide. Many different couplings, controlled by gap width, are required; shown is a 3 dB design.

The same kind of waveguide circuit is used in reverse for combining power in the surface KCS buildings of the 18 or 19 klystrons in each cluster, with the CTO's spaced much more closely and no step tapers until the end. For their power to combine effectively in the passive circuit, the klystrons must all be run at the proper power with the proper relative phases. If power from a single 10 MW source drops out, a similar amount of power will be directed out from its CTO back toward that klystron. Thus a 5 MW isolator is required on each of the klystron outputs. For compactness, power can be fed alternately from klystrons arrayed on either side of the combining network.

Bringing the combined KCS RF power from the surface down to and along the tunnel requires navigating two or three 90° bends. Bending in overmoded waveguide is non-trivial, as modes tend to be coupled. A $TE_{01}$ mode L-band bend, shown in Fig. 3.49, has been designed for this purpose. With ports the same diameter as the CTO's, it consists of a pair of mode converters to a single-polarization $TE_{20}$ mode in a rectangular cross-section on either side of a sweep bend designed to preserve the latter [50]. Tapers to WC1890 are used to minimise losses in the shafts and in any other significant runs.

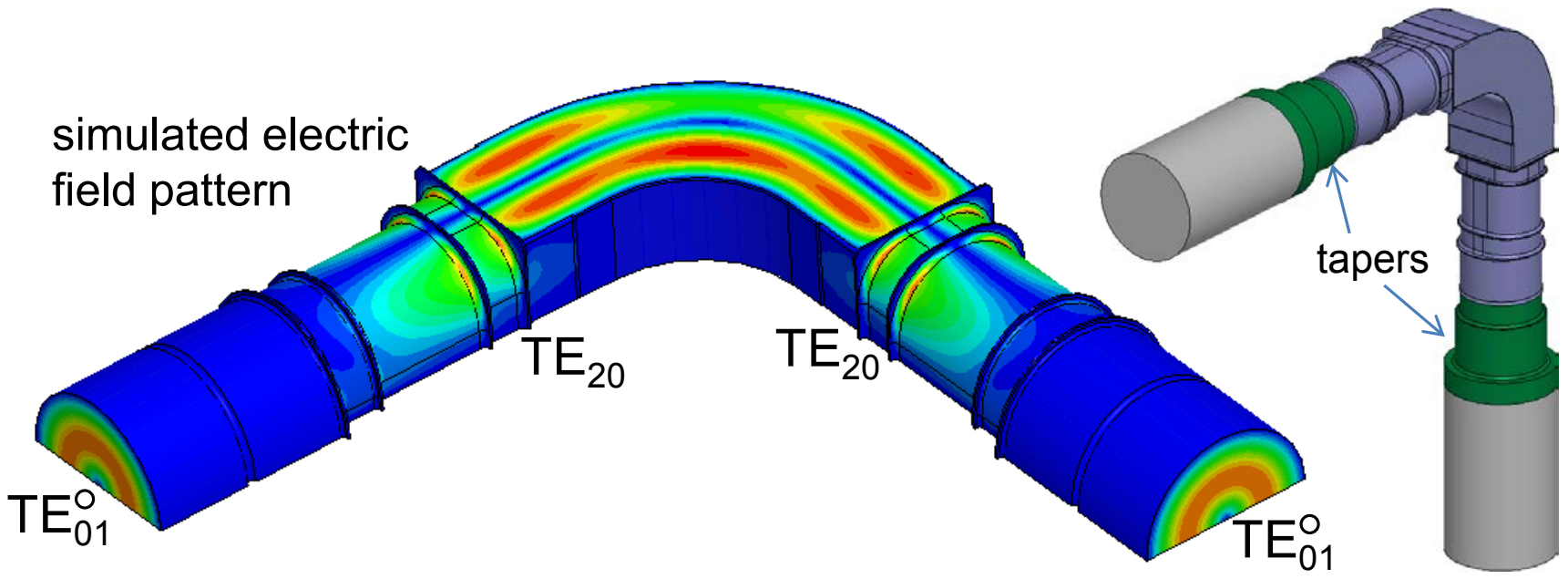

**Figure 3.49** Overmoded bend for the circular $TE_{01}$ mode KCS main waveguide. Mode converting sections allow the actual bending to be done in the rectangular $TE_{20}$ mode. WC1375 ports connect to WC1890 through step-tapers.

Given that each KCS involves a large network of waveguide that includes 18 or 19 combiners and 25 or 26 tap-offs, the RF match of the CTOs and other components (bends and tapers) has to be very good to avoid significant losses. The power losses for KCS listed in Table 3.18 assumes the RF matches, especially those for the CTOs, are very good, and that the klystron RF amplitudes/phases can be precisely controlled to achieve optimal combining (e.g. by minimising reflections back to the klystrons). For the CTOs, this will likely require adding features that allow fine tuning of the match during cold-test setup.

From a circuit standpoint, the RF power the KCS pipe is back-terminated through the klystron isolators and forward-terminated through the cavity isolators, so misdirected power (e.g. from a breakdown) will be readily absorbed. Since the klystrons are isolated, they can be safety turned off when others are running, and as with DKS, the pressurised variable power dividers in the LPDS can be used to zero the power that goes to groups of 4-5 cavities if needed. From a safety standpoint, there are windows on the CTO ports to isolate the $N_2$ so the entire KCS pipe is not vented if there is a vent in the klystron feed lines or in one of the LPDS's.

### 3.9.4 LLRF control for KCS

The principle mechanisms of regulating the vector sum field of many cavities driven from a single RF source apply equally to the KCS configuration as to the RDR, FLASH, and XFEL layouts, and to that of the DKS. However, as will be explained, the unique features of the KCS layout apply some important additional constraints and functional requirements on the LLRF control system. resulting from the unique features very large number of cavities, long distances and correspondingly long cable delays, tapped RF power distribution and the use of a cluster of klystrons as the RF source.

At first glance, the implementation block diagram for KCS LLRF system in Fig. 3.50) is very similar to that of the RDR and the DKS system described in Section 3.8.4. Front-end LLRF controllers contain the analog interfaces, downconverters and digitisers for the field probe, forward and reflected RF power signals from each cavity. They also provide the control and monitoring interfaces to the cavity mechanical tuners, piezo tuners, cavity input coupler, and RF power dividers on the Local Power Distribution System (see figure Fig. 3.1 power dividers, and cavity input couplers). The front-end electronics are located in a radiation-shielded and temperature-regulated rack under each cryomodule.

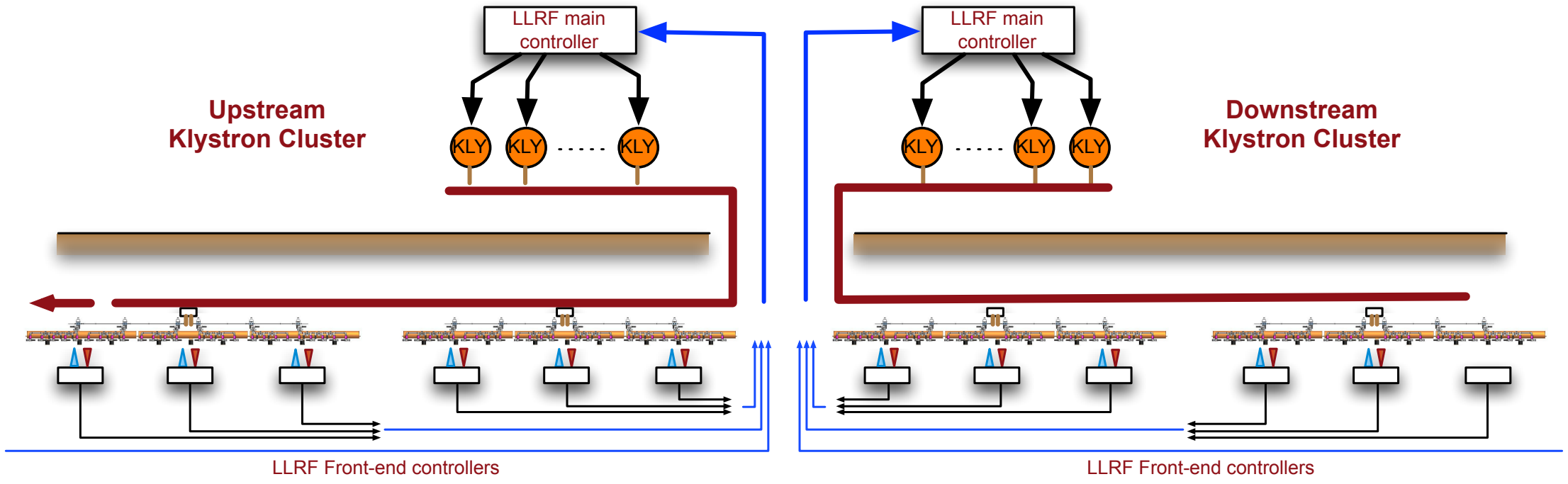

**Figure 3.50.** Implementation block diagram for the DKS LLRF system

Each front-end controller computes in real time a partial cavity field vector sum (i.e. the sum for those cavities in the cryomodule), and sends it to the master LLRF controller located in the surface building for the associated klystron cluster. This is done over dedicated synchronous data links and fast Ethernet. The master controller, located close to the klystrons, performs the full vector-sum addition, klystron drive waveform generation including feedback and feedforward corrections, exception handling, and overall system coordination. The master controller communicates with the ILC global control system and with high-level applications such as linac energy, energy profile management and beam current feedforward (i.e. RF waveform corrections based on the measured current profile in the associated damping ring prior to the pulse).

In addition to computing partial vector sums, the front-end controllers implement the algorithms and control functions associated with individual cavities or crymodules such as the motor control of the power dividers, phase shifters and couplers (see Section 3.7.4).

#### 3.9.4.1 Vector-sum regulation

The very long distance between the klystron cluster and the furthest cryomodule results in a delay of more than 4 μs from the time a change is made to the RF forward power to the time the change is seen by the furthest cavity and it then takes a further 4 μs for the change on the cavity field probe to be detected by the front-end controller and sent back to the LLRF main controller. This 8 μs total round-trip delay reduces closed-loop stability margins by $\sim$3 degrees per 1 kHz, reducing the maximum gain and bandwidth. Transport delays of the beam itself also play a role in the regulator dynamics and are different for the upstream and downstream clusters depending whether the RF power is traveling in the same direction of the opposite direction as the beam. Conversely, transport delays with respect to the closest cryomodules are less than a microsecond, and hence have no significant impact on regulator stability margins.

This range of transport delays over the cavity string results in a rather complex timing relationship between the beam, RF power, and RF signals, and these must be taken into account in the design of the vector-sum controller. A MIMO (multi-input/multi-output) optimal regulator design approach, where each partial vector sum is treated as a separate input, is likely to yield a regulator performance (gain-bandwidth) that is significantly better than would be achieved if the closed-loop performance were entirely dictated by the longest transport delays.

The impact of the reduced regulator performance depends largely on the environmental conditions around the main linacs (ground vibration, power line disturbances, microphonics induced by the flow of cryogenic fluids, etc.). Provided the environment is quiet, then the reduced jitter attenuation will still be sufficient to keep the cavity field stability within specification. Should it prove necessary, fast regulation of the total linac energy could be achieved by configuring a few of the cryomodules at the high energy end of the linacs in the RDR of DKS layout and use them to provide faster fine-tuning of the linac final energy.

#### 3.9.4.2 Cavity-level algorithms

At the cavity and cryomodule level, there is an additional dimension to control of the local power distribution system, which in the case of KCS includes resistive loads to allow some of the power from the CTO to be diverted from the cavities, effectively giving some fine tuning of the total RF power to the sum of all cavities in the three-cryomodule cavity string. This allows fine tuning of the RF power that is fed to the 26 cavities in that local string

#### 3.9.4.3 RF power source control

At the klystron cluster level, additional supervisor control functions are required to monitor and balance the RF power amongst all the klystrons. The total klystron power will be regulated in both the phase and amplitude, the latter which can be achieved by varying only the relative phase between banks of klystrons. There may also be a slow feedback loop on each klystron to optimise the match to the upstream power by minimising the reflected power to the klystron when trying to achieve the maximum beam energy.

At the local power distribution level, the input forward power is determined by the power ratio of the Coaxial Tap-off. An additional control knob on the local power distribution system allows the total power to the 26-cavity string to be fine-tuned by diverting some of the power from the CTO to a separate RF load.

#### 3.9.4.4 Klystron cluster LLRF system tests

While it is impractical to build a dedicated linac facility for testing the KCS, it may be possible to test LLRF control algorithms for the klystron cluster at the European XFEL, which has a similar number of cavities over similar distances as a klystron cluster in the TDR baseline design. The large-scale vector-sum control could be emulated by adding a 'supervisory' LLRF controller that communicates with the XFEL LLRF systems.

# Chapter 4
# Electron source

## 4.1 Overview

The ILC polarized electron source must produce the required train of polarized electron bunches and transport them to the Damping Ring. The nominal train is 1312 bunches of $2.0 \times 10^{10}$ electrons at 5 Hz with polarization greater than 80 %. For low energy ILC operation ($E \leq$150 GeV /beam), the source is required to run at 10 Hz. The beam is produced by a laser illuminating a photocathode in a DC gun. Two independent laser and gun systems provide redundancy. Normal-conducting structures are used for bunching and pre-acceleration to 76 MeV, after which the beam is accelerated to 5 GeV in a superconducting linac. Before injection into the damping ring, superconducting solenoids rotate the spin vector into the vertical, and a separate superconducting RF structure is used for energy compression.

The SLC polarized electron source already meets the requirements for polarization, charge and lifetime. The primary challenge for the ILC source is the long bunch train, which demands a laser system beyond that used at any existing accelerator, and normal conducting structures which can handle high RF power. R&D prototypes have demonstrated the feasibility of both of these systems [60, 61].

## 4.2 Beam Parameters

The key beam parameters for the electron source are listed in Table 4.1.

**Table 4.1**
Electron Source system parameters.

| Parameter | Symbol | Value | Units |
|---|---|---|---|
| Electrons per bunch (at gun exit) | $N_{-}$ | 3×10$^{10}$ | Number |
| Electrons per bunch (at DR injection) | $N_{-}$ | 2×10$^{10}$ | Number |
| Number of bunches | $n_b$ | 1312 | Number |
| Bunch repetition rate | $f_b$ | 1.8 | MHz |
| Bunch train repetition rate | $f_{rep}$ | 5 (10) | Hz |
| FW Bunch length at source | $\Delta t$ | 1 | ns |
| Peak current in bunch at source | $I_{avg}$ | 3.2 | A |
| Energy stability | $\sigma_E/E$ | <5 | % rms |
| Polarization | $P_e$ | 80 (min) | % |
| Photocathode Quantum Efficiency | QE | 0.5 | % |
| Drive laser wavelength | $\lambda$ | 790±20 (tunable) | nm |
| Single bunch laser energy | $u_b$ | 5 | μJ |

## 4.3 System Description

Figure 4.1 depicts schematically the layout of the polarized electron source. The key beam parameters are listed in Table 4.1. Two independent laser systems are located in a surface building. The light is transported down an evacuated light pipe to the DC guns. The beam from either gun is deflected on line by a magnet system which includes a spectrometer, and it then passes through the normal-conducting subharmonic bunchers, travelling wave bunchers and pre-accelerating sections. This is followed by the 5 GeV superconducting linac. The SC linac has 8 10MW klystrons each feeding 3 cryomodules, giving 24 cryomodules, 21 required and 3 spares. The Linac-to-Ring transfer line that brings the beam to the damping rings provides spin rotation and energy compression.

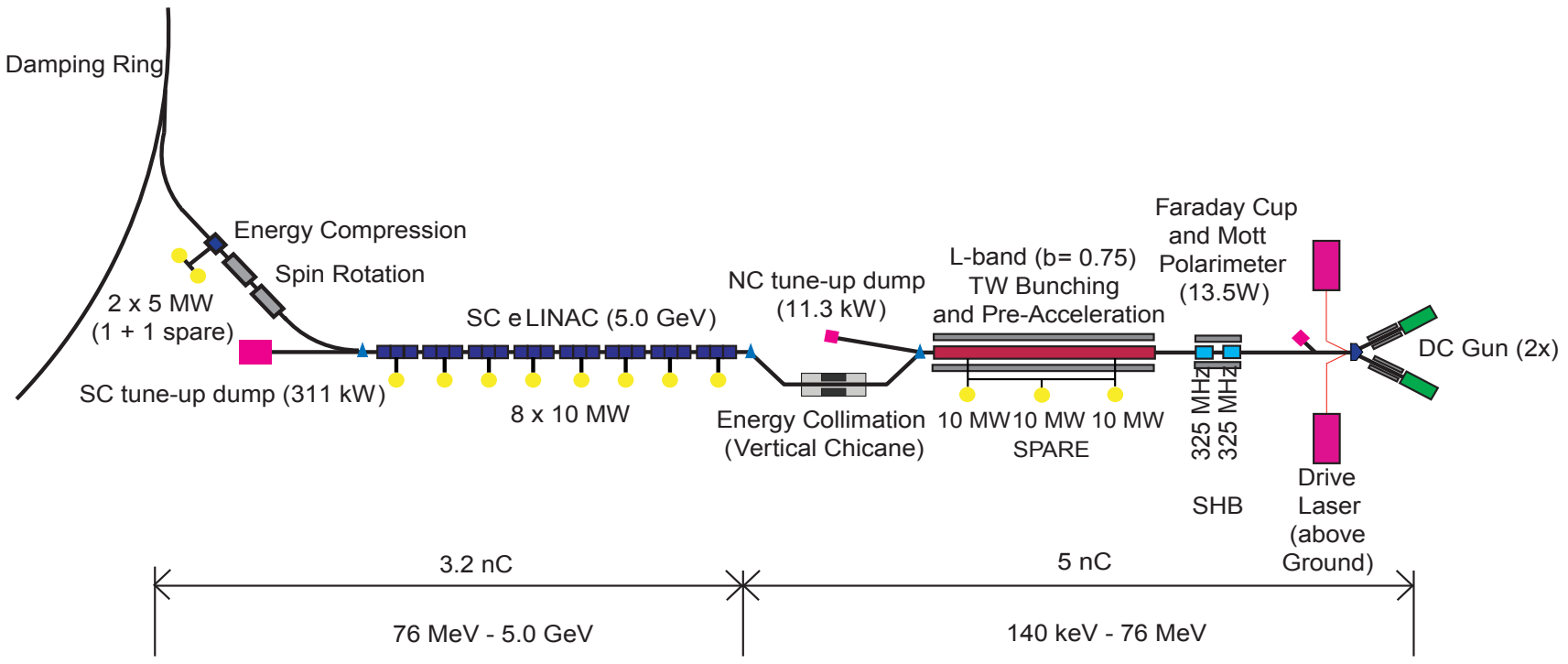

**Figure 4.1** Schematic view of the polarized Electron Source.

### 4.3.1 Photocathodes for Polarized Beams

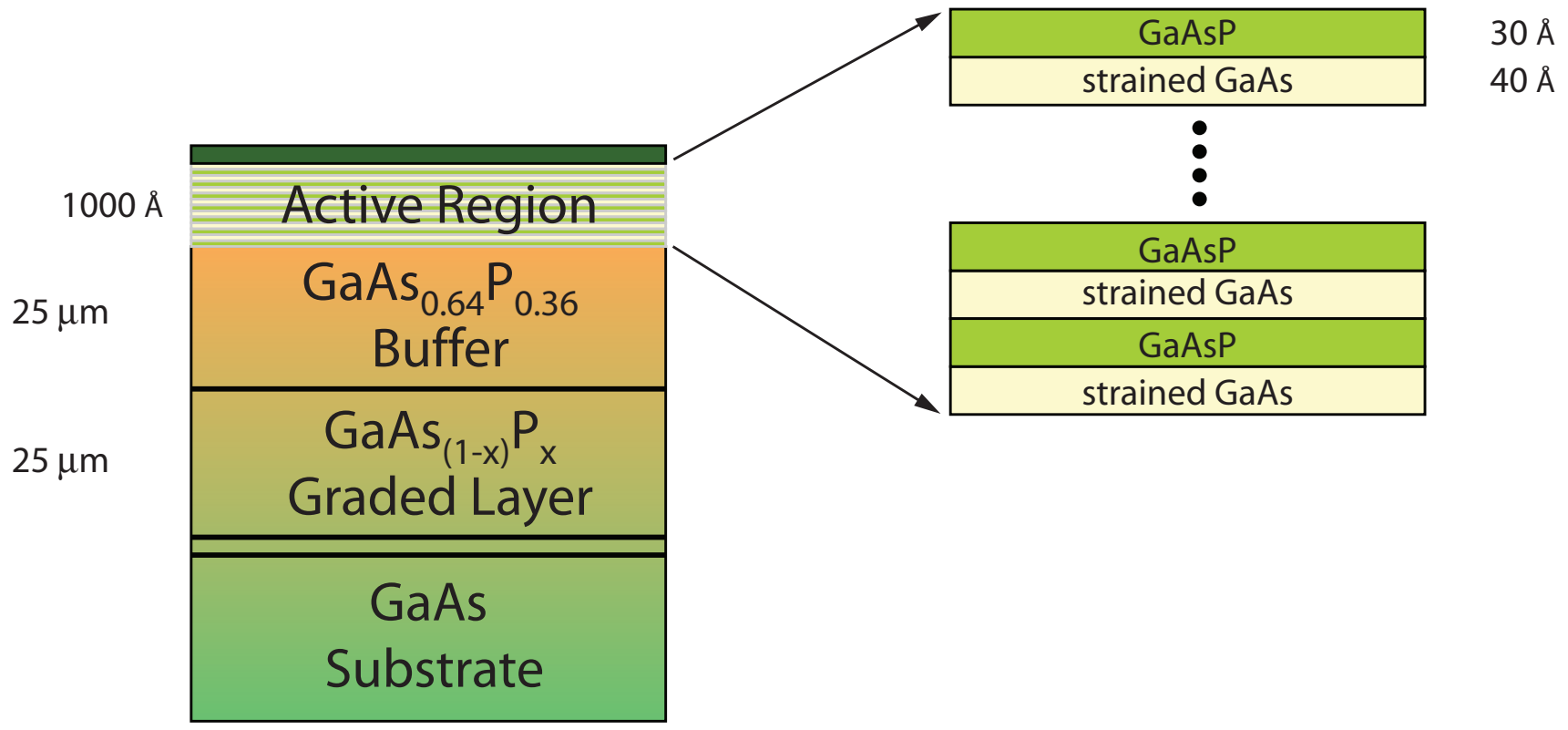

**Figure 4.2** Structure of a strained GaAs/GaAsP superlattice photocathode for polarized electrons.

Photocathode materials have been the subject of intense R&D efforts for more than 20 years. The most promising candidates for the ILC polarized electron source are strained GaAs/GaAsP superlattice structures (see Fig. 4.2). GaAs/GaAsP superlattice photocathodes routinely yield at least 85 % polarization with a maximum QE of ∼1 % (routinely 0.3 to 0.5 %) [62–64]. The present cathodes consist of very thin quantum-well layers (GaAs) alternating with lattice-mismatched barrier layers (GaAsP). Each layer of the superlattice (typically 4 nm) is considerably thinner than the critical thickness (∼10 nm) for the onset of strain relaxation, while the transport efficiency for electrons in the conduction band can still be high [65]. The structures are p-doped using a high-gradient doping technique, consisting of a thin (10 nm), very highly doped ($5 \times 10^{19}\mathrm{cm}^{-3}$) surface layer with a lower density doping ($5 \times 10^{17}\mathrm{cm}^{-3}$) in the remaining active layer(s). A high-surface doping density is necessary to achieve high QE while reducing the surface-charge-limit problem [66, 67]. A lower doping density is used to maximize the polarization [68]. With bunch spacing of ∼500 ns, the surface-charge-limit problem for the ILC is not expected to be a major issue. The optimum doping level remains to be determined. An alternative under study is the InAlGaAs/GaAs strained superlattice with minimum conduction band offset where a peak polarization of 91 % has been observed [69]. Research continues on various cleaning and surface-preparation techniques. Atomic hydrogen cleaning (AHC) is a well-known technique for removing oxides and carbon-related contaminants at relatively low temperatures [70].

### 4.3.2 Polarized Electron Gun

The ILC polarized electron gun is a DC gun producing a 200 keV electron beam based on the design developed at Jefferson Laboratory [71, 72]. Photocathodes for polarized electron production are not viable in an RF gun vacuum environment. DC gun technology for polarized sources has evolved considerably since the SLC [73, 74]. The ILC gun is optimized for a peak current, limited by space charge, of 4.5-5 A (4.5-5 nC/1 ns). This provides overhead to compensate for losses that occur primarily through the bunching system. The gun power-supply provides a cathode bias of 200 kV. An ultrahigh vacuum system with a total pressure $\leq 10^{-10}$ Pa (excluding $H_2$) is required to maintain the negative electron affinity (NEA) of the cathode. During HV operation the electric field on the cathode surface must be kept below 9 MV/m to ensure low dark current ($< 25\,\mathrm{nC}$). Excessive dark current leads to field emission resulting in molecular desorption from nearby surfaces. This process leads to deterioration of the gun vacuum and is destructive to the cathode's NEA surface.

The gun has an integrated cathode preparation and activation chamber and load-lock system. The activation chamber is attached to the gun and both volumes are maintained under high vacuum. The preparation chamber allows the option of local cathode cleaning and activation as well as storage of spare cathodes. Cathodes may be rapidly exchanged between the gun and preparation chamber. Cesiator channels in the preparation chamber are located behind the retractable photocathode. This eliminates the deposition of cesium on electrode surfaces, thereby reducing the dark current of the gun. The load-lock consists of a small rapidly pumped vacuum chamber for transferring cathodes from an external atmospheric source into or out of the preparation chamber without affecting the latter's vacuum.

The gun area is equipped with a Mott polarimeter to measure polarization and a Faraday cup to measure the charge. Several Residual Gas Analyzers (RGAs) characterise the vacuum near the gun. Other special diagnostics for the DC gun include measurement of the quantum efficiency of the cathode (using a cw diode laser integrated into the gun) and a nano-ammeter for dark-current monitoring.

The dominant source of intensity variations and timing jitter is the laser system. A secondary source for intensity variations is the gun power supply and beam dynamics influenced by space charge forces within the gun and the low-energy sections of the injector.

### 4.3.3 ILC Source Laser System

The conceptual layout schematic of the laser system is depicted in Fig. 4.3. To match the bandgap energy of GaAs photocathodes, the wavelength of the laser system must be 790 nm and provide tunability ($\pm$20 nm) to optimise conditions for a specific photocathode. Therefore, the laser system is based on Ti:sapphire technology.

A 1.8 MHz pulse train is generated by a cavity-dumped mode-locked oscillator. After diffractive pulse stretching to 1 ns and temporal pulse shaping, the bunch train is amplified using a multi-pass Ti:sapphire amplifier. The amplifier crystal must be cryogenically cooled to facilitate power dissipation and minimize instabilities caused by thermal lensing induced by the high-power amplifier pump [75]. A cw frequency-doubled Nd:YAG (or similar such as Nd:Vanadate) diode-pumped solid state (DPSS) laser provides the pump power for the Ti:sapphire amplifier. Additional amplification can be supplied by one or multiple flashlamp-pumped Ti:sapphire stages. Final laser pulse energy and helicity control is achieved via electro-optical techniques (Pockels cells, polarizers, and waveplates). This system is also used as a feedback device to compensate for the QE decay of the photocathode between cesiations, to remove slow intensity drifts of laser and/or electron beam, and to maintain the circular polarisation state of the laser beam. Various optical techniques are used to cancel systematic effects caused by an asymmetric laser-beam profile or effects associated with the sign of the helicity of the laser light.

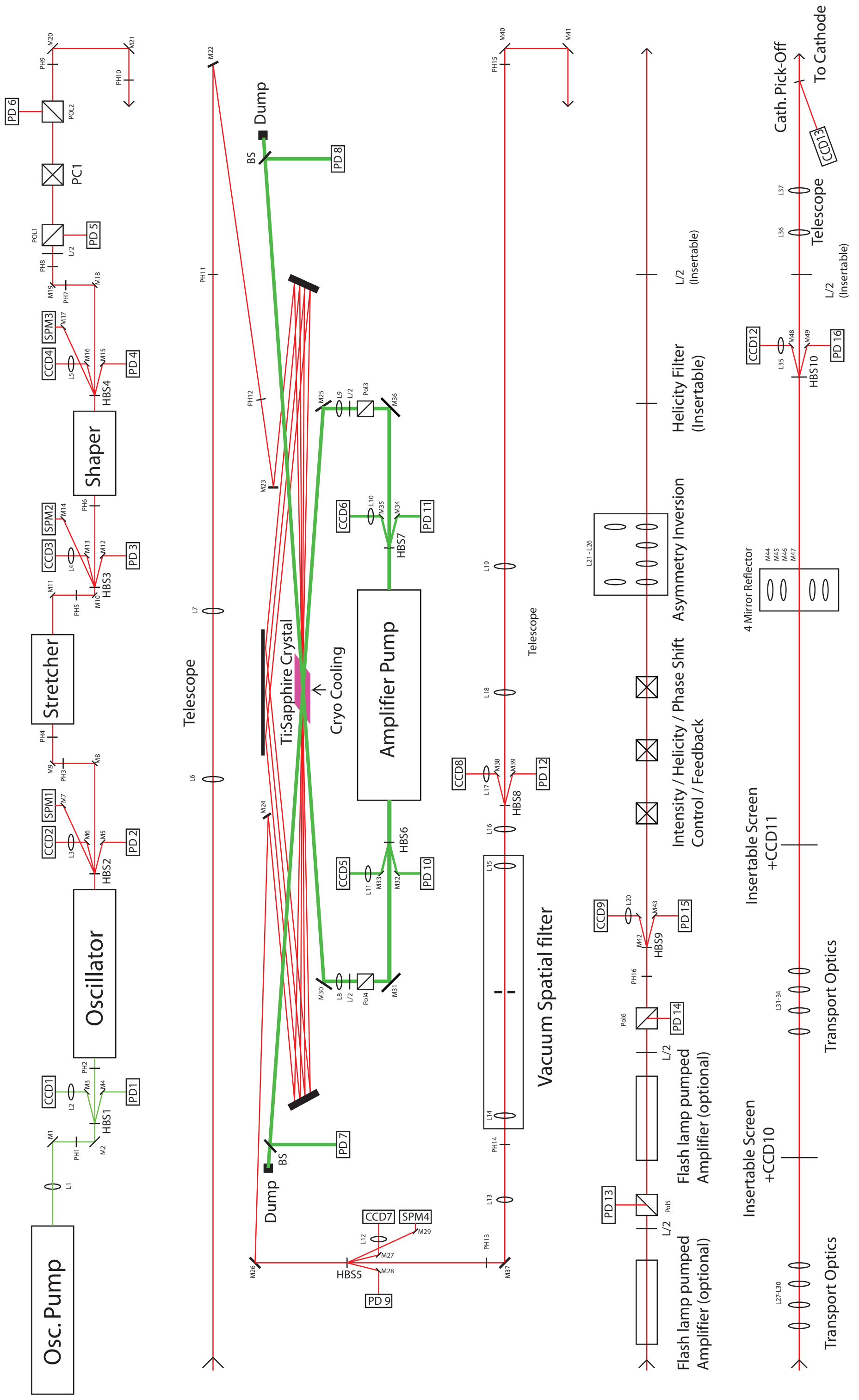

**Figure 4.3.** Schematic view of source drive laser system.

### 4.3.4 Bunching and Pre-Acceleration

The bunching system compresses the 1 ns micro-bunches generated by the gun down to $\sim$20 ps FWHM. It includes two subharmonic bunchers (SHB) and a 5 cell travelling wave $\beta$ =0.75 L-band buncher. Two SHB cavities both operate at 325 MHz. Together they compress the bunch to $\sim$200 ps FWHM. The L-band bunching system is a modification of the TESLA Test Facility [76] design with a travelling-wave buncher to maximize capture efficiency. The buncher has 5 cells with $\beta$ =0.75 and a gradient of 5.5 MV/m and compresses the bunch to 20 ps FWHM. The buncher and the first few cells of the following travelling wave pre-accelerator are immersed in a $7 \times 10^{-2}$T solenoidal field to focus the beam. Two 50 cell $\beta$ =1 normal conducting (NC) TW accelerating sections at a gradient of 8.5 MV/m increase the beam energy to 76 MeV. These structures must withstand very high RF power for the duration of the very long pulse but they are identical to those being developed for the positron source. Further details of the bunching system are summarised elsewhere [77].

### 4.3.5 Chicane, Emittance Measurement and Matching Sections

Immediately downstream of the NC pre-acceleration, a vertical chicane provides energy collimation before injection into the SC booster linac. The chicane consists of four bending magnets and several 90° FODO cells. The initial dipole at the chicane entrance can be used as a spectrometer magnet (see Fig. 4.1). A short beam line leads to a diagnostic section that includes a spectrometer screen. The injector beam emittance is measured by conventional wire scanners downstream of the chicane. Two matching sections connect the chicane and emittance measurement station to the downstream SC booster linac.

### 4.3.6 The 5 GeV Superconducting Pre-Acceleration (Booster) Linac

Twenty-one standard ILC-type SC cryomodules accelerate the beam to 5 GeV and FODO cells integrated into the cryomodules transversely focus the beam. An additional string of three cryomodules is added to provide redundancy (total of 24 cryomodules). The booster linac consists of two sections. In the first section, the $e^-$ beam is accelerated from 76 MeV to 1.7 GeV in cryomodules with one quadrupole per module. In the second section, the $e^-$ beam is accelerated to the final 5 GeV in cryomodules with one quadrupole every other module.

### 4.3.7 Linac to Damping Ring Beamline and Main $e^-$ Source Beam Dump

The Linac To Ring (LTR) beam line transports the beam to the injection point of the damping ring and performs spin rotation and energy compression. The 5 GeV longitudinally polarized electron beam is first bent through an arc. At 5 GeV, the spin component in the plane normal to the magnetic field precesses 90° in that plane for every n $\times$ 7.9° (n: odd integer) of rotation of momentum vector. An axial solenoid field integral of 26.2 T-m rotates the spin direction into the vertical [78]. A 5 GeV tune-up beam dump is installed near the LTR. To dump the 5 GeV beam, the first bend of the LTR is turned off, and the dump bend downstream energized. The dump drift is $\sim$ 12 m long.

## 4.4 Accelerator Physics Issues

Simulations indicate that >95% of the electrons produced by the DC gun are captured within the 6-D damping ring acceptance: $\gamma(A_x+A_y) \leq$0.07 m and ΔE×Δz $\leq$(±3.75 MeV)×(±3.5 cm). The starting beam diameter at the gun is 2 cm, which is focused to a few mm diameter before it is injected into the DR. Calculations in the low-energy regions of the injector ($\leq$ 76 MeV) include space-charge effects and use PARMELA [79]. The beam propagation through the superconducting booster linac and LTR beam line has been optimized using MAD [80] and tracked by the ELEGANT code [81].

### 4.4.1 DC Gun and Bunchers

The DC gun [72] creates a 200 keV electron beam with a bunch charge of 4.5-5 nC with a bunch length of 1 ns and an unnormalized transverse edge emittance at the gun exit of 70 mm-mrad. To minimize longitudinal growth of the bunch, it is desirable to locate the first subharmonic buncher (SHB) as close to the gun as possible. However, the beam lines needed to combine both guns require a distance of ∼1-1.5 m between gun and first SHB. The SHBs capture 92 % of the electrons generated at the gun. The beam parameters after the preaccelerator at 76 MeV (see Section 4.3.4) are summarized in Table 4.2. A plot of the beam envelope from gun up through the bunching system is given in Fig. 4.4.

**Table 4.2**
76 MeV beam parameters after NC bunching and pre-acceleration.

| Parameters | $\beta$ = 0.75 TW Buncher Design |
|---|---|
| Initial charge | 4.5 - 5 nC |
| Transmitted charge | 92 % |
| Phase extension FWHM | 9 deg L-band |
| Energy spread FWHM | <100 keV |
| Normalized rms emittance | 70 µm rad |

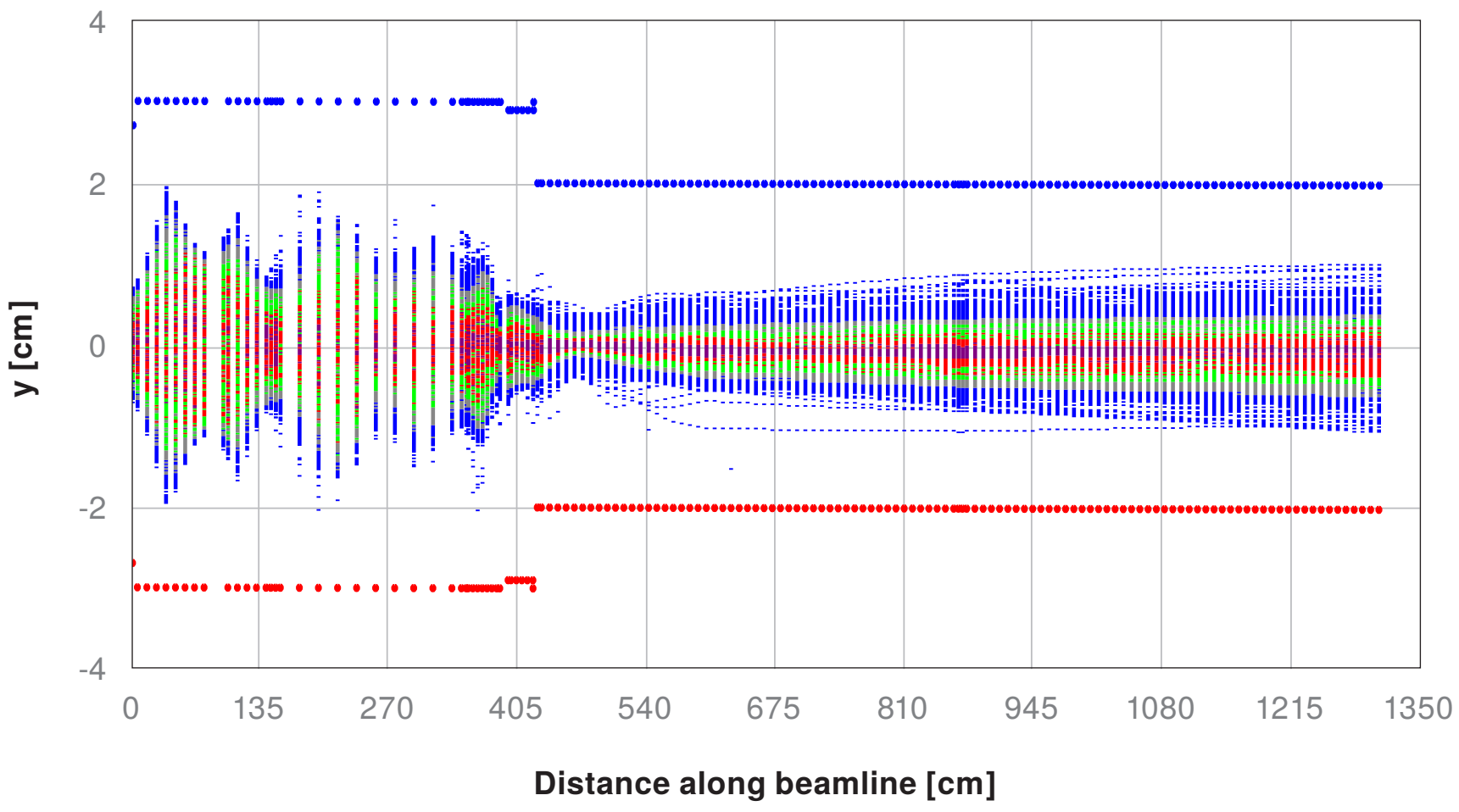

**Figure 4.4**
Beam envelope along the 76 MeV injector.

### 4.4.2 The 5 GeV Booster Linac and Linac to Damping Ring Line (eLTR)

The optics of the superconducting booster linac are shown in Fig. 4.5.

At the dump window, the beam size $\sigma_x/\sigma_y$ is 0.72 cm/1.4 cm and 13.9 cm/1.4 cm for 0 % and $\pm 10\,\%$ energy spread, respectively. These beam sizes are within the dump window specifications. At the profile monitor before the beam dump location, the dispersion dominates the beam size and thus the dump also serves as an energy spectrometer with 0.1 % resolution.

The arc of the eLTR is designed to rotate the spin vector by 90 degrees from longitudinal into a horizontal position before injection into the damping ring and to provide the $R_{56}$ necessary for energy compression. For every 90° of spin rotation, an arc angle of 7.9° is required. The initial LTR arc bending angle is 3 $\times 7.9^\circ = 23.8^\circ$. The $R_{56}$ is adjustable (-0.75 ±0.40 m). The arc is followed by a SC solenoid and a standard SCRF cryomodule. A 8.3-m-long superconducting solenoid with 3.16 T solenoid rotates the horizontal spin vector into the vertical. After the bunch is decompressed by the arc, an RF voltage of 225 MV provided by a 12.3 m-long 9-cavity superconducting linac, rotates the electrons in longitudinal phase space to match the longitudinal DR acceptance. The LTR includes approximately 200 m of additional transport and an optical transformer to match the Twiss parameters at the DR injection line [82]. There are three PPS stoppers each 1 m long in the LTR arc. Two FODO cells upstream of the LTR arc have laser-wire emittance-measurement stations. The optics of the LTR system are shown in Fig. 4.6.

**Figure 4.5**
Optics of the SC electron booster linac.

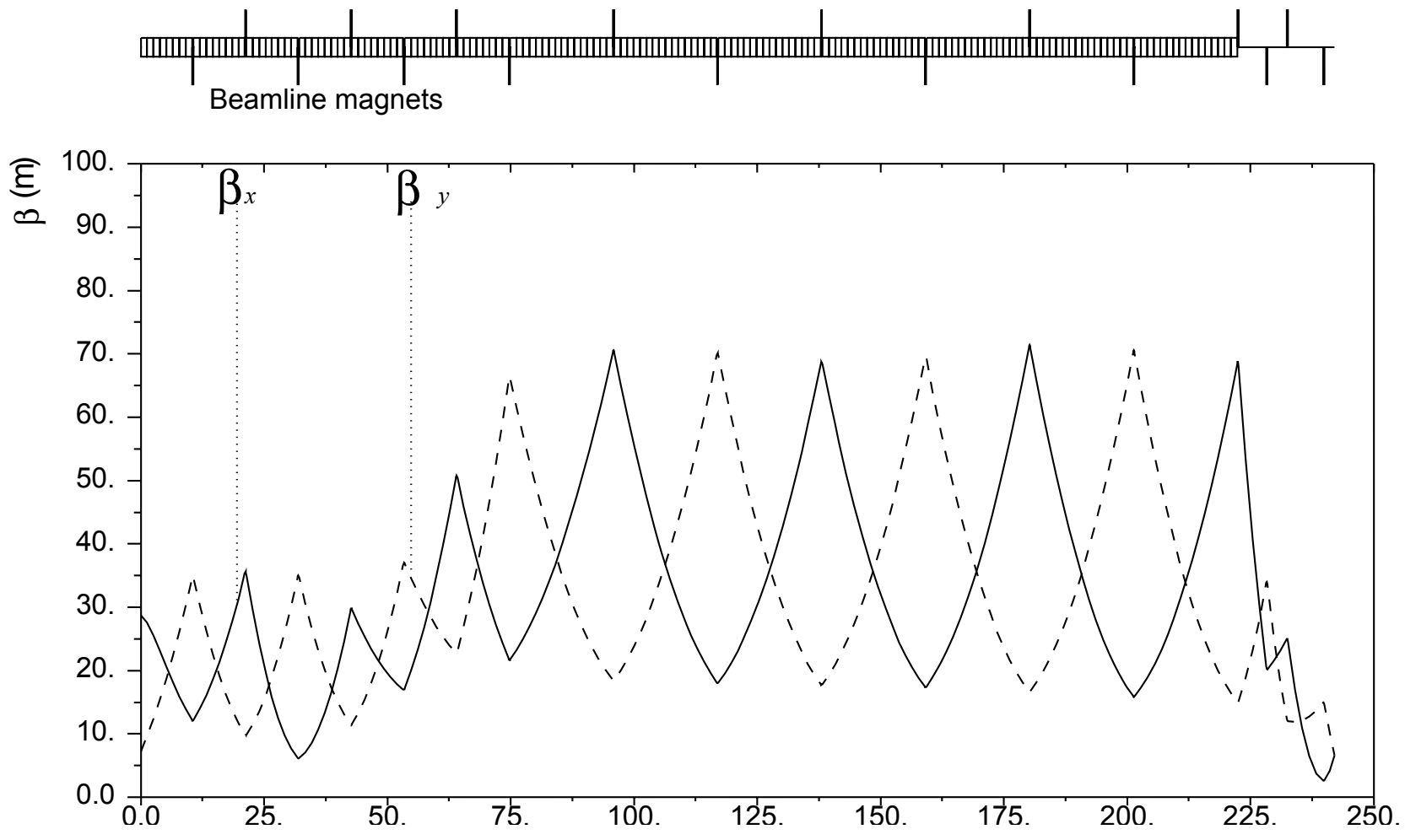

**Figure 4.6**
Optics of the LTR.

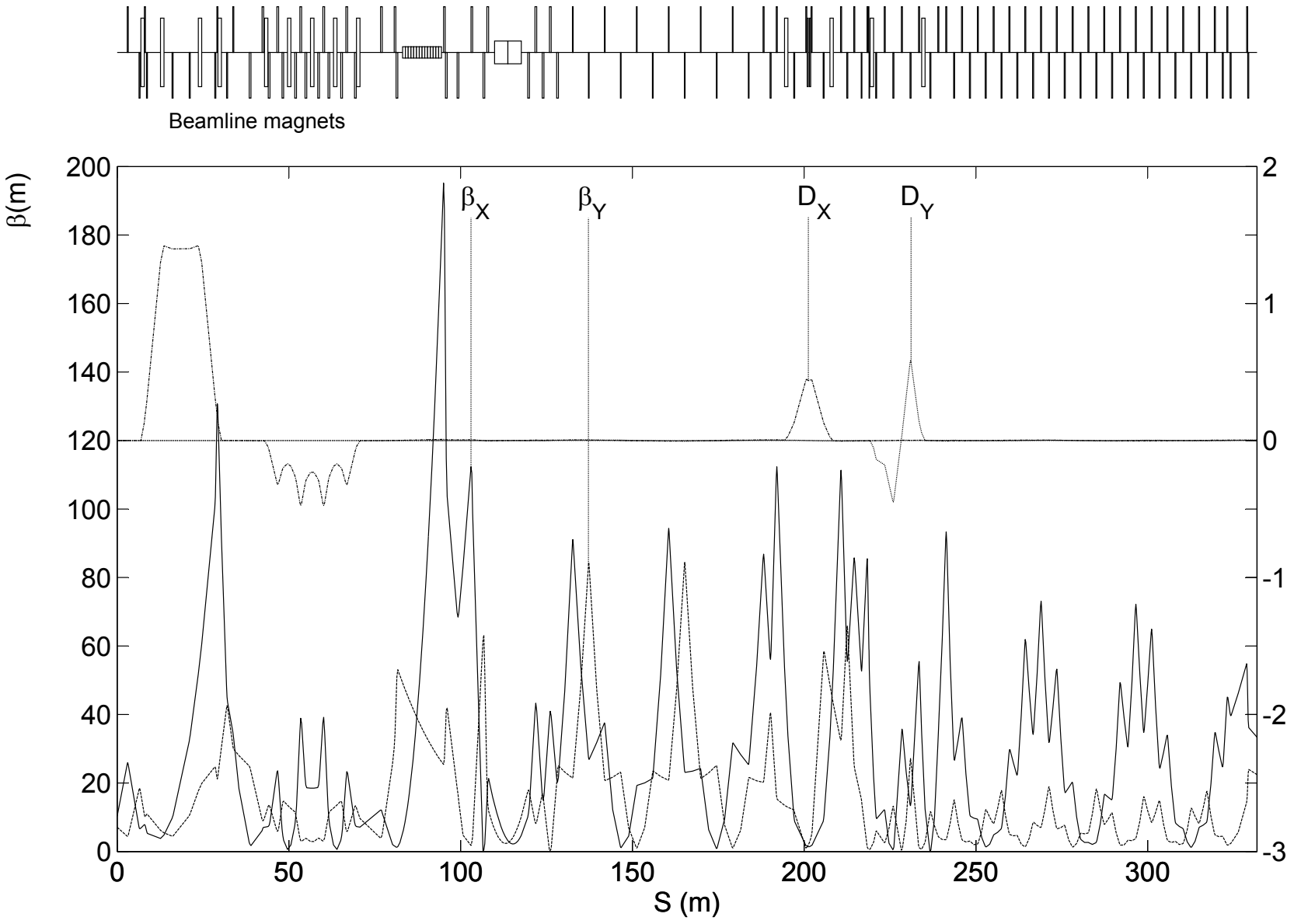

## 4.5 Accelerator Components

### 4.5.1 Table of Parts Count

Table 4.3 lists the major components of the ILC electron source and Table 4.4 the lengths of the various electron source beamlines.

**Table 4.3**
Total number of components for the polarized electron source.

| Magnets | | Instrumentation | | RF | |
|---|---|---|---|---|---|
| Bends | 27 | BPMs | 100 | 325 MHz SHB Cavities | 2 |
| Quads (NC) | 158 | Wirescanners | 4 | 5 Cell L-band buncher | 1 |
| Quads (SC) | 16 | Laserwires | 1 | L-band TW structures | 2 |
| Solenoids(NC) | 12 | BLMs | 5 | 1.3 GHz cryomodules | 24 |
| Solenoids(SC) | 2 | OTRs | 2 | L-band klystrons/modulators | 13 |
| Correctors(SC) | 32 | Phase monitors | 2 | | |

**Table 4.4**
System lengths for the $e^-$ source beamlines.

| Beam Line Section | Length |
|---|---|
| Gun area | 7 m |
| NC beam lines | 14 m |
| Chicane + emittance station | 54 m |
| SC beam lines | 245 m |
| eLTR | 332 m |
| Dumplines | 12 m |
| Total beam line length | 664 m |
| Total tunnel length | 680 m |

# Chapter 5
# Positron source

## 5.1 Introduction

The ILC Positron Source generates the positron beam. The production scheme uses the electron main linac beam passing through a long helical undulator to generate a multi-MeV photon drive beam which then impinges onto a thin metal target to generate positrons in electromagnetic showers. The positrons are captured, accelerated, separated from the shower constituents and unused drive-beam photons and transported to the Damping Rings. The baseline design is for 30 % polarised positrons. There are spin rotators before injection into the damping rings to preserve the polarisation and there is also sufficient beamline space to allow for an upgrade to a polarisation of $\sim 60\,\%$ [83].

The positron source performs several critical functions:

- generation of a high-power multi-MeV photon production beam. This requires suitable short-period, high-K-value helical undulators;
- production of the positron bunches in a metal target that can reliably deal with the beam power and radioactive environment induced by the production process. This requires high-power target systems;
- capture, acceleration and transport of the positron bunch to the Damping Rings with minimal beam loss. This requires high-gradient normal-conducting RF and special magnets to capture the positrons efficiently. The long transport lines also require large aperture magnets to transport efficiently the positron beams which have large transverse emittance.

The Positron Source also has sufficient instrumentation, diagnostics and feedback (feedforward) systems to ensure optimal operation.

## 5.2 Beam parameters

The key parameters of the Positron Source are given in Table 5.1 [84].

The source produces $2 \times 10^{10}$ positrons per bunch at the IP with the nominal ILC bunch structure and pulse repetition rate. It is designed with a 50 % overhead and can deliver up to $3 \times 10^{10}$ at injection into the 0.075 mrad transverse dynamic aperture of the damping ring. The main electron linac beam has an energy that varies between 100 and 250 GeV and passes through $\sim$ 150 m of helical undulator, with a 1.15 cm period and a K value of 0.92. At 150 GeV, the first harmonic cut-off of the photon drive beam is 10.1 MeV and the beam power is $\sim$63 kW. Approximately 4.4 kW of this power is deposited in the target in $\sim$ 1 mm rms. A windowless moving target is required to handle the high beam power and heat deposition.

The Positron Source undulator is long enough to provide adequate yield for any electron beam energy over 150 GeV. For lower energy operation, the electron complex operates at a 10 Hz repetition rate with 5 Hz of 150 GeV electrons used to produce positrons and 5 Hz of electrons at the desired energy for collisions.

**Table 5.1**
Nominal Positron Source Parameters

| Parameter | Symbol | Value | Units |
|---|---|---|---|
| Positrons per bunch at IP | $n_b$ | $2 \times 10^{10}$ | number |
| Bunches per pulse | $N_b$ | 1312 | number |
| Pulse Repetition Rate | $f_{rep}$ | 5 | Hz |
| Positron Energy (DR injection) | $E_0$ | 5 | GeV |
| DR Dynamic Aperture | $\gamma(A_x + A_y)$ | <0.07 | m rad |
| DR Energy Acceptance | $\Delta$ | 0.75 | % |
| DR Longitudinal Acceptance | $A_l$ | 3.4 x 37.5 | cm-MeV |
| Electron Drive Beam Energy† | $E_e$ | 150/175/250 | GeV |
| Undulator Period | $\lambda$ | 1.15 | cm |
| Undulator Strength‡ | $K$ | 0.92/0.75/0.45 | - |
| Undulator Type | - | Helical | - |
| Undulator Length | $L_u$ | 147 | m |
| Photon Energy ($1^{st}$ harm cutoff) | $E_{c10}$ | 10.1/16.2/42.8 | MeV |
| Photon Beam Power | $P_\gamma$ | 63.1/54.7/41.7 | kW |
| Target Material | - | Ti-6%Al-4%V | - |
| Target Thickness | $L_t$ | 0.4 / 1.4 | r.l. / cm |
| Target Absorption | - | 7 | % |
| Incident Spot Size on Target | $\sigma_i$ | 1.4/1.2/0.8 | mm, rms |
| Positron Polarisation | $P$ | 31/30/29 | % |

† For centre-of-mass energy below 300 GeV, the machine operates in 10 Hz mode where a 5 Hz 150 GeV beam with parameters as shown in the table is a dedicated drive beam positron source.
‡ K is lowered for beam energies above 150 GeV to bring the polarisation back to 30 % without adding a photon collimator before the target.

## 5.3 System description

The layout of the electron side of the ILC is shown in Fig. 5.1, including the relative position of the major systems of the positron source. Figure 5.2 is a schematic of the positron source beamlines with dimension indicated, split into two sections [85]. The upper section shows the beamlines from the end of electron main linac to the end of the 400 MeV positron pre-accelerator. The lower section shows the beamlines from the end of the pre-accelerator to the end of the positron-source beamline or the beginning of the damping ring.

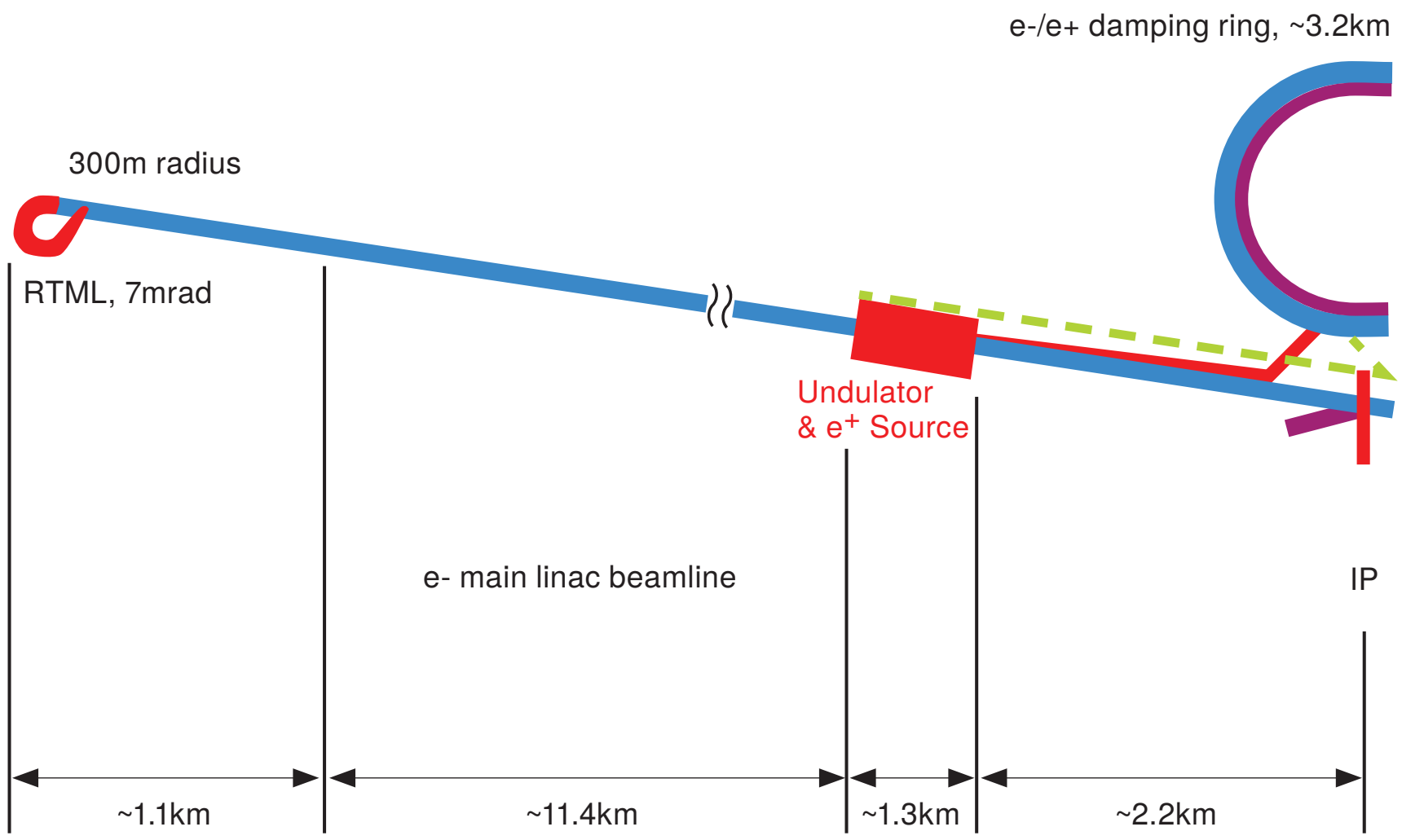

**Figure 5.1**
Layout of Positron system relative to the ILC

The electron beam from the main linac passes through the undulator and a dogleg before continuing to the IP for collisions. These beamlines are labeled as EUPM, EUND and EDOGL in Fig. 5.2. For lower energy operations (CM=200 GeV, 230 GeV and 250 GeV), a dedicated 5 Hz 150 GeV drive beam, alternating with the lower energy beam for physics, is used for positron production. After passing through the EUND beamline to generate photons, this 150 GeV drive beam is then sent to a beam dump in the beamline EPUNDDL.

The photon beam produced by the electron beam drifts through the section UPT and strikes a 1.4 cm thick Ti-alloy target to produce an electromagnetic shower of positrons and electrons. The positrons are then captured within optical matching device (OMD) and then matched into a capture system (labeled PTAPA) consisting of normal conducting (NC) L-band RF cavities and surrounding solenoid. The positron beam is accelerated to 125 MeV before entering the chicane where the positrons, electrons and photons are separated, into beamlines PCAP, PCAPEDL and PCAUPDL respectively. Both electrons and photons are dumped. After the chicane, the positron beam is further accelerated to 400 MeV using a NC L-band RF system with solenoidal focusing (labelled beamline PPA).

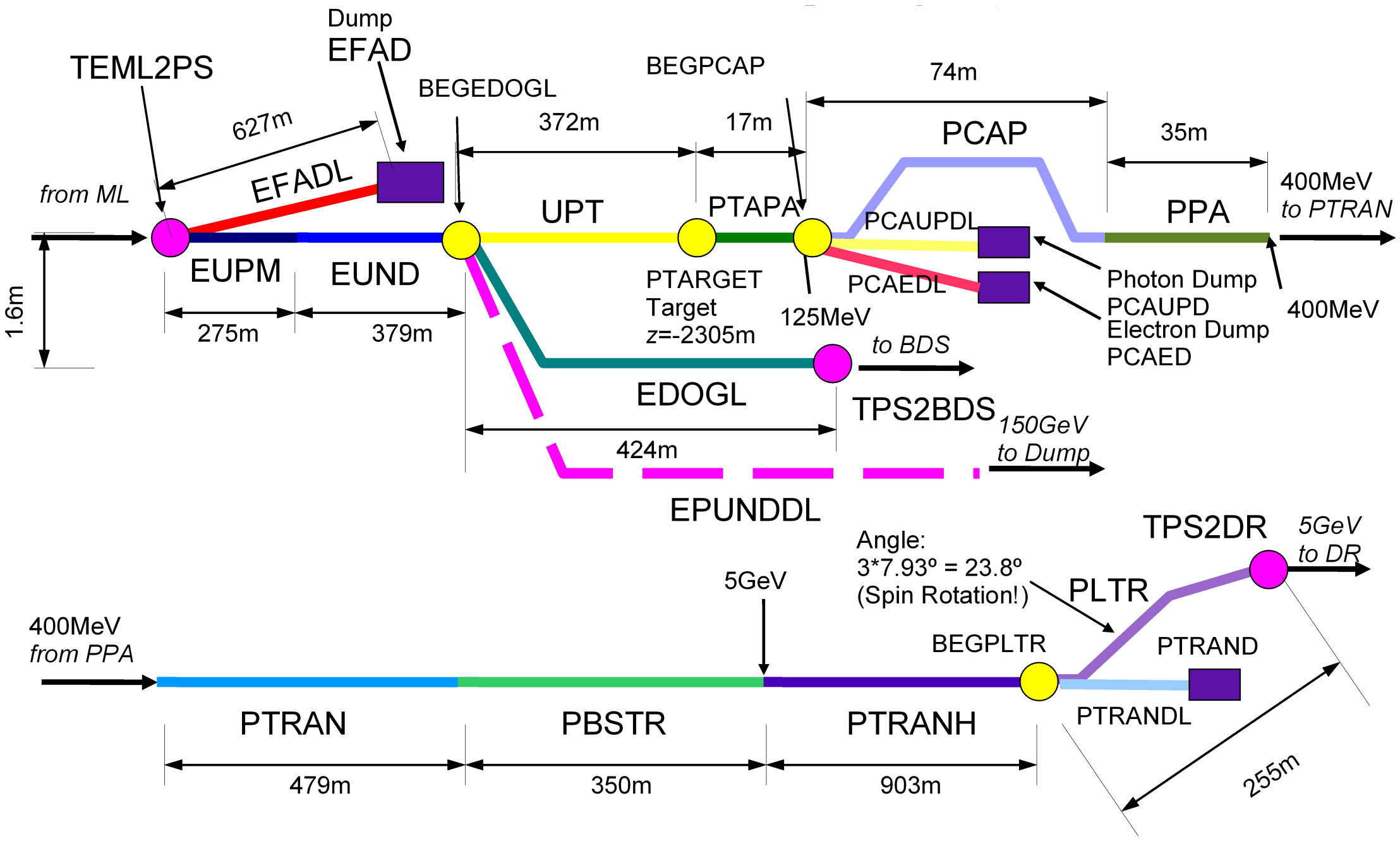

**Figure 5.2.** Positron Source beamlines cartoon

The 400 MeV positron beam is transported for approximately 479 m in beamline PTRAN (400 MeV) to a booster linac (PBSTR) where the beams are further accelerated to 5 GeV using SC L-band RF. Before injection into the damping ring, the beam is transported 903 m in PTRANH before passing through a beamline section (PLTR) that carries out spin rotation and energy compression in order to maximise injection acceptance. Finally, the beam is injected into the positron damping ring at point TPS2DR.

Figure 5.3 shows how the performance of the positron source (yield and polarisation) strongly depends on the main electron-beam energy for the given undulator parameters (K, $\lambda_u$). At higher electron-beam energy, the undulator B field is re-optimized to restore the polarisation to 30 %. The final undulator parameters for a yield of 1.5 at 350 and 500 GeV energy are listed in Table 5.2.

One additional part of the positron-source system is the Auxiliary Source [86]. The current auxiliary source scheme generates a single-bunch low-intensity ($\sim$ 1 % of nominal beam intensity) positron beam which is intended for commissioning. This source uses 500 MeV electron drive beam from a conventional S-band electron accelerator impinging on the same production target as the normal beam to produce positrons which then pass through the capture, acceleration and transport beamlines sections, and subsequently injected into the damping ring. The 500 MeV S-band electron injector has 8 SLAC-type 3 m-long accelerator structures [87] and a microwave photo cathode gun. The KAS is less than 40 m long and is installed along-side the 370 m long undulator photon transport line. The electron injector is powered by 4 S-band RF stations, each with a 100 MW modulator, a 50 MW klystron and a SLED cavity [88].

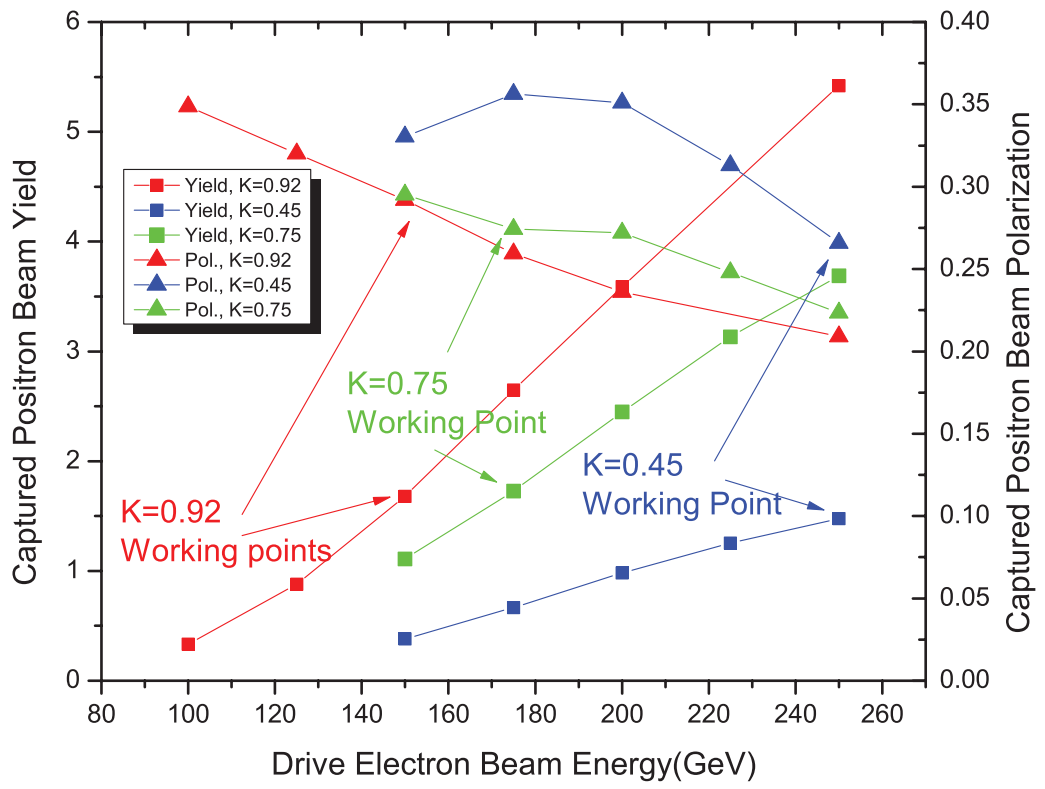

**Figure 5.3**
Simulation results of positron source yield and polarisation as a function of drive-beam energy for 147 m long undulator and $\lambda_u$=1.15 cm using a flux concentrator as OMD.

**Table 5.2**
Parameters for 350 GeV CM and 500 GeV CM.

| Parameter | units | 350 GeV | 500 GeV |
|---|---|---|---|
| Electron beam energy ($e^+$ prod.) | GeV | 178 | 253 |
| Bunches per pulse | N | 1312 | 1312 |
| Photon energy (first harmonic) | MeV | 16.2 | 42.8 |
| Photon openning angle ($=1/\gamma$) | µrad | 2.9 | 2 |
| Undulator length | m | 147 | 147 |
| Required undulator field | T | 0.698 | 0.42 |
| undulator period length | cm | 1.15 | 1.15 |
| undulator K | | 0.75 | 0.45 |
| Electron energy loss in undulator | GeV | 2.6 | 2 |
| Induced energy spread (assume 0% initial) | % | 0.122 | 0.084 |
| Emittance growth | nm | -0.55 | -0.31 |
| Average photon power on target | kW | 54.7 | 41.7 |
| Incident photon energy per bunch | J | 8.1 | 6 |
| Energy deposition per bunch ($e^+$ prod.) | J | 0.59 | 0.31 |
| Relative energy deposition in target | % | 7.20% | 5% |
| Photon rms spot size on target | mm | 1.2 | 0.8 |
| Peak energy density in target | J/ cm$^3$ | 295.3 | 304.3 |
| | J/ g | 65.6 | 67.5 |
| Pol. of Captured Positron beam | % | 30 | 30† |

† Flux concentrator needs to operate at a stronger field.

## 5.3.1 Photon production

Production of an adequate number of positrons requires that the photons hitting the target have both sufficient intensity and high-enough energies to produce ∼1–100 MeV electron-positron pairs that can escape from the target to be captured. In general this means photon energies of at least 10 MeV. The total number of positrons produced must be suffient to allow for losses between the target and the IP.

A helical undulator generates twice the synchrotron radiation power per period than the equivalent (same maximum field) planar undulator, reducing the length required to produce the same number of positrons. Another benefit is that the helical undulator generates circularly polarised photons which in turn generate longitudinally polarised positrons. For the baseline undulator system with a 150 GeV drive beam, the photons produce enough captured positrons but the resulting polarisation is only $\sim 30\,\%$. To achieve higher positron polarisation requires a longer undulator to produce an excess of photons. That allows photons with the wrong polarisation state to be absorbed by photon collimator and still leave adequate photon yield on the target. A polarisation of 60 % can be achieved with an additional 73.5 m long undulator.

The undulator is installed at the end of the electron main linac as shown in Fig. 5.1. Above 150 GeV, the electron beam used in the final collisions at the IP is used as the drive beam, passing through the undulator to generate the required photons. At lower beam energy, the positron yield is too low and a dedicated 150 GeV drive beam is interleaved with the electron beam used in the IP collisions.

### 5.3.2 Positron production & capture

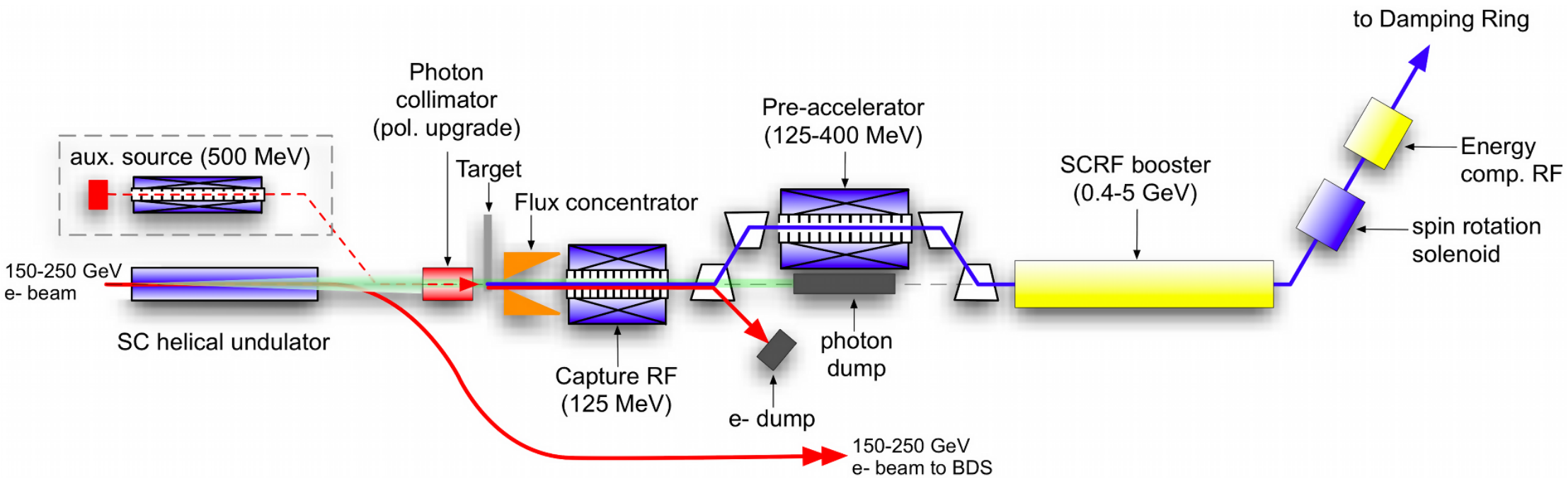

**Figure 5.4.** Schematic layout of Positron Source. Beamline sections are defined in Section 5.3.

Figure 5.4 shows the schematic layout for the positron-beam production, capture and transport to the damping rings. The photon beam generated from the helical undulator is incident on the rim of a rotating titanium target (see Section 5.5.2) with 0.4 radiation-lengths thickness. The incident photon beam has transverse size of ∼1 mm rms and electron and positron emerging from the downstream side of the target are captured in a 0.07 mrad transverse dynamic aperture. The target is followed by a tapered magnet called the Optical Matching Device (OMD) (see Section 5.5.3) which has a field starting from <0.5 T at the target and then quickly ramped to over 3 T in ∼2 cm, and then decays to 0.5 T over 14 cm. This OMD has wide energy acceptance and is used to match the beam phase-space out of the target into the capture L-band RF cavities (TAP). The capture RF cavities are placed directly after the OMD to accelerate the positron beam to 125 MeV. The accelerating RF cavities have an average gradient of 9 MV/m and are located inside 0.5 T solenoids which provide beam focusing.

The target and equipment immediately downstream will become highly activated. A remote-handling system is used to replace the target, OMD and 1.3 m NC RF cavities. The remote handling system is described in detail in Part I Section 4.3.10.

### 5.3.3 Positron transport

After capture, positrons are separated from electrons and photons in the dipole magnet at the entrance of an achromatic chicane which horizontally deflects the positrons by 1.5 m. The chicane includes collimators to remove positrons with large incoming angles and energy far from the nominal value.

The pre-accelerator immediately downstream of the chicane accelerates the positron beam from 125 MeV to 400 MeV. The normal-conducting L-band RF structures are immersed in a constant solenoid field of 0.5 T. The accelerating gradient is ∼ 8 MV/m and the total length is 34.6 m. The transport line is 480 m long and transfers the 400 MeV positron beam to the positron booster linac.

### 5.3.4 5 GeV SC Booster Linac

It accelerates the beam from 400 MeV to 5 GeV using SC L-band RF modules. There are three sections with a periodic FODO lattice. The first low-energy section which accelerates up to 1083 MeV contains four cryomodules with six 9-cell cavities and six quadrupoles. The quadrupole field strength $(\partial B/\partial x) \times L$ varies from 0.8-2.4 T. The second section up to 2507 MeV uses six standard ILC-type cryomodules, each containing eight 9-cell cavities and two quadrupoles. The quad strength ranges from 0.6-1.4 T. The last section up to 5 GeV has twelve standard ILC-type cryomodules, each with eight 9-cell cavities and one quadrupole. The quadrupole field strength ranges from 0.8-1.7 T.

### 5.3.5 Linac to Damping-Ring Beam Line

The linac to damping-ring (LTR) system from the booster linac to the DR injection line has two main functions: to rotate the polarisation into the vertical plane, and to compress the energy spread to match the DR longitudinal acceptance.

The polarisation is preserved through transport and acceleration. The polarisation must be rotated into the vertical plane to preserve the polarisation in the DR. The spin-rotation system contains bending magnets and solenoids, changing the spin of positrons first from the longitudinal to the horizontal plane and then from horizontal to vertical. To produce $n \cdot 90°$ of spin rotation ($n$ is an odd integer) from the longitudinal to horizontal plane at 5 GeV, a total bending angle $\theta_{bend} = n \cdot 7.929°$ is required. To rotate the spin by $90°$ from the horizontal to vertical plane at 5 GeV energy requires a solenoid magnetic-field integral of 26.2 T m. This is achieved with an 8.3 m-long superconducting solenoid with 3.16 T field.

The energy compression uses a combination of booster-linac RF phase, a chicane at the beginning of the LTR and RF voltage. The chicane has a transverse offset of 1.5 m and a nominal R56 of −0.75 m. The first arc of the LTR has a bending angle of $3 \times 7.929° = 23.787°$ to rotate the spin by $90°$. After the first arc, an RF voltage of 225 MV is provided by a 9-cavity ILC cryomodule with no quads. This compresses the positron energy to match into the DR. The rest of the LTR system includes: a section with an additional $9.626°$ horizontal bending, a vertical dogleg to raise the elevation up by 1.65 m, another vertical dogleg to lower the elevation back to 0.35 m and a FODO lattice to transport the beam into a matching section into the DR injection line. Its geometry is shown in Fig. 5.5.

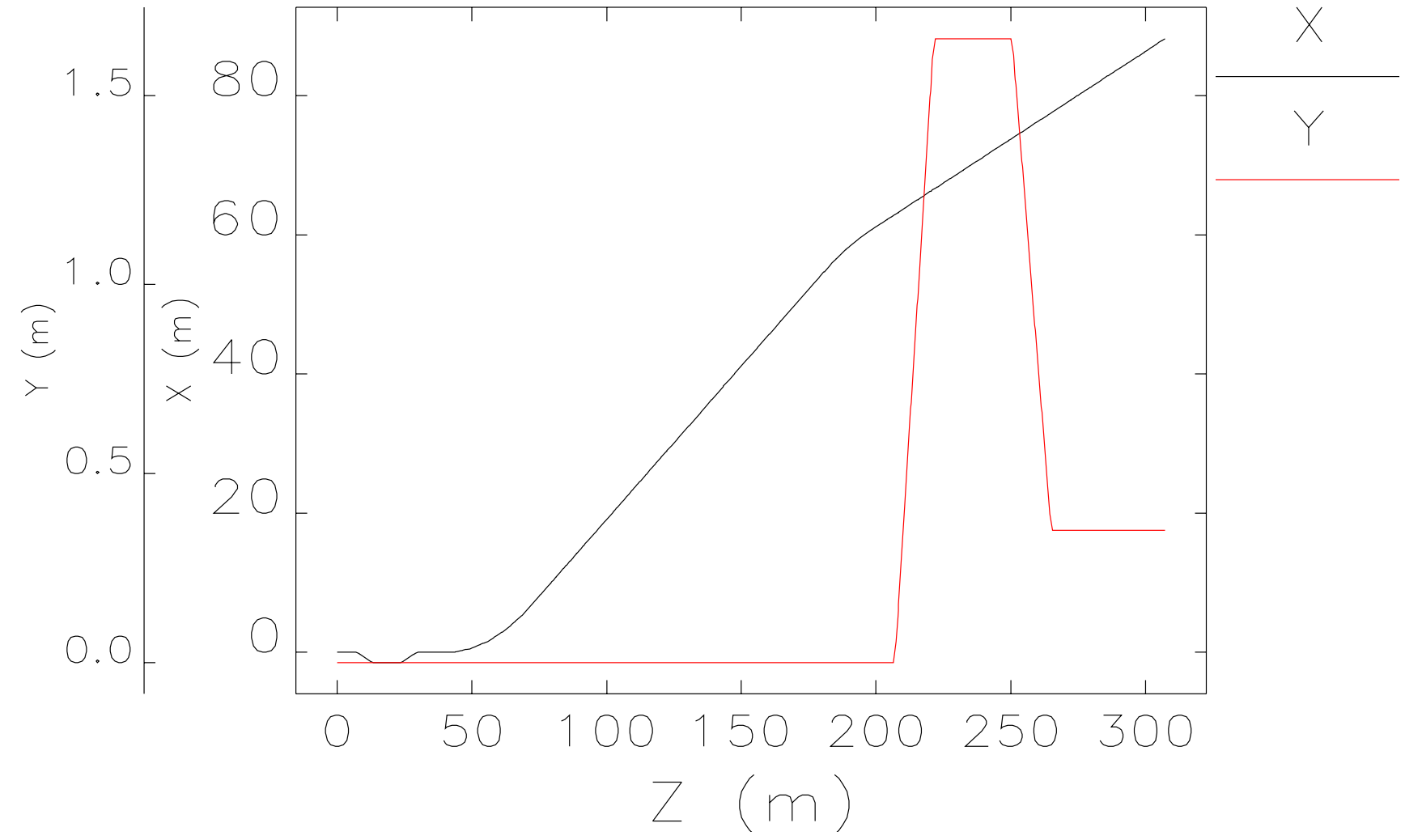

**Figure 5.5** Geometry of LTR beamline. The LTR beamline starts at $z = 0$.

## 5.4 Optics parameters

The optics of the positron-source system starting from the capture section to the DR injection is shown in Fig. 5.6. The lattice is optimized to have maximum transmission and minimum emittance growth.

Multi-particle tracking has been performed from the target to the DR injection, using Elegant [89] to track the large angular divergence and long tails at low energy. Energy compression is required before injection into the DR to accommodate more positron beam within the 6-D acceptance in the DR equal to $A_x + A_y \leq 0.07\,\mathrm{m}$ and $\Delta E \times \Delta x \leq (\pm 3.75\,\mathrm{MeV}) \times (\pm 3.5\,\mathrm{cm})$.

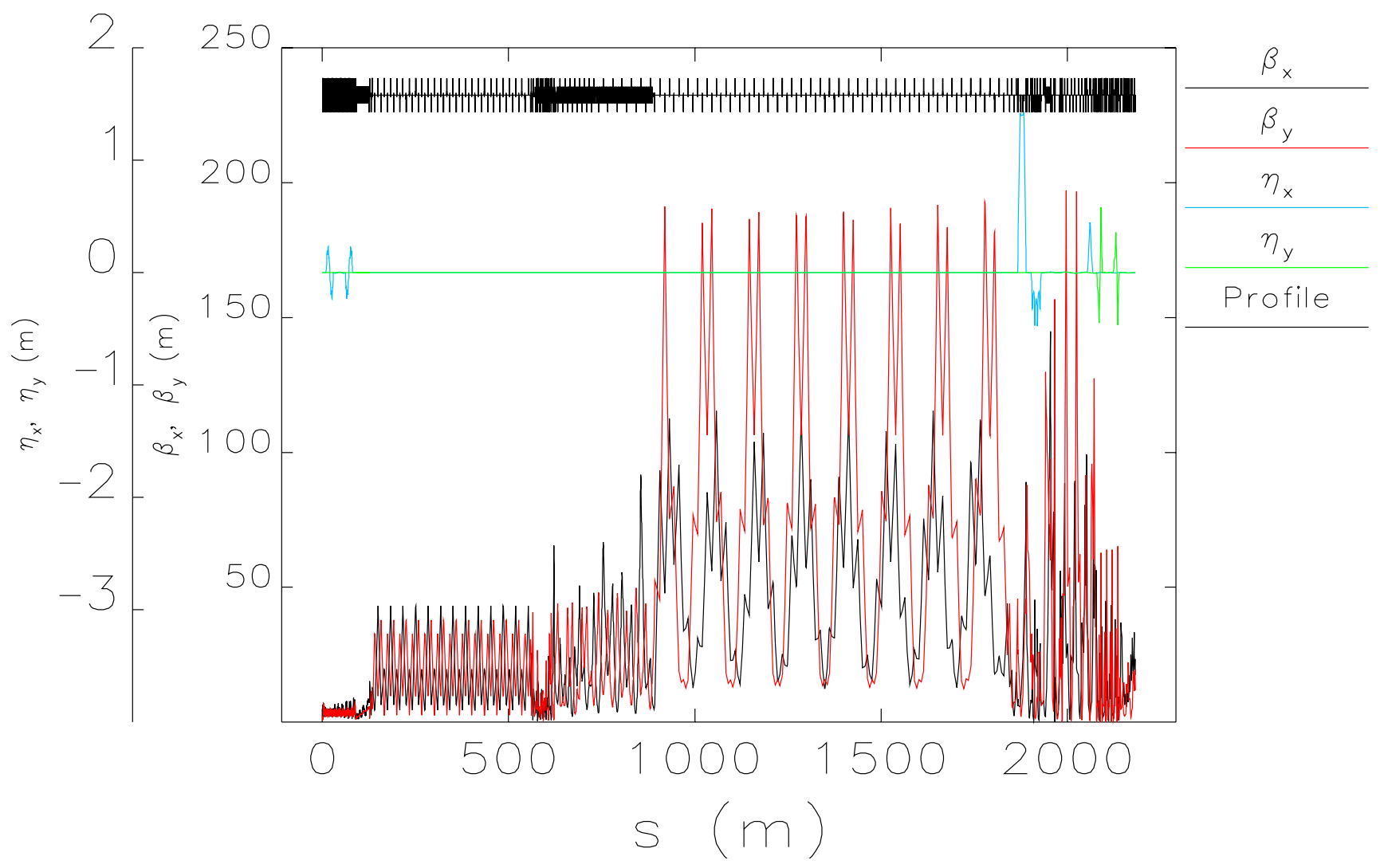

**Figure 5.6**
Optics of positron source

## 5.5 Accelerator components

### 5.5.1 Undulator

The undulator uses superconducting technology to achieve high field with a short period. Two interleaved helical windings of NbTi spaced half a period apart generate the transverse helical field. The large length of the undulator requires that it be built in modular units. Each 4 m-long cryomodule contains two separate undulators with an active undulator length of $\sim$ 3.5 m [90]. The present baseline parameters are given in Table 5.3.

**Table 5.3**
Helical undulator parameters

| Parameter | Value |
|---|---|
| Period (mm) | 11.5 |
| K | 0.92 |
| Field on Axis (T) | 0.86 |
| Beam aperture (mm) | 5.85 |
| First Harmonic Energy (MeV) | 10.1 |
| Nominal Drive Beam Energy(GeV) | 150 |

The undulator vacuum chamber is made of copper and operates at a temperature of 4.2 K. Copper is selected for its high conductivity which alleviates resistive wall effects. Estimates for a 150 µm-long Gaussian bunch containing $1 \times 10^{10}$ electrons (a more demanding case than the ILC nominal parameters of 300 µm and $2 \times 10^{10}$ particles per bunch), interacting with a 200 m-long copper vessel with internal aperture of 5.6 mm, indicate that the resistive wall effect would increase the RMS energy spread from the nominal value of 0.05 % to 0.0505 %. Another advantage of using copper is that excellent surface quality is readily achievable in copper vessels. A pessimistic wakefield model has suggested that for a measured surface roughness (RA value) of $<$ 100 nm, the electron energy spread would only increase from 0.05 % to $< 0.051\,\%$. The resistive-wall wakefield could potentially cause emittance growth, but numerical simulations have shown that there is no effect until the transverse kick strength is $>$5000 times the nominal value [91, 92].

The material between the superconducting windings is soft magnetic iron which serves as an outer yoke to increase the field and to provide additional support. Each cryomodule contains a liquid-helium bath; zero liquid boil off is achieved through the use of in-situ cryocoolers.

Since the electron vacuum vessel is at cryogenic temperatures, each module effectively acts as a long cryopump. Roughing pumps are installed in room temperature sections between cryomodules (approximately every 12 m) but achieving UHV conditions relies on cryopumping. The baseline pressure target of $10^{-8}$ mbar is set to avoid fast-ion instability problems. Vacuum calculations indicate that the cryopumping is adequate provided that the number of photons with energy >10 eV striking the vessel surface is kept sufficiently low. Extensive calculations of the undulator photon output down to these very low energies have been carried out. These indicate that low-power photon absorbers [93] should be placed approximately every 12 m to provide an adequate shadowing of the cold vessel surfaces. These absorbers are in room temperature sections.

The electron-beam transport through the complete undulator system is based on a simple FODO arrangement with quadrupole spacing of ∼ 12 m (in the room temperature sections). There are electron beam-position monitors at every quadrupole and two small horizontal and vertical corrector magnets per cryomodule. Simple electron-beam transport calculations have shown that excellent relative alignment between the quadrupoles and neighboring BPMs is required. In this simple model, quadrupole to BPM misalignment of ∼ 5 µm leads to an emittance growth of ∼2%. It is important to note however that this is not due to the undulator but to the effect of the quadrupoles and is therefore a general problem for the ILC beam transport. Dispersion-free steering-correction algorithms will be required, similar to those used for the Main Linacs (see Part I Section 4.6).

### 5.5.2 Target

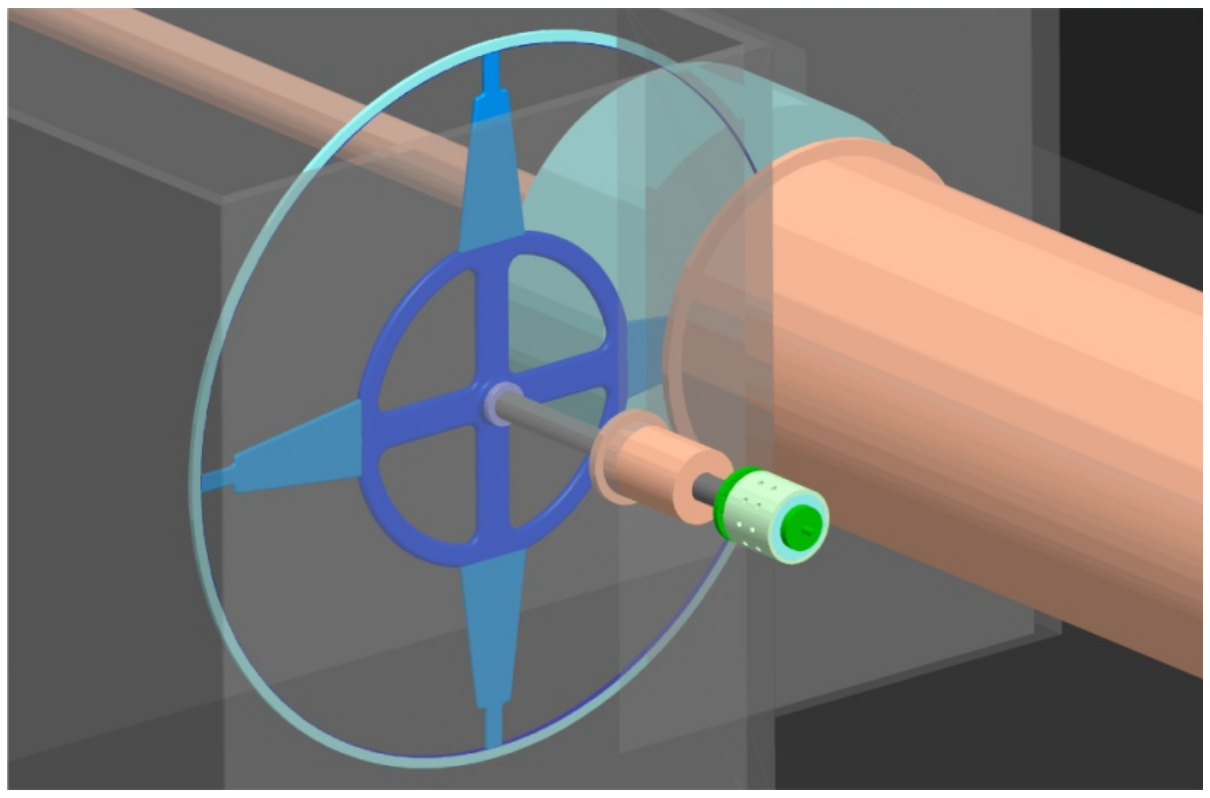

**Figure 5.7**
Overall target layout.

The positron-production target is a rotating wheel made of titanium alloy (Ti6Al4V). The photon beam from the undulator is incident on the rim of the spinning wheel. The diameter of the wheel is 1 m and the thickness is 0.4 radiation lengths (1.4 cm). During operation the outer edge of the rim moves at 100 m/s. The combination of wheel size and speed offsets radiation damage, heating and the shock-stress in the wheel from the ∼ 300 kW photon beam. A picture of the conceptual target layout is shown in Fig. 5.7. The motor is mounted on a single shaft, with a rotating water union to allow cooling water to be fed into the wheel. The beam power is too high to allow a vacuum window downstream of the target. The target wheels sit in a vacuum enclosure at $10^{-8}$ Torr (needed for NC RF operation), which requires vacuum seals for access to the vacuum chamber. The rotating shaft penetrates the enclosure using one vacuum passthrough. The optical matching device (OMD – see Section 5.5.3), is mounted on the target assembly, and operates at room temperature. The motor driving the target wheel is designed to overcome forces due to eddy currents induced in the wheel by the OMD.

The target-wheel assembly is designed for an operational life of two years. In the event that the target fails during a run, the assembly can be replaced by a new assembly in less than a day using a vertically removable target remote-handling scheme [94].

A series of sensors provides information on the target behavior. An infrared camera tracks temperatures on the wheel, to allow for quick shutdown in case of a cooling failure. Flowmeters monitor cooling water flow in and out of the wheel (to check for leaks), and thermocouples check ingoing and outgoing flow temperature. There is a torque sensor on the shaft, and vibration sensors on the wheel to monitor mechanical behavior. Finally, the wheel's rotational speed is monitored.

R&D on the target — and in particular on the rotating vacuum seal — remains on-going, and progress is reported in Part I Section 4.3. While the vacuum specification of the seal has been demonstrated, its lifetime and reliability require further R&D.

### 5.5.3 Optical matching device

The OMD generates a solenoidal magnetic field which peaks in strength at 3.2 T close to the target and falls off to 0.5 T to match the solenoidal field at the entrance of the capture section. The OMD increases the capture efficiency by a factor of 2. The OMD is a normal-conducting pulsed flux concentrator designed and prototyped by LLNL.

The magnetic field of the OMD interacts with the spinning metal of the target to create Eddy currents. The target design must take into account this drag force which produces an increased average heat load, requires a stronger drive motor and possibly causes 5 Hz resonance effects.

### 5.5.4 Normal-conducting RF accelerator system

Due to the extremely high energy deposition from positrons, electrons, photons and neutrons behind the positron-production target, the 1.3 GHz pre-accelerator uses normal conducting structures up to an energy of 400 MeV. Major challenges are achieving adequate cooling with the high RF and particle-loss heating, and sustaining high accelerating gradients during millisecond-long pulses in a strong magnetic field. The current design contains both standing-wave (SW) and travelling-wave (TW) L-band accelerator structures. The capture region has two 1.27 m SW accelerator sections at 15 MV/m and three 4.3 m TW accelerator sections with 8.5 MV/m accelerating gradient. The electrons are then accelerated from 125–400 MeV in the pre-accelerator region, which contains eight 4.3 m TW sections at 8.5 MV/m accelerating gradient. All accelerator sections are surrounded with 0.5 T solenoids. Figure 5.8 shows the schematic layout [61].

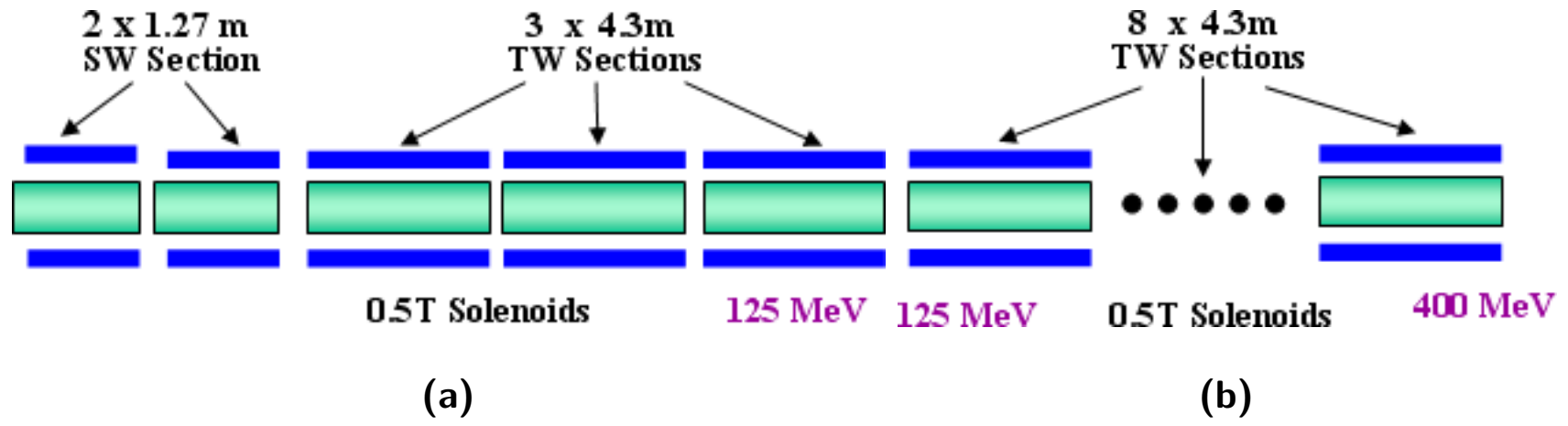

**Figure 5.8**
Layout of the capture region (a) and pre-accelerator region (b).

#### 5.5.4.1 Standing wave accelerator structure for positron capture

The high-gradient (15 MV/m) positron-capture sections are simple $\pi$ mode 11 cell SW accelerator structures. The benefits are an effective cooling system, high shunt impedance with larger aperture (60 mm), low RF pulse heating, simplicity and cost efficiency. The mode and amplitude stability under various cooling conditions have been theoretically verified for this type of structure. Figure 5.9 shows a section view of the SW structure; Table 5.4 gives the important RF parameters.

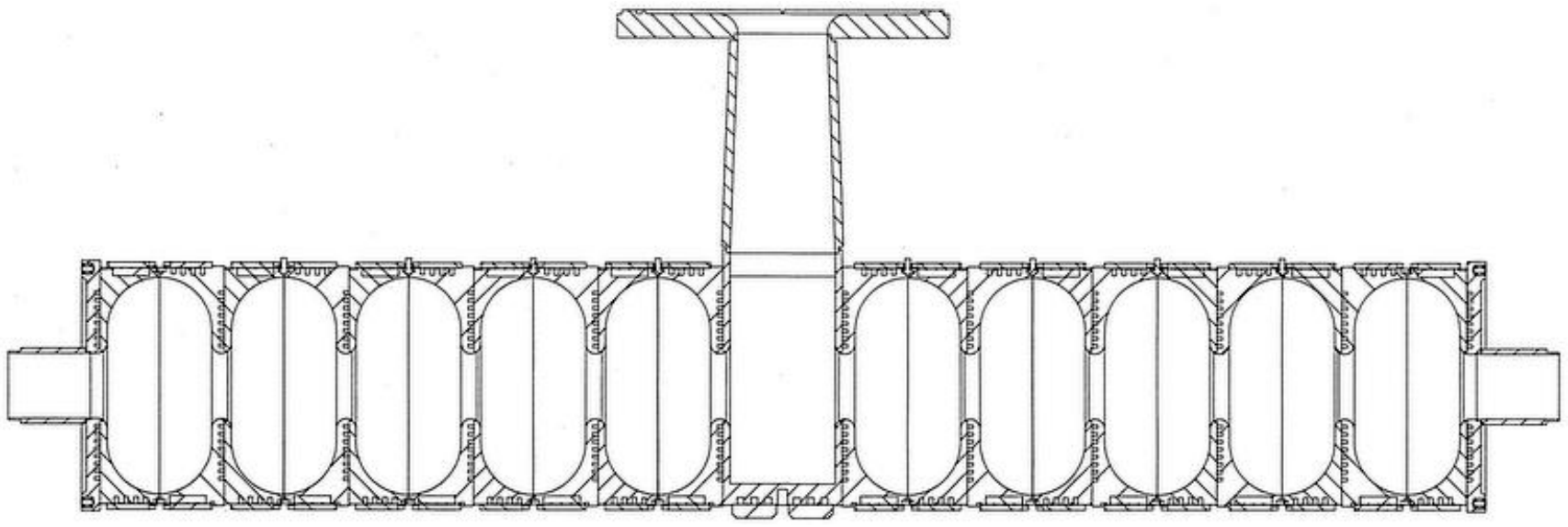

**Figure 5.9**
11–cell SW Structure.

**Table 5.4**
Parameters of SW structure.

| | |
|---|---|
| Structure Type | Simple $\pi$ Mode |
| Cell Number | 11 |
| Aperture 2a | 60 mm |
| Q | 29700 |
| Shunt impedance r | 34.3 MΩ/m |
| E0 (8.6 MW input) | 15.2 MV/m |

#### 5.5.4.2 Travelling wave accelerator structure for pre-accelerator region

All TW sections are 4.3 m long, $3\pi/4$-mode constant-gradient accelerator structures. The "phase advance per cell" was chosen to optimize the RF efficiency for this large-aperture structure. The benefits are a low pulse heating, easy installation for long solenoids, no need to use circulators for RF reflection protection, apparent simplicity and cost efficiency. Table 5.5 gives the important RF parameters.

**Table 5.5**
Parameters of TW structure.

| | |
|---|---|
| Structure Type | TW $3\pi/4$ Mode |
| Cell Number | 50 |
| Aperture 2a | 46 mm |
| Attenuation $\tau$ | 0.98 |
| Q | 24842 – 21676 |
| Group velocity Vg/c | 0.62% – 0.14% |
| Shunt impedance r | 48.60 – 39.45 MΩ/m |
| Filling time $T_f$ | 5.3 µs |
| Power Dissipation | 8.2 kW/m |
| $E_0$ (10 MW input) | 8.5 MV/m |

#### 5.5.4.3 RF systems

Each of these accelerator sections has an individual RF station powered by a 1300 MHz, peak-power 10 MW pulsed klystron. The RF station consists of modulator, RF windows, phase shifters, RF loads, directional couplers and low-level RF system. For the SW structures, RF circulators are needed for reflection protection of the power klystrons.

### 5.5.5 Magnets

The positron-source magnet system has 157 dipoles, 509 quadrupoles and 253 corrector magnets. The large magnet count is a result of the long beamlines connecting the various segments of the source. The magnet designs themselves are quite straightforward. In addition, are used large-aperture DC solenoids, surrounding the L-band capture RF elements, to focus the positrons at low energies. These magnets are normal conducting to withstand the beam loss in the target area. There are also two SC solenoids for spin rotation in the PLTR. The three types of solenoids and their parameters are shown in Table 5.6.

**Table 5.6**
Solenoid Parameters

| Item | Length (m) | ID (cm) | Field Range (kG) | Field, nominal (kG) | N (#) |
|---|---|---|---|---|---|
| Short Solenoid | 1.3 | 36 | 4-8 | 5 | 4 |
| Long Solenoid | 4.3 | 31 | 4-8 | 5 | 23 |
| SC Solenoid | 2.5 | 6 | 52.4 | 52.4 | 2 |

### 5.5.6 Diagnostics

The Positron source has the normal complement of beamline instrumentation to measure orbit, emittance, charge and energy spread. Specialised diagnostics are designed into systems that are unique to the positron source, e.g. target. The largest channel count comes from the BPM system and the long beamlines. Performance specifications for the diagnostics are in most cases more relaxed than the Main Linac or RTML.

### 5.5.7 Electron & photon beam dumps

There are 9 beam dumps, 16 variable-aperture collimators, 1 fixed-aperture collimator and 5 stoppers with burn-through monitors in the positron-source system. Three of the beam dumps must absorb sufficiently large beam power that they require designs with water in the path of the beam. The plumbing required to cool and treat the resulting radioactive water dominates the cost of the dump and collimator technical system in this area of the ILC.

There is a tune-up dump upstream of the undulator (downstream of the Main Linac). It is assumed that this dump is only used with a shortened bunch train (100 bunches) at nominal beam parameters and 5 Hz. At the maximum electron-beam energy of 250 GeV, the tune-up dump must absorb 400 kW. This dump also serves as the abort dump for up to a full train of electrons (1.35 MJ) to protect the undulator. The dump is a 40 cm diameter by 250 cm long stainless-steel vessel filled with 10 mm- diameter aluminum balls through which flows approximately 114 liters per minute of water; it is backed by a short length of solid copper cooled on its periphery by water. Personal access needs to be shielded from the dump by 10 cm of steel and 40 cm of concrete. A service cavern is required to house a heat exchanger, pumps and a system to extract and safely dispose of hydrogen and tritium from the water.

A second dump, technically identical is required for tuning the 5 GeV positrons before injection into the damping ring.

The most challenging dump in the positron-production system is the one that absorbs non-interacting undulator photons from the positron-production target. This dump must absorb 300 kW continuously ($2 \times 10^{17}$ photons/sec of 10 MeV average energy produced with a 3 µrad angular spread.) The primary absorber in this case must be water, contained in a vessel with a thin window. Calculations have shown that, at the nominal distance of 500 m from the middle of the undulator to the positron target and the nominal distance of 150 m from the target to the dump, the power density on a 1 mm Ti window is 0.5 kW/cm$^2$ and the resultant temperature rise after the passage of one bunch train is 425 °C; in the core of the beam, the rise in the water temperature would be 190 °C. The dump is a compact (10 cm diameter by 100 cm long) pressurized (12 bar) water vessel with a Ti window, with a radioactive-water processing system.

The remaining dumps and collimators in the positron system are all based on peripherally cooled solid-metal construction, with the cooling water supplied directly from the accelerator low-conductivity water (LCW) system.

# Chapter 6
# Damping Rings

## 6.1 Introduction

The ILC damping rings include one electron and one positron ring, each 3.2 km long, operating at a beam energy of 5 GeV. The two rings are housed in a single tunnel in the central region of the site, with one ring positioned directly above the other. The damping rings must perform three critical functions:

- accept $e^-$ and $e^+$ beams with large transverse and longitudinal emittances and produce the low-emittance beams required for high-luminosity production;
- damp incoming beam jitter (transverse and longitudinal) and provide highly stable beams for downstream systems;
- Delay bunches from the source to allow feed-forward systems to compensate for pulse-to-pulse variations in parameters such as the bunch charge.

The damping ring system includes the injection and extraction systems, which themselves include sections of transport lines matching to the sources (upstream of the damping rings) and the RTML system (downstream of the damping rings).

This chapter first discusses the parameters and configuration of the damping rings before describing the lattice and various instabilities, most notably the electron-cloud effect, that can affect the beam parameters. The vacuum and RF systems are described, followed by the magnet systems and finally injection and extraction.

## 6.2 Top-level parameters and configuration

The nominal parameters of injected and extracted beams for both the electron and positron damping rings in the baseline configuration are listed in Table 6.1.

**Table 6.1**
Nominal parameters of injected and extracted beams for the baseline configuration.

| Parameter | Electron Beam | Positron Beam |
|---|---|---|
| Train repetition rate [Hz] | 5.0 | |
| Main Linac Bunch separation [ns] | 554 | |
| Nom. # bunches per train | 1312 | |
| Nom. bunch population | $2 \times 10^{10}$ | |
| Required acceptance:† | | |
| Norm. betatron amplitude $(a_x + a_y)_{\max}$ [m rad] | 0.07 | |
| Long. emittance $(\Delta E/E \times \Delta l)_{\max}$ [% × mm] | 0.75×33 | |
| Extraction Parameters: | | |
| Norm. horizontal emittance $\gamma\epsilon_x$ [µm rad] | 5.5 | |
| Norm. vertical emittance $\gamma\epsilon_x$ [nm rad] | 20 | |
| RMS relative energy spread $\sigma_p/p$ [%] | 0.11 | |
| RMS Bunch length $\sigma_z$ [mm] | 6 | |
| Max. allowed transfer jitter $[\sigma_{x,y}]$ | 0.1 | |

† specified for the positron damping ring

The configuration of the damping rings is constrained by the timing scheme of the main linac [95]. In particular, each damping ring must be capable of storing a full bunch train (1312 bunches) and reducing the emittances to the required level within the 200 ms (100 ms in the 10 Hz mode) interval between machine pulses. In addition, the relatively large bunch separation in the main linacs means that the damping rings must be capable of injecting and extracting individual bunches without affecting the emittance or stability of the remaining stored bunches. The exact circumference has been chosen to provide adequate flexibility in the fill pattern allowing different timing configurations in the linac. The bunch trains are separated by gaps to mitigate the fast ion instability in the electron ring.

The superconducting RF system is operated at 650 MHz which is exactly half the linac frequency. To achieve the short damping times necessary to reduce the emittance (by roughly six orders of magnitude in the case of the positron vertical emittance) within the allowed 200 ms interval, superconducting wigglers of total length roughly 100 m are used in each damping ring.

The layout [96] of the damping ring is a racetrack, with long straights [97] to accommodate damping wigglers, RF cavities, phase trombone, injection, extraction, and circumference-adjusting chicane as shown in Fig. 6.1.

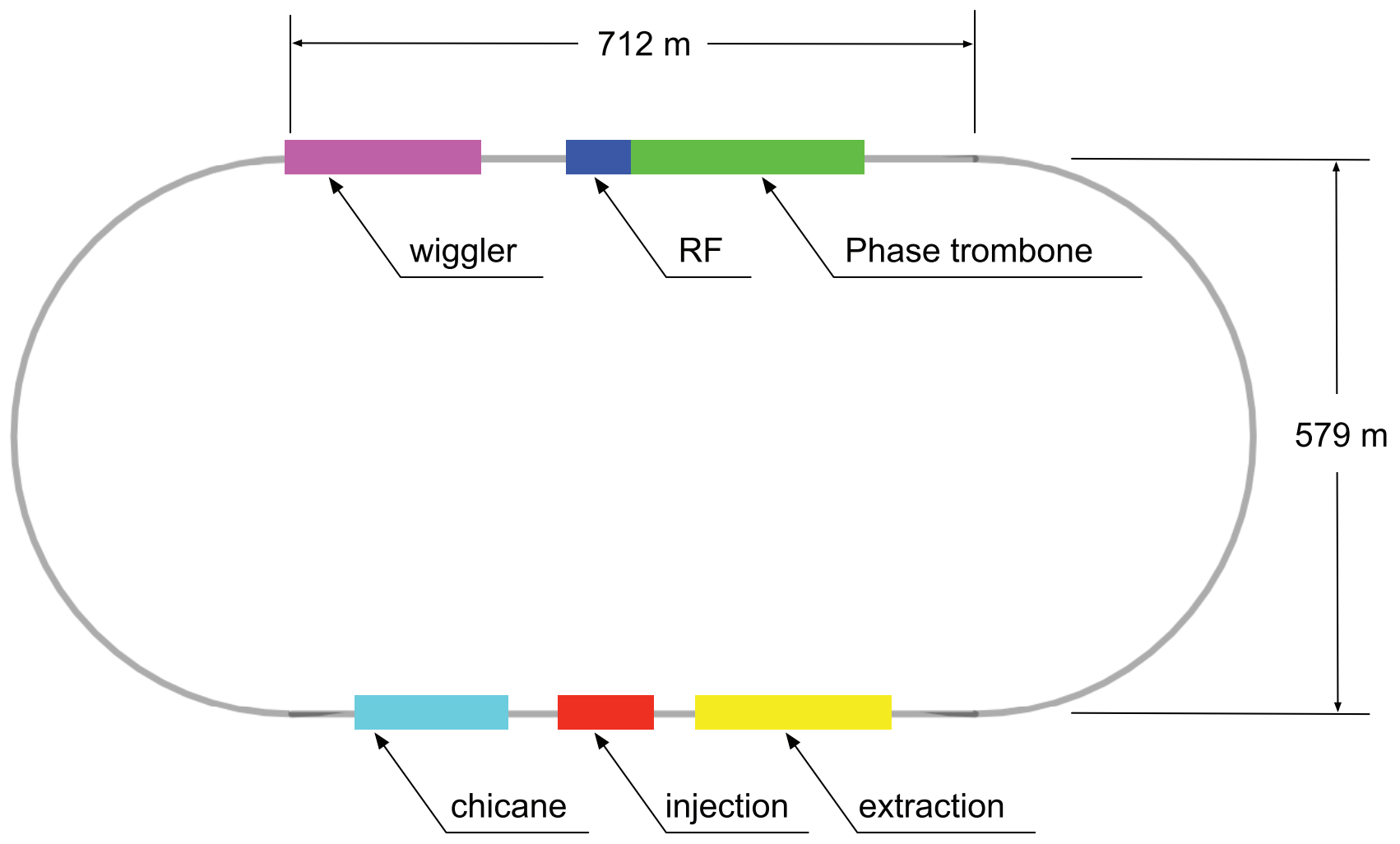

**Figure 6.1**
Damping-ring layout: the circumference is 3238.7 m; the length of each straight is 710.2 m.

Damping-ring parameters for three ILC operating modes, corresponding to four distinct configurations, have been developed. Two of these operating modes utilize a 5 Hz repetition rate: the low-power baseline with 1312 bunches in each ring; and the high-luminosity upgrade with 2625 bunches. The third operating mode is at 10 Hz and has distinct configurations for the operation of the positron and electron rings. The parameters of these operating modes are summarized in Table 6.2 based on the current version of the DR baseline lattice.

In the 10 Hz operating configuration, the electron linac is operated with alternating pulses, a high-energy pulse for positron production followed by a low energy pulse for collisions. Each damping ring has a pulsed time profile with beam injection/extraction times of 1 ms. In the positron ring, full beam current is stored for 100 ms and then extracted, leaving the ring empty for 100 ms before the next injection cycle. A shorter damping time is necessary to achieve the same extracted vertical emittance in half the nominal storage time.

For the high-luminosity upgrade, twice the number of bunches need to be stored in the DR with 3.1 ns bunch spacing. The doubling of the current in the rings poses a particular concern for the positron DR due to the effects of the electron-cloud instability. In the event that the electron-cloud mitigations that have been recommended [99] are insufficient to achieve the required performance for this configuration, allowance had been made for the installation of a second positron ring in the same tunnel. In this scenario, the two positron rings would both operate with the baseline parameters. The third ring would be placed above the electron ring as indicated in Fig. 6.2a and Fig. 6.2b.

**Table 6.2**
Damping ring lattice parameters for 5 Hz *Low Power* (baseline) and *High Luminosity* (upgrade) operating modes and 10 Hz repetition rate (baseline) operation [98].

| Parameter | 5 Hz Mode | | 10 Hz Mode | |
|---|---|---|---|---|
| | Low Power | High Lumi | Positrons | Electrons |
| Circumference [km] | 3.238 | | 3.238 | |
| Number of bunches | 1312 | 2625 | 1312 | |
| Particles per bunch [$\times 10^{10}$] | 2 | 2 | 2 | |
| Maximum beam current [mA] | 389 | 779 | 389 | |
| Transverse damping time $\tau_x, \tau_y$ [ms] | 23.95 | | 12.86 | 17.5 |
| Longitudinal damping time $\tau_z$ [ms] | 12.0 | | 6.4 | 8.7 |
| Bunch length $\sigma_z$ [mm] | 6.02 | | 6.02 | 6.01 |
| Energy spread $\sigma_E/E$ [%] | 0.11 | | 0.137 | 0.12 |
| Momentum compaction factor $\alpha_p$ [$\times 10^{-4}$] | 3.3 | | 3.3 | |
| Normalized horizontal emittance $\gamma\epsilon_x$[μm] | 5.7 | | 6.4 | 5.6 |
| Horizontal chromaticity $\xi_x$ | −51.3 | | −50.9 | −51.3 |
| Vertical chromaticity $\xi_y$ | −43.3 | | −44.1 | −43.3 |
| Wiggler Field [T] | 1.51 | | 2.16 | 1.81 |
| Number of Wigglers | 54 | | 54 | |
| Energy loss/turn [MeV] | 4.5 | | 8.4 | 6.19 |
| RF Specifications: | | | | |
| Frequency [MHz] | 650 | | 650 | |
| Number of cavities | 10† | 12 | 12 | |
| Total voltage [MV] | 14.0 | | 22.0 | 17.9 |
| Voltage per cavity [MV] | 1.40 | 1.17 | 1.83 | 1.49 |
| RF synchronous phase [°] | 18.5 | | 21.9 | 20.3 |
| Power per RF coupler [kW]‡ | 176 | 294 | 272 | 200 |

† The baseline RF deployment for positrons is 12 cavities to support 5 and 10 Hz modes.
‡ Power/coupler computed as (Max. Current) × ($E$ loss/turn)/(No. cavities).

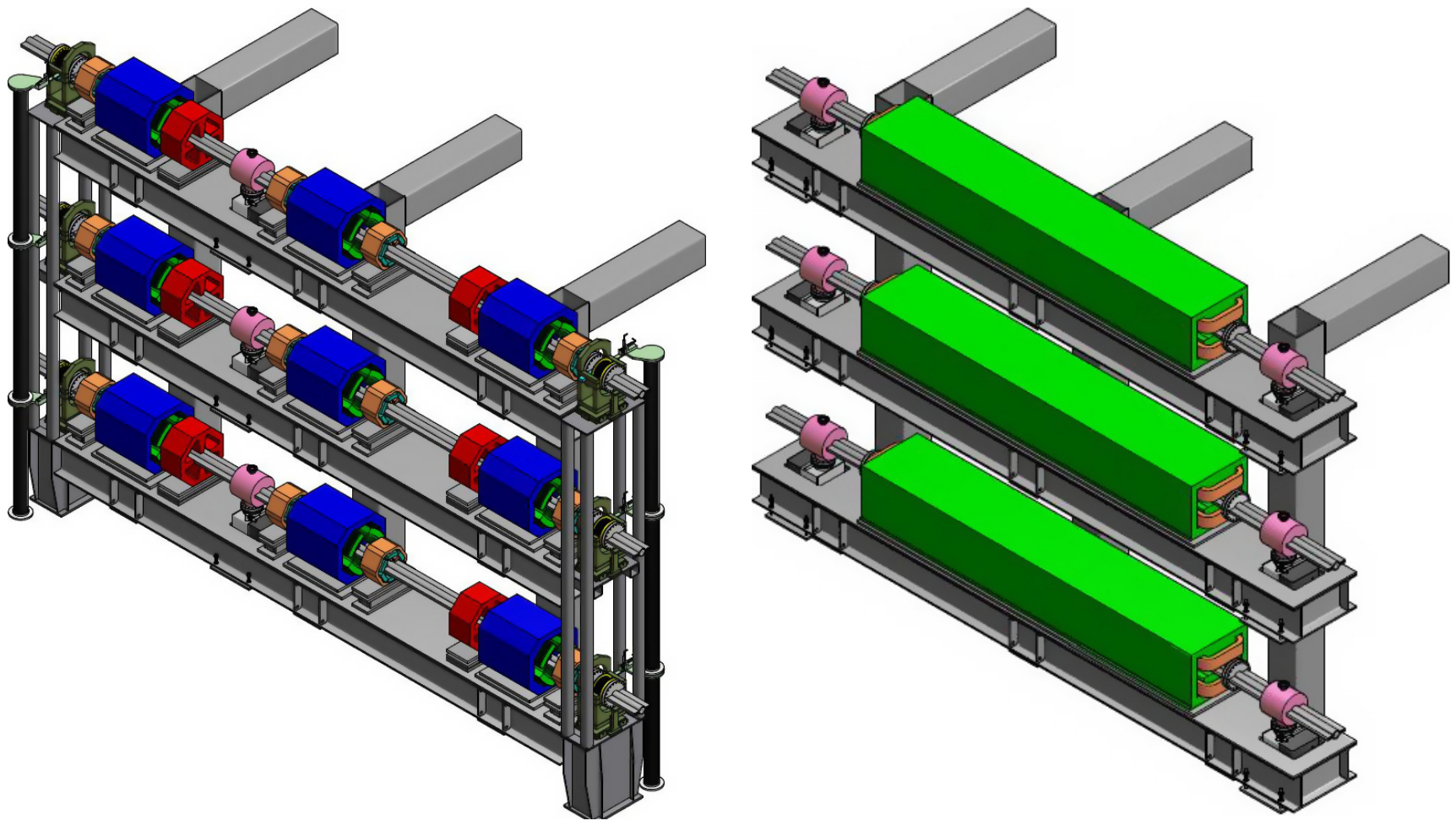

(a) (b)

**Figure 6.2**
Damping-ring arc magnet layout with positron ring at the bottom and electron ring directly above. A second positron ring would be placed above the electron ring if required: (a) quadrupole section layout and (b) dipole section layout.

The superconducting damping wigglers [100] are based on the CESR-c design, with 14 poles and 30 cm period. The peak field of the 54 1.875 m long wigglers is 1.51 T for a 24 ms damping time in the 5 Hz mode and 2.16 T for a 13 ms damping time for 10 Hz operation. The horizontal emittance is near 0.5 nm rad over the range of relevant wiggler fields. There are 10 single-cell 650 MHz superconducting cavities in the baseline configuration. For 5 Hz operation, 8 of these cavities can provide a total of 14 MV for a 6 mm bunch length, even in the event of a single klystron failure. For 10 Hz operation, the number of cavities is increased to 12 and the accelerating voltage to 22 MV for the same 6 mm bunch length. A phase trombone provides for adjustment of betatron tune and a chicane for small variations of the circumference.

## 6.3 Lattice description

Each arc in the DR consists of 75 cells [101], each with one focusing and two defocusing quadrupoles placed symmetrically about a single 3 m bend. Focusing and defocusing sextupoles are located adjacent to the corresponding quadrupoles. Each cell contains one vertical, one horizontal, and one skew quad corrector as well as two beam-position monitors adjacent to the defocusing sextupoles, as shown in Fig. 6.3. Dispersion suppressors, at the ends of the arc, match the finite dispersion in the arcs to zero dispersion in the straights. The dispersion suppressor beam line includes two dipole bending magnets and seven quadrupoles. There is a skew quad corrector at each of the two dipoles.

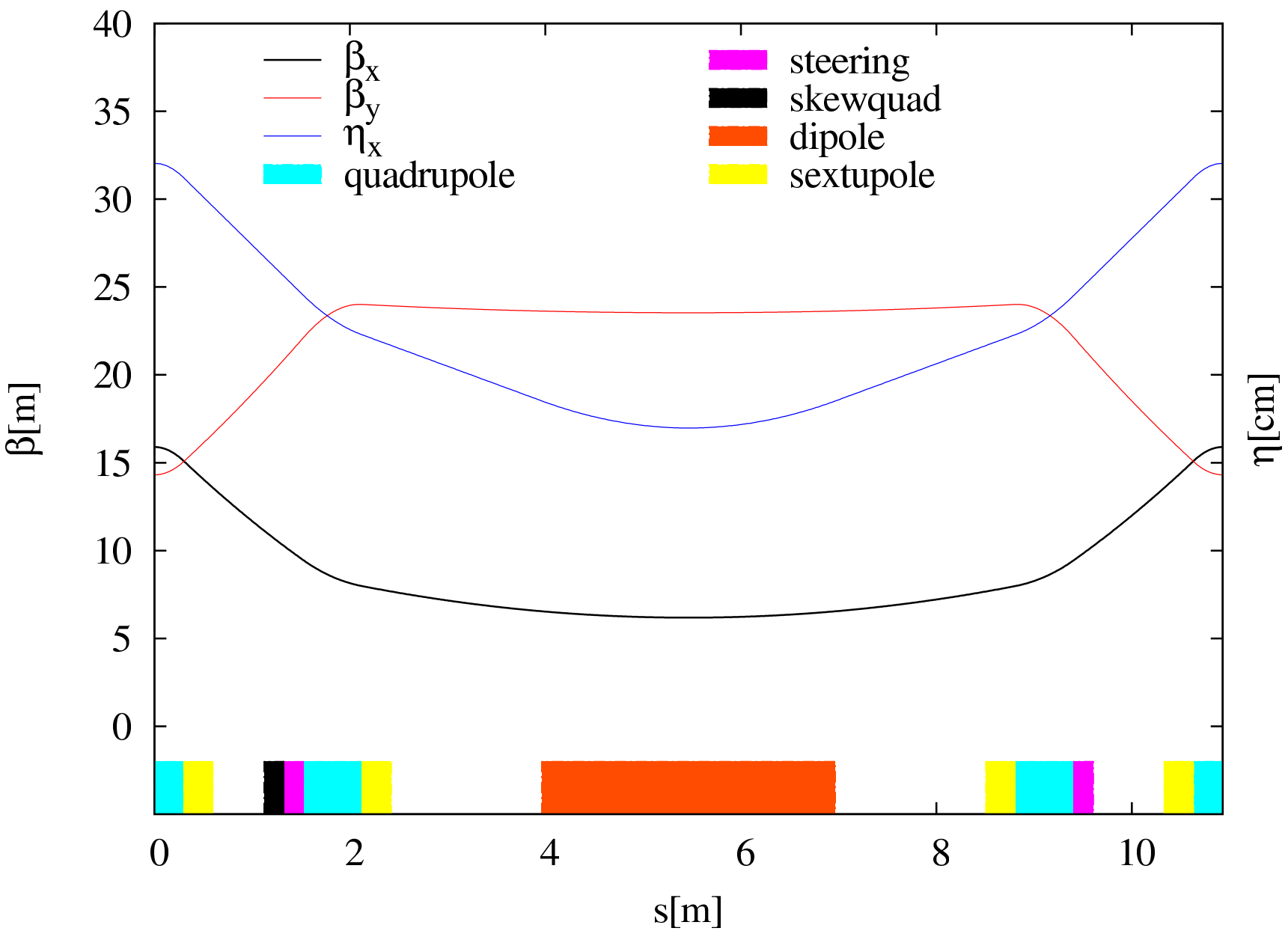

**Figure 6.3** Arc cell. The cell boundaries are at the midpoint of the focusing quadrupole. Each cell contains a vertical and horizontal dipole corrector, a skew quad corrector in each cell, and two beam position monitors adjacent to the defocusing sextupoles.

Acceptable values of the momentum compaction are bounded from below by the single-bunch instability threshold, and from above by the RF voltage required to achieve the requisite 6 mm bunch length. In practice, the dynamic aperture shrinks as the focusing strength is increased to reduce momentum compaction. The design has nonlinear aperture adequate to accept the entire phase space of the injected positrons, and consistent with the specified horizontal emittance. The resulting momentum compaction is $3.3 \times 10^{-4}$. The TME-style arc cell gives somewhat better dynamic aperture than the FODO cell tuned to give comparable emittance and with the same 3 m dipole.

The RF-Wiggler straight provides space for 12 RF cavities at 650 MHz (as well as open space for up to 4 additional cavities) and 54 wigglers (with open space for 6 more). The phase trombone is also located in this straight.

The phase trombone consists of five cells, each constructed from six equally spaced, alternating-gradient quadrupoles. The overall length of the phase trombone is 339 m. The range of the phase trombone is a full betatron wavelength in both horizontal and vertical. There is a single-skew quadrupole corrector in each of the five cells.

The machine circumference is adjusted by varying the field of the chicane dipoles. A 4.4 mm increase in path length is accompanied by an increase in horizontal emittance of about 3 %.

There are horizontal and vertical dipole correctors and a beam-position monitor adjacent to each quadrupole in the straights. The lattice functions for the ring are shown in Fig. 6.4.

The injection is located upstream of the centre of the long injection/extraction straight, and the extraction downstream of the centre as shown in Fig. 6.1.

The injection line grazes the outside of a quadrupole, and is deflected horizontally by a pair of septum bend magnets so the trajectory passes inside the aperture of the next quadrupole. This horizontally defocusing quadrupole makes the trajectory nearly parallel to the stored orbit. At 90° of betatron phase downstream from the septa, where the injection trajectory crosses the stored orbit, a set of kickers deflects the single injected bunch onto the stored orbit.

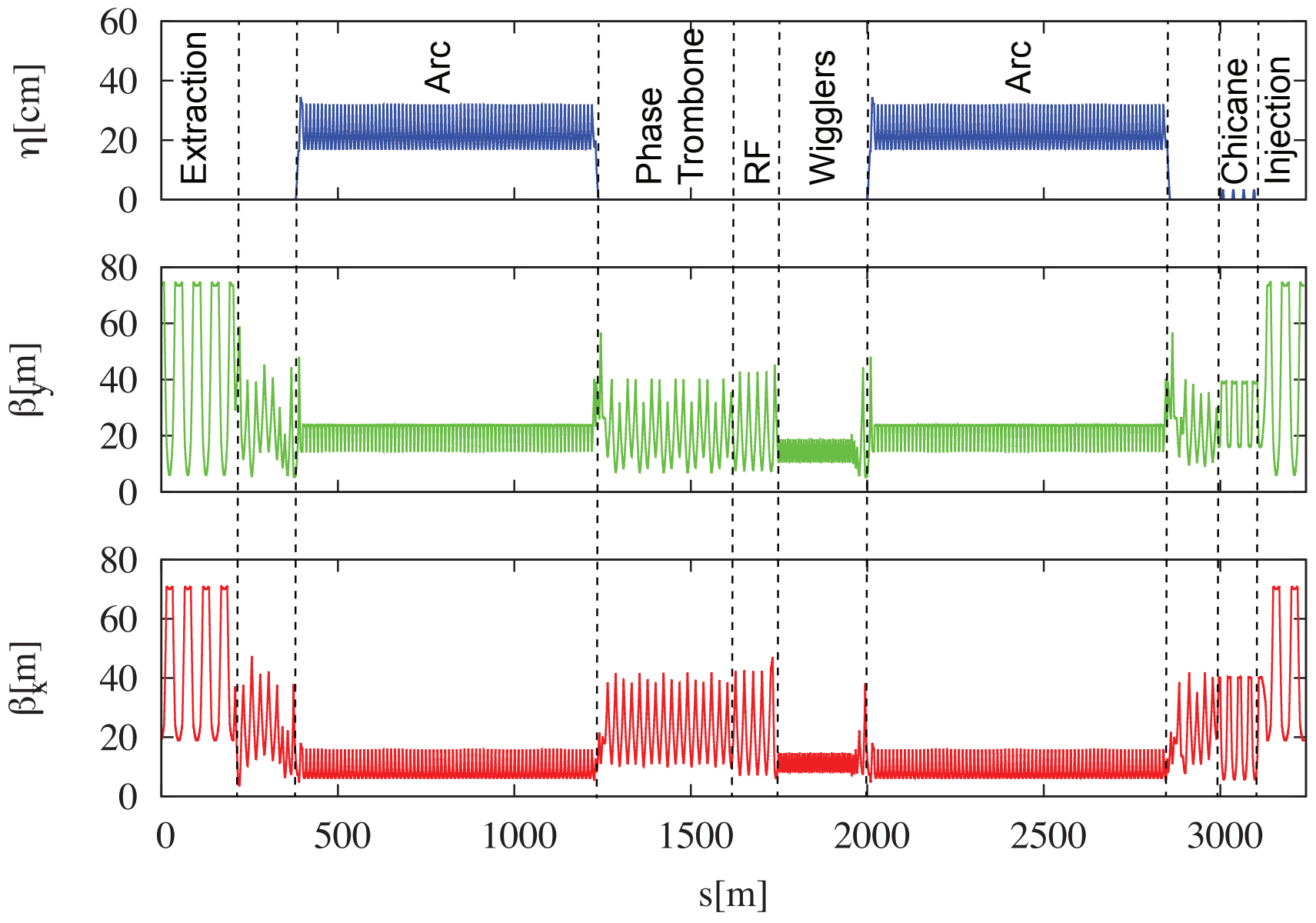

**Figure 6.4** Damping ring lattice functions with the major functional sections (injection, extraction arcs, RF, wigglers, circumference chicane, and phase trombone) labelled.

Extraction is accomplished with a similar set of kickers that deflects a single damped bunch horizontally. A horizontally defocusing quadrupole increases the deflection and at 90° of betatron phase downstream of the kickers, the bunch passes through the bending fields of a pair of septum magnets. These deflect the trajectory further horizontally, so it passes outside of the next focusing quadrupole and into the extraction-line optics. The stored orbit is located in the nominally field-free region of the septum magnets and is not significantly affected.

## 6.4 Beam Dynamics

### 6.4.1 Emittance Tuning

An emittance-tuning procedure based on that developed at CesrTA [102] has been used to compensate for misalignments and field errors in the DR design. The procedure has three basic steps:

1. Measure and correct the closed orbit errors using all BPMs and all dipole correctors;
2. Measure betatron phase and coupling by resonant excitation and correct errors, using all quadrupoles and skew-quadrupole correctors;
3. Repeat measurement of orbit and coupling, and measure dispersion by resonant excitation of the synchrotron motion, and then fit simultaneously using vertical dipole correctors and skew quadrupoles.

The tuning algorithm depends for its effectiveness on the accuracy of the beam-position monitors. The BPM specification is given in Table 6.3. The alignment tolerances are given in Table 6.3, as are the multipole errors measured at SLAC for the SPEAR and PEPII dipoles, quadrupoles and sextupoles [103]. The results of the emittance-tuning procedure for 100 lattices with a randomly chosen Gaussian distribution of multipole and alignment errors are summarised in Fig. 6.5a and Fig. 6.5b. The tuning procedure consistently achieves the specified geometric vertical emittance of 2 pm rad.

**Table 6.3**
BPM and magnet alignment tolerances.

| Parameter | RMS |
|---|---|
| BPM Differential resolution† | 2 µm |
| BPM Absolute resolution | 100 µm |
| BPM Tilt | 10 mrad |
| BPM differential button gain | 1% |
| Quad & Sextupole Offset (H&V) | 50 µm |
| Quadrupole Tilt | 100 µrad |
| Dipole Roll | 100 µrad |
| Wiggler vertical Offset | 200 µm |
| Wiggler - Roll | 200 µrad |

† Reproducibility of single pass measurement

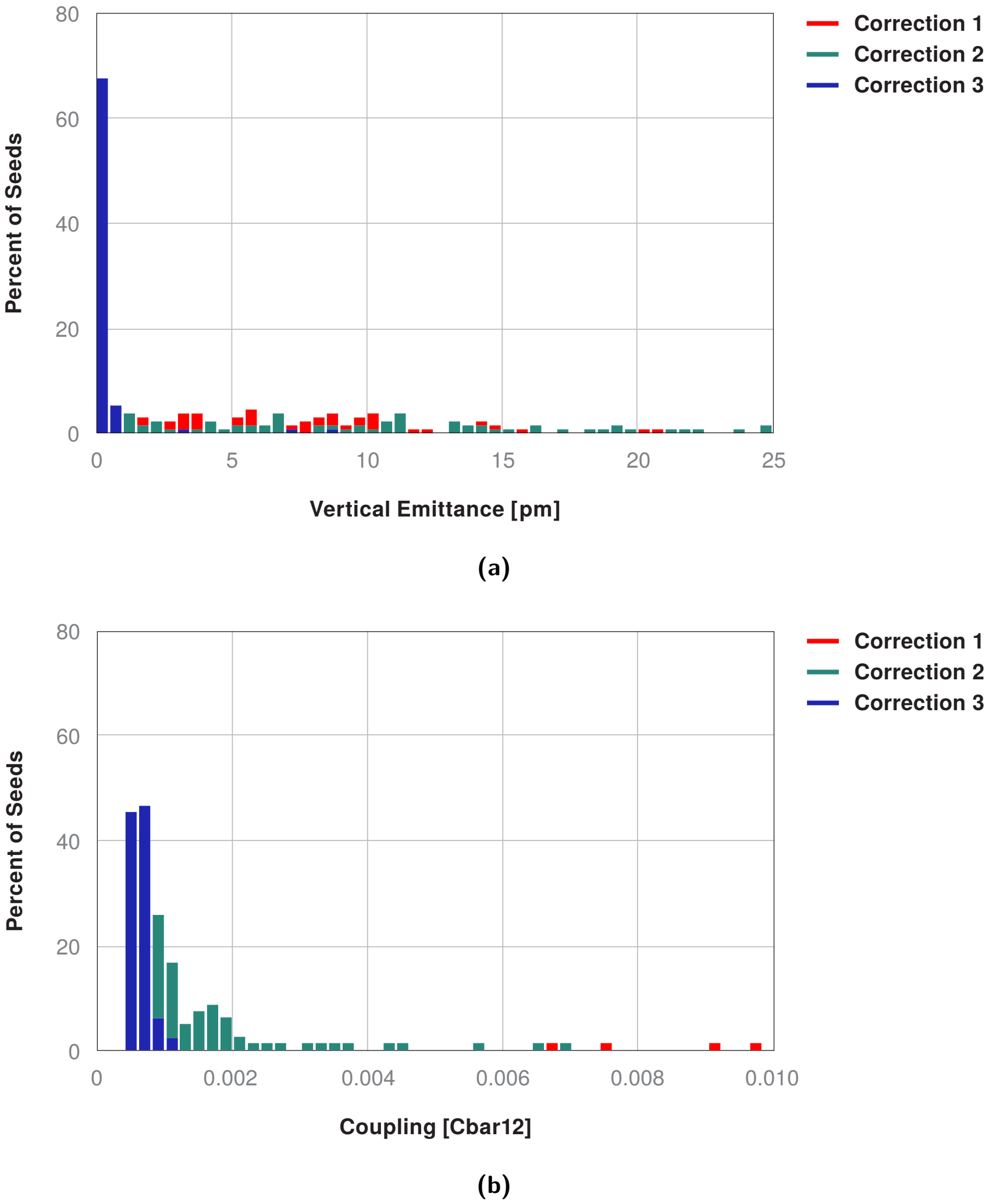

**Figure 6.5**
Histogram of the (a) vertical emittance and (b) rms coupling ($\overline{C}_{12}$) at the conclusion of each step in the low-emittance tuning procedure for 100 lattice files with randomly chosen misalignments and multipole errors.

## 6.4.2 Dynamic aperture

The specification for the phase space distribution of the injected positron bunch is an amplitude of $A_\mathrm{x} + A_\mathrm{y} \leq$0.07 m rad (normalized) and an energy spread of $\Delta E/E \leq$0.75%. The dynamic aperture for a lattice with the specified misalignments and multipole errors, and wiggler nonlinearities including those due to end effects and finite pole width, is shown in Fig. 6.6. In order to evaluate the effect of the wiggler nonlinearities on dynamic aperture, the wiggler field is computed with a finite-element code (Vector Fields), and fit to an analytic form as a Fourier expansion that automatically satisfies Maxwell's equations. A symplectic tracking algorithm ensures that the phase space is not distorted by numerical noise. None of the injected particles are lost in these simulations.

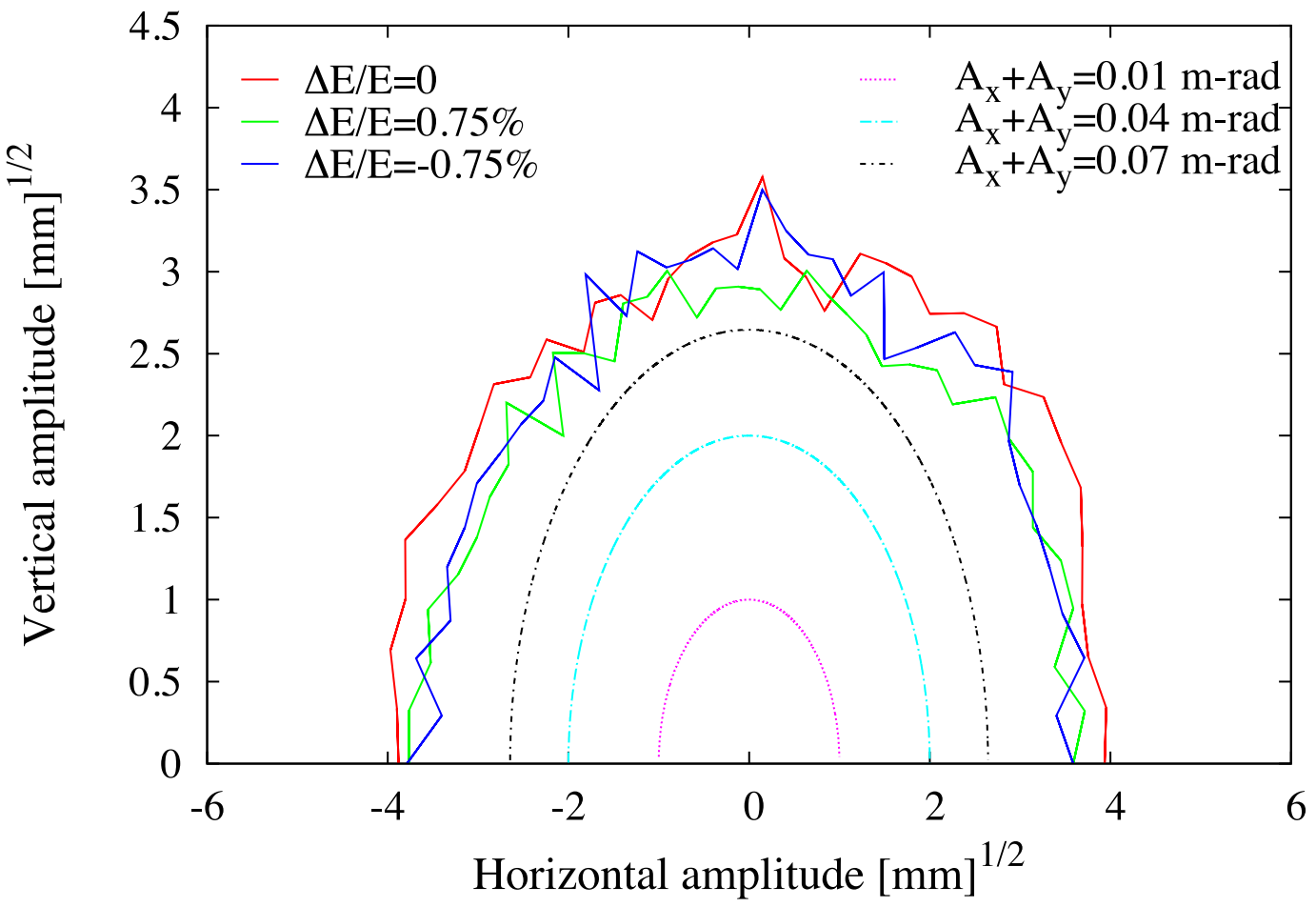

**Figure 6.6**
Dynamic aperture including multipoles, wiggler nonlinearities and magnet misalignments. Defined as the largest stable amplitude after tracking 1000 turns.

## 6.4.3 Collective Effects

The many collective effects that may affect beam quality in the damping rings were examined during the original baseline configuration studies [104]. These include impedance-driven instabilities, intrabeam scattering, space-charge effects, electron cloud effects in the positron ring and ion effects in the electron ring. The beam's Touschek lifetime is expected to be much longer than the nominal 200 ms spent in the DR, however, obtaining suitable lifetimes for commissioning and tuning will be important. The largest sources of emittance dilution were found to be the Electron Cloud (EC) instability in the positron DR and the Fast Ion Instability (FII) in the electron DR.

## 6.4.4 Electron Cloud

Electron Cloud (EC) induced instabilities and emittance growth are critical issues for the positron damping ring. The electron cloud develops quickly as photons striking the vacuum chamber wall knock out photoelectrons that are then accelerated by the beam, gain energy, and strike the chamber again, producing secondary electrons. The peak secondary electron yield (SEY) of typical vacuum chamber materials is $>1$, even after surface treatment, leading to electron amplification of the cascade. Once the cloud is present, coupling between the electrons and the circulating beam can cause a single-bunch (head-tail) instability and incoherent tune spreads that may lead to increased emittance, beam oscillations, or even beam losses. The threshold electron cloud density, beyond which there is emittance growth and the onset of instabilities, has been determined using measurements at CESRTA and calculations with the simulation code CMAD. The electron density is computed with codes that simulate cloud growth for various bunch configurations and vacuum chamber geometries and surface properties.

The build-up of the EC in the DR vacuum chambers has been modeled using the EC mitigations specified by the ILC Electron Cloud Working Group [99] (see Table 6.4). The simulations employ the vacuum system conceptual design described in Section 6.5.

**Table 6.4**
EC mitigations specified for the positron DR.

| Magnetic Region | Primary Mitigation | Secondary Mitigation |
|---|---|---|
| Drift | TiN Coating | Solenoid Windings |
| Dipole | Grooves with TiN Coating | Antechamber |
| Wiggler | Clearing Electrodes | Antechamber |
| Quadrupole | TiN Coating | — |

#### 6.4.4.1 Photon Transport Model

The distribution of absorption sites for synchrotron radiation photon around the ring can be used to predict the sources of the photoelectrons which seed the electron cloud. This distribution has been computed for the DR lattice using a newly developed photon-tracking simulation code, Synrad3D [105]. This code computes the synchrotron-radiation photons per electron generated by a beam circulating in the magnetic lattice, and simulates in three dimensions the propagation of the photons as they scatter off, or are absorbed by, the vacuum chamber. The design vacuum-chamber geometry (see Section 6.5), including details such as antechambers and photon stops, is used in the calculation. Both specular and diffuse photon scattering are included in the simulation. For the scattering calculation, the surface material is approximated as aluminum with a thin carbon coating, and the surface parameters are representative of a typical technical vacuum chamber, namely an rms roughness of ~100 nm and a correlation length of ~5000 nm.

#### 6.4.4.2 EC Buildup

The EC buildup in each of the principal magnetic field regions of the damping ring has been modeled. In the dipole field regions, the modeling code POSINST [106] was employed. Simulations with both POSINST and ECLOUD [107] were carried out in the quadrupole, sextupole and drift regions. The CLOUDLAND [108] package was used for simulations in the wiggler region.

Simulations of the EC buildup in the dipole chambers were based on:

- SEY model parameters for a TiN surface based on the most recent data obtained at CesrTA;
- Photon distributions around the vacuum chambers based on Synrad3d modeling of the ILC DR vacuum system;
- A quantum efficiency of 0.05, independent of photon energy and incident angle.

The SEY model corresponding to the above-mentioned fits yields a peak SEY value of 0.94 at an incident electron energy of 296 eV. In addition to this, the simulations have also been run with the SEY set to 0 (meaning that any electron hitting the chamber walls gets absorbed with unit probability) in order to isolate the contribution to the EC density $\mathrm{N}_e$ from photoemission. The cloud densities in the dipoles are expected to be between these two limits since the simulation does not directly model the reduction in the effective SEY from having grooved top and bottom chamber surfaces.

Cloud densities averaged over the full vacuum chamber as well as those averaged over a $20\sigma_{\mathrm{x}} \times 20\sigma_{\mathrm{y}}$ elliptical cross-sectional area were calculated. The modelling statistical uncertainties are at the level of less than 30 %.

Simulations of the EC buildup in the quadrupoles and sextupoles in the arc and wiggler regions and in the drift regions of the wiggler sections for the DR lattice utilized the same photoelectron and SEY model parameters as were used for the dipole region. Representative field strengths of 10 T/m (70 T/m$^2$ )were used for the quadrupoles (sextupoles). Trapping effects were evident in the beam-pipe-averaged cloud densities, which had not yet reached equilibrium during the eight trains simulated, but since the trapping does not occur in the beam region, the 20$\sigma$ densities prior to the passage of each bunch were stable after just a couple of trains.

The simulations for the field-free regions were repeated with a solenoidal magnetic field of 40 G, as recommended during the ECLOUD10 workshop [99]. Such a field was shown to reduce the cloud buildup in the beam region to negligible levels.

Table 6.5 shows the 20$\sigma$ EC density estimates obtained from the above simulations.

The EC buildup in the wiggler is simulated using the CLOUDLAND code [109]. The ring length occupied by wigglers is 118 m. The simulation assumes a peak SEY of 1.2 for the copper surface of a wiggler chamber. The energy at the peak SEY is 250 eV. The photon flux used in the simulation is 0.198 photon/m/positron with a uniform azimuth distribution. The quantum efficiency is 0.1 and the beam size $\sigma_x/\sigma_y = 80\mu/5.5\mu$. The peak wiggler field for the simulation is set to 2.1 T. The beam chamber of the wiggler section includes an antechamber with 1 cm vertical aperture. A round chamber with diameter of 46 mm is used, which is a good approximation since most electrons accumulate near the horizontal centre due to multipacting. The CLOUDLAND calculation shows that a beam with bunch population of $2 \times 10^{10}$ and bunch spacing of 6 ns can excite strong multiplication near the horizontal centre. The peak electron density seen by the last bunch along the bunch train is about $1.2 \times 10^{13} \mathrm{m}^{-3}$. However, the wiggler vacuum design includes clearing electrodes. Application of a modest positive voltage (of a few hundred volts) has been shown to reduce the electron density near the beam by more than three orders of magnitude, ie, to levels well below those at which beam instabilities are expected to develop.

**Table 6.5**
POSINST and ECLOUD modelling results for the $20\sigma$ density estimates, $N_e$ ($10^{11}$ $m^{-3}$), just prior to each bunch passage with the baseline lattice.

| | Field-free | Field Free w/solenoid | Dipole | Quad | Sext |
|---|---|---|---|---|---|
| Arc Region | 2.5 | 0 | 0.4 | 1.6 | 1.35 |
| Wiggler Region | 40 | 0 | | 12 | |

#### 6.4.4.3 EC Instability

The above estimates of cloud density place an upper limit on the ring-averaged density of about $4 \times 10^{10} \mathrm{m}^{-3}$, about a factor of three below the expected single bunch instability threshold [110]. Thus operation in the baseline configuration is expected to have negligible emittance dilution from the EC. This operating margin should also minimize the possibility of any adverse impacts from sub-threshold emittance growth on the positron beam. For the high-luminosity upgrade, a second positron ring may be added if insufficient operating margin remains with a single ring.

### 6.4.5 Fast Ion Instability

A significant concern for the electron damping ring is the fast ion instability (FII). In contrast to the more familiar ion-trapping effect, where ions oscillate stably for long periods in the potential well of the stored beam, the fast ion instability is associated with ions created in the beam path by interaction with the circulating beam during a single turn. Ions created at the head of the bunch train move slowly, and remain in the beam path, influencing the motion of subsequent bunches. The resultant ion-induced beam instabilities and tune shifts are critical issues due to the ultra-low vertical emittance. A low base vacuum pressure at the $1 \times 10^{-7}$Pa level is essential to reduce the number of ions formed. To mitigate bunch motion, there are also bunch-by-bunch feedback systems with a damping time of ≈0.1 ms.

To reduce further the core ion density, short gaps are introduced into the electron-beam bunch train by omitting a number of successive bunches. The use of such *mini-gaps* in the train significantly mitigates the FII by reducing the core ion density and by inducing tune variation along the train. The bunch patterns selected for the DR provide these mini-gaps for any of the proposed main-linac train structures.

The dependence of FII growth rates on gaps in the bunch trains is evident in simulations. Two sets of simulations have been carried out to study this effect in the DR. The simplest meaningful simulation is based on a weak strong model. The circulating bunch is represented as a single macro particle, and is sensitive only to centroid dipole motion. The ion cloud consists of multiple macro particles, that are free to move transversely in the vacuum chamber. This method can be used to characterise the growth of the vertical oscillation amplitude [111].

For a pressure of $1 \times 10^{-7}$Pa with CO as the only gas species, and a single long train of 1312 bunches, with $2 \times 10^{10}$ particles/bunch and $4\lambda_{\mathrm{RF}}$ bunch spacing, the vertical amplitude versus turn is shown in Fig. 6.7a. The oscillation amplitude is greater than the beam size in only 6 turns.

If the 1312 bunches are distributed into 41 trains of 32 bunches, with a train gap of $43\lambda_{\mathrm{RF}}$, then the growth time is 26 turns as shown in Fig. 6.7b.

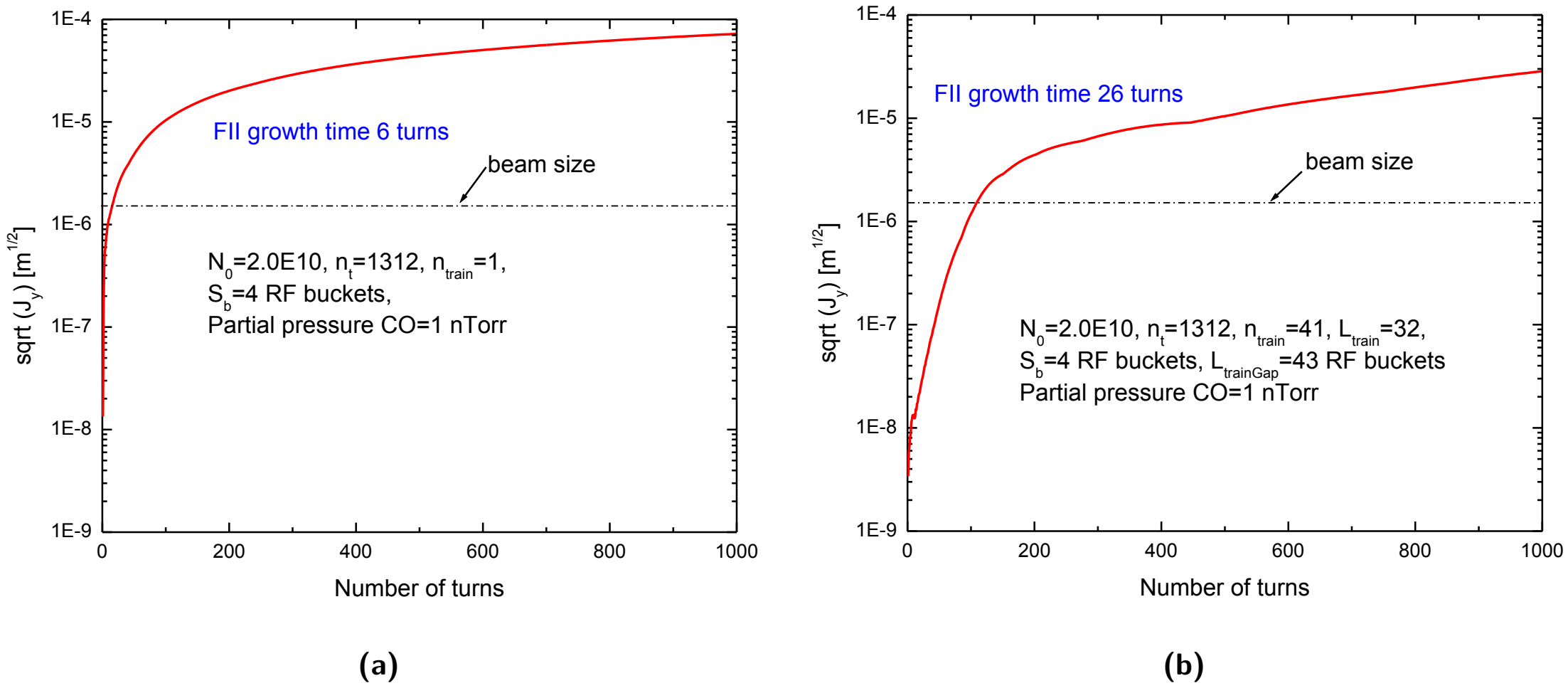

**Figure 6.7.** Simulated vertical amplitude vs number of turns for (a) 1312 bunches in a single train at $1 \times 10^{-7}$Pa CO and (b) for 41 trains of 32 bunches.

A second simulation [112] has been carried out using vacuum parameters based on those observed at SPEAR3 (0.5 $\times$ 10$^{-7}$ Pa pressure with a composition of: 48 % $H_2$ , 5 % $CH_4$, 16 % $H_2O$, 14 % $CO$, and 17 % $CO_2$). The inclusion of multiple gas species is expected to contribute some Landau damping due to the spread in ion frequencies. A uniform pressure along the ring is assumed and the bunch patterns used in the baseline DR configuration as well as in the high-luminosity upgrade were explored. The electron bunch was divided into roughly 11 slices to allow for the possibility of a more complex single-bunch instability than the simple dipole motion in the simulation described above [113]. The code has been benchmarked with a SPEAR3 experiment [114] where there is a good agreement.

Modeling of two possible bunch patterns in the DR gives the fastest exponential growth times of 56 turns and 84 turns (see Table 6.1) during operation in the baseline configuration and 37 turns for the high-luminosity configuration. Figure 6.8 shows the unstable modes within the first of these bunch patterns. A broad-band spectrum is exhibited.

**Figure 6.8**
Simulated vertical amplitude versus time for the bunch pattern with the shortest growth time. The different lines in the plots correspond to individual bunches. The vertical instability growth time is 56 turns. Note (b) uses a logarithmic scale.

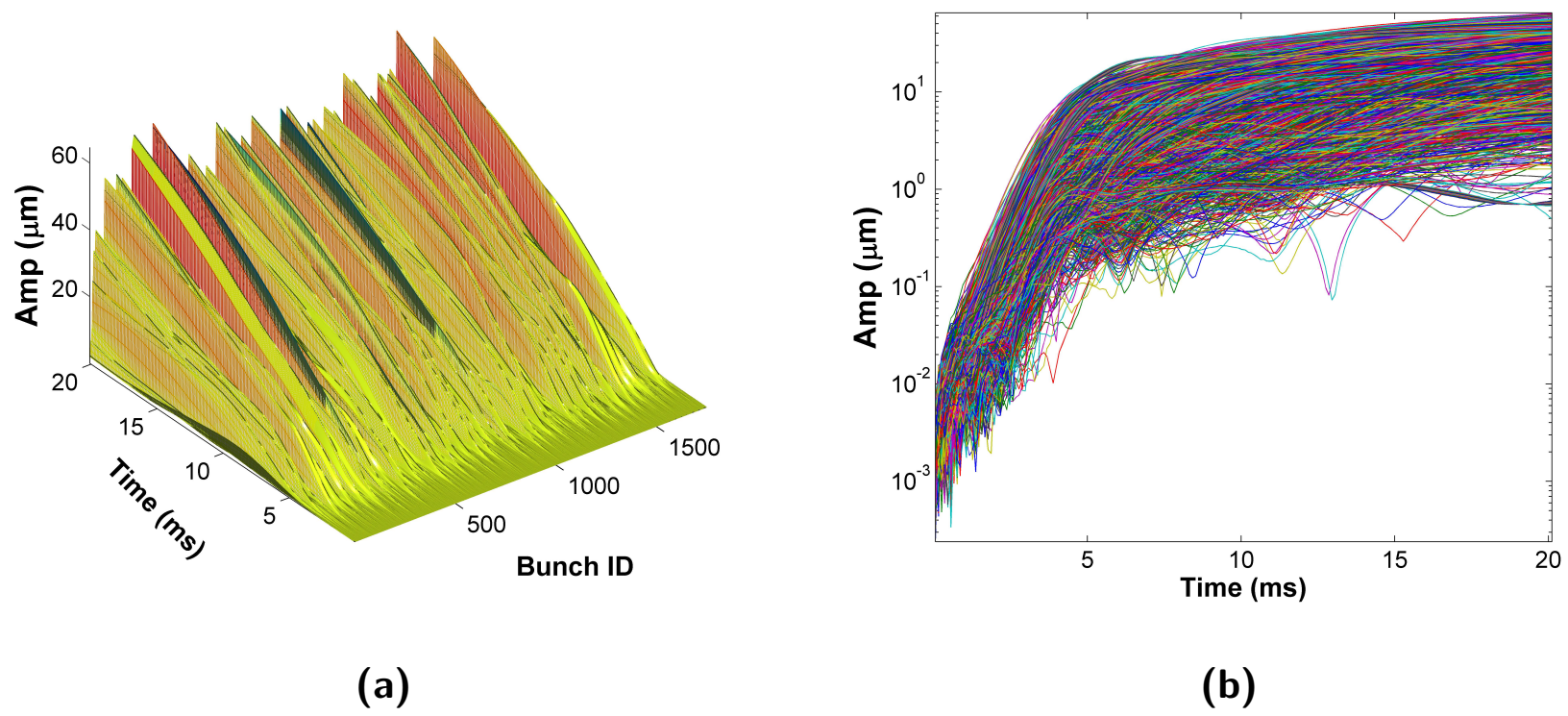

The dependence on pressure and bunch spacing in the two simulations is consistent and growth rates are comparable. The radiation damping time is approximately 2000 turns so feedback is essential to stabilize the beam-ion interaction. Experience at KEK-B and DAΦNE suggests that feedback systems with damping times of order 20 turns, which can suppress the FII, are practical [115].

## 6.5 Vacuum System

The vacuum system conceptual design incorporates EC-mitigation techniques enumerated in Table 6.4, and targets the vacuum performance required to suppress the FII in the electron ring as described in Section 6.4.5. The present conceptual design [116] draws on previous design work [117–119] and incorporates inputs from the lattice designers, magnet engineers, and electron-cloud-dynamics simulation group.

Dipole chambers, shown in Fig. 6.9a, incorporate three EC mitigation techniques: antechambers with radially sloped outside walls are used to minimize scattered photons entering the main beam aperture; a titanium nitride (TiN) coating is applied to the inside surface of the chamber to reduce secondary electron yield (SEY); and grooves on the top and bottom of the vacuum chamber further reduce the number of secondary electrons that enter the central region of the vacuum chamber near the beam [109]. The inside of the antechamber contains non-evaporable Getter (NEG) strips to provide distributed pumping. Explosion bonded transition pieces are used on the ends of the chambers to allow the use of stainless steel flanges.

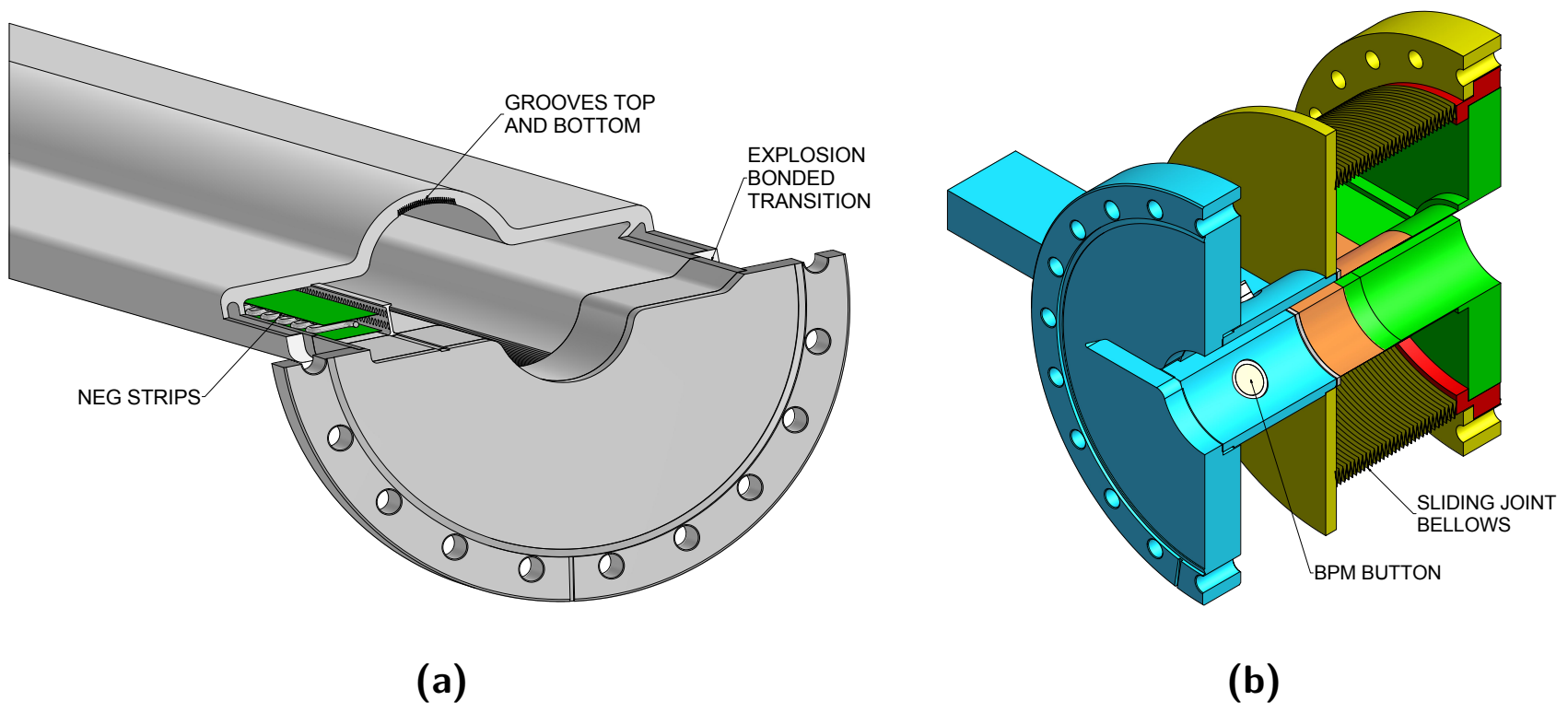

**Figure 6.9** (a) Dipole Chamber with grooved top and bottom surfaces, radially inside antechamber with NEG strips, and radially outside antechamber with sloped wall. (b) BPM and sliding joint assembly. [116].

The remainder of an arc cell consists of two short drift chambers on either side of the dipole, and one chamber extending through three quadrupoles, four sextupoles, and three corrector magnets. These chambers have the same profiles and TiN coating as the dipole chamber, but without the grooves on top and bottom. BPM assemblies as shown in Fig. 6.9b are located at each end of the chamber extending through the quadrupoles.

The wiggler-region vacuum chambers shown in Fig. 6.10a are made from copper to provide good thermal conductivity in this high-power region. The copper also minimizes the rate of scattered photons that escape the vacuum chamber to deposit energy into the cold mass and coils of the superconducting wigglers. The wigglers are grouped in pairs and a single vacuum chamber runs through two wigglers as well as the quadrupole magnet between them. The long vacuum chamber traversing each wiggler pair has a 46 mm beam aperture and 20 mm tall antechambers, including in the quadrupole. The choice of 20 mm-tall antechambers was based on photon-tracking simulation results in Synrad3D [109]. NEG Strips are recessed into the upper wall of the antechambers to act as distributed pumping and are shielded from beam-induced heating by means of a perforated aluminum strip. Most synchrotron radiation (SR) passes through the wiggler antechambers and is trapped in the photon stops located at the end of each cell. A tungsten clearing electrode is deposited via thermal spray on the bottom of the chamber as the primary EC mitigation technique [120, 121].

The other drift and quadrupole chambers in the wiggler section are copper chambers with TiN coating. They also have a 46 mm aperture and incorporate 20 mm tall antechambers to match the wiggler chambers and minimize impedance issues. The design is based on those previously specified for an earlier lattice variant [117, 118]. These chambers have gradually sloping, grooved antechambers are shown in Fig. 6.10b to dilute power density striking the photon stop. The gap between the sloping

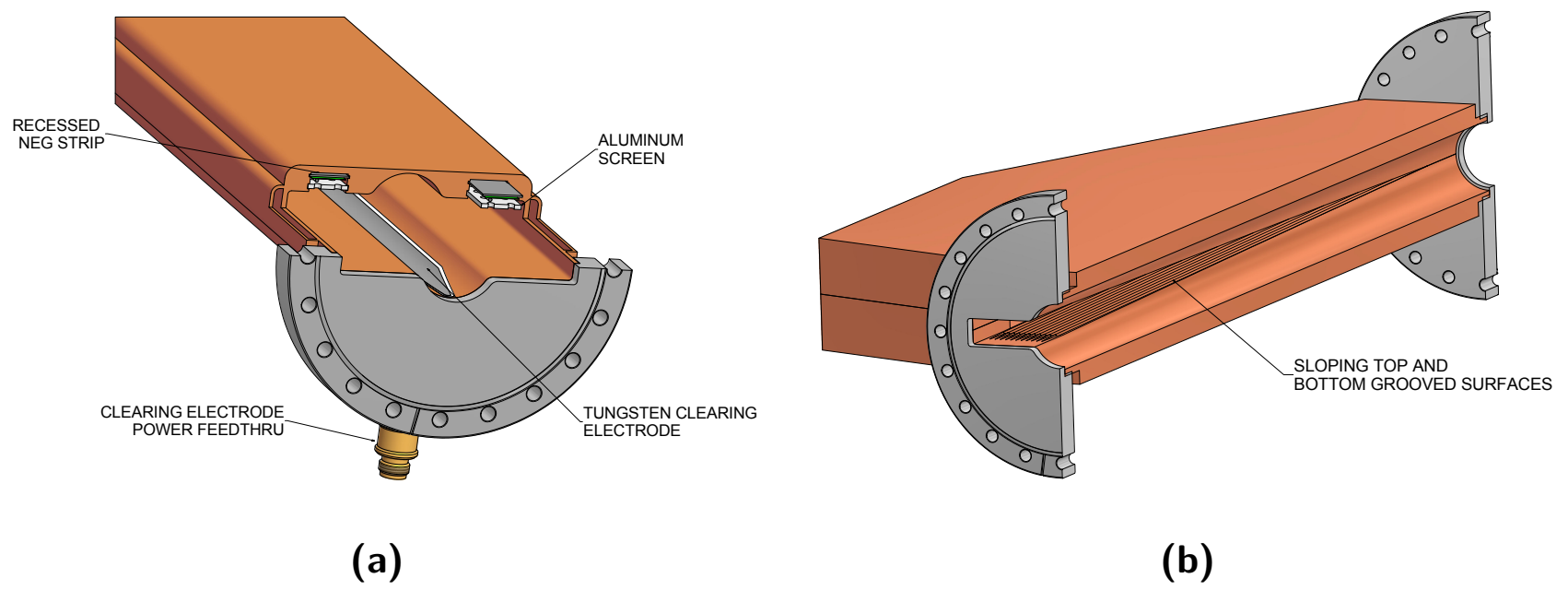

**Figure 6.10** (a) Wiggler vacuum chamber with clearing electrode and 20 mm tall antechambers with recessed NEG strips. (b) Wiggler section photon stop showing sloping and grooved photon-absorbing walls [116].

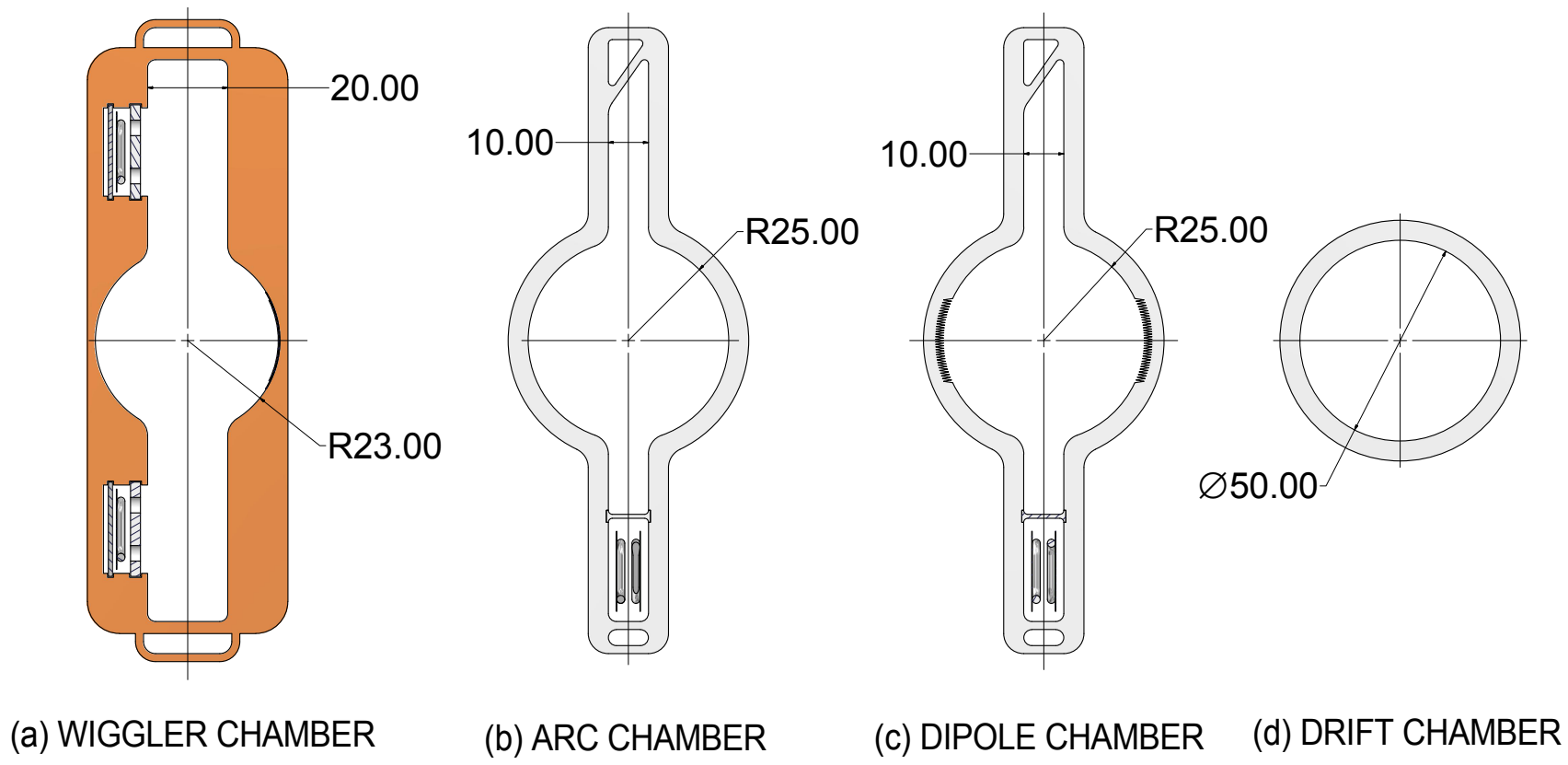

**Figure 6.11** Side-by-side comparison of ILC DR vacuum chamber profiles. Dimensions are in millimeters [116].

surfaces opens to antechambers pumped with an ion pump and Titanium sublimation pumps through ducts. An additional photon stop is required at the first bending magnet of the arc after the wiggler straight to intercept the forward SR component.

The remaining straight sections have simple round aluminum vacuum chambers with 50 mm aperture and TiN coating as shown Fig. 6.11. Aluminum-stainless-steel explosion-bonded transitions on the ends allow welding to stainless steel flanges. At the ends of the straight drift sections, tapered chambers match to the sections with antechambers.

Solenoid windings cover all accessible drift sections throughout the DRs to further reduce the number of secondary electrons approaching the beam axis.

Beam-position monitors (BPMs) are located near the majority of quadrupoles in the DR. There are no BPMs near the quadrupole trapped between the wiggler pairs due to lack of space, nor is there a BPM near the centre quadrupole in each arc cell. The BPM near the central quad in the arc cell is omitted because simulations indicate that, with the support and alignment scheme of the arc cell magnets, one BPM at the beginning and end of the magnet girder is sufficient [122]. The BPM blocks are paired with a sliding joint on one side as shown in Fig. 6.9b to allow them and the chamber they are connected to through the quadrupole to float. This maintains the absolute position of the BPM as steady as possible and allows movement of the BPM to be monitored. The sliding joint also allows for expansion and contraction of the surrounding vacuum chambers.

A sufficient number of ultra-high vacuum pumps, both localized (lumped) and distributed, are installed in the vacuum system to maintain the required average gas pressure ($\sim 10^{-9}$ torr) at the design beam current. The installed pumping system has enough pumping speed and capacity to allow vacuum system conditioning in a reasonably short duration during the initial accelerator commissioning, and after installation of new vacuum components for upgrades and/or repairs. Typical pumps are sputter-ion pumps (noble-diode style), non-evaporable Getters (NEGs) and titanium sublimation pumps (TiSPs). As illustrated for the dipole and wiggler chambers, NEG strips are inserted into the ante-chambers to provide distributed pumping. The final design must provide adequate pumping speed and capacity to handle the SR-induced gas load. Lumped ion pumps are installed periodically, with a typical spacing of 5 m. These ion pumps assist initial pump down and beam conditioning of the DR vacuum chambers, and handle any non-Getterable gases (such as CH4 and trace Ar) in the vacuum system.

The vacuum system is divided into sectors by RF-shielded gate valves to facilitate staged installation, upgrades, maintenance, and repairs. A typical length for each vacuum sector is 50 m. Cold-cathode ion gauges (CCGs) are installed periodically throughout the vacuum system to monitor performance and for trouble shooting. Each vacuum sector is equipped with at least one residual gas analyzer (RGA). Numerous thermocouples monitor local temperatures of vacuum components. Monitoring and interlock functions are integrated into the central control system.

## 6.6 RF systems

The damping-ring RF frequency of 650 MHz has a simple relationship with the main linac RF (1.3 GHz) to accommodate varying bunch patterns. While high-power 650 MHz RF sources are not commercially available, several klystron manufacturers can develop them by modifying 500 MHz klystrons of equivalent power level. Similarly, the RF cavity units can be designed by scaling from existing 500 MHz superconducting module designs currently in operation at CESR, KEK, [123–125] and elsewhere. New designs are required for the input coupler because the power handling capability must be kept at a level of about 300 kW CW, as well as for the HOM dampers and cryostats. The parameters presented in Table 6.6 are scaled from the 500 MHz units developed by industry and being operated in various laboratories. The RF-cryomodules are 3.5 m in length and 1.5 m in diameter.

**Table 6.6**
RF system parameters for the 3 different operating configurations [98].

| Parameter | Unit | Nominal 5 Hz | 10 Hz mode $e^+$ ring | Luminosity upgrade |
|---|---|---|---|---|
| Frequency | MHz | | 650 | |
| Total RF voltage | MV | 14 | 22 | 14 |
| Overvoltage factor | | 2.94 | 2.49 | 2.94 |
| Active cavity length | m | | 0.23 | |
| $R/Q$ | Ω | | 89 | |
| $Q_0$ at operating gradient | $10^9$ | 1 | 0.6 | 1 |
| Number of cavities/ring | | 10 | 12 | 12 |
| Cavity RF voltage | MV | 1.4 | 1.83 | 1.17 |
| Cavity average gradient | MV/m | 6.1 | 8.0 | 5.1 |
| Beam power per cavity | kW | 185.5 | 287 | 309 |
| Input coupler $Q_{\text{ext}}$ | | | $68 \times 10^3$ | |
| Cavity tuning | | stationary | fixed | stationary |
| RF reflected power | % | 8.0 | 11.4 | 2.6 |
| HOM Power | % | | 5 | |
| Total RF power | MW | 2.00 | 3.83 | 3.80 |
| Number of klystrons/ring | | 5 | 6 | 6 |
| Klystron peak power (10% overhead) | kW | 441 | 703 | 698 |
| Operating temperature | K | | 4.5 | |
| RF cryogenic losses per cavity | W | 15 | 50 | 15 |
| Number of SC modules per ring | | 10 | 12 | 12 |
| Static cryo losses at 4.5 K | W | | 30 | |
| Total cryo losses per ring | W | 450 | 960 | 540 |

### 6.6.1 Baseline

For the nominal baseline configuration, a beam current of 0.4 A is stored in 1312 bunches. The value of the RF voltage is chosen in order to achieve a 6 mm bunch length. The beam power and the total RF voltage for each ring is shared among 10 superconducting cavities in 5 RF stations. The 10 modules ensure adequate energy and beam-power margin in case of an RF-station fault, and permit continued operation at full performance with the 8 remaining units by increasing their RF output power. The cavities of the faulty station can still contribute to the bunch longitudinal focusing in this case, being passively excited by the 650 MHz spectral harmonic of beam current. The stations are located in the RF straight section, roughly 100 m long, upstream of the wigglers. The section has space for up to 16 cavities. Waveguides connect pairs of cavities to klystrons housed in a centrally located alcove which has access shafts to the surface. Each distribution system has magic-tees for power splitting and circulators for protecting the klystron against reflected power.

The possibility to add 2nd harmonic cavities in order to increase the flexibility of the system and reduce the cost for the nominal baseline configuration has been considered. This would require a new design to adapt the cells of the 1.3 GHz cavities as 2nd harmonic cavities. The beam is powered only through the fundamental cavities and the harmonic cavities are used to control the bunch length, allowing the same bunch length with less fundamental voltage or a shorter bunch with the same voltage.

### 6.6.2 10 Hz operation

For the positron damping ring in the 10 Hz operating mode the radiated energy is doubled to achieve the required shorter damping time. As shown in Table 6.6, this requires twice the beam power and two more RF cavities. In this configuration the damping ring has a pulsed time profile with beam injection/extraction times of 1 ms. Full beam current is stored for 100 ms and then extracted, the ring is then empty for the next 100 ms before the next injection cycle. This is a concern for the operation of the superconducting cavities whose tuning actuators have limited speed and excursion, so that it is quite difficult to follow, in real time, the rapidly changing beam-loading conditions. The simplest approach to overcome this difficulty requires keeping the cavities tuned at a certain fixed resonant frequency. The RF system can be optimized for this operating mode [126]. The main RF parameters for the positron ring are listed in Table 6.6. It is assumed that both rings will be operated identically in the 10 Hz mode, keeping the cavities tuned at a fixed resonant frequency. The only difference between the electron and positron ring is that the damping time required for the electrons is longer and the power required for the RF system is lower: 10 cavities assure adequate beam power.

### 6.6.3 Luminosity upgrade

For the luminosity upgrade there are 2625 bunches per main linac pulse, corresponding to a 0.8 A damping-ring current. The beam power required for the RF system is doubled with respect to nominal and 12 RF cavities are needed in order to keep the power per cavity at a level of about 300 kW. The parameters are listed in Table 6.6.

As with the baseline operation, in case of the failure of one klystron the system can be retuned to exploit the two unpowered cavities as passive, beam-excited devices, providing the same RF gradient across the bunch. The power to restore the beam losses will be provided by the ten active cavities. To guarantee a sufficient power margin to operate in the various configurations, including with a klystron fault, the maximum klystron power is 0.8 MW CW.

# 6.7 Magnets and Power Supplies

## 6.7.1 Superconducting Wigglers

The superferric wiggler design [100] provides the necessary operating flexibility over the range of peak fields required for the various DR operating modes, while maintaining the requisite field quality. Table 6.7 compares the parameters of the CESR-c wiggler, the ILC RDR wiggler and the recent redesign necessitated by the 10 Hz operating mode. The ten central poles of the magnet, each of 15 cm length, utilize coils of 660-turns carrying 93 kA. The poles at each end taper successively to 3/4- and 1/2-length as was used in the CESR-c design. The end poles have been simplified, omitting trim coils used to tune the second integral. Instead, the number of turns in the end pole coil has been adjusted to limit residual horizontal orbit displacement for 5 GeV electrons incident on axis to about 50 µm. There are 158 turns in the end-pole coils in this design. Each wiggler is powered by an individual AC-to-DC power supply.

**Table 6.7**
Superferric Wiggler Comparison

| Parameter | Unit | CESR-c | ILC RDR | ILC TDR |
|---|---|---|---|---|
| Peak Field | T | 2.10 | 1.67 | 2.16 |
| No. Poles | | 8 | 14 | 14 |
| Length | m | 1.3 | 2.5 | 1.875 |
| Period | m | 0.40 | 0.40 | 0.30 |
| Pole Width | cm | 23.8 | 23.8 | 23.8 |
| Pole Gap | cm | 7.6 | 7.6 | 7.6 |
| $\Delta B/B\|_{x=10\,\mathrm{mm}}$ | % | 0.0077 | 0.0077 | 0.06 |
| Coil Current | A | 141 | 112 | 141 |
| Beam Energy | GeV | 1.5–2.5 | 5 | 5 |

The superconducting damping wigglers are 4.5 K devices with static heat loads of 2 W/m or less, based on CESR-c experience [100]. To avoid a significant dynamic heat load, care must be taken to minimize the scattered synchrotron radiation that reaches the cold mass. The wigglers are co-located in the RF/Wiggler straight with the superconducting RF cavities in order to concentrate the cryogenic infrastructure.

## 6.7.2 Conventional Magnets

The damping ring has conventional electromagnets for the dipole, quadrupole, sextupole, and corrector magnets. This technology choice offers flexibility for tuning and optimizing the rings as well as for adjusting the operating beam energy by a few percent around the nominal value of 5 GeV. Magnet counts are shown in Table 6.8. Table 6.9 gives the key magnet parameters and maximum higher-order harmonic content specifications.

**Table 6.8**
Magnet types and counts for a single ILC Damping Ring using the baseline lattice. These counts do not include magnets, kickers, and septa associated with the damping ring abort beam dump located in the RTML (see Chapter 7). Wiggler magnets are superconducting, all others are normal conducting. In the engineering style designation, which is of the form *MxxLyyy*, *M* indicates the magnet type, *xx* indicates the bore diameter in millimetres, and *yyy* indicates the physical length in millimetres [127].

| Magnet | Type | Eng. Style | Qty | Power Method |
|---|---|---|---|---|
| Dipoles: | Corrector | D60L250 | 304 | Individual |
| | Chicane | D60L940 | 28 | String |
| | Disp. Supp. | D60L1940 | 10 | String |
| | Arc | D60L2940 | 150 | String |
| Quadrupoles: | Arc | Q60L480 | 482 | Individual |
| | Straight | Q60L700 | 121 | Individual |
| | Wig/Inj/Ext | Q85L309 | 50 | Individual |
| | Wiggler | Q85L600 | 30 | Individual |
| Skew Quads | Corrector | Q60L250 | 158 | Individual |
| Sextupoles | — | SX60L250 | 600 | Individual |
| Wigglers | — | WG76L2100 | 54 | Individual |
| Kickers | Inj/Ext | Striplines | 42 | Individual |
| Thin Pulsed Septa | Inj/Ext | — | 2 | Individual |
| Thick Pulsed Septa | Inj/Ext | — | 2 | Individual |

**Table 6.9**
Target field tolerances used for error simulations at a reference radius of 30 mm for damping-ring magnets. For the wigglers, the field quality is specified by the observed roll-off for a horizontal displacement from the beam axis by the indicated distance. The maximum KL-value specifies the nominal strength of the strongest magnet of each magnet type.

| Type | Unit | Max Field Max $KL$ | Error |
|---|---|---|---|
| Dipoles | mrad | 41 | $2 \times 10^{-4}$ |
| Quadrupoles | $m^{-1}$ | 0.35 | $2 \times 10^{-4}$ |
| Sextupoles | $m^{-2}$ | 1.23 | $2 \times 10^{-4}$ |
| H correctors | mrad | 2 | $5 \times 10^{-3}$ |
| V correctors | mrad | 2 | $5 \times 10^{-3}$ |
| Skew quads | $m^{-1}$ | 0.03 | $3 \times 10^{-3}$ |
| Wigglers | | – | $3 \times 10^{-3}$ |

### 6.7.3 Power Supply System

All quadrupoles, sextupoles, wigglers and corrector magnets (dipole, skew quadrupole, and possibly other multipoles) have individual power supplies. Individual control of the quadrupole and sextupole magnets significantly enhances the ability to tune and locally correct the machine optics in a ring with very aggressive operating parameters. Individual power supplies for the wigglers offer simplified control in the event of a magnet quench by eliminating the power system coupling between magnets. Alcoves used by the power system are located at the junctions between each straight and the arcs. Because of the long distances between individually powered magnets and the alcoves, the power supply system uses bulk supplies located in the main alcoves that power a master "bus" from which DC-to-DC converters supply power to individual magnets. This design minimizes cable heat loads in the ring and provides for an efficient distribution of power. For the arc dipole magnets, one-half of each arc is powered as a string from the nearest alcove. The pulsed power supplies for the stripline kickers require short cable runs to preserve the necessary timing synchronization, and are housed in small secondary alcoves near each group of kickers.

## 6.8 Instrumentation and Feedback Systems

### 6.8.1 Diagnostics and Instrumentation

The principal ring instrumentation required consists of systems whose performance is fairly standard for light sources or which has been demonstrated as part of the R&D program. This complement includes the following:

1. beam position monitors with turn by turn capability and precision as in Table 6.3;
2. "tune tracker" that tracks three normal modes of a single bunch and drives the bunch at those tunes via feedback kickers or equivalent;
3. visible and/or x-ray synchrotron light monitors for measuring transverse bunch dimensions and streak camera for bunch length
4. toroid current monitor and bunch-bunch current monitor;
5. Beam-loss monitors (based on ion chambers, photomultiplier and scintillator).

### 6.8.2 Fast Feedback systems

With over a thousand bunches circulating in the ring, wakefields induced in vacuum chamber components can give rise to coupled-bunch instabilities that cause bunch jitter and/or emittance growth. To combat this, the rings have fast bunch-by-bunch feedback systems in all three oscillation planes (longitudinal, horizontal and vertical) [128, 129]. Modern commercial FPGAs (Field Programmable Gate Arrays), with many digital signal processor units on a single chip, can easily manage the requirements of the feedback systems in terms of speed and number of bunches. The bandwidth of the fast feedback system must be at least $f_{\mathrm{RF}}$ (that is, 650 MHz in the DR). This means that every block of the system must have the capability to manage the full bandwidth except for the power section (amplifiers and kickers), where half bandwidth is sufficient. The main elements of each system are the analog front end, the digital processing unit, the analog back end, amplifiers and kicker(s).

The pickups can be 4-button monitors (two or three for each ring) with at least full bandwidth and dynamic range of the order ∼90 dB. The analog front ends are capable of extracting the oscillation signals from the monitors in each of the three planes (L, H, V) and giving them to the digital sections with a swing in the range of ∼0.5 V (typical of many analog-to-digital converters).

To minimize the quantization noise and have an adequate dynamic range, the digital units are based on a 14- or 16-bit signal processing system. The processing must compute the correction signal for all buckets (including the empty ones) to decouple the feedback behavior from the fill pattern. This means that all feedback systems have the capability to process, in real time, 7,022 input/output channels, although the real bunches are in, at most, 2,625 buckets. The digital unit sampling frequency is 650 MHz. A real-time FIR (finite impulse response) filter (with ≥50 taps, corresponding to an individual memory for each bunch of ≥50 acquisitions) provides the correction signal in terms of synchrotron or betatron phase advance using only one pickup for each system. The feedback setup is designed to be easily configurable using software tools. A down-sampling feature is also needed to manage very low-frequency oscillations.

The analog back-end systems adapt the output correction signals to the power section. The longitudinal kicker (an over-damped c cavity) works with a centre frequency between 800 and 2000 MHz, whereas the transverse kickers (striplines) operate at baseband (from 10 kHz up to half the bandwidth of the fast feedback system). Each power section has four 250 W amplifiers (1 kW total), with the bandwidth required by the kicker.

## 6.9 Injection and Extraction systems

The bunch separation in the main linacs is much longer than in the damping rings, so individual bunches must be injected and extracted without affecting the emittance or stability of the remaining stored bunches. For this to be the case, the kicker field must be negligible for any stored bunch upstream or downstream of the injected or extracted bunch, requiring that the effective kicker pulse width be less than twice the bunch spacing.

Individual bunch injection and extraction is accomplished in the horizontal plane using a fast kicker with rise/fall time ∼3 ns. The injection septum and injection kicker are separated by a horizontal phase advance of $\pi/2$ (as are the extraction septum and extraction kicker) and inserted in long drifts with low phase advance and high horizontal beta function, ∼70 m. If the DR is filled with 1312 bunches separated by 4 DR RF buckets in a train, the extraction kicker pulses with a repetition rate of up to 1.8 MHz (3 MHz is needed for the luminosity upgrade, 2625 bunches) to provide the specified uniform bunch spacing in the extraction line. The injection and extraction are performed simultaneously to reduce variations in beam loading, but the injection kicker starts to pulse a few turns after the beginning of the extraction. Thus, injected bunches fill the gaps vacated by extracted bunches in the same order as the bunches were extracted.

The kicker modules are 50 Ω stripline structures inside the vacuum pipe, each 30 cm long with a 30 mm gap. The required kick angle to extract the damped low emittance (∼0.5 nm rad) bunch is ∼0.6 mrad and nearly twice that for the large ($\sim 7 \times 10^{-6}$ mrad) injected bunch. Based on experience with ultrafast pulsers at the KEK-ATF, the design provides ±10 kV pulses on opposite electrodes. Thus a total complement of 42 kickers is required to handle injection and extraction for each ring. This corresponds to 84 high voltage pulsed power supplies.

The 30 cm stripline gives a 2 ns contribution to the kicker pulse width, leaving less than 10 ns for the electrical pulse width at the nominal ring bunch spacing of 6 ns. The kickers pulse about every 554 ns during the linac pulse of about 1 ms. For the high luminosity parameters, the ring bunch spacing is 3 ns, requiring an electrical pulse width of less than 4 ns and a pulse about every 366 ns. The tolerance on horizontal beam jitter of the extracted beam is 0.1—0.2$\sigma$, which requires the extraction kicker amplitude stability to be less than 5–10 $\times$ $10^{-4}$. A similar tolerance applies to the kicker amplitude for bunches before and after the target bunch.

The septum magnets are modeled after the Argonne APS injection septa. The thin (2 mm) septum magnet has a 0.73 T field, and the thick (30 mm) septum magnet has a 1.08 T field. Each magnet has an effective length of 1 m. Both magnets are pulsed once per linac cycle to reduce power dissipation, with eddy currents in the septum shielding the circulating beam. A half-sine pulse of about 10 ms width is used, and post-regulation is required to produce a 1 ms plateau flat to $10^{-4}$.

# Chapter 7
# Ring to Main Linac

## 7.1 Overview

The ILC Ring to Main Linac (RTML) is responsible for transporting and matching the beam from the Damping Ring to the entrance of the Main Linac. The RTML must perform several critical functions:

- transport of the electron and positron beams from the damping rings, at the center of the ILC accelerator complex, to the upstream ends of their respective linacs ("geometry matching");
- collimation of the beam halo generated in the damping ring to $\leq 10^{-5}$, based on SLC experience;
- rotation of the spin polarisation vector from the vertical to any arbitrary angle required at the IP;
- compression of the long Damping-Ring bunch length by a factor of 20–30 to provide the short bunches required by the Main Linac and the IP.

In addition, the RTML must provide sufficient instrumentation, diagnostics and feedback/ feedforward systems to preserve and tune the beam quality.

This chapter is organised as follows. The first two sections define the beam parameters and give a detailed description of the operation of the differing parts of the system. Important beam dynamics considerations relevant to the operation of the system are discussed in Section 7.4. The final section gives a manifest and definition of the various elements required to construct the system.

## 7.2 Beam Parameters

The key beam parameters of the RTML are listed in Table 7.1. Parameters are shown for the different operation modes, namely the low-energy 5/10 Hz configuration, nominal 5 Hz and luminosity upgrade.

**Table 7.1**
Basic beam parameters for the RTML.

| Parameter | Unit | Nominal $e^-/e^+$ | Low energy $e^-/e^+$ |
|---|---|---|---|
| Repetition rate | Hz | 5 | 5/10 |
| Initial energy | GeV | 5.0 | 5.0 |
| Initial energy spread | % | 0.11 | 0.12/0.137 |
| Initial norm. hor. emit. | µm | 5.5 | 5.9/6.0 |
| Initial norm. ver. emit. | nm | 19.6/20.0 | 20.1/20.9 |
| Initial hor. beam jit. | $\sigma_{x,y}$ | 0.1 | 0.1 |
| Initial bunch length | mm | 6.0 | 6.0 |
| Final bunch length | mm | 0.3 | 0.3 |
| Final energy | GeV | 15.0 | 15.0 |
| Final energy spread | % | 1.5 | 1.5 |
| Final hor. beam jitter | $\sigma_{x,y}$ | 0.1 | 0.1 |
| Norm. hor. emit. budget | µm | 0.9 | 0.9 |
| Norm. ver. emit. budget | nm | 6.5 | 6.5 |

## 7.3 System Description

### 7.3.1 Layout

Figure 7.1 depicts schematically the general layout of the ILC with emphasis on the various sub-beamlines and components of the RTML relative to Damping Rings, Main Linacs and Beam Delivery System (BDS). The RTML includes the short transfer line from the Damping Ring (DR) extraction to the main tunnel and the long low-emittance transport from the DR. This is followed by a $180^\circ$ turn-around, the spin-rotation and the two-stage bunch-compression sections. The beamlines upstream of the turn-around are collectively known as the "upstream RTML," while those from the turn-around to the start of the main linac are collectively known as the "downstream RTML". In order to accommodate the different damping-ring elevations and linac lengths, the electron and positron RTMLs have minor differences in their short transport sections, but are otherwise identical. The Twiss functions along the positron RTML are shown in Fig. 7.2. The electron RTML is almost identical and is not shown here.

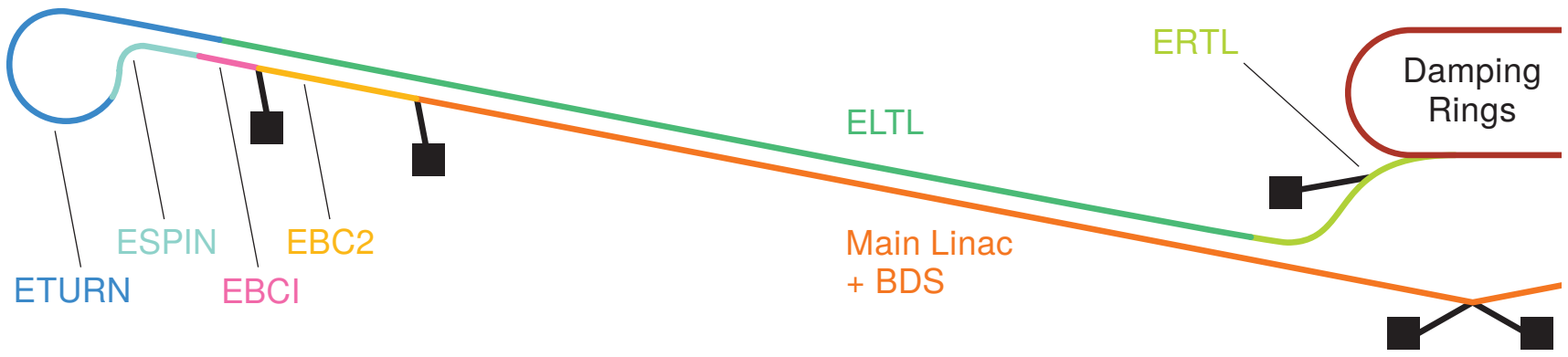

**Figure 7.1**
Schematic of the RTML, indicating the various functions described in the text.

Each of the key functions of the RTML listed in Section 7.1 is supported by several of the sub–beamlines shown in Fig. 7.1.

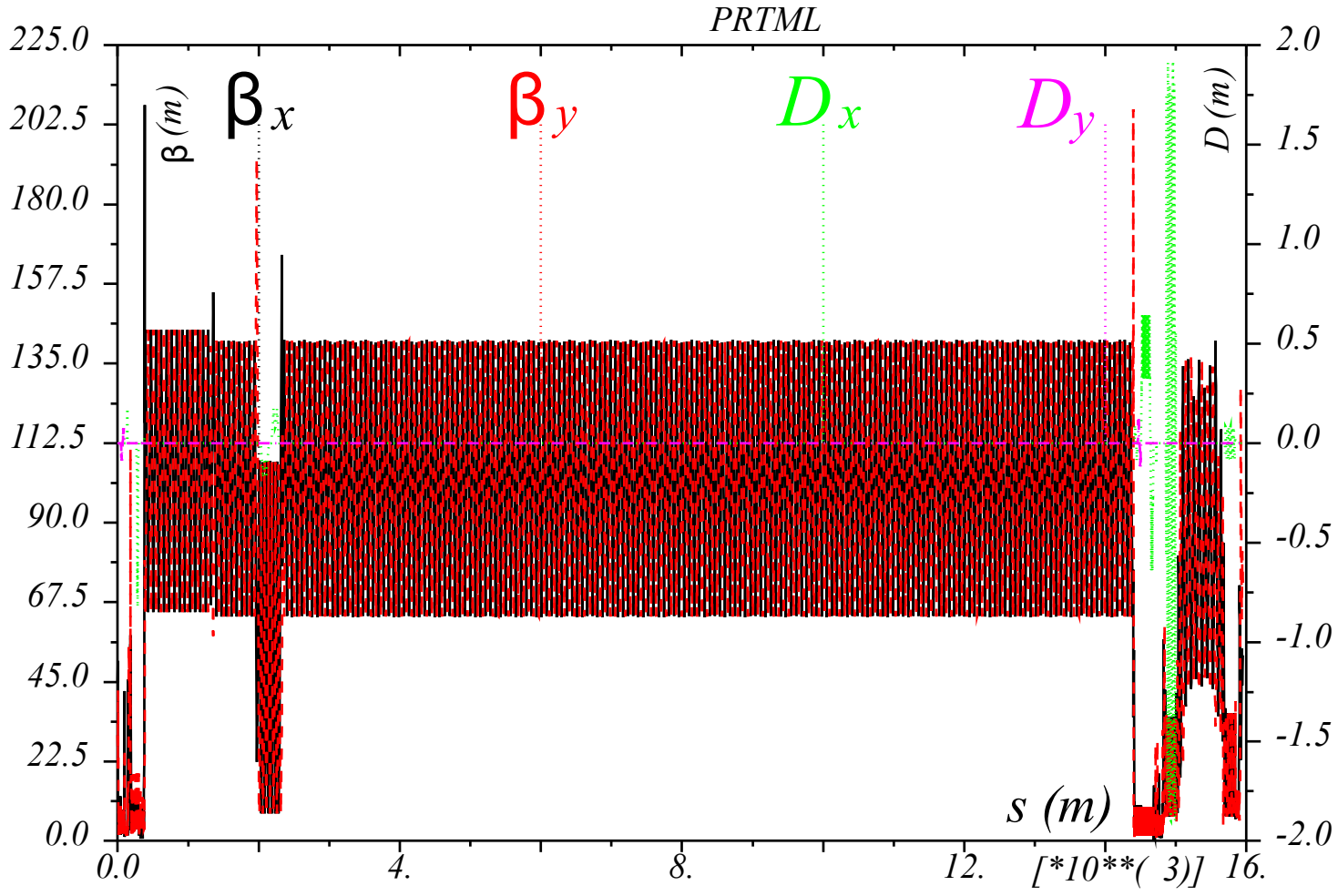

**Figure 7.2**
Twiss functions of the positron RTML.

The name abbreviations used for RTML sub-beamlines are the following:

- ERTL (PRTL) for the Electron (Positron) "Ring-To-Line" beamline from the damping ring to main tunnel, including Dump Lines (EC_DL);
- ELTL (PLTL) for Electron (Positron) "Long-Transfer-Line" or "Return Line";
- ETURN (PTURN) for Electron (Positron) TURN-around beamline;
- ESPIN (PSPIN) for Electron (Positron) SPIN-rotation system;
- EBC1 (PBC1) and EBC2 (PBC2) for the first and second stage of the Electron (Positron) Bunch Compressor, including their Dump Lines BC1_DL and BC2_DL.

### 7.3.2 Geometry Match

The RTML geometry and design is largely determined by the requirements of other areas, for example the length of the linac and its required curvature in the vertical plane, the positioning of the Damping Rings and their diameter, etc. The exact coordinates and angles of the connection points between Damping Rings and RTML and RTML and BDS/Main Linac are specified [130, 131]. Following extraction from the Damping Rings, the beams follow the lines ERTL and PRTL located in the Central Region tunnel and are injected into the long transfer line, parallel to the axis of the Main Linac. The PRTL contains a vertical dogleg which brings the positron beam from the height of the positron Damping Ring to the height of the PLTL in the main tunnel. The ELTL and PLTL (Return Lines) have an Earth-curved geometry along the Main Linac and a straight-line geometry elsewhere, except for areas near the connection to the Main Linac and BDS, where the beam geometry is adjusted using horizontal doglegs. In addition, small vertical and horizontal doglegs at the upstream end of the Turn-around change the beam elevation from the ceiling of the linac tunnel to the nominal linac elevation, and adjust the horizontal position between the ELTL (PLTL) line axis and the main linac axis.

### 7.3.3 Sub-systems

This subsection describes the functionality and specification of the subsystems of the RTML, starting from the damping ring and working out to the turn-arounds and then back towards the Interaction Point, as shown in Fig. 7.1.

#### 7.3.3.1 Extraction from the DR (ERTL/PRTL)

Figure 7.3a shows a plot of survey data for the PRTL line using the global Cartesian coordinate system $x, y, z$ with origin at the interaction point. In order to specify a complicated geometry, the ERTL and PRTL lines have been divided onto four logical sub-lines: horizontally straight section B containing a vertical dogleg, horizontal arc C, straight section D, and horizontal arc E. (Section A is the extraction line from the Damping ring.) Section B of PRTL consists of a matching section followed by regular FODO cells and a vertical dogleg to change the elevation of the positron beam. This plot shows circles for the given coordinates of the connection points between the sections along with solid-lines from the beamline survey. The nominal values of the coordinate displacements $x, y, z$ and angles the corresponding spherical coordinates $\Theta, \Phi, \Psi$ for the connection points are given [131]. The ERTL is identical to the PRTL except for an extra vertical dogleg.

For the Luminosity Upgrade configuration, there are 2 positron Damping Rings and two vertical doglegs instead of one. They merge the two positron beams coming from lower and upper Damping Rings into a single beamline (see Fig. 7.4).

The Twiss functions of the PRTL are shown in Fig. 7.3b where the boundaries of sections B, C, D and E are marked by blue lines.

#### 7.3.3.2 Return Line (ELTL/PLTL)

The ELTL and PLTL lines follow the earth's curvature in the Main Linac tunnel and have straight line geometry in other locations except near the positron production where there is a horizontal dogleg. The first section of each line contains a system of skew quadrupoles for coupling correction, a beam diagnostics section with 4 laser wires, a magnetic chicane for emittance measurements, and a collimation section to remove beam halo. Since the first part of the ELTL and PLTL share the BDS tunnel, a horizontal dogleg must be inserted at the junction between the BDS and Main Linac [132], corresponding to the dogleg after the undulator, to follow the geometry of the main beam line in this area. The positron source is located at the end of the electron linac where it joins the BDS; and the

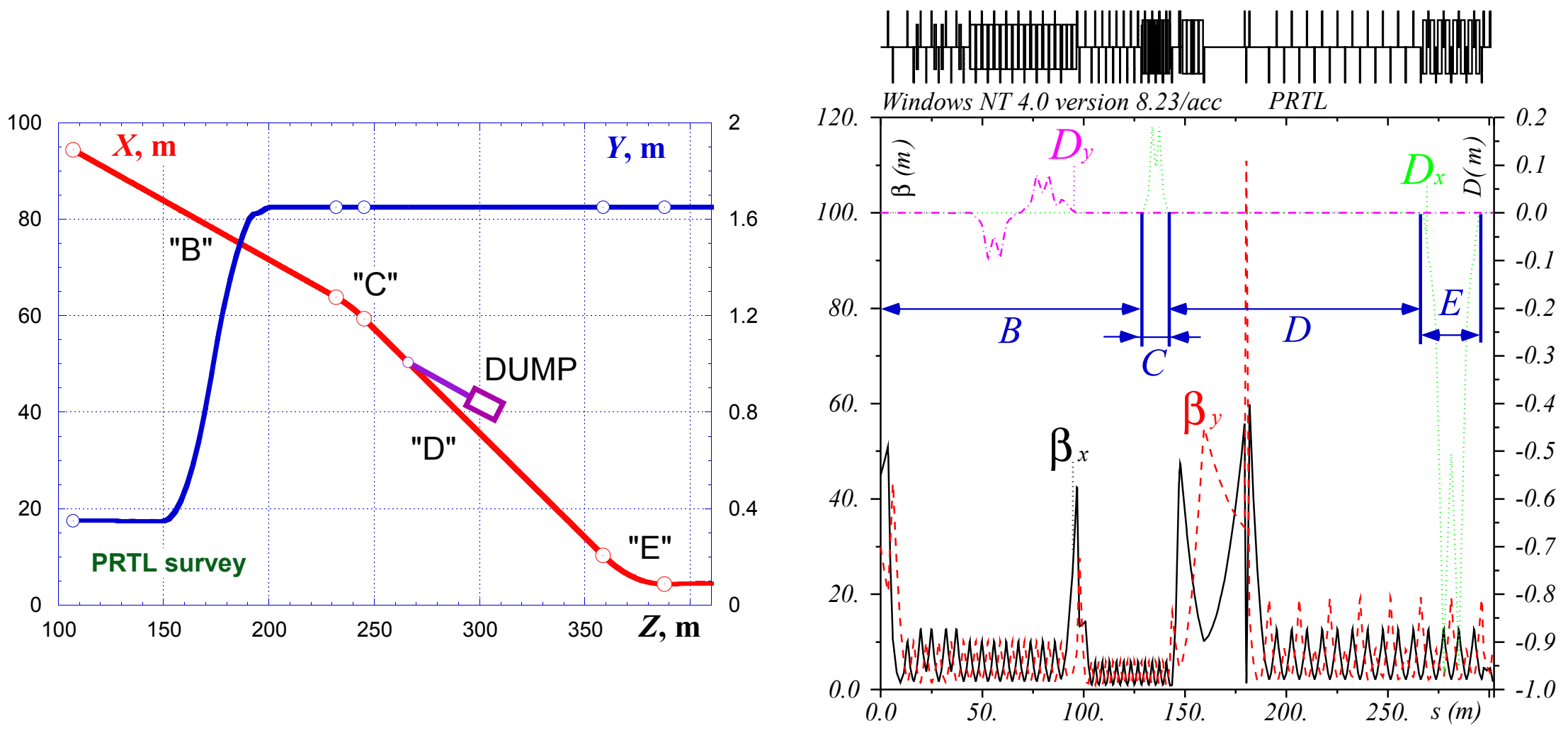

**(a)** Survey plot for PRTL

**(b)** PRTL lines

**Figure 7.3.** Survey plot (left) and Twiss functions (right) for PRTL line.

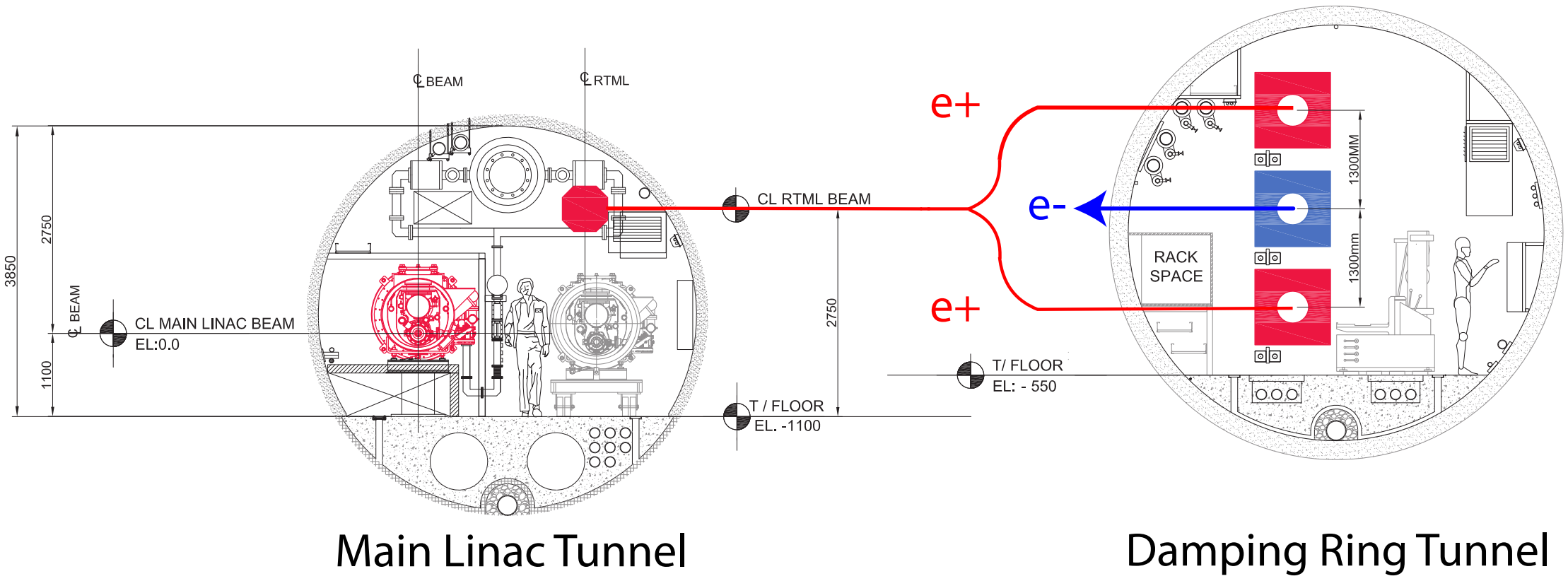

**Figure 7.4.** Layout of the vertical Doglegs in PRTL.

electron beam performs a dogleg around the positron target. The ELTL follows the same path in the opposite direction (see Fig. 7.5). Because of the high radiation at the positron target, no nearby beamlines has magnets in that section of beamline. The PLTL line also includes a dogleg at the junction between the Main Linac and the BDS, but without the complication of the positron target and radiation.

Because the Main Linac tunnel follows the curvature of the Earth, the ELTL and PLTL have vertical correctors at each quadrupole of the FODO system in the Main Linac tunnel. Each of these correctors gives the beam the vertical kick necessary for the beam to follow the curvature of the Earth. These correctors also generate a small dispersion that is propagated periodically along the FODO system cells. To match this vertical dispersion to the straight sections, there are an additional 4 vertical correctors before and after the curved sections.

#### 7.3.3.3 Turn-around

As well as changing the direction of the beams, the turn-around copes with a change in elevation and a change in horizontal offset to get from the return line location at the top of the tunnel to the main linac orientation. The Turn-around accomplishes this change in geometry as follows:

- The Turn-around is achieved by 29 cells with 108°/108° phase advance per cell; there are 18 "forward" arc cells, 3 "reverse" arc cells, and 8 cells for matching or suppressing dispersion.

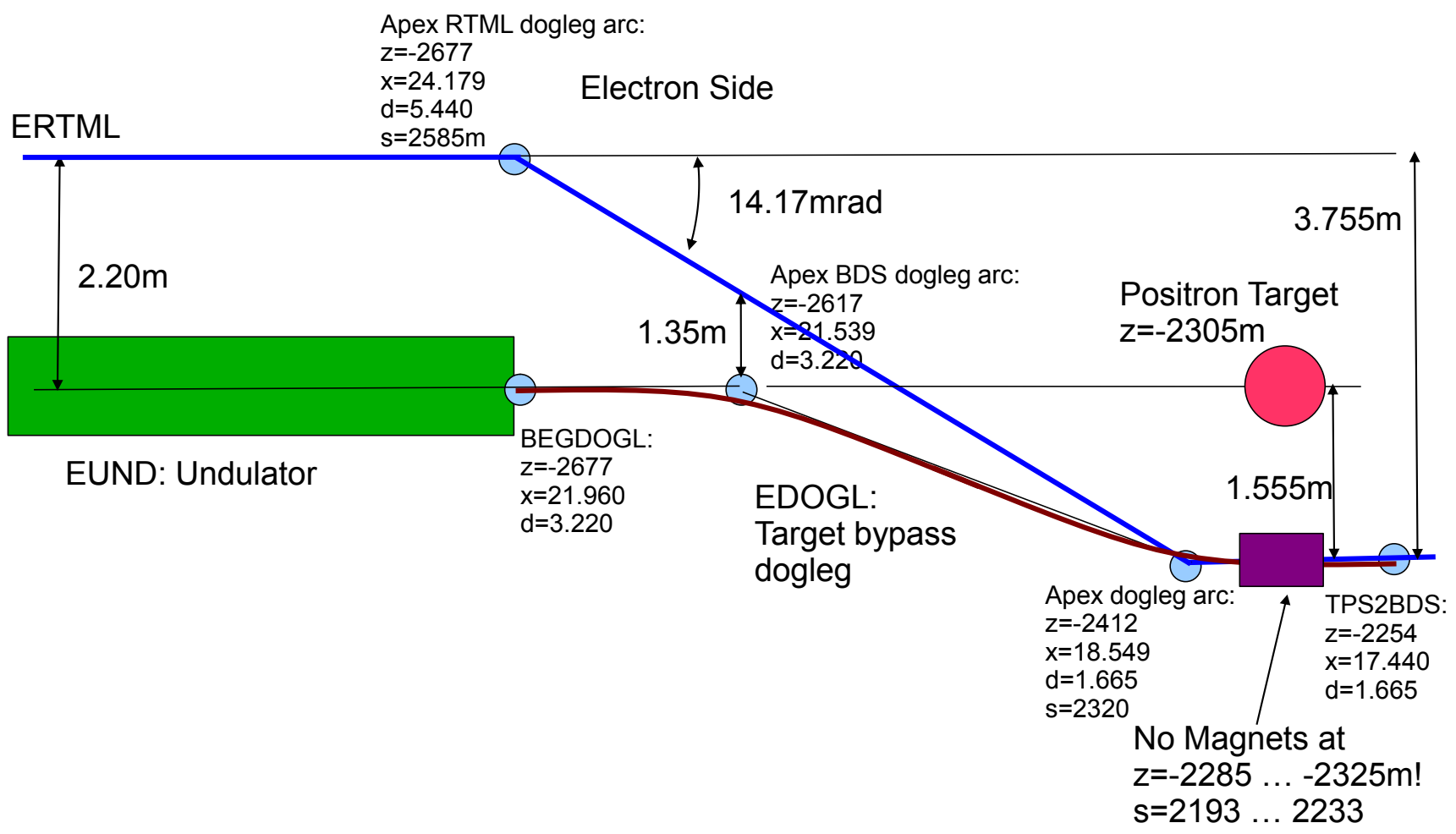

**Figure 7.5** Layout of the horizontal dogleg on the electron side of the RTML.

This is a total of 180° minus the 7.9° needed in the Spin Rotator, and also includes most of the needed horizontal offset;

- upstream of the turnaround is a vertical dogleg which produces the 1.65 m offset in 8 cells of 90°/90° phase advance;
- upstream of the vertical dogleg is a horizontal dogleg which produces a small horizontal offset in 8 cells of 90°/90° phase advance.

The radius of the arc is about 30 m. This is a compromise between the cost, which implies a short Turn-around, and the emittance growth, which implies relatively small pole-tip fields in the bending magnets. The dispersion is corrected entirely by adjusting the strengths of the quadrupoles in the vertical dogleg and in the main Turn-around arc. The Turn-around also includes a feedforward correction system, which corrects residual bunch-by-bunch orbit errors from the DR extraction. The BPMs for the feedforward correction are near the end of the Return Line, separated by 2 cells (for 90 degree coverage). The time delay of the beam through the turn-around is 600 ns, which is adequate for applying corrections. The $R_{56}$ of the turn-around is 2.37 m.

#### 7.3.3.4 Spin Rotation

The beam polarisation is vertical in the damping rings; this polarisation is transported with negligible loss or precession to the end of the Turn-around. Before entering the linac, the spin orientation should be set to any direction required by the experiments. This is accomplished for both the electrons and positrons by a spin-rotation system composed of a pair of 5 T superconducting solenoids, followed by an arc with a net 7.9° bend angle, followed by another pair of solenoids. The spin direction is selected by adjusting the excitation in the solenoid pairs. In order to rotate the spin without introducing undesirable $x - y$ coupling, the solenoid-based rotators each use a pair of identical superconducting solenoids separated by a quadrupole lattice which introduces a $+I$ transformation in the horizontal plane and a $-I$ transformation in the vertical plane [133], the net effect of which is to cancel the cross-plane coupling.

#### 7.3.3.5 Bunch Compression

There is a two-stage bunch-compression system in order to achieve the required factor of 20 [134]. Compared with a single-stage compression [135], the 2-stage compressor reduces the energy spread of the beam throughout the RTML and allows more flexibility to reduce the bunch length below 0.3 mm for the Energy Upgrade configuration. Table 7.2 summarises the important parameters for both the first stage (BC1) and second stage (BC2) of the Bunch Compressor.

**Table 7.2**
Key parameters for the two-stage bunch compressor in the nominal and low-energy configurations, assuming compression to 0.3 mm RMS final bunch length.

| Parameter | Unit | BC1 Nominal $e^-/e^+$ | BC1 Low Energy $e^-/e^+$ |
|---|---|---|---|
| Repetition rate | Hz | 5 | 10/5 |
| Initial energy | GeV | 5.0 | 5.0/5.0 |
| Initial energy spread | % | 0.11 | 0.12/0.137 |
| Initial bunch length | mm | 6.0 | 6.0 |
| RF voltage | MV | 465 | 465 |
| RF phase | ° | −115 | −115 |
| Wiggler $R_{56}$ | mm | −372 | −372 |
| Final energy | GeV | 4.8 | 4.8 |
| Final energy spread | % | 1.42 | 1.42 |
| Final bunch length | mm | 0.9 | 0.9 |
| **Parameter** | **Unit** | **BC2 $e^-/e^+$** | **BC2 $e^-/e^+$** |
| RF voltage | GV | 11 | 11 |
| RF phase | ° | −24 | −25.3 |
| Wiggler $R_{56}$ | mm | −55 | −55 |
| Final energy | GeV | 14.9 | 14.8/14.6 |
| Final energy spread | % | 1.12 | 1.17/1.24 |
| Final bunch length | mm | 0.3 | 0.3 |

The two-stage bunch compressor also allows some flexibility to balance longitudinal and transverse tolerances by adjustment of the wiggler magnet strengths, RF voltages, and RF phases. The nominal compressor configurations ease tolerances on damping-ring extraction phase, damping-ring bunch length, and bunch-compressor phase stability, at the expense of tightening the tolerances on the transverse alignment of accelerator components. Alternate configurations are possible that loosen transverse alignment tolerances but tighten the longitudinal (i.e. phase) tolerances.

The linacs in both compressor stages use standard SCRF cryomodules and an RF power-unit configuration similar to that of the Main Linac (i.e. one klystron driving three cryomodules). The first-stage compressor has a single RF unit, with 8 cavities and one quadrupole in each of its 3 cryomodules; the second-stage compressor uses 16 RF units which are identical to the main-linac configuration (i.e. 26 cavities and 1 quad per 3 cryomodules). The stronger focusing in the first stage is necessary to mitigate the higher wakefields and cavity-tilt effects resulting from the longer bunch length in the compressors.

Each bunch compressor stage includes a 150 m lattice of wiggler magnets which provides the required momentum compaction.

Figure 7.6 shows the longitudinal phase space after compression from 6 mm to 0.3 mm RMS length.

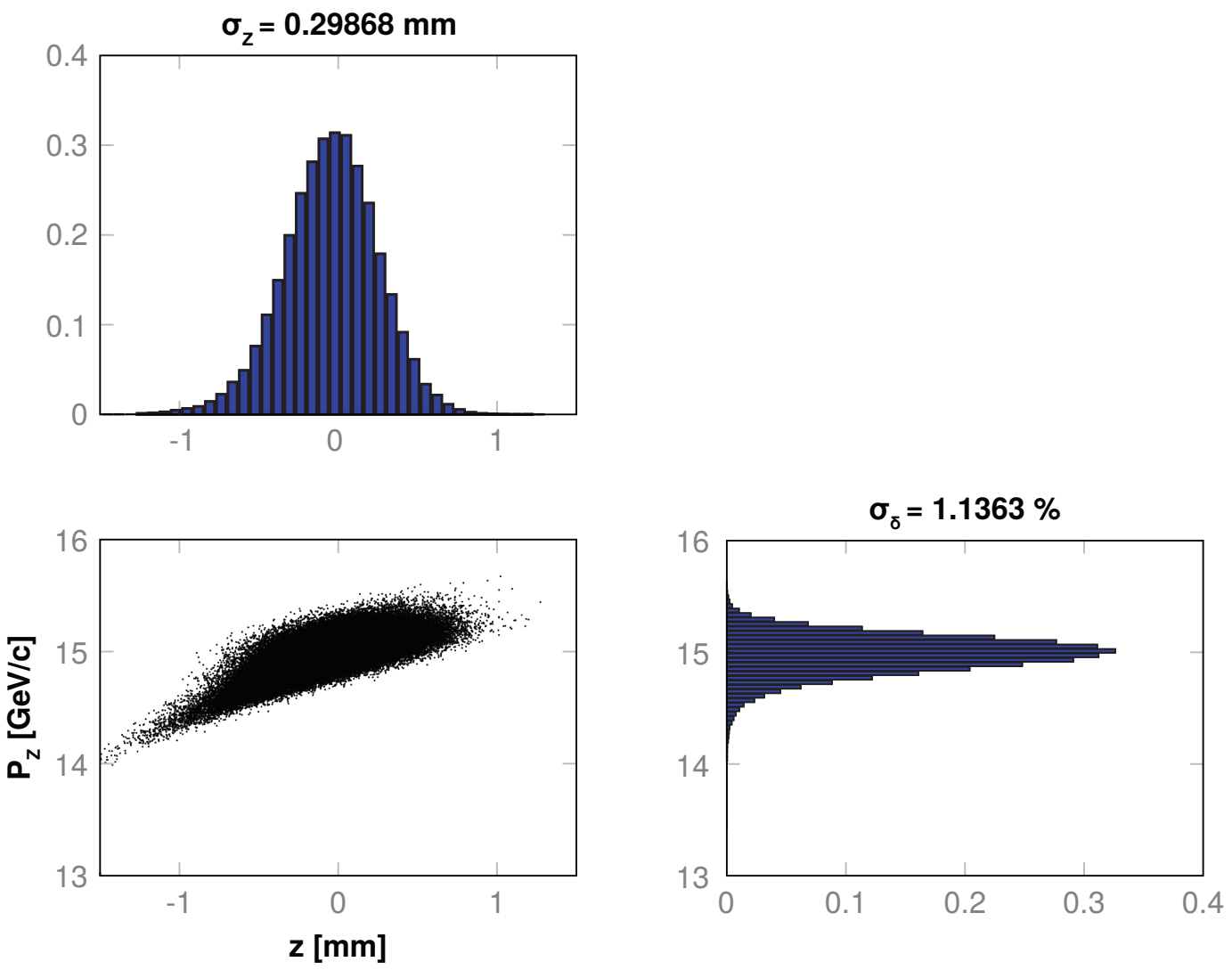

**Figure 7.6**
Longitudinal phase space of the compressed bunch.

## 7.3.4 Collimation and diagnostics

The betatron collimation section is downstream of the Central Region, at the beginning of the Return Line. It consists of two sets of thin spoiler and thick absorber pairs, placed 90° apart in betatron phase. This is sufficient to reduce the halo density by 3-4 orders of magnitude. The thin spoilers are needed to protect the absorbers from a direct hit from an errant beam [136]. There are additional collimators for energy collimation in the Turn-around and in the wigglers of the Bunch Compressor. There are both adjustable-jaw collimators with a rectangular aperture, and fixed-aperture collimators with a cylindrical geometry.

### 7.3.4.1 Adjustable-Aperture Collimators

All of the adjustable-aperture, rectangular collimators (RCOLs) are of the same design: two jaws in the same plane ($x$ or $y$), with a 0.6 RL titanium "spoiler" and no water cooling. The amount of collimated halo is about 0.1 % of the beam power of 220 kW, corresponding to 220 W; since the energy deposited in the titanium is a small fraction of the total halo power, water cooling is not required. The nominal betatron collimation depth is $10\sigma_{\mathrm{x}}$ and $60\sigma_{\mathrm{y}}$. Each RTML has two $x$ and two $y$ betatron spoilers, for a total of eight adjustable spoilers for betatron collimation. Each RTML also has 6 sets of adjustable spoilers for longitudinal collimation: two in the horizontal dogleg portion of the turn-around; two in the BC1 wiggler; and two in the BC2 wiggler. The number of two-jaw, single-plane adjustable collimators in each RTML is 10, giving 20 in total.

### 7.3.4.2 Fixed-Aperture Collimators

There are 28 fixed-aperture collimators with circular geometry (ECOLs): 8 circular collimators in each collimation section before the Return Line, 2 in each turn-around (in the horizontal dogleg), 2 in each BC1, and 2 in each BC2. The collimators in the Return Lines and turn-arounds have 6.5 mm half-gaps, the ones in BC1 have 30 mm half-gaps, and the ones in BC2 have 5 mm half-gaps. The circular collimators absorb the beam power which is scattered from the adjustable-aperture spoilers. To accomplish this, they are 20 RL lengths thick. These collimators are water cooled, and can handle a CW power of about 200 W. The circular collimators are "shadowed" by the spoilers, so that for a particle or bunch to hit a circular collimator it is generally necessary that the particle or bunch pass through a spoiler first. Table 7.3 gives the type, number and location of the collimators in the RTML.

**Table 7.3**
Type, number and location of collimators in RTML.

| Type | Aperture X×Y mm$^2$ | Budget W | Cooling | Location | Number |
|---|---|---|---|---|---|
| Rectangular | 3.43×10 | ≪ 220 | — | ELTL/PLTL | 4 |
| Rectangular | 10×1 | ≪ 220 | — | ELTL/PLTL | 4 |
| Circular | 6.5 | 200 (CW) | water | ELTL/PLTL | 16 |
| Rectangular | 1×10 | ≪ 220 | — | ETURN/PTURN | 4 |
| Circular | 6.5 | 200 (CW) | water | ETURN/PTURN | 4 |
| Rectangular | 18×20 | ≪ 220 | — | EBC1/PBC1 | 4 |
| Circular | 30 | 200 (CW) | water | EBC1/PBC1 | 4 |
| Rectangular | 4×10 | ≪ 220 | — | EBC2/PBC2 | 4 |
| Circular | 5 | 200 (CW) | water | EBC2/PBC2 | 4 |

### 7.3.4.3 Diagnostics

At the entrance to each Return Line before the collimation section, there is a skew quadrupole system for coupling correction, a beam-diagnostics section with multiple laser-wire stations and a magnetic chicane for emittance measurements. The lattices for these sections are taken from the RDR design rematched into the new line [137].

Figure 7.7 shows the layout of the electron RTML with the location of collimators and diagnostic sections. The positron side is identical.

**Figure 7.7**
Location of collimator and diagnostic sections in the RTML.

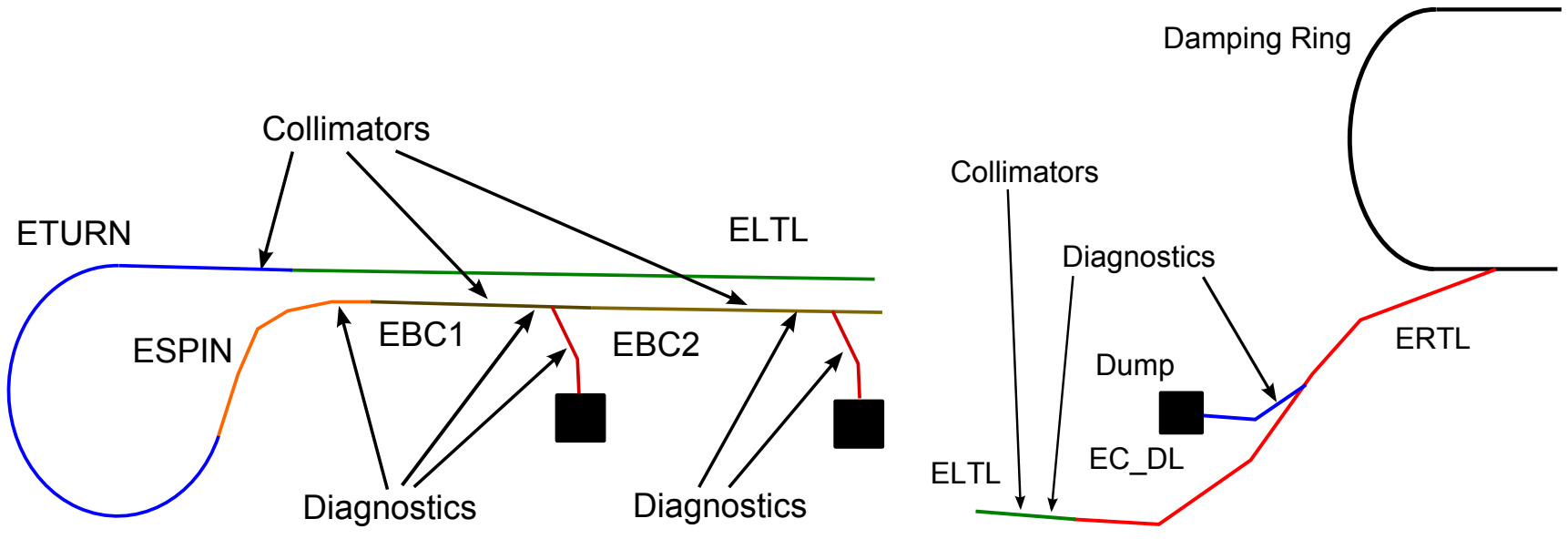

**(a)** ELTL, ETURN, ESPIN, EBC1 and EBC2 lines. **(b)** ERTL and ELTL lines.

## 7.3.5 Tuning, Correction, and Operations

The diagnostic, correction, and operational requirements of the RTML have been carefully integrated into the design of the entire beamline and are described in detail below.

### 7.3.5.1 Global Dispersion Correction

The Arc, the BC1 wiggler, and the BC2 wiggler contain normal and skew quads in regions of horizontal dispersion which are used to tune any residual dispersion due to misalignments and errors. The quads are arranged in pairs, with an optical $-I$ transform between the two quads in a pair; this permits tuning of the dispersion without introducing any betatron coupling or beta beats. The dispersions in the turn-around is adjusted by tuning normal quads in the horizontal and vertical doglegs at the upstream end of the turn-around.

### 7.3.5.2 Global Coupling Correction

There are two decoupling regions: the first is immediately downstream of the Arc, and the second is immediately downstream of the Spin Rotator. Each decoupling region contains 4 orthonormal skew quads in regions of zero dispersion, which allow complete and independent control of the 4 betatron coupling terms. The first station is conceptually intended to correct the coupling introduced by the damping-ring extraction system, while the second corrects coupling generated by errors in the spin-rotation system, as well as the remaining betatron coupling from small rotation errors on the RTML quads.

#### 7.3.5.3 Emittance Measurements

There are three emittance measurement stations: the first is between the first decoupling section and the first collimation section, the second is between the second decoupling station and the bunch compressor, and the third is between the bunch compressor and the linac. Each of these stations contains 4 laser-wire scanners embedded in a FODO lattice with 45° betatron phase; each station can therefore measure the projected $x$ and $y$ emittances of the beam. The first station can be used to tune the Arc dispersion and the skew quads in the first decoupler; the second station can be used to tune the Turn-around dispersion and the skew quads in the second decoupler; the third station can be used to tune the dispersion correction in the Bunch Compressor wigglers. Although none of the systems have the capability to measure normal-mode emittances and coupling parameters directly, the optics of the first two stations are compatible with a later upgrade if needed.

#### 7.3.5.4 Beam-Position Monitors

There are cavity-type beam-position monitors (BPMs) with horizontal and vertical readout at each quadrupole, with additional units close to the laser wires, at high-dispersion regions in the Bunch-Compressor wigglers, and at other critical locations. The BPMs in the room-temperature sections of the RTML almost all operate in the 6 GHz frequency band ("C-band"), while the BPMs in the cryomodules and at a handful of other locations use the 1 GHz frequency band ("L-band"). At the nominal bunch charge of 3.2 nC, these BPMs can achieve sub-micron single-bunch resolution. The standard RTML BPM requires high precision and stability of the BPM's offset with respect to the device's mechanical centre; a few of the BPMs have other requirements, such as high bandwidth or low latency.

#### 7.3.5.5 Longitudinal Diagnostics

Each stage of the Bunch Compressor contains arrival-time (phase) monitors, beam position monitors at dispersive locations, X-ray synchrotron-light monitors, and two types of bunch-length monitors (a passive device based on measuring the RF spectrum of the bunch, and an active device based on transverse deflecting cavities [138]). The active bunch-length monitor can also measure the correlation between energy and longitudinal position within a bunch.

#### 7.3.5.6 Feedback and Feedforward

The RTML is not expected to require any intra-train trajectory feedback systems, although there are a number of train-to-train (5 Hz) trajectory feedbacks. In addition, the beam energy at BC1 and BC2 is controlled by a 5 Hz feedback, as is the electron-positron path-length difference through their respective bunch compressors (see Section 7.4). There is also a trajectory feed-forward that uses BPMs at the end of the Return line to make bunch-by-bunch orbit measurements, which are fed forward to a set of fast correctors downstream of the Turn-around. The speed-of-light travel time between these two points is about 600 ns, and the actual distance between them is on the order of a few tens of meters; the resulting delay of the beam relative to the propagated signal is more than adequate for a digital low-latency orbit correction system [139].

#### 7.3.5.7 Machine Protection

*Intermediate Extraction Points:* There are 3 locations where the beam in the RTML may be directed to a beam dump: upstream of the first collimation section, downstream of BC1, and downstream of BC2 [140]. At each of these locations, there are both pulsed kickers and pulsed bends for beam extraction. The kickers are used when an intra-train extraction is required, for example during a machine protection fault, while the bends are used to send entire trains to their beam dumps. The pulsed bends can also be energized by DC power supplies if a long period of continual dump running is foreseen. All 3 dumps are capable of absorbing 220 kW of beam power. This implies that the first 2 dumps, which are at 5 GeV, can absorb the full beam power, while the third dump, at 15 GeV, can absorb only about 1/3 of the nominal beam power. Full trains can be run to this dump at reduced repetition rate, or short trains at full rate.

## 7.4 Accelerator Physics Issues

A number of beam dynamics issues were considered in the design and specifications of the RTML.

### 7.4.1 Incoherent (ISR) and Coherent (CSR) Synchrotron Radiation

Current estimates indicate that the horizontal emittance growth from ISR is around 90 nm (1.1 %) in the Arc, 380 nm (4.8 %) in the Turn-around, and 430 nm (5.4 %) in the Bunch Compressor in its nominal configuration. Vertical emittance growth from ISR in the vertical dogleg is negligible.

Studies of the ILC Bunch Compressor indicate that there are no important effects of coherent synchrotron radiation, primarily because the longitudinal emittance of the beam extracted from the damping ring is so large [141].

### 7.4.2 Stray Fields

Studies have found that fields at the level of 2.0 nT can lead to beam jitter at the level of $0.2\sigma_y$ [142]. This is considered acceptable since the orbit feed-forward corrects most of this beam motion. Measurements [143] indicate that 2 nT is a reasonable estimate for the stray-field magnitude in the ILC. Emittance-growth considerations also place limits on the acceptable stray fields, but these are significantly higher.

### 7.4.3 Beam-Ion Instabilities

Because of its length and its weak focusing, the electron Return line has potential issues with ion instabilities. To limit these to acceptable levels, the base pressure in the Return line must be smaller than 2 μPa [144].

### 7.4.4 Static Misalignments

The main issues for emittance growth are: betatron coupling introduced by the Spin Rotator or by rotated quads; dispersion introduced by rotated bends, rotated quads in dispersive regions, or misaligned components; wakefields from misaligned RF cavities; and time-varying transverse kicks from pitched RF cavities.

Studies of emittance growth and control in the region from the start of the Turn-around to the end of the second emittance region have shown that a combination of beam steering, global dispersion correction, and global decoupling can reduce emittance growth from magnetostatic sources to negligible levels, subject to the resolution limits of the measurements performed by the laser wires [145, 146]. Although the upstream RTML is much longer than the downstream RTML, its focusing is relatively weak and as a result its alignment tolerances are actually looser. Studies have shown that the same tuning techniques can be used in the upstream RTML with the desired effectiveness [147]. The tolerances for RF cavity misalignment in the RTML are large (0.5 mm RMS would be acceptable) because the number of cavities is small and the wakefields are relatively weak [148]. Although in principle the RF pitch effect is difficult to manage, in practice it leads to a position-energy correlation which can be addressed by the Bunch Compressor global dispersion correction [149]. A full and complete set of tuning simulations have not yet been performed, but it is expected that the baseline design for the RTML can satisfy the emittance preservation requirements. Table 7.4 summarises the alignment errors estimated for the RTML components.

**Table 7.4**
Standard local alignment error in RTML.

| Error | Unit | Cold Sections | Warm Sections |
|---|---|---|---|
| Quad. offset | µm | 300 | 150 |
| Quad. roll | µrad | 300 | 300 |
| RF cavity offset | µm | 300 | — |
| RF cavity tilt | µrad | 300 | — |
| BPM offset | µm | 300 | 100 (w.r.t. magnet) |
| CM offset | µm | 200 | — |
| CM pitch | µrad | 20 | — |
| Bend offset | µm | — | 300 |
| Bend. roll | µrad | — | 300 |

## 7.4.5 RF Phase and Amplitude Jitter

Phase and amplitude errors in the bunch compressor RF lead to energy and timing jitter at the IP, the latter directly resulting in a loss of luminosity. Table 7.5 shows the RMS tolerances required to limit the integrated luminosity loss to 2 %, and to limit growth in IP energy spread to 10 % of the nominal energy spread [150].

**Table 7.5**
Key tolerances for the two-stage bunch compressor.

| Parameter | Arrival Time | Energy Spread |
|---|---|---|
| Correlated BC phase errors | 0.24° | 0.35° |
| Uncorrelated BC phase errors | 0.48° | 0.59° |
| Correlated BC amplitude errors | 0.5 % | 1.8 % |
| Uncorrelated BC amplitude errors | 1.6 % | 2.8 % |

The tightest tolerance which influences the arrival time is the relative phase of the RF systems on the two sides: in the nominal configuration, a phase jitter of the electron and positron RF systems of 0.24° RMS, relative to a common master oscillator, results in 2 % luminosity loss. The tight tolerances are met through a three-level system:

- Over short time scales, such as 1 second, the low-level RF system is required to keep the two RF systems phase-locked to the level of 0.24° of 1.3 GHz;
- Over longer time periods, the arrival times of the two beams are directly measured at the IP and a feedback loop adjusts the low-level RF system to synchronize the beams. This system is required to compensate for drifts in the low-level RF phase-locking system which occur over time scales long compared to a second;
- Over a period of many minutes to a few hours, the arrival time of one beam is "dithered" with respect to the arrival time of the other beam, and the relative offset which maximizes the luminosity is determined. This offset is used as a new set-point for the IP arrival-time feedback loop, and serves to eliminate drifts which arise over time scales long compared to a minute.

### 7.4.6 Halo Formation from Scattering

Halo formation is dominated by Coulomb scattering from the nuclei of residual gas atoms, and it is estimated that $10^{-5}$ Pa base pressure in the downstream RTML will cause approximately $9 \times 10^{-7}$ of the beam population to enter the halo [151]. A similar calculation was performed for the upstream RTML, which indicates that $2 \times 10^{-6}$ Pa base pressure causes approximately $2 \times 10^{-6}$ of the beam population to enter the halo. This is well below the requirement of $10^{-5}$.

### 7.4.7 Space-Charge effects

In the long, low-energy, low-emittance transfer line from the damping ring to the bunch compressor, the incoherent space-charge tune shift is on the order of 0.15 in the vertical, the impact of which will be the subject of further studies.

### 7.4.8 Wake field in SRF cavities and collimators

Assuming collimation of the beam extracted from the damping ring at $10\sigma_{\mathrm{x}}$, $60\sigma_{\mathrm{y}}$, and $\pm 1.5\,\%$ ($10\sigma_\delta$) in momentum, the worst-case jitter amplification for untapered, "razor-blade" spoilers is expected to be around 10 % in $x$, around 75 % in $y$, and the contribution to $x$ jitter from energy jitter is expected to be negligible [152, 153]. The vertical jitter amplification figure can be brought to an acceptable level by the use of spoilers with modest longitudinal tapers. The other collimator wakefield "figures of merit" are acceptable even assuming untapered spoilers.

### 7.4.9 Emittance preservation

Preservation of the vertical emittance in the RTML is a challenging task, which cannot be achieved without dedicated beam-based alignment algorithms. Simulations have studied one-to-one correction, kick minimisation, dispersion bumps and coupling-correction algorithms to achieve small emittance growth in the entire RTML, excluding the Bunch compressor. The results are shown in Fig. 7.8. After corrections, the growth of normalised emittance is 5.3 nm rms (9.94 nm for 90 % of seeds). In the two-stage bunch compressor the biggest effect comes from cavity misalignments, tilts and asymmetries from the geometry of the main power and HOM couplers. In addition to the other BBA algorithms, the simulations applied a "girder optimisaton" (or tilting of cryomodules) algorithm to minimize emittance growth to 1.09 nm rms (1.48 for 90 % of seeds).

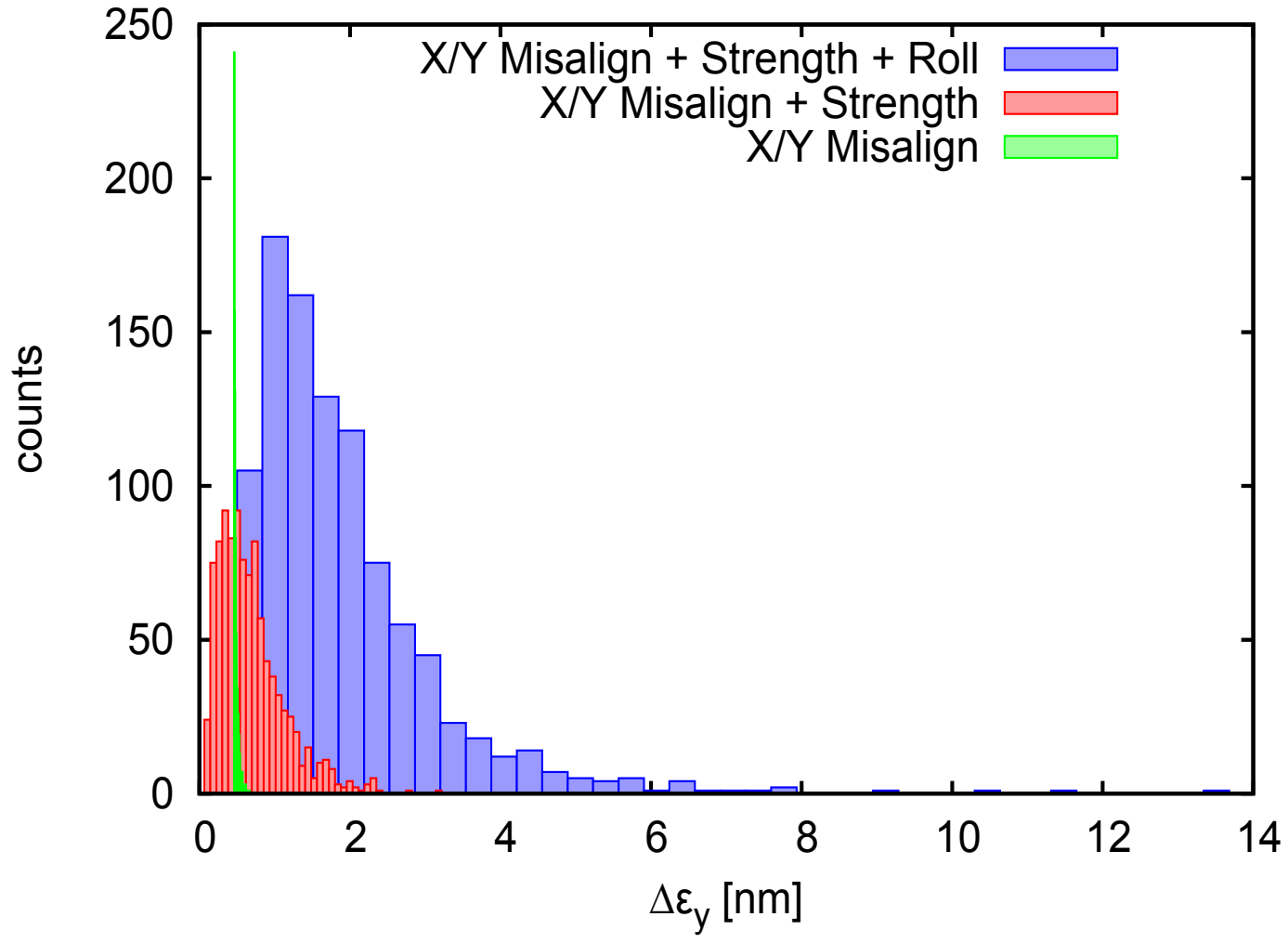

**Figure 7.8**
Histogram of the final emittance growth for 1000 seeds in RTML, excluding BC.

## 7.5 Accelerator Components

### 7.5.1 Magnets, Pulsed elements

Table 7.6 shows the total number of components of each type in each RTML. The number of quadrupoles, dipole correctors, and BPMs is larger in the electron RTML than in the positron RTML due to the longer electron Return line; for these 3 component classes, the different totals for each side are shown. Detailed information about the magnet families used in the RTML can be found elsewhere [154]. Each quadrupole and dipole has its own power supply, while other magnets are generally powered in series with one power supply supporting many magnets [155]. The cost estimate for the S-band dipole-mode structures was developed based on recent experience with accelerator-structure construction at IHEP.

**Table 7.6**
Total number of components in each RTML. Where 2 totals are shown, the larger number refers to the longer electron-side RTML, the smaller number refers to the shorter positron-side RTML.

| Magnets | | Instrumentation | | RF | |
|---|---|---|---|---|---|
| Bends | 336/356 | BPMs | 782/752 | Cavities | 440 |
| Quads | 825/793 | Wires | 12 | Cryo-Mod. | 51 |
| Dipoles | 1229/1157 | BLMs[†] | 2 | RF sources | 17 |
| Kickers | 18 | OTRs | 5 | S-band struct. | 2 |
| Septa | 14 | $\Phi$ monitors | 5 | S-band sources | 2 |
| Pulsed bends | 3 | Xray SLMs[‡] | 1 | | |
| Extr. bends | 12 | Rect. Coll. | 10 | | |
| Rasters | 6 | Circ. Coll. | 14 | | |
| Solenoids | 4 | | | | |

[†] Bunch Length Monitor
[‡] Synchrotron Light Monitor

Table 7.7 shows the system lengths for the RTML beamlines.

**Table 7.7**
System lengths for each RTML beamline. The larger number refers to the longer electron-side RTML, the smaller number refers to the shorter positron-side RTML.

| Beamline | Length |
|---|---|
| ERTL/PRTL | 302 m |
| ELTL/PLTL | 15 302 /14 109 m |
| ETURN/PTURN | 275 m |
| ESPIN/PSPIN | 123 m |
| EBC1/PBC1 | 231 m |
| EBC2/PBC2 | 908 m |
| Dumplines (E/P) | 182 m |
| Total | 17 323 /16 130 m |
| Total excluding dumplines | 17 141 /15 948 m |
| Footprint | 30 456 m |

### 7.5.2 Vacuum Systems

The base-pressure requirement for the downstream RTML is set by limiting the generation of beam halo to tolerable levels, while in the upstream RTML it is set by the necessity of avoiding beam-ion instabilities. As described in Section 7.4, the base pressure requirement for the downstream RTML is 10 µPa, while in the upstream RTML it is 2 µPa. Both upstream and downstream RTML vacuum systems are stainless steel with 2 cm OD; the upstream RTML vacuum system is installed with heaters to allow *in situ* baking, while the downstream RTML vacuum system is not. The bending sections of the turn-around and bunch compressors are not expected to need photon stops or other sophisticated vacuum systems, as the average beam current is low, and the fractional power loss of the beam in the bending regions is already small to limit emittance growth from ISR.

### 7.5.3 Cryogenics

Each RTML includes 51 cryomodules in the RF system of BC1 and BC2 and 4 superconducting solenoids in the Spin Rotator. Solenoids are cooled by a local cryocooler system operating at 4.2 K. The cryocooler requires only a small volume of liquid helium which is recondensed in the system. The RTML cryomodules are the same as used in the Main Linac. Liquid helium at 2 K needed for cryomodule cooling is transported from the ML area by a transfer line.

### 7.5.4 Service tunnels and Alcoves

In the area of the two-stage Bunch Compressor, there is a service tunnel that runs parallel to the beam tunnel. All of the power supplies, RF sources, and rack-mounted instrumentation and controls equipment needed for the bunch compressor are installed in the service tunnel. This configuration allows repairs and maintenance to be performed while minimizing disruption to the accelerator itself.

# Chapter 8
# Beam Delivery System and Machine Detector Interface

## 8.1 Introduction

The ILC Beam Delivery System (BDS) is responsible for transporting the $e^+/e^-$ beams from the exit of the high-energy linacs, focusing them to the sizes required to meet the ILC luminosity goals ($\sigma_x^*$ =474 nm, $\sigma_y^*$ =5.9 nm in the nominal parameters, see Section 2.2), bringing them into collision, and then transporting the spent beams to the main beam dumps. In addition, the BDS must perform several critical functions:

- measure the linac beam and match it into the final focus;
- protect the beamline and detector against mis-steered beams from the main linacs[1];
- remove any large amplitude particles (beam-halo) from the linac to minimize background in the detectors;
- measure and monitor the key physics parameters such as energy and polarization before and after the collisions.

The BDS must provide sufficient instrumentation, diagnostics and feedback systems to achieve these goals.

## 8.2 Parameters and System Overview

Tables 8.1 and 8.2 show the key BDS parameters.

**Table 8.1**
Key parameters of the BDS [12]. The range of $L^*$, the distance from the final quadrupole to the IP, corresponds to values considered for the existing SiD and ILD detector concepts.

| Parameter | Value | Unit |
|---|---|---|
| Length (start to IP distance) per side | 2254 | m |
| Length of main (tune-up) extraction line | 300 (467) | m |
| Max. Energy/beam (with more magnets) | 250 (500) | GeV |
| Distance from IP to first quad, $L^*$, for SiD / ILD | 3.51 / 4.5 | m |
| Crossing angle at the IP | 14 | mrad |
| Normalized emittance $\gamma\epsilon_x$ / $\gamma\epsilon_y$ | 10 000 / 35 | nm |
| Nominal bunch length, $\sigma_z$ | 300 | µm |
| Preferred entrance train to train jitter | <0.2–0.5 | $\sigma_y$ |
| Preferred entrance bunch to bunch jitter | $< 0.1$ | $\sigma_y$ |
| Typical nominal collimation aperture, $x/y$ | 6-10 / 30-60 | beam sigma |
| Vacuum pressure level, near/far from IP | 0.1 / 5 | µPa |

[1] This applies to the positron side of the BDS; on the electron side the protective fast abort extraction is located upstream of the positron source undulatory section.

**Table 8.2.** Energy-dependent parameters of the Beam Delivery System [84].

| | | Center-of-mass energy, $E_{\rm cm}$ (GeV) | | | | | | | |
|---|---|---|---|---|---|---|---|---|---|
| | | Baseline | | | | Upgrades | | | |
| Parameter | | 200 | 250 | 350 | 500 | 500 | 1000 (A1) | 1000 (B1b) | Unit |
| Nominal bunch population | $N$ | 2.0 | 2.0 | 2.0 | 2.0 | 2.0 | 1.74 | 1.74 | $\times 10^{10}$ |
| Pulse frequency | $f_{\rm rep}$ | 5 | 5 | 5 | 5 | 5 | 4 | 4 | Hz |
| Bunches per pulse | $N_{\rm bunch}$ | 1312 | 1312 | 1312 | 1312 | 2625 | 2450 | 2450 | |
| Nominal horizontal beam size at IP | $\sigma^*_{\rm x}$ | 904 | 729 | 684 | 474 | 474 | 481 | 335 | nm |
| Nominal vertical beam size at IP | $\sigma^*_{\rm y}$ | 7.8 | 7.7 | 5.9 | 5.9 | 5.9 | 2.8 | 2.7 | nm |
| Nominal bunch length at IP | $\sigma^*_{\rm z}$ | 0.3 | 0.3 | 0.3 | 0.3 | 0.3 | 0.250 | 0.225 | mm |
| Energy spread at IP, $\rm e^-$ | $\delta E/E$ | 0.206 | 0.190 | 0.158 | 0.124 | 0.124 | 0.083 | 0.085 | % |
| Energy spread at IP, $\rm e^+$ | $\delta E/E$ | 0.190 | 0.152 | 0.100 | 0.070 | 0.070 | 0.043 | 0.047 | % |
| Horizontal beam divergence at IP | $\theta^*_{\rm x}$ | 57 | 56 | 43 | 43 | 43 | 21 | 30 | µrad |
| Vertical beam divergence at IP | $\theta^*_{\rm y}$ | 23 | 19 | 17 | 12 | 12 | 11 | 12 | µrad |
| Horizontal beta-function at IP | $\beta^*_{\rm x}$ | 16 | 13 | 16 | 11 | 11 | 22.6 | 11 | mm |
| Vertical beta-function at IP | $\beta^*_{\rm y}$ | 0.34 | 0.41 | 0.34 | 0.48 | 0.48 | 0.25 | 0.23 | mm |
| Horizontal disruption parameter | $D_{\rm x}$ | 0.2 | 0.3 | 0.2 | 0.3 | 0.3 | 0.1 | 0.2 | |
| Vertical disruption parameter | $D_{\rm y}$ | 24.3 | 24.5 | 24.3 | 24.6 | 24.6 | 18.7 | 25.1 | |
| Energy of single pulse | $E_{\rm pulse}$ | 420 | 526 | 736 | 1051 | 2103 | 3409 | 3409 | kJ |
| Average beam power per beam | $P_{\rm ave}$ | 2.1 | 2.6 | 3.7 | 5.3 | 10.5 | 13.6 | 13.6 | MW |
| Geometric luminosity | $L_{\rm geom}$ | 0.30 | 0.37 | 0.52 | 0.75 | 1.50 | 1.77 | 2.64 | $\times 10^{34}$ cm$^{-2}$ s$^{-1}$ |
| – with enhancement factor | | 0.50 | 0.68 | 0.88 | 1.47 | 2.94 | 2.71 | 4.32 | $\times 10^{34}$ cm$^{-2}$ s$^{-1}$ |
| Beamstrahlung parameter (av.) | $\Upsilon_{\rm ave}$ | 0.013 | 0.020 | 0.030 | 0.062 | 0.062 | 0.127 | 0.203 | |
| Beamstrahlung parameter (max.) | $\Upsilon_{\rm max}$ | 0.031 | 0.048 | 0.072 | 0.146 | 0.146 | 0.305 | 0.483 | |
| Simulated luminosity (incl. waist shift) | $L$ | 0.56 | 0.75 | 1.0 | 1.8 | 3.6 | 3.6 | 4.9 | $\times 10^{34}$ cm$^{-2}$ s$^{-1}$ |
| Luminosity fraction within 1 % | $L_{1\%}/L$ | 91 | 87 | 77 | 58 | 58 | 59 | 45 | % |
| Energy loss from BS | $\delta E_{\rm BS}$ | 0.65 | 0.97 | 1.9 | 4.5 | 4.5 | 5.6 | 10.5 | % |
| $\rm e^+e^-$ pairs per bunch crossing | $n_{\rm pairs}$ | 45 | 62 | 94 | 139 | 139 | 201 | 383 | $\times 10^3$ |
| Pair energy per B.C. | $E_{\rm pairs}$ | 25 | 47 | 115 | 344 | 344 | 1338 | 3441 | TeV |

On the electron side, the BDS starts at the end of the target bypass dogleg of the positron source; on the positron side, it begins at the exit of the machine-protection system of the positron main linac [131]. The main subsystems of the beam delivery are [156]: the fast extraction and tuneup beam line; the betatron and energy collimation; the final focus; the interaction region; and the extraction line. A diagnostic section to determine the beam properties is located at the end of the main linacs. The layout of the beam delivery system is shown in Fig. 8.1. The BDS is designed for 500 GeV centre-of-mass energy but can be upgraded to 1 TeV with additional magnets.

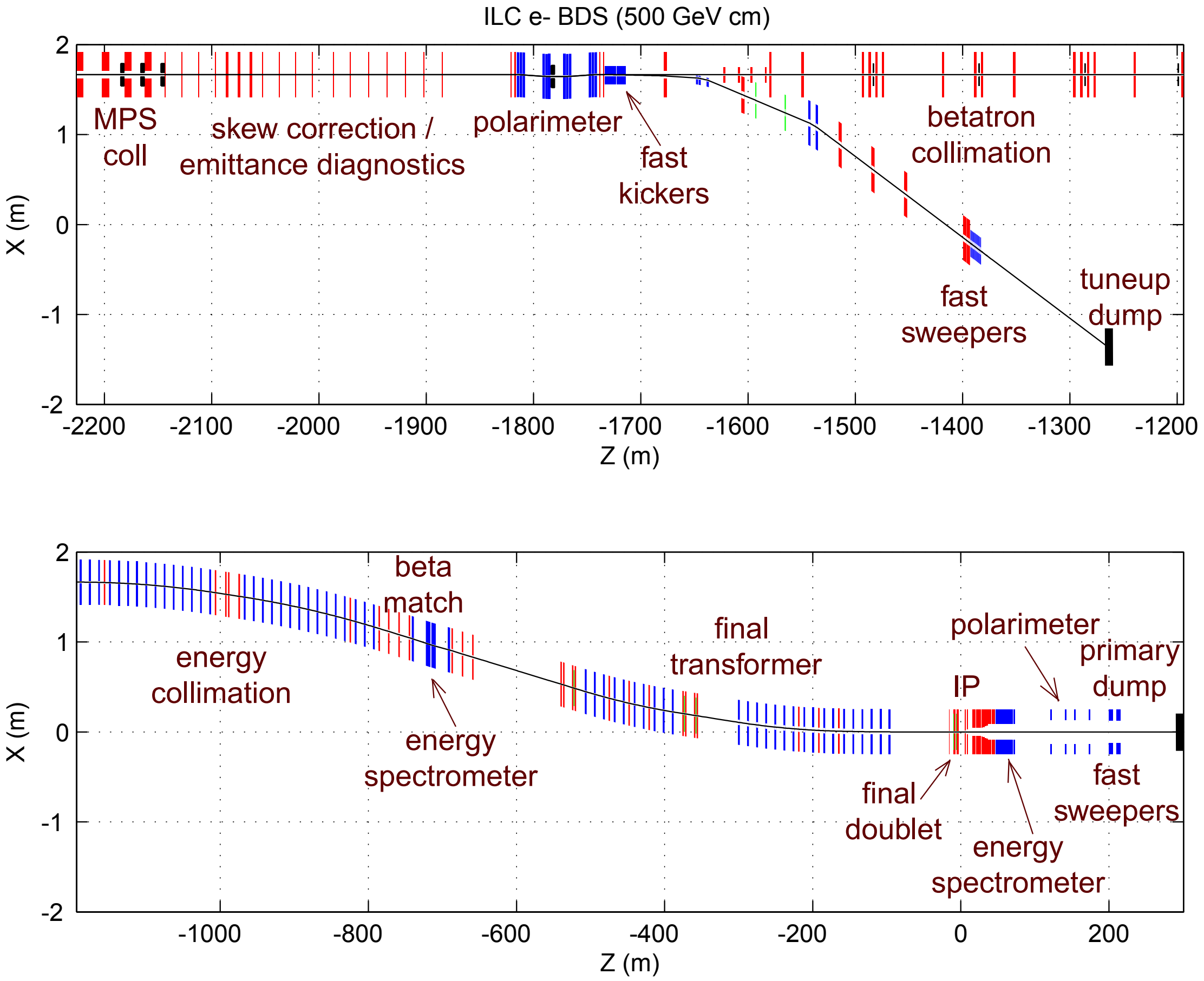

**Figure 8.1.** BDS layout showing functional subsystems, starting from the linac exit; X – horizontal position of elements, Z – distance measured from the IP.

There is a single collision point with a 14 mrad beam-crossing angle. To support future energy upgrades the beam-delivery systems are in line with the linacs and the linacs are also oriented at a 14 mrad angle. The 14 mrad geometry provides space for separate extraction lines and requires crab cavities to rotate the bunches into the horizontal for head-on collisions. There are two detectors in a common IR hall which alternately occupy the single collision point, in a so-called "push-pull" configuration. This approach, which is significantly more challenging for detector assembly and operation than a configuration with two separate interaction regions, has been chosen for economic reasons.

## 8.3 Lattice description

The BDS lattice [157] starts 2254 m away from the Interaction Point; on the electron side, the BDS follows the target bypass section of the positron source, while on the positron side it starts after the Machine Protection and Collimation section of the Main Linac [131].

### 8.3.1 Diagnostics, Tune-up dump, Machine Protection

The initial part of the BDS, from the end of the main linac to the start of the collimation system, is responsible for measuring and correcting the properties of the beam before it enters the Collimation and Final-Focus systems. In addition, errant beams must be detected here and safely extracted in order to protect the downstream systems. Starting at the exit of the main linac, the system includes the skew-correction section, emittance-diagnostic section, polarimeter with energy diagnostics, fast-extraction/tuning system and beta-matching section.

#### 8.3.1.1 Skew Correction

The skew correction section contains 4 orthonormal skew quadrupoles which provide complete and independent control of the 4 betatron-coupling parameters. This scheme allows correction of any arbitrary linearised coupled beam.

#### 8.3.1.2 Emittance Diagnostics

The emittance diagnostic section contains 4 laser wires which are capable of measuring horizontal and vertical RMS beam sizes down to 1 µm. The wire scanners are separated by 45° in betatron phase to allow a complete measurement of 2D transverse phase space and determination of the projected horizontal and vertical emittances.

#### 8.3.1.3 Polarimeter and Energy Diagnostics

A magnetic chicane used for Compton polarimetry and auxiliary beam-energy measurement is situated after the emittance-diagnostic section, directly after the branch-off of the tune-up extraction line [158]. At the center of the chicane is the interaction point for Compton scattering and two BPMs to monitor relative beam-energy changes and the angle. The length of the chicane is set to limit horizontal emittance growth due to synchrotron radiation to less than 1 % with a 500 GeV beam. The detector for the Compton-scattered electrons is placed behind the last chicane magnet.

#### 8.3.1.4 Tune-up and Emergency Extraction System

The pulsed extraction system is used to extract beams in the event of an intra-train Machine Protection System (MPS) alarm. It is also used at any time when beams are not desired in the collimation, final-focus, or IR areas, for example during commissioning of the main linacs. The extraction system includes both fast kickers which can rise to full strength in the 300 ns between bunches, and pulsed bends which can rise to full strength in the 200 ms between trains. These are followed by a transfer line with $\pm 10$ % momentum acceptance which transports the beam to a full-beam-power water-filled dump. There is a 125 m drift which allows the beam size to grow to an area of $2\pi$mm$^2$ at the dump. A set of rastering kickers sweep the beam in a 6 cm-radius circle on the dump window. By using the nearby and upstream BPMs in the polarimeter chicane and emittance sections, it is possible to limit the number of errant bunches which pass into the collimation system to 1–2.

## 8.3.2 Collimation System

Particles in the beam halo produce backgrounds in the detector and must be removed in the BDS collimation system. One of the design requirements for the ILC BDS is that no particles are lost in the last several hundred meters of beam line before the IP. Another requirement is that all synchrotron radiation passes cleanly through the IP to the extraction line. The BDS collimation must remove any particles in the beam halo that do not satisfy these criteria. These requirements define a system where the collimators have very narrow gaps and the system is designed to address the resulting machine protection, survivability and issues related to beam-emittance dilution.

The collimation system has a betatron-collimation section followed by energy collimators. The downstream energy collimators help to remove particles whose energy has been degraded that originate in the betatron-collimation section but are not absorbed there. The betatron-collimation system has two spoiler/absorber $x/y$ pairs located at high-beta points, providing single-stage collimation at each of the final doublet (FD) and IP betatron phases. The energy-collimation section has a single spoiler located at the central high-dispersion point (1530 µm/%). Dedicated studies [159] show that two additional quadrupoles between the collimators may be beneficial to tune the phase advance between the collimators and the interaction point. All spoilers and absorbers have adjustable gaps. Protection collimators (PC) are located throughout the system to provide local protection of components and additional absorption of scattered halo particles.

The spoilers are 0.5–1 $X_0$ (radiation length) thick, the absorbers are 30 $X_0$, and the protection collimators are 45 $X_0$. The betatron spoilers as well as the energy spoiler are "survivable", i.e. they can withstand a hit of two errant bunches of 250 GeV/beam, matching the emergency-extraction design goal. With 500 GeV beam, they would survive only one bunch, and would therefore require more effective MPS or the use of a collimator pre-radiator.

The collimation apertures required are approximately $\sim$6–9$\sigma_{\mathrm{x}}$ in the $x$ plane and $\sim$40–60$\sigma_{\mathrm{y}}$ in the $y$ plane. These correspond to typical half-gaps of the betatron spoiler of $\sim$1 mm in the $x$ plane and $\sim$0.5 mm in the $y$ plane.

### 8.3.2.1 Beam Energy Measurement

Following the energy-collimation section is another magnetic chicane for the beam-energy spectrometer. The chicane consists of four dipoles which introduce a fixed dispersion of $\eta = 5$ mm at the centre. Its length is chosen to limit horizontal emittance growth due to synchrotron radiation to less than 1 % with a 500 GeV beam. Before, after and at the centre of the chicane, the beam line is instrumented with cavity BPMs mounted on translation systems. When operating the spectrometer with a fixed dispersion over the whole energy range, a BPM resolution of 0.5 µm is required.

### 8.3.2.2 Muon suppression

Electromagnetic showers created by primary beam particles in the collimators produce penetrating muons that can easily reach the collider hall [161]. The muon flux through the detector is reduced by a 5 m-long magnetised iron shield 330 m upstream of the collision point that fills the cross-sectional area of the tunnel and extends 0.6 m beyond the ID of the tunnel, as shown in Fig. 8.2 [162]. The shield has a magnetic field of 1.5 T, with opposite polarities in the left and right halves of the shield such that the field at the beam line is zero. The shield also provides radiation protection for the collider hall during access periods when beam is present in the linac.

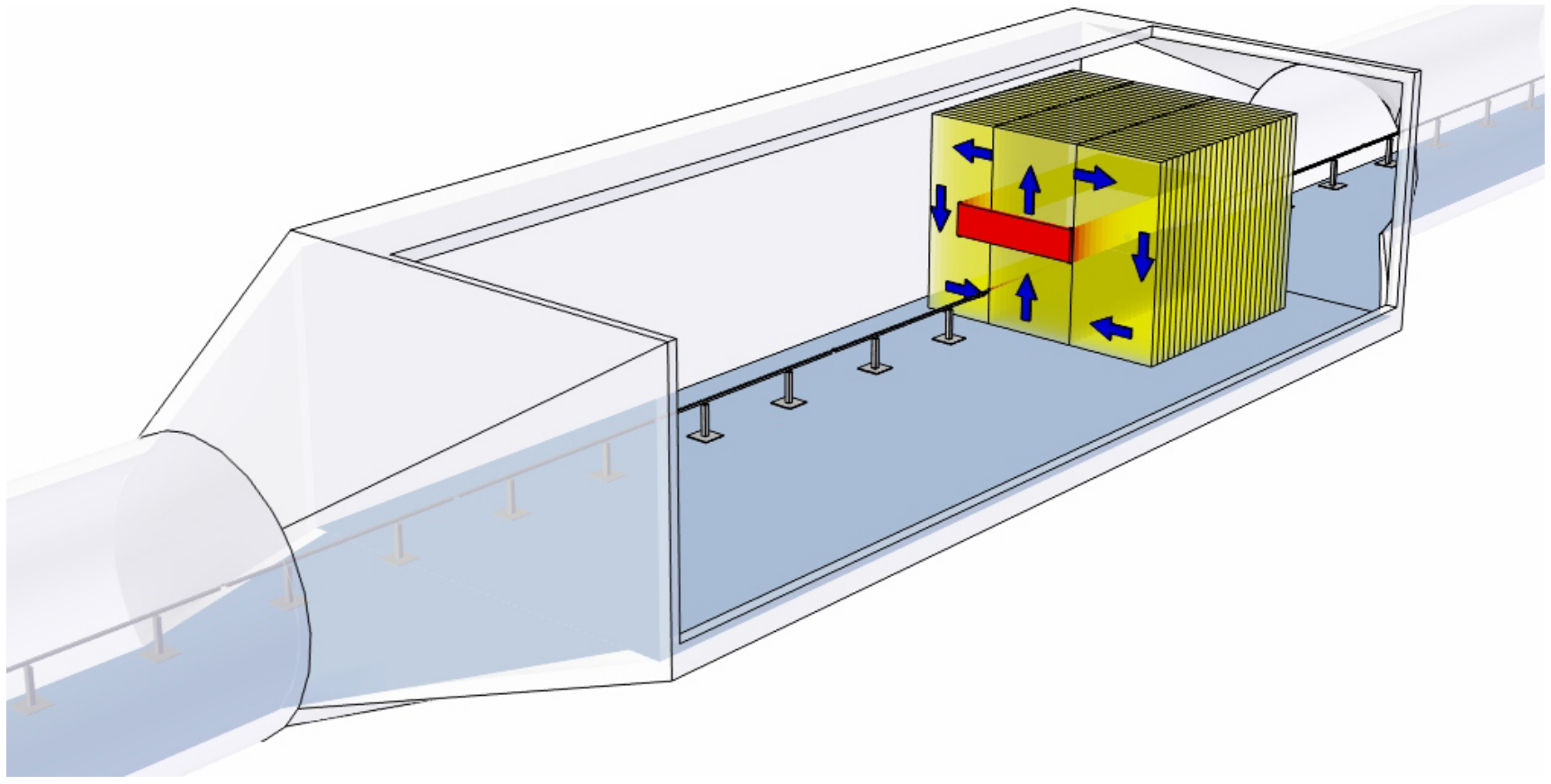

**Figure 8.2**
Schematic of the 5 m-long magnetised muon shield installed in a tunnel vault which is configured to accommodate a possible upgrade to a 19 m-long shield. The coil is shown in red, and blue arrows indicate the direction of the magnetic field in the iron.

#### 8.3.2.3 Halo-power handling

The power-handling capacity of the collimation system is set by two factors: the ability of the collimators to absorb the incident beam power and the ability of the muon-suppression system to reduce the muon flux through the detector. In the baseline design, the muon-suppression system presents the more restrictive limitation, setting a tolerance of $1 - 2 \times 10^{-5}$ on the fraction of the beam collimated in the BDS. With these losses and the 5 m wall, the number of muons reaching the collider hall would be a few muons per 150 bunches (a reduction by more than a factor of 100). Since the actual beam-halo conditions are somewhat uncertain, the BDS includes caverns large enough to increase the muon shield from 5 m to 19 m and to add an additional 9 m shield downstream. Filling all of these caverns with magnetized muon shields would increase the muon suppression capacity of the system to $1 \times 10^{-3}$ of the beam. The primary beam spoilers and absorbers are water cooled and capable of absorbing $1 \times 10^{-3}$ of the beam continuously.

#### 8.3.2.4 Tail-folding octupoles

The final focus includes two superconducting octupole doublets [163]. These doublets use nonlinear focusing to reduce the amplitude of beam-halo particles while leaving the beam core untouched [164]. This “tail-folding” would permit larger collimation amplitudes, which in turn would dramatically reduce the amount of beam power intercepted and the wakefields. In the interest of conservatism, the collimation system design described above does not take this tail folding into account in the selection of apertures and other parameters.

### 8.3.3 Final focus

The role of the final-focus (FF) system is to demagnify the beam to the required size (~474 nm horizontal and ~5.9 nm vertical) at the IP. The FF optics creates a large and almost parallel beam at the entrance to the final doublet (FD) of strong quadrupoles. Since particles of different energies have different focal points, even a relatively small energy spread of ~0.1 % significantly dilutes the beam size, unless adequate corrections are applied. The design of the FF is thus mainly driven by the need to cancel the chromaticity of the FD. The ILC FF has local chromaticity correction [165] using sextupoles next to the final doublets. A bend upstream generates dispersion across the FD, which is required for the sextupoles to cancel the chromaticity. The dispersion at the IP is zero and the angular dispersion is about $\eta'_x$ ~0.009, i.e. small enough that it does not significantly increase the beam divergence. Half of the total horizontal chromaticity of the whole final focus is generated upstream of the bend in order for the sextupoles to cancel the chromaticity and the second-order dispersion simultaneously [166].

The horizontal and the vertical sextupoles are interleaved in this design, so they generate third-order geometric aberrations. Additional upstream sextupoles in proper phase with the FD sextupoles partially cancel the third-order aberrations. The residual higher-order aberrations can be minimised further with octupoles and decapoles. The final-focus optics is shown in Fig. 8.3.

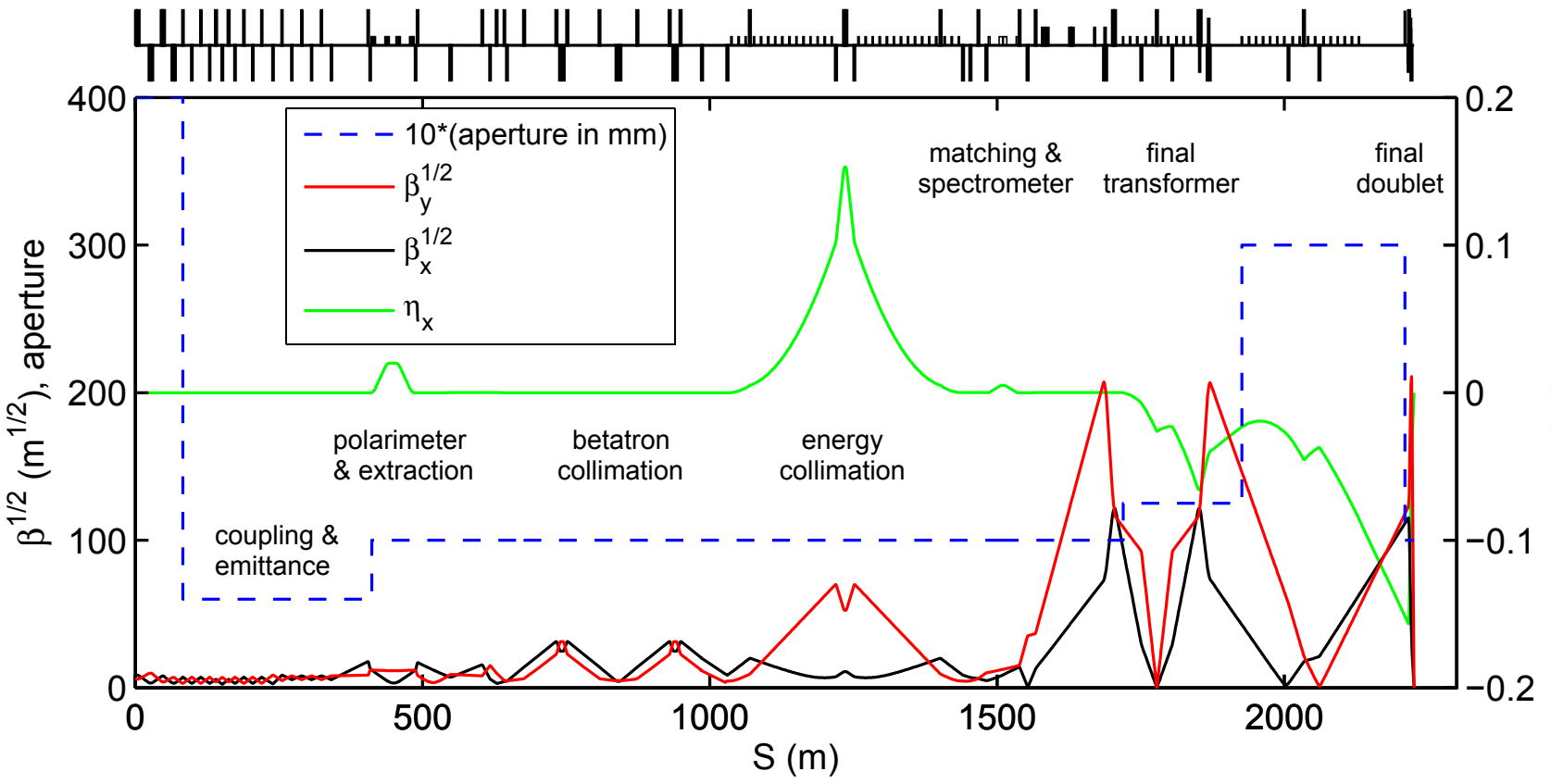

**Figure 8.3** BDS optics, subsystems and vacuum chamber aperture; $S$ is the distance measured from the entrance.

Synchrotron radiation from the bending magnets causes emittance dilution, so it is important to maximize the bending radius, especially at higher energies. The FF includes bending magnets for 500 GeV centre-of-mass energy and space for additional bending magnets that are necessary at higher energies. At 500 GeV, every fifth bending magnet is installed, leading to an emittance dilution of 0.5 %; at 1 TeV, with all bending magnets implemented, the figure is 1 %.

In addition to the final-doublet and chromaticity-correction magnets, the final focus includes: an energy spectrometer (see Section 8.7.2.1); additional absorbers for the small number of halo particles that escape the collimation section; tail-folding octupoles (see Section 8.3.2); the crab cavities (see Section 8.9); and additional collimators for machine protection or synchrotron-radiation masking of the detector.

### 8.3.4 Extraction line

The ILC extraction line [167, 168] has to transport the beams from the IP to the dump with acceptable beam losses, while providing dedicated optics for beam diagnostics. After collision, the beam has a large angular divergence and a huge energy spread with very low-energy tails. It is also accompanied by a high-power beamstrahlung photon beam and other secondary particles. The extraction line must therefore have a very large geometric and energy acceptance to minimise beam loss.

The optics of the ILC extraction line is shown in Fig. 8.4. The extraction line can transport particles with momentum offsets of up to 60 % to the dump. There is no net bending in the extraction line, which allows the charged-particle dump to also act as a dump for beamstrahlung photons with angles of up to 0.75 mrad.

The first quadrupole is a superconducting magnet 5.5 m from the IP, as shown in Fig. 8.7. The second quadrupole is also superconducting, with a warm section between their cryostats. The downstream magnets are normal conducting, with a drift space to accommodate the crab cavity in the adjacent beamline. The quadrupoles are followed by two diagnostic vertical chicanes for the energy spectrometer and Compton polarimeter, with a secondary focal point in the centre of the latter. The horizontal angular amplification ($R_{22}$) from the IP to the Compton IP is set to $-0.5$ so that the measured Compton polarisation is close to the luminosity weighted polarisation at the IP. The lowest-energy particles are removed by a vertical collimator in the middle of the energy chicane. A large chromatic acceptance is achieved through the soft D-F-D-F quadruplet system and careful optimization of the quadrupole strengths and apertures. The two SC quadrupoles are compatible with up to 250 GeV beam energy, and the warm quadrupoles and the chicane bends with up to 500 GeV.

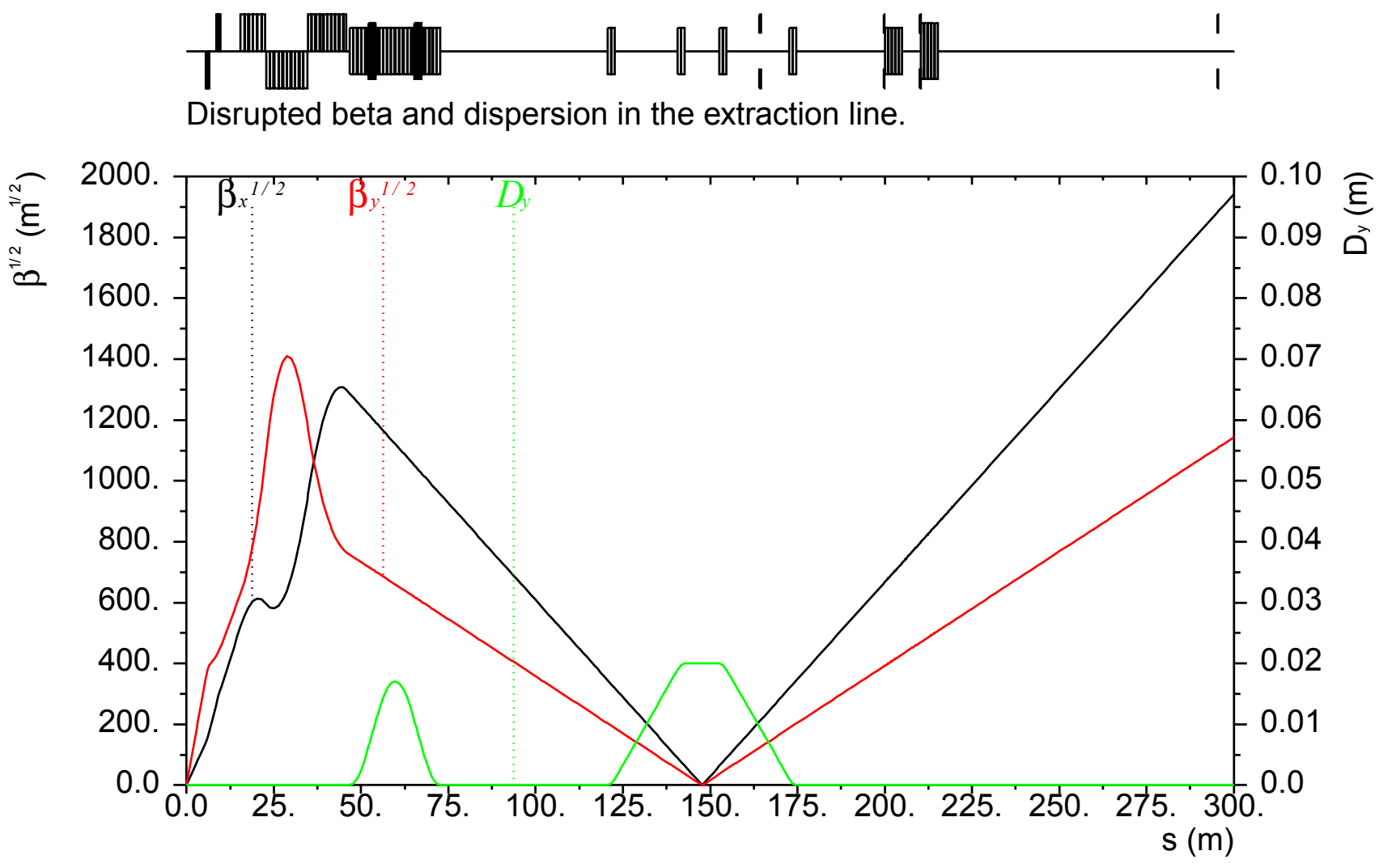

**Figure 8.4**
Disrupted $\beta$-functions and dispersion in the extraction line for the nominal 250 GeV beam.

The diagnostic section is followed by a 100 m-long drift to allow adequate transverse separation (>3.5 m) between the dump and the incoming line. It also allows the beam size to expand enough to protect the dump window from the small undisrupted beam. A set of rastering kickers sweep the beam in a 3 cm circle on the window to avoid boiling the water in the dump vessel. They are protected by three collimators in the 100 m drift that remove particles that would hit outside the 15 cm-radius dump window.

Extraction beam loss has been simulated for realistic 250 GeV GUINEA-PIG beam distributions [169], with and without beam offset at the IP. No primary particles are lost in the SC quadrupoles, and all particles above 40 % of the nominal beam energy are transmitted cleanly through the extraction magnets. The total primary loss on the warm quadrupoles and bends is a few Watts, while the loss on the protection collimators is a few kW for the nominal beam parameters. Figure 8.5 shows that even for an extreme set of parameters, with very high beamstrahlung energy loss, the radiation deposition in the magnet region is manageable.

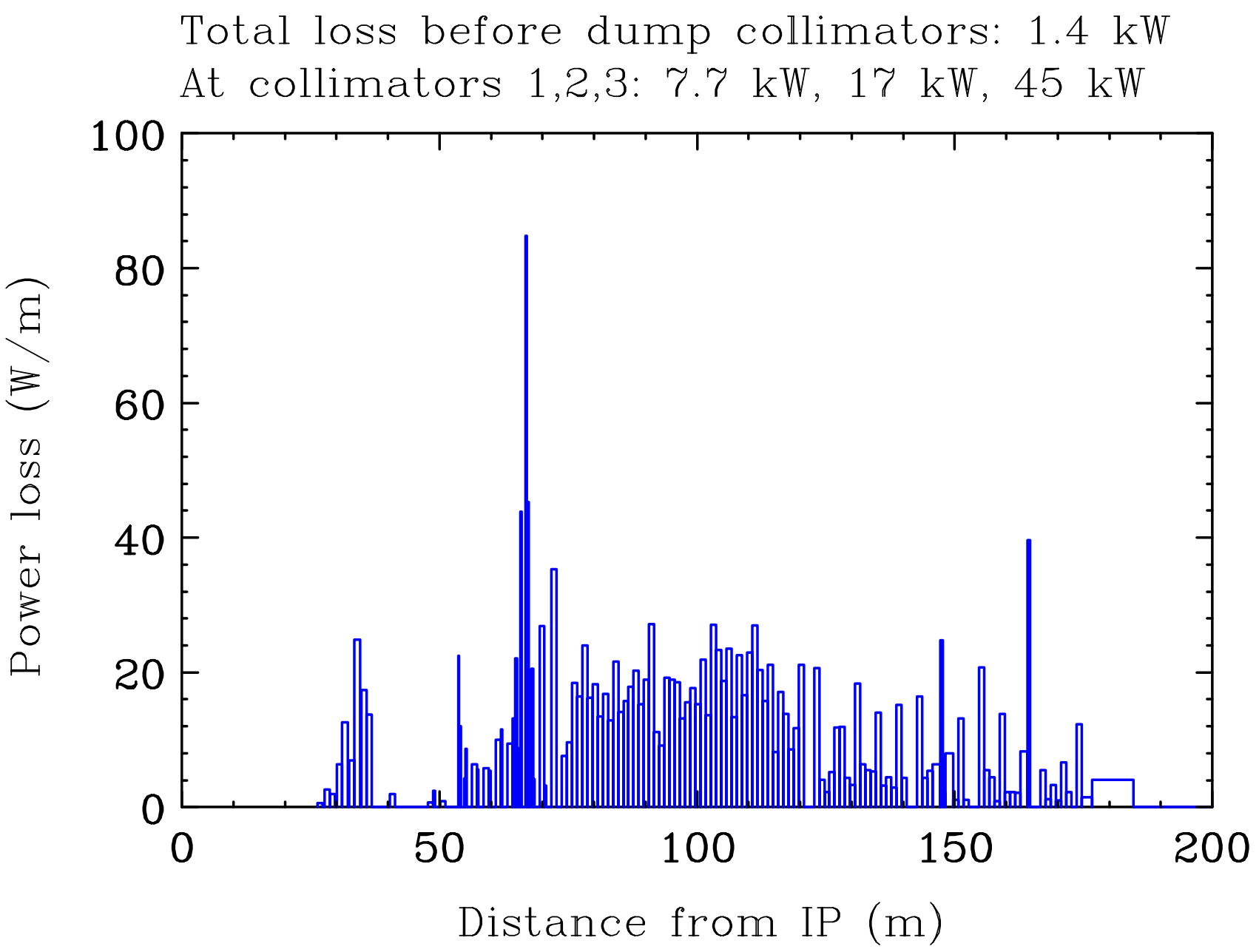

**Figure 8.5**
Power loss density in the magnet region for disrupted beam at 250 GeV, for high-luminosity operation.

### 8.3.5 Beam dynamics and emittance growth

Wakefield calculations for the BDS spoilers and absorbers give IP jitter amplification factors [153] of $\mathcal{A}_{\mathrm{x}}$ =0.14 and $\mathcal{A}_{\mathrm{y}}$ =1.05 for an assumed collimation depth of $9\sigma_x$ and $65\sigma_y$ in the horizontal and vertical planes respectively. Estimated as $\delta\varepsilon/\varepsilon = (0.4\, n_{\mathrm{jitter}}\, \mathcal{A})^2$, these parameters give an emittance dilutions of 0.08 % and 4.4 % in the $x$ and $y$ planes respectively, assuming a $0.5\,\sigma$ incoming beam jitter. The current ILC collimation depth is still being re-evaluated and is likely to be smaller, in particular for lower centre-of-mass energy operations. A more stringent requirement on the pulse-to-pulse jitter of $0.2\sigma_y$ will likely be required at the entrance of the BDS (specifically at the collimators), but this should be achievable using the fast intra-train feedback system located at the exit of the linac. Energy jitter at the collimators also amplifies the horizontal jitter at the IP. An energy jitter of 1 % produces a horizontal emittance growth of 2.2 %.

## 8.4 Interaction-Region Layout and Machine-Detector Interface

### 8.4.1 Requirements and boundary conditions

The ILC is configured to have two detectors that share one interaction point with only one detector in data-taking position at a time, so-called "push-pull" operation mode. The time spent to roll detectors in and out needs to be as short as possible to maximise the time available for data taking.

The need for high efficiency sets specific requirements and challenges for many detector and machine systems, in particular the IR magnets, the cryogenics, the alignment system, the beamline shielding, the detector design, and the overall integration. The minimal functional requirements and interface specifications for the push-pull IR have been successfully developed and published [170]. This constrains all further IR design on both the detectors and machine. The developments lead to a detailed design of the technical systems and the experimental area layout that follow detailed engineering specifications [171].

### 8.4.2 The push-pull system

The detector motion and support system is designed to ensure reliable push-pull operation, allowing a hundred moves over the life of the experiment, while preserving the internal alignment of the detector's internal components and ensuring accuracy of detector positioning. The motion system preserves the structural integrity of the collider-hall floor and walls. Moreover, the motion and support system must be compatible with the vibration stability requirements of the detector, which are at the level of tens of nanometers. In regions with seismic activity, the system must also be compatible with earthquake-safety standards.

The detectors are placed on platforms that preserve the detector alignment and distribute the load evenly onto the floor (see Fig. 8.6). The platform also carries some of the detector services like electronic racks. Cables and supply lines are routed to the platform in flexible cable chains that move in trenches underneath the platform itself. In combination with a simple indexing mechanism, the platform with the detector can be positioned quickly within the required precision of 1 mm with respect to the beam axis.

An engineering study on a possible platform design has concluded that the flexure of the platform and the distortion of the cavern invert would total less than $\pm 2$ mm [172]. Two different types of transport systems are under study for the platform, air pads and Hilman rollers. In both cases, the platforms are jacked onto the transport system to allow for the movement of a slightly undulating surface. The platform with the detector can be positioned within approximately six hours. In parking or beam position, the platforms are lowered onto permanent supports. Trenches in the hall floor for the cable chains also provide access to the platform undercarriage for maintenance.

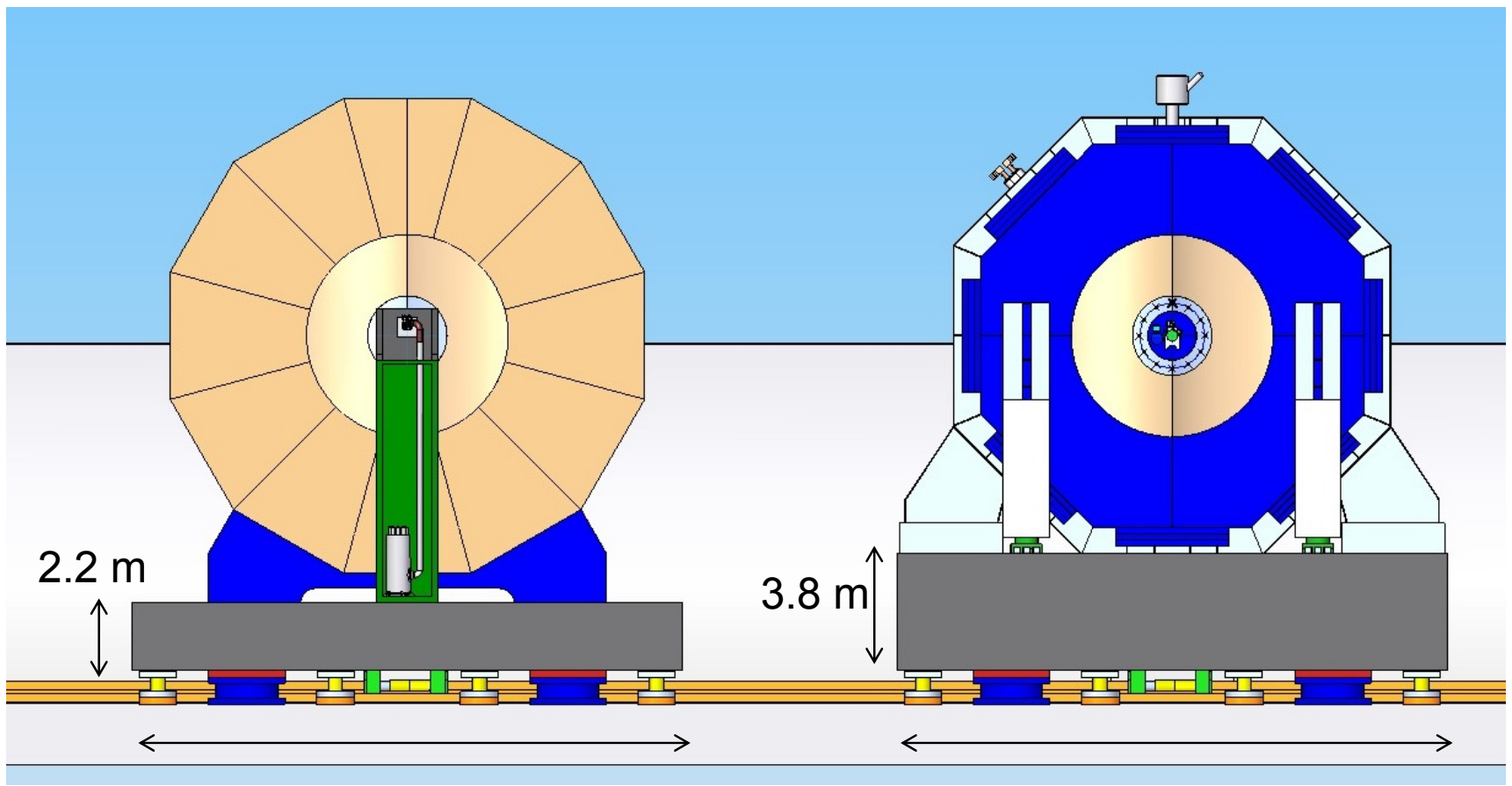

**Figure 8.6**
Platform support concept for the push-pull system. Left - ILD; right - SiD

#### 8.4.2.1 SiD in a push-pull configuration

As the half-height of SiD is 1.7 m less than that of ILD, an extra thick support platform is required. With the magnetic field on and the endcap doors sucked into the central barrel, SiD is very stiff. The last quadrupole lens package, QD0, rests on a 5 d.o.f. magnetically insensitive mover system which in turn rests on cylindrical cutouts in the doors which are only marginally larger than the diameter of the QD0 cryostat. This design emphasises maximal hermeticity and rapid push-pull detector exchange. The forward-calorimeter package (LumiCal, BeamCal and masking) is logically a cantilevered extension of the QD0 cryostat. An alignment system based on Frequency Scanning Interferometry (FSI) aligns the opposing QD0/FCAL packages to the tunnel-mounted QF1 cryostats that complete the final doublet telescope and ensure precision positioning of the LumiCal with respect to the IP. The same FSI system guarantees vertex- and tracking-detector alignment after each push-pull operation without the need to reacquire beam-based alignment data. This design requires that all mechanical systems mounted on the detector be vibration free. The IP vacuum is assumed to be achievable via QD0 cryo-pumping without external or NEG pumps.

#### 8.4.2.2 ILD in a push-pull configuration

The ILD detector is somewhat larger than SiD and is also designed to be assembled from slices in a similar way to the CMS detector at LHC. The detector placement on the platform preserves detector alignment and distributes the load evenly onto the floor. The platform also carries some of the detector services like electronic racks. The ILD slices have their own motion system based on air pads and grease pads. In the parking position, the detector can be opened for maintenance by moving the yoke slices on air pads from the platform. The QD0 magnets of ILD are supported by an external pillar that couples the magnet directly to the platform floor. In the barrel of the detector, the QD0 magnets are stabilised by a tie-rod system. This arrangement allows the detector end caps to open to some extent without removing the quadrupoles. An FSI system ensures the alignment of the quadrupoles to each other and to the beam line that is defined by the stationary QF1 magnets.

### 8.4.3 Final focus

The ILC final focus uses independently adjustable compact superconducting magnets for the incoming and extraction beam lines. The adjustability is needed to accommodate beam-energy changes and the separate beam line allows optics suitable for post-IP beam diagnostics. The BNL direct-wind technology is used to produce closely spaced coil layers of superconducting multi-strand cable. The design is extremely compact and the coils are almost touching inside shared cold-mass volumes. Cooling is provided by superfluid helium at 2 K.

To facilitate "push-pull" at a shared IP, the superconducting final-focus magnets are arranged into two groups so that they can be housed in two separate cryostats as shown in Figure 8.7, separated by only warm components and vacuum valves. The first cryostat grouping in Fig. 8.7 moves with the detector during switchover, while the second remains fixed on the beam line.

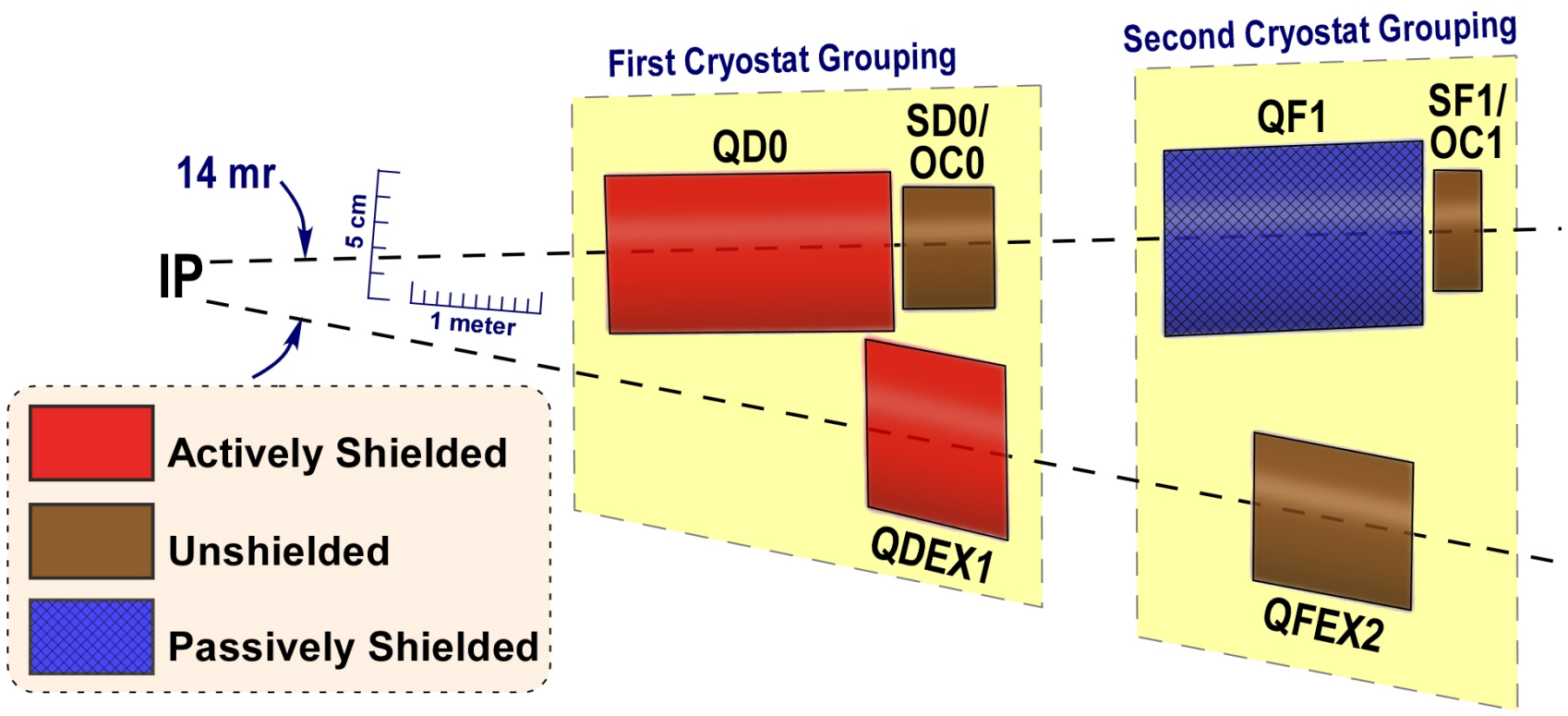

**Figure 8.7** Schematic layout of magnets in the IR.

Figure 8.8 shows the engineering model of the magnets that are in the detector-mounted cryostat: the QD0 quadrupole; the sextupole package; and the extraction line quadrupole. In the current design, the QD0 magnet is split into two coils. This allows for higher flexibility in running at lower energies.

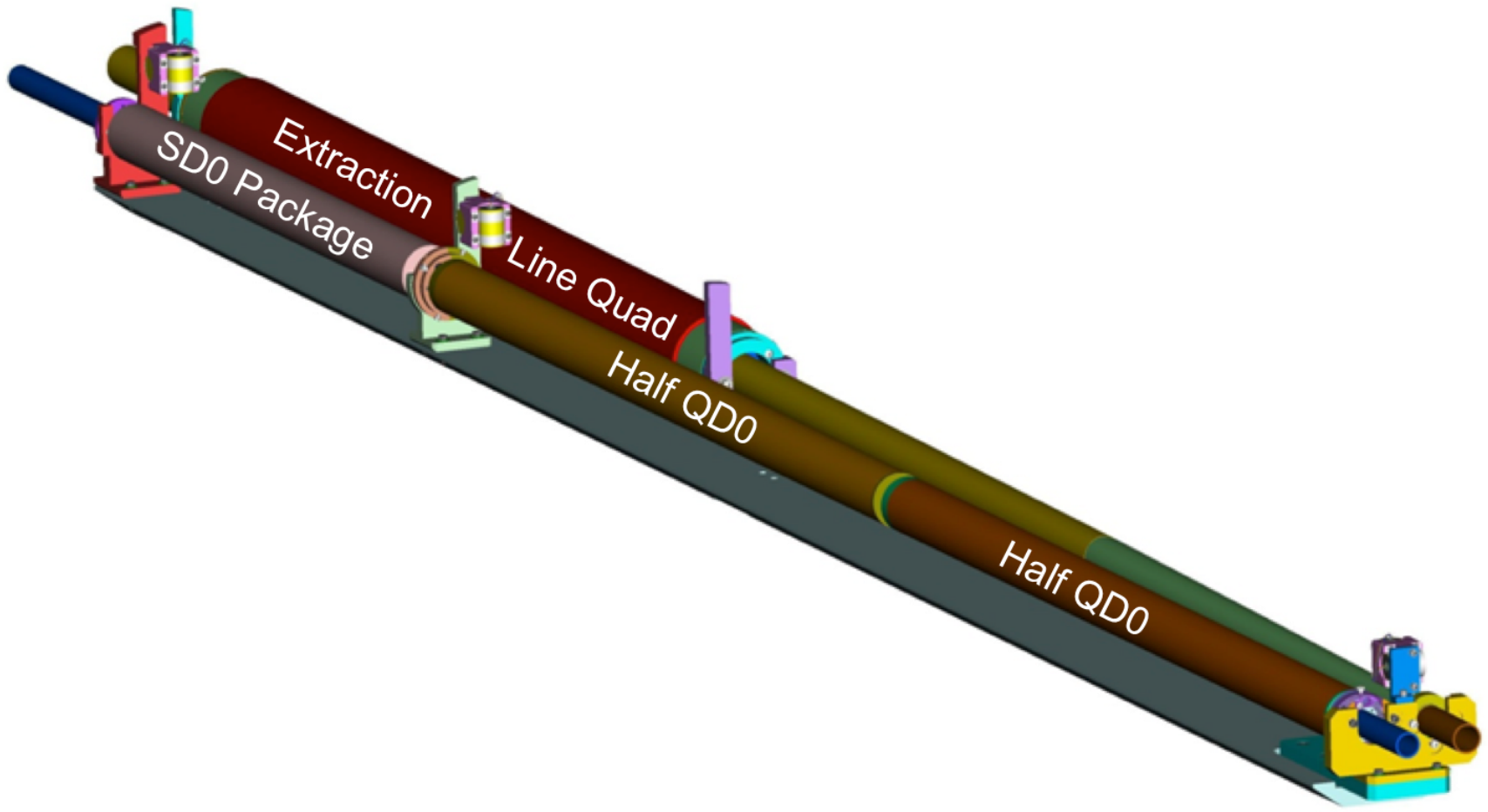

**Figure 8.8** Engineering model of the detector-mounted final-focus magnets [173]. The QD0 magnet is split into two coils to allow for energy flexibility.

The technology of the superconducting final-focus magnets has been demonstrated by a series of short prototype multi-pole coils. The schematic layout of magnets in the IR is shown in Fig. 8.7. The quadrupoles closest to the IP are actually inside the detector solenoidal field and therefore cannot have magnetic-flux-return yokes; at the closest coil spacing, the magnetic cross talk between the two beam lines is controlled by using actively shielded coil configurations and by use of local correction coils, dipole, skew dipole and skew quadrupole or skew sextupole, as appropriate.

An additional optical element is required in the IR to compensate the effects of the detector's solenoidal field interacting with the accelerator IR magnets. The so-called large-aperture Detector Integrated Dipole (DID) [175] reduces detector backgrounds at high beam energies, while minimising orbit deflections at low energies.

The vertical position of the centre of the incoming-beam-line quadrupole field must be stable to order of 50 nm, in order to stay within the capture range of the intra-train-collision feedback (see Section 8.7.1 and references [170, 176]). This requirement is well beyond experience at existing accelerators and is being addressed in a world-wide R&D program.

### 8.4.4 Experimental-area layout and infrastructure

The design layout of the experimental areas – at the surface and underground – needs to fulfil the requirements of both detectors and the machine while at the same time minimising the cost. As the boundary conditions for flat topography sites and for mountain sites for the ILC are very different, two different solutions have been developed.

#### 8.4.4.1 Flat topography sites

At the flat topography ILC sites, e.g. in the US or in Europe, the access to the detector cavern is provided by vertical shafts of ≈ 100 m length. The detectors are pre-assembled and tested in large sections in surface buildings, similar to what was done for CMS. Only late in the construction phase, about one year before the machine start-up, can the detector parts be lowered via a large shaft of 18 m diameter into the cavern. This procedure decouples to a large extent the time lines of the civil construction and the machine and detector installation work. In addition, the space in the underground cavern is minimised as no lengthy detector-installation procedures need to be done there.

Figure 8.9 shows the layout for these sites. The hall floor layout follows a z-shape that allows for two maintenance positions where detector parts could be moved away from the push-pull platform. The platforms run along the straight section of the hall, perpendicular to the beam line. Access is provided by a set of five vertical shafts. The largest, with 18 m diameter, is only used in the installation phase of both detectors. It is located directly over the IP, so that the heavy detector parts with masses of up to 3500 t can be lowered directly onto the respective platforms. Two shafts of 8 m and 10 m diameter are located in the maintenance parts of the hall. They provide independent access to each maintenance region so that one detector can always take data undisturbed at the beam position. These shafts allow transport of material for maintenance and upgrades and contain service lines (power, data, cooling, etc.) into the hall. Two small 6 m additional shafts are needed for personnel transport and safety egress.

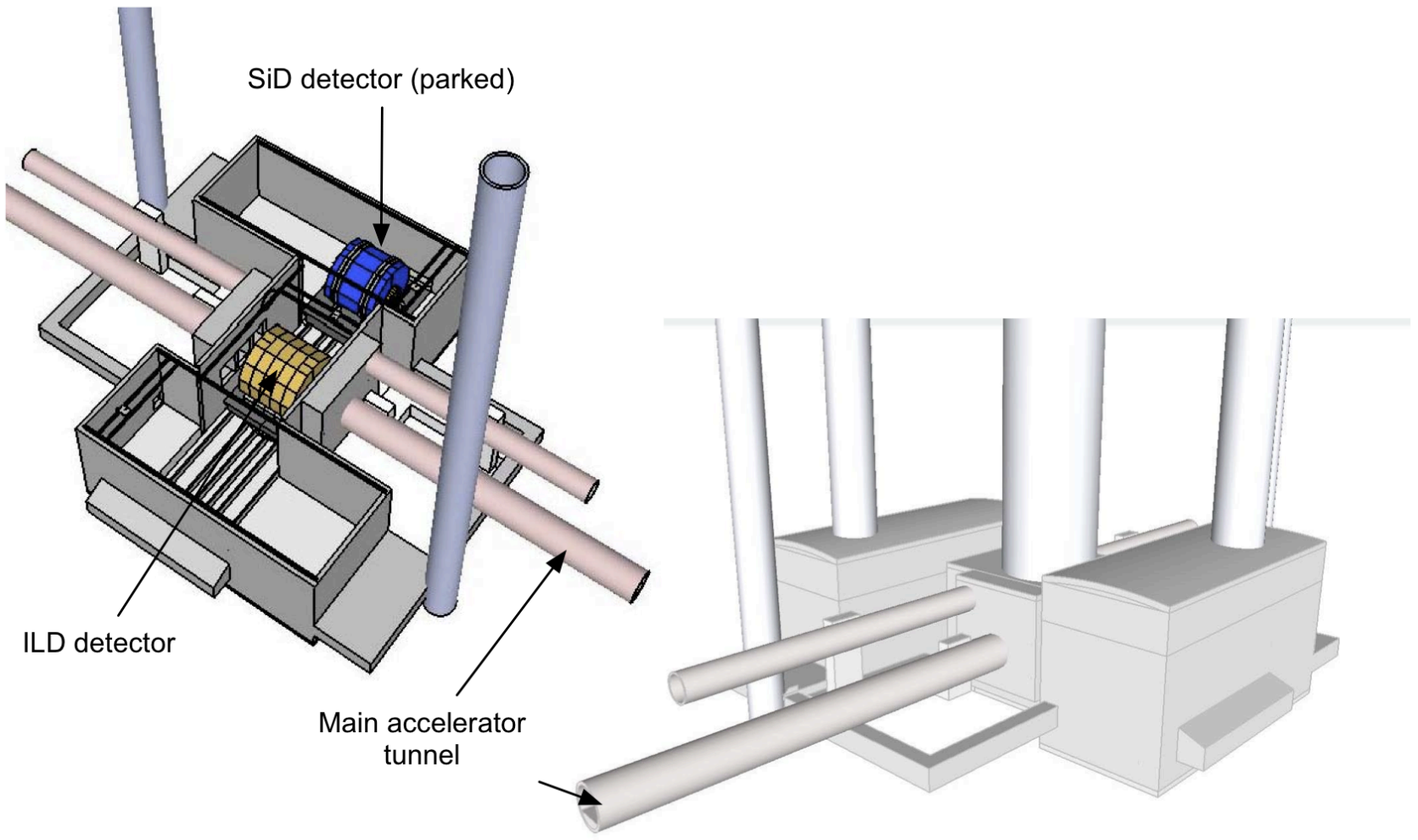

**Figure 8.9** Layout of the detector cavern for the American region.

#### 8.4.4.2 Mountainous sites

At the mountainous sites that are under study in Japan, no vertical access will be possible to the detector cavern. All material and personnel needs to be brought into the hall via an access tunnel of ≈ 1 km length that may have a slope of up to 10 %. The diameter of this tunnel and the capacity of the transport system limits the masses and sizes of the detector parts that can be brought into the hall. This forbids the application of the CMS-type detector-assembly scheme as described above.

A modified scheme is needed, where most of the detector assembly is done inside the underground cavern. The largest parts that cannot be assembled in situ are the superconducting coils of the detector solenoids. They define the diameter of the access tunnel to be ≈ 11 m.

Figure 8.10 shows the underground cavern for the Japanese sites. The access tunnel on the right extends beyond the cavern to the central ILC region with the damping rings. The larger entrance into the hall is used for ILD, the slightly smaller rear entrance for SiD. The SiD coil is smaller and fits into the tapered tunnel that passes underneath the ILC beam line. The main cavern has alcoves that extend the parking positions of the detectors to allow the unslicing and maintenance operations. The assembly phase of the detectors in this arrangement takes much longer inside the cavern (>3 y compared to 1 y in the flat topography case) and needs careful planning of the use of the underground space and the transport capacity in the access tunnel.

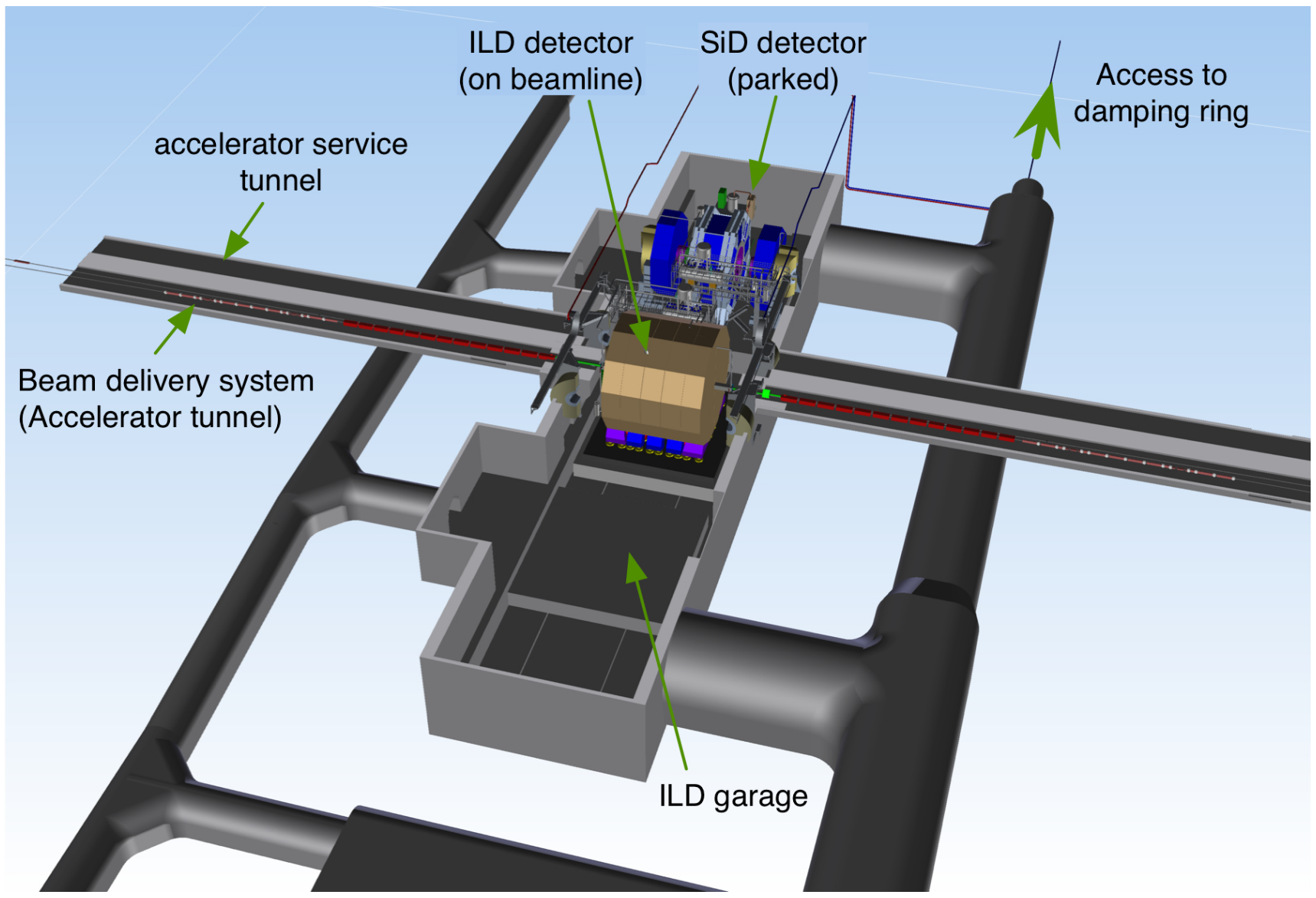

**Figure 8.10** Layout of the detector cavern for mountainous sites.

## 8.4.5 Shielding

### 8.4.5.1 Radiation

The ILC detectors are self-shielding with respect to ionising radiation stemming from maximum-credible-beam-loss scenarios [177]. Additional shielding in the hall is necessary to fill the gap between the detector and the wall in the beam position. The design of this beam-line shielding needs to accommodate both detectors, SiD and ILD, which are significantly different in size. A common 'pac-man' design has been developed, where the movable shielding parts are attached to the wall of the detector hall and matched to interface pieces on the experiments (see Fig. 8.11).

### 8.4.5.2 Magnetic fields

The magnetic stray fields outside the iron return yokes of the detectors need to be small enough to cause no disturbance to the other detector during operation or maintenance. The magnetic-field limit has been set to 5 mT at a lateral distance of 15 m from the beam line [170]. This allows the use of standard iron-based tools at the other detector. The design of the detector return yokes has been verified in simulations for the design fringe fields.

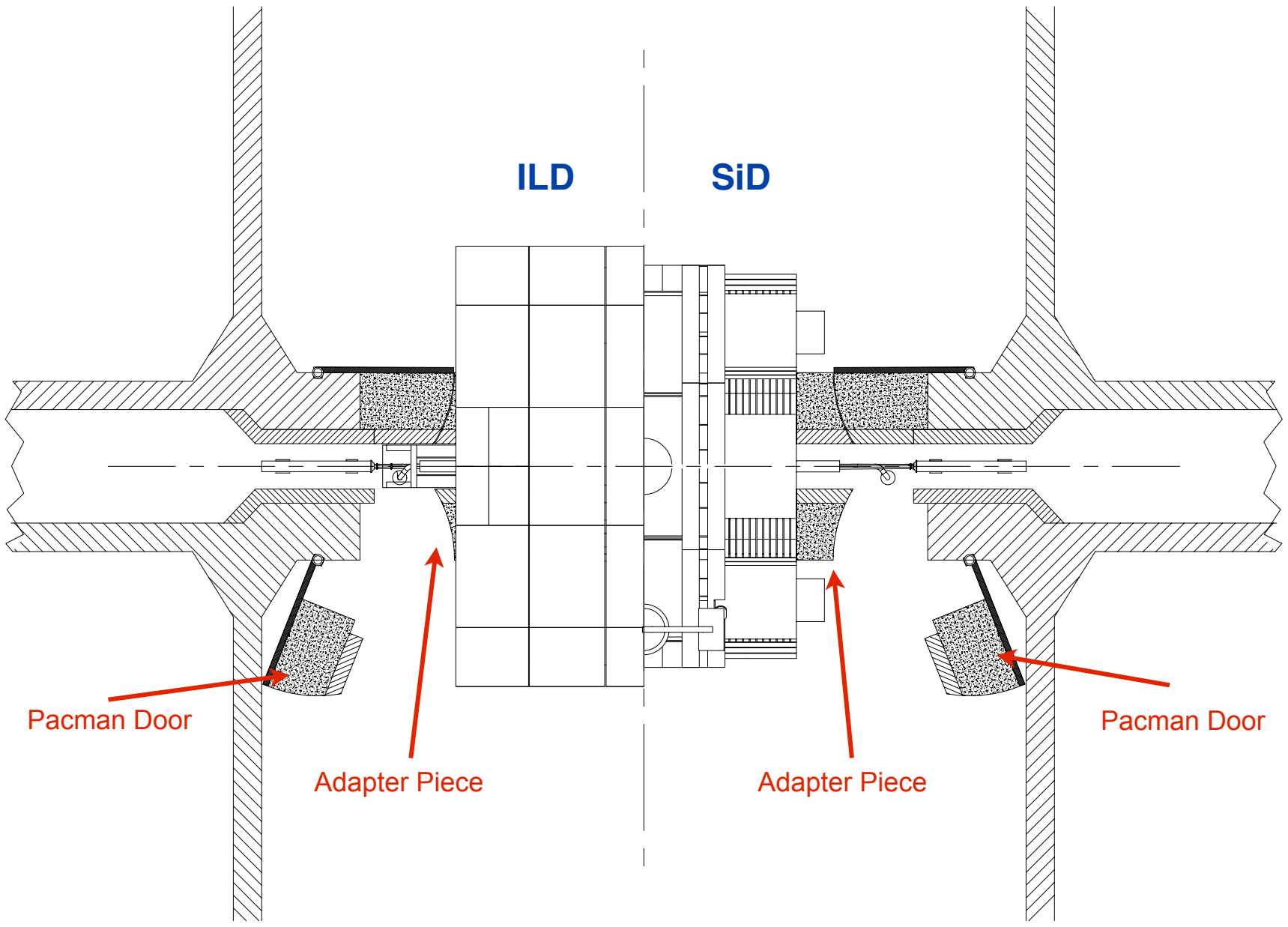

**Figure 8.11**
Design of the beamline shielding compatible with two detector of different size.

## 8.4.6 Detector services

Services are required for the operation and maintenance of both detectors, with different requirements on their proximity to the detector. Primary services are located on the surface above the experimental hall (in the flat-topography sites) or in nearby service caverns (in the mountainous sites). There are usually large and sometimes noisy facilities such as water chillers, high-voltage transformers, auxiliary power supplies, Helium storage and compressors and gas-storage systems. Secondary services are placed in the underground cavern in dedicated service areas. Examples are cooling-water distributions, power supplies, gas-mixture systems, power converters, and parts of the cryogenic system for the detectors. As the detectors are not disconnected during push-pull exchange, all supplies that go directly to the detector are run in flexible cable chains. The detectors carry on-board those services that need short connections, e.g. front-end electronics, patch panels, electronic containers.

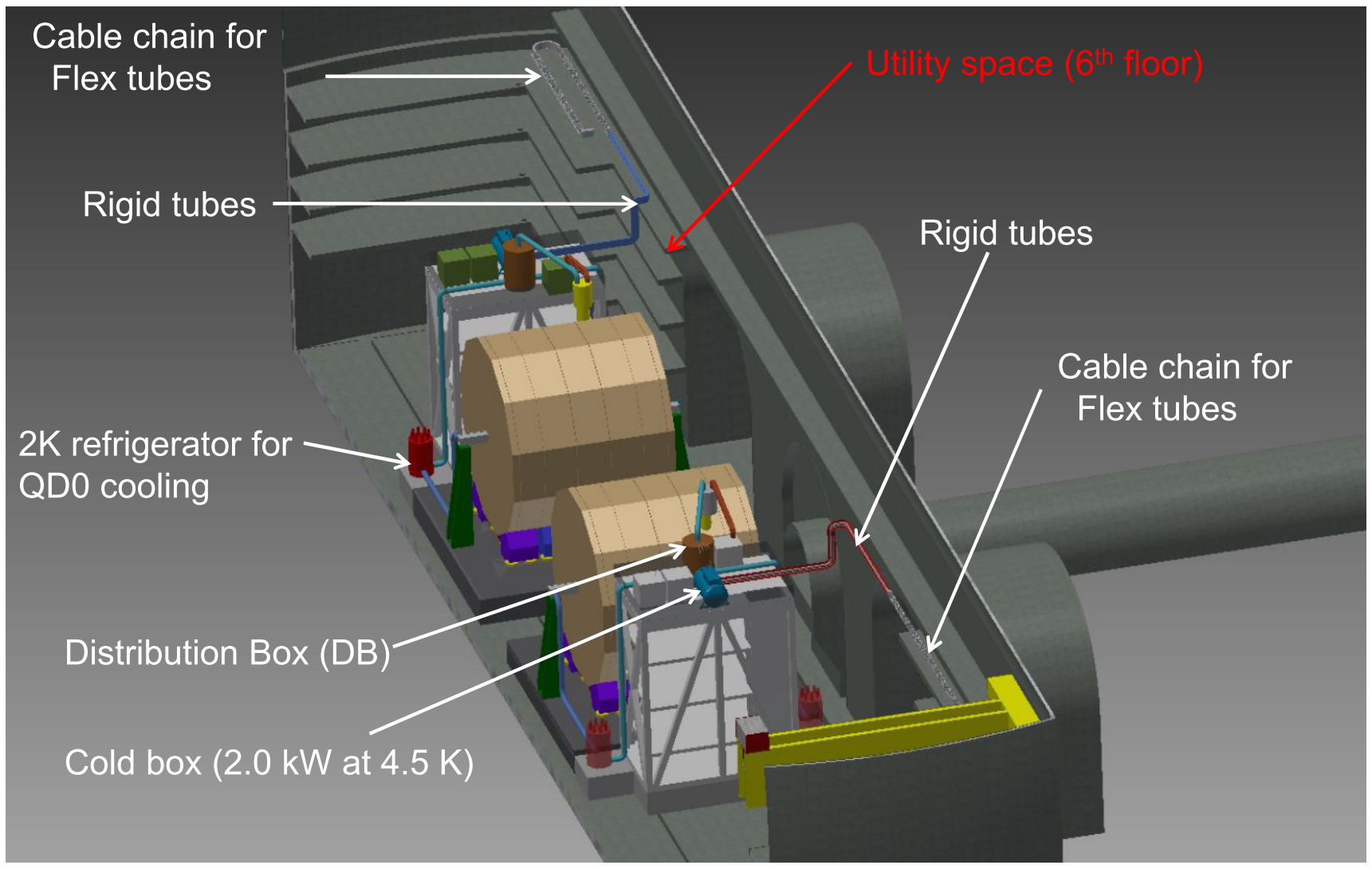

**Figure 8.12**
Common detector cryogenic system (study) with the cold boxes placed on service racks close to the detectors.

Cryogenic Helium for the superconducting solenoids and the QD0 magnets is supplied by a common system for both detectors. Two solutions are currently under study. In one, the liquid He is brought to the detectors via flexible cryogenic lines (see Fig. 8.12), and the cold boxes are in service areas at the cavern walls. The second solution places the cold boxes close to the detectors while gaseous He is supplied via flexible lines to the detector platforms. In each case, a re-cooler is placed on the platform of each detector for the 2 K He supply of the QD0 magnets.

## 8.5 Magnets and power supplies

The BDS has a wide variety of different magnet requirements, and the most distinct magnet types (67) of any ILC area, even though there are only 636 magnets in total. Of these, 86 are superconducting magnets clustered into 4 cryostats close to the IP, as described in Section 8.4.3, and the tail-folding octupoles described below. There are 64 pulsed magnets: 5 types of abort kickers, sweepers and septa. These are used to extract the beams to a fast extraction/tuning dump and to sweep the extracted beam in a 3 cm circle on a dump window.

The remaining 474 magnets are conventional room-temperature magnets, mostly with low-carbon steel cores and water-cooled coils of hollow copper conductor. The bending magnets in the final focus have fields of less than 0.5 kG to minimize synchrotron radiation that would cause beam dilution; they use solid wire coils. The quadrupoles and sextupoles have straight-forward designs adequate for up to 500 GeV beam. The extraction-line magnets have large apertures, e.g. over 90 mm and up to 272 mm, to accommodate the disrupted beam and the photons emerging from the IP. These magnets must fit in alongside the incoming beamline.

The main technical issue with the BDS magnets is their positional stability. All the incoming beam line quadrupoles and sextupoles sit on 5 degree-of-freedom magnet movers with a smallest step size of 50 nm. Data on the relative position of each magnet with respect to the beam is provided by BPMs inserted in the magnet bores so that the magnets can be moved if necessary. The absolute field strength of the BDS magnets has a tight tolerance, requiring power supplies with stability of a few tens of ppm for the tightest tolerances, although most are looser. Magnet-temperature changes lead to strength and position variations so the ambient temperature in the tunnel must be controlled to a relative temperature of about 0.5 °C and the cooling water to within 0.1 °C.

### 8.5.1 Tail-Folding Octupoles

The tail-folding octupoles are the only superconducting magnets in the BDS (other than the FD and extraction quadrupoles) and have the smallest (14 mm ID) clear working aperture in order to reach the highest practical operating gradient. The magnets are energised via NbTi conductor cooled to 4.5 K. With such a small aperture, the beam pipe must have high conductivity to minimise the impact of wakefields. This can be achieved with a beam pipe at 4.5 K made of either aluminium or stainless-steel with a high-conductivity coating. Because these magnets are isolated in the BDS, being far from either the IP or the linac, cryocoolers are used to provide standalone cooling.

## 8.6 Vacuum system

While the aperture of the BDS vacuum chamber is defined by the sizes of the beam, its halo and other constraints, the design of the chambers and vacuum level are governed mainly by two effects: resistive and geometric wakes and the need to preserve the beam emittance; beam-gas scattering and minimisation of detector background.

### 8.6.1 Wakes in vacuum system

The resistive-wall (RW) wakefield of the BDS vacuum system and the geometric wakefield of the transitions in the beam pipe may cause emittance growth due to incoming (transverse) jitter or drift, or due to beam-pipe misalignment. In order to limit these effects to tolerable levels, the inner surface of the BDS vacuum chamber is coated with copper, the vacuum chambers are aligned with an RMS accuracy of $\sim$ 100 µm [178], and incoming beam jitter is limited to 0.5 $\sigma_y$ train-to-train and 0.25 $\sigma_y$ within a train, to limit the emittance growth to 1–2 %.

### 8.6.2 Beam-gas scattering

The specification for the pressure in the BDS beam pipe is driven by detector tolerance to beam-gas-scattering background. Studies have shown that electrons scattered within 200 m of the IP can strike the beam pipe within the detector and produce intolerable backgrounds, while electrons which scatter in the region from 200 to 800 m from the IP are much more likely to hit the protection collimator upstream of the final doublet and produce far less severe detector backgrounds [179]. Based on these studies, the vacuum in the BDS is specified to be 0.1 µPa within 200 m of the IP, 1 µPa from 200 m to 800 m from the IP, and 5 µPa more than 800 m from the IP.

In the extraction lines, the pressure is determined by backgrounds from beam-gas scattering in the Compton polarimeter located about 200 m from the IP. Here the signal rates are large enough that 5 µPa would contribute a negligible background in the detectors.

### 8.6.3 Vacuum-system design

The BDS vacuum is a standard UHV system. The main beam pipes are stainless steel, copper coated to reduce the impedance, with the option of an aluminum-alloy chamber. In locations where there is high synchrotron-radiation (SR) power ($\geq$ 10 kW/m) (e.g. in the chicanes or septa regions), the beam pipe is copper with a water-cooled mask to intercept the photons. The beam pipes are cleaned and baked before installation. There is no *in situ* baking required except possibly for the long drift before the IP.

The required maximum pressure of 5 µPa ($N_2$/CO equivalent) can be achieved by standard ion pumps located at appropriate intervals. The beam pipe near the IP must have pressure below 0.1 µPa for background suppression, and may be baked *in situ* or NEG-coated.

## 8.7 Instrumentation and feedback systems

### 8.7.1 Feedback systems and Stability

Maintaining the stability of the BDS is an essential prerequisite to producing luminosity. Since the beams have RMS vertical sizes of roughly 6 nm at the IP, vertical offsets of about 1 nm will noticeably reduce the luminosity. In addition, especially for parameter sets with higher disruption, the beam-beam interaction is so strong that the luminosity is extremely sensitive to small variations in the longitudinal shape of the bunch caused by short-range wakefields. Finally, the size of the beam at the IP is sensitive to the orbit of the beam through the final-doublet quadrupoles, the sextupoles and other strong optical elements of the BDS. Care must be taken to minimise thermal and mechanical disturbances, by stabilising the air temperature to 0.5 °C and the cooling water to 0.1 °C, and by limiting high-frequency vibrations from local equipment to $\sim$ 100 nm.

Beam-based orbit-feedback loops are used to maintain the size and position of the beam at the IP. All of the feedback loops use beam-position monitors with at least micron-level (and in some cases sub- micron) resolution to detect the beam position and dipole magnets or stripline kickers to deflect the beam. There are two basic forms of feedback in the BDS: train-by-train feedbacks, which operate at the 5 Hz repetition rate of the ILC, and intra-train feedbacks, which can apply a correction to the beam between bunches of a single train.

#### 8.7.1.1 Train-by-train feedback

A train-by-train feedback with five correctors controls the orbit through the sextupoles in the horizontal and vertical planes, where the optical tolerances are tightest. Additional correctors throughout the BDS help reduce long-term beam-size growth. The orbit control feedback can maintain the required beam sizes at the IP over periods from a few hours to several days depending on details of the environment. On longer timescales, IP dispersion and coupling knobs need to be applied.

#### 8.7.1.2 Intra-Train IP position and angle feedbacks

The intra-train feedbacks use the signals detected on early bunches in the train to correct the IP position and angle of subsequent bunches. The offset of the beams at the IP is determined by measuring the deflections from the beam-beam interaction; this interaction is so strong that nanometre-level offsets generate deflections of tens of microradians, and thus BPMs with micron-level resolution can be used to detect offsets at the level of a fraction of a nanometer. Corrections are applied with a stripline kicker located in the incoming beam line between SD0 and QF1. The angle of the beams at the IP is determined by measuring the beam positions at locations 90° out of phase with the IP; at these locations the beam is relatively large so micron resolution is sufficient to measure the beam position (and hence the IP angle) directly to a small fraction of its RMS size. A stripline kicker for the angle correction is located at the entrance to the final focus, causing a latency of about four bunch spacings.

The position-feedback BPM is located near the IP in a region where electromagnetic backgrounds or particle debris from the collisions are a concern. Results from simulations and from a test-beam experiment indicate that backgrounds are an order of magnitude too small to cause a problem [180].

#### 8.7.1.3 Luminosity feedback

Because the luminosity may be extremely sensitive to bunch shape, the maximum luminosity may be achieved when the beams are slightly offset from one another vertically, or with a slight nonzero beam-beam deflection. After the IP position and angle feedbacks have converged, the luminosity feedback varies the position and angle of one beam with respect to the other in small steps to maximize the measured luminosity.

#### 8.7.1.4 Hardware Implementation for intra-train feedbacks

High-bandwidth, low-latency ($\sim$ 5 ns) signal processors for stripline and button BPMs have been tested at the NLCTA, ATF [181] and ATF2 [182]. The feedback processor has been prototyped using fast state-of-the-art FPGAs. A complete system prototype has been demonstrated with a total latency of $\sim$ 140 ns [183].

### 8.7.2 Energy, Luminosity and polarization measurements

#### 8.7.2.1 Energy measurements

Absolute beam-energy measurements are required by the ILC physics program to set the energy scale for particle masses. An absolute accuracy better than 200 ppm is required for the centre-of-mass energy, which implies a requirement of 100 ppm on determination of the absolute beam energy. The intra-train relative variation in bunch energies must be measured with a comparable resolution. Measurements of the disrupted energy spectrum downstream of the IP are also useful to provide direct information about the collision process. It is important that the energy spectrometers be able to make precision energy measurements between 45.6 GeV ($Z$-pole) and the highest ILC energy of 500 GeV. A precise measurement at $Z$-pole energies is of particular importance since it defines the absolute energy scale.

To achieve these requirements, there are two independent and complementary measurements for each beam [158]. About 700 m upstream of the IP, a spectrometer similar to the one employed at LEP-II [184] is capable of making high-precision bunch-to-bunch relative measurements in addition to measuring the absolute beam-energy scale. A four-magnet chicane in the energy-spectrometer region provides a point of dispersion which can be measured using triplets of high-precision RF BPMs. The nominal displacement of the beam is 5 mm and must be measured to a precision of 100 microns. Precision movers keep the beam nearly centred in the BPMs in order to achieve this accuracy.

About 55 m downstream from the IP is a synchrotron radiation spectrometer [185]. A three-magnet chicane in the extraction line, shown in Fig. 8.13, provides the necessary beam deflection, while the trajectory of the beam in the chicane is measured using synchrotron radiation produced in wiggler magnets imaged $\sim 70\,\mathrm{m}$ downstream at a secondary focus near the polarimeter chicane. The synchrotron light produced by the wigglers also comes to a vertical focus at this point, and position-sensitive detectors in this plane arrayed outside the beam pipe measure the vertical separation between bands of synchrotron radiation.

The energy spectrum of the beam after collision contains a long tail as a result of the beam-beam disruption in the collision process. This disrupted beam spectrum is not a direct measure of the collision energy spectrum, but it is produced by the same physical process, and direct observation of this disrupted tail serves as a useful diagnostic of the collision process. The position-sensitive detector in the spectrometer is designed to measure this beam-energy spectrum down to 50 % of the nominal beam energy.

### 8.7.2.2 Luminosity measurements

The ILC luminosity can be measured with a precision of $10^{-3}$ or better by measuring the Bhabha rate in the polar-angle region from 30–90 mrad. Two detectors are located just in front of the final doublets. The LumiCal covers the range from 30–90 mrad and the BeamCal covers the range from 5–30 mrad. At 500 GeV centre-of-mass energy, the expected rate in the LumiCal region is $\sim 10$ Bhabhas per bunch train, which is too low to permit its use as an intra-train diagnostic for tuning and feedback. At smaller polar angles of 5-30 mrad the rate or energy deposition of beamstrahlung $\mathrm{e}^+\mathrm{e}^-$ pairs can be measured and provides a fast luminosity diagnostic. The expected rate in this region is 15 000 pairs (and 50 TeV energy deposition) per bunch crossing. Furthermore, the spatial distributions of pairs in this region can be used to determine beam-collision parameters such as transverse sizes and bunch lengths.

### 8.7.2.3 Polarisation measurements

Precise polarimetry with 0.25 % accuracy is needed to achieve the ILC physics goals [186]. Compton polarimeters [158] are located both $\sim$ 1800 m upstream of the IP, as shown in Fig. 8.1, and downstream of the IP, as shown in Fig. 8.13, to achieve the best accuracy for polarimetry and to aid in the alignment of the spin vector.

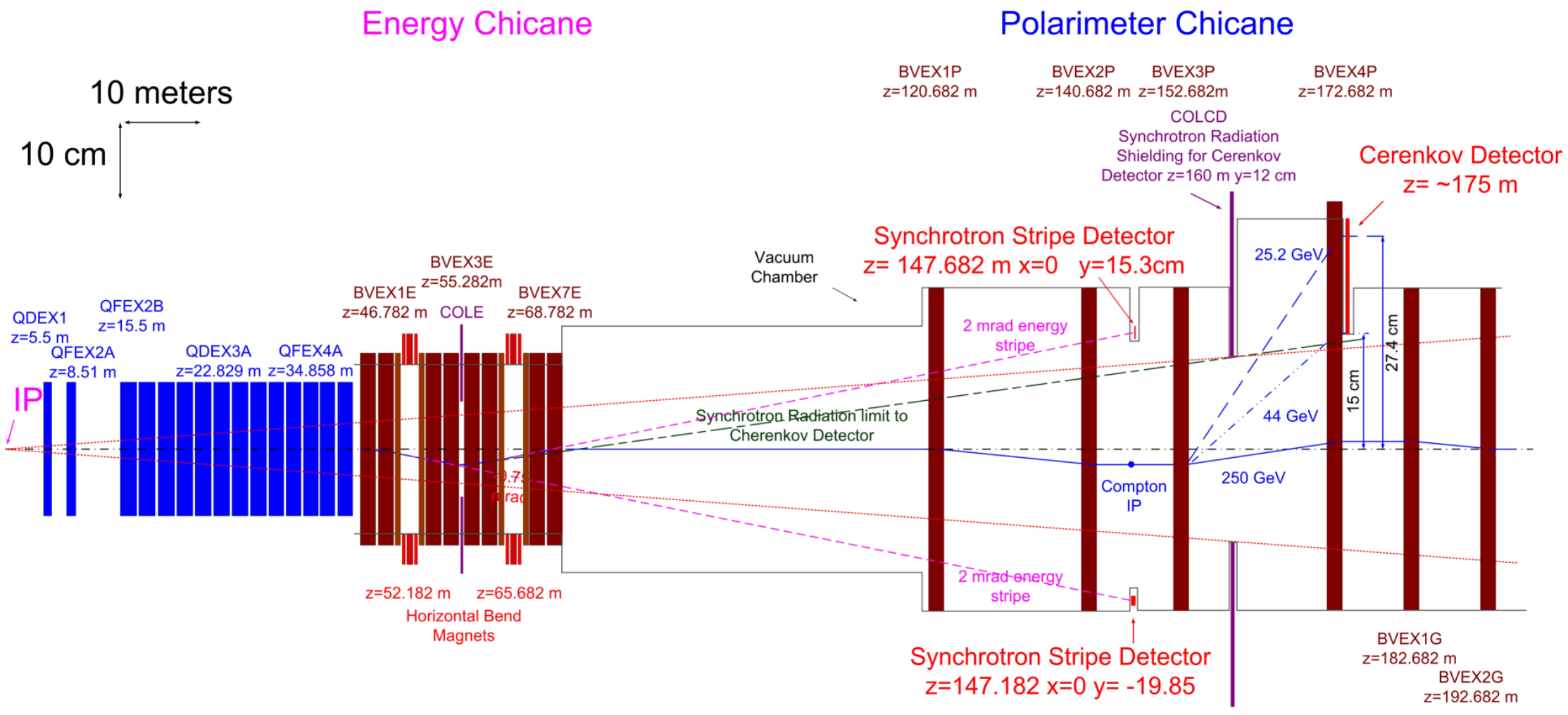

**Figure 8.13.** Schematics of energy and polarimeter chicanes in the 14 mrad extraction line, shown in a configuration with two additional bends at the end. Longitudinal distances are given from the IP. Also shown is the 0.75 mrad beam stay-clear from the IP.

The upstream polarimeter measures the undisturbed beam before collisions. It consists of a dedicated 4-bend horizontal chicane with the Laser-Compton IP in the middle and a detector for the Compton-scattered electrons at the end, as shown in Fig. 8.14. The length of the chicane is chosen such that the total emittance growth due to synchrotron radiation stays below 1%, even at the highest beam energy of 500 GeV. The relatively clean environment allows a laser system that measures every single bunch in the train and a large lever arm in analysing power for a multi-channel detector, which facilitates internal systematic checks. The good field region of the individual dipoles is wide enough to accommodate all beam energies from 500 GeV down to 45.6 GeV.

**Figure 8.14**
Schematics of upstream polarimeter chicane.

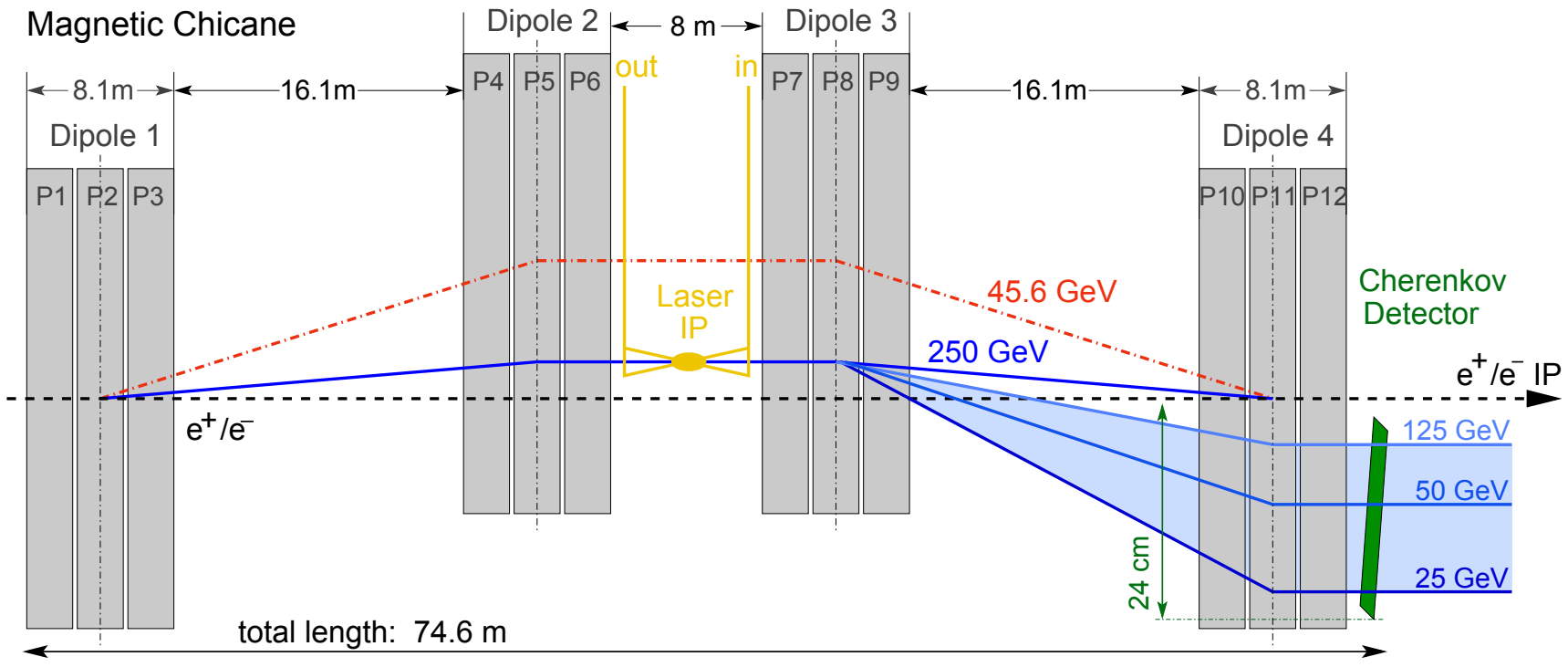

The downstream polarimeter measures the polarisation of the outgoing beam after collision. The estimated average depolarisation for colliding beams is 0.3 %, and for the outgoing beam 1 %. A schematic drawing of the extraction line is shown in Fig. 8.13. In the high background environment of the disrupted beam, the required high laser power allows measurement of only a few bunches out of each train. The chicane of the downstream polarimeter consists of six vertical bends to maximise the analysing power and to deflect the Compton-scattered electrons out of the synchrotron-radiation fan [187].

Both polarimeters are designed to meet the physics requirements at all energies from the $Z$ pole to the full energy of the ILC.

## 8.7.3 Diagnostic and Correction devices

Each quadrupole, sextupole, and octupole magnet in the incoming BDS beam lines is placed on a 5 degree-of-freedom mover, and has an associated BPM. There are also several tens of correctors in the incoming beam lines for 5 Hz feedback, and in the extraction lines, where there are no movers. The BPMs in the incoming beam line are RF cavities, either S, C or L-band, depending on the beam line aperture. Long chains of bends or kickers have sparsely placed BPMs. BPMs in the extraction lines are button or strip-line design.

Additional instrumentation in the BDS includes a deflecting cavity to measure vertical-time correlation, ion-chamber and PMT loss monitors, transverse profile monitors for horizontal synchrotron light, OTR monitors, current monitors, pickup phase monitors, etc.

## 8.8 Beam dumps and Collimators

### 8.8.1 Main Dumps

The beam-delivery system contains two tune-up dumps and two main beam dumps. These four dumps are all designed for a peak beam power at nominal parameters of 18 MW at 500 GeV per beam, which is also adequate for the 14 MW beam power of the 1 TeV upgrade. The dumps consist of 1.8 m-diameter cylindrical stainless-steel high-pressure (10 bar) water vessels with a 30 cm diameter, 1 mm-thick Ti window and also include their shielding and associated water systems (Fig. 8.15). The design [188] is based on the SLAC 2.2 MW water dump [189, 190].

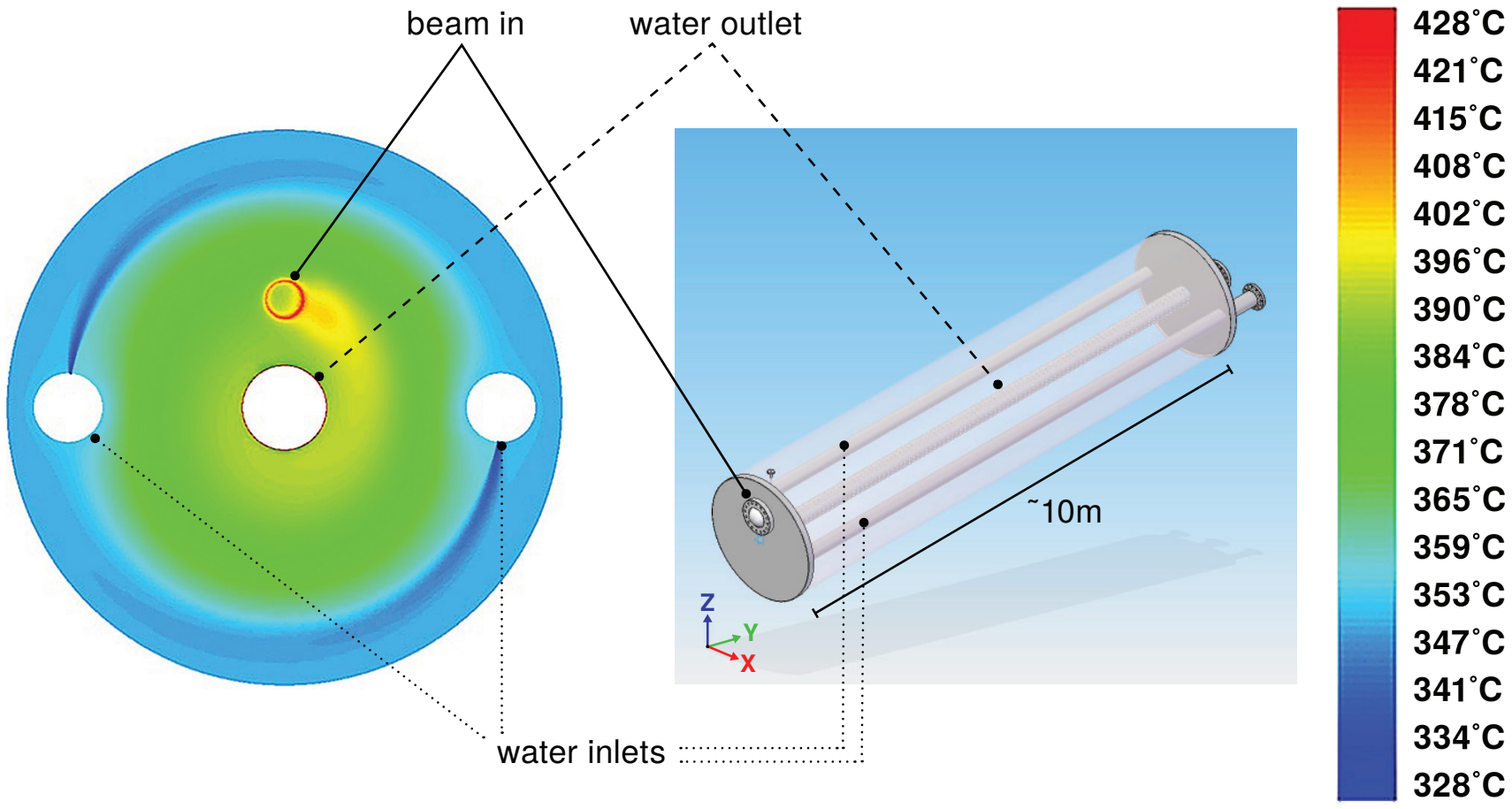

**Figure 8.15** Temperature distribution at the shower maximum of the beam in the main 18 MW dump just after passage of the beam train (left). (The geometry of the dump is also shown on the right.) The colour bar shows temperature in kelvin; the maximum temperature is 155 °C [191].

The dumps absorb the energy of the electromagnetic shower cascade in 11 m $(30\,X_0)$ of water. Each dump incorporates a beam-sweeping magnet system to move the charged beam spot in a circular arc of 6 cm radius during the passage of the 1 ms-long bunch train. Each dump operates at 10 bar pressure and also incorporates a vortex-flow system to keep the water moving across the beam. In normal operation with 500 GeV beam energy, the combination of the water velocity and the beam sweepers limits the water temperature rise during a bunch train to 155 °C [191]. The pressurisation raises the boiling temperature of the dump water; in the event of a failure of the sweeper, the dump can absorb up to 250 bunches without boiling the dump water.

The integrity of the dump window, the processing of the radiolytically evolved hydrogen and oxygen, and containment of the activated water are important issues for the full-power dumps. The dump service caverns include three-loop pump-driven 145 L/s heat-exchanger systems, devices to remotely exchange dump windows during periodic maintenance, catalytic $H_2$-$O_2$ recombiners, mixed-bed ion-exchange columns for filtering of $^7$Be, and sufficient storage to house the volume of tritiated water during maintenance operations.

#### 8.8.1.1 Ensuring the integrity of the dump and dump window

The main vessel is welded using low-carbon stainless steel (316L) and all welds are radiographed to ensure quality; the 10 bar radioactive-water cooling system is closed but communicates with the atmosphere via a small diameter tube from the gas space on top of the surge tank to avoid it being classified as a nuclear pressure vessel. Several materials are under consideration for use in the dump window: 316L stainless, Ti-6Al-4V, and Inconel (A601,718,X750). All of these materials have been extensively used in nuclear reactors; their mechanical properties, thermal properties, and reaction to radiation damage have been thoroughly studied. As described above, the bunches in each train are swept in a circle to reduce further the thermal stress and radiation damage to the dump windows; the windows also have additional water cooling from multiple water jets in a separate cooling loop from the main vessel. Each dump incorporates a remote-controlled mechanism for exchanging the highly activated windows on a regular schedule driven by integrated specific dose, along with local temporary storage for all tritiated water. As a final backup to guarantee environmental safety in the event of a failure of the dump body or dump window, the dump enclosure is air tight and incorporates adequate sump volume and air drying capacity to prevent the release of tritiated water even in the case of catastrophic dump failure. Since a failure of the window could create a catastrophic water-to-vacuum leak with highly radioactive tritiated water, a pre-window, with peripheral and gas cooling, isolates the beamline vacuum system and provides secondary containment. Storage space for a damaged dump and a removable cavern wall are provided for dump replacement.

#### 8.8.1.2 Mitigation of water-activation products

Activation products are primarily the result of photo-spallation on $^{16}$O, primarily $^{15}$O, $^{13}$N, $^{11}$C, $^{7}$Be and $^{3}$H (tritium). The first three radionuclides have short half lives and decay after $\sim$ 3 hours. The $^{7}$Be is removed from the system by filtering it out in a mixed-bed ion-exchange column located in the dump-support cavern. Tritium, a $\sim$ 20 keV $\beta$ emitter with a half life of 12.3 years, builds up in the water to some equilibrium level; the tritium is contained by the integrity of the dump system and the backup measures described in the preceding section.

#### 8.8.1.3 Radiolysis and hydrogen and oxygen evolution

Hydrogen is produced via the reaction $H_2O \rightarrow H_2 + H_2O_2$ at the rate of 0.3 L/ MW s, or 5.4 L/s at 18 MW beam power. The lower explosive limit (LEL) of hydrogen in air is $\sim$ 4 %. Experience at SLAC [192] indicates that a catalyst consisting of a high-nickel stainless-steel ribbon coated with platinum and palladium, in the form of a 46 cm diameter 6.4 cm-thick mat, will reduce the $H_2$ concentration to 25 % of the LEL in one pass. Other types of higher-density catalyst are also available. The gases released in a surge tank are heated to 65 °C and are pumped through the catalyst, which does not need replacement or servicing.

#### 8.8.1.4 Shielding and protection of site ground water

Assuming a dry rock site, as in the baseline configuration, 50 cm of iron and 150 cm of concrete shielding are needed between the dump and other areas of the tunnel enclosure to protect equipment from radiation damage. If the chosen site is not dry, the area surrounding the dump must be enveloped by an additional 2 m-thick envelope of concrete to prevent tritium production in the ground water.

### 8.8.2 Collimators

The beam-delivery system contains 32 variable-aperture collimators and 32 fixed aperture collimators. The devices with the smallest apertures are the 12 adjustable spoilers in the collimation system. To limit their impedance to acceptable levels, these 0.6–1.0 $X_0$ Ti spoilers have longitudinal Be tapers. Figure 8.16 shows a collimator design suitable for the ILC [160, 193].

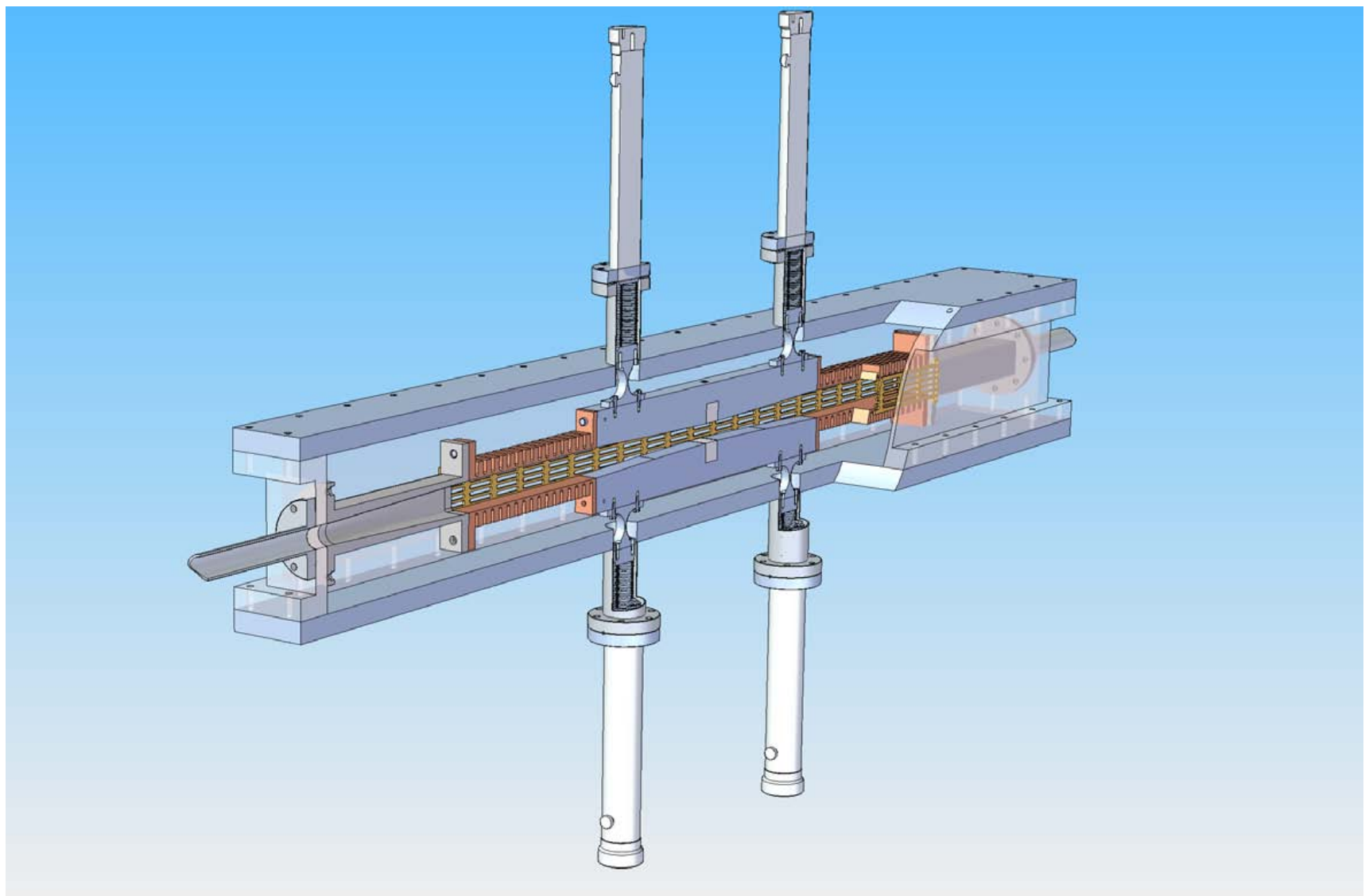

**Figure 8.16**
Tentative spoiler candidate design [160, 193].

## 8.9 Crab cavity system

Crab cavities are required to rotate the bunches from a 14 mrad crossing angle so they collide head on. Two 3.9 GHz SC 9-cell cavities in a 2–3 m long cryomodule are located 13.4 m from the IP. The cavities are based on the Fermilab design for a 3.9 GHz $\mathrm{TM}_{110}$ $\pi$-mode 13-cell cavity [194, 195]. The ILC has two 9-cell versions (see Fig. 8.17) of this design operated at 5 MV/m peak deflection. This provides enough rotation for a 500 GeV beam and 100 % redundancy for a 250 GeV beam [196, 197].

The most challenging specification of the crab-cavity system is on the uncorrelated phase jitter between the incoming positron and electron cavities which must be controlled to 61 fsec to maintain optimised collisions [199]. A proof-of-principle test of a 7-cell 1.5 GHz cavity at the JLab ERL facility [200] has achieved a 37 fsec level of control, demonstrating feasibility. The higher- and lower-order modes of the cavity must be damped effectively to limit unwanted vertical deflections at the IP, as must the vertical polarization of the main deflecting mode.

Couplers with lower $Q_{\mathrm{ext}}$ and greater power-handling capability are required to handle beam loading and LLRF feedback for off-axis beams. The crab cavity needs $\sim$ 3 kW per cavity for about 10 msec, with a $Q_{\mathrm{ext}}$ of $\sim 10^6$ [196–198, 201]. The crab cavity is placed in a cryostat with tuner, $x - y$ and roll adjustment which provides proper mechanical stability and microphonic rejection. The cryostat also accommodates the beam pipe of the extraction line which passes about 19 cm from the centre of the cavity axis.

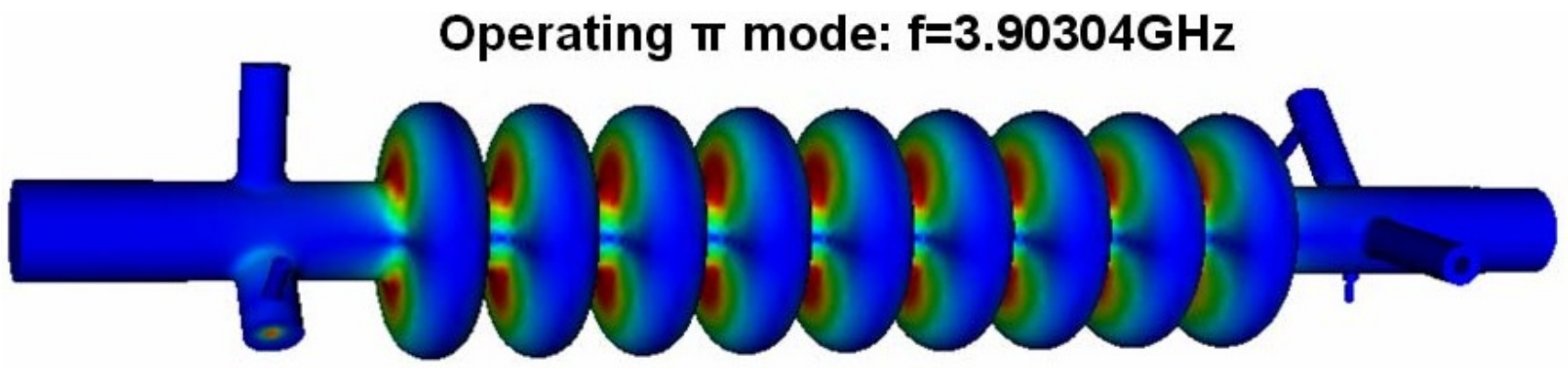

**Figure 8.17**
Field distribution for the operating mode of the 3.9 GHz crab cavity [198].

## 8.10 Accelerator Components

The total counts of the BDS accelerator components are summarized in Table 8.3.

**Table 8.3**
BDS components, total counts.

| Magnets | | Instrumentation | | Dumps & Collimators | |
|---|---|---|---|---|---|
| Warm dipoles | 190 | BPMs C-band | 262 | Full power dumps | 4 |
| Warm quads | 204 | BPMs L-band | 42 | Insertable dumps | 2 |
| Warm sextupoles | 10 | BPMs S-band | 14 | Adjustable collim. | 32 |
| Warm octupoles | 4 | BPMs stripline/button | 120 | Fixed apert. collim. | 32 |
| SC quads | 32 | Laser wire | 8 | Stoppers | 14 |
| SC sextupoles | 12 | SR transv. profile imager | 10 | | |
| SC octupoles | 14 | OTR screens | 2 | Vacuum | |
| Muon spoilers | 2 | Crab & deflection cavities | 4 | Pumps | 3150 |
| | | Loss monit. (ion chamb., PMT) | 110 | Gauges | 28 |
| Warm correctors | 64 | Current monitors | 10 | Gate valves | 30 |
| SC correctors | 36 | Pick-up phase monitors | 2 | T-connections | 10 |
| Kickers/septa | 64 | Polarimeter lasers | 3 | Switches | 30 |

# Chapter 9
# Global Accelerator Control Systems

## 9.1 Overview

Rapid advances in electronics and computing technology in recent decades have had a profound effect on the performance and implementation of accelerator control systems. These advances will continue through the time of ILC construction, when network and computing capabilities will far surpass that of equipment available today. Nevertheless, a machine of the scope of an ILC presents some unique control system challenges independent of technology, and it is important to set out functional requirements for the ILC control system.

This section discusses the control-system requirements for the ILC, and describes a functional and physical model for the system. In several places implementation details are described, but this has been done largely as a means to describe representative technologies, and in particular, to establish a costing model. Regardless of the final technology implementation, the control system model described in this chapter contains a number of architectural choices that are likely to survive.

## 9.2 Requirements and Technical Challenges

The broad-scope functional requirements of the ILC control system are largely similar to those of other modern accelerator control systems, including control and monitoring of accelerator technical systems, remote diagnostics, troubleshooting, data archiving, machine configuration, and timing and synchronisation. However, several features of the ILC accelerator push implementation beyond the present state of the art. These are described below.

### 9.2.1 Scalability

The ILC has an order of magnitude more technical system devices than other accelerators to date. The primary challenges of scalability in relation to existing accelerator control systems are the physical distances across the accelerator, the large number of components and number of network connections, and the implied network bandwidth. Real-time access to control-system parameters must be available throughout the site, and by remote access. These challenges are also present in the commercial domain, notably in telecommunication applications, and lessons learned there are almost certainly applicable to the ILC control system.

### 9.2.2 High Availability

Requirements for high availability drive many aspects of the ILC control system design and implementation. These requirements were derived from accelerator-wide availability simulations. The control system as a whole is allocated a 2500 hour MTBF and 5 hour MTTR (15 hours downtime per year). This translates to control system availability between 99% and 99.9% (2-nines and 3-nines). A detailed analysis of how control system availability relates to beam availability is complicated. However, a coarse analysis shows that if the control system comprises some 1200 controls shelves (electronics crates), then each shelf must be capable of providing 99.999% (5-nines) availability. Such availability is routinely implemented in modern telecom switches and computer servers, but has not been a requirement of present accelerator control systems.

### 9.2.3 Support extensive automation and beam-based feedback

A very complex series of operations is required to produce the beams and deliver them to the collision point with the required emittance. The control system must provide functionality to automate this process. This includes both getting beam through the entire chain and also tune-up procedures to maximise the luminosity. Beam-based feedback loops are required to compensate for instabilities and time-dependent drifts in order to maintain stable performance. Inter-pulse feedback should be supported in the control-system architecture to minimise development of custom hardware and communication links. The automation architecture should have some built-in flexibility so procedures can easily be changed and feedback loops added or modified as needed. Automation and feedback procedures should incorporate online accelerator models where appropriate.

### 9.2.4 Synchronous Control-System Operation

The ILC is a pulsed machine operating at a nominal rate of 5 Hz. Sequences of timing events must be distributed throughout the complex to trigger various devices to get beam through the accelerator chain. These events are also used to trigger acquisition of beam instrumentation and other hardware diagnostic information so that all data across the machine can be properly correlated for each pulse.

### 9.2.5 Precision RF-Phase-Reference Distribution

The control system must generate and distribute RF phase references and timing fiducials with stability and precision consistent with the RF system requirements.

### 9.2.6 Standards and Standardisation, Quality Assurance

A critical aspect of implementing a high-availability control system is the use of consistent ("best") work practices and a level of quality assurance process that is unprecedented in the accelerator-controls environment. Additional technical solutions to high availability will rely on this foundation of work practices and quality-assurance processes. Commercial standards should be used wherever they can meet the requirements, for such things as hardware packaging and communication networks.

The control system must specify standard interfaces between internal components and to all other systems. This makes integration, testing, and software development easier and more reliable. Standard interfaces allow parts of the system to be more easily upgraded if required for either improved performance or to replace obsolete technologies.

### 9.2.7 Requirements on Technical Equipment

Technical equipment comprises field hardware such as power-supply controllers, vacuum equipment, beam instrumentation, and motion-control devices. These systems are the responsibility of the technical groups. However, they must interface to the control system in a coherent way to allow equipment to be accessed via a common interface for application programming, data archiving, and alarms. In order to meet the very stringent requirements for overall system reliability, as well as provide for more efficient R&D and long-term maintenance, standards must be applied to the technical equipment for packaging, field bus, communication protocol, cabling, and power distribution.

### 9.2.8 Diagnostic Interlock Layer

A Diagnostic Interlock Layer (DIL) complements normal self-protection mechanisms built into technical equipment. The DIL utilises information from diagnostic functions within the technical equipment to monitor the health of the equipment and identify anomalous behaviour indicative of impending problems. Where possible, corrective action is taken, such as pre-emptive load balancing with redundant spares, to avert or postpone the fault before internal protective mechanisms trip off the equipment.

## 9.3 Impact of Requirements on the Control-System Model

In order to meet the high-availability requirements of the ILC, a rigorous failure-mode analysis must be carried out in order to identify the significant contributors to control system downtime. Once identified, many well-known techniques can be brought to bear at different levels in the system, as well as system wide, and at different time scales (i.e. bunch-to-bunch, macro pulse, process control) to increase availability. The techniques begin with relatively straightforward, inexpensive practices that can have a substantial impact on availability. A careful evaluation and selection of individual components such as connectors, processors, and chassis are crucial. Administrative practices such as QA, agile development methodology, and strict configuration management must also be applied. Other techniques are much more complex and expensive, such as component redundancy with automatic detection and failover [202]. The control system must be based on new standards for next-generation instrumentation that:

1. are modular in both hardware and software for ease in repair and upgrade;
2. include inherent redundancy at internal module, module assembly, and system levels;
3. include modern high-speed, serial, inter-module communications with robust noise-immune protocols;
4. include highly intelligent diagnostics and board-management subsystems that can predict impending failure and invoke evasive strategies.

The Control System Model incorporates these principles through the selection of the front-end electronics packaging standard and component redundancy.

In addition to its intrinsic availability, the control system is responsible at the system level for adapting to failures in other technical systems. For example, the feedback system is responsible for reconfiguring a response matrix due to the loss of a corrector, or switching on a spare RF unit to replace a failed station.

Scalability requirements are met through a multi-tier hierarchy of network switches that allow for the flexible formation of virtual local area networks (VLANs) as necessary to segment network traffic. Control system name-servers and gateways are utilised extensively to minimise broadcast traffic and network connections. These software components manage the otherwise exponential growth of connections when many clients must communicate with many distributed control points.

Automation and flexible pulse-to-pulse feedback algorithms are implemented by a coordinated set of software services that work together through global coordination and distributed execution. The distributed execution is synchronised with the machine pulse rate via the timing event system which can produce software interrupts where needed. The network backbone accommodates the distribution of any sensor value to any feedback computation node. This distribution can be optimised to allow for efficient local as well as global feedback.

## 9.4 Control System Model

The model of the ILC control system is presented here from both functional and physical perspectives. This model has served as a basis for the cost estimate, as well as to document that the control-system requirements have been satisfied. Functionally, the control-system architecture is separated into three tiers, as shown in Fig. 9.1. Communication within and between these tiers is provided by a set of network functions. A physical realization, as applied to the Main Linac, is shown in Fig. 9.2. The remainder of the section describes the functional and physical models in more detail.

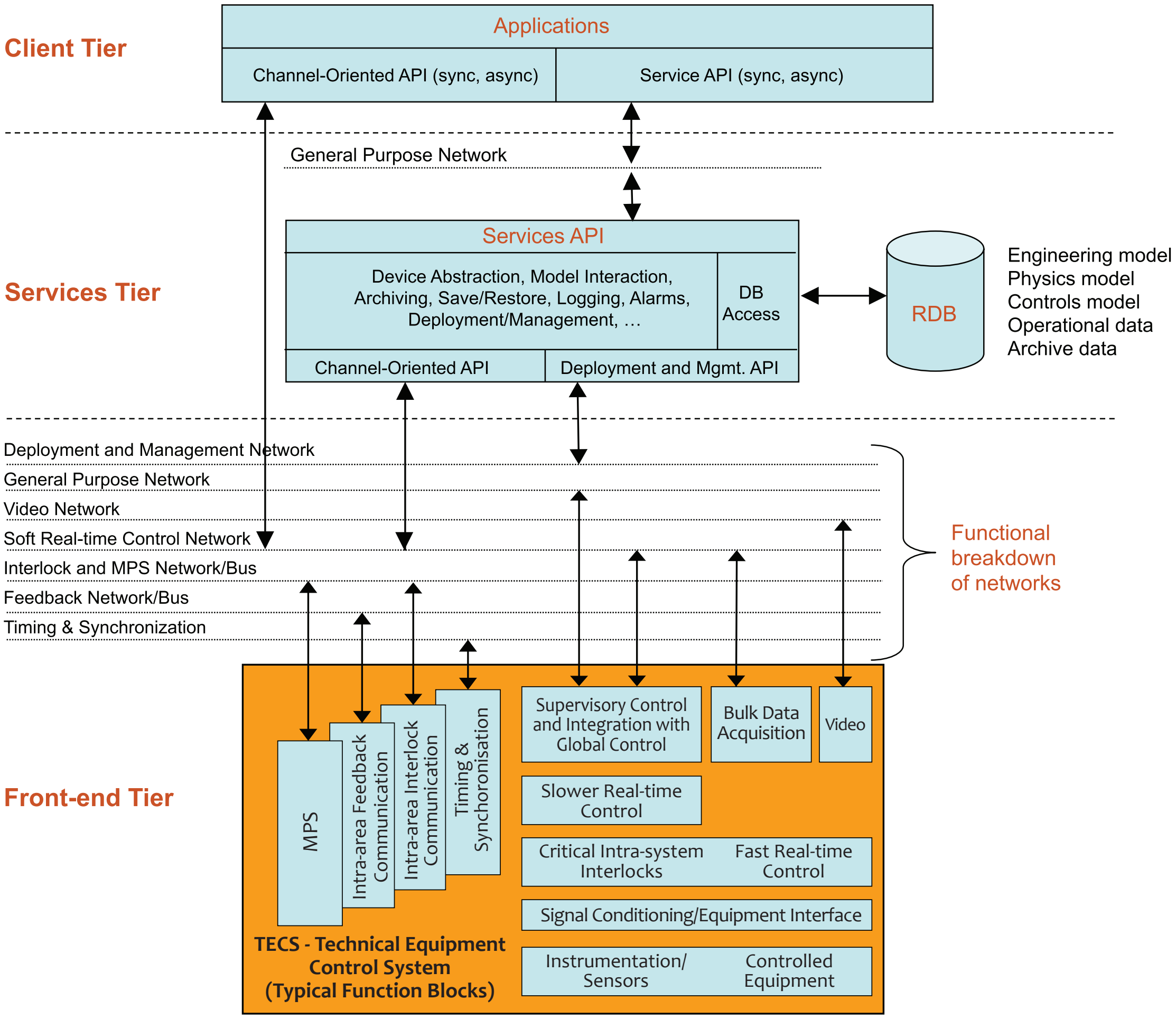

**Figure 9.1.** Control-system functional model.

### 9.4.1 Functional Model

The control-system model is functionally composed of three distinct tiers, as shown in Fig. 9.1. The 3-tier model includes a middle tier that implements significant portions of the logic functionality through software services that would otherwise reside in the client tier of a 2-tier system [203]. The three tiers are described in more detail below:

**Client Tier**: Provides applications with which people directly interact. Applications range from engineering-oriented control consoles to high-level physics control applications to system configuration-management applications. Engineer-oriented consoles are focused on the operation of the underlying accelerator equipment. High-level physics applications require a blend of services that combine data from the front-end tier and supporting data from the relational database in the context of high-level device abstractions (e.g. magnets, BPMs).

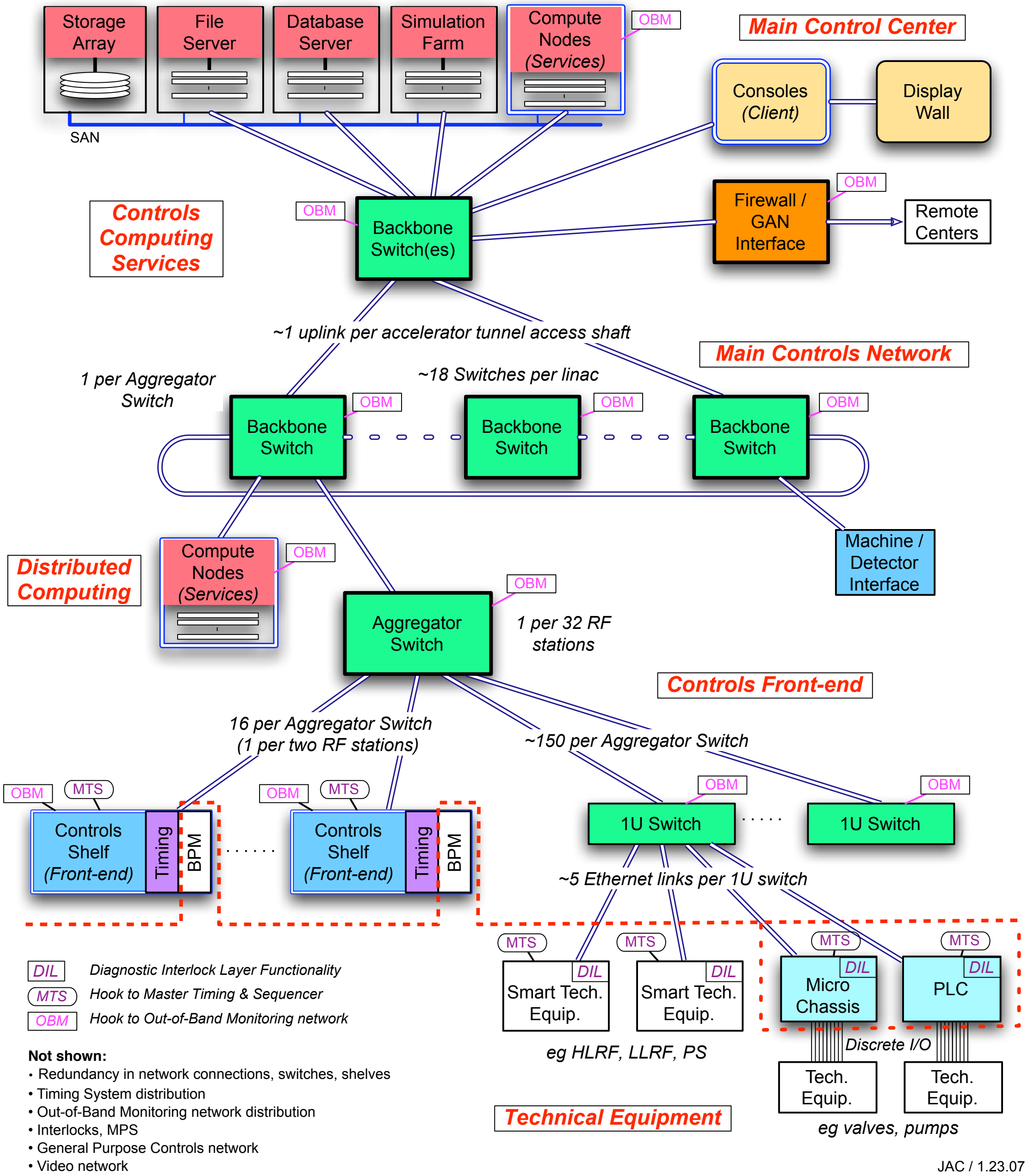

**Figure 9.2.** Control-system physical model.

**Services Tier**: Provides services that coordinate many activities while providing a well-defined set of public interfaces (non-graphical). Device abstractions such as magnets and BPMs that incorporate engineering, physics, and control models are represented in this tier. This makes it possible to relate high-level machine parameters with low-level equipment settings in a standard way. For example, a parameter save/restore service can prevent two clients from simultaneously attempting to restore a common subset of operational parameters. This centralisation of control provides many benefits in terms of coordination, conflict avoidance, security, and optimisation.

**Front-end Tier**: Provides access to the field I/O and underlying dedicated fast feedback systems. This tier is configured and managed by the services tier, but can run autonomously. For example, the services tier may configure a feedback loop in the front-end tier, but the loop itself runs without direct involvement. The primary abstraction in this tier is a channel, or process variable, roughly equivalent to a single I/O point.

### 9.4.2 Physical Model

The ILC control system must reliably interact with more than 100,000 technical system devices that could collectively amount to several million scalar and vector Process Variables (PVs) distributed across the many kilometres of beam lines and facilities at the ILC site. Information must be processed and distributed on a variety of timescales from microseconds to several seconds. The overall philosophy is to develop an architecture that can meet the requirements, while leveraging the cost savings and rapid evolutionary advancements of commercial off-the-shelf (COTS) components.

#### 9.4.2.1 Main Control Centre

The accelerator control room contains consoles, servers, displays, and associated equipment to support operations of the ILC accelerator from a single location. Operators and technical staff run the accelerator and interact with technical equipment through Client Tier applications that run in the Main Control Centre.

#### 9.4.2.2 Controls Computing Services

Conventional computing services dedicated to the control system include storage arrays, file servers, and compute nodes. A separate simulation farm is anticipated for offline control-system modelling and simulation, and for potentially performing model-reference comparisons to dynamically detect unusual conditions. Enterprise-grade relational databases act as a central repository for machine-oriented data such as physics parameters, device descriptions, control system settings, machine model, installed components, signal lists, and their relationships with one another.

#### 9.4.2.3 Controls Networks and Distributed Computing

9.4.2.3.1 Main Controls Network Data collection, issuing and acting on setpoints, and pulse-to-pulse feedback algorithms are all synchronised to the pulse repetition rate. The controls network must therefore be designed to ensure adequate response and determinism to support this pulse-to-pulse synchronous operation, which in turn requires prescribing compliance criteria for any device attached to this network. Additionally, large data sources must be prudently managed to avoid network saturation.

For example, in the Main Linac, waveform capture from the LLRF systems is likely to dominate linac network traffic. Full-bandwidth raw waveforms from individual RF stations could be required for post-event analysis and therefore must be captured on every pulse. However, only summary data is required for archiving and performance verification. By grouping multiple RF stations together (notionally into groups of 32), full-bandwidth waveforms can be locally captured and temporarily stored, with only summary data sent on.

Dedicated compute nodes associated with each backbone network switch run localised control-system services for monitoring, data reduction, and implementing feedback algorithms.

9.4.2.3.2 Other Physical Networks To accommodate communication functions that are not compatible with the Main Controls Network, several other physical networks are envisioned, namely: a *General-purpose controls network* for general controls network access, including wireless access and controls network access to non-compliant devices; an *Out-of-band monitoring network*: to provide independent means to access and configure all Network switches and Controls Shelves; a *Video network* to distribute video data streams facility-wide. A *Technical Equipment Interlock Network* provides a means to distribute interlock signals. Functionally, this has similarities with the Machine Protection System described elsewhere. Technical equipment may report equipment or sensor status for use by other systems or utilise status information provided by other technical systems.

Based on initial assessments, commodity-computing equipment (e.g. 10-GB redundant Ethernet) is adequate to meet the requirements for all the networks.

#### 9.4.2.4 Controls Front-end

The control-system model front-end comprises the following three main elements:

**1U Switch**: Aggregates the many Ethernet-controlled devices in a rack or neighbourhood of racks. Some of these devices speak the controls protocol natively, while others have proprietary protocols that must be interfaced to the control system. It is assumed these 1U switches reside in many of the technical equipment racks.

**Controls Shelf**: Consists of an electronics chassis, power supplies, shelf manager, backplane switch cards, CPUs, timing cards, and instrumentation cards (mainly BPMs). The Controls Shelf serves several purposes: (1) to host the protocol gateways, reverse gateways, and name servers to manage the connections required for clients to acquire controls data; (2) to run the core control system software for managing the various Ethernet device communication protocols, including managing any instrumentation (BPM) cards in the same shelf; (3) to perform data reduction, for example, so that full-bandwidth RF/BPM waveforms need not be sent northbound in the control system. The control-system physical model references the commercial standard AdvancedTCA (ATCA) for the Controls Shelves. This is a specification that has been developed for the telecommunications industry [204], and has applicability for the ILC control-system in part because of its high-availability feature set.

**Aggregation Switch**: Aggregates network connections from the 1U switches and Controls shelves and allows flexible formation of virtual local-area networks (VLANs) as needed.

#### 9.4.2.5 Technical Equipment Interface

It has been common practice at accelerator facilities for the control system to accommodate a wide variety of interfaces and protocols, leaving the choice of interface largely up to the technical system groups. The large scale of the ILC accelerator facility means that following this same approach would almost certainly make the controls task unmanageable, so the approach must be to specify a limited number of interface options. For the purpose of the conceptual design and for the costing exercise, two interface standards were chosen: a Controls-shelf compliant electronics module for special sensor signals and specific beam-instrumentation applications such as BPM electronics; a controls compliant redundant network for all *smart* technical systems. While not explicitly part of the control-system model, it is assumed that discrete analog and digital I/O can be provided through micro-controller chassis or PLCs.

In addition to conventional interfaces for controls purposes, the control system provides functionality for remote configuration management of technical equipment for micro- controllers, PLCs, application-oriented FPGAs, etc.

### 9.4.3 Pulse-to-Pulse (5 Hz) Feedback Architecture

Many of the beam-based feedback algorithms required for ILC apply corrections at the relatively low machine pulse rate (nominally 5 Hz). This low correction rate and the distributed nature of many of the monitors and actuators make it desirable to use the integrated controls infrastructure for these feedback systems.

Using the integrated control-system architecture to implement the feedback algorithms offers many advantages, including:

- simpler implementation, since dedicated interfaces are not required for equipment involved in feedback loops;
- higher equipment reliability, since there are fewer components and interfaces;

- greater flexibility, since all equipment is inherently available for feedback control, rather than limited to predefined equipment;
- simplified addition of ad-hoc or un-anticipated feedback loops with the same inherent functionality and tools. This could significantly enhance the commissioning process and operation of the ILC.

Referring to Fig. 9.2, feedback algorithms are implemented as services running in both distributed and centralised compute nodes. Design and implementation of feedback algorithms is enhanced through high-level applications such as Matlab [205] integrated into the Services Tier shown in Fig. 9.1.

Implementing feedback at the machine pulse rate demands synchronous activity of all involved devices and places stringent compliance criteria on technical equipment, control system compute nodes, and the main controls network.

## 9.5 Remote Access – Remote Control

It is becoming commonplace for accelerator-based user facilities to provide means for technical experts to access remotely machine parameters for troubleshooting and machine-tuning purposes. This requirement for remote access is more critical for the ILC because of the likelihood that expert personnel are distributed worldwide.

## 9.6 Timing and RF-Phase Reference

Precision timing is needed throughout the machine to control RF phase and time-sampling beam instrumentation [206]. The timing system emulates the architecture of the control system, with a centrally located, dual-redundant source distributed via redundant fibre signals to all machine sector nodes for further local distribution. Timing is phase-locked to the RF system.

### 9.6.1 RF-Phase-Reference Generator

The RF-phase-reference generator is based on dual phase-locked frequency sources for redundancy. It includes fiducial generation (nominally at 5 Hz) and line lock. The macro-pulse fiducial is encoded on the distributed phase reference by a momentary phase shift of the reference signal. Failure of the primary frequency source can be detected and cause an automatic failover to the backup source.

### 9.6.2 Timing and RF-Phase-Reference Distribution

The phase reference is distributed via dual redundant active phase-stabilised links. Figure 9.3 shows an overview of dual redundant phase-reference transmission and local, intra-sector distribution.

The Phase Comparator unit detects failures in the primary phase-reference link and automatically fails over to the secondary link. Both the Phase Comparator unit and the Sector Timing Control units are fault tolerant. A local DRO or VCXO is phase-locked to the phase reference to develop a local reference with low phase noise for distribution within an RF sector of the main linac.

Figure 9.4 shows a block diagram of a single active phase-stabilised link. A portion of the optical signal is reflected at the receiving end. The phase of the reflected optical signal is compared with the phase of the frequency source. The resulting error signal controls the temperature of the shorter series section of fibre to compensate for environmentally induced phase shifts [207].

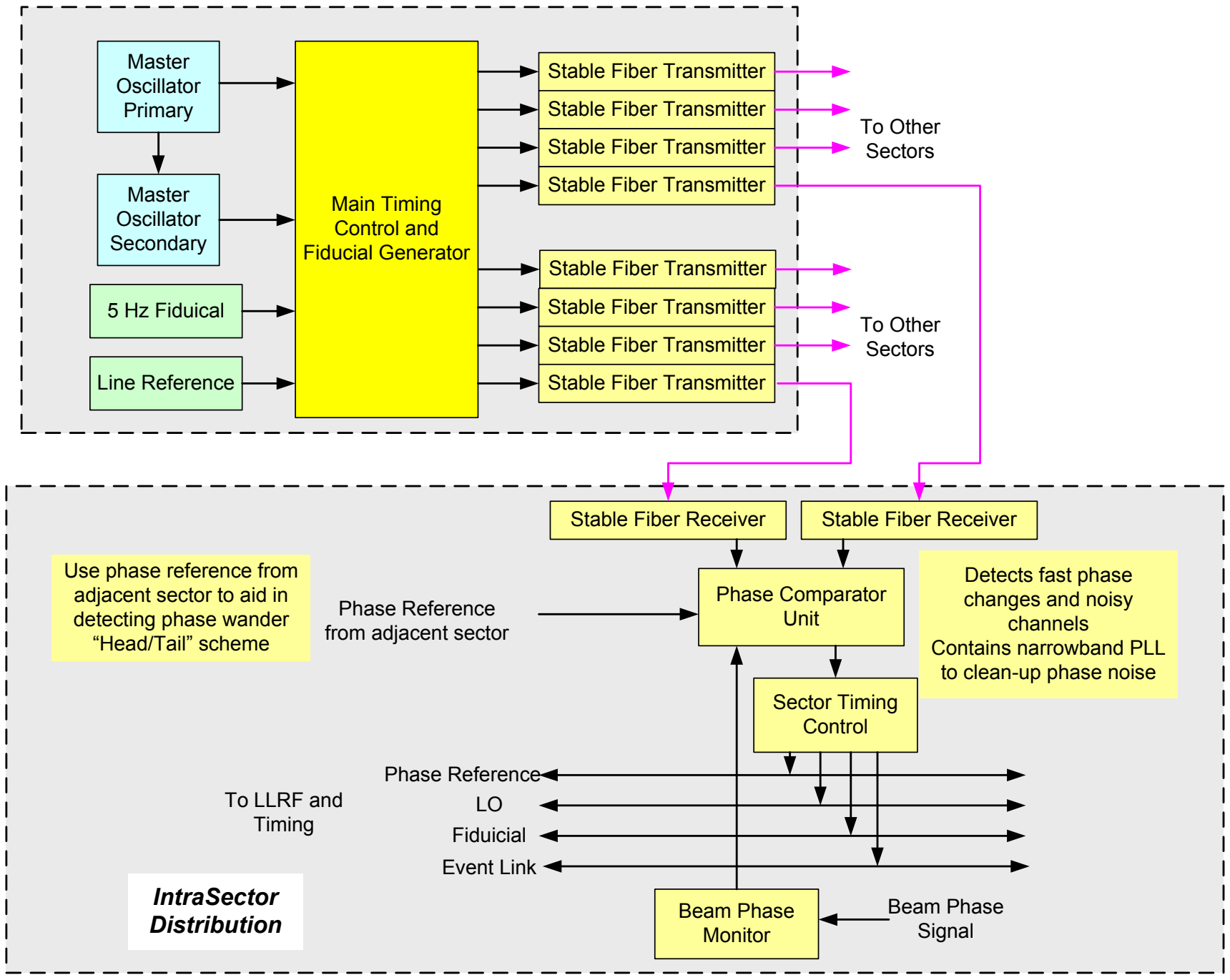

**Figure 9.3** Timing-system overview showing redundant phase-reference distribution and local intra-sector timing distribution.

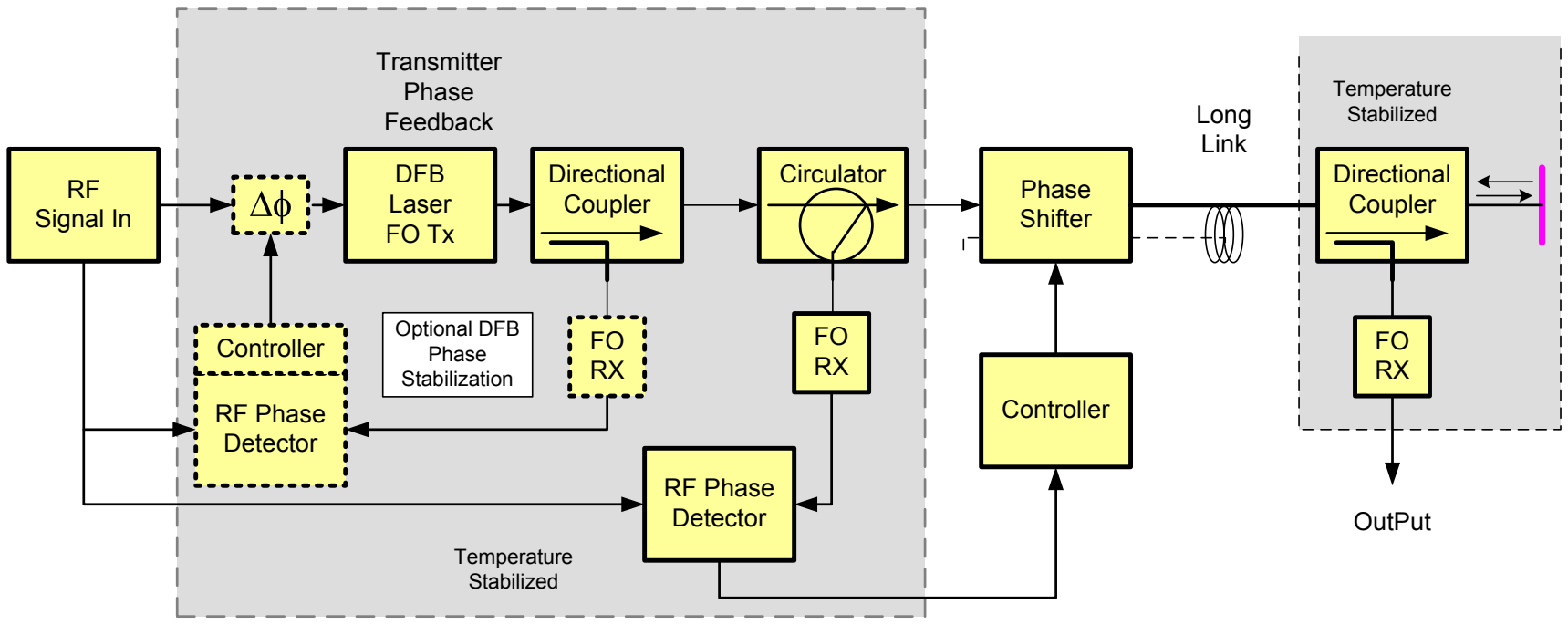

**Figure 9.4** Phase stabilised reference link.

## 9.6.3 Timing and Sequence Generator

An event stream is distributed via dual redundant links in a star configuration. The system automatically fails over to the redundant link upon detection of a failure. The event system provides a means for generating global and local sequences, synchronising software processing to timing events, and generating synchronous time stamps.

# 9.7 Beam-based Feedback

Beam-based dynamical feedback control is essential for meeting the high performance and luminosity needs of the ILC. Feedback systems stabilise the electron and positron trajectories throughout the machine, correct for emittance variations, and provide measurement and correction of dispersion in the Main Linac. Two timescales of beam-based feedback are anticipated, namely pulse-to-pulse feedback at the 5 Hz nominal pulse-repetition rate, and intra-train feedback that operates within the bunch train.

### 9.7.1 Architecture for Intra-Bunch Feedback Systems

Unlike pulse-to-pulse feedback, which is implemented through the control system, dedicated systems are required for intra-bunch feedback. These must operate at the bunch rate of $\sim$3 MHz, and include the RTML turnaround trajectory feed-forward control and intra-bunch trajectory control at the IP. Orbit feedback in the damping ring is synchronised to the damping-ring revolution frequency.

Local input/output processors acquire beam position, cavity fields, beam current, and other local beam parameters at the full 3 MHz bunch rate and distribute that information through a fast synchronous network. Local interconnections with the low-level RF systems provide opportunities for local feedback loops at the full 3 MHz bunch rate. Dedicated processing crates provide both dedicated real-time bunch-to-bunch control, and dispersion-free steering, while additional uncommitted crates could provide spare capacity and flexibility.

### 9.7.2 Hardware Implementation

Most of the feedback-processing requirements described in this section, including dynamic orbit control in the damping ring can be met using commercial hardware. Custom hardware solutions are used in cases where low latency or unique capabilities are required, such as for the RTML turnaround trajectory feed-forward and the IP intra-bunch trajectory feedback. High-availability solutions are implemented as appropriate, using the same standards and approach as for other instrumentation and control-system equipment.

# Chapter 10
# Availability, Commissioning and Operations

## 10.1 Overview

The ILC is a complex machine with hundreds of thousands of components most of which must be tuned with exquisite precision to achieve the design luminosity. This high luminosity must be maintained routinely in order to deliver the required integrated luminosity. Great care must be taken at all stages of the design to ensure that the ILC can be commissioned rapidly and operate efficiently with minimal downtime. Some of the critical design issues are:

- high-availability components and redundancy to minimize downtime;
- ease of commissioning;
- separation of regions to allow beam in one region while another is in access;
- Machine Protection System (MPS) to prevent the beam from damaging the accelerator;
- ensuring automated rapid recovery;
- feedback systems and control procedures to maintain optimum performance.

Many of these issues are mentioned elsewhere but are presented here as an integrated package to emphasise their importance to the ILC and the need for a powerful state-of-the-art control system.

## 10.2 Availability

### 10.2.1 Importance of Availability

The important figure of merit for the ILC is not the peak luminosity but the integrated luminosity recorded by the experiments. The integrated luminosity of the accelerator is the average luminosity multiplied by the uptime of the accelerator. Having surveyed the uptime fraction (availability) of previous accelerators, a goal of 75 % availability has been chosen for the ILC. This is comparable to HEP accelerators whose average complexity is much less than that of the ILC. As such, it should be a challenging but achievable goal. This goal is made even more challenging by the fact that all ILC subsystems must be performing well to generate luminosity. In contrast, a storage ring has an injector complex that can be offline between fills without impacting performance. Because it has more components and all systems must be working all the time, attaining the target availability for the ILC requires higher-availability components and more redundancy than previous accelerator designs. High availability must be an essential part of the design from the very beginning. A methodology is in place to apportion the allowed downtime among various components and hence arrive at availability requirements for the components.

### 10.2.2 Methodology

A simulation has been developed that calculates accelerator availability based on a list of parts (e.g. magnet, klystron, power supply, water pump). Input includes the numbers of each component, an estimate of its mean time between failure (MTBF) and mean time to repair (MTTR), and a characterisation of the effect of its failure (e.g. loss of energy headroom, minor loss of luminosity, or ILC down). The simulation includes extra repair time for components in the accelerator tunnel (for radiation cool-down and to turn devices off and on), repair of accessible devices while the accelerator is running, repair of devices in parallel to overlap their downtimes, and extra time to recover the beam after repairs are completed. It also allows repairs to be made in one region of the ILC while beam is used for accelerator physics studies in an upstream region. The inputs to the simulation were varied to test different machine configurations and different MTBFs/MTTRs to develop a machine design that had a calculated downtime of 15 %. The ILC design goal is $>$ 75 % uptime, but 10 % downtime was reserved as contingency for things that are missing from the simulation or for design errors. More details of the availability simulation model and its application to the ILC can be found in [3, 208].

### 10.2.3 Availability Studies

Simulations have been used to evaluate the impact of proposed design changes during the Technical Design Phase (TDP) [209]. The largest design change with impact on overall availability was going from a twin tunnel to a single tunnel for the two main linacs. This was part of the SB2009 change proposals. This is a rather complete analysis of the impact on alternate RF system designs which would be required to maintain a constant availability in a single tunnel design as a function of the installed energy overhead. The cases considered included KCS (Klystron Cluster System), DRFS (Distributed RF System with many smaller klystrons) along with a Central Region which contained the electron, positron sources, the DR's and the BDS which have second tunnel for support equipment that is accessible during beam running.

The results of a typical example simulation run giving the desired 15 % "downtime" are shown in Fig. 10.1 for a KCS with 4 % minimum overhead.

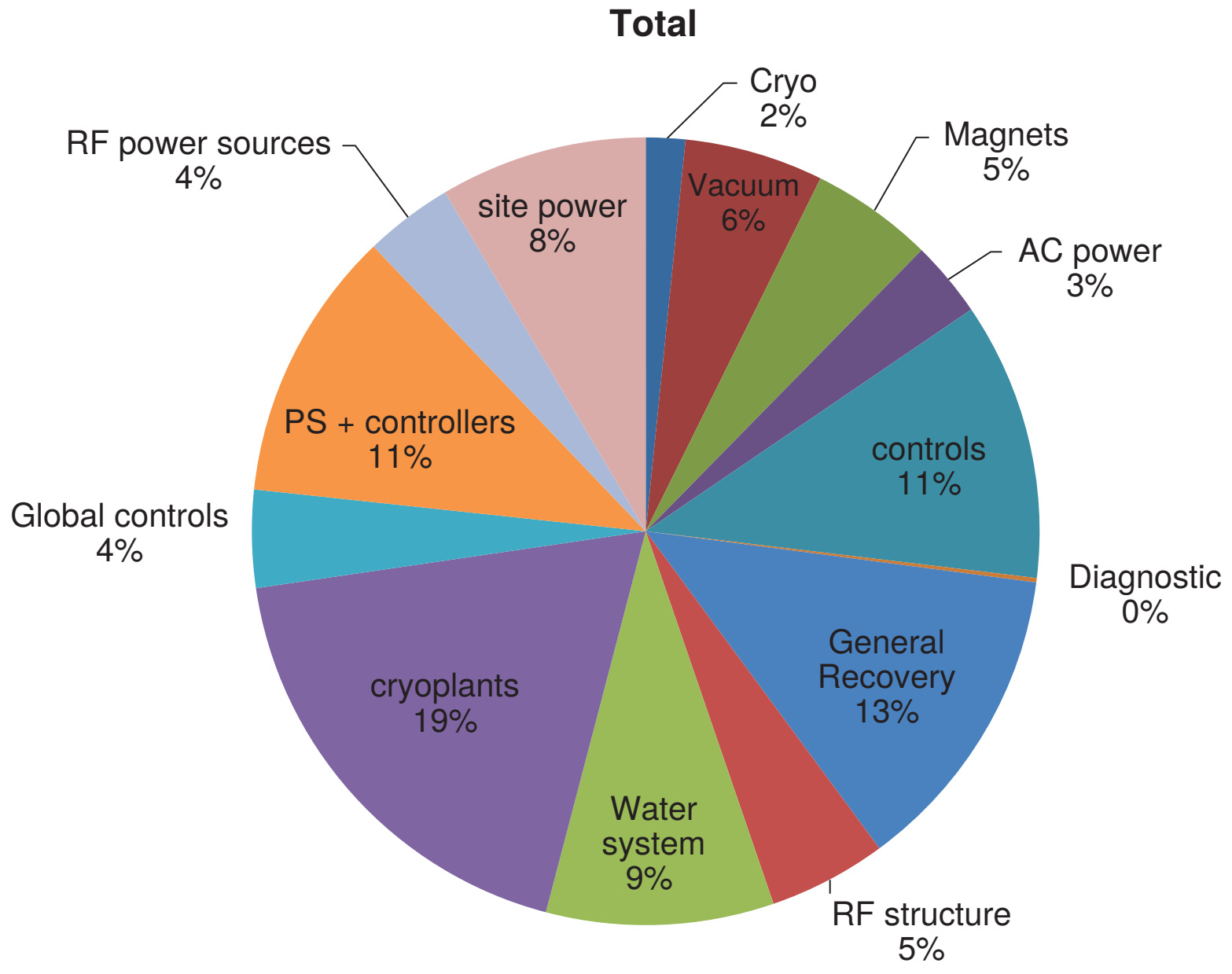

**Figure 10.1**
This figure shows the distribution of the downtime by area of the accelerator for a typical simulation run (KCS with 4% energy overhead). The downtime fractions shown are percent of the total downtime of about 15 %. So the actual downtime caused by the cryoplants is 19 % of 15 % = 2.8 %. 'General Recovery' is the excess, (beyond time nominally allotted), time spent recovering from scheduled maintenance days and is lumped because it cannot be directly attributed to a particular area [209].

The energy overhead varies with the actual operating energy and will be much larger as one operates the ILC below the installed energy capability but the downtime associated with the linac does not get much lower than 1 % as there always remain some non-RF accelerator components, e.g. RTML systems, in the accelerator tunnel.

## 10.3 Bunch Timing and Path-length Considerations

In order to extract the bunches in the damping ring one by one and inject into the main linac, there are certain constraints to satisfy among the DR circumference, number of bunches, RF frequencies and bunch distances in the DR and main linac [95]. The present parameters satisfy these constraints and allow for a flexible choice of bunch patterns as required for best operating performance. In addition, there is another constraint due to the fact that the positrons are generated by electrons from the previous pulse. For the most flexible operation, it is highly desirable that the difference in path-length travelled by the positrons (from target, through DR, RTML, Linac and BDS to IP) and the electrons (from target location direct to IP) is an integer multiple of the DR circumference. Other solutions involving pulse-to-pulse variation of the timing of electron injection are possible but less flexible. Because of this constraint, the exact location of the injector complex and the layout of the transport lines is a subject that can be fixed only after the final component lengths and the site details are decided.

## 10.4 Commissioning

This section describes general ideas on commissioning. The actual implementation will evolve with the schedule for construction of the conventional facilities, the installation of services and technical components and the availability of access to regions of the accelerator. These schedules will be site dependent but a typical example is shown in Fig. 10.2.

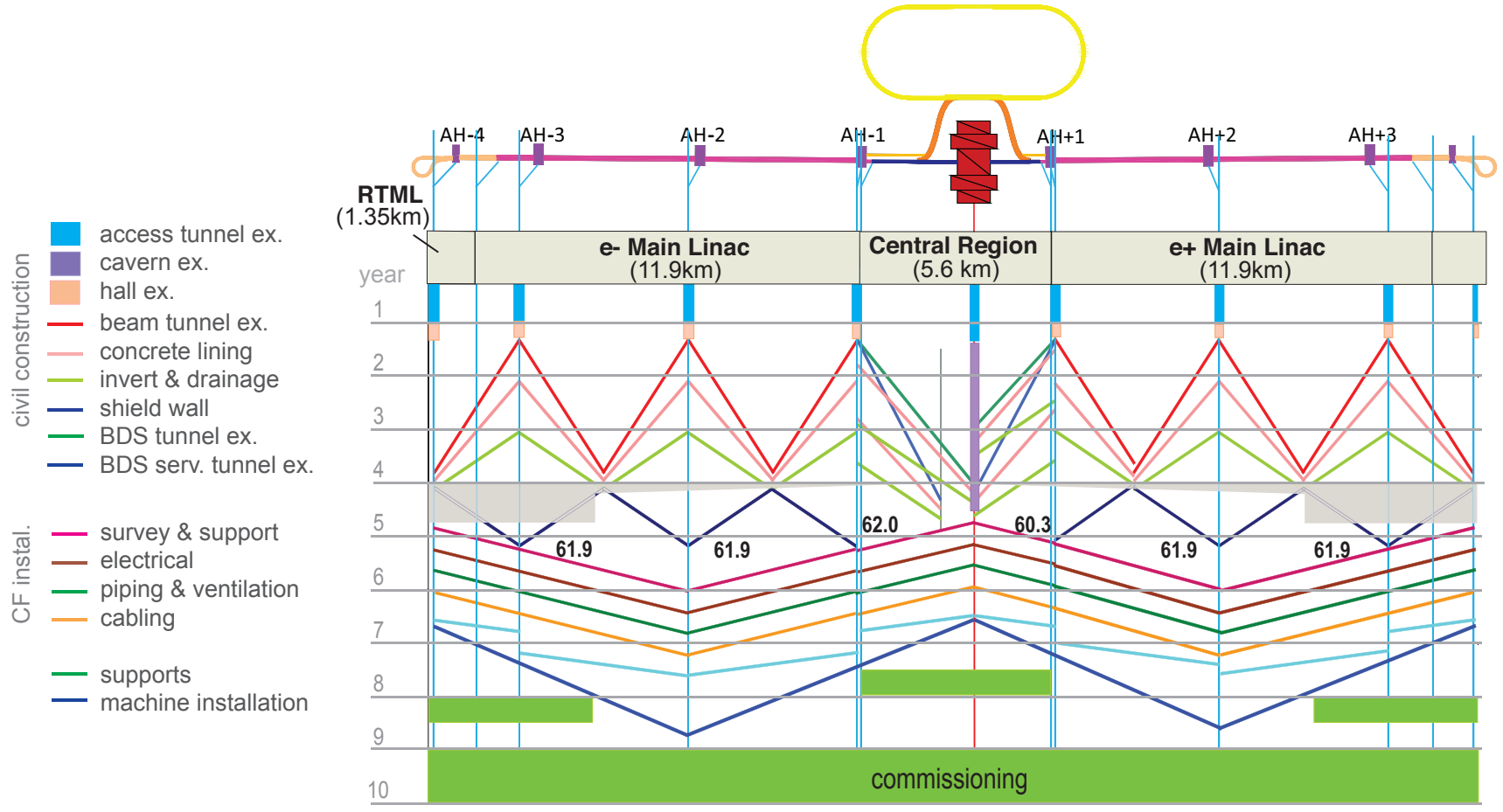

**Figure 10.2**
An example schedule which shows civil construction, installation of services and accelerator components followed by testing and beam commissioning of ILC systems (green bars).

### 10.4.1 Phased Commissioning

To minimize the time from completion of construction of the ILC to operation for high luminosity, it is desirable to complete upstream regions of the accelerator early. Commissioning can then start on these regions while construction continues downstream. This is called phased commissioning. In particular, it would be beneficial to complete the injectors and damping rings in time to allow one or two years of commissioning while construction of the linacs and BDS continues. The central region of the ILC includes the Interaction Region hall and facilities whose construction schedule, combined with the construction and commissioning of the detectors, is a critical path element to begin operation for physics. These drive the general scheduling philosophy of starting in the central region and progressing outwards in both directions along the two 11 km linacs.

A large amount of hardware validation and alignment and beam commissioning studies are necessary to produce low-emittance beams with good stability and availability. Consequently, it is important to allocate a sufficient amount of commissioning time at an early stage. A major function of the DR commissioning period is to achieve the alignment of optical components and to establish a small beam emittance. In addition, there are issues related to beam intensity that need to be checked and high-intensity beams are needed for vacuum chamber scrubbing. The use of the damping rings obviously necessitates functional beam source systems. Since both DRs are in the same tunnel, a schedule optimization has to be done to determine if it is best to install both DRs at the same time or if the $e^-$ ring should be installed and commissioned followed by the $e^+$ ring. The example construction schedule shown in Fig. 10.2 assumes that both rings are installed together and commissioned in parallel. The electron ring can be commissioned to full current with the standard electron source whereas the positron ring will use electrons from the auxiliary source or positrons produced by the auxiliary source on the target systems. The construction of the experiment is likely to consume the largest contiguous amount of time. It is recognized that construction of the underground detector hall is a major undertaking which cannot be completed until several years after groundbreaking of ILC. To mitigate the schedule impact, most of the sub-assemblies of ILC detector facilities are built on the surface and later installed into the hall in large pieces.

## 10.5 Radiation shielding and Personnel Protection System zones

To enable efficient operation and commissioning, the personnel protection system (PPS) is designed to allow personnel access in one region while beam is in another upstream region. As an example, the main-linac beam tunnel can be accessed while there is beam in the damping ring. Those parts of the accelerator accessible to people could have radiation levels that exceed the levels allowed for the public. Therefore, the radiation shielding and PPS zones described here are designed for radiation workers.

For the single main-linac tunnel design there is no service tunnel and therefore no need for detailed evaluation of radiation levels during the operation. The Japanese 'Kamaboko' design is an exception to this. The 'Kamaboko' tunnel design allows the main-linac tunnel to be divided into two regions, one for the main linac and the other for RF sources with utilities (service side). Personnel access to the service side during main-linac beam operation is crucial for long-term continuous operation of the main linac. Therefore, the separation wall thickness should be designed to allow such access.

### 10.5.1 Summary of Regions' Radiation Requirements

Maximum-allowable radiation levels for radiation and non-radiation workers in several scenarios have been examined and the most conservative case has been used for shielding calculations for non-site-specific design. This is referred to as the "maximum credible beam-loss condition" where all active limiting systems are off (system failure); radiation levels for radiation workers must be less than "250 μSv/hr" or "30 mSv/event". For the separation wall in the Japanese 'Kamaboko' main-linac tunnel, site-specific regulatory limits have been applied, i.e, radiation level for occupied area must be kept below "1 mSv/week" under normal operation condition. Radiation shielding and PPS devices must be designed to satisfy these criteria under the ILC operating beam loss scenarios.

### 10.5.2 Summary of the Radiation Safety Design for the Main Linac

Induced activity in air is estimated for 1 W/m continuous beam loss in the main-linac ventilation unit which is 5000 m long, between access tunnels. The ventilation system is designed to replace the entire air in the unit within 3 hours, therefore the air is irradiated with neutrons and photons for a maximum of 3 hours. The exhausted air from the main-linac tunnel passes through a vertical shaft. The induced activities were calculated based on Swanson's parameter with 1 W/m beam loss, 1 % fraction of deposited energy per beam loss and 2 m average path length of photons passing through the air. The nuclide, $^{3}$H, $^{7}$Be, $^{11}$C, $^{13}$N, $^{15}$O, $^{38}$Cl and $^{39}$Cl, were obtained by this manner. In addition to these,

$^{40}$Ar production is estimated from thermal neutron flux in the section. The highest activation comes from 13N, $5.8 \times 10^{-4}$ Bq/m$^3$, which is low in comparison with the airborne activation limit.

Induced activities in cooling water were estimated in the same way as for air. The amount of cooling water in the main-linac section is assumed to fill two 2-inch-diameter, 5000 m-long, water channels for each ventilation unit. The nuclides, $^3$H, $^7$Be, $^{11}$C, $^{13}$N, $^{15}$O, are produced. Highest value is 162 Bq/m$^3$ for $^{15}$O, but its short half-life means that this is unimportant.

The conclusion of the radiation-safety study for the main-linac tunnel is that the beam loss from normal operation with 1 W/m continuously produced is acceptable from the radiation-safety viewpoint. In actual operation, hardware systems and operation procedures to maintain beam losses at less than design value are quite important, as are tunnel design and installation. Radiation-safety design was performed for typical main-linac tunnel design concerning the tunnel separation wall, induced activities in air and cooling water assuming 1 W/m continuous beam loss.

The separation wall in the 'Kamaboko tunnel' design, should be designed considering the following items: the radiation level on the service side should be less than 1 mSv/week; the wall should have through holes every 20 m; and the tunnel should have horizontal emergency passage way every 500 m. An example of such a wall that satisfies these items has the following features: 3 m thickness for the normal section with ordinary concrete (2.3 g/cm$^3$); a 5 m-thick region of heavy concrete (3.0 g/cm$^3$) region with 8 m long non-modulator area for the emergency passageway; a reduced thickness part above 3 m from floor level.

### 10.5.3 PPS Zones

The personnel protection system (PPS) prevents people from being in the accelerator tunnel when the beam is on. A system of gates and interlocks turn off the beam before allowing access to the accelerator housing. Access to the service tunnel is not part of the PPS system. The ILC is divided into different regions (PPS zones) with tune-up dumps and shielding to allow beam in one region while there is access in another region. The PPS zones are the injectors, DR, main linac and BDS. Entrance gates for PPS zones are monitored and dump the beam when opened.

The ILC PPS zones are long and it would be burdensome to search the full region after each permitted access. To ameliorate this problem, they are divided into multiple search zones separated by fences with gates that are also monitored. The search zones are up to several hundred meters long. For example, in the linac a search zone is 500 m long and is separated by gates midway between each cross tunnel passageway or safety vault. Personnel access from a service area (service tunnel, shaft, detector hall etc.) to an accelerator area is controlled by PPS gates, as is the access from one accelerator region (PPS zone) to another accelerator region. Fences, doors, or moving shields are used for these gates and they have redundant gate-closed status switches for PPS monitoring. They are locked to prevent careless access but have an unlocking mechanism for emergencies. Information and communication systems are provided at the gates to show the operational status and allow communication between a person at the gate and an operator granting permission to go through the gate.

There are personnel-access passages between accelerator area and service area at the main linac, shafts, alcoves and the detector hall with PPS gates near each end. Since the passageways are used as emergency exits, heavy moving doors are avoided if possible. PPS gates between the accelerator areas and the service areas (including the access passageway) need to restrict the flow of activated air from the accelerator tunnel to the service area.

### 10.5.4 Shielding between PPS Zones

Shielding between PPS zones is designed to allow beam in the upstream zone while people are in the downstream zone. The upstream beam is deflected into a tune-up dump and there are triply redundant beam stoppers between the beam and the accessed region to ensure the beam does not enter the accessed region.

## 10.6 Machine-Protection System

The task of the machine-protection system, MPS, is to protect the machine components from being damaged by the beam when equipment failure or human error causes the beam to strike the vacuum envelope. The MPS design must take into account all types of failures that may occur and the damage they could produce.

### 10.6.1 Overview

Both the damage caused by a single bunch and the residual radiation or heating caused by small (fractional) losses of many bunches are important for MPS. The MPS consists of:

- a single bunch damage mitigation system;
- an system to limit the average beam loss;
- a series of abort kickers and dumps;
- a restart ramp sequence;
- a fault-analysis recorder system;
- a strategy for limiting the rate with which magnetic fields (and insertable device positions) can change;
- a sequencing system that provides for the appropriate level of protection depending on machine mode or state;
- a protection collimator system.

The systems listed must be tightly integrated in order to minimize time lost to aberrant beams and associated faults.

### 10.6.2 Single-Pulse Damage

Single-pulse damage is mitigated by systems that check the preparedness of the machine before the high-power beam passes. Single pulse damage control is only necessary downstream of the damping ring. Three basic subsystems are involved:

1. a beam-permit system that surveys all appropriate devices before damping-ring beam extraction begins and provides a permit signal if each device is in the proper state;
2. an abort system that stops the remaining bunches of a train if a bunch does not arrive at its intended destination;
3. spoilers upstream of devices (typically collimators) to expand the beam size enough that several incident bunches do not cause damage.

In addition, some exceptional devices (damping-ring RF and extraction kickers for example) have fast-monitoring systems and redundancy. Spoilers or sacrificial collimators are placed before the bunch compressors, in the undulator chicane, at the beginning of the BDS system and in the collimator section of the BDS. Locations with dispersion downstream of an accelerator section have spoilers to intercept off-energy beam caused by klystron faults or phase errors before the beam can hit a downstream collimator or beam pipe. The spoilers are designed to survive the number of incident bunches that hit before the abort system can stop the beam. The use of a *pilot bunch* is also being kept as an option. A pilot bunch is one percent of nominal current and is spaced 10 μs ahead of the start of the nominal train. If it does not arrive at its intended destination, the beam abort system is triggered to prevent full-intensity bunches from hitting the spoiler.

Studies have shown that for many failure scenarios such as quadrupole errors or klystron phase errors, the beam is so defocused by the time it hits the linac aperture that it does not cause damage. For this reason, no spoilers or extra beam-abort kickers are included in the linac.

The beam-abort system uses BPMs and current detectors to monitor the beam trajectory and detect losses. On a bunch-by-bunch basis, the system checks for major steering errors or loss of beam. When a problem is detected, it inhibits extraction from the damping ring and fires all abort kickers upstream of the problem. The abort kickers cleanly extract the beam into dumps, protecting downstream beam lines. In the few milliseconds before the start of the pulse train, the beam-permit system checks the readiness of the modulators and kicker pulsers, and the settings of many magnets before allowing extraction of beam from the damping rings.

## 10.6.3 Average Beam-Loss-Limiting System

Average beam loss is limited, throughout the ILC, by using a combination of radiation, thermal, beam intensity and other special sensors. This system functions in a manner similar to other machines, such as SLC, LHC, SNS and Tevatron. If exposure limits are exceeded at some point during the passage of the train, damping ring extraction or source production $(\mathrm{e}^+/\mathrm{e}^-)$ are stopped. For stability, it is important to keep as much of the machine as possible operating at a nominal power level. This is done by segmenting it into operational MPS regions. There are 7 of these regions, as noted in Table 10.1. Beam rate or train length can be limited in a downstream region while higher rate and train lengths are maintained in upstream regions. The maximum power-handling capabilities of the beam dumps, as shown in Table 10.1, vary with the location, beam energy and the operating requirements.

**Table 10.1** Maximum power handling capabilities of the beam dumps.

| | | | | | |
|---|---|---|---|---|---|
| E-1 | SC Tune-up Dump | 311 kW‡ | E+1 | SC Tune-up Dump | 311 kW‡ |
| E-2 | EDRX Tune-up Dump | 220 kW | E+2 | PDRX Tune-up Dump | 220 kW |
| E-3 | RTML Tune-up Dump | 220 kW | E+3 | RTML Tune-up Dump | 220 kW |
| E-4 | BDS Tune-up Dump | 14 MW | E+4 | BDS Tune-up Dump | 14 MW |
| E-5 | Primary E- Dump | 14 MW† | E+5 | Primary E+ Dump | 14 MW† |
| E-6 | RTML Tune-up Dump | 220 kW | E+6 | RTML Tune-up Dump | 220 kW |
| E-7 | E- Fast Abort Dump | 250 kW | E+7 | E+ Target Dump | 200 kW† |

† Always ON
‡ 45 kW always ON

## 10.6.4 Abort Kickers and Dumps

Abort systems are needed to protect machine components from single-bunch damage. It is expected that a single-bunch impact on a vacuum chamber will leave a small hole, roughly the diameter of the beam. Each abort system uses a fast kicker to divert the beam onto a dump. The kicker rise time must be fast enough to produce a guaranteed displacement of more than the beampipe radius in an inter-bunch interval.

There are abort systems at the end of each linac, before the undulator entrance, and one at the entrance to the BDS on the positron linac.

There will be many meters of fast kickers needed at each dump and megawatts of peak power from pulsers. R&D will be needed to optimize the final system and ensure its reliability.

### 10.6.5 Restart Ramp Sequence

Actual running experience is needed to exactly define the restart ramp sequence. For that reason the sequencer must be flexible and programmable. Depending on the beam dynamics of the long trains, it may be advisable to program short trains into a restart sequence. There may also be single bunch, intensity dependent effects that require an intensity ramp. The system must be able to determine in advance if the beam loss expected at the next stage in the ramp sequence is acceptable. Given the number of stages and regions, the sequence controller must distribute its intentions so that all subsidiary controls can respond appropriately and data-acquisition systems are properly aligned. It may be necessary to have a pilot bunch mode with the nominal intensity but large emittance. The initial stages of the sequence can be used to produce 'diagnostic' pulses to be used during commissioning, setup and testing.

### 10.6.6 Fault-Analysis Recorder System

A post-mortem analysis capability is required that captures the state of the system at each trip. This must have enough information to allow the circumstances that led to the fault to be uncovered. Data to be recorded on each fault include: bunch-by-bunch trajectories, loss-monitor data, machine-component states (magnets, temperature, RF, insertable-device states), control system states (timing system, network status) and global system status (sequencer states, PPS, electrical, water and related sensors). The fault-analysis system must automatically sort this information to find what is relevant.

### 10.6.7 Rapidly Changing Fields

In addition to the above, there are critical devices whose fields (or positions) can change quickly, perhaps during the pulse, or (more likely) between pulses. These devices need 1) special controls protocols, 2) redundancy or 3) external stabilization and verification systems.

1. Depending on the state of the machine, there are programmed (perhaps at a very low level) ramp-rate limits that keep critical components from changing too quickly. For example, a dipole magnet is not allowed to change its kick by more than a small fraction of the aperture (few percent) between beam pulses during full power operation. This may have an impact on the speed of beam-based feedbacks. Some devices, such as collimators, are effectively frozen in position at the highest level of beam power. There may be several different modes, basically defined by beam power, that indicate different ramp-rate limits.

2. There are a few critical, high-power, high-speed devices (damping-ring kicker and RF, linac front-end RF, bunch-compressor RF and dump magnets) that need some level of redundancy or extra monitoring in order to reduce the consequence of failure. In the case of the extraction kicker, this is done by having a sequence of independent power supplies and stripline magnets that have minimal common-mode failure mechanisms.

3. There are several serious common-mode failures in the timing and phase distribution system that need specially engineered controls. This is necessary so that, for example, the bunch compressor or linac common phase cannot change drastically compared to some previously defined reference, even if commanded to do so by the controls, unless the system is in the benign beam-tune-up mode.

### 10.6.8 Sequencing System Depending on Machine State

The ILC is divided into segments delineated by beam stoppers and dump lines. There may be several of these in the injector system, two beam dumps in each RTML, and 2 (or 3) in the beam delivery and undulator system. In addition, the ring-extraction system effectively operates as a beam stopper assuming the beam can remain stored in the ring for an indefinite period. This part of the MPS assumes that the beam power in each of these segments can be different and reconfigures the protection systems noted above accordingly.

### 10.6.9 Protection Collimators

The entire ILC requires protection collimators and spoilers that effectively shadow critical components. These devices must be engineered to withstand innumerable single-pulse impacts.

## 10.7 Operability

To ensure high average luminosity it is important that the ILC have many features built in to make its operation mostly automatic and efficient. These features include:

- accurate, reliable, robust diagnostics;
- monitoring, recording, and flagging of out-of-tolerance readings of all parameters that can affect the beam, some of which must be checked milliseconds before each pulse train so beam can be aborted if there is a problem;
- beam-based feedback loops to keep the beam stable through disturbances like temperature changes and ground motion;
- automated procedures to perform beam-based alignment, steering, dispersion correction, etc.;
- automatic recovery from MPS trips starting with a low-intensity, high-emittance beam and gradually increasing to nominal beam parameters.

### 10.7.1 Feedback systems

Transporting the beam through the ILC while maintaining a small emittance requires a large number of feedback systems.

These feedback systems include measurements from various beam-position monitors, from laser wires scanning the beam profile and other diagnostics. The feedback loops must be carefully designed to be orthogonal and to maintain corrections that are within the device ranges. The feedback systems must avoid trying to compensate for large deviations of the beam due to component failure. It is hence necessary to use flexible setups for the control loops such as provided by MATLAB tools and analysis techniques (see Section 9.7).

# Chapter 11
# Conventional Facilities and Siting

## 11.1 Introduction

In the RDR, a generic CFS design was developed and used in each of three regional sample sites. This resulted in very similar overall layouts using a twin Main Linac (ML) tunnel configuration and common designs for supporting mechanical and electrical utility systems. The current design is tailored to accommodate local site conditions and incorporates the results of value engineering and tunnel configuration studies and detailed site-specific designs for conventional facilities and mechanical and electrical utility systems. For the Americas and European regions, a single ML tunnel constructed by tunnel-boring machines is preferred; for the Asian region, where both candidate sites are in mountainous regions and there is great experience with tunneling in mountainous regions, the drill-and-blast New Austrian Tunneling Method (NATM) is preferred. This means that a larger single tunnel is the preferred method of construction, since NATM can cost-effectively produce large tunnels. This larger tunnel can then be divided into two with a shielding wall to allow klystrons etc to be separated from the running accelerator. Another major development associated with these tunnel geometries is the introduction of the Klystron Cluster high-level RF system (KCS) for the Americas and European regions and the Distributed Klystron high level RF system (DKS) for the Asian region.

The designs that have been developed for the Americas and European regions are very similar. The Americas design has been based on the Fermilab site in northeastern Illinois. The European design has been developed for a site near the CERN Laboratory in Switzerland. A preliminary evaluation of a second site near the Joint Institute for Nuclear Research in Dubna, Russia has also been performed. In all cases a single ML tunnel is used with the KCS, which places all of the klystrons and related equipment in surface buildings at the tops of vertical shafts. From these klystron buildings, waveguides distribute the microwave power through the vertical shafts and ML tunnel.

Two candidate sites have been identified in the Asian region, both in mountainous areas of Japan. Access considerations preclude the use of vertical shafts so that inclined tunnels are used for access to the main tunnels and IR Hall. There are some surface buildings at the entrances to the access tunnels, but surface facilities are minimized to limit environmental impact. Detector construction and assembly methods are different between a mountain site and one with relatively uniform surface elevations. This alone has a direct impact on the integrated construction, installation and commissioning schedule and possibly overall project cost.

After a sketch of the overall layout of the ILC, (Fig. 11.1), and some other general considerations related to common design criteria and general site considerations, this chapter sets out the detailed conventional facility designs for first the Asian region and then the European and Americas regions. Finally, common issues of handling and the installation plan and estimate of effort required are discussed.

## 11.2 Overall Layout and Common Design Criteria

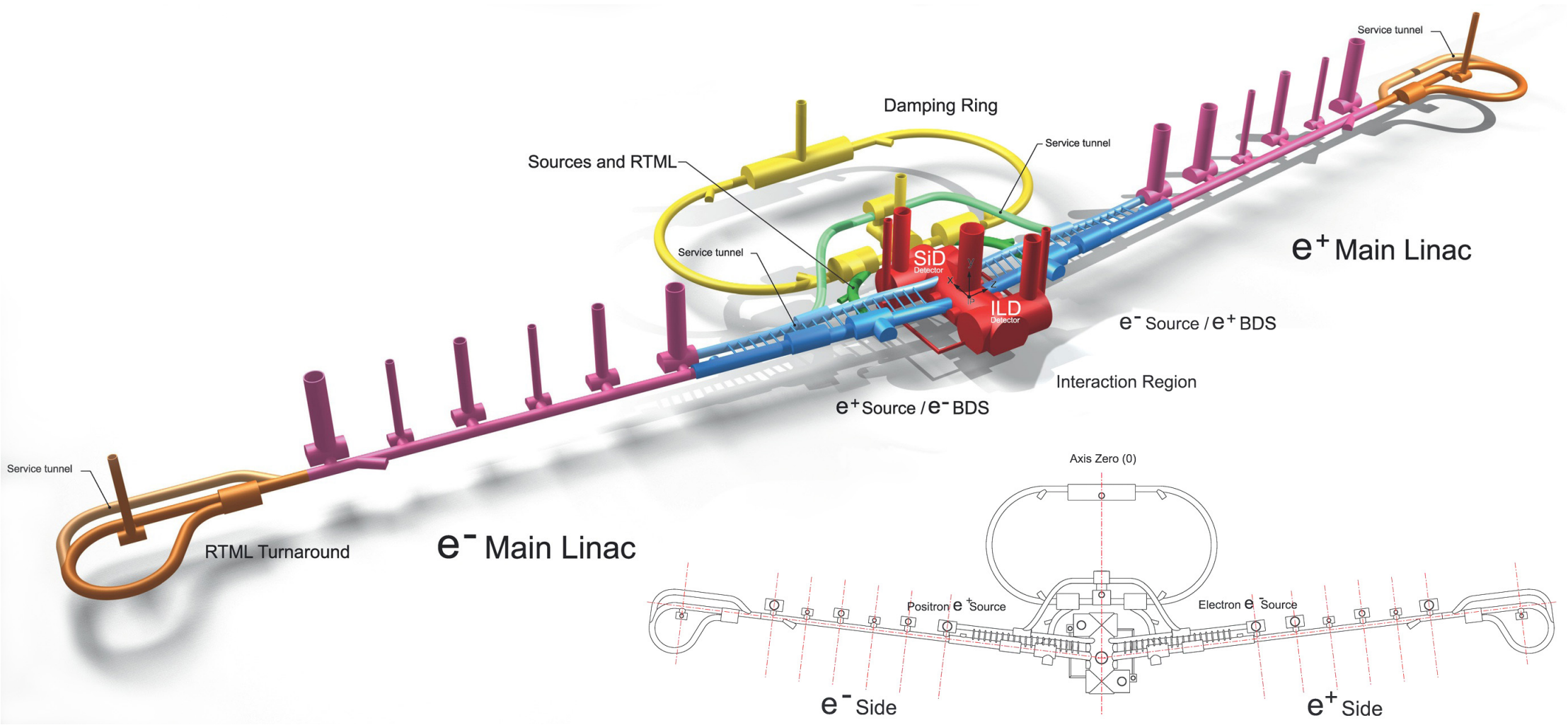

**Figure 11.1.** ILC tunnel schematic for KCS showing accelerator systems, IR hall and support tunnels.

The ML is housed in a single tunnel; in the Asian design this is a wide tunnel with parallel galleries, one containing the beamlines and one accessible by personnel. The ML beam tunnel also houses the RTML 5 GeV transport line supported from the ceiling and positioned towards the center of the tunnel. The DR has a single tunnel large enough to contain an electron ring, a positron ring and a possible future second positron ring.

The Central Region area, from the IR Hall to the ends of the MLs, has both a beam tunnel and a parallel service tunnel. The beam tunnel houses multiple beamlines including the $e^-$ and $e^+$ sources, the BDS, the RTML and beam abort and dump lines. This region also includes the short segments that route beamlines to and from the DR. All tunnels have been sized for the respective equipment and its installation, transport and replacement, as well as personnel egress. The beam and service tunnels are widened as needed to maintain the same aisle width as in the ML.

The service and beam tunnels are separated by sufficient material to provide structural stability and radiation shielding for workers in the service tunnel while the accelerator is operating. Penetrations between tunnels have been sized and configured to provide the required radiation shielding, as have the V-shaped personnel passageways between the two tunnels.

The IR Hall is sized to support two detectors in a 'push-pull' configuration. Each detector garage area is connected to the beam tunnel, and to the egress elevator through a passageway.

The beamline configuration and the arrangement and operational requirements of the IR Hall and detectors are site invariant. Each detector will be constructed on a moveable platform which must have the capability to accurately move efficiently into and out of the interaction point. However the design and construction of the enclosures and tunnels that house the beamlines and related equipment must conform to local geological conditions.

While the KCS and DKS differ in configuration and equipment layout, the electrical power and cooling systems are very similar, though local conditions and climate will have a direct influence on their design.

## 11.3 General site requirements

The site must accommodate the initial 31 km overall length, as well as the upgrade to 50 km length, and the area adjacent to the IR for the DRs. Requirements for tunnel access, support equipment and surface buildings must be included.

Alignment and stability are very important for reliable accelerator operation. Even more critical is the stability of the IR floor. Two detectors , with respective weights of approximately 15 kt and 10 kt, will be supported on concrete platforms each weighing approximately 2 kt. The geology at any proposed site must be able to accommodate the detector movements and allow their repositioning without unsatisfactory deflection or settlement over time.

Electrical power requirements are substantial. Operation at 500 GeV (1 TeV) will require approximately 161 MW (285 MW) respectively. These requirements are almost certainly a significant addition to any existing local electrical-grid power capability. In addition a reliable and ample water source for process cooling will be needed.

Suitable access will be needed during both the construction phase, during which a great deal of excavated material will be removed, and the operational phase. Trucking routes and deposit locations will need to be identified. For the installation of components, shipping by road is likely to be the main delivery option and roads to the site must be able to accommodate both the length and weight requirements of the major components. Rail, air and/or seaport access may be required for specific components and convenient access to a major airport is essential for a fully international project.

## 11.4 Asian region (Mountain topography)

### 11.4.1 Siting studies

#### 11.4.1.1 Location and properties of Asian sample sites

The Asian region currently has two candidate sites, both in Japan (Fig. 11.2), which were selected after several years of study:

- Kitakami site: located in Iwate prefecture (Tohoku district);
- Sefuri site: located in Fukuoka & Saga prefecture (Kyushu district).

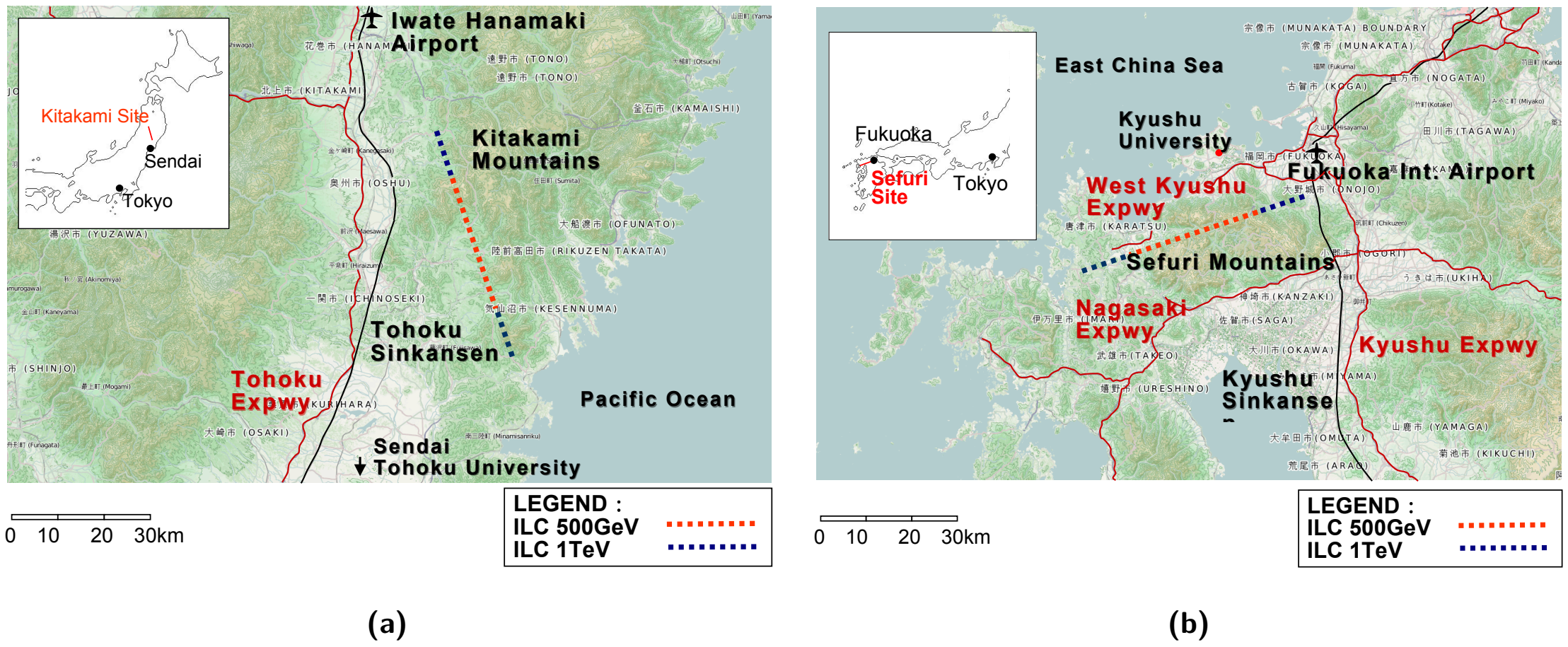

**Figure 11.2.** Candidate sites in (a) Kitakami and (b) Sefuri.

They are favoured because of their geographical and geological characteristics, as well as the strong support of the local government and residents. The common geographical and geological features of the two sites include:

- although mountainous, the region is not particularly steep.

- location in bedrock suitable for construction of a 31 km accelerator tunnel and a large IR Hall cavern;
- potential to extend to 50 km tunnel length;
- small ground vibrations;
- no active faults near the tunnel so no danger of nearby earthquakes;
- no man-made vibration source nearby.

Additional favourable characteristics include a suitable climate, stable source of 300 MW electric power, adequate cooling water supply, possibility for adequate groundwater treatment, tunnel access capability, and suitable road access for construction and delivery vehicles.

#### 11.4.1.2 Land features

Although the sites are located in mountainous regions largely covered by forest, the base of the mountains is more gently sloping terrain, sparsely populated with small clusters of houses and comparatively small-scale agriculture and dairy farms. Access to the underground tunnels would be located in this more accessible terrain. These areas would also serve as a base for construction, and provide access to the experimental facility after completion.

#### 11.4.1.3 Climate

The Kitakami site has a slightly cold climate with a mean air temperature in the coldest (hottest) month of $-4.8\,°C$ ($+28.8\,°C$) respectively. The mean annual rainfall is 1,318 mm. The Sefuri site has a mild climate with a mean air temperature in the coldest (hottest) month of $+3.0\,°C$ ($+32.1\,°C$) respectively. The mean annual rainfall is 1,612 mm. In both sites, there is occasional light snowfall in January and February.

The Kitakami site is located in the Tohoku district which was hit by a massive earthquake in March 2011, although there was little damage at the site itself. The Sefuri site is located in the Kyushu district, which, although hit by typhoons every year, suffers little damage.

#### 11.4.1.4 Geology and tunnel structures

The tunnels at both sites would be built in hard granite bedrock (Fig. 11.3). If the ML tunnel is extended to 50 km, one end would extend beyond the granite into sedimentary rock, which, however, would also provide a stable base for construction.

The ML tunnel is located at a depth of between 50 m and 400 m; access is through a sloped tunnel with a grade of no more than 10 %. The access tunnel for the IR Hall has a maximum grade of 7 %.

The ML tunnel has a ‘Kamaboko’ shape. Rock-bolt reinforcement is not usually required in stable granite, but may be required in wider sections. The tunnel interior is lined with concrete to provide waterproofing so that external groundwater can be processed by normal drainage. The access tunnels do not need to be waterproof and the interiors are of sprayed concrete. Rock anchor or bolt reinforcement is required for the IR Hall.

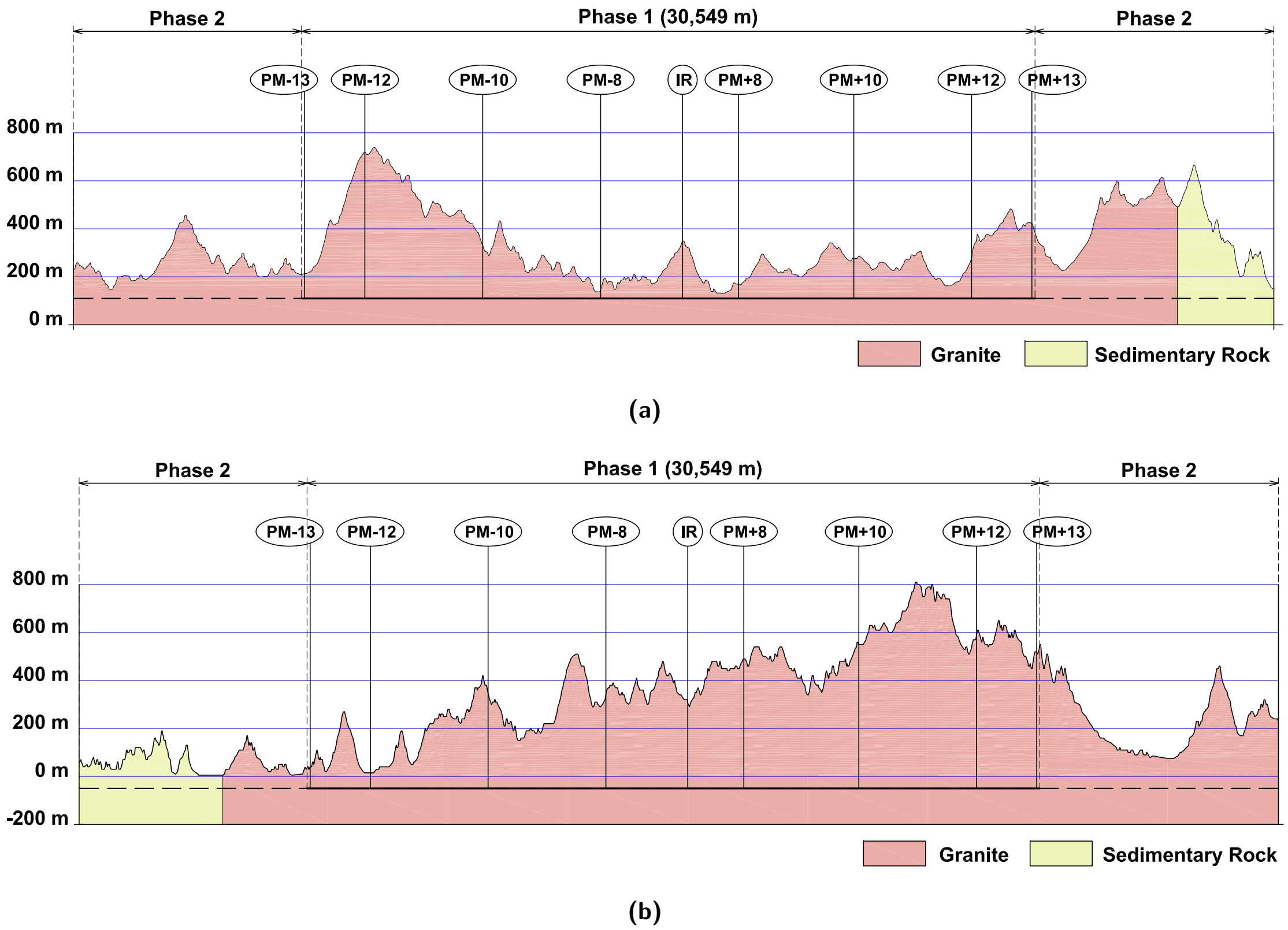

**Figure 11.3.** Elevation of the sites (a) Kitakami and (b) Sefuri. The 500 GeV baseline tunnel (phase 1) is shown, as well as the extent of the 1 TeV upgrade (phase 2). For Phase 1 the access tunnel spacing is shown.

#### 11.4.1.5 Power Availability

Both sites have sufficient electric power to meet requirements. Kitakami has a 275 kV line and Sefuri a 345 kV line. Electric power stability should be adequate for accelerator operation.

#### 11.4.1.6 Construction Methods

The ML tunnel is 11.0 m wide and 5.5 m high and is excavated using NATM. After construction it is divided into two parallel galleries by a central concrete wall The access tunnels are also excavated by NATM, except where they penetrate the 10 m to 20 m thick surface soil layer, where steel reinforcement is required. The IR Hall is excavated from the top down, starting from a top-heading tunnel connected to the access tunnel using a bench-cut construction method. As the excavation progresses it is reinforced by rock-bolts into the cavern wall.

### 11.4.2 Civil construction

Because the Japanese sites are deep underground, they have some unique requirements:

- the ML RF power is supplied via DKS with the RF sources in the service gallery;
- due to the capacity of the cryogenic plants, the underground structures are separated into seven zones, each with a maximum span of ±2.5 km from the access point (Fig. 11.4);
- the sloped access tunnels dictate a particular design for the underground enclosures as well as a particular installation method;
- electrical, mechanical and cryogenic utilities are located in underground caverns.

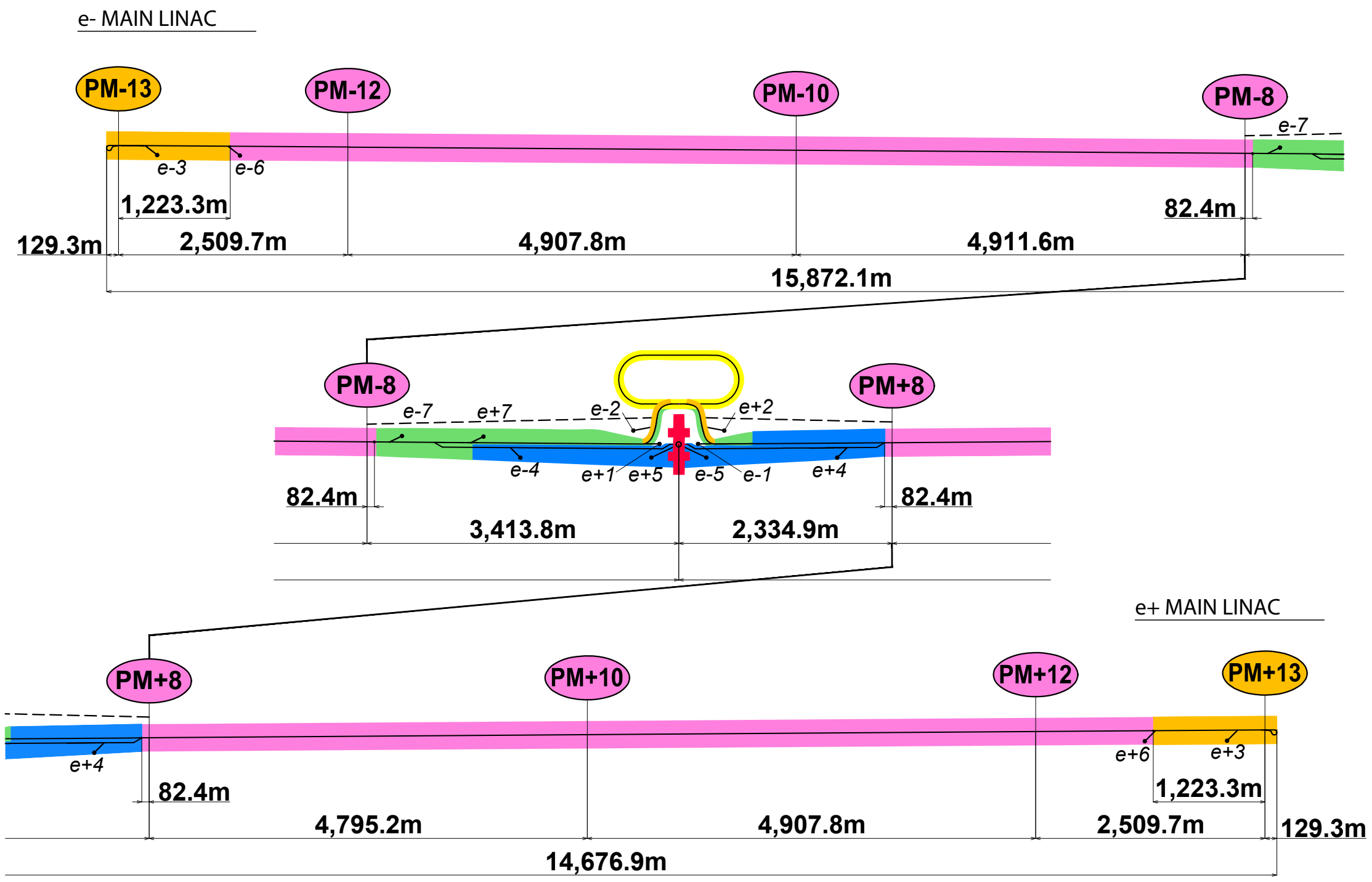

**Figure 11.4.** Asian region overall site layout.

## 11.4.2.1 Overall site layout

The overall site layout is shown in Fig. 11.4. Because of the cryogenic-string length, the ML, shown schematically in Fig. 11.5, is slightly longer with DKS than with KCS: the electron arm is 11,188 m, while the positron arm is 11,072 m. The number of access points is the minimum consistent with the cryogenic plant layout.

## 11.4.2.2 Underground enclosures

Tables 11.1, 11.2 and 11.3 show the extent of the required underground construction for the different accelerator systems. The features of the major underground enclosures are described below.

**Table 11.1**
Tunnel lengths and volumes

| Accelerator section | Length(m) | Volume ($m^3$) |
|---|---|---|
| $e^-$ source (beam) | 368 | 17,757 |
| $e^-$ source (service) | 223 | 4,881 |
| $e^+$ source (beam) | 1,678 | 67,364 |
| $e^+$ source (service) | 1,523 | 33,351 |
| Damping Ring | 3,239 | 120,352 |
| RTML | 3,305 | 200,237 |
| Main Linac | 22,425 | 1,395,754 |
| BDS (beam) | 3,847 | 184,019 |
| BDS (service) | 3,102 | 67,915 |
| TOTAL | 39,710 | 2,091,630 |

**Table 11.2**
Cavern summary and volumes. The six large Main Linac caverns house the helium compressors, cryogenic facilities, electrical substations, cooling-water systems, and plumbing systems.

| Accelerator section | Qty | Volume ($m^3$) |
|---|---|---|
| $e^-$ source | 0 | 0 |
| $e^+$ source | 0 | 0 |
| Damping Ring | 4 | 21,151 |
| RTML | 2 | 15,522 |
| Main Linac | 6 | 293,687 |
| BDS | 0 | 0 |
| IR | 1 | 189,381 |
| TOTAL | 13 | 519,741 |

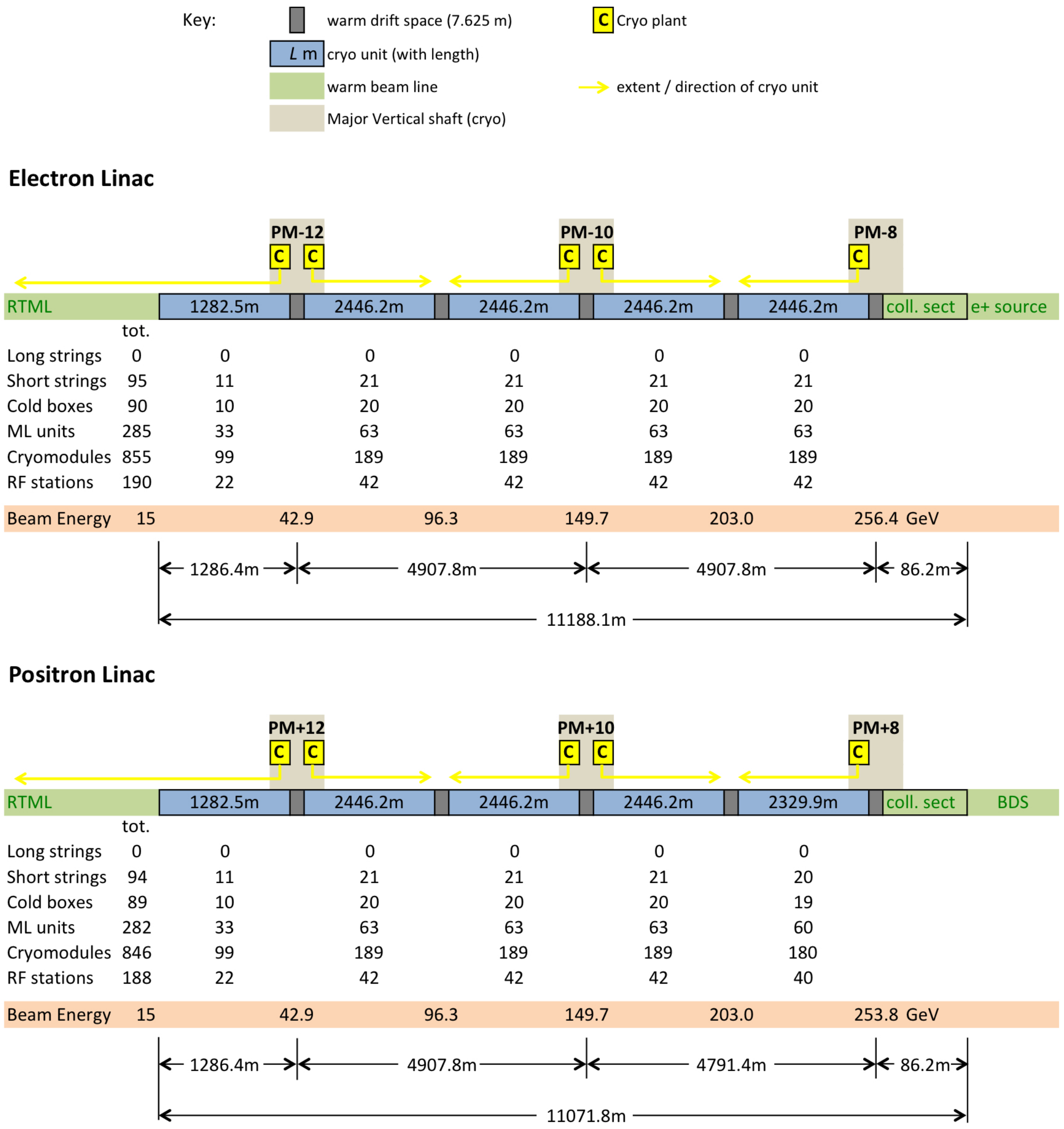

**Figure 11.5.** Main Linac schematic showing the cryostrings, cryogenic fluid control cold boxes, 3-cryomodule ML units, cryomodules and HLRF generation stations. The beams energy increases from the left side of the figure for both the electron and positron linac diagrams.

**Table 11.3**
Access tunnel lengths and volumes

| Accelerator section | Qty | Volume ($m^3$) |
|---|---|---|
| $e^-$ source | 0 | 0 |
| $e^+$ source | 0 | 0 |
| Damping Ring | 1,320 | 88,335 |
| RTML | 2,000 | 117,186 |
| Main Linac | 6,000 | 351,558 |
| BDS | 0 | 0 |
| IR | 1,772 | 155,914 |
| TOTAL | 11,092 | 712,993 |

11.4.2.2.1 ML tunnel A comparison of various construction methods indicated that NATM would be the most cost effective for mountainous sites (see Part I Section 5.2.2.2) The slower excavation speed ($\sim 100\,\mathrm{m}/$month) is compensated by greater flexibility with short construction zones and more parallel excavation.

The cross-sectional layout of the ML tunnel with centre wall is shown in Fig. 11.6. Both beam and service gallery have functional zones for equipment installation, survey, conveyance, and human egress. Water pipes are installed in the lower part of the tunnel and electric power lines are installed in shielded racks on the ceiling.

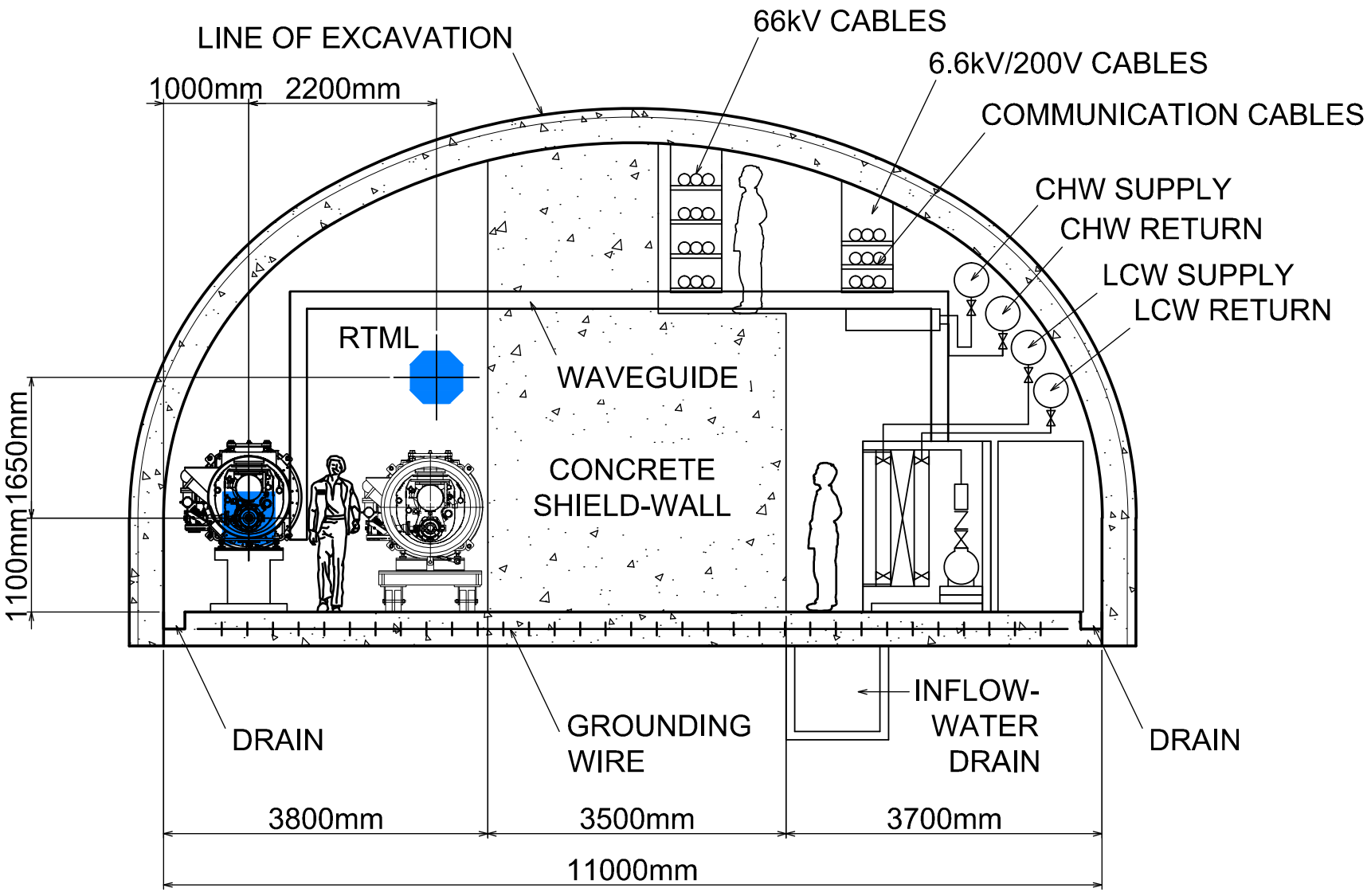

**Figure 11.6** Equipment layout in the ML tunnel

At its base, the centre wall is 3.5 m thick to provide radiation shielding and the upper side thickness is 2.0 m. At intervals of 12 RF units, there is a connection passage between the beam and service galleries, which can be used for evacuation in case of emergency. At each connection, the centre wall thickness is increased by 2 m in the downbeam direction to provide more shielding. To allow efficient excavation, the tunnel height must be at least 5 m, based on the passage of standard excavation machines ($\sim$3.5 m high), plus the sliding-form for the concrete lining ($\sim$1 m thick), and the concrete liner itself 30 cm thick. The tunnel floor is 40 cm thick. The ML tunnel is nominally aligned along a geoid surface. A small slope of no more than $\pm$0.8 % is acceptable for the main-linac cryomodule and may be introduced to minimise total access-tunnel length for a given specific site.

Previous constructions have shown that grout can limit inflow to no better than $\sim$1 l/min/m at 100 m depth underground.The inflow water rate must be confirmed by geologic studies before construction, but any inflow beyond this will be completely isolated by the concrete liner and drained to a ditch that will be sized assuming the inflow water for 5 km is gathered to one access point.

11.4.2.2.2 Access halls (AH) An“access hall” at the Asian site (Fig. 11.7) corresponds functionally to a “shaft-base cavern” at the other sites. The six halls for the ML/RTML/BDS areas are located alongside the main underground tunnel (Fig. 11.3). They provide an entrance to the tunnel as well as a local center for electricity, air, cooling water, and liquid He infrastructure. Each AH includes:

- an electrical substation with two 30 MW 66 kV/6.6 kV transformers, an incoming panel, and a distribution panel for cryogenics, accelerator supplies, and service equipment;
- a mechanical station with the second-loop heat exchangers with pumps which isolate it from the first loop, which handles the water-pressure differential due to the depth;
- a liquid He cryogenic station with 4 K cold boxes with dewars, cold compressors, 2 K cold boxes, and He distribution system;
- a warm compressor whose location must take into account its vibration and noise impacts.

11.4.2.2.3 IR Hall The IR Hall (Fig. 11.8) consists of a main hall (142 m long by 25 m wide by 42 m high) that has enough space for assembly of the two detectors, as well as space for the operating detector that sits on the beamline. It also has several work areas on either side and a tunnel loop for egress. All of the central region beamline equipment, including the DR equipment, and the detector components are carried in through the IR Hall access tunnel.

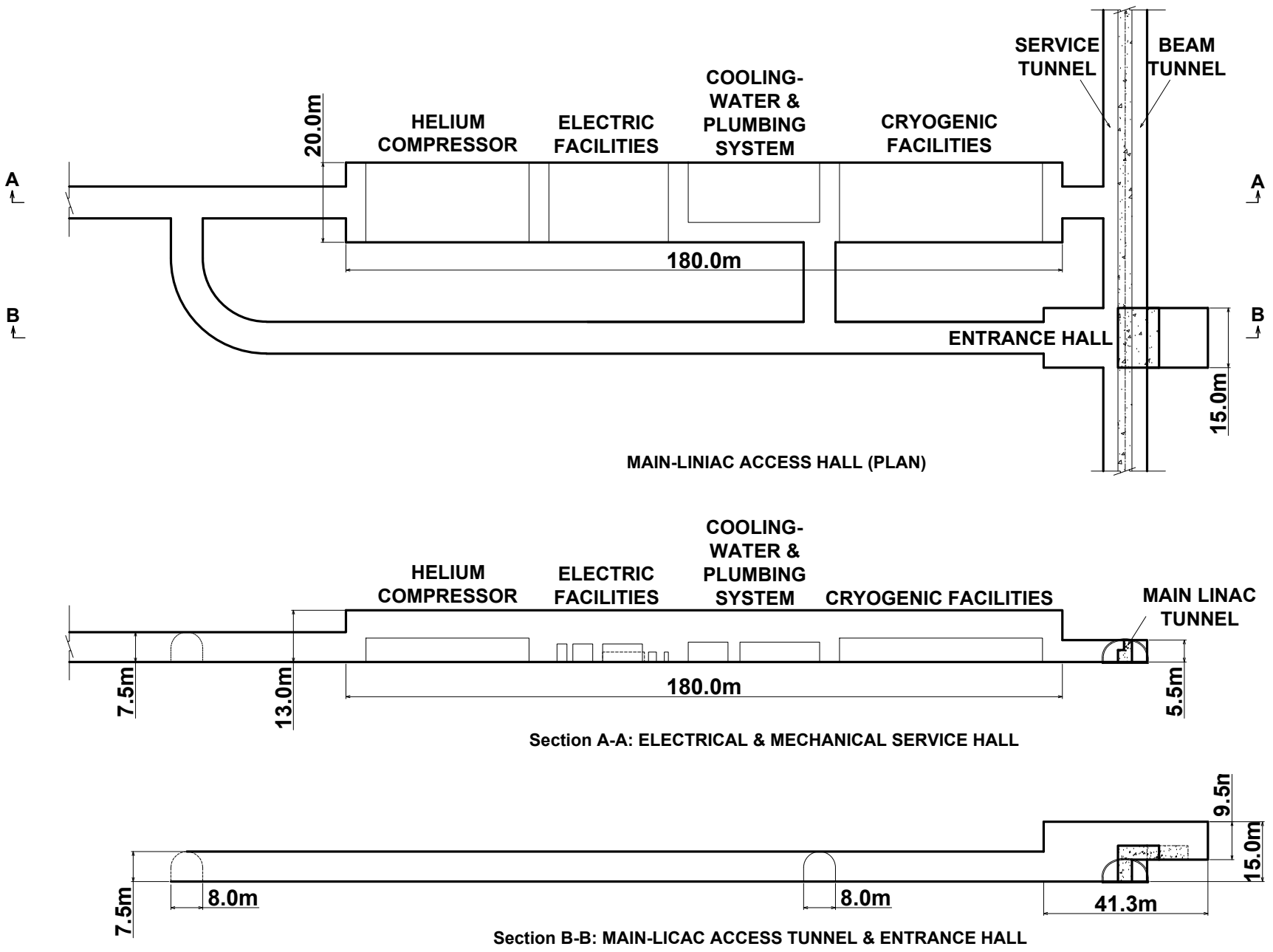

**Figure 11.7**
Typical Access Hall Plan and Elevation View

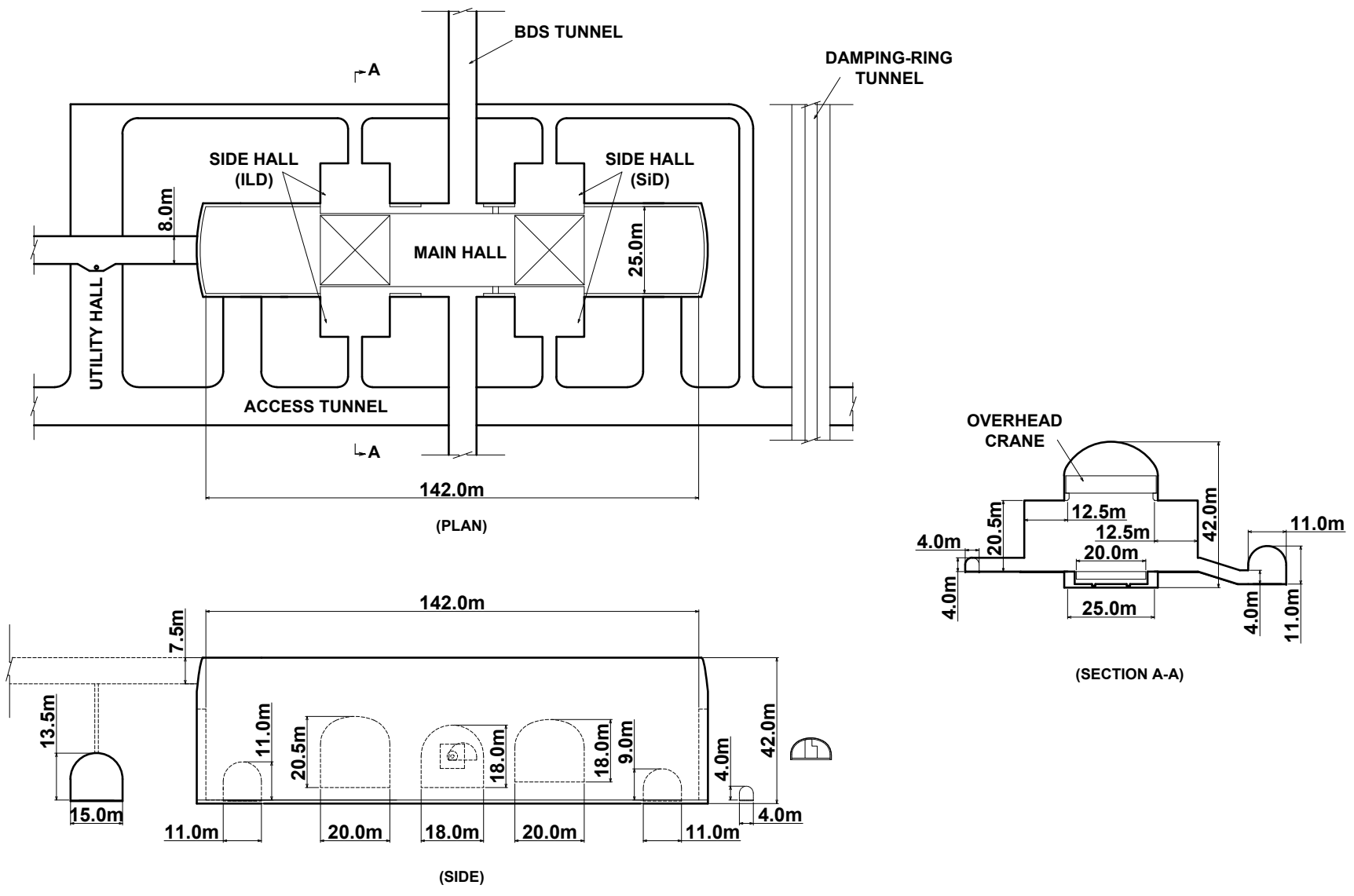

**Figure 11.8**
IR Hall

### 11.4.2.3 Access to Underground Areas

Inclined tunnels provide access to the underground facilities. There are 10 access tunnels; six for the main linacs, two for both ends of ILC, and two for the central region. A great advantage of access tunnels over vertical shafts is that vehicles can be used to transport people and equipment. A disadvantage is the long distances involved which affect the size of cooling/ventilation pipes and pumps; an alternative option is to use small-bore vertical shafts, which can be excavated by a boring machine.

The eight access tunnels at PM-13, -12, -10, -8, +8, +10, +12, +13 along the main linac and an access tunnel for the damping ring, have a vaulted section with an inner width of 8 m and height of 7.5 m (Fig. 11.9 left). The tunnel width is wide enough that two large trucks can pass each other and leave a human escape zone, and the height is sufficient to accommodate large pipes for cooling water and air ventilation. The IR Hall access tunnel (Fig. 11.9 right) is larger to allow transport of the detector solenoids from the detector assembly building on the surface. Its width is 11.0 m and its height is 11.0 m. A damping ring access tunnel is also constructed. Assuming the lengths of access tunnel, which will be known after a real site and the tunnel routes are fixed, to be 1 km in average, the total length and excavation volume of the ten access tunnels are 10 km and 643,000 $m^3$ respectively.

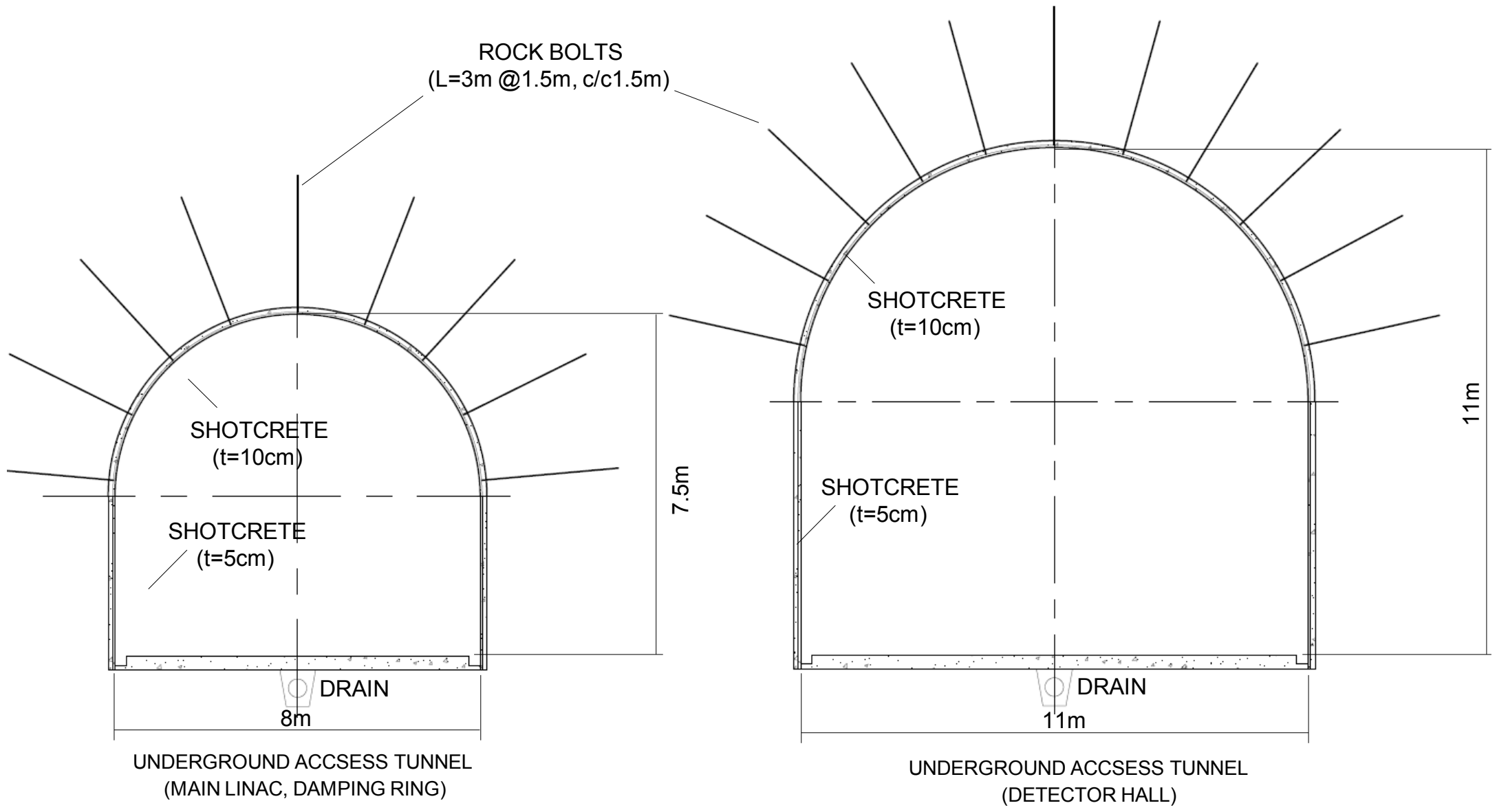

**Figure 11.9.** Underground access tunnels.

The surface entrances of the access tunnels are located near existing roads. The surface sites around the entrances support construction and are later utilised for facilities such as cooling towers. The tunnel excavation starts from the surface which is assumed to be soil or soft rock down to a depth of ~20 m. The tunnel walls are reinforced by rock bolts and finished with sprayed concrete ("shotcrete") of ~20 cm thick. The tunnel floor is 30 cm thick.

#### 11.4.2.4 Surface facilities

In these mountainous sites, some facilities that would otherwise be on the surface must be located underground. Table 11.4 summarises the area of the surface facilities. Neither of the two sites is close to an existing accelerator facility so provision must be made for general purpose buildings, accounting for roughly half of the total in Table 11.4. The remaining surface facility area is roughly 60 % of that at the Americas site.

**Table 11.4**
Asian site surface facilities. The IR surface facilities include general purpose buildings.

| Accelerator section | Qty | Area ($m^2$) |
|---|---|---|
| $e^-$ source | 0 | – |
| $e^+$ source | 0 | – |
| Damping Ring | 0 | – |
| RTML | 0 | – |
| Main Linac | 65 | 22,375 |
| BDS | 10 | 3,650 |
| IR | 28 | 65,250 |
| TOTAL | 103 | 91,275 |

### 11.4.3 Mechanical services

The main aspects of the mechanical design are:

The location and quantity of the ML heat loads (Table 11.5) are based on the DKS. ~90 % of the heat load is cooled by processed water at a temperature of ~34 °C. About 10 %of the heat loads such as air conditioning and racks are cooled by chilled water at ~9 °C. Cryomodules are in a 9-cryomodule string and RF is fed by klystrons and modulators installed in the service gallery. As a staged approach, in the baseline one klystron feeds 4.5 modules located every 54 m. Later, for the luminosity upgrade, more klystrons are added and each klystron feeds 3 modules located every 36 m. Chilled water is used to cool instrument racks and gallery air.

The cooling-water plants are located next to the cryogenic plants in the underground access halls. Heat is transferred through the access tunnels and released to the air by cooling towers on the surface near the access tunnel entrance. Cryogenic warm compressors are distributed in the underground access halls.

**Table 11.5** Summary of DKS heat loads (MW) by Accelerator section. Heat loads generated by the utilities themselves (pumps, fan motors, and etc) are listed as 'Conventional'.

| Accelerator section | load to LCW | load to Air | Conventional | Cryo (Water load) | Total |
|---|---|---|---|---|---|
| $e^-$ sources | 1.40 | 0.70 | 1.87 | 0.80 | 4.77 |
| $e^+$ sources | 5.82 | 0.64 | 2.27 | 0.59 | 9.32 |
| DR | 10.92 | 0.73 | 2.69 | 1.45 | 15.79 |
| RTML | 4.16 | 0.76 | 2.02 | part of ML cryo | 6.94 |
| Main Linac | 42.17 | 5.57 | 16.89 | 32.00 | 96.63 |
| BDS | 9.20 | 1.23 | 1.68 | 0.41 | 12.52 |
| Dumps | 14.00 | | 1.12 | | 15.12 |
| IR | 0.40 | 0.76 | 1.79 | 2.65 | 5.60 |
| TOTALS | 88.1 | 10.4 | 30.3 | 37.9 | 167 |

## 11.4.3.1 Processed water

The heat loads are distributed up to ±2.5 km from the nearest access hall. Considering both construction costs and operational safety, the cooling-water system is based on 3 loops (Fig. 11.10). The first loop includes surface cooling towers, pumps, and piping underground. The second loop provides processed cooling water out to ±2.5 km in both directions along the ML service gallery. The heat exchanger protects underground equipment against high water pressure from the surface. The third loop provides low conductivity water (LCW) to the local heat loads.

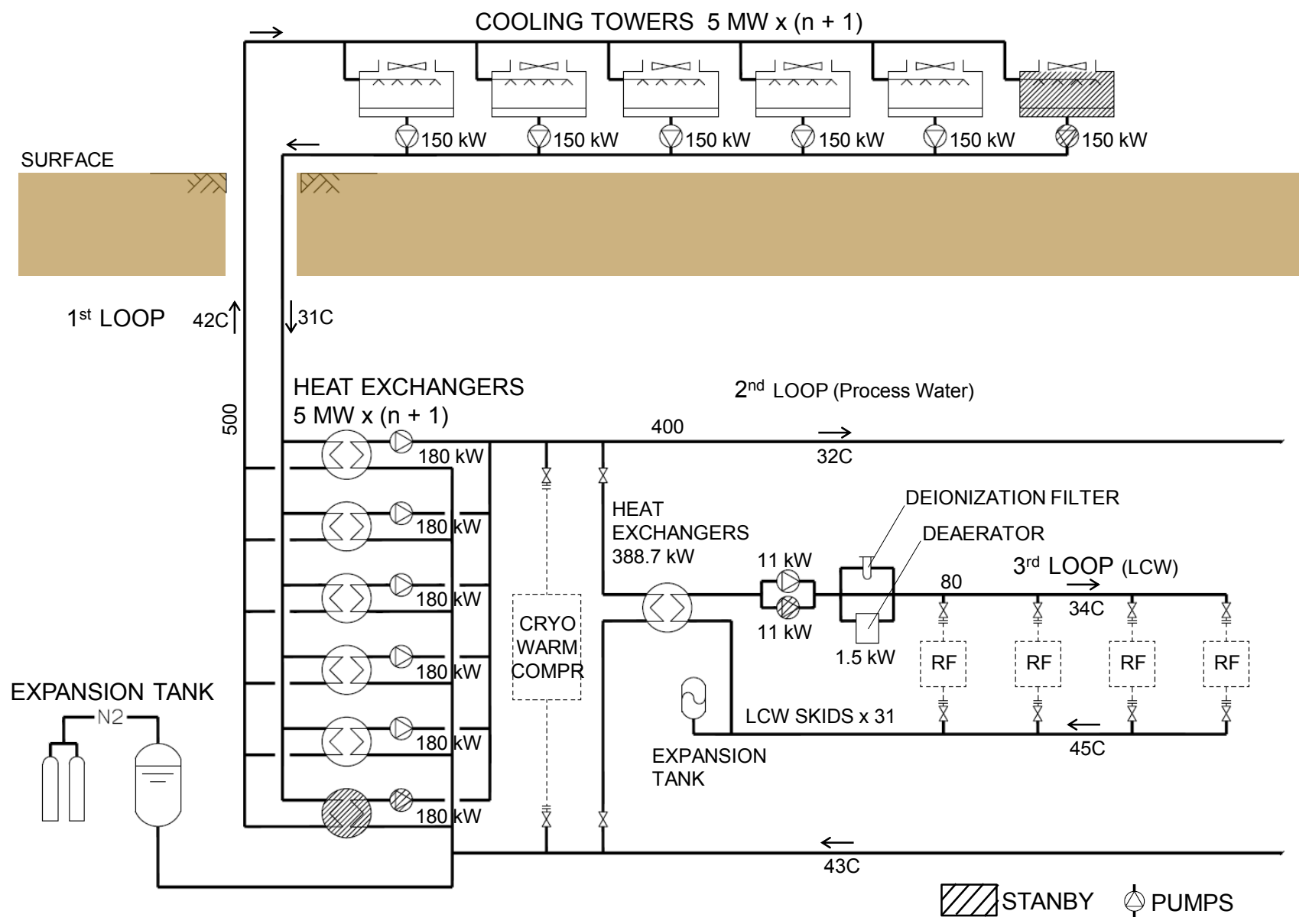

**Figure 11.10** The cooling water system in access hall 2.

11.4.3.1.1 Cooling towers The cooling towers are open-water type because of the advantages of lower construction cost, smaller footprint, and lower noise. The evaporation rate of 600 m$^3$/h for cooling 200 MW of heat load can be compensated with water flowing into the tunnel, which would otherwise need to be disposed of. A group of cooling towers with one stand-by tower is located at each access tunnel entrance, supplying cooling water of 31 °C and returning water of 42 °C.

11.4.3.1.2 Underground cooling-water loops The second-loop water temperature is 32 °C in supply and 43 °C in return water. The second loop has also a group of heat exchangers and pumps with one back-up. The third loop, which finally cools the accelerator technical equipment, needs to supply deoxygenated and demineralised water via stainless-steel pipes. It feeds four RF units using a compact cooling-water unit with heat exchanger, pump, de-aerator, and de-ionizer. The water temperature is 34 °C in supply and 45 °C in return.

11.4.3.1.3 Chilled-water system The chilled-water system (Fig. 11.11) is similar to the cooling - water system except it includes a refrigerator. Chilled water is produced by "Inverter-Turbo"-type refrigerators which have high efficiency and small $CO_2$ gas emission. The system configuration is also three loop. The third loop covers four RF units and the water temperature is 7 °C at the supply and 18 °C at the return.

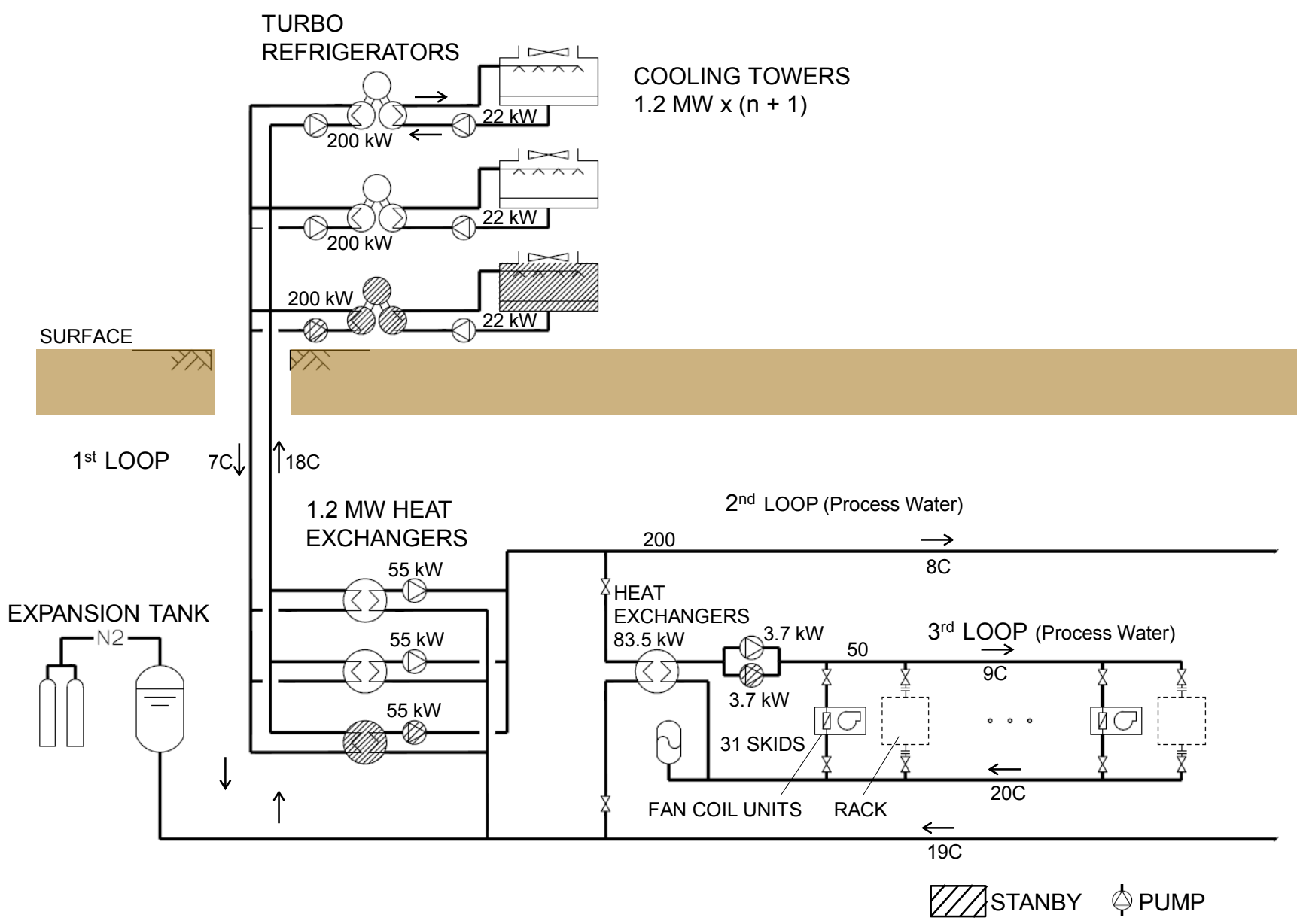

**Figure 11.11**
The chilled-water system (access hall 3).

## 11.4.3.2 Piped utilities

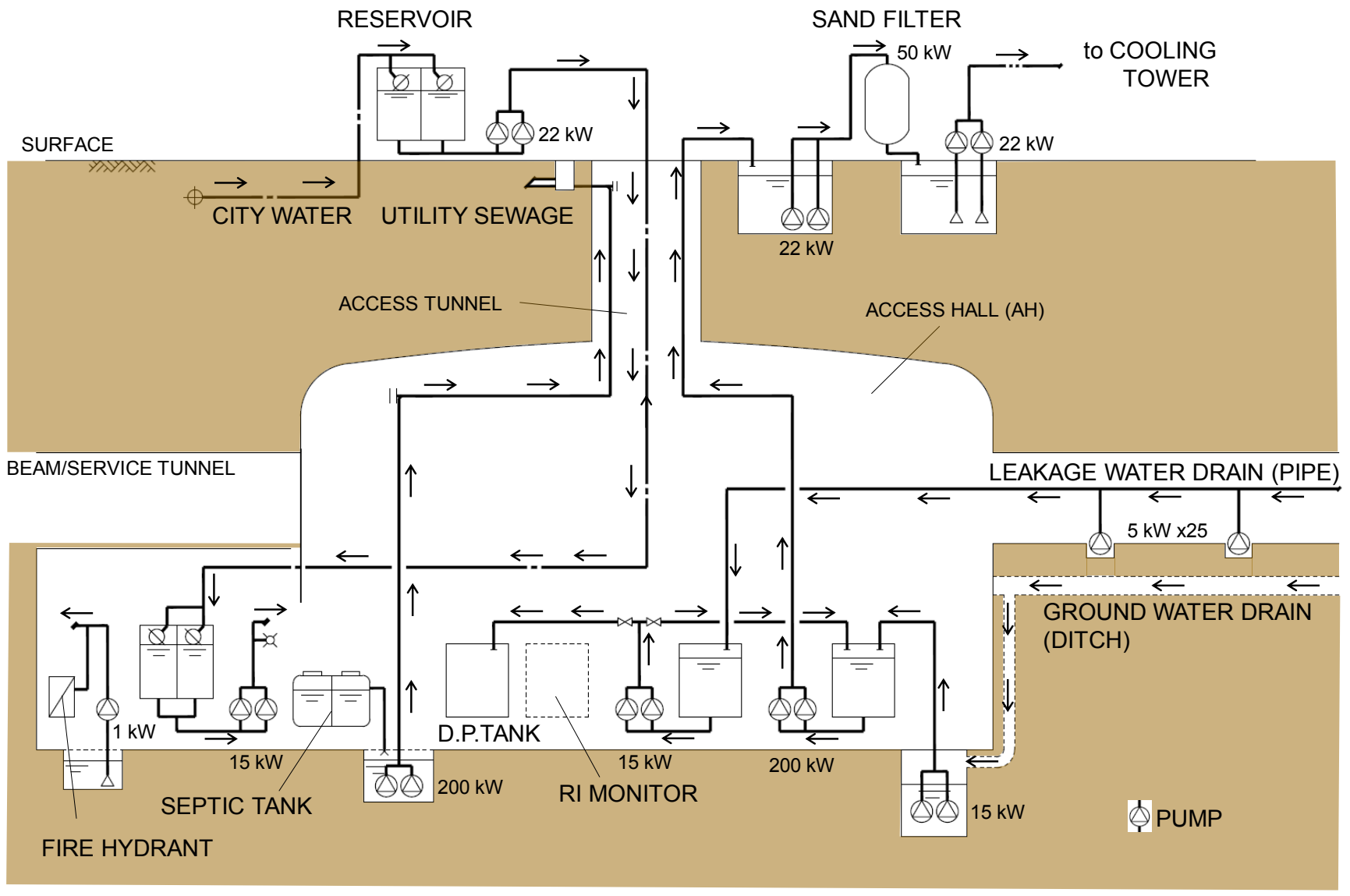

**Figure 11.12**
Piped utilities.

Figure 11.12 shows a flow chart of the piped utilities. The municipal water system is used for potable water. It is stored in tanks both on the surface and underground. Sewage water is processed and sent to a drain sewer on the surface.

The inflow water outside the thick tunnel lining is collected in a tank at each access hall. Water leaking into the tunnel is collected to pits located at intervals and pumped to the access hall tank. This water is monitored for activation, and if activated, stored in a holding tank. Otherwise, it is merged into the inflow water and pumped to the surface. Part of the water is sand-filtered and utilised for the cooling-tower makeup water.

#### 11.4.3.3 Air treatment

Fresh ambient air is treated by air-conditioning equipment on the surface (Fig. 11.13). The air is cooled and dehumidified in the summer and heated in the winter, and supplied to the underground structures by large-bore ducts installed in the access tunnels. The air blows in the tunnel without ducts at a flow rate of ∼0.5 m/s. The tunnel temperature is 29 °C and the humidity is 35 %. The service tunnel is cooled by fan-coil units using chilled water. The air is exhausted to the surface. The atmospheric pressure is controlled by dampers in the ducts so that the pressure of the service tunnel is slightly higher than the beam tunnel. The exhaust duct is also the smoke exhaust in the case of fire. Helium leakage is vented through the small-bore survey shafts located every ∼2.5 km (Fig. 11.13).

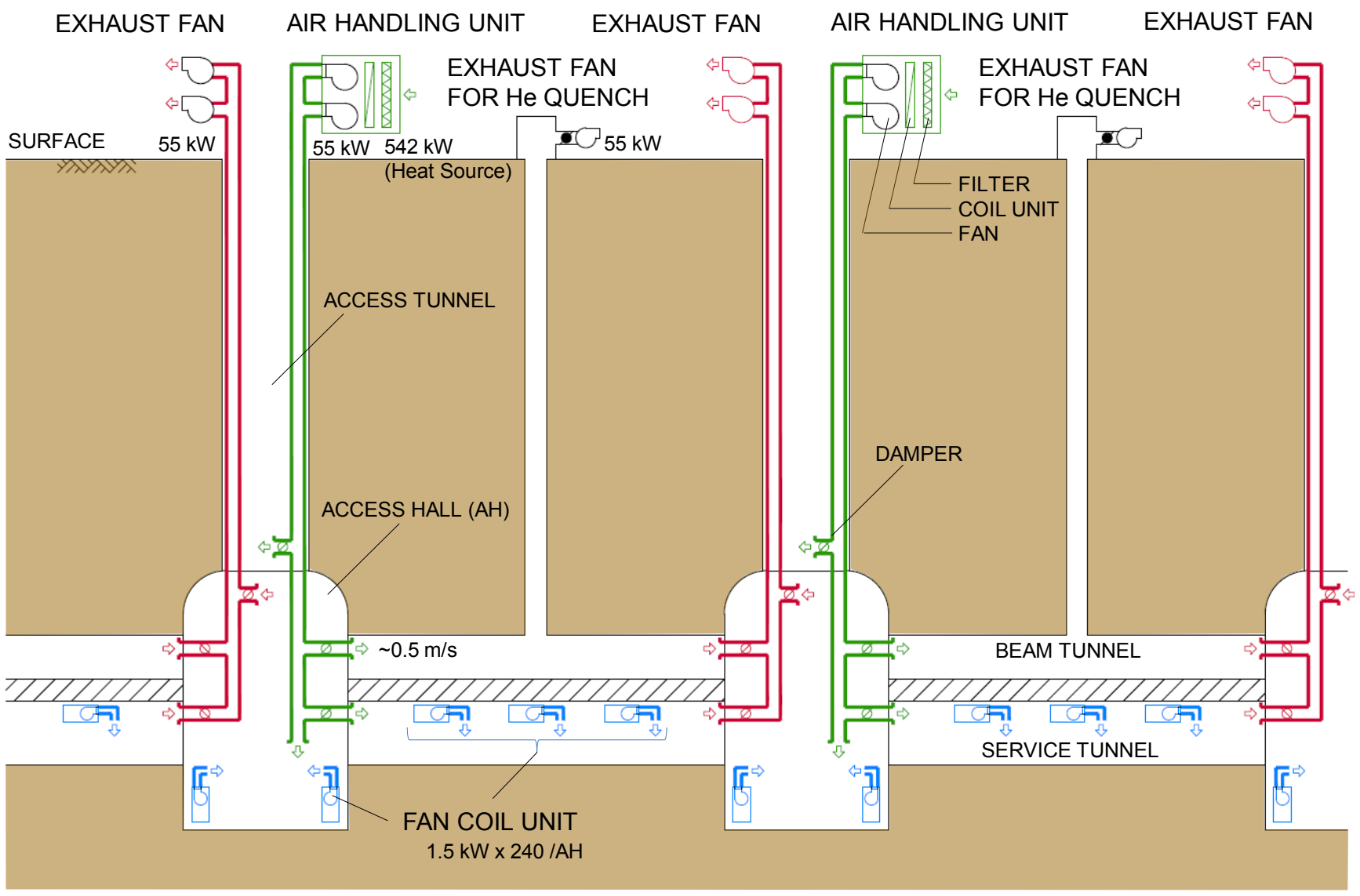

**Figure 11.13** Main Linac tunnel ventilation scheme.

### 11.4.4 Electrical

This section describes the electrical power requirements for the ILC site. A summary of electrical power loads is given in Table 11.6.

**Table 11.6** Estimated DKS power loads (MW) at 500 GeV centre-of-mass operation. 'Conventional' refers to power used for the utilities themselves. This includes water pumps and heating, ventilation and air conditioning, (HVAC). 'Emergency' power feeds utilities that must remain operational when main power is lost.

| Accelerator section | RF Power | Racks | NC magnets | Cryo | Conventional | | Total |
|---|---|---|---|---|---|---|---|
| | | | | | Normal | Emergency | |
| $e^-$ sources | 1.28 | 0.09 | 0.73 | 0.80 | 1.47 | 0.50 | 4.87 |
| $e^+$ sources | 1.39 | 0.09 | 4.94 | 0.59 | 1.83 | 0.48 | 9.32 |
| DR | 8.67 | | 2.97 | 1.45 | 1.93 | 0.70 | 15.72 |
| RTML | 4.76 | 0.32 | 1.26 | | 1.19 | 0.87 | 8.40 |
| Main Linac | 52.13 | 4.66 | 0.91 | 32.00 | 12.10 | 4.30 | 106.10 |
| BDS | | | 10.43 | 0.41 | 1.34 | 0.20 | 12.38 |
| Dumps | | | | | 0.00 | 1.21 | 1.21 |
| IR | | | 1.16 | 2.65 | 0.90 | 0.96 | 5.67 |
| TOTALS | 68.2 | 5.2 | 22.4 | 37.9 | 20.8 | 9.2 | 164 |

### 11.4.4.1 Electrical power distribution

The electrical power is distributed in three stages:

- the site electric power is stepped down from local-district high voltage (150-500 kV) to 66 kV in the main substation and distributed to the 6 access hall and the IR Hall substations;
- the 66 kV electricity is further stepped down to 6.6 kV at each substation and distributed inside the areas;
- the electric loads such as RF modulators and cryogenic warm compressors are powered directly at 6.6 kV and other local loads are fed at lower voltages stepped down in local substations distributed along the accelerator.

### 11.4.4.2 Main substation and 66 kV power distribution

A primary voltage of 275 kV was assumed for the site. The single-line diagram of the main substation is shown in Fig. 11.14. The primary-line configuration is a two-way system including a stand-by line. The power capacity is designed to be 300 MW and space is reserved for an additional 200 MW for the future 1 TeV upgrade.

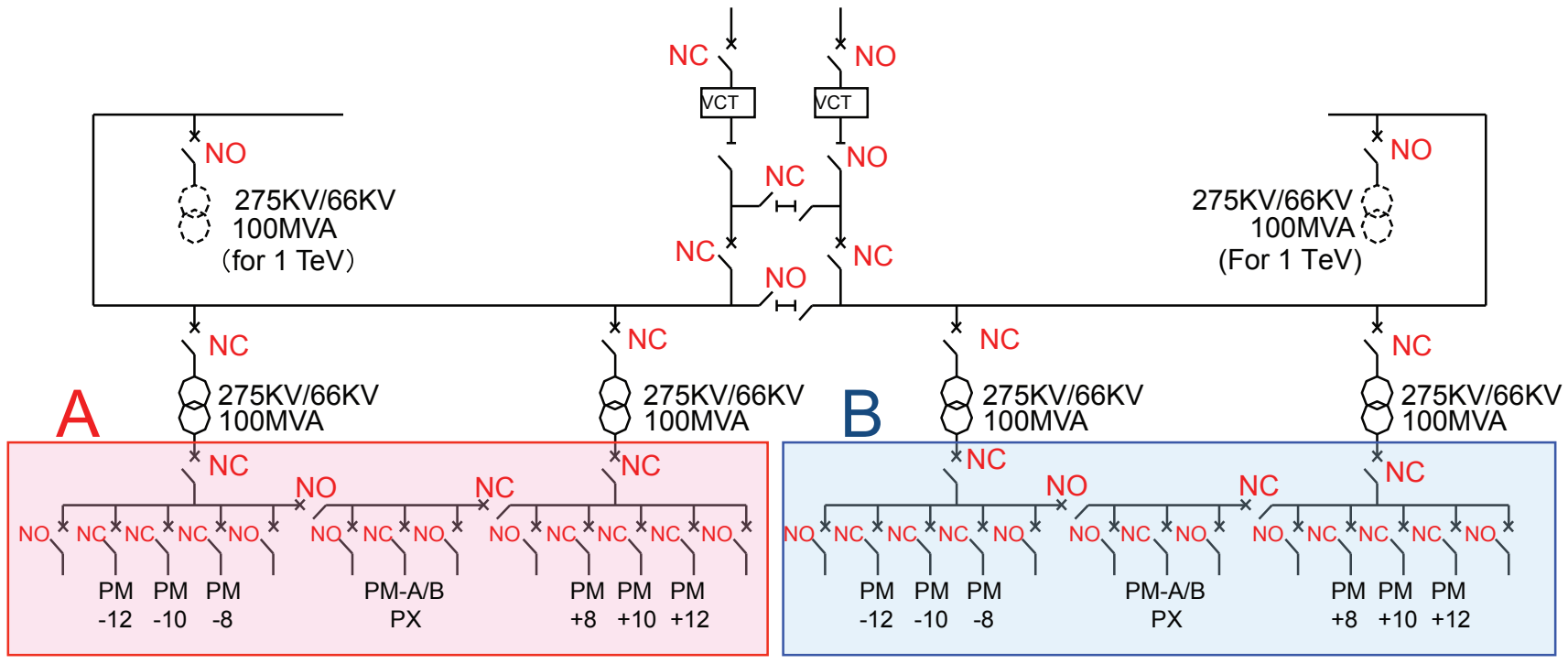

**Figure 11.14** Single-line diagram for the main substation.

The main transformers have an n+1 redundancy scheme and use four banks of 100 MW transformers. The switching gears are gas insulation type. They are located in an outside yard of area $\sim$4000 m$^2$. The secondary voltage is 66 kV and the power is distributed through the BDS and ML service galleries and access halls with two pairs of three-phase cables.

### 11.4.4.3 Access hall substations

With a power range between 28 MW and 44 MW, two 30 MVA 66 kV/6.6 kV transformers are required at each substation, allowing more than a half of the operational power to be maintained in case of a transformer fault. There is one spare transformer at the main substation, with capacitors to improve power efficiency. The single-line diagram and the equipment layout in the hall are shown in Fig. 11.15.

### 11.4.4.4 Local substations

The local distribution board diagram is shown in Fig. 11.16. There are 6.6 kV boards for the modulators at an interval of every four RF units, and cryogenic compressors in the access halls. The local substations step down 6.6 kV to lower voltages at an interval of every four RF units.

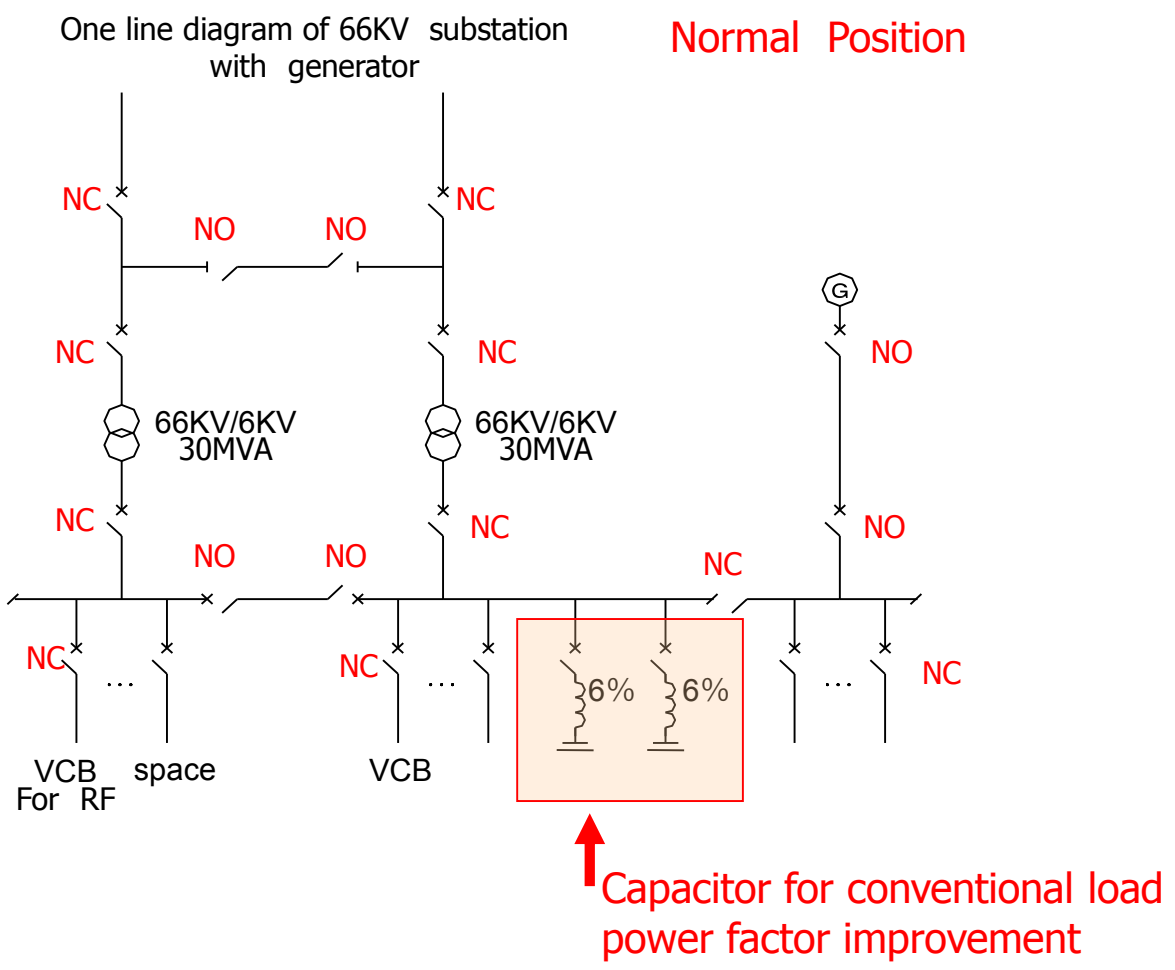

**Figure 11.15**
Access hall substation single-line diagram and the equipment layout.

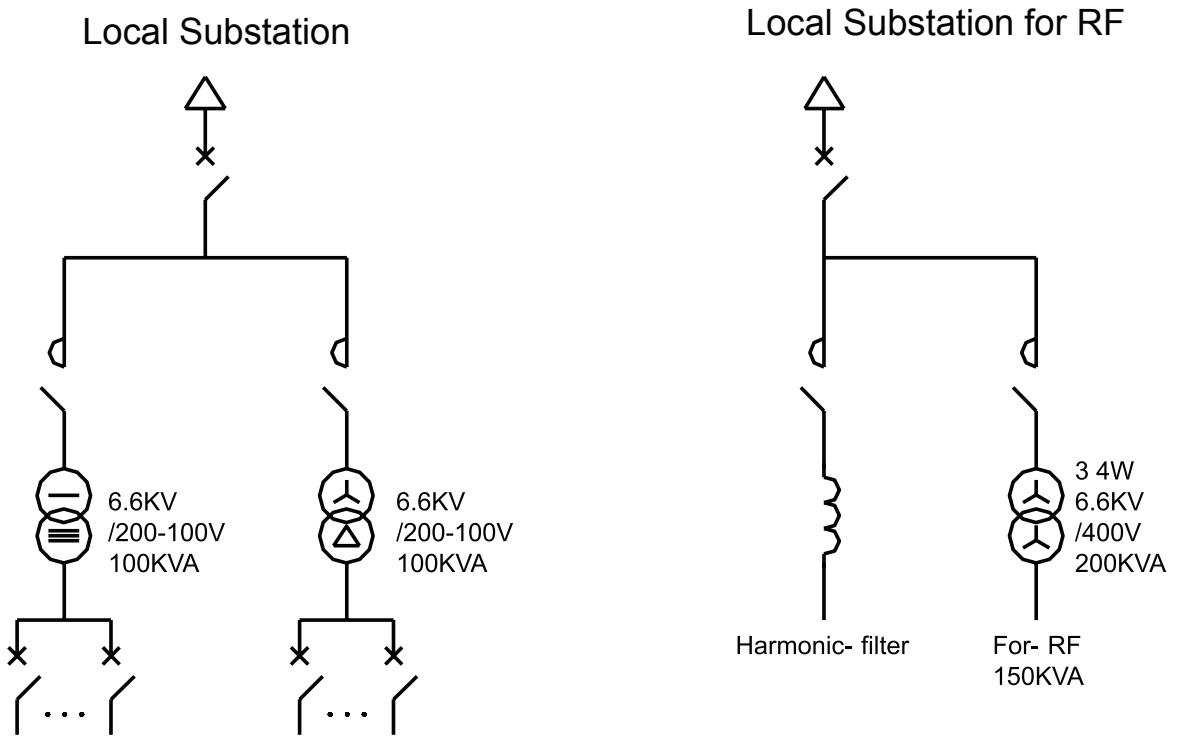

**Figure 11.16**
Local distribution panels.

### 11.4.4.5 Emergency and backup electrical power

There are emergency generators at each of the seven 66 kV substations. The emergency power generators are adequate for fire-fighting and to maintain minimal functioning of the building and the compressor for He-gas storage during an electricity outage. Each of seven diesel-engine generators installed at the surface yards supplies the underground 66 kV substation with ~1 MVA power.

DC power supplies are used for the substation control system and emergency lights. They are installed in seven AHs and the IR Hall. Chargeable batteries are used for tunnel emergency lights, evacuation lights, and local substations. The equipment is a cubicle system and valve-regulated, sealed-cell lead-acid batteries are used.

To provide backup power for critical systems, UPSs are installed in each control room beside the substations. Technical equipment includes its own UPS where necessary.

## 11.4.5 Life safety and egress

### 11.4.5.1 Fire safety

There are no existing laws and design guidelines in Japan which specify safety and disaster prevention measures for deep underground tunnels. A special committee established by the Japan Society of Civil Engineers is currently reviewing the basic policy proposal on the disaster prevention design for the ILC underground facilities.

Of primary importance in an underground tunnel is safe refuge when a fire breaks out. However, the distance to the surface via an access hall can be as long as 5-6 km and a secondary evacuation route is required. This is provided by access passages located every 500 m along the ML that connect the two galleries so the other gallery can provide an escape route (see Section 10.5.3).

Evacuation from the tunnel to the surface is via the access hall and the access tunnel every 5 km. Even with rapid egress, it can take up to 1 hour to reach the surface. If a fire is detected, the partition door and damper for the access passage will close automatically, and will prevent smoke reaching the escape route.

Each of the two galleries is separately ventilated from the access halls. There is no separate emergency smoke-control system. The main ventilation system switches to a smoke-exhaust function automatically in case of a fire.

There is no installed fire-extinguishing sprinkler system to avoid possible water damage to the accelerator machine and experimental equipment. The ML tunnel is equipped with the following standard emergency equipment:

- smoke detector and fire detector;
- fire alarm system;
- emergency lighting system;
- emergency illuminated exit signs;
- emergency exit guide lights;
- fire extinguishers.

#### 11.4.5.2 Safety for Helium

Since there is a large quantity of liquid helium in the ML tunnel, oxygen deficiency monitoring is required throughout. When the oxygen concentration drops below an acceptable level, emergency measures are taken and an alarm sounds. The main ventilation system switches to emergency mode and the helium gas from the upper part of the tunnel is discharged outside by exhaust shafts in the access tunnels.

## 11.5 European region (Flat topography)

### 11.5.1 Siting studies

Two European sites were considered: the Geneva region (deep-site study), along the French-Swiss border and the Dubna region (shallow-site study) in Russia. The European design is based on the KCS RF concept developed by the Americas Region. This assumes that as much as possible of the technical equipment is housed on the surface in order to minimise the underground enclosure volumes.

#### 11.5.1.1 Geneva region (deep site)

11.5.1.1.1 Location This site is set in the North-Western part of the Geneva region, near the CERN laboratory (Fig. 11.17). Since no real discussions with local authorities have taken place, this position is only indicative. The IR is fully located within existing CERN land at the Prevessin Campus. The new underground structures will mostly be constructed at a depth of 100-150 m in stable Molasse rocks in an area with moderate seismic activity.

All necessary infrastructure to accommodate the project is available in the Geneva area. This includes the possibility of accommodating specialists for the accelerator construction period, storage and assembly of equipment, and the provision of project-production support during manufacturing of the special-purpose equipment. Excellent transport and communication networks already exist.

11.5.1.1.2 Land Features The proposed location is within the Swiss midlands embedded between the high mountain chains of the Alps and the lower mountain chain of the Jura. CERN is situated at the foot of the Jura mountain chain in a plain slightly inclined towards Lake Geneva. The absolute altitude of the surface ranges from 430 m to 500 m with respect to sea level.

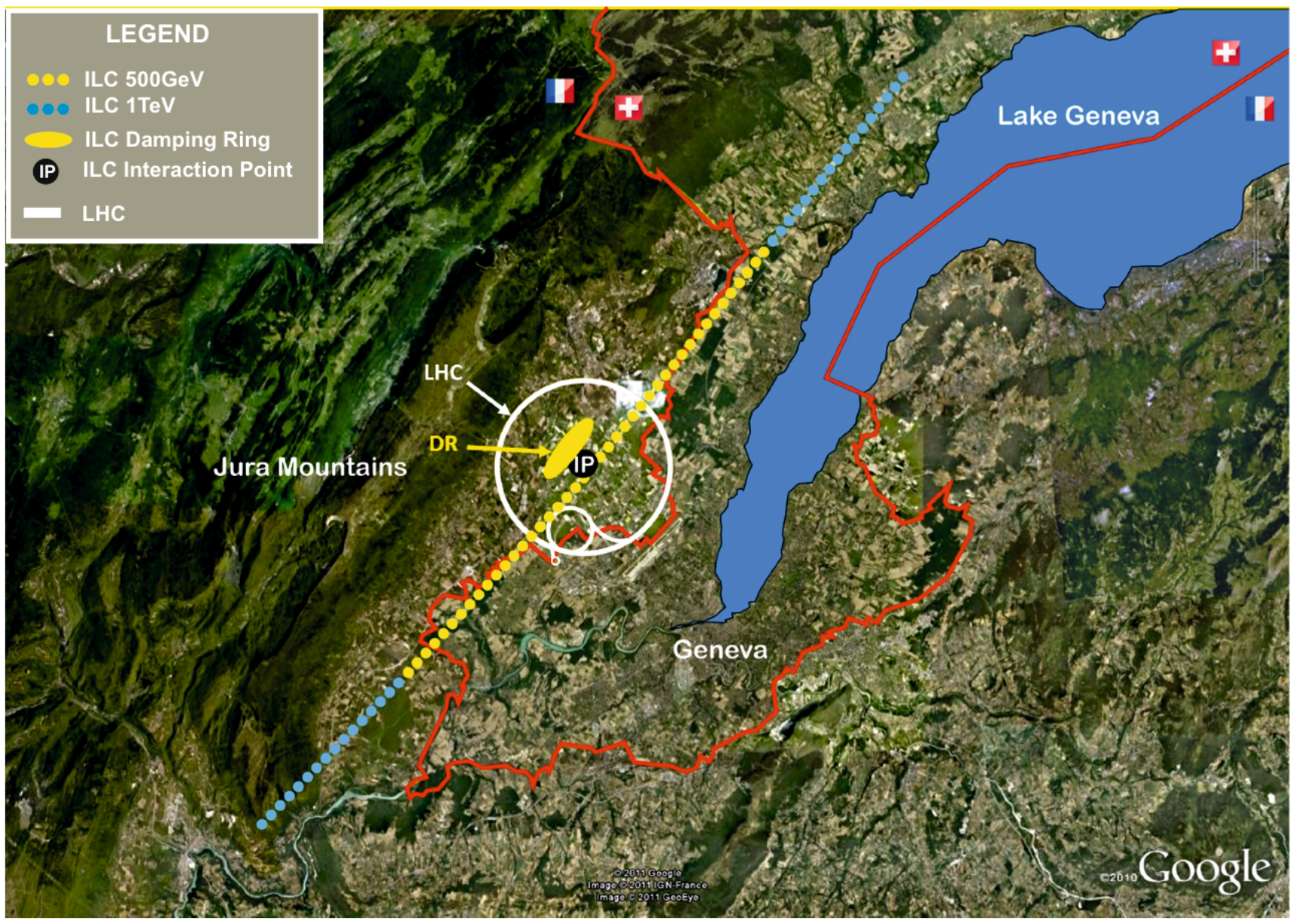

**Figure 11.17** The potential location of ILC in the Geneva region.

11.5.1.1.3 Climate The climate of the Geneva region is temperate, with mild winters and warm summers. The mean annual air temperature is 9.6 °C, with a maximum temperature of 25.7 °C in July and a minimum temperature of −1.9 °C in January. The mean annual relative humidity is 75%. Precipitation is well-distributed throughout the year, with a mean annual precipitation of 954 mm. An average of 42.5 cm of snow falls in the period November to March.

11.5.1.1.4 Geology Most of the proposed path of the ILC is situated in the Geneva Basin, a sub-basin of the large North Alpine Foreland (or Molasse) Basin. Characterized as stable and impermeable, the Molasse rock is considered to be very suitable for underground constructions. A simplified geological profile of the region is shown in Fig. 11.18.

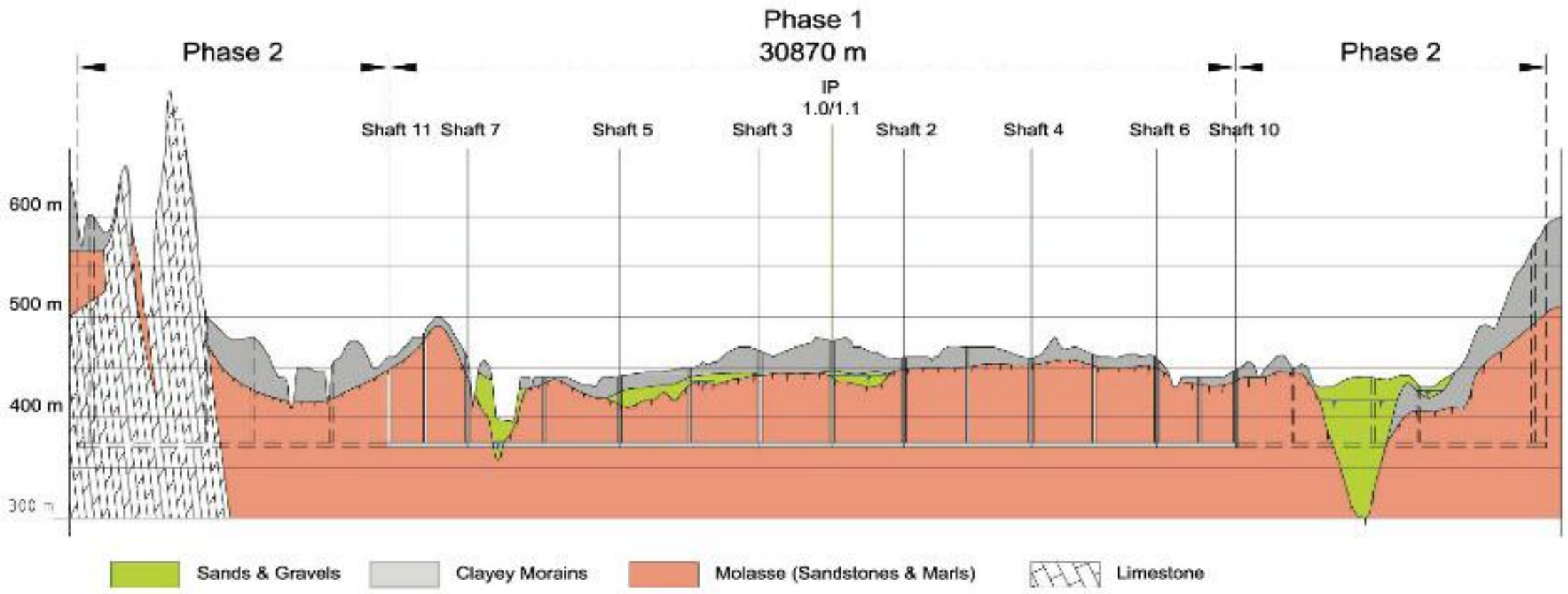

**Figure 11.18.** Simplified geological profile. ILC is mostly housed in the Molasse Rock.

11.5.1.1.5 Power Availability The CERN region has a very well developed electrical supply with a 400 kV line coming into the Prevessin Site on the French side of the campus and a 130 kV line arriving on the Meyrin Site on the Swiss side. The existing CERN networks carry electrical energy to major sub-stations via 66 kV and 18 kV underground links. Final loads are supplied at 18 kV, 3.3 kV or 400 V.

11.5.1.1.6 Construction Methods For the upper parts of the shafts, located in dry moraines up to 50 m depth, traditional excavation means are foreseen. Where water-bearing units are encountered the ground-freezing technique will be used to allow safe excavation of the shafts under dry conditions. This involves freezing the ground with a primary cooling circuit using ammonia and a secondary circuit using brine at −23 °C, circulating in vertical tubes in pre-drilled holes at 1.5 m intervals. Besides creating dry conditions, the frozen ground acts as a retaining wall.

When the underlying rock (sandstone) is reached the shafts and caverns will be excavated using rock breakers and road headers. A temporary lining will be set in place using rock bolts, mesh and shotcrete, after which the walls and vaults will be sealed with waterproof membranes and covered with cast in-situ reinforced concrete.

The underground enclosures have diameters of 5.2 m, 6.0 m, 8.0 m and up to 12.0 m. For the Molasse rock, it is estimated that it is cheaper to excavate these tunnels using a TBM with 8.0 m diameter for the entire length of the BDS tunnels, with some local cavern enlargements using roadheaders in a second phase.

### 11.5.1.2 Dubna Region (shallow site)

11.5.1.2.1 Location The Dubna area provides a potential shallow tunnel site. The Joint Institute for Nuclear Research (JINR) is an International Intergovernmental Organization and has experience of organizing and realizing large-scale research projects based on international cooperation among scientific centers and industrial enterprises. Figure 11.19 shows the site location.

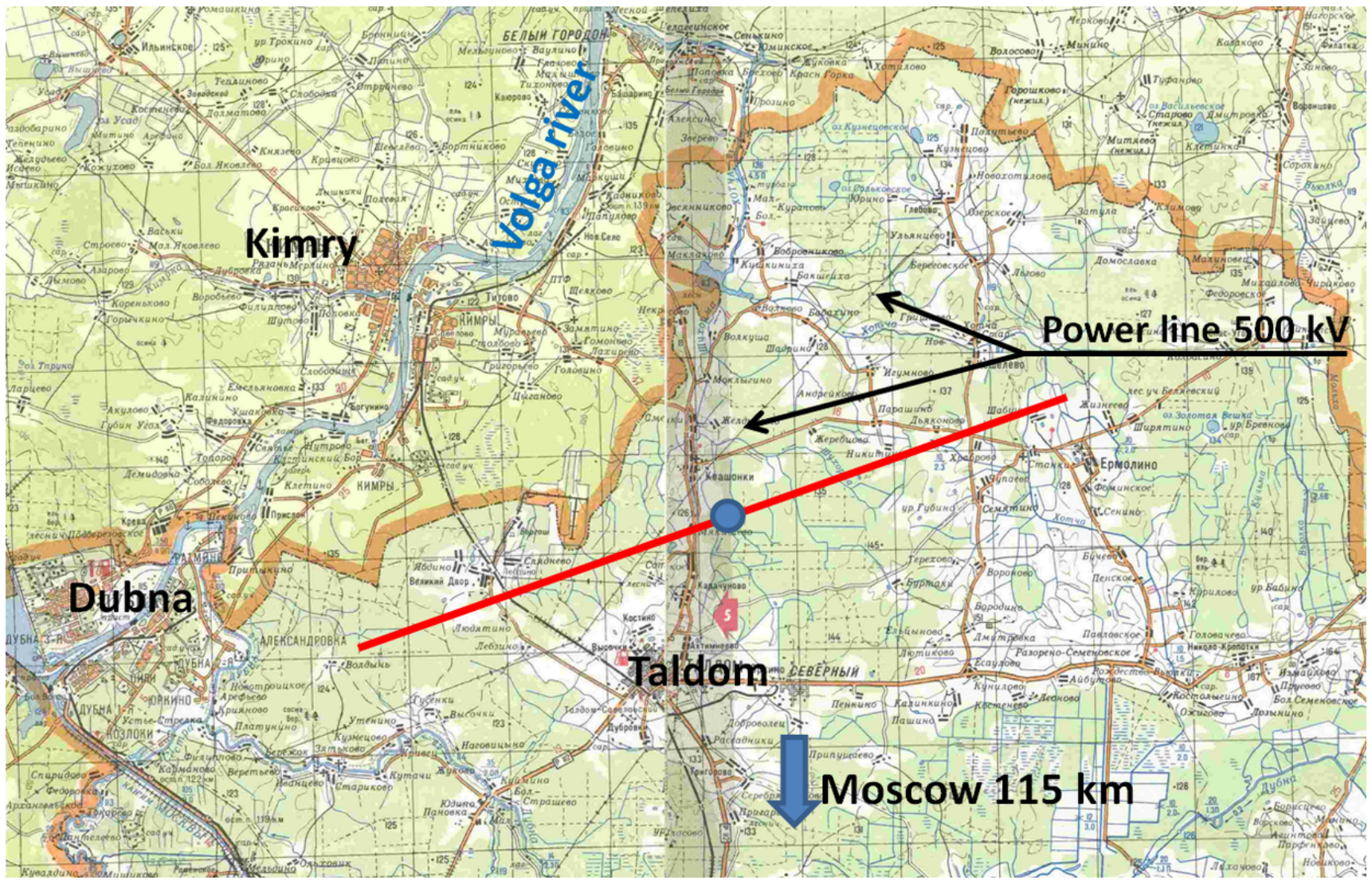

**Figure 11.19** The proposed path of the accelerator construction (indicated as a red line) in the Dubna region

Due to the special economic zone established in Dubna in December 2005, preferential terms for the development and manufacturing of high technology products are provided. Furthermore, the prevalent legal conditions in the Dubna region provide the opportunity to acquire land free of charge, as has been the case for JINR, with the agreement of the Russian Federation government.

11.5.1.2.2 Land features The main feature of the proposed location is a flat topography, with an altitude ranging between 125 m and 135 m above sea level. The relief increases away from the site as the plain changes into smoothly sloping separate hills. The area is swampy with potential waterlogged conditions. During floods of the Dubna River, the groundwater level increases by up to between 0.6 m and 0.9 m, and a high percentage of the area is flooded. The territory is sparsely populated and practically free of industrial structures. The region around the accelerator path is mainly covered with forests containing small inclusions of agricultural land. The accelerator path traverses two small settlements and a railway with light traffic between Taldom and Kimry. Construction will not affect national parks or religious and historical monuments. Infrastructure and communication systems are in place.

11.5.1.2.3 Climate The region is characterized by a moderate continental climate with long and relatively cold winters and warm summers. The average annual air temperature is +3.10 °C, with a maximum of +36.0 °C and a minimum of -43.0 °C. The average maximum air temperature of the hottest and coldest months is +22.7 °C, and -19.0 °C respectively. The average monthly relative air humidity in the region during the coldest and the hottest months is 84% and 57% respectively. The annual rainfall is 630 mm, of which 447 mm precipitates during the warm period (April-October) and 183 mm during the cold period (November-March). Snow cover typically starts in November, with an average snow depth of 30-40 cm in open places during the winter period.

11.5.1.2.4 Geology The site is situated within the Russian Plate, which is a part of the ancient East-European platform. The area is located in the southern part of a very gently sloping saucer-shaped structure, called the Moscovian syncline. The top layer consists of alluvial deposits, i.e. fine water-saturated sands with a varying thickness of 1 m to 5 m. These deposits cover the underlying semi-solid moraines of the Moscovian glaciation, which contains inclusions of detritus and igneous rocks. The thickness of the moraine deposits is between 30 m and 40 m. The moraines cover the fluvio-glacial saturated sands and loams of the Dnieper glaciation. Jurassic clays and carboniferous limestones are located at a depth of 50 m–60 m. The region has low seismic activity.

As the ILC is proposed to be placed in the moraines, at a depth of 20 m, an impermeable soil layer should be present under the tunnel to prevent water inflow from underlying water-bearing units (see Fig. 11.20). Overall, the available data show that the geological, hydrological and geotechnical conditions are favourable.

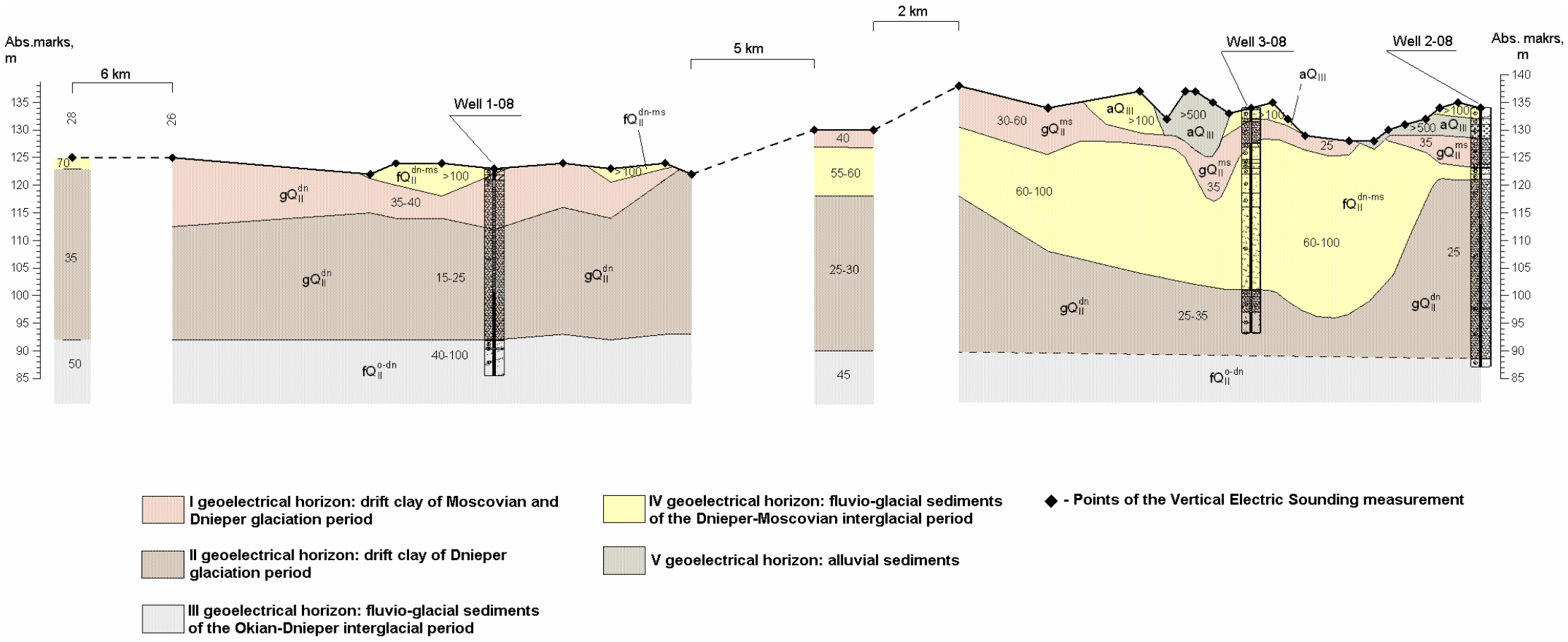

**Figure 11.20.** A geology study near Dubna

11.5.1.2.5 Power availability The northern part of the Moscow region, as well as the neighboring regions, has a developed electrical energy generation, transmission and distribution network. Two trunk transmission lines with voltages of 220 kV and 500 kV pass through the Dubna territory. The proposed ILC path is deliberately placed nearby and parallel to these power lines.

11.5.1.2.6 Construction methods A one-tunnel solution for the accelerator structures is possible at the Dubna site. A communication tunnel will be placed directly above the accelerator tunnel near the ground surface at a depth of 3 m-4 m. This tunnel is necessary for power supplies, RF power sources, data storage devices, electronic and control systems, etc. Near sub-surface buildings would be constructed by an open pit method and the tunnel could be constructed using TBMs, although 'cut and cover' construction techniques are possible over nearly the whole length.

### 11.5.2 Civil construction

The European design is developed to fit the local geological and environmental constraints of the Geneva area. Studies performed by external consultants [210–212] have been aimed at minimising the infrastructure costs, such as civil engineering, which are main cost-drivers of the project. This section describes the technical designs for the civil engineering for the 500 GeV baseline.

#### 11.5.2.1 Overall site layout

Figure 11.21 shows a schematic layout of the civil engineering complex. Key characteristics of the ILC baseline layout are:

- tunnel footprint of approximately 31 km, positioned at 100-150 m depth;
- horizontal tunnel following a geoid surface;
- IR and injection complex fully located on the Prevessin Campus;
- ML housed within a single tunnel with an internal diameter of 5.2 m;
- two turn-around tunnels connected to the ML with a bending radius of 30 m in the horizontal plane;
- a service tunnel, linking the ML with the IR Hall and the DR;

  two additional RTML tunnels are planned for the central injector region, connecting the DR and the sources.
- two independent caverns for detector assembly and maintenance linked via a transfer tunnel;
- Shafts and surface installations approximately every 2 km along the ML.

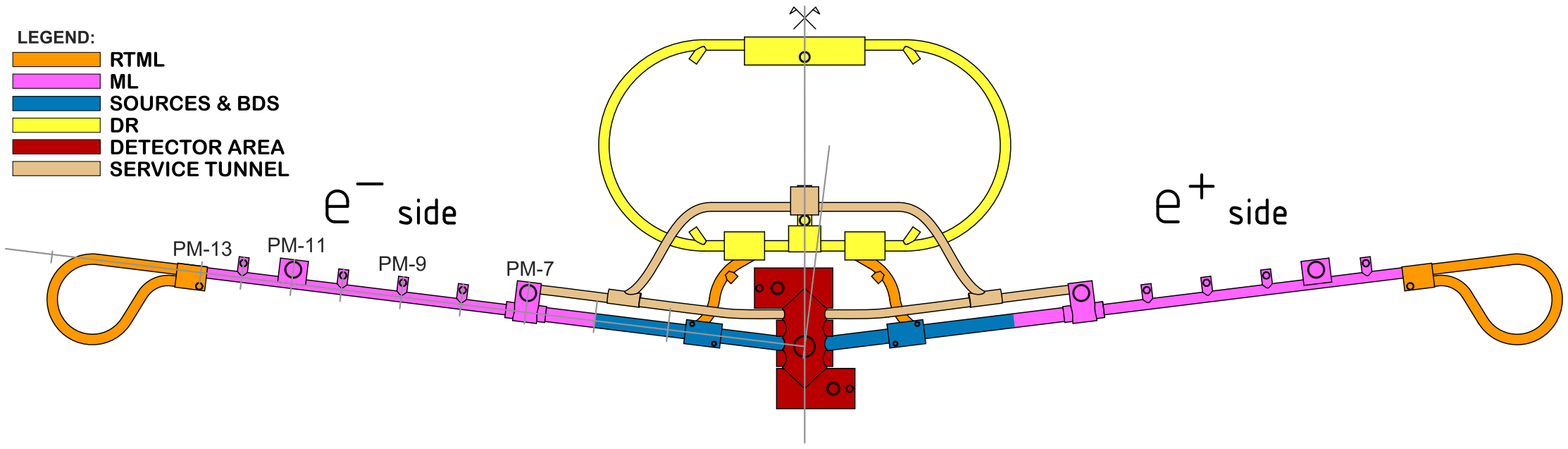

**Figure 11.21.** Schematic layout of the civil engineering complex

#### 11.5.2.2 Underground Construction

Table 11.7, Table 11.8, and Table 11.9 summarize the European underground construction tunnels, caverns and access shafts, respectively. Of all the deep-tunnel sites studied, the soft 'molasse' sandstone near CERN is the most soft and weakest. For this reason, the cavern ceilings are dome-shaped leading to a larger overall excavation.

**Table 11.7**
Tunnel lengths and volumes by Accelerator section

| Accelerator section | Length (m) | Volume ($m^3$) |
|---|---|---|
| $e^-$ source (beam) | 368 | 18,522 |
| $e^-$ source (service) | 618 | 9,828 |
| $e^+$ source (beam) | 1,678 | 84,329 |
| $e^+$ source (service) | 2,203 | 35,038 |
| Damping Ring | 3,239 | 91,571 |
| RTML (beam) | 3,305 | 74,546 |
| RTML (service) | 1,955 | 31,090 |
| Main Linac | 22,168 | 470,782 |
| BDS (beam) | 3,847 | 193,379 |
| BDS (service) | 3,847 | 61,183 |
| TOTAL | 43,228 | 1,070,268 |

**Table 11.8**
Cavern summary and volumes

| Accelerator section | Qty | Volume ($m^3$) |
|---|---|---|
| $e^-$ source | 1 | 2,029 |
| $e^+$ source | 1 | 6,715 |
| Damping Ring | 6 | 59,604 |
| RTML | 10 | 20,312 |
| Main Linac | 12 | 41,280 |
| BDS | 6 | 26,292 |
| IR | 5 | 127,100 |
| TOTAL | | 283,332 |

**Table 11.9**
Shaft depths and volumes

| Accelerator section | Depth (m) | Volume ($m^3$) |
|---|---|---|
| $e^-$ source | 100 | 0 |
| $e^+$ source | 100 | 0 |
| Damping Ring | 100 | 12,723 |
| RTML | 100 | 5,655 |
| Main Linac | 100 | 69,665 |
| BDS | 100 | 707 |
| IR | 100 | 39,584 |
| TOTAL | | 128,334 |

### 11.5.2.3 ML tunnel

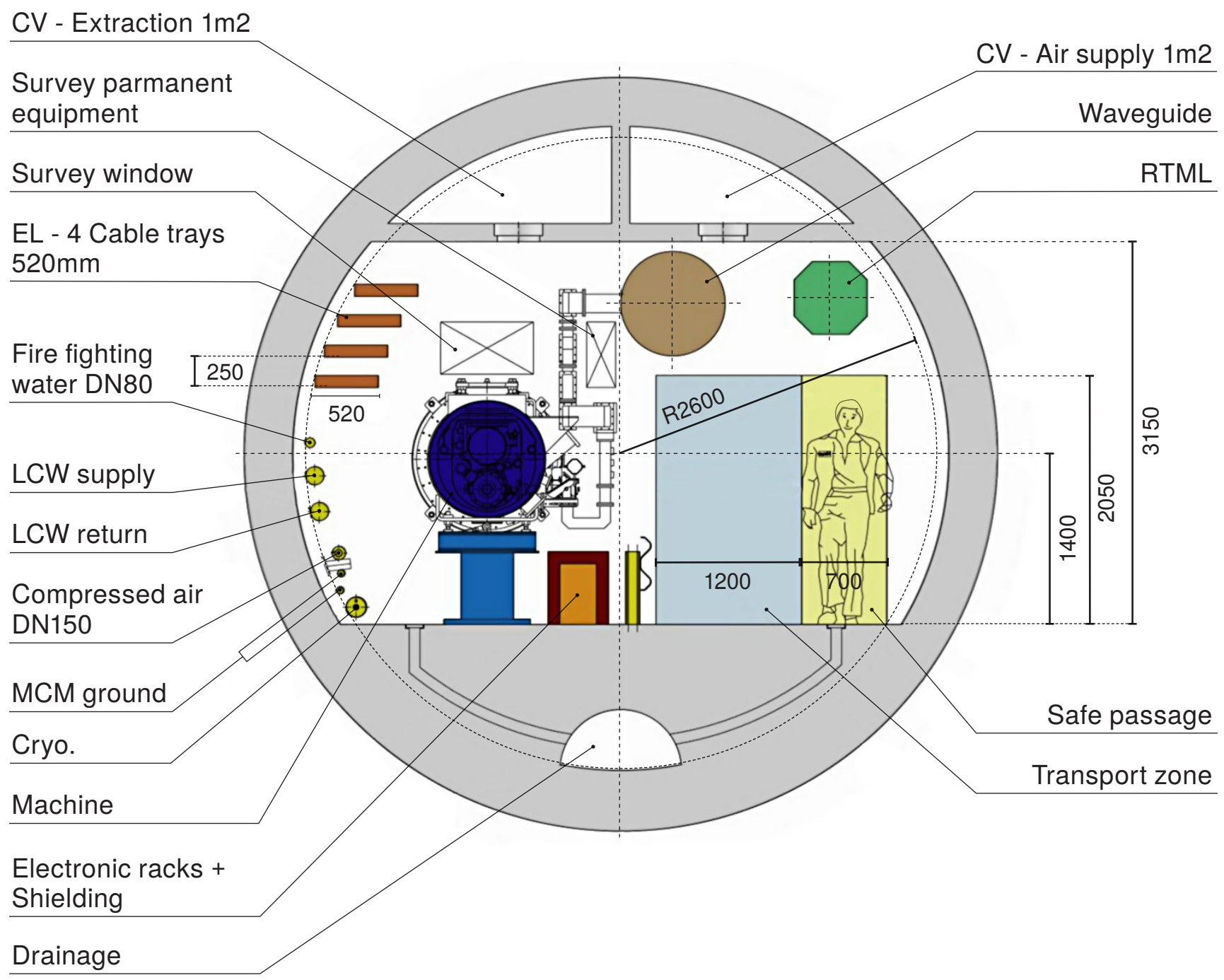

**Figure 11.22**
Typical tunnel cross section.

Figure 11.22 shows a typical cross section of the ML tunnel. The diameter has been optimized through 3D modelling of the accelerator and its services. The diameter is within the common range of TBMs used for metro transportation tunnels; machinery and spare parts are easily found on the market.

A driving factor of the tunnel size is the ventilation concept, adopted mainly for safety reasons. This differs from the LHC, which has a longitudinal ventilation scheme. Cryo-modules are attached directly to the tunnel floor, which minimizes ground movement and allows for easy access.

### 11.5.2.4 Central injection region

The central injection complex (Fig. 11.23) consists of DRs, polarised electron and positron sources and the electron and positron 5 GeV SCRF injector linacs.

**Figure 11.23**
Model of the central injection region.

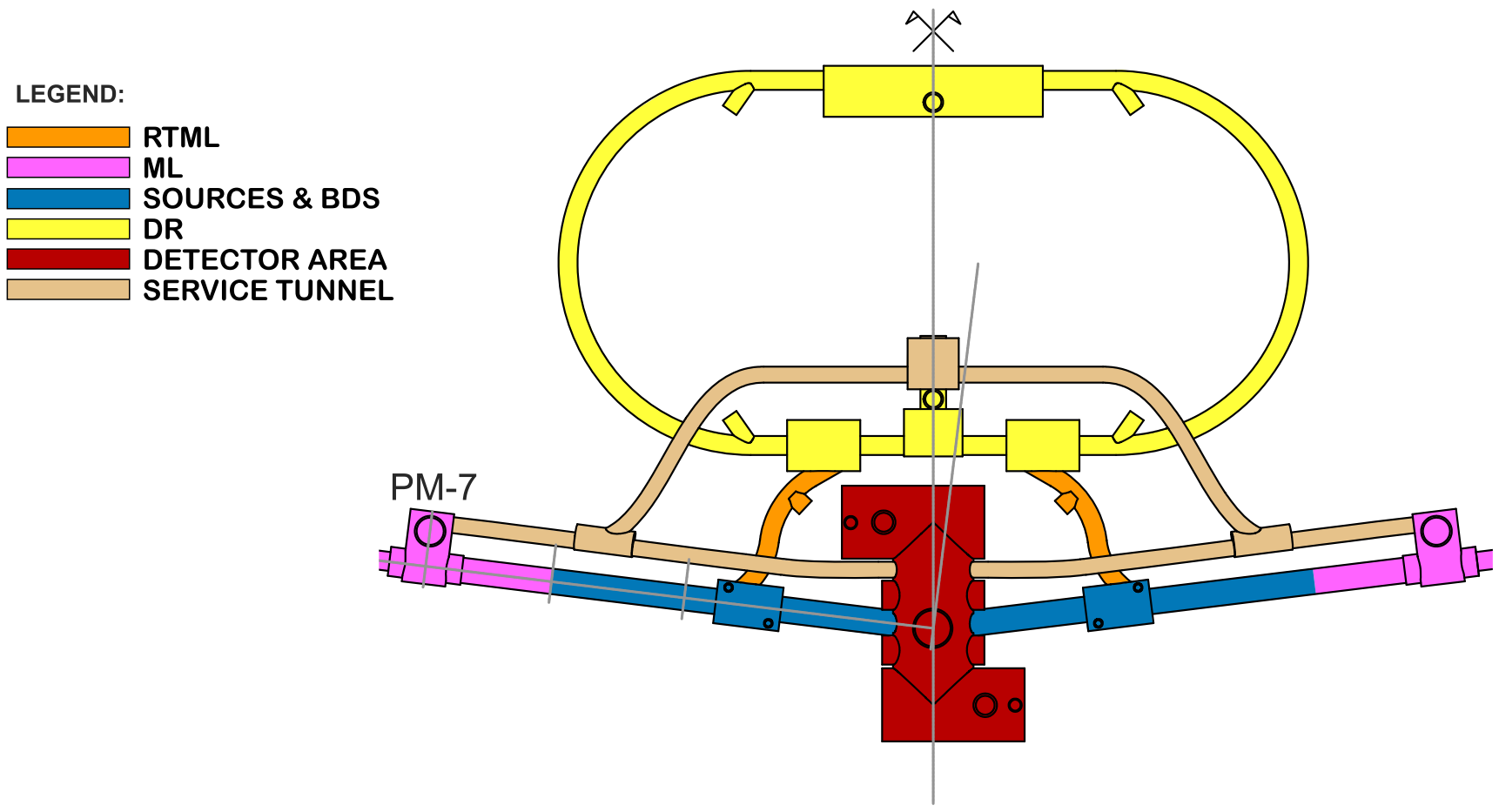

The DR complex is an approximately 3 km-long quasi-circular tunnel with an internal diameter of 6 m and containing 4 alcoves. There are two 9 m-diameter shafts, one in the middle of each long straight section. It is connected to the ML through two 250 m long RTML tunnels, the ELTR and PLTR transfer tunnels, with an internal diameter of 6 m. The electron and positron injector linacs are located in tunnels of 8 m internal diameter. The sources are housed in 7 m-diameter tunnels connected at their ends to the ML. A 4.5 m-diameter service tunnel passes over the DR and connects the Ring to the IR Hall and ML.

### 11.5.2.5 IR and BDS

The IR and BDS facilities are situated in the middle of the complex (Fig. 11.24). The IR Hall (Fig. 11.25) houses the two detectors in two 60 m caverns on either side of the interaction point, each of which has an 8 m-diameter vertical access shaft. Both detector caverns are connected to a sub-cavern, with a 6 m-diameter shaft. Escape tunnels connect each of the detector caverns with a safety shelter located in the other detector cavern. A survey gallery allows the alignment of magnets located in the beam tunnels on both sides of the IR.

Before being lowered underground, the detectors will be assembled and tested in a surface building. An 18 m-diameter shaft connects this surface building to the transfer cavern which in turn connects the two detector caverns. Travelling cranes will have to be installed to allow the assembly and servicing of the detectors. The surface building is equipped with a temporary 4000 t gantry crane and the transfer cavern is equipped with a 40 t crane.

The detector platform allows the sliding of each detector into on-beam position through a push-pull system. The geotechnical and structural behavior of the ground-detector complex interface was studied using existing local geological data and known geotechnical rock characteristics available at CERN. A 3D model was developed for understanding the stress conditions of the underground cavern complex at the IR. The analysis identified the in-situ stress development across the IR and has shown that the current orientation of the cavern alignment is preferred.

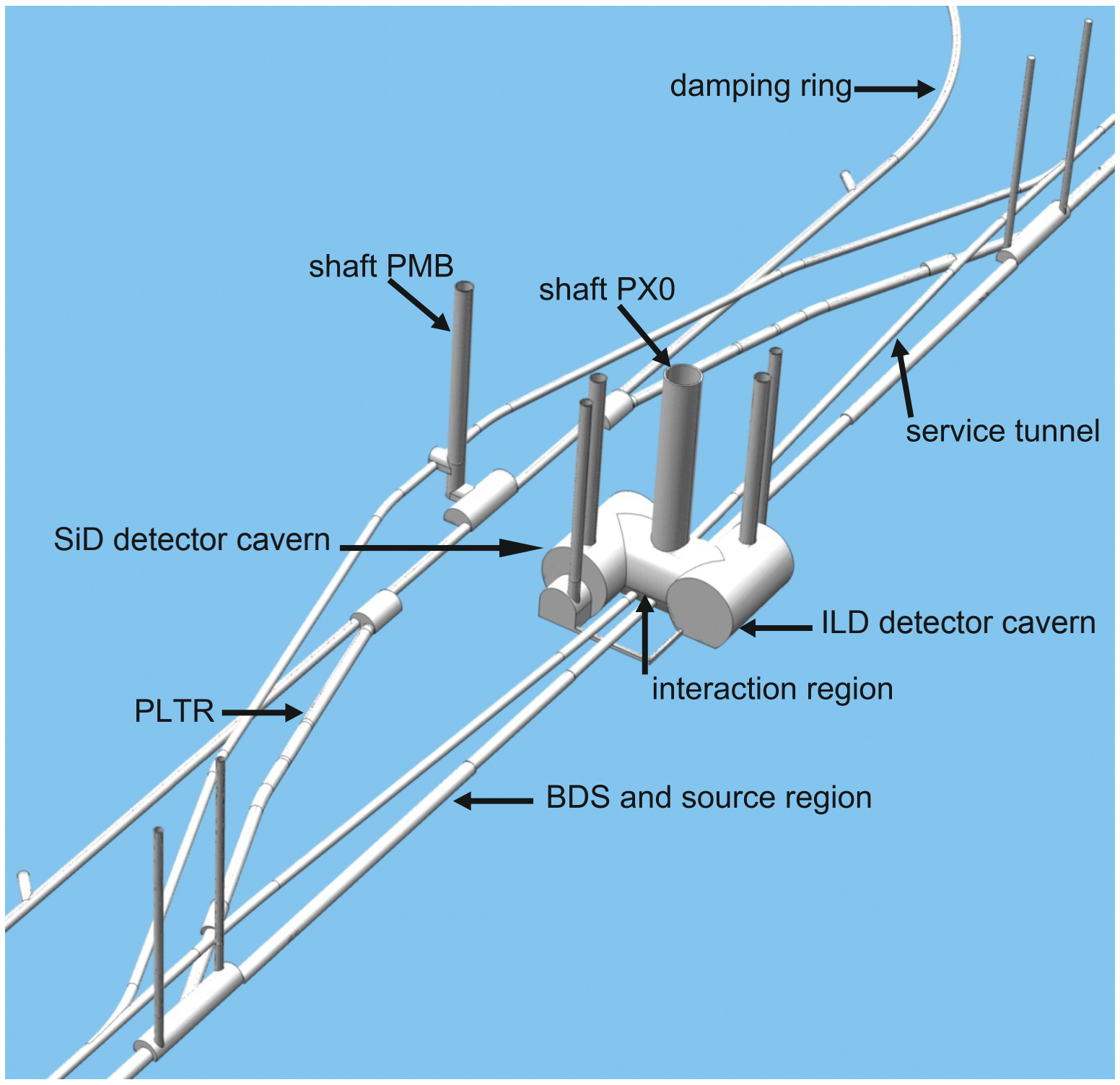

**Figure 11.24**
Model of the Central Injection Region.

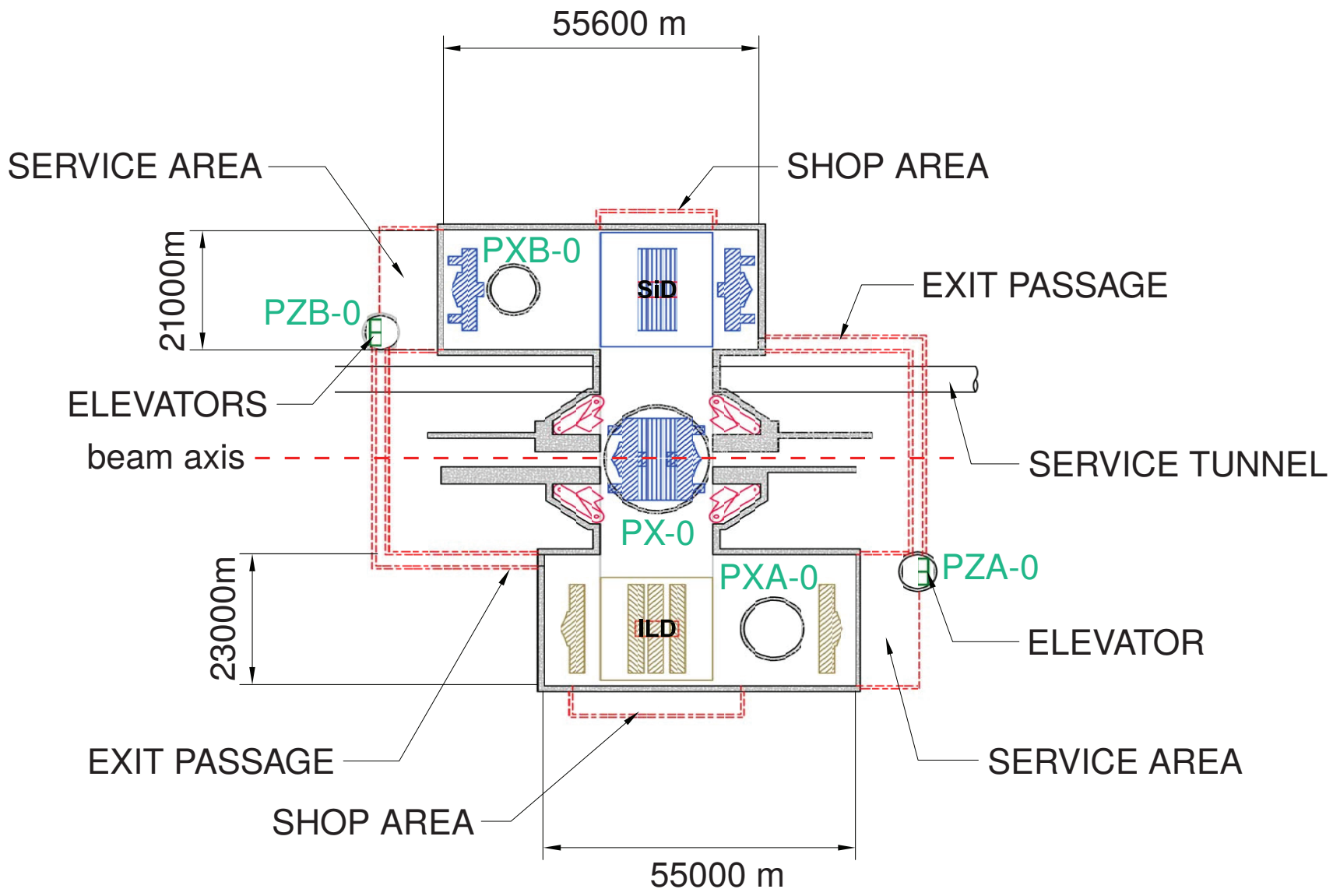

**Figure 11.25**
IR Hall plan view.

The BDS handles the incoming and outgoing beams in and out of the IR. It houses several beam dump caverns ($e^-$ and $e^+$ tune-up dump, $e^-$ and $e^+$ fast-abort dump, photon dump), positron-capture chicanes, target-bypass 'dog-leg' areas, undulator areas and service caverns for equipment storage. The beam-dump facilities are located at both sides of the IR in caverns accommodating water dump tanks at high pressure.

#### 11.5.2.6 Surface facilities

Surface buildings are foreseen approximately every 2 km along the machine length in a rural environment with easy access for large vehicles. This includes equipment buildings, cooling towers and pump stations, cryo buildings, shaft-head buildings, storage areas, and assembly areas. As local workshops and technical offices are already in place at CERN, these are not considered. A large fraction of the buildings are expected to be located at the Prevessin Campus, near the IR.

Stations with klystron clusters and cryo-plants have to be located roughly every 4 km on the surface. Klystron clusters without cryo-plants are located approximately every 2 km. Hybrid installations are foreseen at the outer end of the tunnels, where a single klystron cluster powers the first 1.25 km of the ML [213].

### 11.5.3 Mechanical

The European mechanical design is based on the America's KCS mechanical design (see Section 11.6.3). However, a major difference between the two concepts lies in the ventilation systems, which for Europe consists of an overhead ventilation scheme in the main tunnel. This scheme has been adopted for CLIC, mainly due to fire safety constraints, and its design is readily applicable to the ILC complex. For further details, see the CLIC CDR [214].

### 11.5.4 Electrical

This is based on the Americas design, see Section 11.6.4.

### 11.5.5 Life safety and egress

#### 11.5.5.1 Introduction

A detailed life-safety study has been conducted for CLIC. From a fire-safety point of view, the ILC single-tunnel complex is comparable with CLIC. Therefore the CLIC life safety and egress study can be applied to the ILC facility. For further details, see the CLIC CDR [215].

#### 11.5.5.2 Fire Risk Assessments and scenarios

Detailed fire risk assessments and scenarios will have to be made for every specific area i.e. tunnels, experiment caverns, alcoves for equipment, linking galleries, once more information is available on the layouts and their interconnects through ventilation systems.

#### 11.5.5.3 Fire Prevention strategy

Fire prevention measures at every possible level of functional design need to be implemented to ensure that large adverse events are only possible in the very unlikely event of multi-level safety barrier failure.

#### 11.5.5.4 Fire Safety Measures

The tunnels can be split into compartments with solid doors and fire walls, with internal longitudinal passages. An example of such a firewall is shown in Fig. 11.26.

The action of splitting the facility into compartments needs to be accompanied by a coherent design of the ventilation and smoke-handling systems. The ventilation system in the tunnel should be capable of creating a lower pressure in the compartment affected by the fire and an over pressure in the areas at the sides, as shown in Fig. 11.27. The smoke-handling system should withstand the thermal impact of fire and ensure the continuity of its functioning to prevent smoke propagation from one compartment to another.

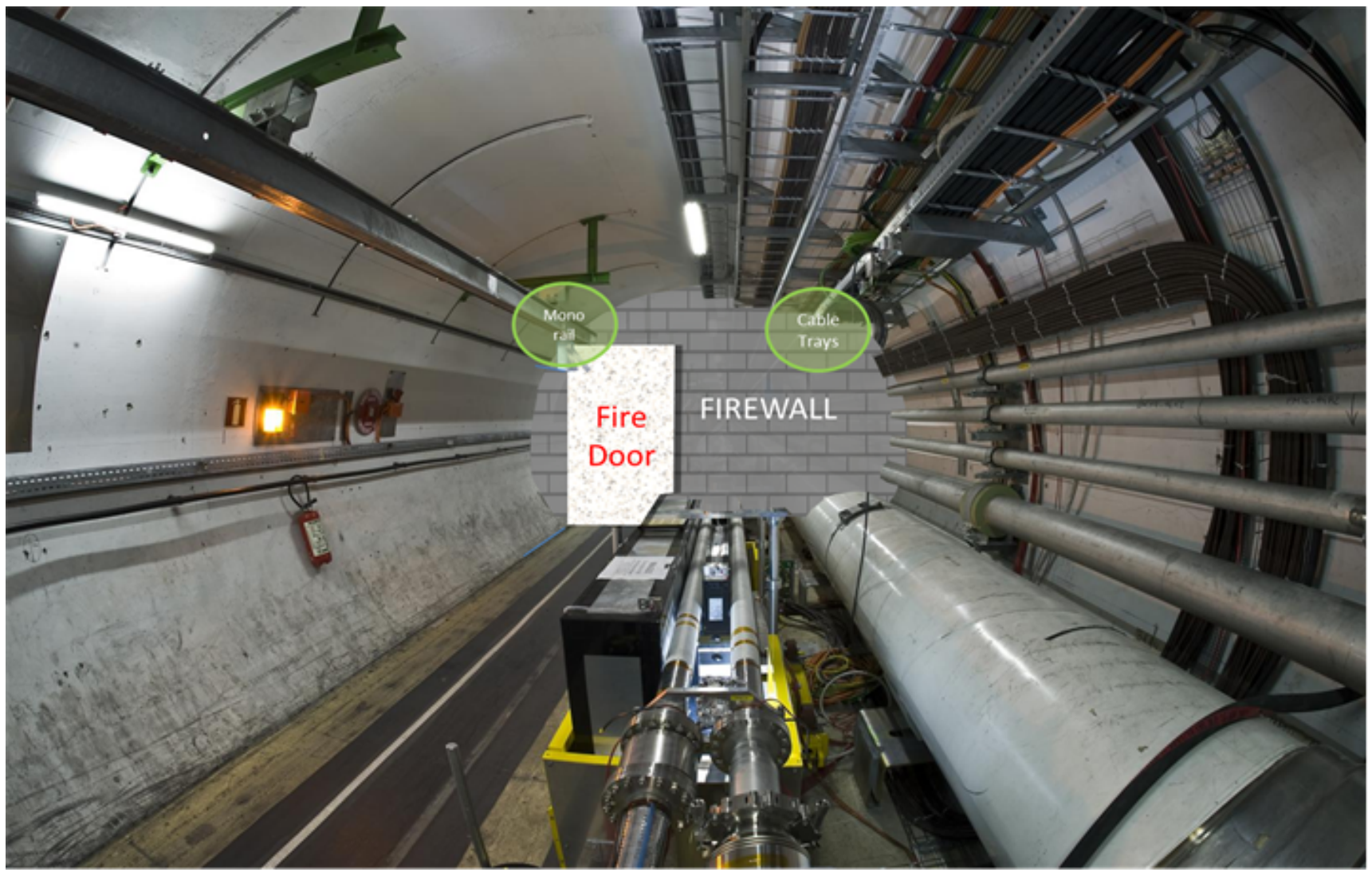

**Figure 11.26** Conceptual representation of a tunnel firewall.

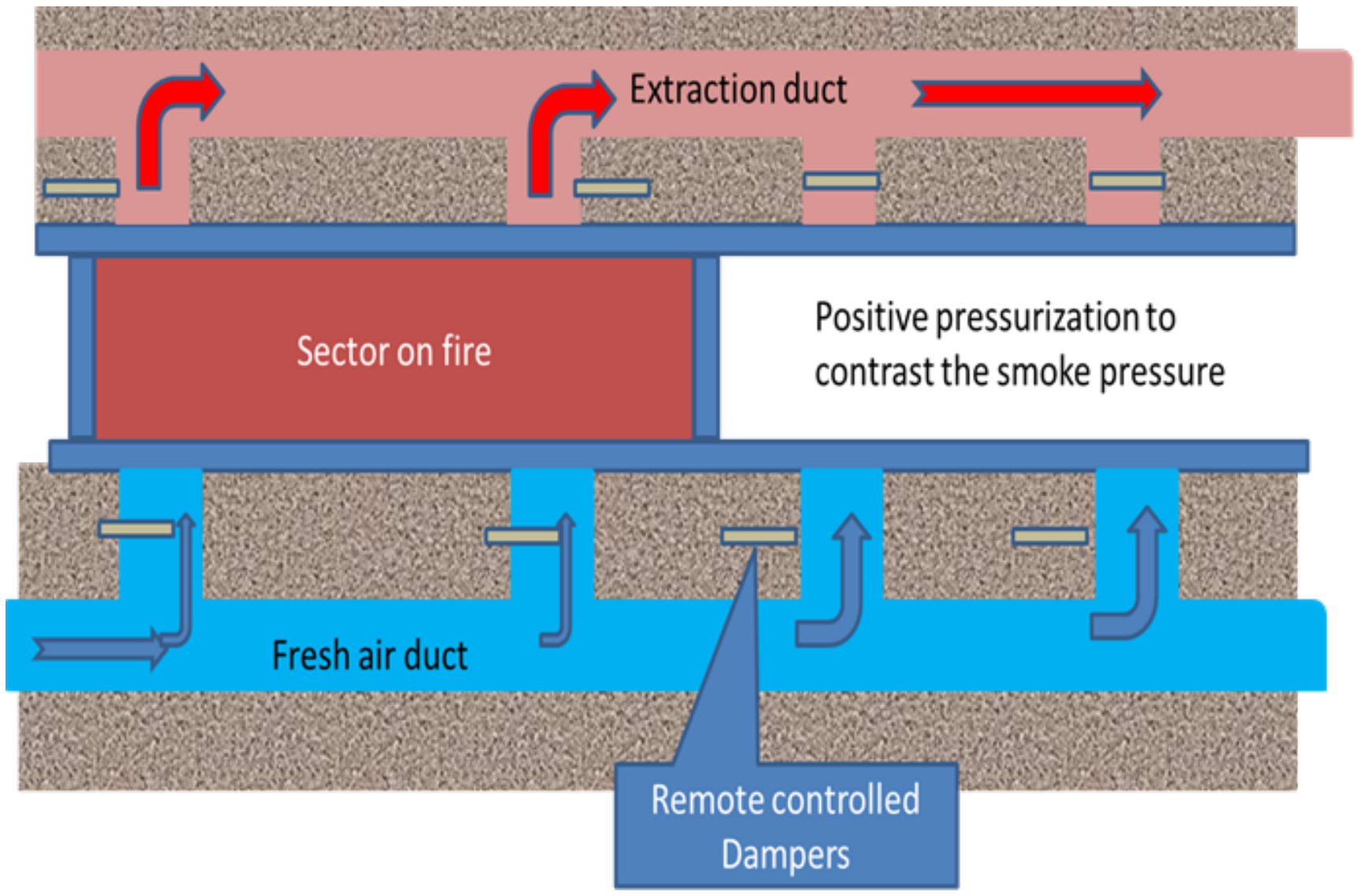

**Figure 11.27** Schematic representation of the pressurization of a sector adjacent to the sector on fire

## 11.6 Americas region (Flat topography)

### 11.6.1 Siting studies

#### 11.6.1.1 Location

The Americas sample site is in Northern Illinois, with a north-south orientation roughly centered on Fermilab. The central campus and IR are located on the Fermilab site, located approximately 35 miles west of downtown Chicago. The surrounding area has a medium population density supported by robust utilities and transportation infrastructure. While the routing requires the tunnel to pass below residential areas, the shafts can be located in non-residential areas.

#### 11.6.1.2 Land Features

The surface of northern Illinois is primarily flat, with surface elevations ranging from 200 m to 275 m above sea level. Much of the eastern half of northern Illinois is developed with many commercial, residential and industrial complexes. The 2751 hectare (6800-acre) Fermilab site is also relatively flat with less than 15 m of fall from northwest to southeast.

### 11.6.1.3 Climate

The climate is typical of the Midwestern United States, with four distinct seasons. In summer, temperatures ordinarily reach between 26°C and 33°C and humidity is moderate. Yearly precipitation averages 920 mm. Winter temperatures average -2°C during the daytime, and -10°C at night. Temperatures can be expected to drop below -18°C for on average 15 days throughout the winter season.

### 11.6.1.4 Geology

Geologic information has been obtained from previous underground construction at Fermilab and in northeastern Illinois, and not from ILC-specific investigations. The tunnels are located in a dry, uniform and massive dolomitic limestone deposit (Fig. 11.28). An overlying layer of shale provides a hydrogeologic barrier between upper aquifers and the dolomite. These geologic conditions should provide a relatively dry tunnel, during both construction and operations, but it is expected that some grouting will be required.

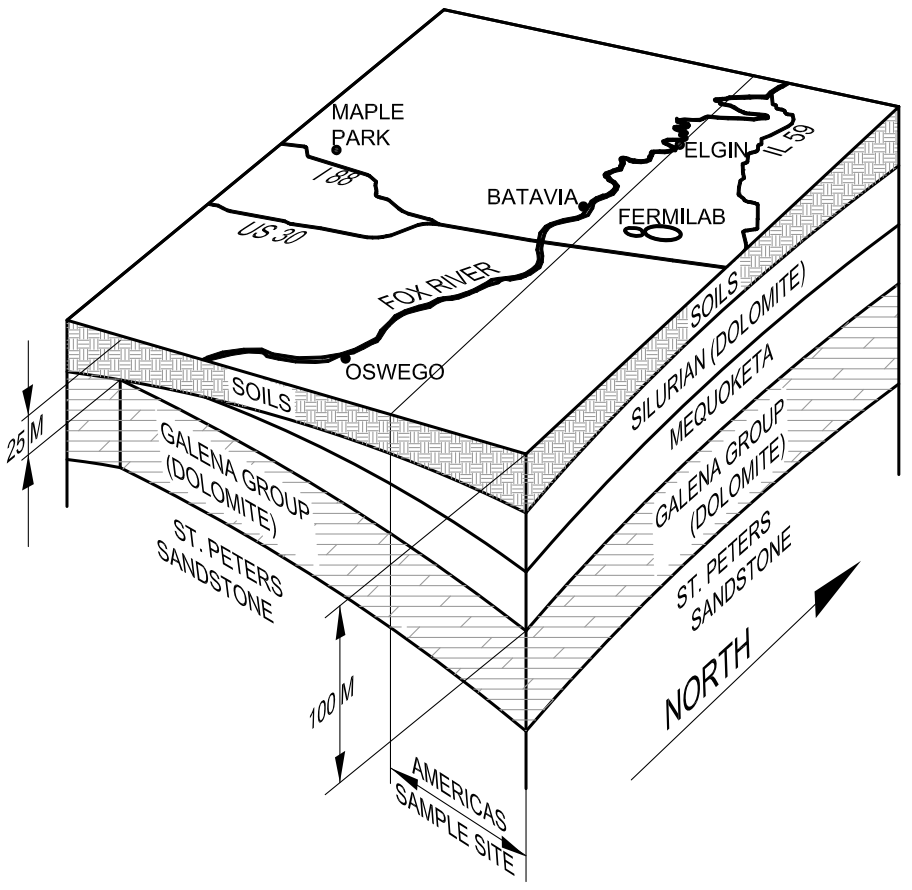

(a) North/South and East/West geology sections near the Americas sample site

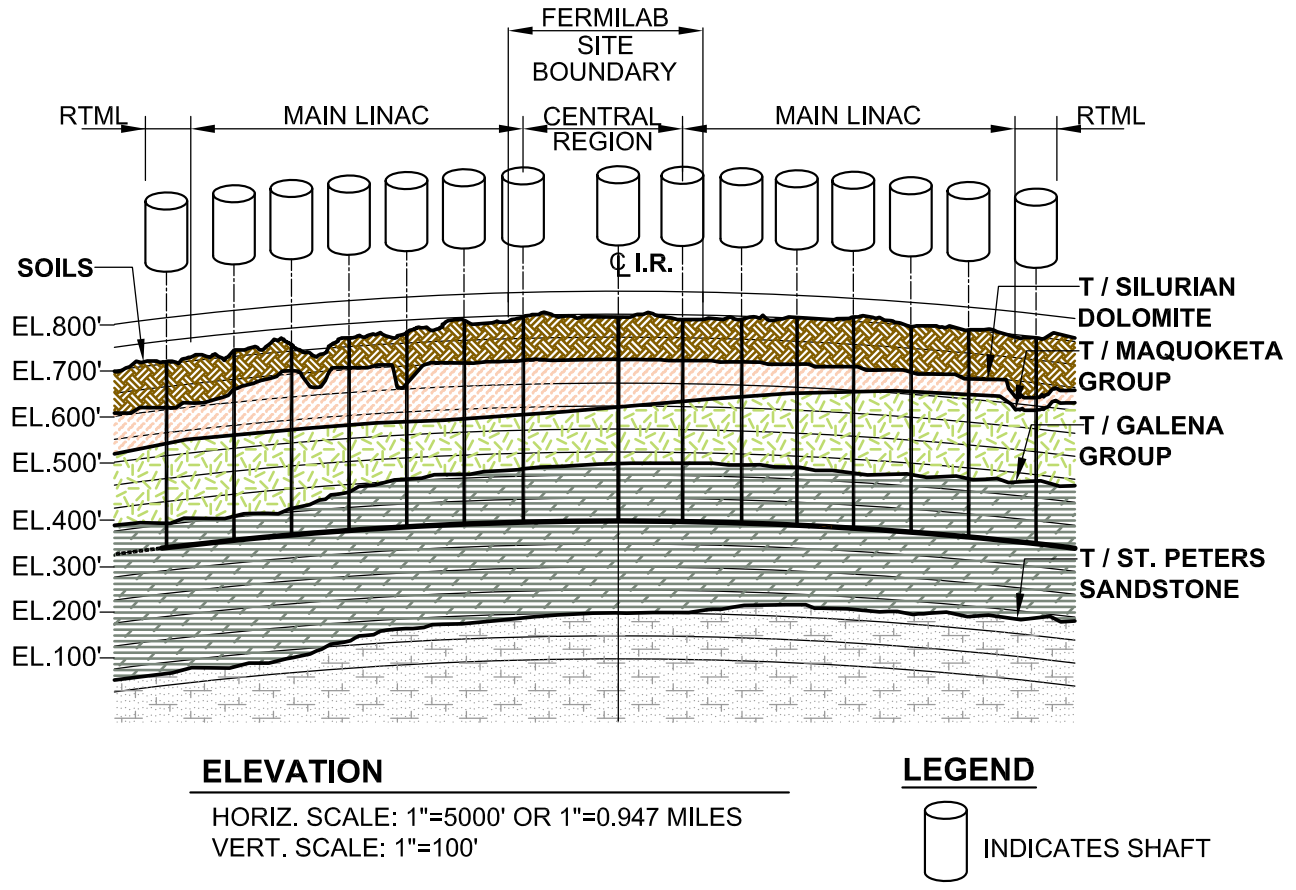

(b) Geological profile of the Americas sample site showing the layers of hard limestone

**Figure 11.28.** Geology of the Americas sample site

### 11.6.1.5 Power Distribution System

The investor-owned utility, Commonwealth Edison Company, services the Northern Illinois area with a capacity of more than 22,000 MVA. This capacity is made available through both fossil fuel and nuclear power generating stations. The electrical transmission infrastructure in Northern Illinois is very strong. The local power grid is capable of tying to three other national power generating sources. Transmission lines with voltage at 365 kV currently serve Fermilab along the eastern boundary of the site.

### 11.6.1.6 Construction Methods

The tunnels are excavated with TBMs and lined with a cast concrete invert. Widened portions and caverns are excavated using drill and blast. Temporary supports are required for the largest spans, permanent support is provided by rock bolts. Shaft overburden is excavated using standard earth excavators and muck boxes, supported by ring beams and timber lagging, keyed into the underlying rock. Excavation through the limestone and shale to the final depth uses conventional drill and blast methods. Support is provided by resin encapsulated rockbolts and the shaft is reinforced and concrete lined.

## 11.6.2 Civil construction

### 11.6.2.1 Underground enclosures

Figure 11.29 shows the overall plan view.

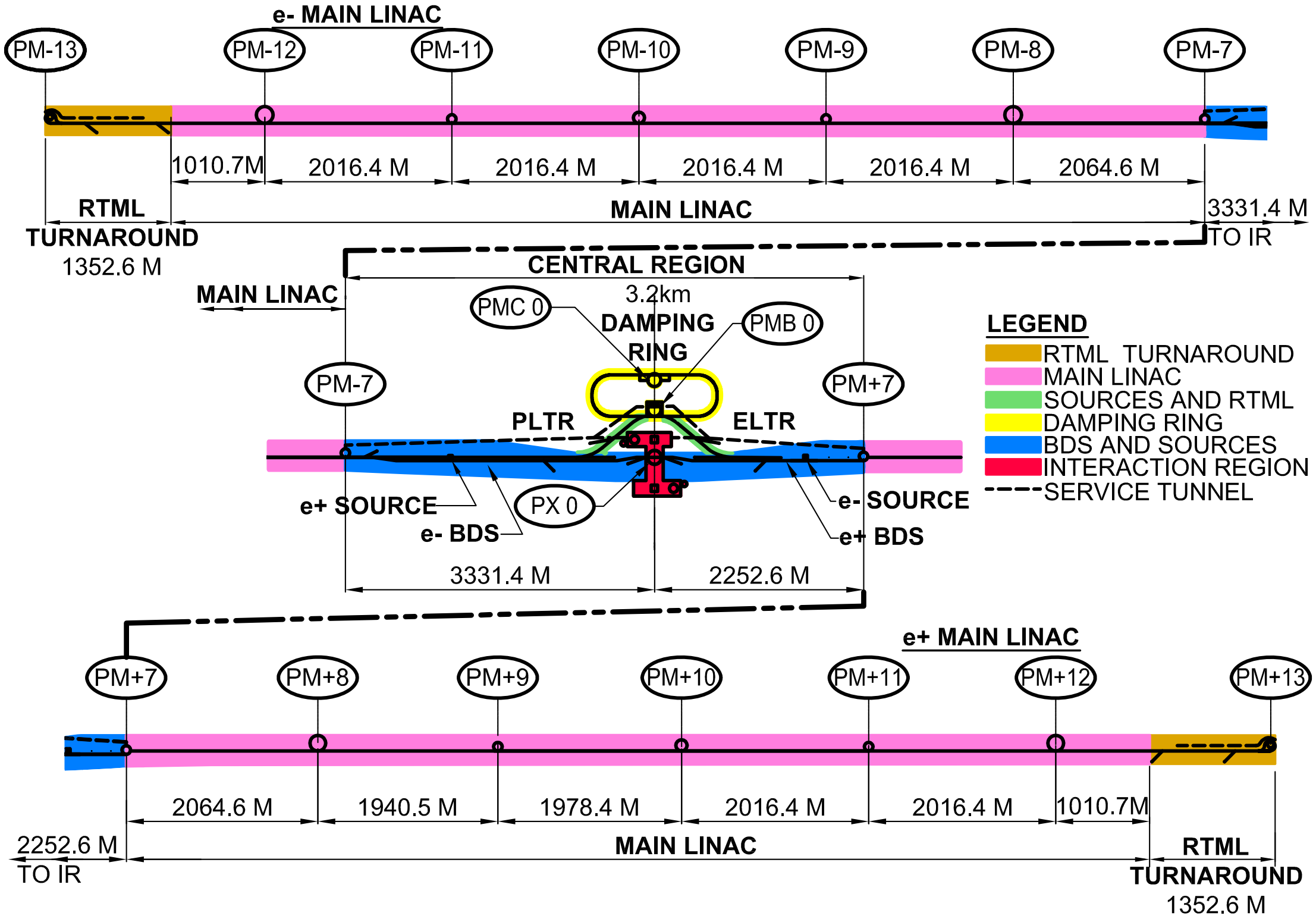

**Figure 11.29.** Americas tunnel layout. The colors reflect the Accelerator sections locations.

11.6.2.1.1 Tunnels The accelerator complex contains a combined total tunnel length of $\sim 44\,\mathrm{km}$, where the breakdown by accelerator section is listed in Table 11.10. Where two accelerator sections share the same tunnel segment, the length of tunnel is apportioned according to the fraction of the tunnel length occupied by each respective accelerator section. So for example, the ELTR and PLTR are divided equally between the Sources and RTML, while the tunnel between the end of the Main Linac and the IR Hall is apportioned 54 % to 46 % between the BDS and the $e^+$ source on the electron side and 90 % to 10 % between the BDS and $e^-$ source on the positron side.

**Table 11.10**
Tunnel lengths and volumes by Accelerator section

| Accelerator section | Length (m) | Volume (m$^3$) |
|---|---|---|
| $e^-$ source (beam) | 368 | 8,064 |
| $e^-$ source (service) | 618 | 11,584 |
| $e^+$ source (beam) | 1,678 | 33,770 |
| $e^+$ source (service) | 2,203 | 36,922 |
| Damping Ring | 3,239 | 76,945 |
| RTML (beam) | 3,305 | 68,619 |
| RTML (service) | 1,955 | 31,090 |
| Main Linac | 22,168 | 435,264 |
| BDS (beam) | 3,847 | 141,440 |
| BDS (service) | 3,847 | 61,183 |
| TOTAL | 43,228 | 904,881 |

Figure 11.30 shows a typical cross-section through the ML tunnel. The cryomodule waveguides are located on the aisle side of the cryomodule and are fed from circular over-moded waveguide on the tunnel ceiling. The circular over-moded waveguide comes from the Klystron Service Building located at each of the ML Shafts. Space is reserved for survey lines of sight.

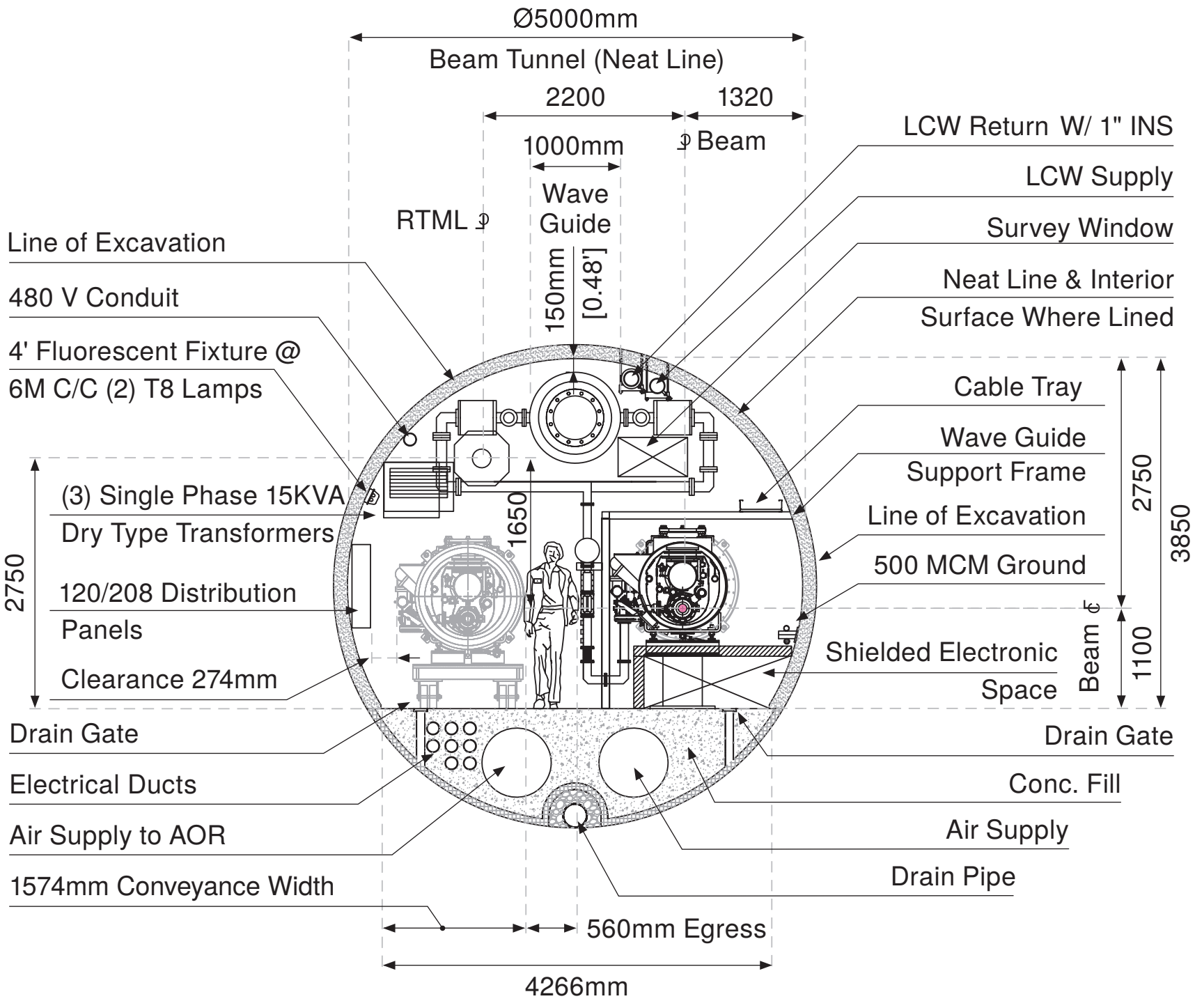

**Figure 11.30** Typical ML tunnel cross section

11.6.2.1.2 Cost Allocation of Underground Tunnels between Area Systems The costs for the tunnel that are attributed to each of the area systems are based on the percentage of beam length for each Area System in a specific tunnel segment. As shown in Table 11.11, the ELTR and PLTR Beam and adjacent Service tunnels are divided equally between the Sources and RTML. The Beam and adjacent Service tunnels between the end of the Main Linac and the IR Hall are proportioned 54 % to the $\mathrm{e}^-$ BDS and 46 % to the $\mathrm{e}^+$ source, on the electron side, and 90 % to the e+ BDS and 10 % to the $\mathrm{e}^-$ source, on the positron side. The RTML Turnarounds, Main Linacs and Damping Ring tunnel lengths are considered 100 % part of those Area Systems respectively. The RTML Service tunnel at the turnarounds is considered 100 % part of the RTML.

**Table 11.11** Tunnel Cost Allocations

| | Beam Tunnel | Adjacent Service Tunnels Americas and European Region | Adjacent Service Tunnels or Service side of Tunnel in Asian Region |
|---|---|---|---|
| RTML Turnarounds | 100% RTML | 100% RTML | 100% RTML |
| Main Linacs | 100% Main Linac | NA | 100 Main Linac |
| End of -e Main Linac to IR Hall | 54% e- BDS<br>46% e+Source | 54% e- BDS<br>46% e+Source | 54% e- BDS<br>46% e+Source |
| PLTR | 50% RTML<br>50% e+ Source | 50% RTML<br>50% e+ Source | 50% RTML<br>50% e+ Source |
| Damping Ring | 100% Damping Ring | NA | NA |
| ELTR | 50% RTML<br>50% e- Source | 50% RTML<br>50% e- Source | 50% RTML<br>50% e- Source |
| End of +e Main Linac to IR Hall | 90% e+ BDS<br>10% e- Source | 90% e+ BDS<br>10% e- Source | 90% e+ BDS<br>10% e- Source |

11.6.2.1.3 IR Hall, caverns, and alcoves There are underground caverns and alcoves along the tunnels, in addition to the central IR Hall; the IR Hall is described in Section 11.5.2.5. Caverns are located at the base of each shaft, and alcoves provide safe havens in emergencies and also house equipment. The caverns and alcoves, summarised in Table 11.12 are sized for:

- the amount and nature of equipment to be housed: cryogenic, electrical, cooling and ventilation,

**Table 11.12**
Cavern summary and volumes

| Accelerator section | Qty | Volume ($m^3$) |
|---|---|---|
| $e^-$ source | 0 | - |
| $e^+$ source | 1 | 324 |
| Damping Ring | 8 | 26,821 |
| RTML | 6 | 12,575 |
| Main Linac | 49 | 57,165 |
| BDS | 8 | 36,869 |
| IR | 1 | 135,703 |
| TOTAL | | 269,458 |

water distribution, electronics, etc;

- connecting services between access shafts and tunnels;
- lowering, assembly, and commissioning of TBMs for the excavation work (at those caverns where excavation starts or ends).

The caverns have moveable steel-concrete shielding doors moving on air-pads or rails, which can be opened for equipment transfer into the beamline area.

#### 11.6.2.2 Underground access

There are a total of 14 vertical shafts (Fig. 11.29): two 6 m-diameter shafts for the RTML, four 14 m-diameter major equipment shafts, two 9 m-diameter shafts and six smaller 6 m-diameter shafts for the ML, summarised in Table 11.13. The shafts allow movement of equipment and personnel, and provide accessways for services such as cooling water, potable water, compressed air, cryo-fluids, electrical supply, and controls. The over-moded waveguide also uses these shafts. Two shafts service the ML and Sources/BDS areas. There are two access shafts serving the DR tunnel. The 9 m-diameter shafts are situated at opposite sides of the DR at the midpoint of the straight sections. In the Central Region there are four 1.5 m-diameter shafts that supply utilities to the high-power-beam-abort caverns. The IR Hall has an 18 m-diameter shaft used for lowering major detector segments from the surface-assembly building. There are also two 8 m-diameter shafts for lowering smaller equipment into the hall, one for each detector, and two 6 m-diameter shafts for utilities and personnel egress.

**Table 11.13**
Shaft depths and volumes

| Accelerator section | Depth (m) | Volume ($m^3$) |
|---|---|---|
| $e^-$ source | 125 | 0 |
| $e^+$ source | 125 | 0 |
| Damping Ring | 125 | 15,896 |
| RTML | 125 | 7,065 |
| Main Linac | 125 | 114,021 |
| BDS | 125 | 883 |
| IR | 125 | 54,950 |
| TOTAL | | 192,815 |

#### 11.6.2.3 Surface structures

The KCS RF feeds the ML at 2.5 km intervals. The ML surface infrastructure installations are spaced every 2.5 km at the heads of service shafts to the tunnel. The surface installations every 5.0 km also have cryogenic cooling plants (see Fig. 11.31). At the central region end of the ML, there are hybrid installations where a single Klystron cluster powers the last 1.25 km section.

The numbers and sizes of buildings associated with each accelerator section are listed in Table 11.14. The Americas proposed sample site centers the alignment of the ILC on the Fermilab site. Spaces normally required in a central campus such as office space, tech space, machine shops, storage, and cafeteria are considered to be existing on the Fermilab Site and suitable for the needs of the ILC project, and therefore not included in the table. Electrical power and other associated infrastructure are also considered to be part of these existing buildings. Therefore no additional surface support facilities are included in the Americas Region TDR design.

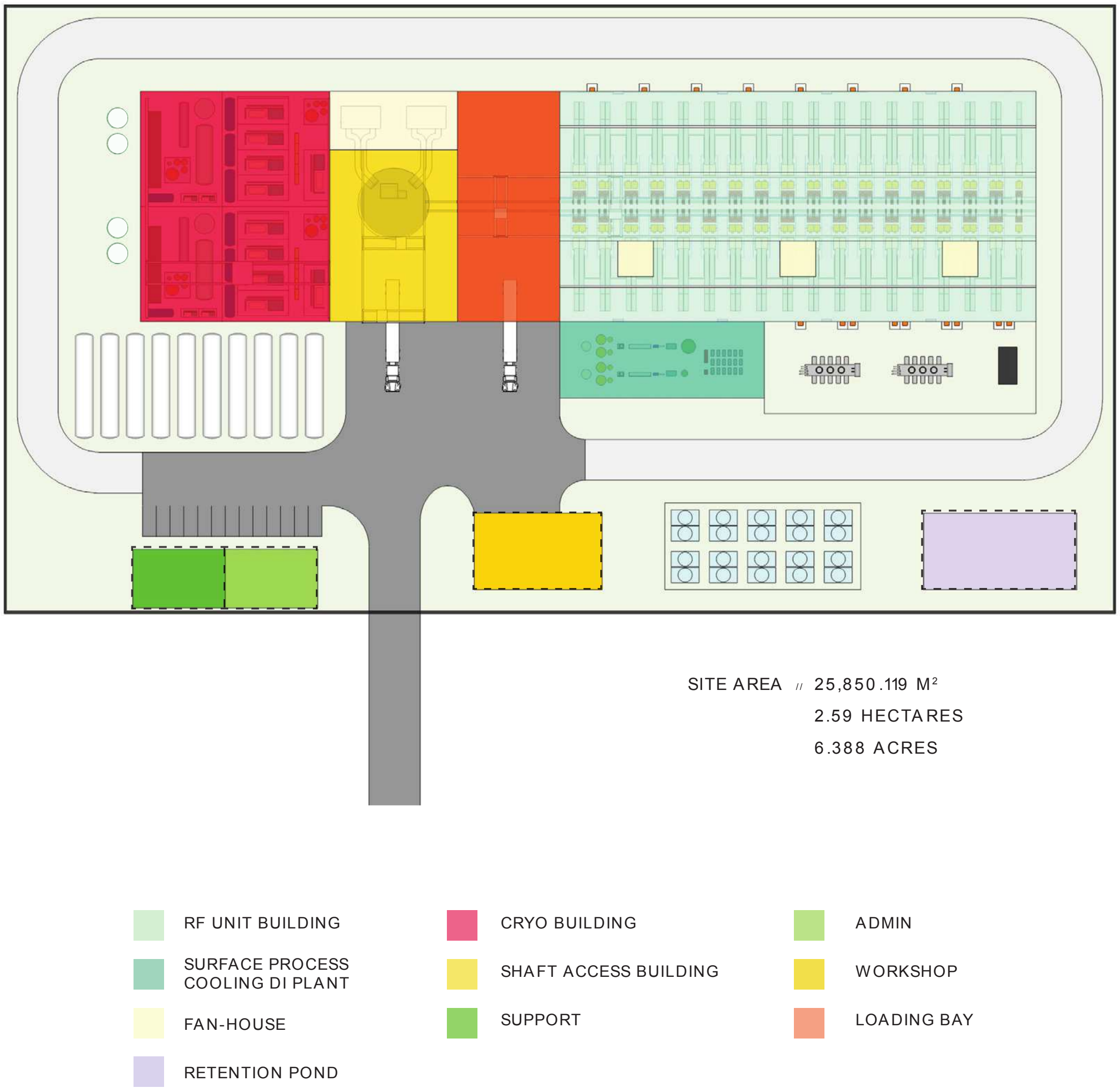

**Figure 11.31.** Typical KCS surface facility layout

**Table 11.14**
Surface structures by Accelerator section

| Accelerator section | Qty | Area (m$^2$) |
|---|---|---|
| e$^-$ source | 0 | – |
| e$^+$ source | 0 | – |
| Damping Ring | 3 | 5,294 |
| RTML | 2 | 1,410 |
| Main Linac | 18 | 60,200 |
| BDS | 0 | – |
| IR | 1 | 7,695 |
| TOTAL | 24 | 74,599 |

## 11.6.3 Mechanical

### 11.6.3.1 Processed water

Thermal heat loads were tabulated for each accelerator section. Design specifications were developed [216–225]. The ML accounts for about 60 % of the total load. Tables 11.15 and 11.16 show the distribution of heat loads by component (above and below ground) and accelerator sections.

**Table 11.15.** Main Linac KCS RF Heat Load (TDR Baseline Low Power)

| | | | To Low Conductivity Water | | | | | | | | to CHW | to AIR |
|---|---|---|---|---|---|---|---|---|---|---|---|---|
| | Q.ty | Average heat load (kW) | Heat load to LCW water (kW) | Max all. temp. (°C) | Supply temp. (°C) | Delta temp. (°C delta) | Water flow (l/min) | Max all. press. (bar) | Typical (wtr) press. drop (bar) | Accept. temp. variation (delta °C) | Racks heat load (kW) | Heat load to air (kW) |
| COMPONENTS IN THE SURFACE (listed per RF unit) | | | | | | | | | | | | |
| RF components x (413) | | | | | | | | | | | | |
| RF charging supply | 413/ML | 2.39 | 1.67 | | 40 | 8.5 | *2.84* | 18 | 5 | 10 | NA | 0.72 |
| Switching power supply | 413/ML | 5.5 | 3.3 | | 35 | 6.25 | 7.6 | 13 | 5 | 10 | NA | 2.2 |
| Filament transformer | 413/ML | 0.79 | 0.6 | 60 | 35 | 0.40 | 20 | | 1 | n/a | NA | 0.2 |
| Marx modulator | 413/ML | 4.96 | 3.0 | | 35 | 2.14 | 20 | 10 | 5 | n/a | NA | 2.0 |
| Klystrn scket tank / gun | 413/ML | 0.99 | 0.79 | 60 | 35 | 1.14 | 10 | 15 | 1 | n/a | NA | 0.2 |
| Focusing coil (solenoid ) | 413/ML | 1.68 | 1.6 | 80 | 55 | 2 | 10 | 15 | 1 | n/a | NA | 0.1 |
| Klystron collector | 413/ML | 38.43 | 37.1 | 87 | 38† | 14 | 37 | 15 | 0.3 | n/a | NA | 1.29 |
| Klystron body & windows | 413/ML | 3.37 | 3.4 | 40 | 25 to 40 | 5 | 10 | 15 | 4.5 | ±2.5 °C | NA | |
| CTOs & combining loads/circulators | 2/klstrn | 11.71 | 9.36 | | | 6.04 | 22.28 | | *(80 psid)* | | | 2.3 |
| Relay racks (Instrument racks) | | 3.0 | 0 | N/A | N/A | N/A | | N/A | N/A | None | 3 | 0.0 |
| Subtotal surface heat load to LCW water | | | 60.74 | Total surface (heat to water and air) | | | | 69.82 | | | 3.0 | 9.1 |
| COMPONENTS IN THE TUNNEL (listed per RF unit) | | | | | | | | | | | | |
| RF components (x 567) | | | | | | | | | | | | |
| RF pipe in shaft (shaft & bends) | | 1.89 | 1.70 | | | 10 | *2.445* | | *(80 psid)* | | | 0.2 |
| Relay racks (instrument racks) | | 5 | 5 | N/A | N/A | N/A | N/A | N/A | N/A | None | | 0.0 |
| Main tunnel wvgde & local wvgd | | 12.23 | 11.62 | | | 12 | *13.9* | | *(80 psid)* | | | 0.6 |
| Distribution end loads & Cavity reflection loads | | 31.80 | 31.3 | | | 20 | *20.54* | | *(80 psid)* | *±2.5 °C* | | 0.5 |
| Subtotal tunnel heat load to LCW water | | | 49.62 | Total tunnel (heat to water and air) | | | | 50.9 | | | | 1.3 |

† (inlet temp 25 to 63)

**Table 11.16** Summary of Heat Loads (MW) by Accelerator section. Heat loads generated by the utilities themselves (pumps, fan motors, and etc) are listed as 'Conventional'.

| Acc. section | load to LCW | load to Air | Conventional | Cryo (Water Load) | Total |
|---|---|---|---|---|---|
| $e^-$ sources | 1.40 | 0.70 | 0.80 | 0.80 | 3.70 |
| $e^+$ sources | 5.82 | 0.64 | 1.51 | 0.59 | 8.56 |
| DR | 10.92 | 0.73 | 1.79 | 1.45 | 14.89 |
| RTML | 4.16 | 0.76 | 0.68 | part of ML cryo | 5.59 |
| Main Linac | 46.5 | 5.53 | 5.32 | 32.0 | 89.34 |
| BDS | 9.2 | 1.23 | 3.23 | 0.41 | 14.07 |
| Major Dumps | 14 | | 0.05 | | 14.05 |
| IR | 0.4 | 0.76 | 0.10 | 2.65 | 3.91 |
| Total | 92.4 | 10.4 | 13.5 | 37.9 | 154 |

There are two types of water-cooling system: the first uses a chiller to provide cool supply water (chilled water/LCW), and the second uses a cooling tower that provides somewhat warmer water (process water/LCW). The chilled water/LCW-type system is used in the DR, IR and Central Region (which includes $e^+$ source, $e^-$ source, and BDS). This provides tight air-temperature stability in these areas. The ML, RTML and the Main Dump use the process/LCW water. Figures 11.32 and 11.33 show typical schematic diagrams of the process water and chilled-water support utility systems.

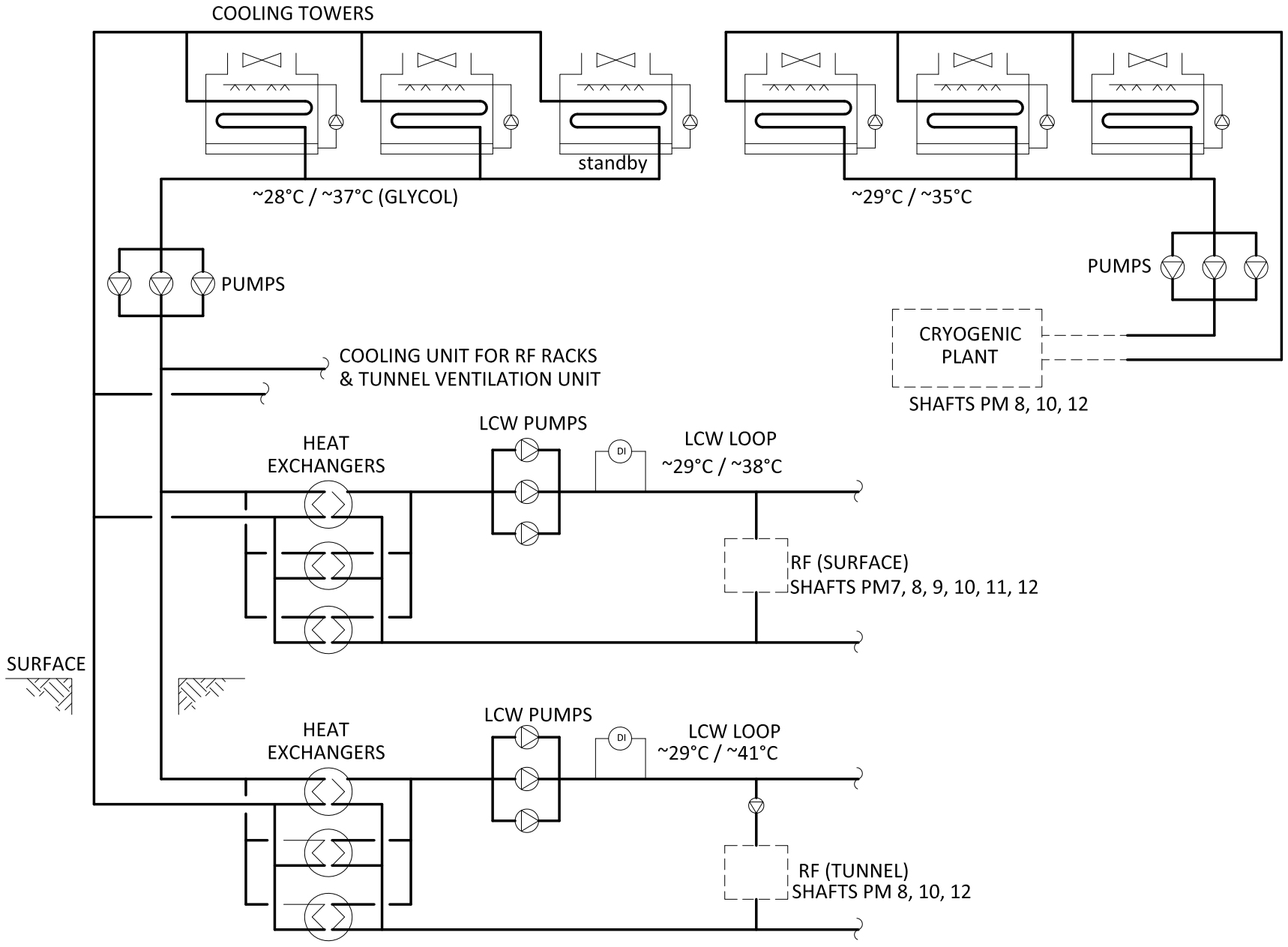

**Figure 11.32** Typical process water schematic.

For both systems, the cooling tower type is a closed-circuit evaporative cooler, using closed-loop glycol water as is customary in cold climates. This type of tower conserves water and treatment chemicals as compared to an open tower system. All surface plants are provided with n+1 redundancy. The make-up water to the system is supplied from individual wells or municipal water supply at each surface plant.

For the ML and RTML process water/LCW system, cooling towers provide a maximum 28.3 °C cooling water supply to the LCW heat exchangers. The heat exchanger supplies about 29.4 °C LCW to the loads. About 60 % of the heat loads from the ML are located on the surface. The load-to-air component of the tunnel heat loads is minor and handled by the tunnel ventilation system. At the surface, the ML surface heat loads to air from the RF components are dissipated using ambient air-ventilation systems. The HLRF for the RTML is located in a short support tunnel adjacent to the accelerator and therefore requires fan coils for conditioning its relatively large heat load to air. The Main Dumps near the IR have a dedicated process water system.

For the DR, tunnel fan coils use a cooler 10 °C supply water to maintain a tunnel temperature closer to the mean temperature of the magnet loads and to provide for better air-temperature stability [226]. The rest of the loads in the damping ring and central region, such as magnets, power supplies, and RF, are provided with approximately 18 °C LCW supply. The chiller-system design includes a waterside economiser that would automatically provide free cooling using the cooling tower if the ambient conditions are adequate.

**Figure 11.33** Chilled Water/LCW System at Central Region.

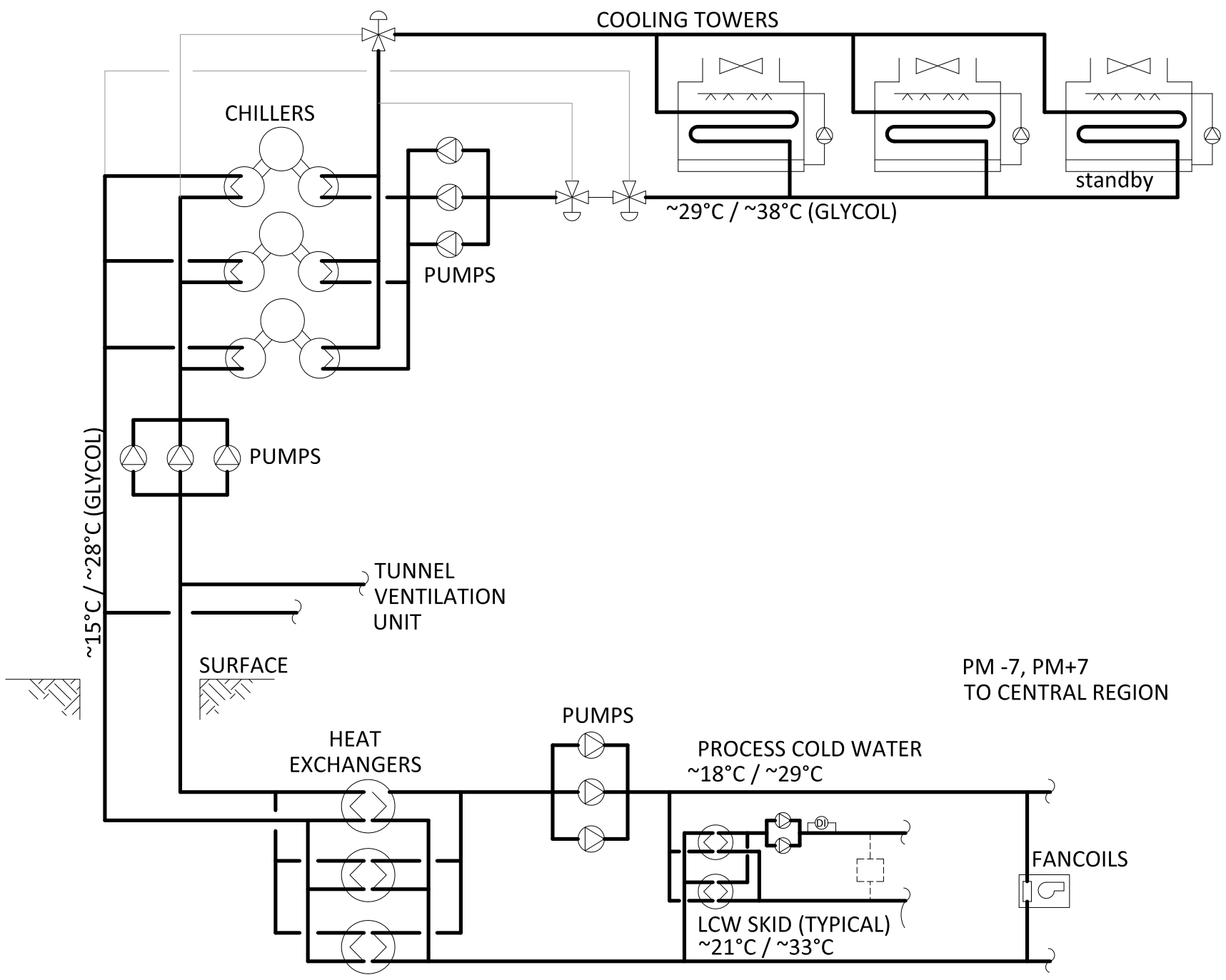

The main distribution of the cooling-water system follows the location of the shafts [227]. There are nineteen surface water plants, twelve for the ML and RTML, two for the central region, two for the IR, two for the DR, and one for the main dumps.

#### 11.6.3.2 Piped utilities

Groundwater inflow and condensate drainage for all underground areas is estimated to be 21 $m^3$/h/km. The total number of duplex sump pumps required are 132 in the ML, 121 in the Central Region and IR, and 32 in the DR. Groundwater duplex lift pumps and collection tanks are provided at every major shaft location. Each groundwater lift station has three pumps, any of which can pump the entire inflow volume. Water discharge is piped up the shaft through separate and protected piping systems.

#### 11.6.3.3 Air treatment

There are two ventilation systems, one for the Areas of Refuge (AOR) and the other for the general tunnel ventilation. Both systems have individual separate supply air ducts through the shafts from the surface ventilation units down to the cavern floor. They use the tunnel floor for further distribution along the length of the tunnel as well as into the AOR. Each unit is sized to 20% overcapacity to provide some redundancy in case one surface-unit fails. The general tunnel ventilation is conditioned to provide neutral temperature dehumidified air, while the AOR ventilation unit is non-conditioned raw outside air to be used only when the AOR is occupied. Return air from the general tunnel ventilation system is ducted up from the caverns to the surface units. In general the heat is removed from the tunnel areas by separate fan coils, except in the ML area where the heat load is minor and the tunnel ventilation is adequate. The tunnel ventilation provides 0.45 m/s air speed and an air change rate of approximately 2 per hour. The temperature in the tunnels is a maximum of 29 °C in the ML/RTML tunnel area and 29 °C in the central region, Damping Ring, IR, and service tunnels/caverns. Figure 11.34 shows a typical schematic diagram of the ventilation system.

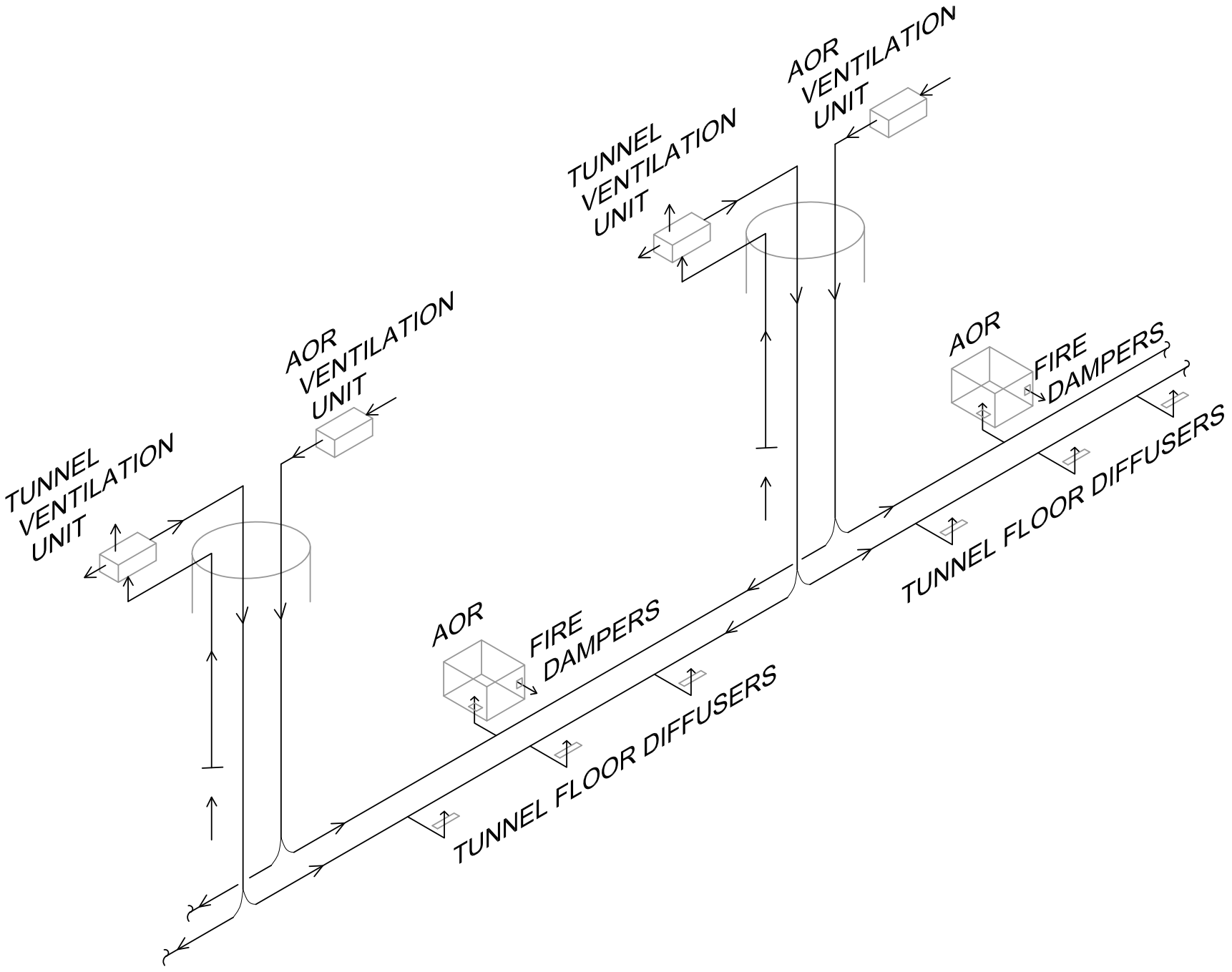

**Figure 11.34** Typical ventilation scheme.

## 11.6.4 Electrical

Electrical load tables were compiled for each area and the systems designed. The ML has about 70 % of the total loads. The conventional loads are from the components associated with running support facilities for the experimental equipment and facilities, such as pumps, fans and other mechanical/electrical systems not provided by the experiment. The power-factor value used for equipment sizing is the actual expected, if given, or a 90 % for all other equipment. Table 11.17 shows a summary of the power loads distributed by component and Accelerator section.

**Table 11.17** Summary of power loads (MW) by Accelerator section. 'Conventional' refers to power used for the utilities themselves. This includes water pumps and heating, ventilation and air conditioning, (HVAC). 'Emergency' power feeds utilities that must remain operational when main power is lost.

| Accelerator section | RF Power | RF Racks | NC magnets & Power supplies | Cryo | Conventional | | Total |
|---|---|---|---|---|---|---|---|
| | | | | | Normal load | Emergency load | |
| $e^-$ source | 1.28 | 0.09 | 0.73 | 0.80 | 1.02 | 0.16 | 4.08 |
| $e^+$ source | 1.39 | 0.09 | 4.94 | 0.59 | 2.19 | 0.35 | 9.56 |
| Damping Ring | 8.67 | | 2.97 | 1.45 | 1.84 | 0.14 | 15.08 |
| RTML | 4.76 | 0.32 | 1.26 | part of ML cryo | 0.12 | 0.14 | 6.59 |
| Main Linac | 58.1 | 4.9 | 0.914 | 32 | 8.10 | 5.18 | 109.16 |
| BDS | | | 10.43 | 0.41 | 0.24 | 0.28 | 11.36 |
| Dumps | | | | | 1 | | 1.00 |
| IR | | | 1.16 | 2.65 | 0.09 | 0.17 | 4.07 |
| Total | 74.2 | 5.4 | 22.4 | 37.9 | 14.6 | 6.4 | 161 |

The electrical power supply is divided into major systems by function:

- supply: 345 kV large overhead interconnect with the local Utility transmission grid;
- transmission: 69 kV and 34.5 kV main feeders serving local substations;
- medium voltage distribution: 34.5 kV distribution lines from local substations to service transformers distributed throughout the project;
- medium voltage standby power distribution: 4.16 kV distribution lines from generators to dedicated power transformers that serve standby loads;
- low voltage distribution: 480 and 208/120 V local distribution lines that directly serve loads;
- low voltage standby power distribution: 480 and 208/120 V local distribution lines that directly serve standby power loads.

### 11.6.4.1 Supply system

The Supply system consists of a 345 kV overhead line from the local utility grid to the central campus substation. The interconnect point with the local utility serves as the ownership demarcation point with a switching device and revenue metering. The local Utility has a switching device at this point to manage services to the project. All loads and losses beyond this point are included in the electrical power bill. Due to the large power requirements, the electrical system is designed to be independent and standalone from the local electrical utility infrastructure at the highest possible level.

The electrical power system for the project originates at the Central Campus Substation, which includes two 345 kV to 69 kV transformers and two 345 kV to 34.5 kV transformers. Each transformer serves a specific part of the project through switchgear. The 69 kV switchgear is outdoor rated, SF6 gas insulated switchgear (GIS) that enables a compact reliable installation at this voltage class. The 34.5 kV switchgear type is enclosed bus with vacuum circuit breakers that provide a compact reliable installation at this voltage class. Figure 11.35 illustrates the Central Campus Substation.

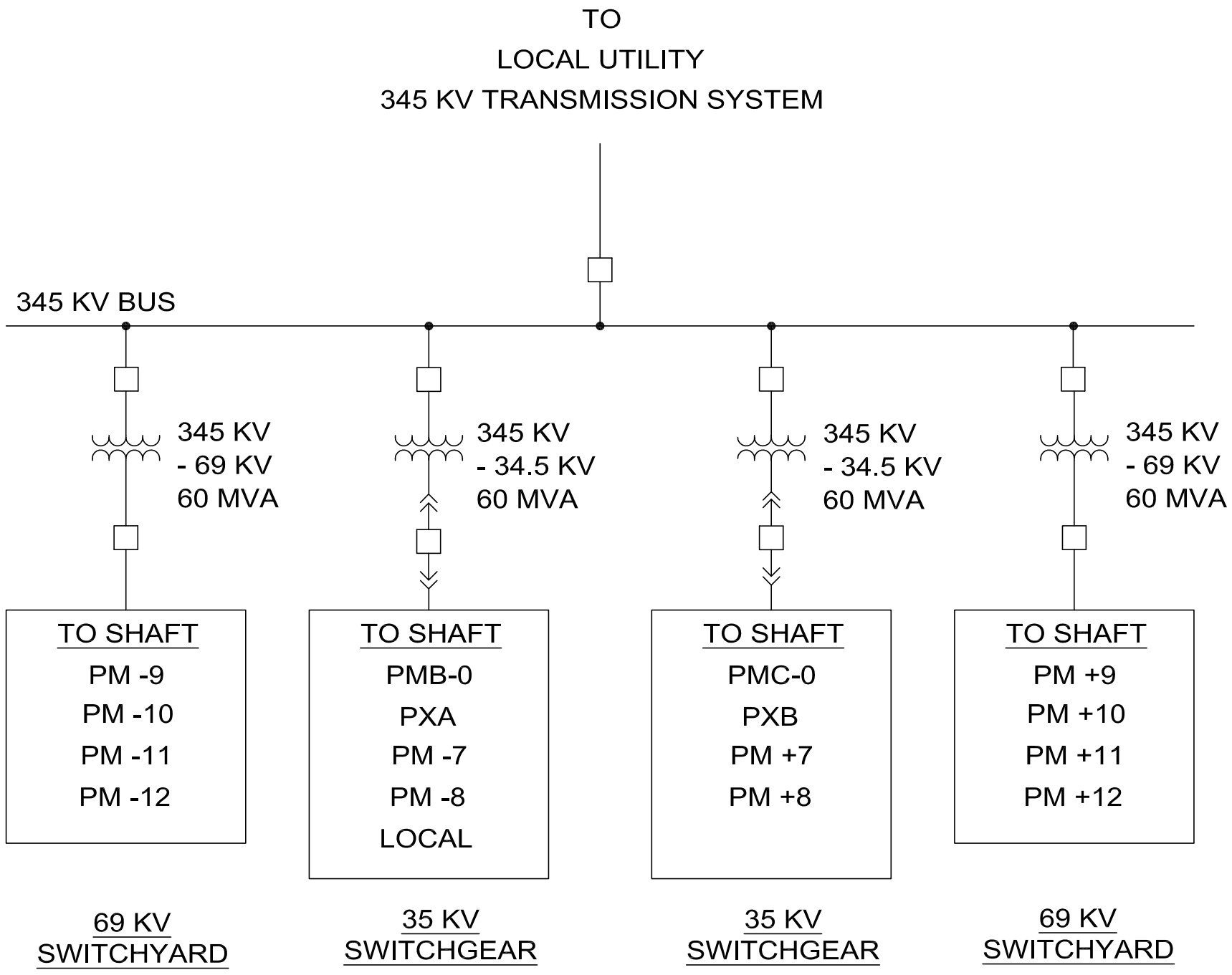

**Figure 11.35**
Electrical transmission system.

### 11.6.4.2 Transmission system

The 69 kV and 34.5 kV transmission system provides the required power to each local substation or switching station generally located at the top of each shaft. The system is a combination of substations, switching stations, 69 kV feeder lines and 34.5 kV feeder lines. The architecture of the system is a single-feed radial configuration extending from the central campus to the ends of the accelerator tunnels and the far side of the DR.

The 34.5 kV transmission feeders originating in the Central Substation serve the DR and near shafts, PM-7, PM+7, PM-8 and PM+8 local substations. The local Central Region loads are served directly from the substation switchgear while 34.5 kV feeders are routed through the tunnel to other shafts. The feeder that serves shafts PM-7 and PM-8 extends down the tunnel to shaft PM-7 where 34.5 kV switchgear provides service to local Medium-Voltage Distribution and a feed through to shaft PM-8 for local distribution. Similarly a feeder configuration is included for PM+7 and PM+8. The 34.5 kV transmission voltage to these locations enables direct Medium-Voltage Distribution through switchgear without the installation of local substation transformers at the DR, PM±7 and PM±8.

The 69 kV transmission feeders originating in the Central Substation serve the shafts PM±9 to the end of the tunnels. The 69 kV voltage level is used to minimize the number and size of conductors and conduits installed in the tunnel. The 69 kV feeders are extended from the Central Substation GIS to shafts PM+/-9 and PM±11. 69 kV to 34.5 kV substations are located at each of these shafts to provide local Medium Voltage Distribution and 34.5 kV feed to subsequent shafts, PM±10 and PM±12. No substation transformers are required at shafts PM±10 and PM±12.

### 11.6.4.3 Medium-Voltage Distribution system

The Medium-Voltage Distribution system provides power to each distribution transformer that serves a load in the tunnel or on the surface. The distribution feeder system is a radial configuration from the local substation switchgear to the distribution transformers. On the surface, transformers serve specific loads such as RF units, cryogenics or conventional facilities. In the tunnel, a distribution transformer is located in the base cavern to serve all conventional loads. The technical loads in the central region that include a service tunnel are served by a separate local transformer. Figure 11.36 illustrates a 34.5 kV distribution switchgear that serves both local loads and provides the origin of the transmission feeders to other shaft substations.

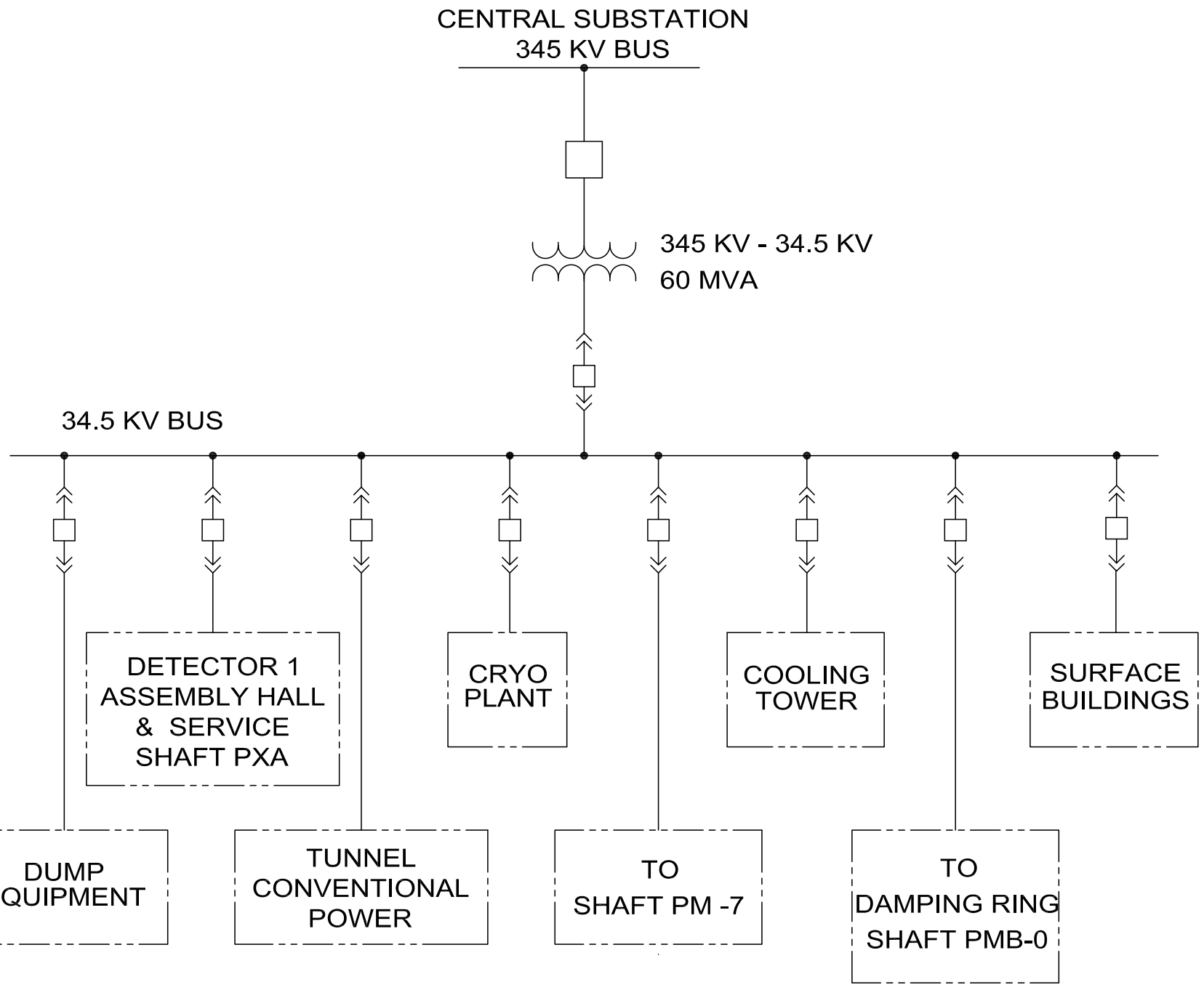

**Figure 11.36** Central Region 35 kV Switchgear One-Line Diagram

### 11.6.4.4 Standby power

Standby power generation is provided at each shaft location to support life-safety facilities when normal power is not available. The standby power-distribution system automatically generates electricity for selected facilities when called upon using diesel generators. The generators are rated at 4.16 kV and sized for the load served. The 4.16 kV voltage is needed due to the length of the distribution feeders. On the surface and in the tunnel, a dedicated standby power transformer is provided to serve the standby power loads.

The electrical lines are installed in the underground tunnels and enclosures in conduits that are either in concrete-encased duct banks or embedded in the tunnel floor and routed up to the surface at each shaft. Cable installation and splicing is accommodated with vaults spaced at approximately 522 m along the length of the main tunnel.

## 11.6.5 Life safety and egress

The life-safety requirements and fire-protection systems for the single-tunnel design concept are based upon the National Fire Protection Association (NFPA) 520, Standard for Subterranean Spaces, 2005 Edition. In addition, Hughes Associates, Inc. was retained to assess the feasibility of the single-tunnel design by analysing different fire scenarios in the beam tunnel and damping ring. A fire analysis of the single-tunnel portions of the ILC installation was conducted using the Fire Dynamics Simulator (FDS) computational fluid-dynamics program developed by USA National Institute of Standards. Models of different fire sizes were constructed for the ML tunnel, base caverns, and DR. Anticipated combustible fuel loadings in the single tunnels were evaluated and it was determined that pool/spill fire scenarios involving transformer oil were the most demanding fire scenarios. Table 11.18 summarises fire-size limitations for various size tunnels based on spill area, rates and volumes.

**Table 11.18**
Summary of Tunnel Fire Modelling Results

| Tunnel type | Tunnel diameter (m) | Limiting fire size (kW) | Confined spill area ($m^2$) | Continuous spill rate (l/min) | Unconfined spill volume (l) |
|---|---|---|---|---|---|
| Main Linac/ Straight Damping Ring | 4.5 | 750 | 0.8 | 1.28 | 1.25 |
| | 5.0 | 1,000 | 1.0 | 1.70 | 1.57 |
| | 5.5 | 1,100 | 1.1 | 1.87 | 1.69 |
| | 6.5 | 1,500 | 1.4 | 2.55 | 2.17 |
| Base cavern | 4.5 | 3,000 | 2.4 | 5.10 | 3.86 |
| | 5.0 | 4,000 | 3.0 | 6.81 | 4.95 |
| | 5.5 | 4,500 | 3.3 | 7.66 | 5.48 |
| | 6.5 | 6,000 | 3.6 | 10.21 | 7.08 |
| Curved Damping Ring | 4.5 | 2,500 | 2.0 | 4.25 | 3.31 |
| | 5.0 | 3,250 | 2.5 | 5.53 | 4.13 |
| | 5.5 | 4,000 | 3.0 | 6.81 | 4.95 |
| | 6.5 | 5,000 | 3.6 | 8.51 | 6.02 |

The findings support the single-tunnel concept and prove that the life-safety requirements and fire-protection system requirements of NFPA 520 will allow occupants in the single-tunnel portions to evacuate safely during a fire, provided that the maximum anticipated fire size in a tunnel can be restricted to the limiting fire sizes for each tunnel type and diameter established in the analysis. In addition, the analysis concluded that it is not necessary for ventilation systems in the tunnel to shut down during a fire event, provided air velocities supplied during the fire are less than 1 m/s.

### 11.6.5.1 Personnel egress

At the base of each ML access shaft is a cavern that contains oil-filled electrical equipment, water pumps, motors and other utility equipment. This equipment has the highest risk for fire. The prevailing codes require the containment of such areas through the use of fire-rated walls and doors. In addition, the elevators located in each shaft are also isolated by fire-rated walls and doors. Once these areas are properly isolated, the main linac (or DR) tunnel can be used for personnel travel to the shaft exit in the event of an emergency incident. Due to the overall tunnel length, it is also required to have a fire-protected AOR located at the midpoint between shafts to provide an intermediate safe area for injured personnel or to await emergency assistance. In areas where a service tunnel is located adjacent to the main tunnel, such as the RTML and BDS, crossover labyrinths are provided for passage between the two tunnels. These crossover labyrinths are located such that the travel distance to the crossover does not exceed 120 m from any point in either tunnel. The crossover labyrinths are separated by 2 h fire rated construction. The DR is an extension of the single portion of the tunnel and is provided with two vertical exits. These exits are separated from the common space by 2 h fire-rated construction. The following provisions for the required emergency fixtures are included:

- emergency lighting;

- illuminated exit signage;
- illuminated exist direction signs;
- a check in and check out system.

#### 11.6.5.2 Fire suppression

Automatic sprinkler protection is required throughout the facility. It is a class I standpipe system with 2.5-inch fire department hose valves spaced approximately 100 m apart. Portable fire extinguishers are also provided.

#### 11.6.5.3 Fire detection

Addressable fire detection and voice alarm is provided. Manual pull stations are spaced approximately 120 m apart. Smoke detection is provided at caverns and other sensitive areas. A voice/alarm system is capable of transmitting voice instructions from the fire command station located at the surface buildings. A two-way fire department communication system is provided and operated from the Fire Command Station. The two-way communication jacks are spaced 130 m apart.

## 11.7 Handling equipment

### 11.7.1 Introduction

This section covers the handling equipment used for on-site transport and installation of components. The on-site handling and transport operations start with unloading of components following delivery by their supplier to the site and finish when the components are installed in their final positions in the accelerator tunnels and service buildings.

ILC Handling equipment can be split into two main categories:

- "installed handling equipment" that is permanently installed in buildings or underground structures, such as cranes, elevators, hoists, and the external gantry used to lower experiment modules to the underground area;
- "mobile handling equipment" that can move between buildings or underground structures, such as road transport and handling equipment, industrial lift trucks, tractors and trailers,and custom-designed vehicles for transport and installation of equipment underground.

For the underground transport and installation of cryomodules and magnets, special equipment is needed so as to fit within the tunnel cross section, taking account of cost and installation timescale considerations. The mobile equipment used on the surface and in the tunnels is essentially the same for the Americas, European and Asian sites.

Equipment used to move detector components before lowering is not discussed in this section. The Americas and European handling equipment solutions that are based on the use of vertical access shafts are described. Inclined access tunnels are used in the Asian design. In this case, a fleet of goods and passenger vehicles is used for equipment and personnel transit between the surface and underground areas. The fleet is defined and operated to ensure adequate throughput as required by the installation schedule and also to ensure safe exit for personnel working underground in the event of fire or accident. Vehicles equipped with internal combustion engines are used for the inclined access tunnels; these are not suitable for use in the rest of the underground areas. To allow transfer of equipment from the inclined tunnel access vehicles to the tunnel transport and installation vehicles, junction caverns equipped with overhead travelling cranes provide the interface between the underground accelerator areas and the sloping access tunnels.

## 11.7.2 Transport operations

### 11.7.2.1 Initial delivery to site

Delivery of equipment to site is covered by the supply contracts for each item of equipment. This means delivery to assembly halls, storage areas or tunnel-access points as appropriate.

### 11.7.2.2 Surface transport and handling on and between sites

Surface transport operations include transfers inside and between buildings on the main laboratory site as well as transfers between the main laboratory site and the tunnel access points. These operations are carried out using a fleet of road transport vehicles. Vehicle unloading is carried out by industrial lift trucks or overhead travelling cranes.

### 11.7.2.3 Transfer between surface and underground via vertical access shafts

Lowering of equipment from the surface to the underground areas is carried out via vertical access shafts equipped with elevators and overhead travelling cranes. Shafts of different diameters are used along the length of the accelerator; four 14 m diameter shafts are available for lowering of cryomodules.

The surface buildings above the access shafts are equipped with overhead travelling cranes with sufficient lift heights to lower equipment to the caverns at the base of the shafts via handling openings reserved in the shaft cross section. In addition goods/personnel lifts allow personnel access and are also used to lower equipment. The cross section of a 9 m machine-access shaft with crane-handling opening and lift shaft (European site version) is shown in Fig. 11.37.

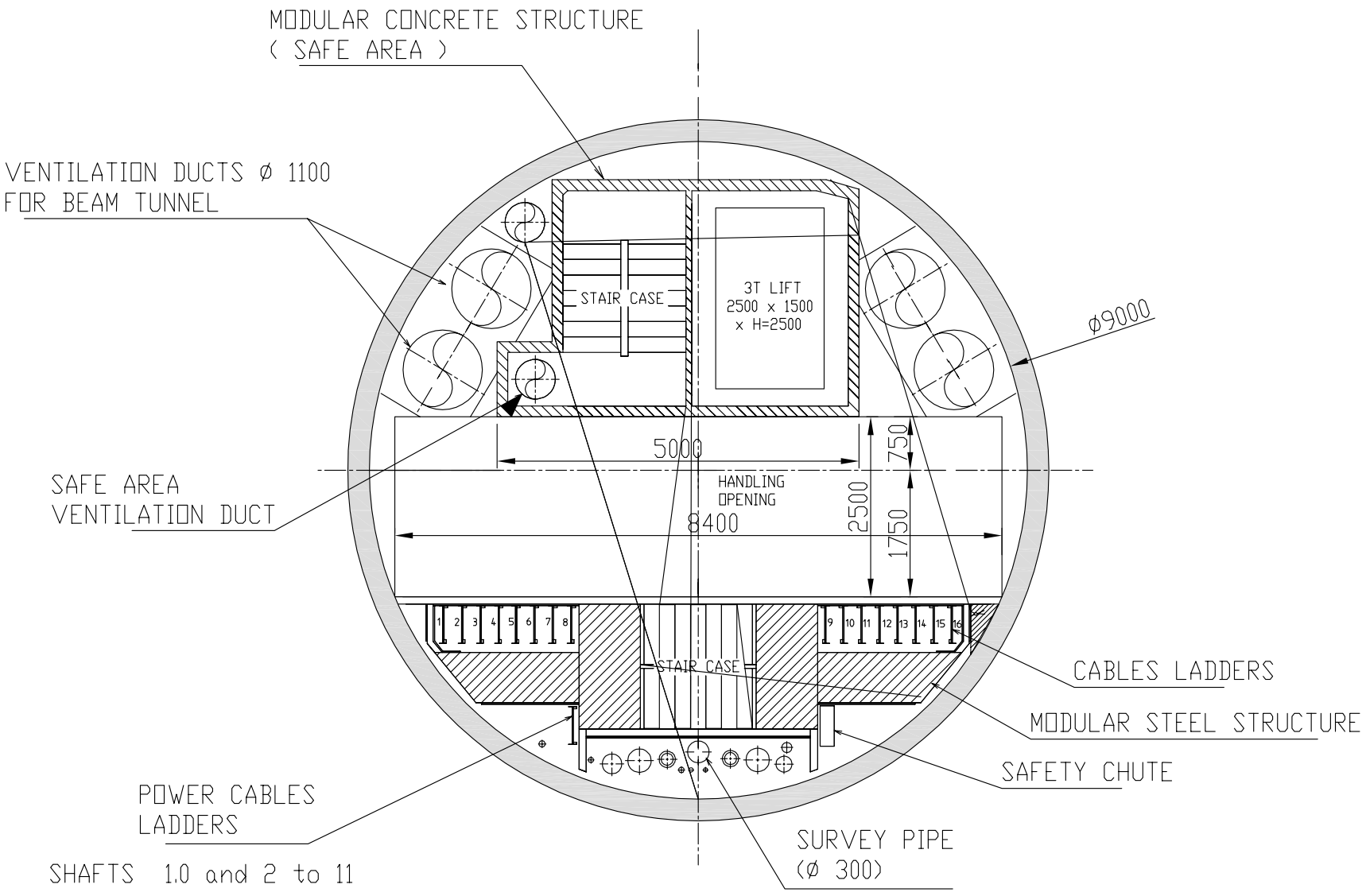

**Figure 11.37** Cross section of access shaft

## 11.7.3 Installed handling equipment

### 11.7.3.1 Elevators

The elevators for the European scheme are listed in Table 11.19. The European scheme is based on one elevator per access shaft with an adjacent stairwell built into the concrete modules that are used to build up the elevator shaft. The Americas scheme is based on the use of twin elevators in the access shafts; the elevator shafts are separated by a fire-resistant wall.

**Table 11.19** Elevators (European scheme)

| Location | | Area | | | |
|---|---|---|---|---|---|
| Shaft Diameter | Shaft/location | DR | RTML | Main Linac | Experiment |
| 14m | PM-12, PM-8, PM+8, PM+12 | | | 4× 3 t | |
| 9m | PM-10, PM+10 | | | 2× 3 t | |
| 6m | PM-13, PM-11, PM-9, PM-7, PM+7, PM+9, PM+11, PM+13 | | 2× 3 t | 6× 3 t | |
| 9m | PMB-0, PMC- 0 | 2× 3 t | | | |
| 6m | PZB-0, PZA-0 | | | | 2× 3 t |
| | Control room | | | | 2× 1.6 t |
| | Detector caverns | | | | 2× 1.6t |

### 11.7.3.2 Cranes

11.7.3.2.1 Shaft transfer and underground area cranes For the Americas and European sites the transfer of heavy loads between the surface and the underground areas is carried out using overhead travelling cranes installed in the surface buildings above shafts. Cranes are used for handling of loads underground in the experimental detector caverns and interaction region as well as in the beam-dump and positron-source caverns. Table 11.20 lists the cranes and hoists.

**Table 11.20.** Cranes and hoists

| | A1 $e^-$ source | A2 $e^+$ source | A3 DR | A4 RTML | A5 Main Linac | A6 BDS | E Experiment |
|---|---|---|---|---|---|---|---|
| Surface Buildings | | | | | | | |
| Detector assembly | | | | | | | 1× 4000 t+ 2× 20 t+ 2×400 t 40 t aux |
| Cooling Towers | | | 2× 5 t | | 12× 5 t | | 1× 5 t |
| Cooling ventilation | | | 2× 15 t | | 12× 15 t | | 1× 15 t |
| Shaft access | | | 2× 20 t | 2× 20 t | 12× 20 t | | |
| Cryo Compressors | | | 1× 20 t | | 6× 20 t | | 1× 20 t |
| KlyCluster Building | | | | | 12× 10 t | | |
| Miscellaneous surface hoists | | | 2× 5 t | 2× 5 t | 12× 5 t | | 4× 5 t |
| Underground Structures | | | | | | | |
| Detector Caverns | | | | | | | 3× 40 t |
| Beam Dumps | | | | | | 4× 5 t | |
| Sources Facilities | 1× 20 t | | | | | | |
| Miscellaneous cavern hoists | | | 2× 5 t | 2× 5 t | 12× 5 t | 6× 5 t | 6× 5 t |

In addition to the cranes installed in surface buildings above access shafts, cranes are installed in the service buildings to carry out installation and maintenance of plant.

### 11.7.3.3 Hoists

Hoists are installed in surface buildings and underground areas for various installation and maintenance activities.

### 11.7.3.4 External gantry used to lower experiment modules to the underground area

An external gantry of 4000 t capacity is used to lower assembled detector modules from the surface to underground. This gantry is rented from an industrial supplier for the period scheduled for lowering the modules; the supply contract includes its assembly, operation then dismantling and removal from site.

## 11.7.4 Mobile handling equipment

### 11.7.4.1 Underground transport and handling

11.7.4.1.1 Schedule and space considerations Initially the full width of the accelerator tunnels are available for installation of services, allowing the use of standard industrial lift trucks, tractors and trailers. The available space for transport narrows once the beam-line equipment starts to be installed. For tunnel construction cost reasons the transport passage is kept to a minimum; this means that cryomodule transport vehicles, for instance, are not able to pass each other in the tunnel.

11.7.4.1.2 Cryomodules The space required for module transport and installation in the tunnel has a major influence on the cross section of the main linac tunnel. The large number of cryomodules to be transported and installed means that it is important to optimise the whole sequence of cryomodule transport to allow rapid transport and installation.

The design of the cryomodule transport vehicle design minimises the width of the reserved transport volume. It is capable of transport along the tunnel as well as transfer from the transport zone onto the support jacks. The vehicle (Fig. 11.38) is based on that used to install conventional magnets for the LHC. The vehicles are equipped with an automatic guidance system. The operator is required to ensure that the floor is clear of personnel and equipment. The vehicle can be configured for different loads and can therefore also be used for transport of other items.

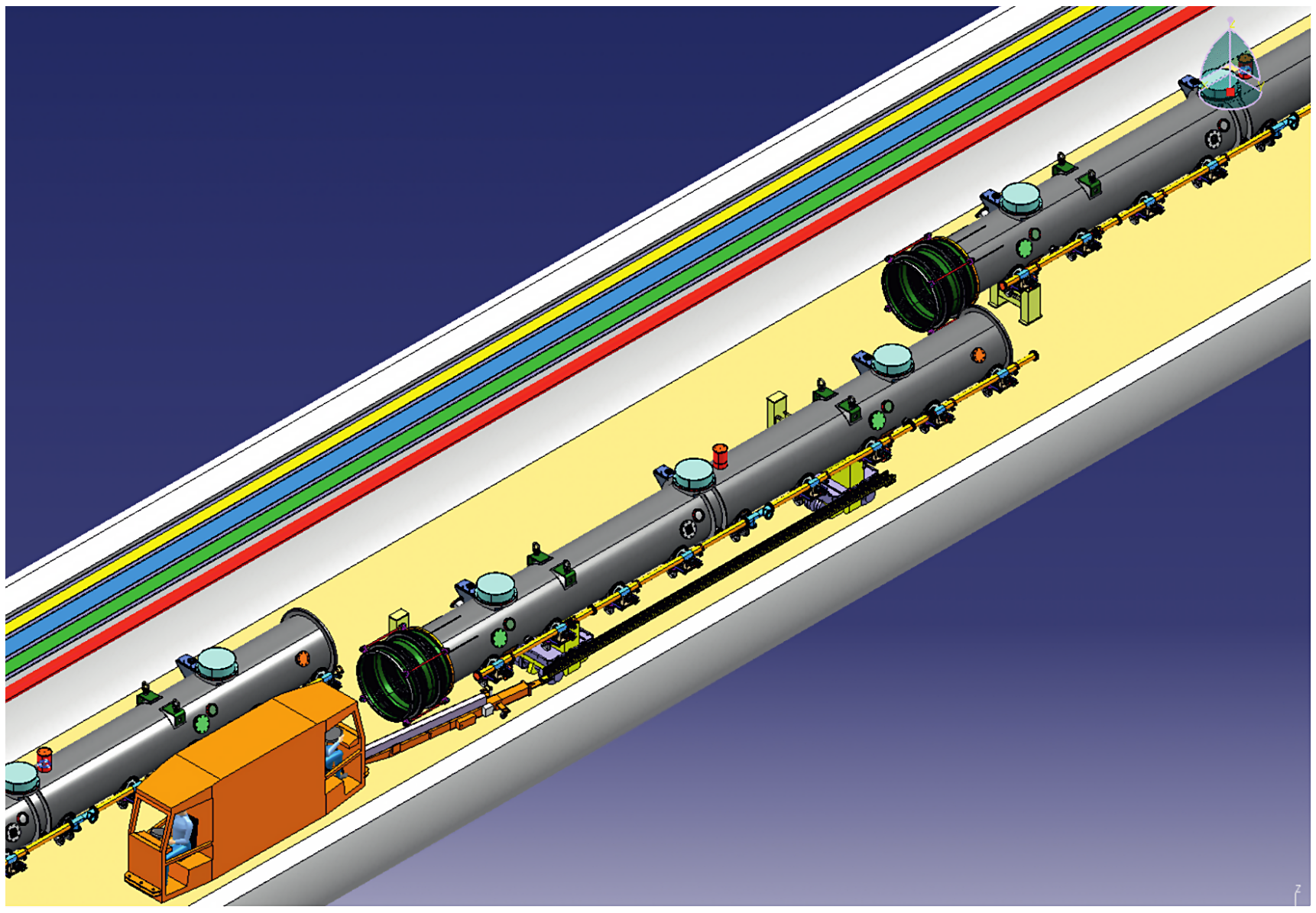

**Figure 11.38** Cryomodule transport vehicle during transfer onto supports (Case shown is installation between two previously installed cryomodules)

Although module installation logistics aims for sequential installation, the installation process allows installation of cryomodules between two previously installed cryomodules in the event of supply delays or if sorting of modules is required. In addition the system is able to remove a previously installed cryomodule if major repairs are needed.

The interconnections between cryomodules are installed after the cryomodules have been positioned on their floor supports – this gives a clearance between modules of over 150 mm during their transfer onto their supports which allows rapid lateral transfer under manual control with minimum risk of damage to the adjacent cryomodule.

The cryomodule design includes lifting points and support points to allow the whole sequence of transport and handling operations. These are needed during the phases of module assembly, testing, storage, road transport to access points, lowering, tunnel transport and installation. The cryomodule design includes the transport restraints and special lifting beams used when handling fully assembled cryomodules during the installation process.

11.7.4.1.3 Magnets Specially designed vehicles (Fig. 11.39) are used for magnet transport along the tunnel followed by their installation. The use of vehicles combining transport, lifting and transfer avoids the need to transfer the load between different items of equipment and results in optimised installation times.

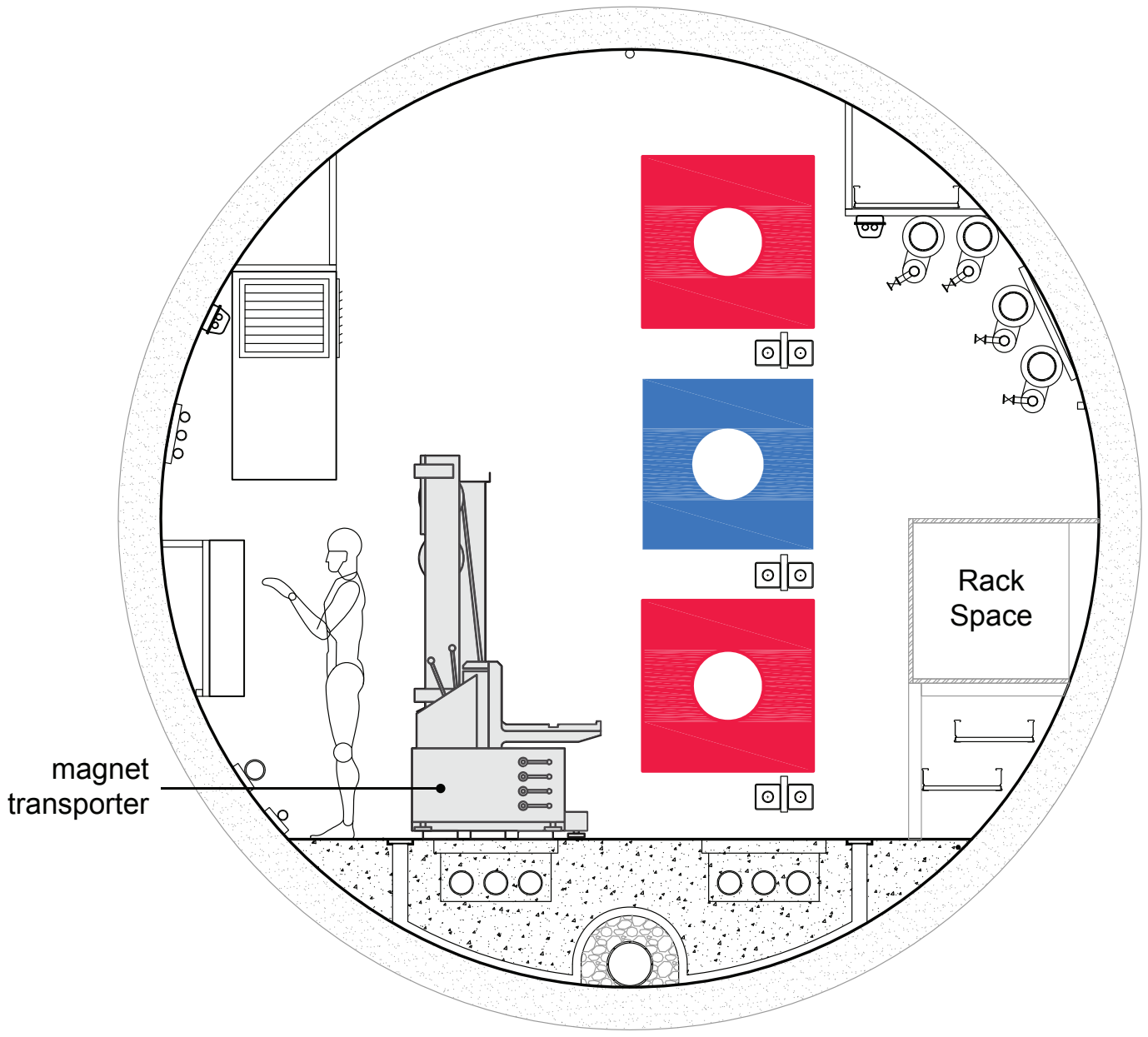

**Figure 11.39**
Special vehicle for magnet installation – shown in damping rings tunnel.

11.7.4.1.4 RF equipment RF equipment installation requires transport along the tunnel followed by precise positioning at a range of heights. The solution is to use an adaptation of the magnet transport and installation vehicle.

11.7.4.1.5 Other accelerator equipment Standard industrial handling equipment such as forklift trucks, electrical tractors and trailers are used to transport and install equipment other than cryomodules, magnets and RF in the tunnel. Where optimal this installation is carried out before cryomodule and magnet installation.

11.7.4.1.6 Personnel transport Personnel transport in the underground areas is by means of small electrical tractors or bicycles.

### 11.7.4.2 Surface transport and handling equipment

Standard road-going trucks and trailers are used for surface transport between sites. Standard industrial handling equipment such as forklift trucks, are used for material handling where it is not feasible to use overhead travelling cranes.

## 11.8 Alignment and survey

Survey and alignment covers a very broad spectrum of activities, starting from the conceptual design of the project, through the commissioning of the machines, to the end of operations. The cost estimate developed covers the work necessary until successful completion of the machine installation. It includes equipment needed for the tasks to be performed, and equipment for a dedicated calibration facility and workshops. It also includes the staff that undertake the field work, and the temporary manpower for the workshops. Full time, regular staff is considered to be mainly dedicated to organisational, management, quality control, and special alignment tasks. The cost estimate is mostly based on scaling the equivalent costs of the LHC to the ILC scope.

### 11.8.1 Calibration Facility

A 100 m long calibration facility is needed for the calibration of all the metrological instruments. The facility is housed in a climate controlled and stable building. Due to the range limit of current day commercial interferometers against which the instruments are to be compared the facility has been restricted to 100 m. A mechanical and an electronic workshop are also needed during the preparation phase and throughout the entire project. They are used for prototyping, calibration, and maintenance of the metrological instruments.

### 11.8.2 Geodesy and Networks

A geodetic reference frame is established for use across the whole site, together with appropriate projections for mapping and any local 3D reference frames appropriate for guaranteeing a coherent geometry between the different beam lines and other parts of the project. An equipotential surface in the form of a geoid model is also established and determined to the precision dictated by the most stringent alignment tolerances of the ILC. The geodetic reference frame consists of a reference network of approximately 80 monuments that cover the site. These monuments are measured at least twice, by multi-satellite GPS for horizontal coordinates, and by direct levelling for determining the elevations. The first determination is used for the infrastructure and civil engineering tasks. The second, and more precise determination, is used for the transfer of coordinates to the underground networks prior to the alignment of the beam components. A geodetic reference network is also installed in the tunnel and in the experimental cavern. For costing purposes it is assumed that the reference points in the tunnel are sealed in the floor and/or wall (depending on the tunnel construction) every 50 m. In the experimental cavern, the reference points are mostly wall brackets. The underground networks are connected to the surface by metrological measurements through vertical shafts or horizontal access ways. The distance between two consecutive shafts can exceed 2.5 km in some cases and some additional small diameter shafts may be required.

### 11.8.3 Civil Engineering Phase

The layout points which define the tunnel locations and shapes are calculated according to the beam lines in the local 3D reference frame. The tunnel axes are controlled as needed during the tunnel construction. All tunnels, including profiles, are measured in 3D using laser scanner techniques when the tunnels are completed. The same process is applied to the experiment cavern(s) and other underground structures. The buildings and surface infrastructure are also measured and the as-built coordinates are stored in a geographical information system (GIS).

### 11.8.4 Fiducialisation

Systematic geometrical measurements are performed on all beamline elements to be aligned prior to their installation in the tunnels. The alignment of elements installed on common girders or in cryomodules is first performed, and the fiducial targets used for the alignment in the tunnels are then installed on the girders (cryomodules) and all individually positioned elements. The positional relation between the external markers and the defining centrelines of the elements are then measured. For this report, an estimated 10,000 magnetic elements were assumed to need referencing. It is also assumed that most corrector magnets do not need fiducialisation. This number does not account for instrumentation, collimators, or other special beam elements.

### 11.8.5 Installation and Alignment

The trajectories of all the beamlines are defined in the local 3D reference frame which covers the entire site. The location of reference markers at the ends of each beam line element to be aligned are defined in this reference system, together with the roll angle giving a full 6 degrees of freedom description of element location and orientation. Likewise the position of all geodetic reference points is determined in this reference frame.

Prior to installation, the beamlines and the positions of the elements are marked out on the floors of the tunnels. These marks are used for installing the services, and the element supports. The supports of the elements are then aligned to their theoretical position to ensure that the elements can be aligned whilst remaining within the adjustment range of the supports.

After installation of services such as LCW and cable trays, the tunnels are scanned with a laser scanner. The point clouds are then processed, and the results inserted into a CAD model. A comparison with theoretical models is used by the integration team to help identify any non-conformity and prevent interference with the subsequent installation of components. The current requirements for the one sigma tolerances on the relative alignment of elements or assemblies are given in Table 11.21.

**Table 11.21**
Components and required alignment tolerances.

| Area | (km) | Nb of beam | Error of misalignment on the fiducials ($1\sigma$) |
|---|---|---|---|
| $e^-$ source | 2.3 | 1 | 0.1 mm rms over 150 m |
| $e^+$ source | 3.3 | 1 | 0.1 mm rms over 150 m |
| 2 DRs | 6.6 | 2 | 0.1 mm rms over 150 m |
| RTML | 1.7 | 1 | 0.1 mm rms over 150 m |
| Main linac | 23.9 | 1 | 0.2 mm rms over 600 m |
| BDS | 6.5 | 1 | 0.02 mm rms over 200 m |

The components are aligned in two steps:

- A first alignment is performed to allow connection of the vacuum pipes or interconnection of the various devices. This is done using the underground geodetic network as reference.
- After all major installation activities are complete in each beamline section, a final alignment, or so-called smoothing, is performed directly from component to component in order to guarantee their relative positions over long distances.

To reach and maintain the positioning tolerances of the final doublets in the BDS IR, a 150 m straight reference line is set up as close as possible to the beam components. This line, consisting of lasers or stretched wires and hydrostatic levels and allows for the geometrical connection between the beam lines and the detectors. The IR hall with movable detectors will require an extensive network with monuments and markers on the floors and walls, at several levels, for the use of laser trackers to develop and to maintain a 3D network which is coupled to the in-tunnel networks. This network will continue to evolve during the assembly of the detectors and into operation.

### 11.8.6 Information Systems

The theoretical positions of all the components to be aligned on the beam lines is managed in a dedicated database. This database is also used for managing all the geodetic and alignment measurements and the instrument calibrations. All measurement data are captured and stored electronically and subsequently transferred to the database. Pre-processing of the measurements are carried out in the database and then dedicated software for data analysis is used to calculate the best fit position of the elements and components. These results are also stored in the database where they can be accessed for further post-processing, analysis and presentation. A geographic information system (GIS) is set up for managing all location data.

## 11.9 Installation

### 11.9.1 Scope

This section covers activities required to prepare, coordinate, integrate, and execute a detailed plan for the installation of the ILC components, including the associated site-wide logistics. It includes all labor, materials and equipment required to receive, transport, situate, affix, accurately position, interconnect, integrate, and check out all components and hardware from a central storage or subassembly facility to its operational location within the beam and service tunnels as well as the surface service buildings where applicable. The premise is that the installation group receives fully tested assemblies certified for installation. Fabrication, assembly, component quality control and commissioning, as well as the basic utilities provided by conventional facilities, such as ventilation, air conditioning, fire prevention, high voltage electrical, chilled water and low-conductivity water distribution are described elsewhere.

### 11.9.2 Methodology

The installation WBS is broken down into two major categories, general installation and technical system installation. General installation includes all common activities and preparations and associated logistics. It is further broken down into logistics management, engineering support, equipment, vehicles, shipping and receiving, warehousing, and transportation. Technical system installation includes all efforts required for complete installation of the technical components underground, and in the surface buildings where applicable. General Installation is further broken down into logistics management, engineering support, equipment, vehicles, shipping-receiving, warehousing, and transportation. Technical System Installation covers the six machine areas, viz. electron sources, positron source, damping ring, RTML, main linac and beam delivery. Each element of the WBS for both general and technical systems is then extended two levels further and populated with required labor as well as incidental material and equipment costs. Table 11.22 shows the top-level Installation WBS.

**Table 11.22**
Top-Level WBS Installation.

| WBS | | | | | Component |
|---|---|---|---|---|---|
| 1 | 7 | 3 | 1 | | General installation |
| | | | | 1 | Logistics management |
| | | | | 2 | Engineering support |
| | | | | 3 | Equipment |
| | | | | 4 | Vehicles |
| | | | | 5 | Shipping and receiving |
| | | | | 6 | Warehousing |
| | | | | 7 | Surface transport |
| 1 | 7 | 3 | 2 | | Technical-system installation |
| | | | | 1 | Electron source |
| | | | | 2 | Positron source |
| | | | | 3 | Damping Ring |
| | | | | 4 | RTML |
| | | | | 5 | Main Linac |
| | | | | 6 | Beam Delivery |

The installation estimates made for the RDR are used as a starting point. The scope of major changes impacting the installation work was identified and used to scale the RDR installation estimates accordingly.

General installation accounts for 18 % of the effort in Japan and 16 % at the US/CERN sites; the remaining effort is required for the accelerator systems.

Figure 11.40 indicates the split of effort among accelerator systems.

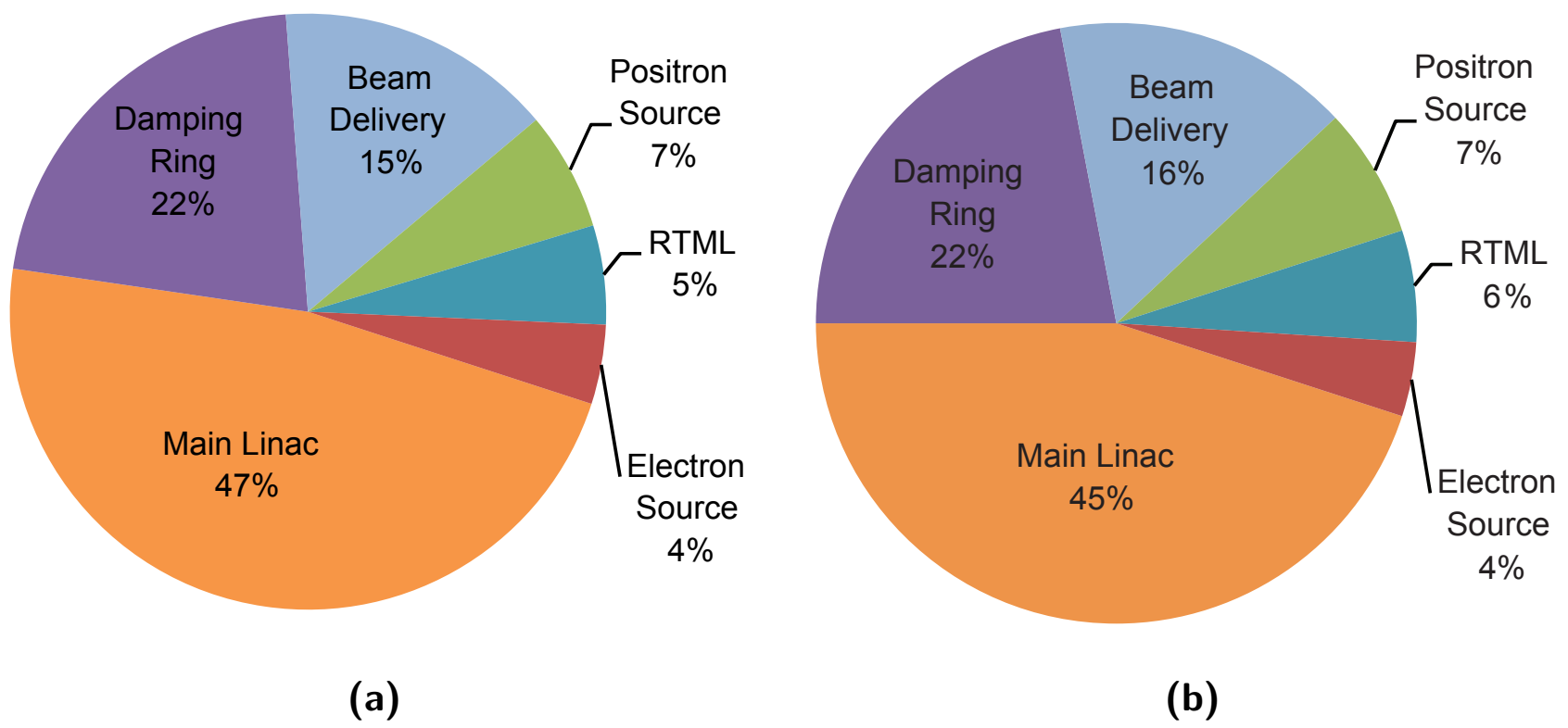

**Figure 11.40** Relative labor effort for the various accelerator systems for (a) the Japan sites and (b) the US/CERN sites.

## 11.9.3 Model of Main-Linac Installation

The complete ILC main linac requires the installation of 1,840 Cryomodules, over 11,130 magnets and approximately 480 high-level RF stations. Since the main linac is a major cost driver, the installation of cryomodules and RF sources is modelled in detail.

The installation rate is one RF unit (Fig. 11.41) and associated services per day for each crew, which includes the following steps:

- tunnel preparation for installation;
- move, place, adjust and fix cryomodule supports;
- install, adjust, fix and prepare section of cryo and beam pipe connections;
- complete cryogenic and vacuum connections, leak check;
- cryomodule sleeve coupling connection.

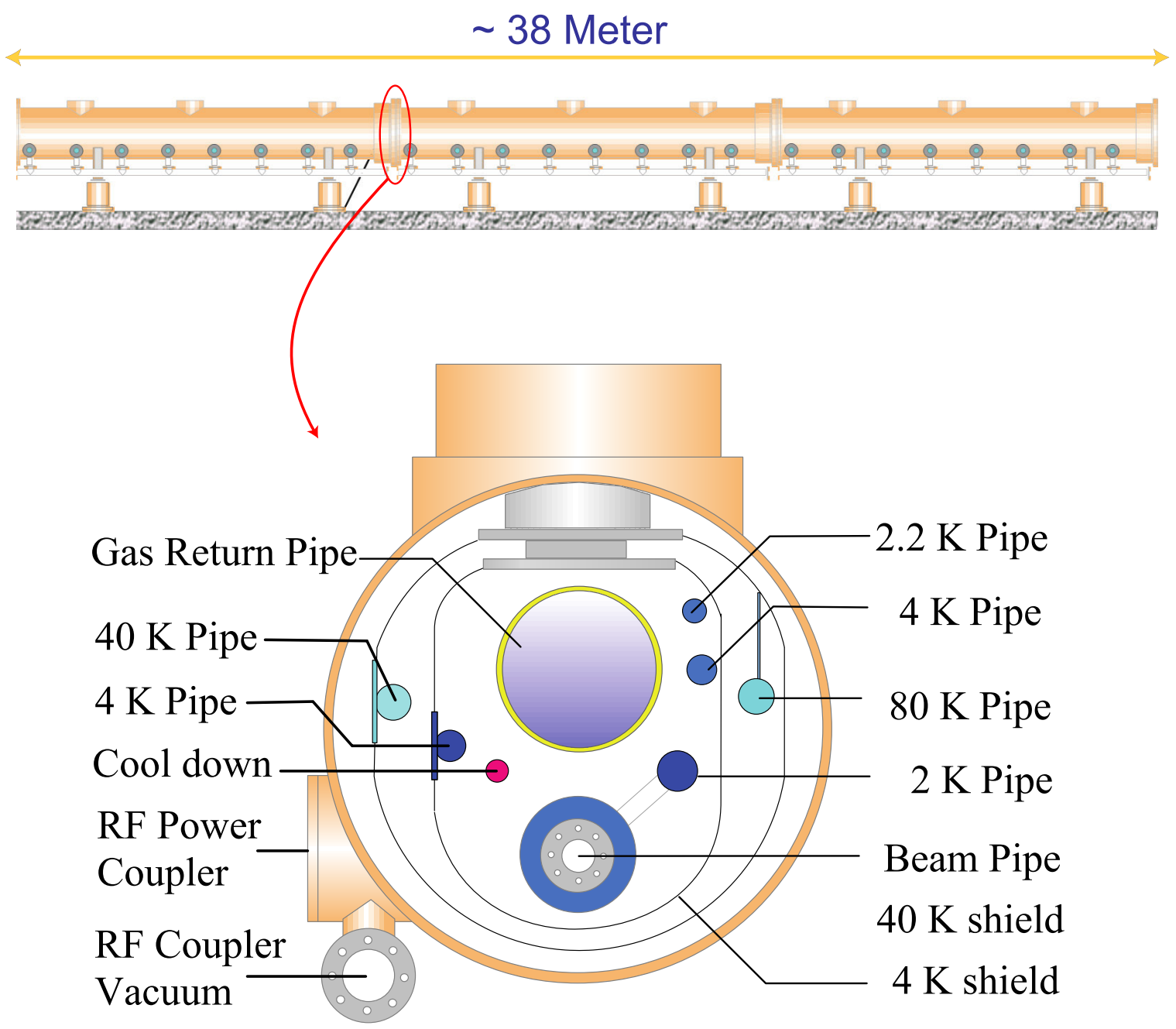

**Figure 11.41** The installation of one of the Main Linac RF units. The cryomodule cross section indicates various pipe interconnects that need to be made.

Labor productivity is taken to be 75 %, or 6 hours per shift, given transport distances and handling difficulty. The model includes the number and size of the equipment pieces, distances to installation, speed of transportation and estimates of number of staff and hours for each task. To estimate the installation effort to assemble an entire RF unit, the Japan site using the DKS is used as a model. The entire RF unit includes Marx modulator, klystron, control racks, cable trays, control cables and complex waveguides. The installation of one RF unit (three cryomodules) is estimated to take a total of 72 person-days.

#### 11.9.3.1 Installation planning in underground segment

To create a cost effective, timely and safe installation plan, certain facility conditions are assumed to exist prior to the start of installation. Some examples include, but are not limited to, the availability of utilities, communication, above-ground warehousing and equipment staging areas. Once these and details of the technical components are known, a very general model, both in time and 3-D, can be developed. Figure 11.42 shows a schematic installation plan for the main linac. The 72-man crew is working in a (moving) 1 km section of the tunnels at the rate of one RF unit per day. Similar activities and crews will be working in other sections of the linac tunnels as they become available. This is also true for the central complex of injectors and damping ring.

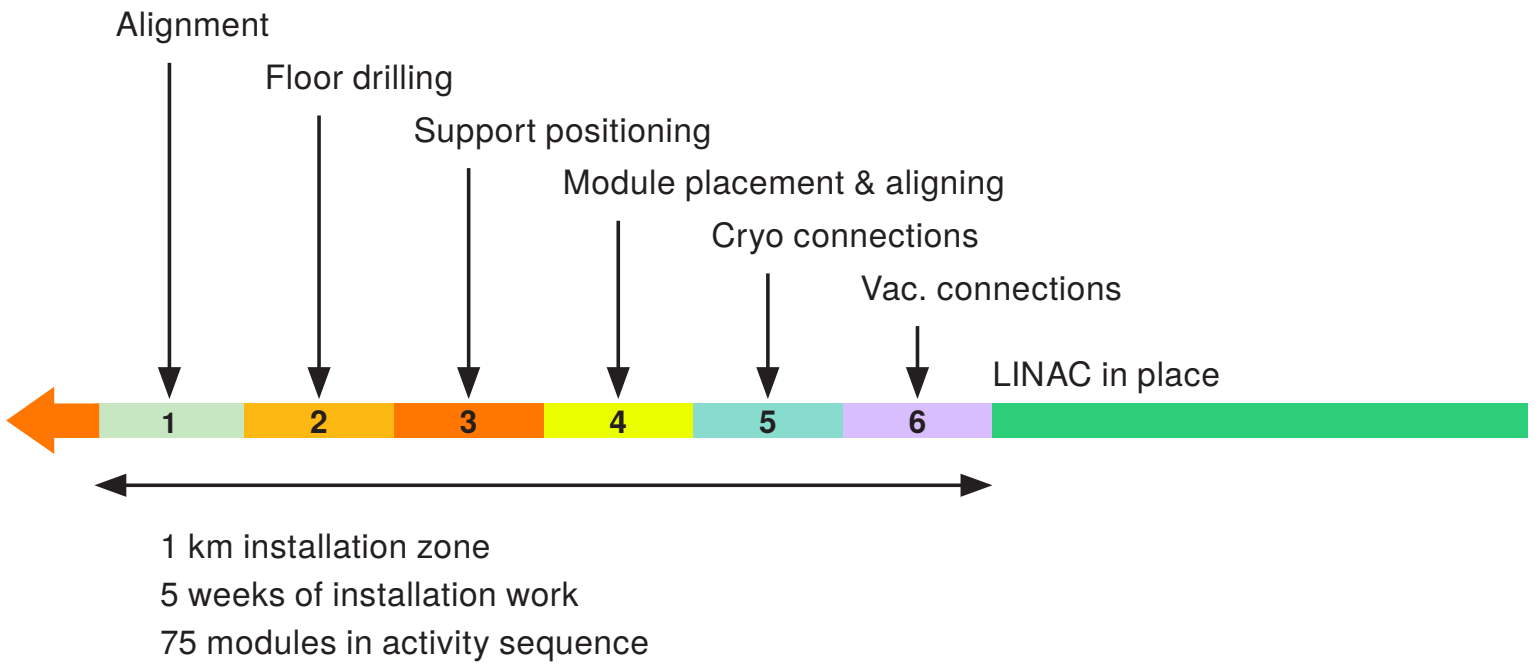

**Figure 11.42**
Installation Model for main-linac components in an underground segment.

This estimate assumes a 2-year installation schedule, a six-month period of ramp-up and on the job training, and 75 % efficiency. In-tunnel activities are concentrated on a day shift, with transport and staging on swing shift. Based on this multi-shift model, the total manpower to fit all the installation activities into the 2-year peak period comprises over 450 people on day shift and another 250 on swing shift in various parts of the tunnel. There are also about 100 people involved in surface logistics.

A detailed plan for installation of the ILC must await the choice of a site. However, an outline plan has been drawn up which is illustrated in Fig. 11.43. This shows the relative effort required to install each accelerator systems, including the main linac and is the basis for the costing of the ILC installation in Chapter 15.

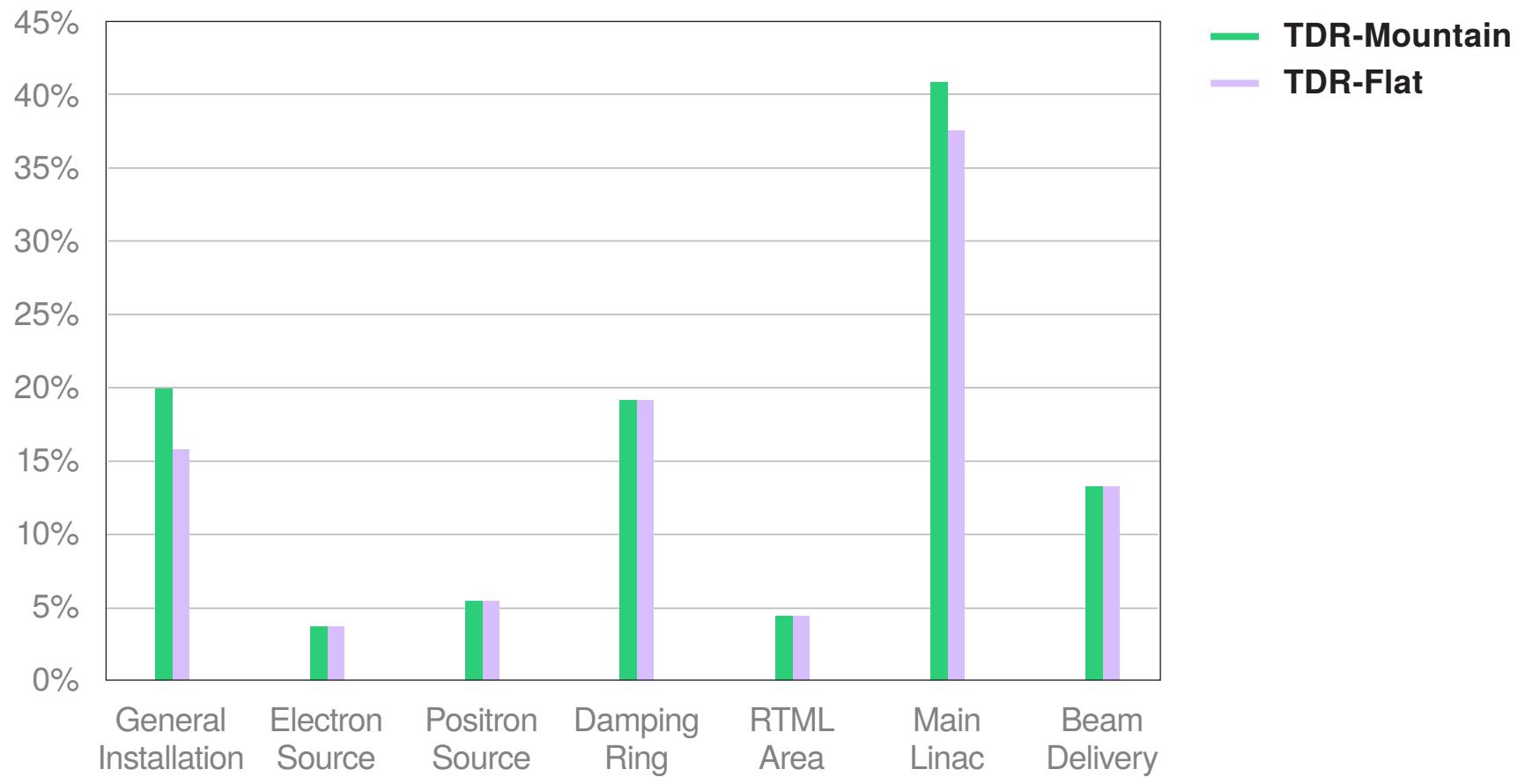

**Figure 11.43**
Relative effort required to install the accelerator system indicated on the $x$ axis.

# Chapter 12
# Possible upgrade and staging options

## 12.1 Introduction

The physics performance requirements for the ILC [228] specify a continuous centre-of-mass energy range from 200 GeV to 500 GeV, with the possibility of an optional upgrade to a centre-of-mass energy of 1 TeV after some years of running. The GDE has focused on providing a mature and robust design and cost estimate for the 200–500 GeV baseline machine, which has been the subject of the previous chapters of this report. The design represents a solution that is cost-performance optimised for that energy range. This chapter presents scenarios for the energy upgrade to 1 TeV and a luminosity upgrade of the baseline 500 GeV machine by up to a factor of two. Furthermore, with the recent discovery of a Higgs boson by the LHC [7, 8] at approximately 125 GeV, a further scenario for a staged approach to the baseline machine is presented, starting at an initial centre-of-mass-energy range of 200–250 GeV.

The level of design detail of the staging and upgrade scenarios is significantly less mature then the baseline. In particular, the TeV upgrade parameters and associated conceptual design are a relatively simple and straightforward scaling of the baseline machine, based on forward-looking assumptions of higher achievable operational parameters for the SCRF technology of 45 MV/m average accelerating gradient with $Q_0 = 2 \times 10^{10}$; achieving these values requires further R&D beyond the baseline technology, but the extrapolation seems realistic. It is assumed that this R&D will continue in parallel to both construction and operation of the baseline machine, such that the extension to the main linacs required for the energy upgrade would benefit from the improved technology (see Part I Section 2.3.4). Both the luminosity upgrade and the low-energy staging are based on the existing technology and require no additional R&D. However, no attempt has been made at this time to study engineering and potential cost trade-offs. The initial 250 GeV stage ("Light Higgs Factory") in particular could well benefit for a re-evaluation of machine parameters that may lead to further cost-performance optimisation at that energy.

The remainder of this chapter deals with the top-level parameters for the staging and upgrade scenarios, and the implications for the machine sub-systems. For the TeV upgrade, an approach to construction that has the minimum impact on ILC operation is discussed. The two TeV upgrade parameter sets presented (so-called low and high Beamstrahlung) were arrived at after careful consideration of the physics impact after discussion with the physics and detector community. Rough cost estimates for the upgrades based on a direct scaling of the TDR baseline costs are provided in Section 15.8.

## 12.2 Parameters

Table 12.1 shows the main parameters for a possible first-stage 250 GeV centre-of-mass-energy machine, the 500 GeV luminosity upgrade, and two possible parameter sets for the TeV upgrade. In the remainder of this section, the parameter sets for the luminosity and TeV upgrade will be discussed. The parameters for the first-stage 250 GeV machine are identical to the baseline parameter set for that energy (see Section 2.2) with the exception of the AC power which will be discussed in Section 12.5.

**Table 12.1.** Primary parameters for a proposed 250 GeV centre-of-mass-energy first stage, the luminosity upgrade for the 500 GeV baseline machine, and the two parameter sets for the TeV upgrade: low Beamstrahlung (A) and high Beamstrahlung (B). The baseline 500 GeV parameters are included for reference.

| | | | Baseline | 1st Stage | L Upgrade | TeV Upgrade A | TeV Upgrade B |
|---|---|---|---|---|---|---|---|
| Centre-of-mass energy | $E_{CM}$ | GeV | 500 | 250 | 500 | 1000 | 1000 |
| Collision rate | $f_{rep}$ | Hz | 5 | 5 | 5 | 4 | 4 |
| Electron linac rate | $f_{linac}$ | Hz | 5 | 10 | 5 | 4 | 4 |
| Number of bunches | $n_b$ | | 1312 | 1312 | 2625 | 2450 | 2450 |
| Bunch population | $N$ | $\times 10^{10}$ | 2.0 | 2.0 | 2.0 | 1.74 | 1.74 |
| Bunch separation | $\Delta t_b$ | ns | 554 | 554 | 366 | 366 | 366 |
| Pulse current | $I_{beam}$ | mA | 5.79 | 5.8 | 8.75 | 7.6 | 7.6 |
| Average total beam power | $P_{beam}$ | MW | 10.5 | 5.9 | 21.0 | 27.2 | 27.2 |
| Estimated AC power | $P_{AC}$ | MW | 163 | 129 | 204 | 300 | 300 |
| RMS bunch length | $\sigma_z$ | mm | 0.3 | 0.3 | 0.3 | 0.250 | 0.225 |
| Electron RMS energy spread | $\Delta p/p$ | % | 0.124 | 0.190 | 0.124 | 0.083 | 0.085 |
| Positron RMS energy spread | $\Delta p/p$ | % | 0.070 | 0.152 | 0.070 | 0.043 | 0.047 |
| Electron polarisation | $P_-$ | % | 80 | 80 | 80 | 80 | 80 |
| Positron polarisation | $P_+$ | % | 30 | 30 | 30 | 20 | 20 |
| Horizontal emittance | $\gamma\epsilon_x$ | μm | 10 | 10 | 10 | 10 | 10 |
| Vertical emittance | $\gamma\epsilon_y$ | nm | 35 | 35 | 35 | 30 | 30 |
| IP horizontal beta function | $\beta_x^*$ | mm | 11.0 | 13.0 | 11.0 | 22.6 | 11.0 |
| IP vertical beta function (no TF) | $\beta_y^*$ | mm | 0.48 | 0.41 | 0.48 | 0.25 | 0.23 |
| IP RMS horizontal beam size | $\sigma_x^*$ | nm | 474 | 729 | 474 | 481 | 335 |
| IP RMS veritcal beam size (no TF) | $\sigma_y^*$ | nm | 5.9 | 7.7 | 5.9 | 2.8 | 2.7 |
| Luminosity (inc. waist shift) | $L$ | $\times 10^{34}$ cm$^{-2}$s$^{-1}$ | 1.8 | 0.75 | 3.6 | 3.6 | 4.9 |
| Fraction of luminosity in top 1% | $L_{0.01}/L$ | | 58.3% | 87.1% | 58.3% | 59.2% | 44.5% |
| Average energy loss | $\delta_{BS}$ | | 4.5% | 0.97% | 4.5% | 5.6% | 10.5% |
| Number of pairs per bunch crossing | $N_{pairs}$ | $\times 10^3$ | 139.0 | 62.4 | 139.0 | 200.5 | 382.6 |
| Total pair energy per bunch crossing | $E_{pairs}$ | TeV | 344.1 | 46.5 | 344.1 | 1338.0 | 3441.0 |

### 12.2.1 Luminosity upgrade

The luminosity upgrade is achieved by a straightforward doubling of the number of bunches per pulse from the baseline number of 1312 to 2625[1], resulting in a doubling of the average beam power and hence luminosity. All other single-bunch parameters are assumed unchanged from their original baseline values. The bunch spacing is reduced from 554 ns to 366 ns resulting in an increase in beam current from 5.8 mA to 8.8 mA. The beam pulse length increases from 714 μs to 961 μs. The choice of bunch spacing is consistent with both the damping ring harmonic number and the main linac RF pulse length (see Section 12.3).

[1]The number in the original 2007 *Reference Design Report* nominal parameter set.

### 12.2.2 Energy upgrade

The choice of beam parameters and ultimately the luminosity for the TeV energy upgrade is also based on a direct scaling from the baseline parameter set, but is more constrained by additional considerations from the higher energy and average beam power:

- the total AC power required by the upgraded machine should be kept below some realistic limit (assumed to be 300 MW);
- the beam current and pulse length should be compatible with the injectors, damping rings and main linac of the baseline design;
- the energy loss due to Beamstrahlung $\delta_{BS}$ ($\propto \frac{N^2}{\sigma_x^2 \sigma_z}$) should be kept low and the maximum pair-production angle ($\propto \sqrt{\frac{N}{\sigma_z}}$) constrained while maximizing the luminosity per bunch crossing.

The total AC power constraint requires the reduction of the repetition rate from 5 Hz to 4 Hz, while the need to keep the RF pulse length in the original main linac to approximately 1.6 ms and the choice of damping ring harmonic number constrains the number of bunches to 2450 (see Section 12.4).

The Beamstrahlung limits tend to be physics dependent, therefore two parameter sets were proposed to the physics and detector groups for study: a low Beamstrahlung parameter set with $\delta_{BS}$ ~5% and a luminosity of $3.6 \times 10^{34}\,\mathrm{cm^{-2}s^{-1}}$ equal to the luminosity-upgrade value for the 500 GeV baseline, and a second, high-Beamstrahlung set with $\delta_{BS} \sim 10\%$ and a correspondingly higher luminosity of $\sim 4.9 \times 10^{34}\,\mathrm{cm^{-2}s^{-1}}$. Both of these parameter sets are based on a reduced single-bunch charge ($\sim 1.7 \times 10^{10}$), shorter bunch lengths (250 µm and 225 µm for low and high $\delta_{BS}$ respectively) and an increased horizontal beam size to control the Beamstrahlung and pair-production angle, while the vertical beta-function at the interaction point (IP) is further reduced to increase the luminosity per bunch crossing [229]. The bunch lengths and IP beta-functions are within the range of the bunch compressor and final-focus systems respectively. It is relatively straightforward to adjust the machine parameters between these two Beamstrahlung parameter sets.

## 12.3 Scope of the luminosity upgrade

A doubling of the average beam power requires the installation of additional RF power (klystrons and modulators) for the main linacs, as well as significant modifications to the damping rings. The baseline designs for other sub-systems (electron and positron sources, bunch compressors, beam delivery and in particular the high-power dump systems) are already specified to cope with the higher beam power (larger number of bunches per pulse). The reduced bunch spacing in the main linac (366 ns) is consistent with the required bunch patterns in the damping rings with a harmonic number of 7044 [95], and a maximum RF pulse length of 1.65 ms.

The following sections briefly describe the impact and necessary modifications to each accelerator system.

### 12.3.1 Main linacs

The upgraded main linac parameters are given in Table 12.2. The doubling of the number of bunches per pulse (1312 to 2625) and the reduction of the bunch spacing (554 ns to 366 ns), results in a ~50 % increase in beam current (5.8 mA to 8.8 mA). The higher current reduces the matched external Q and thus the fill time by the same factor, resulting in an overall slight shortening of the RF pulse length, and an increase in the RF-beam power efficiency from 44 % to 61 %. Hence a doubling of the average beam power only requires an increase of approximate 44 % in the RF power source (number of klystrons), while the power dumped to the RF loads (reflected power) does not change.

The approach to adding the required additional klystrons, modulators, charging supplies and conventional facilities support differs significantly for the two site-dependent variants considered. For the *flat topography* utilising the *Klystron Cluster Scheme* (KCS, see Section 3.9), the additional RF power sources are added to the surface clusters. Each of the 22 clusters requires an additional 10 klystrons and modulators, or an increase in the total number from 413 to 633. This can most easily be done by adding to the combining waveguide system at the downstream (high power) end, between it and the vertical shaft, if space is left (see Section 3.9). Power extracted by each Coaxial Tap-Off (CTO) along the linac then increases from ∼5.8 MW to ∼8.8 MW. All the RF power-distribution systems, including the main long high-power overmoded waveguide and CTO's, are already specified for the higher power and do not need upgrading. An advantage of the KCS approach is that the additional klystrons can be added adiabatically in parallel to operations, since minimal installation work is require in the accelerator tunnel itself. The majority of the additional water cooling (for the klystrons and modulators) is also primarily localised the surface buildings, making the upgrade relatively straightforward. (Note that there is in principle no additional load in the tunnel itself, with the exception of the water cooling associated with the waveguides.) With the *Distributed Klystron Scheme* (DKS, see Section 3.8) used for the *mountainous topography*, the additional klystrons and modulators must be installed in the tunnel and require modification of the local power distribution systems, since each klystron now drives 26 cavities as opposed to 39 in the baseline. Figure 12.1 shows the approach to upgrading the PDS.

**Table 12.2**
The main linac RF parameters for the luminosity upgrade (the baseline numbers are including for comparison).

| | | Baseline | L upgrade |
|---|---|---|---|
| Gradient | MV/m | 31.5 | 31.5 |
| Bunch spacing | ns | 554 | 366 |
| Bunch charge | nC | 3.2 | 3.2 |
| Bunches per pulse | | 1312 | 2625 |
| Beam current | mA | 5.8 | 8.8 |
| Beam pulse length | μs | 727 | 961 |
| $Q_{ext}$ (matched) | $\times 10^6$ | 5.5 | 3.6 |
| Fill time | μs | 927 | 613 |
| RF pulse length | ms | 1.65 | 1.57 |
| RF to beam power eff. | | 44 % | 61 % |

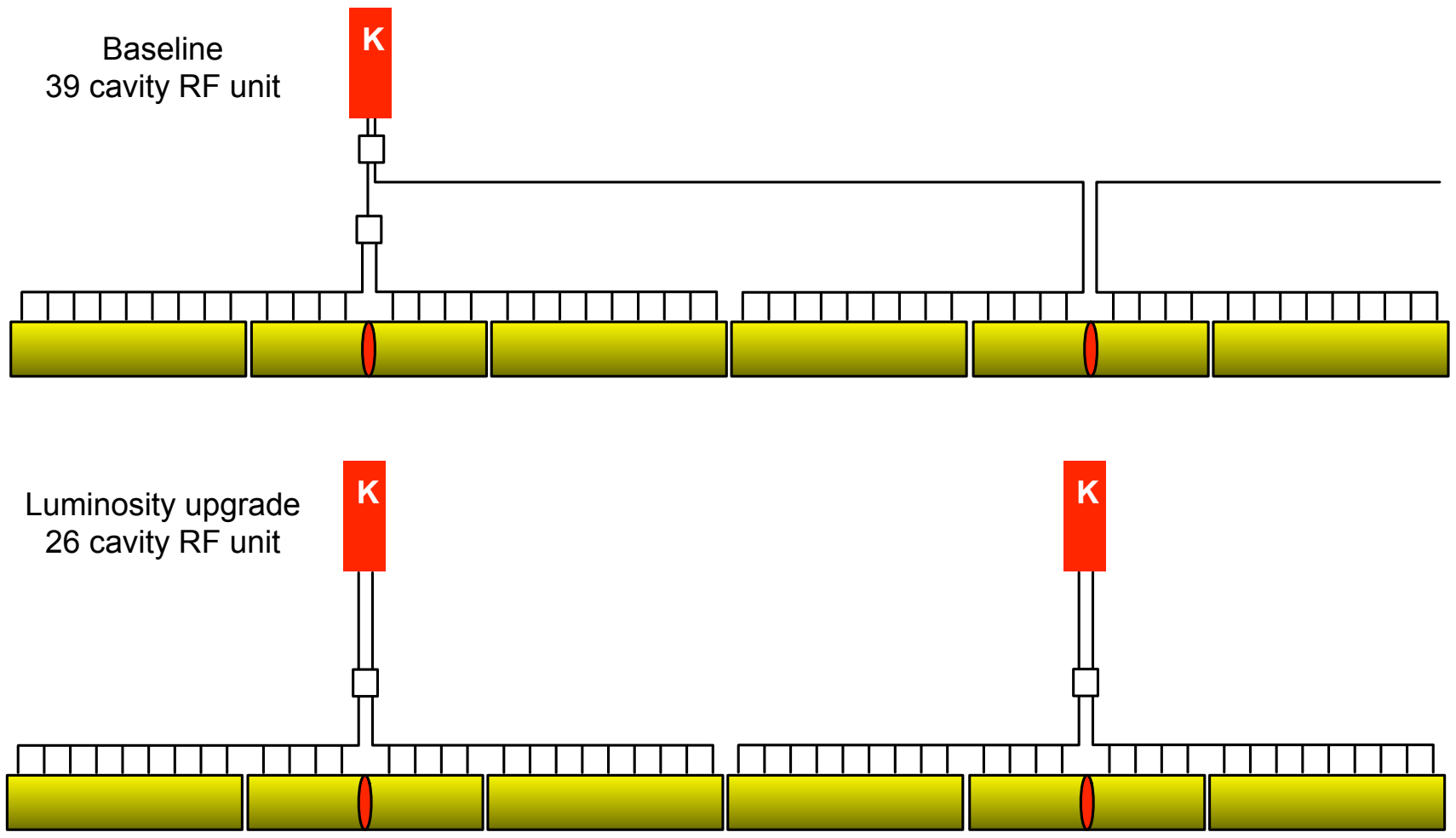

**Figure 12.1**
Scheme for adding klystrons for the luminosity upgrade for the Distributed Klystron Scheme (DKS).

Unlike KCS, the invasive nature of the installation work requires a shutdown during which all the additional RF power would need to be installed. This would also include the additional water cooling and AC power required, although pipe sizes are already specified for the additional load in the baseline, and would not need upgrading. In all other respects, the main linacs would not require modification. In particular, the ∼25 % increase in cryogenic load (dominated by losses from the high-power coupler and HOM losses due to the higher current) is within the baseline specification. All beam-position monitors (and other instrumentation) are compatible with the shorter bunch spacing. Beam dynamics issues (multibunch effects) are also acceptable, and the high-power couplers and HOM couplers/absorbers are specified in the baseline for the higher beam currents (power).

### 12.3.2 Damping Rings

For the high-luminosity upgrade, twice the number of bunches need to be stored in the damping rings, requiring a 3.1 ns bunch spacing. The doubling of the current in the rings poses a particular concern for the positron ring due to the effects of the electron-cloud instability. In the event that the electron-cloud mitigations that have been recommended (see Section 6.5) are insufficient to achieve the required performance for this configuration, the baseline damping ring tunnel and associated underground vaults have been designed to allow the possibility of installing a second positron ring (three rings in total, see Fig. 12.2). The two positron rings would both operate with the baseline parameters (i.e. 1312 bunches per ring). Space has also been provided for the additional power supplies and klystrons in the respective caverns, and the injection and extraction lines (part of the positron RTML system) are designed to accommodate pulsed vertical separation/combination of the positron beam pulse into/from the two rings.

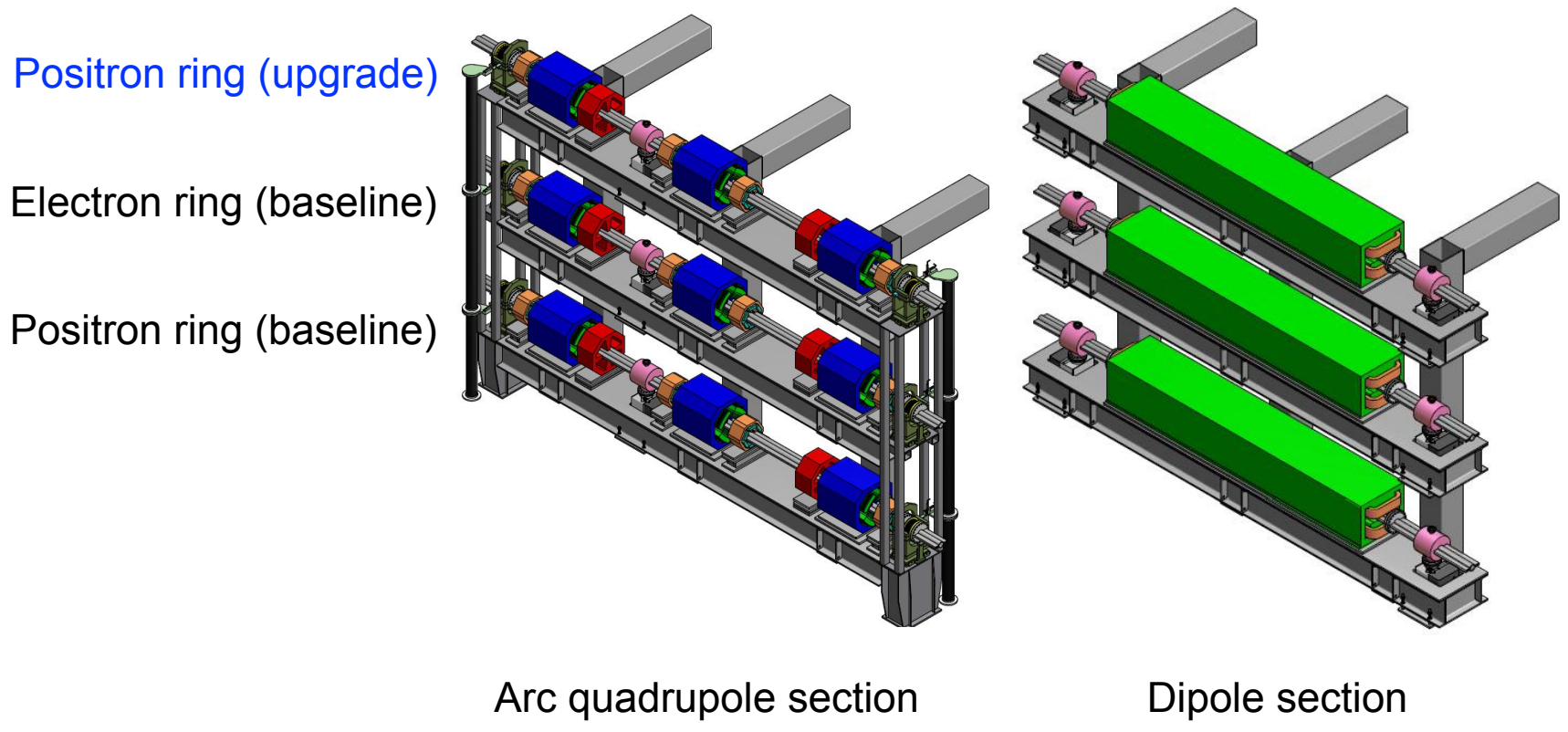

**Figure 12.2** Damping ring configuration, showing the location of the (optional) positron ring for the luminosity upgrade.

For a single ring storing the complete 2625 bunches (corresponding to an average beam current of 0.78 mA and a 200 ms store time at 5 Hz operation), 12 RF cavities are required, supplying 294 kW per cavity to the beam (total of 3.53 MW). However, this is compatible with the power required by the 10 Hz baseline mode (1312 bunches, 0.39 mA). Running the 10 Hz mode with the higher beam current would in principle require a factor of two higher RF power, requiring a doubling of the number of cavities to 24 (since the power coupler is assumed to be limited to $\sim$300 kW). Currently there is space foreseen for an additional 4 cavities, giving a total of 16; thus the number of bunches would be limited to 1750 for low centre of mass running. Although provision is made for a second positron ring should the electron cloud effect at the higher number of bunches prove prohibitive, it is assumed that the electron ring will be able to run with the higher beam current. The principle beam instability issue is the fast ion instability (FII), which becomes more critical at the shorter bunch spacing. However, with the baseline pressure of 1 nTorr, simulations indicate that the FII will be manageable, although the faster growth rate will prove challenging for the multibunch feedback systems (see Section 6.4.5). The vacuum system and in particular the photon stops in the wigglers are all specified for the higher synchrotron-radiation load.

### 12.3.3 Electron and positron sources

The baseline designs for both the electron and positron sources are specified for the production of the higher number of bunches required for the upgrade (see Chapter 4 and Chapter 5 respectively). This includes the DC gun and CW laser systems for the electron source, the photon target for positron production, radiation shielding (positron source), power handling, and the room-temperature RF capture and pre-accelerator sections for both sources. The 5 GeV SCRF booster linac for the positron source will require an additional three 10 MW klystrons and a minor reconfiguration of the power-distribution system. (The electron booster linac, due to its different configuration, has significant RF power margin and requires no additional klystrons for the upgrade.)

### 12.3.4 RTML (bunch compressors)

The RTML — and in particular the SCRF RF linac sections for the bunch compressors — are already compatible with the higher number of bunches (beam current). In particular the RF power configuration has enough overhead to accommodate the increased beam power.

### 12.3.5 Beam Delivery System

All systems in the BDS are specified for the higher beam power (shorter bunch spacing) and no additional modifications are required. In particular the factor of two higher average beam power is well within the specification of the main beam dumps, which are designed to handle the beam power associated with the TeV upgrade parameters.

## 12.4 Scope of energy upgrade to 1 TeV centre-of-mass energy

The upgrade of the beam energy will require extending the main SCRF linacs to provide the additional 250 GeV per beam. The beam current for the TeV upgrade (7.6 mA) is higher than the baseline parameter (5.8 mA) but less than that for the luminosity upgrade (8.8 mA), which assumes some level of the modifications outlined in Section 12.3. Assuming that the luminosity upgrade occurs first, then the injectors (sources and damping rings) will be reused unchanged. The bunch compressor sections will be relocated to the beginning of the extended linacs, as will the 180° turn-around of the RTML (see Section 7.3); the 5 GeV long-transfer line from the damping ring to the turn-around will also need be extended. The undulator-based positron source will remain located at the end of the electron main linac (central campus), but the undulator will need to be replaced with one more suited to the 500 GeV electron beam energy (see Section 12.4.1). The Beam Delivery System will require the installation of additional dipoles to provide the required higher integrated field strength.

The cost and schedule for the upgrade is completely dominated by the extension of the main linacs. One key cost-related consideration is the choice of the accelerating gradient. It is assumed that the current R&D into high-gradient SCRF will continue in parallel with construction and operation of the baseline machine — a period of more than a decade. With this in mind, both a higher gradient and quality factor are assumed for the upgrade linac technology. The actual choice of these parameters will clearly depend on the state-of-the-art at the time the upgrade is approved. However, for the purposes of this discussion, an average accelerating gradient of 45 MV/m with a $Q_0 = 2 \times 10^{10}$ will be assumed. Although significant R&D is required to achieve these ambitious parameters, they are considered realistic (see Part I, Section 2.3.4 and Chapter 6). Assuming that the unit cost of the higher-performance cavities (cryomodules) does not significantly change, the cost of the additional linacs would be reduced by approximately 20% over the baseline (based on the current TDR estimates).

Three scenarios for the upgrade are described in Table 12.3. Scenario A represents a straightforward doubling of the existing main-linac technology, based on the current gradient specifications of 31.5 MV/m average accelerating gradient. Scenario B assumes that the baseline linac is maintained as is (base), but that the additional linac (upgrade) is based on 45 MV/m technology. Finally, scenario C assumes the entire linac is replaced with the higher-gradient technology.

Scenario C would require a complete refurbishing and re-installation of the existing SCRF main linacs. For the linac hardware this is likely to be the most expensive option. However, it would require only an additional 6 km of linac tunnel (and one to two shafts or horizontal access ways) and associated conventional facilities support, and has the smallest overall footprint. Scenario B takes a more conservative approach, and assumes the maximum reuse of the existing baseline infrastructure. Approximately 9 km of additional tunnel (two to three vertical shafts/horizontal access ways) per linac are required (a total of an additional 18 km to the overall footprint). While not as space efficient as scenario C, the assumption of the higher gradient still reduces the overall footprint by $2\times4$ km as compared with a straightforward doubling of the baseline linacs (scenario A). Since scenario B is the less disruptive of the existing hardware, it also opens the possibility of significantly reducing interruption to physics running by allowing the construction and installation of the upgrade linac to occur in parallel with operations. Fig. 12.3 shows a possible scenario for parallel construction based on scenario B.

**Table 12.3** Comparison of main linac upgrade scenarios (gradient). Approximate cavity numbers and linac lengths assume the same cavity length and packing fraction (64%) as the current baseline linac design.

| | | 500 GeV | TeV Upgrade | | | |
|---|---|---|---|---|---|---|
| | | Baseline | Scenario A | Scenario B | | Scenario C |
| | | | | upgrade | base | |
| Energy range | GeV | 15–250 | 15–500 | 15–275 | 275–500 | 15–500 |
| Gradient | MV/m | 31.5 | 31.5 | 45 | 31.5 | 45 |
| Num. of cavities | | 7400 | 15 280 | 8190 | 7090 | 10 700 |
| | | | | total cavities: 15280 | | |
| Linac length | km | 12 | 25 | 9.5 | 11.5 | 17.5 |
| | | | | total length: 21.0 | | |

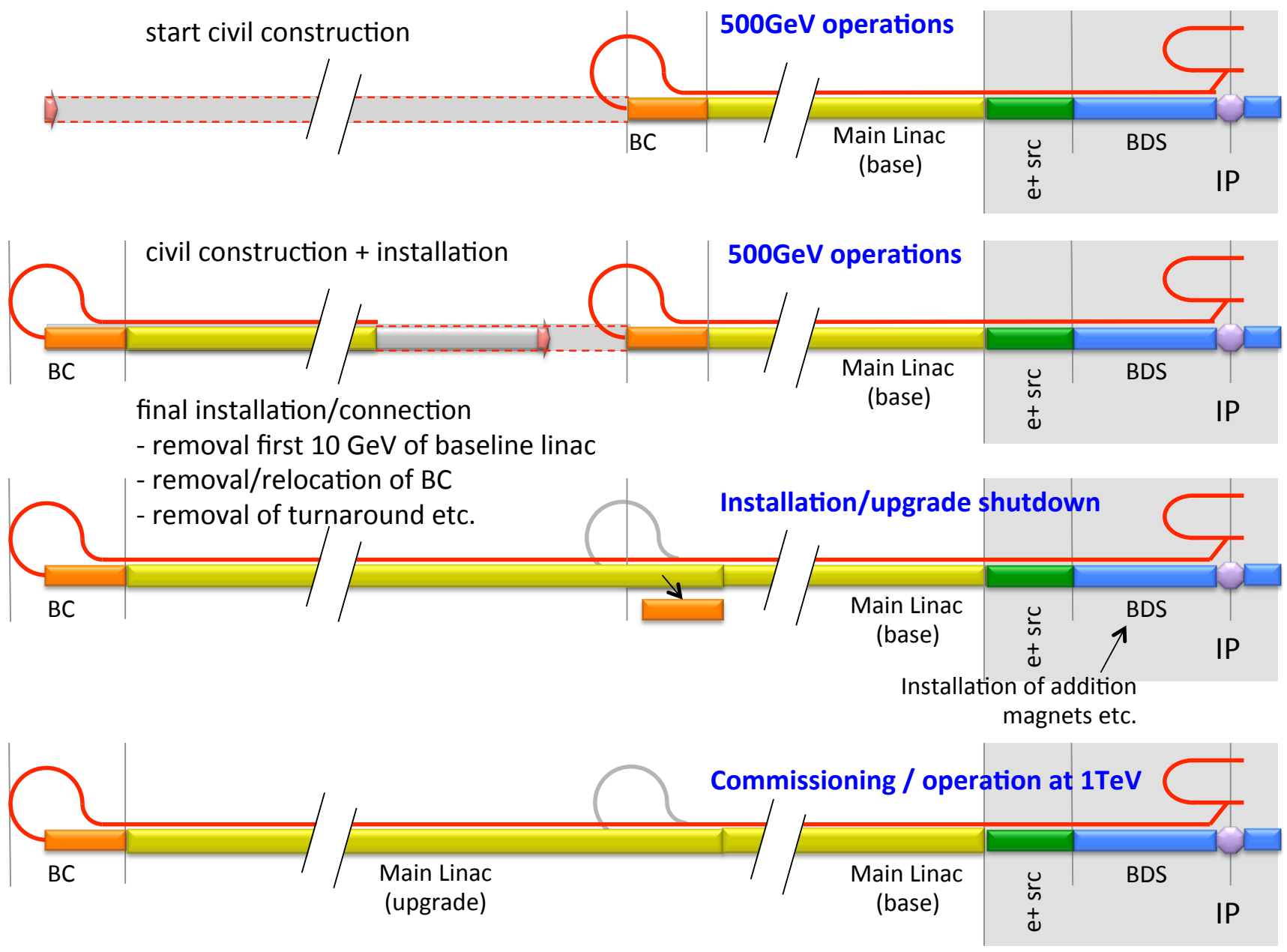

**Figure 12.3** Parallel construction stages for the TeV upgrade (scenario B). Construction of the main linac (yellow) extensions occurs in parallel to 500 GeV operations, requiring a minimum interruption to make the final connections and necessary installation work in the RTML (orange), positron source (green) and BDS (blue). Note that the serial approach shown for the main linac extension construction is oversimplified, and sections of tunnel would likely be constructed and installed in parallel.

Making use of the existing baseline linac in this way has three key implications for the upgrade:

- The beam current and pulse length must be compatible with the existing RF installation and cryogenic cooling capacity.
- The existing linac lattice — which is initially designed to transport a beam energy from 15–250 GeV — must now transport a beam energy of 265 GeV to 500 GeV. This will require the replacement of the first 10 GeV of original linac, since these quadrupoles will not be capable of transporting the higher energy beam (265 GeV to 500 GeV as opposed to 15 GeV to 250 GeV). The remainder of the original linac will use a FoFoDoDo lattice as opposed to the baseline FoDo lattice, resulting in weaker focusing and larger beta functions. Simulations of the beam dynamics have indicated that the vertical-emittance growth can be contained within acceptable limits (see Part I Section 4.6).

- The higher-gradient technology is likely to be based on a cavity shape that has a higher impedance than the current baseline design, potentially resulting in higher wakefield effects that could impact the emittance growth and energy spread of the beam exiting the linacs.

The second and third bullet points would favour replacing the existing linac with the new higher-gradient technology, moving the existing cryomodules upstream. The new quadrupoles could be designed to accommodate the higher beam energy, and the effects of higher wakefields would be reduced. However, this scenario would require a much longer interruption to physics operation than that depicted in Fig. 12.3. For this reason, scenario B is currently seen as more attractive. The approximate numbers given in Table 12.3 for scenario B also assume that the upgrade linac with 45 MV/m will also be used to replace the first 10 GeV of the original baseline linac to provide the stronger quadrupoles (second bullet point). While clearly not the only scenario, this is likely to be the most straightforward and least time consuming. The thirty-five 31.5 MV/m cryomodules removed per linac could in principle be refurbished and used as spares. The principle parameters for the main-linac SCRF for scenario B are given in Table 12.4.

**Table 12.4**
Key main linac parameters for the TeV upgrade (scenario B) compared to the 500 GeV luminosity upgrade parameters. The relative dynamic cryoload gives the total estimated (scaled) dynamic cryogenic load for the linacs relative to the baseline linac.

| | | Baseline (L upgrade) | TeV upgrade upgrade | TeV upgrade base |
|---|---|---|---|---|
| Acceleration | GeV | 15–250 | 15–275 | 275–500 |
| Repetition rate | Hz | 5 | 4 | |
| Gradient | MV/m | 31.5 | 45 | 31.5 |
| $Q_0$ | $\times 10^{10}$ | 1 | 2 | 1 |
| Beam current | mA | 8.8 | 7.6 | |
| Beam pulse length | ms | 0.96 | 0.90 | |
| Fill time | ms | 0.6 | 1.0 | 0.7 |
| RF pulse length | ms | 1.6 | 1.9 | 1.6 |
| Rel. dyn. cryoload | | 1 | 1.2 | 0.8 |

The beam parameters are chosen to keep the RF pulse length for the baseline linac to ∼1.6 ms in accordance with the first bullet point above. In principle this particular constraint could be relaxed if new RF power sources were considered. The RF pulse length in the higher-gradient upgrade linac is ∼2 ms (longer fill time), which will require R&D for the upgrade linac klystron and modulator technology. Finally the lower repetition rate (4 Hz) and assumption of the higher $Q_0$ ($2 \times 10^{10}$) compensate the average dynamic (RF and beam related) cryogenic losses per cavity for the higher gradient, as compared to the 500 GeV luminosity-upgrade parameters.

For the injector systems, it is assumed that the modifications for the luminosity upgrade described in Section 12.3 have been made. For the TeV upgrade parameters, the reduced bunch charge, number of bunches and repetition rate relax the requirements for the sources and damping rings. In particular no modification is in principle required for the electron source or damping rings, the latter of which benefits from the reduced repetition rate to achieve a longer damping time. However, modifications are required for the positron source, RTML and Beam Delivery Systems, which will be briefly described below, followed by a summary of the AC power requirements.

### 12.4.1 Positron source

The undulator-based positron source must be made compatible with an initial electron beam energy of 500 GeV. The solution is to replace the baseline helical undulator with one which is shorter and has a longer period and a lower field (Table 12.5). The upgrade undulator provides a photon beam similar enough to the baseline that the same target and capture arrangement can be used without modification [230]. One important consideration is the photon opening angle ($\sim 1/\gamma_e$) which is reduced by a factor of two for the higher beam energy; this makes photon collimation for polarisation more challenging. Currently a conservative estimate of 20 % polarization is considered feasible, but higher values could be possible provided a suitable solution for the smaller aperture photon collimation can be found [234]. The baseline design geometry of the target-bypass chicane for the high-energy electron beam already accommodates the 500 GeV beam transport (higher synchrotron radiation) with a few percent horizontal emittance growth [231], although additional dipole magnets will need to be installed.

**Table 12.5**
Helical undulator (and other) parameters for the TeV upgrade positron source, compared to the 500 GeV luminosity upgrade parameters. The critical target parameters (and yield) are kept the same for the higher-energy electron beam by replacing the helical undulator [230].

| | | | 500 GeV L upgrade | TeV upgrade |
|---|---|---|---|---|
| Electron beam energy | | GeV | 250 | 500 |
| Positron pulse production rate | | Hz | 5.0 | 4.0 |
| Bunch population | $N$ | $\times 10^{10}$ | 2.0 | 1.7 |
| Effective undulator length | $L_{und}$ | m | 147 | 132 |
| Effective undulator field | $B_{und}$ | T | 0.4 | 0.2 |
| Undulator period | $\lambda_u$ | cm | 1.2 | 4.3 |
| Photon energy ($1^{st}$ harmonic) | | MeV | 42.8 | 27.6 |
| Photon opening angle ($= 1/\gamma$) | | µrad | 2.0 | 1.0 |
| Electron energy loss in undulator | $\Delta E_{und}$ | GeV | 2.6 | 2.4 |
| Average photon power | | kW | 79.3 | 65.5 |
| Peak energy density in target | | J/cm$^{-3}$ | 456 | 475 |

### 12.4.2 RTML

The two-stage bunch-compressor system needs to be "moved" to the new upstream location. The scenario outlined in Fig. 12.3 assumes that a new two-stage compressor will be installed, together with a new turnaround and long transport line. During the shutdown for the final installation work, the baseline warm wiggler sections and cryomodules will be removed together with the first 10 GeV of the baseline Main Linac and replaced with the upgrade linac. The original turnaround would be disconnected and bypassed by the new long transfer line. It is likely that space between the original and upgrade linacs would also be used for additional diagnostics and dump systems, including an emergency extraction dump for machine protection, similar to the one at the exit of the linac (entrance to the BDS).

### 12.4.3 Beam Delivery System (BDS)

The BDS geometry (length and average bending radii) are already compatible with transporting a 500 GeV beam, with acceptable emittance growth generated by synchrotron radiation [230]. Additional dipoles are required (and associated power supplies and cooling) which will be installed in the drift spaces provided in the baseline lattice. The main high-power dumps are already specified for the higher average beam power, to avoid having to replace them for the upgrade (the dumps will be radioactive after several years of operation).

### 12.4.4 AC Power requirements

A comprehensive requirements analysis of the electrical power loads was made for the baseline design. A similar level of engineering analysis was not practicable for the upgrade scenarios presented here. Instead, an extrapolation of the top-level baseline loads (Chapter 11) has been made, based on simplistic scaling laws for main linacs. Table 12.6 gives the approximate loads for the TeV energy upgrade (scenario B). The estimated site power is ~300 MW, compared to ~160 MW for the baseline, and ~210 MW for the luminosity upgrade.

**Table 12.6**
Rough estimate of the power requirements (in MW) for the 1 TeV upgrade (scenario B), based on extrapolation of the baseline design parameters using simple scaling laws [232].

| | RF Power | RF Racks | NC Magnets | Cryo | Conventional load Normal | Conventional load Emergency | Total |
|---|---|---|---|---|---|---|---|
| $e^-$ source | 1.3 | 0.1 | 0.7 | 0.8 | 1.0 | 0.2 | 4.1 |
| $e^+$ source | 1.4 | 0.1 | 4.9 | 0.6 | 2.2 | 0.4 | 9.6 |
| DR | 12.8 | | 4.5 | 1.5 | 2.6 | 0.1 | 21.5 |
| RTML | 7.2 | 0.3 | 2.1 | 2.0 | 0.1 | 0.1 | 11.8 |
| ML (base) | 59.2 | 7.4 | 0.9 | 28.3 | 7.8 | 5.2 | 108.8 |
| ML (upgrade) | 74.2 | 7.4 | 0.7 | 25.1 | 10.2 | 3.9 | 121.3 |
| BDS | | | 16.1 | 0.4 | 0.2 | 0.3 | 17.1 |
| Dumps | | | | | 1.0 | | 1.0 |
| IR | | | 1.2 | 2.7 | 0.1 | 0.2 | 4.2 |
| Total | 156 | 15 | 31 | 61 | 25 | 10 | 300 |

## 12.5 A Light Higgs Factory (250 GeV centre-of-mass) as a possible first-stage.

Following the discovery of a Higgs boson with a mass of ∼125 GeV [7–10], it is useful to discuss an initial-stage project which would start with a centre-of-mass energy of 250 GeV, which could then be later upgraded to 350 GeV for top physics, and then still later to the full 500 GeV, or even directly to a higher energy should the physics case prove compelling.

A first stage 250 GeV centre-of-mass-energy machine would require the installation of approximately half the linacs of the 500 GeV baseline machine. There are two possible scenarios for the civil construction and conventional facilities:

1. Only the tunnel required for the 250 GeV machine is constructed and installed. The next energy stages would then require additional civil construction.
2. The complete tunnel and support shafts (access ways) for the 500 GeV machine is constructed at the beginning, and only half filled with linac, the remaining tunnel housing a beam transfer line to the central region. Staging the energy then simply requires additional linac and associated conventional facilities to be installed.

The first scenario is conceptual the same as that proposed for the 1 TeV upgrade, although half the scale. It is likely to represent the minimum cost for the initial phase machine. The second scenario requires greater investment for the initial phase (for the civil construction), but increasing the centre-of-mass energy then becomes relatively straightforward, and opens up the possibility for a more adiabatic approach to increasing the energy. A very rough scaled estimate suggests approximately 65% and 70% of the 500 GeV baseline cost (Section 15.8) for both scenarios respectively. Since a strong physics case exists for a staged approach up to (or slightly above) 500 GeV centre-of-mass energy, scenario 2 is the preferred option, and will be the focus of the remainder of this discussion. Extension beyond the baseline 500 GeV machine would then require additional civil construction, as already outline in Section 12.4.

The primary machine parameters (including luminosity) are assumed to be the same as those specified for the 500 GeV baseline machine at 250 GeV centre-of-mass energy (see Section 2.2) and are repeated in Table 12.1. This effectively means the injector systems (electron and positron sources, damping rings, bunch compressors) remain unchanged from the baseline. The beam-delivery system could in principle be further optimised for the lower-energy beam, but the overall geometry is still assumed to be consistent with the TeV energy upgrade. For positron production, the 10 Hz mode is currently assumed, again consistent with the approach adopted for the baseline machine (see Section 2.2.2). However, this has additional ramifications for the shorter (electron) linac now running at the full gradient (31.5 MV/m):

- The electron linac must be capable of accelerating the positron production pulse to the nominally required 150 GeV; this now requires an additional 25 GeV of electron linac than would otherwise be required for 250 GeV centre-of-mass-energy running.

- The 10 Hz operation of the electron linac will require a doubling (per unit length) of the average RF power, cryogenic cooling, and associated conventional facilities capacity as compared to the baseline 500 GeV machine.

Thus the electron main linac requires approximately 100% of both the average RF and cryogenic cooling power of the full 500 GeV centre-of-mass-energy baseline linac, while the positron linac would require approximately 50%. The overall scaling is approximately 80% of the AC power load of the baseline machine for the first-stage 250 GeV machine (~129 MW) [232]. The need for the 10-Hz mode at 250 GeV centre-of-mass energy operation could be removed by increasing the length of the superconducting helical undulator from the baseline length of 147 m to approximately 250 m. The electron linac would now only require an additional 3.5 GeV to drive the undulator and only needs to run at 5 Hz. This would reduce the AC power to the ~100 MW level. Another option is the possibility of an independent but unpolarised conventional positron source [233], but this requires further detailed design study, and the loss of polarised positrons should be discussed with the physics community.

The impact on the construction schedule remains to be studied in detail. However, the dominant schedule drivers are likely to be the central region, interaction region hall and the construction of the detectors themselves. While there will be some saving in the overall time required, it is unlikely to be more than 12-18 months based on the current baseline schedule (Chapter 14). The impact on the main-linac component production schedules requires study in order to ascertain the bests cost optimum scenario. If the schedule is indeed constrained by the central region and detectors, a lower production rate could be considered, which may have cost benefits. Furthermore the timescales considered before an upgrade to the second-stage energy would also influence the approach to manufacturing: if a more continuous adiabatic upgrade over several years is considered, this would favour extending the existing linac component manufacturing beyond the initial construction period, possibly at a reduced rate; if the first-stage physics programme requires several years, then it may be necessary or cost beneficial to shutdown and then re-start industrial manufacture. A more detailed study will require a better model for the staging from the physics perspective, as well as an assessment of the most cost-effective approach to dealing with the component manufacture over the longer time scales.

## 12.6 Photon Collider Option

The possibility of developing a gamma-gamma collider option from an $e^+e^-$ or $e^-e^-$ collider, has been extensively discussed over many years. The principle approach has always been to backscatter intense laser beams from the strongly focused charge particle beams close the IR, producing two focused and colliding gamma beams with energies close to that of the particle beams.

To obtain sufficient gamma-gamma luminosity, one requires very high-power lasers, with optical cavities to further enhance the photon intensity, and optical-path designs that can fit around the detectors with photon-bunch timing that matches a possible charged-particle timing pattern. R&D on suitable lasers and optical cavities is ongoing. [235, 236]

With the parameters which give adequate luminosity, the charged-particle beams are severely disrupted and a large crossing angle is required to cleanly extract the beams after collision. Studies suggest that a minimum crossing angle of 25 mrad is required (compared to 14 mrad in the $e^+e^-$ baseline design) and to implement such a layout would require modification of the civil design of the IR hall and the horizontal displacement of the interaction point by a few meters [237].

Given future developments, in lasers and optical cavities and in physics from the LHC and the ILC, a gamma- gamma collider could be considered as a future option for the ILC.

## 12.7 Summary

This chapter has considered staged and upgrade options other than the 500 GeV baseline design described in relative detail in the previous chapters of report, and demonstrates the great flexibility in the design and options of the ILC machine. The baseline design already contains a minimum support for a straightforward luminosity upgrade by doubling the average beam power (increasing the average RF power by $\sim$50 %). Parameters and scope for a future upgrade to 1 TeV centre-of-mass energy have been presented, based on extending the main linacs with a minimum impact on the existing (baseline) machine. Construction of the extended machine could in principle go in parallel with physics running, with only a minimum interruption for connection of the baseline and upgrade linacs and subsequent machine commissioning. The physics parameters (luminosity) for the TeV upgrade represent a trade-off between the physics requirements of the beam-beam (limiting Beamstrahlung and pair-production angle), and a desire to cap the total AC power requirement to approximately 300 MW. Finally, a staged approach to the baseline machine, starting with a 250 GeV centre-of-mass energy first stage has been briefly outlined, where only half the baseline linac would be installed, but the full tunnel and associated civil engineering for the 500 GeV machine would be constructed.

None of these scenarios have been studied in detail, but they represent realistic scaling from the existing baseline design and can be considered as possible example approaches. Other scenarios can certainly be considered as the LHC physics results continue to become available and the physics case for the linear collider becomes further refined.

# Chapter 13
# Project Implementation Planning

In the early 2000s, several study reports [238, 239] were issued by American, Asian and European regional bodies representing the relevant high-energy physics communities on possible organisational structures for the project management of a linear collider (LC). The Consultative Group on High-Energy Physics of the Organisation for Economic Cooperation and Development (OECD) also issued a report [240] on their consensus, concurrently with these regional reports.

All these reports agreed that a high-energy electron-positron LC should be the next major facility on the roadmap of international high-energy physics, and that this project would require a hitherto unknown scale of global collaboration, calling for special attention by the world's research, administrative and political sectors. Together, these reports laid the foundations for an international organisation for the design and development stage of an LC, leading to establishment of the Global Design Effort (GDE) for the International Linear Collider (ILC). Following the International Committee for Future Accelerators (ICFA) decision to base the design of a global linear collider on superconducting radiofrequency (SCRF) technology, the GDE mandate of coordinating the worldwide R&D programme and developing a technical design for a 0.5 TeV linear collider was established. This mandate is completed with the publication of this TDR. Based on various physics studies, ICFA gave the GDE guidance on the accelerator performance to be achieved. In creating the baseline design presented in this report, close attention was also paid as to how best to implement such a global project in order to make it as realistic as possible. The results of these deliberations have been collected as a stand-alone document on Project Implementation Planning (PIP) [11]. In this chapter the guiding principles underlying the PIP are outlined and a brief synopsis of the contents of each section are provided.

It is clear that the international HEP community cannot usurp the role of government or officials in tackling the concomitant intergovernmental issues. Therefore, the PIP focusses on making statements from the standpoint of the primary executor of the research and on presenting the GDE's preferences from the scientific and technical viewpoints in order to inform the debate as much as possible.

Large-scale research undertakings cannot be realised without firm commitments by the nations/regions that undertake them. Moreover, when the scale of a research project goes beyond what can be readily sustained by a single nation/region, its guiding principles have also to expand. One such principle that needs to be underlined is "openness to the world". High-energy physics has been characteristically international in nature since its inception.

The opportunity for research in particle physics has been, and must be, equally open to all scientists in the field, as formulated in the ICFA guidelines. The ILC project is a novel and unique opportunity to realise internationalisation and cooperation in particle physics on a global scale with numerous positive implications for science, technology and education. This is perhaps one of the most important ways in which the ILC can be popularly perceived as making a valuable global contribution to society.

Several different organisational models are conceivable for managing the construction, commissioning and operation of the ILC. Irrespective of the specific details of such models, a clear legal status needs to be defined for an organisation to manage execution of the ILC project. The adequacy of that organisation and its management needs to be assessed from the standpoint of how its legal structure can address the following points effectively: as a scientific project, it is open to participation by any nation/region that is prepared to make a significant contribution; it is driven by in-kind contributions from multiple participants; solid accountability is ensured in both the scientific/technical and budget/financial aspects.

The organisation needs to be able to implement a mechanism that provides long-term stability in terms of maintaining the productivity and continuity of the project, together with the agility to address short-term problems in project execution, in both technical and financial contexts.

The ILC project will go through a number of evolutionary steps towards construction and operation. The early stage of the ILC organisation cannot be completely static because the participating countries/regions may or may not be able to negotiate the necessary approval processes simultaneously. Successful project execution requires a predictable budget with good stability. The ILC project, including construction, will have a life span of 20 years or longer.

The **governance** of a large international science project is a very complex endeavour. There are no precedents for a truly global project without a strong host laboratory. It is crucially important to determine how decisions are made on design and technical issues, who appoints key staff, and the responsibilities of the host when implementing such a project.

A study of other recent major science projects, including ALMA, XFEL, ITER and the LHC was carried out to inform the PIP. Lessons learned from these projects have helped to identify the key considerations for effective governance of the ILC. In developing the ILC *Technical Design Report* (TDR), the importance of defining the responsibilities of the host, having a well established and agreed-to scheme for in-kind contributions, an adequate common fund, etc. were all recognized as important issues. The conclusions and key recommendations of this study with respect to governance have been reported to FALC, ILCSC and publicly at ICHEP 2010. The key points are discussed in the section on governance.

Various **funding models** for a globally supported ILC have been considered in order to understand how it could be built, the responsibilities of the host, etc. Earlier models for the ILC were based on equal sharing among three regions of the world, the Americas, Asia and Europe. Such a model no longer seems viable; instead, a funding model similar to that used in both XFEL and ITER is recommended, namely a “share” system where the “major” partners should contribute a minimum, perhaps 10%, and others would join as members of regional consortia or by making particular contributions. The host nation would contribute a significantly larger share of the construction costs. Running and decommissioning costs need to also be considered and agreed at the time the project is funded.

The responsibilities and the authority of the **project management** and project team need to be determined in advance and must be sufficient to make the team effective. This central management team will be responsible for finalising the design, carrying configuration management, a formal change control process, making technical decisions, maintaining schedules and other responsibilities of project management.

Certain **host responsibilities** are crucial to the success of a global project. The host will need to provide a variety of services similar to those provided by CERN, a successful example of a multi-country large collaborative laboratory. In addition to the necessary contributions to the infrastructure, construction and operations, the host will be expected to achieve status for the ILC laboratory as an international organisation within its local legal system.

**Siting** is a major issue, from selecting the site to dealing with the configuration and site-dependent aspects of the design and implementation. Technical issues, such as seismic conditions, will need to be considered and a site-dependent design, taking the conditions of a particular site into consideration, will need to be developed by modifying the original generic design. Matters such as access, providing infrastructure, safety, etc. will need to be considered issue by issue in developing the site-dependent design to be implemented. The design will evolve from the configuration-controlled ILC design produced by the global design team; the site-dependent changes will be done through a formal change control process.

It is assumed that the majority of contributions from countries to the ILC will be in the form of **in-kind contributions**. This has the substantial advantage that most resources for the construction, other than civil construction, can be made within the collaborating countries. This is important for political reasons, as well as to build technical capacity within the collaborating countries. However, this scheme comes with major challenges in terms of managing the different deliverables, integrating them, maintaining schedules, dealing with unforeseen cost increases for specific items, etc.

The issues discussed above and options to solve them have been carefully studied. A **flexible** form of in-kind contribution, for example one employing a form of '*juste retour*', is preferable (i.e. each member state receives a guaranteed fraction of the industrial contracts in proportion to the value of their contribution). This gives the central management some flexibility to place the work where it will be the most effective while spreading the work and resources equitably. A very important additional lesson from projects which have in-kind contributions is that sufficient central resources must be made available to effectively coordinate and integrate the project through the central management.

An implementation topic unique for the ILC is **the industrialisation and mass production of the SCRF linear accelerator components**. A model for this production that involves multiple vendors worldwide and a globally distributed model based on the "hub laboratory" concept has been developed. Basically, the proposed cost-effective scheme will use industry for what they do best, large-scale manufacturing, and the participating high-energy laboratories for what they do best, integration and carrying the technical risk for performance.

The overall **project schedule** for ILC construction and commissioning has been analysed; it is dominated by the time to construct the conventional facilities as well as by the time required to construct, install and commission long-lead-time technical components such as the SCRF system. An 8 to 10 year construction, installation and commissioning schedule appears necessary.

Finally, the **future technical activities** that will help continue to advance the ILC towards construction have been analysed. Overall, the project implementation planning has served as an integrated element in developing a technical design for the ILC that can smoothly evolve into a final design and implementation plan for the ILC project once it is approved and funded.

One of the most important, problematic and difficult areas is the transition between the current LC organisation and a fully fledged ILC laboratory with an agreed site, specification and budget. In order to separate these considerations, which necessarily change with time, from the more general principles that pertain to a final organisation, only the structure of the final ILC laboratory is discussed in the PIP.

The subjects outlined above are best analysed assuming a specific roadmap. This is particularly important given the evolutionary nature of the ongoing R&D and the steps to follow when a laboratory organisation for the ILC is formed; some must be done in parallel, some in series, some in national and others in international contexts. Figure 13.1 shows a high-level overview of a possible roadmap towards realisation of the ILC.

One important consideration that should be noted is the separation of technical/scientific and political aspects. Without doubt, the final negotiations and decisions concerning the legal agreements, budget sharing and site selection for the ILC will have to be made by the appropriate government agencies of the interested nations/regions. On the other hand, the technical contexts and resultant boundary conditions or specifications for the project (such as the base performance parameters and/or the technical specifications for possible sites) should be dictated by the scientific requirements. It is thus important to identify where the responsibilities of scientists end, and where those of the government officers and statesmen begin.

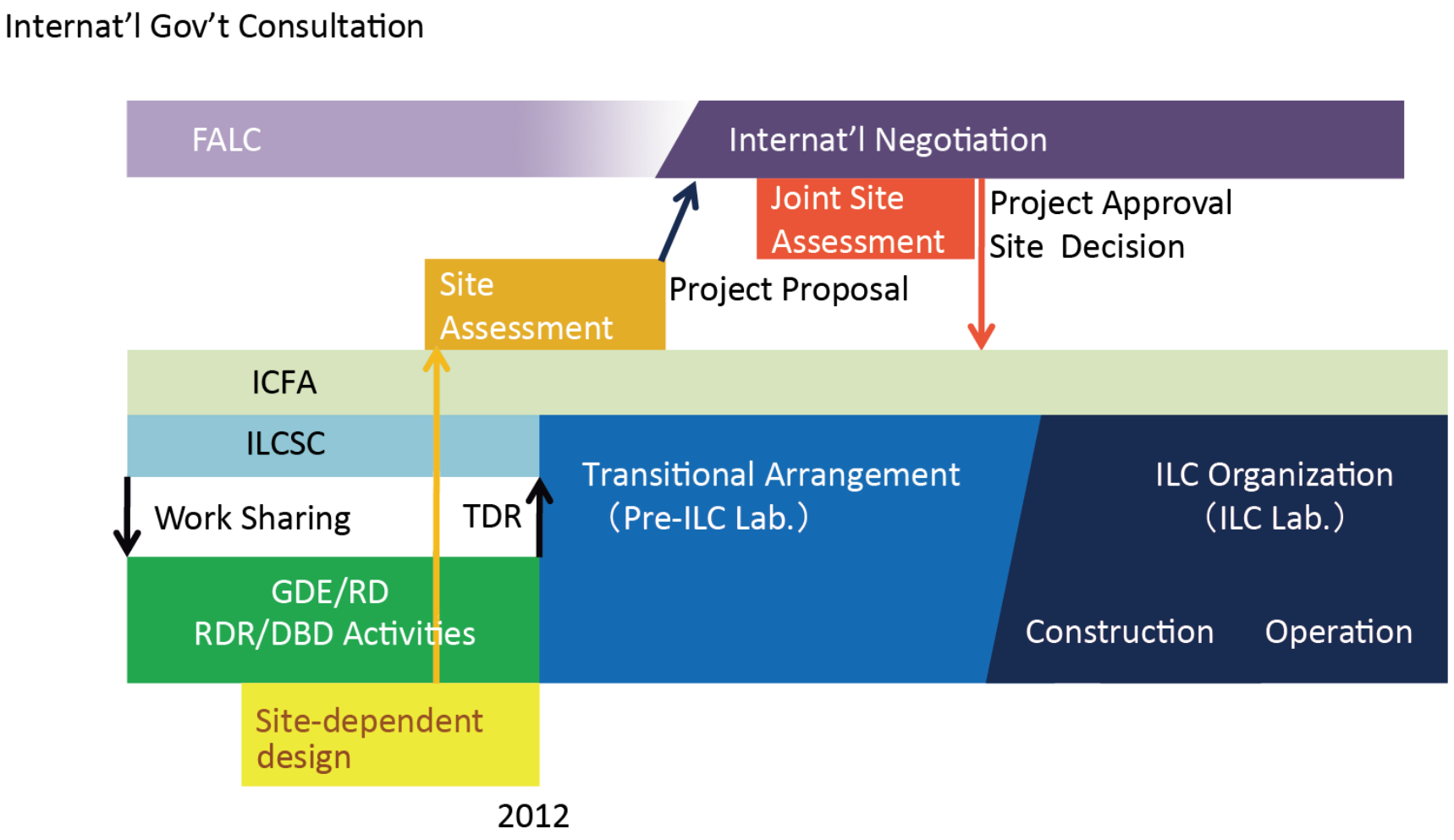

**Figure 13.1** Possible roadmap towards realisation of the ILC

In the area of government-level discussion regarding the ILC's future development, the Funding Agencies for Large Colliders (FALC) currently holds regular meetings. Once the project is ready to be formally proposed, a suitable forum to discuss the necessary arrangements should be formed.

# Chapter 14
# Construction schedule

## 14.1 Introduction

This chapter provides an overview of possible ILC construction schedules for the flat and mountain topography design variants. It updates the study published in the RDR and goes further to cover all major steps from construction to final commissioning. It provides a list of level-1 milestones (top level, a few for the whole project) that can be used to compare various scenarios and later to track project progress. All three sample regions (Europe, Americas and Asia) are considered.

Because of the assumed in-kind nature of the ILC project, this construction schedule should be regarded a target baseline schedule that can be used to plan the pace of various activities, the need for multiple shifts for particular activities, etc., rather than a solid schedule estimate. A more accurate schedule can only be developed after a specific site has been selected and the in-kind contributions to the project have been explicitly defined. *Component or system delivery schedules, with appropriate contingencies, will be an important part of negotiations and agreements between collaborators and the project management.*

## 14.2 Scope and assumptions

The scheduling exercise that is presented in this section focuses on three major steps: construction; installation; and commissioning. A first attempt is made at considering the constraints on the high-tech component mass-production schedule. Other important activities like the R&D programs or procurement processes are not included. The scope of this exercise is to show how the ILC, including the detectors, could be built and delivered for operation at the selected sample sites.

The origin of time considered in what follows is the start of construction work. The mobilisation of equipment and manpower referred to as "site set-up" is omitted. Taking the CERN LEP project as benchmark, this is an activity that can take up to 6 months. It includes building the personnel-support facilities (changing rooms, rest rooms, catering areas) throughout the building sites and the completion of the construction of access roads. Storage facilities required to launch the first steps of the construction, such as parking lots, warehouses, and tip yards, are also assumed to be available.

The acquisition of the land on which the ILC is to be built is also assumed to be complete. This is a step that can take a significant amount of time depending on the location chosen. It is affected by environmental studies, the final layout of the facility and the local context (density of population, accessibility, etc.). For the European sample site, this process will take at least a year.

In what follows, the focus is the critical path of the work sequence. At a later stage the remaining activities that can be carried out in parallel will need to be added. For monitoring purposes, it might be considered to develop an Earned Value Management tool to track the progress of all activities and not only the ones on the critical path. This was successfully used during the construction of the LHC at CERN.

The time estimates that are used in this scheduling exercise are the result of an assessment based on past and on-going projects. The LHC, XFEL and CMS projects have been used most extensively as references. They are all recent and relevant scientific projects. A motorway built in Japan has also been used as it is located in a mountainous region and requires excavation techniques relevant to the ILC project.

Table 14.1 shows reference projects for those relevant areas where ILC time estimates are needed.

**Table 14.1**
ILC time estimates and reference projects

| Activities | Shaft excavation | Tunneling | Cavern construction | Accelerator components installation | Detector installation | Accelerator commissioning | Cooling and ventilation installation | Electrics installation |
|---|---|---|---|---|---|---|---|---|
| LHC | | | × | × | | × | × | × |
| CMS | | | | | × | | × | |
| XFEL | | | | × | | × | | |
| LEP | × | × | | | | | | |
| JP motorway | | × | | | | | | |

Progress rates depend heavily on the organisation of working time. In some cases, workers should be asked to work in shifts. In other cases, having too many workers in the same area is detrimental to efficient working conditions. When using valuable pieces of equipment that are energy and manpower intensive, a three-shift system is recommended. This is most particularly the case with Tunnel Boring Machines or TBMs. In this study, it is assumed that the progress rates stated for TBMs are for three shifts and can therefore not be speeded up. Table 14.2 sums up the progress rates that have been used for the main linac to reach the corresponding time estimates.

This study considers two types of topography. The flat topography (FT) applies to the sample sites located in European and Americas. The mountainous region (MR) applies to the sample site located in Asia.

**Table 14.2**
ILC time estimates and progress rates

| Activity in Main Linac | Region | Progress rates (m/week) | For x shifts |
|---|---|---|---|
| Tunnelling using 8m ⌀TBM | FT<br>MR | 100<br>n/a | 3 |
| Tunnelling using 5.2 m or 6 m ⌀TBM | FT<br>MR | 150<br>n/a | 3 |
| Tunnelling using 6 m–8 m ⌀road header or 'Drill and Blast' (NATM) | FT<br>MR (NATM) | 30<br>20 | 3<br>3 |
| Concreting, invert and tunnel finishing | FT<br>MR | 50<br>Concrete lining 25<br>Invert, drainage 45 | 3<br>3 |
| Ventilation ceiling-ducts installation | FT (Europe only)<br>MR | 50<br>n/a | 3 |
| Survey and set out of components supports | All | 120 | 1 |
| Electrics General Services | All | 120 | 1 |
| Piping and ventilation | All | 120 | 1 |
| Cabling | All | 120 | 1 |
| Installation of supports for machine components | All | 250 | 1 |
| Installation of machine components | All | 100 | 1 |

The number of teams working in a given location is a parameter that can have great impact on the overall progress rate. It might appear tempting to inject more manpower in a particular activity. However, experience with the LHC has proven that this temptation should be resisted. Particularly for space-consuming activities, it is preferable to leave plenty of space for a team to work. When appropriate, this study looks at the impact of doubling the number of teams in action in a sector (section of tunnel between two shafts). In addition to space management, it is important to also balance the benefits of injecting more resources against the required delivery date of a sector. For example, if the commissioning of a sector requires the availability of several facilities, the number of teams should be chosen in order to minimise the time spent waiting for the last facility to become available. This consideration was used when optimising the workflow.

Working time is another parameter that can have a significant impact on progress rates. In this study we assume that work will be carried out five days a week. No public holidays have been taken into account.

Fulfilling the commissioning objectives is a driving force behind this scheduling study. An attempt is made at describing how to go from an installed facility to an operational one, which requires consideration of not only accelerator facilities but also the detectors.

This study is based primarily on the latest ILC European and Asian layouts. At this stage, it is reasonable to assume that the Americas' layout will not introduce major changes in the schedule compared to the European one. The Asia ILC layout is significantly different from the European and American ones and required a specific study. The naming convention used to refer to the various parts of the accelerator is the one designed for the European region (see Fig. 14.1).

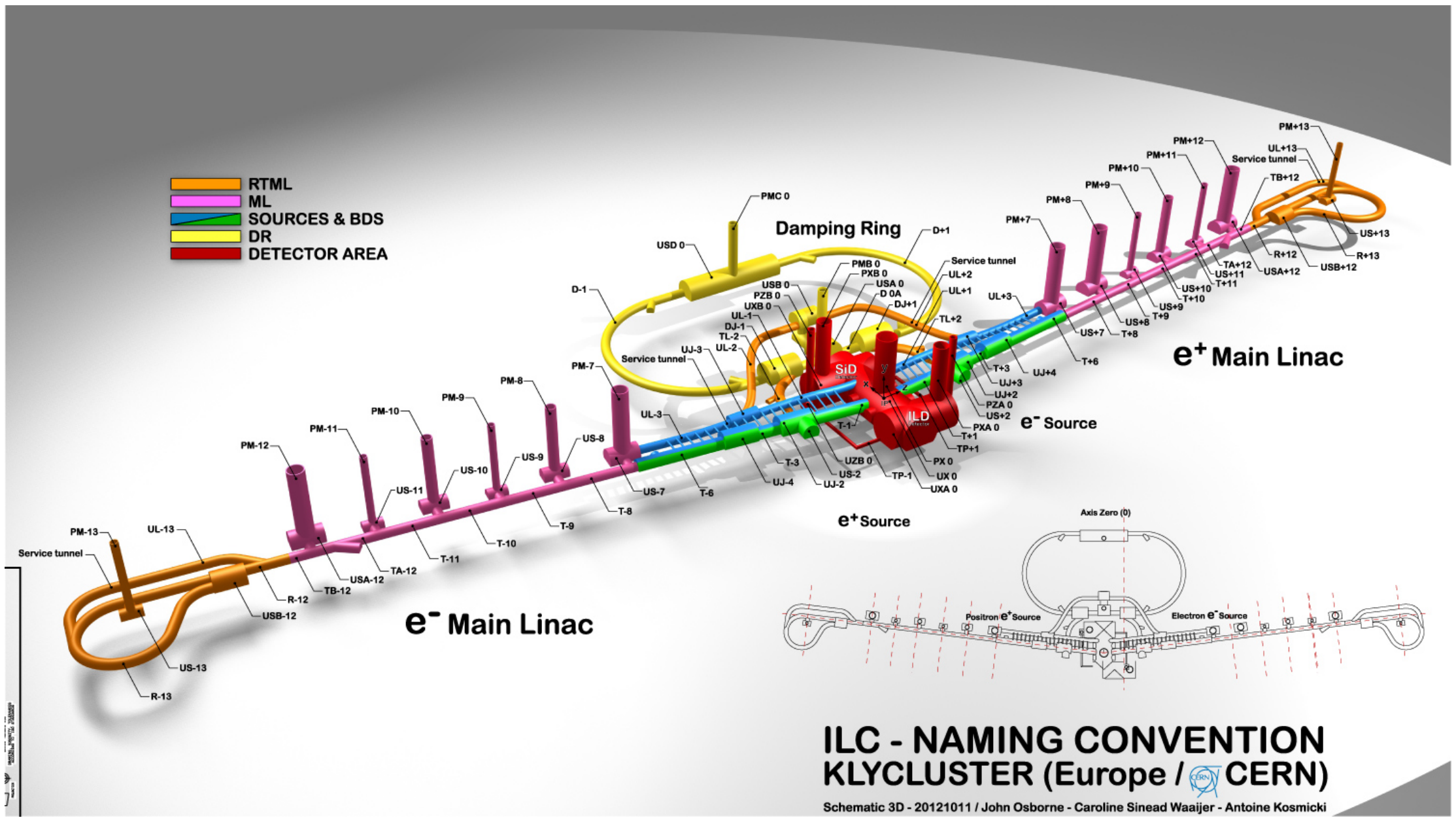

**Figure 14.1.** European ILC layout

The three following sections present the construction of the accelerator complex and high-tech mass-production plans, the commissioning plans and the detector installation and will refer to these figures.

## 14.3 Constructing the accelerator complex

The schedule of the three main phases involved in the delivery of the accelerator complex are discussed below: civil engineering; installation of common facilities; and accelerator component installation. A graphical representation for the complete construction and commissioning schedules for the flat and mountain topography design variants is given in Fig. 14.2 and Fig. 14.3, respectively. These figures will be referred to in the following sections.

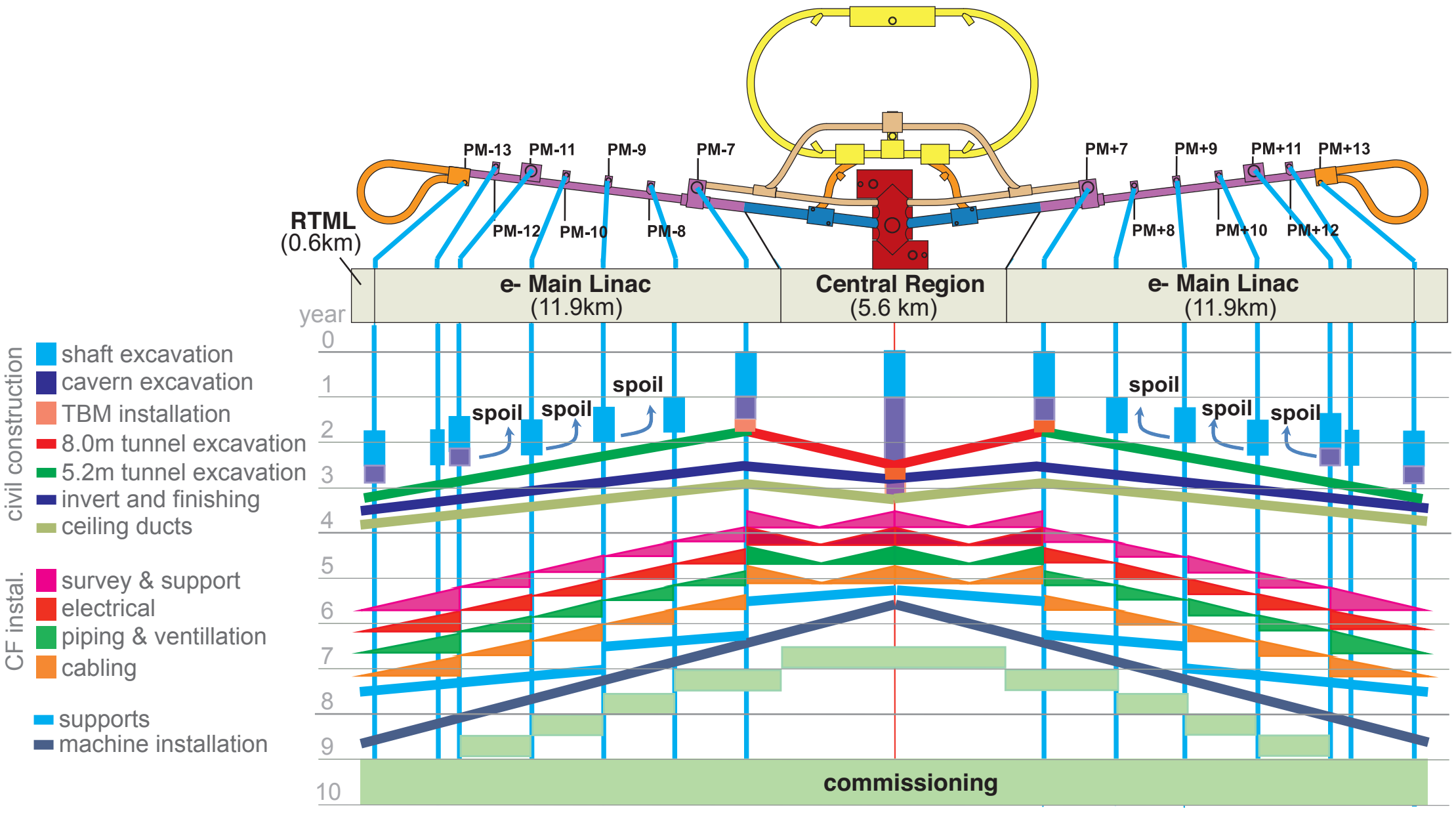

**Figure 14.2.** The construction and commissioning schedule for the flat topography design variant. Years after construction start are represented vertically, while construction progress along the machine footprint is indicated horizontally (not to scale). The vertical lines represent the locations of shafts.

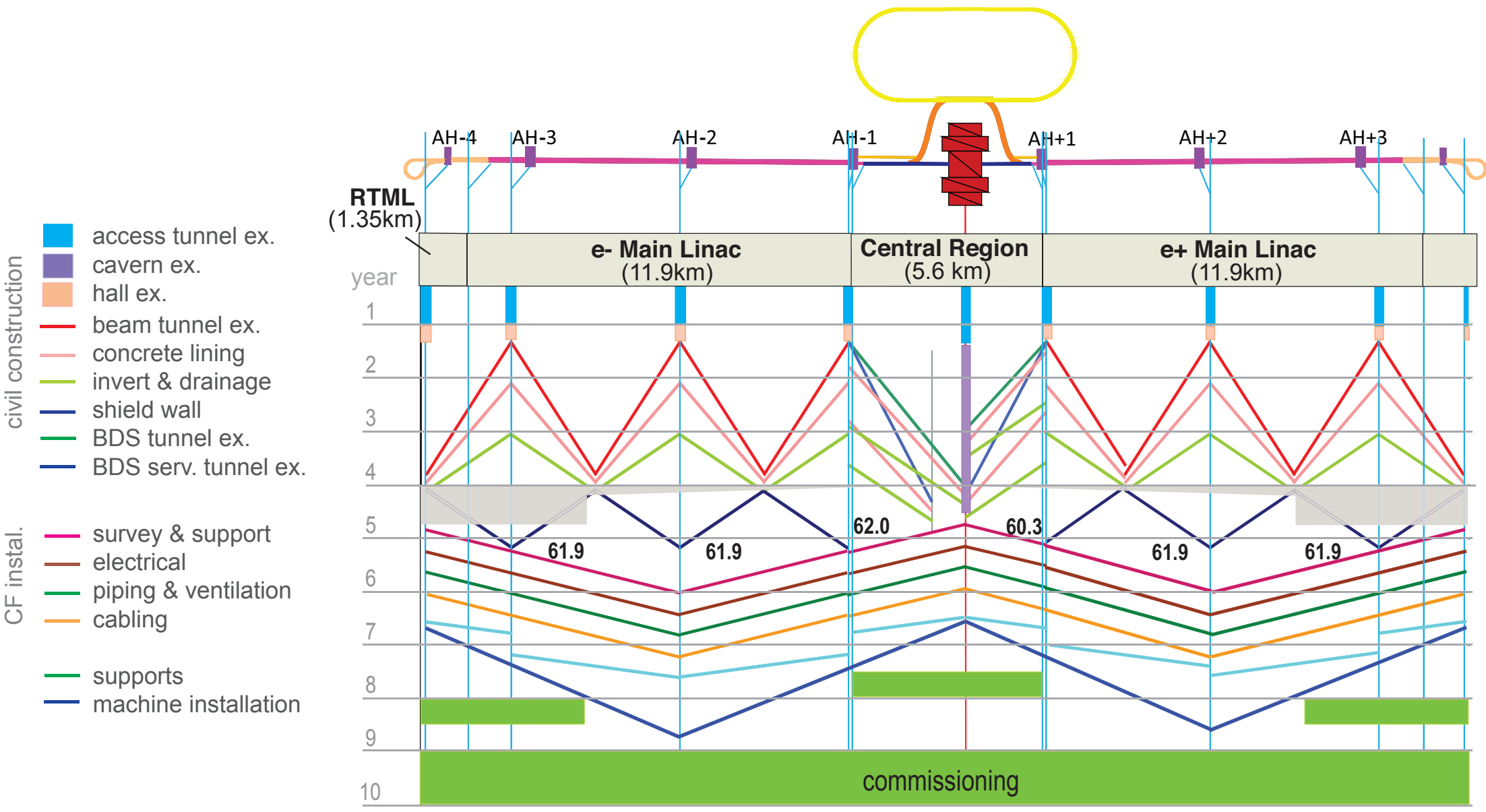

**Figure 14.3.** The construction and commissioning schedule for the mountain topography design variant. See Fig. 14.2 caption for details.

## 14.3.1 Civil engineering

### 14.3.1.1 Flat-Topography Sites

The ILC layouts that are being considered in this study are significantly different from the one presented in the RDR. The Main Linac and BDS consist of a single tunnel of varying diameter. For the FT sites, it was decided that using two types of TBMs with respective diameter of 8 m and 5.2 m would facilitate the construction. Figure 14.4 shows where each type of TBM is to be used.

The civil construction phase is expected to be complete in the first four years of of the construction schedule (Fig. 14.2 years 1–4). The first step in the civil engineering (CE) phase is to excavate the access shafts that will be used to launch the TBMs and start excavating the caverns in the interaction region. Experience from LHC implies one year is necessary to deliver a fully equipped shaft.

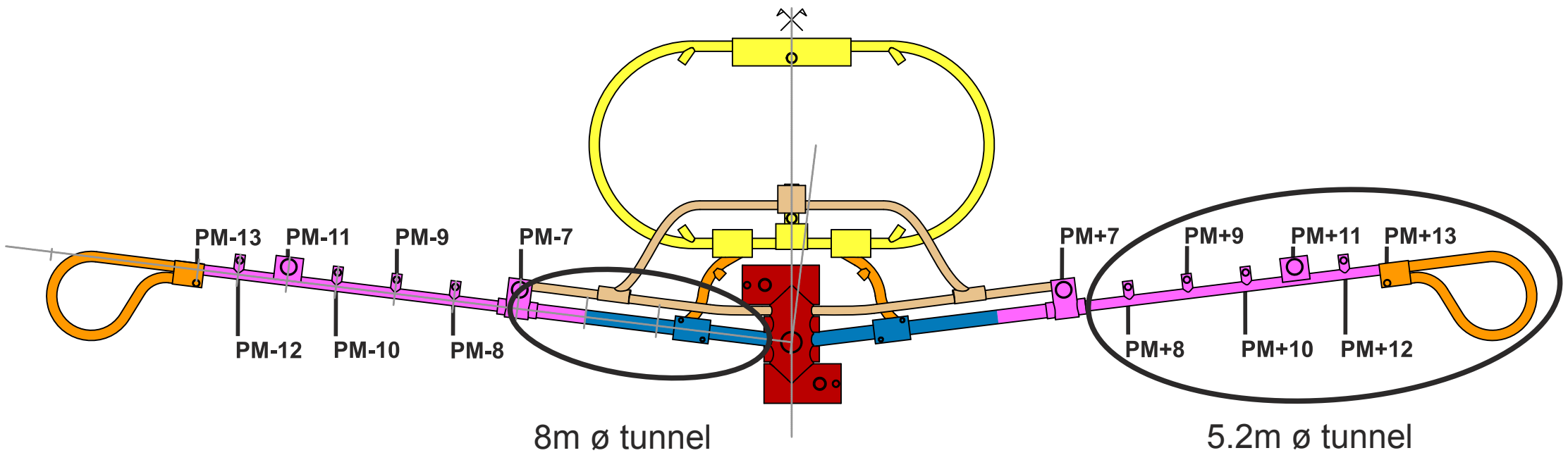

**Figure 14.4.** Choice of TBMs

Results of the ARUP/J Osborne studies [172] recommend minimising stress concentration on the interaction region by excavating and finishing the interaction cavern before tackling the tunnels and service caverns. This leads to the choice of PM7 as the location to launch the TBMs. By the time they reach the IR, the caverns will be excavated and finished.

Once a tunnel section is excavated, the next step is to build the invert and complete the finishing. The progress rate for these steps has been stated in Section 14.2. In the case of the European design, concrete ventilation ducts will be formed on the ceiling. A light-green line in Fig. 14.2 year 4 represents this activity.

#### 14.3.1.2 Mountainous Region site

Years 1–5 of Fig. 14.3 shows the schedule for the civil-engineering phase of the construction schedule for the MR site. The MT site requires excavating horizontal access tunnels as opposed to shafts. Fourteen of these tunnels and the interaction cavern have to be excavated concurrently. Once a long enough section of an access tunnel is made available, the concrete lining activity should start. The next step consists in constructing the invert and drainage system. Finally, the shielding wall has to be erected inside the Main Linac tunnel. This activity will take a full year as the progress rate is 45 m/week.

At this stage. a first set of level-1 milestone can be extracted and are shown in (Table 14.3).

**Table 14.3**
The first set of level-1 milestones.

| Milestone | Flat topography | Mountainous region |
|---|---|---|
| Civil Engineering work complete | Y4, Q4 | Y5, Q1 |

### 14.3.2 Conventional facilities installation

This phase of the construction schedule studies the installation of the supporting infrastructure such as survey and setout supports for accelerator components, electrical general services (cable trays, cables, sockets), infrastructure for cooling and ventilation (pipes, ducts), and accelerator cables.

#### 14.3.2.1 FT sites

Figure 14.2 years 4–7 shows the schedule for the conventional facilities installation. It has been chosen to exclude 2 activities of a different nature sharing the same tunnel section. The number of teams deployed has a significant impact on the completion date of this phase. In what follows it has been chosen to inject 4 teams in the BDS regions and only 2 in the main linac. This choice is justified by the progress rate of the subsequent activities.

#### 14.3.2.2 MR site

The progress rates and the installation sequence for all the activities considered in this phase are the same as for the FT sites. However, due to the shielding-wall partition in the tunnel, it has been chosen to allow activities of different natures to unfold in the same location. (see Fig. 14.3 years 5–7.) A set of level-1 milestones for this phase is shown in Table 14.4.

**Table 14.4**
The second set of level-1 milestones.

| Milestone | Flat topography | Mountainous region |
|---|---|---|
| Civil Engineering work complete | Y4, Q4 | Y5, Q1 |
| Common Facilities installed | Y7, Q3 | Y8, Q2 |

### 14.3.3 Accelerator-Components Installation

This phase consists in first installing accelerator components and their supports. At that stage, the high-tech mass-manufacturing process has to provide the required components. In order to allow for an early commissioning exercise to take place, the schedule has been designed to install accelerator components in the Central Region first. The estimated progress rates are the same for the MR and FT sites.

#### 14.3.3.1 FT sites

Based on experience at the LHC, only two teams are deployed for the installation of the accelerator components and their supports. The installation phase is shown in years 6–9 of Fig. 14.2.

#### 14.3.3.2 MR site

For reasons already given in the previous section, four teams are deployed to install the accelerator components. The installation rates are the same as for the FT. However to reduce potential overcrowding and increase the productivity and efficiency, installation activities for the MT site will be spread over 3 shifts per day. The machine installation can be seen in years 7–9 in Fig. 14.3. The level-1 set of milestones for this phase are given in Table 14.5.

**Table 14.5**
The third set of level-1 milestones.

| Milestone | Flat topography | Mountainous region |
|---|---|---|
| Civil Engineering work complete | Y4, Q4 | Y5, Q1 |
| Common Facilities installed | Y7, Q3 | Y8, Q2 |
| Accelerator ready for early commissioning (BDS and ML up to PM7/AH1) | Y7, Q2 | Y8, Q2 |
| ILC ready for full commissioning (whole accelerator available) | Y9, Q4 | Y9, Q4 |

### 14.3.4 High-tech Mass-Production Plans

Figures 14.5 and 14.6 are a first attempt to match the manufacturing plans of the accelerator parts with the construction schedule. Each figure shows the mass-production plans in the background. The time window dedicated to installation of the accelerator components for the FT and MR is depicted by a coloured rectangle and has been superimposed. The resulting figures show the time constraints of both activities.

For both types of sample site it appears that the components will be ready on time to start the installation of the accelerator. The Asian-region schedule allows for a longer production time of the accelerator parts.

A more detailed study is needed to show how the production, storage and installation rates can be optimised.

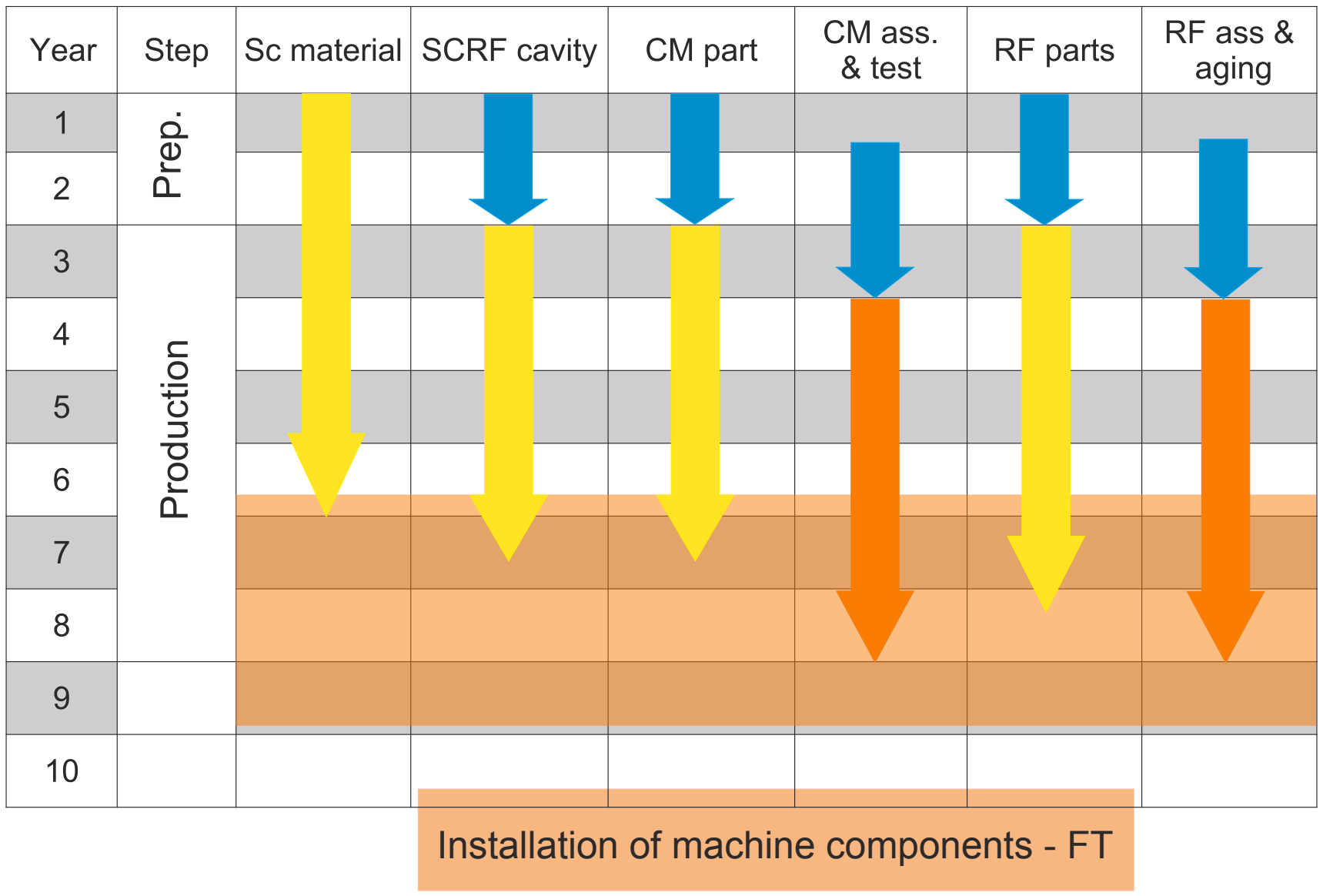

**Figure 14.5** Production plans for FT.

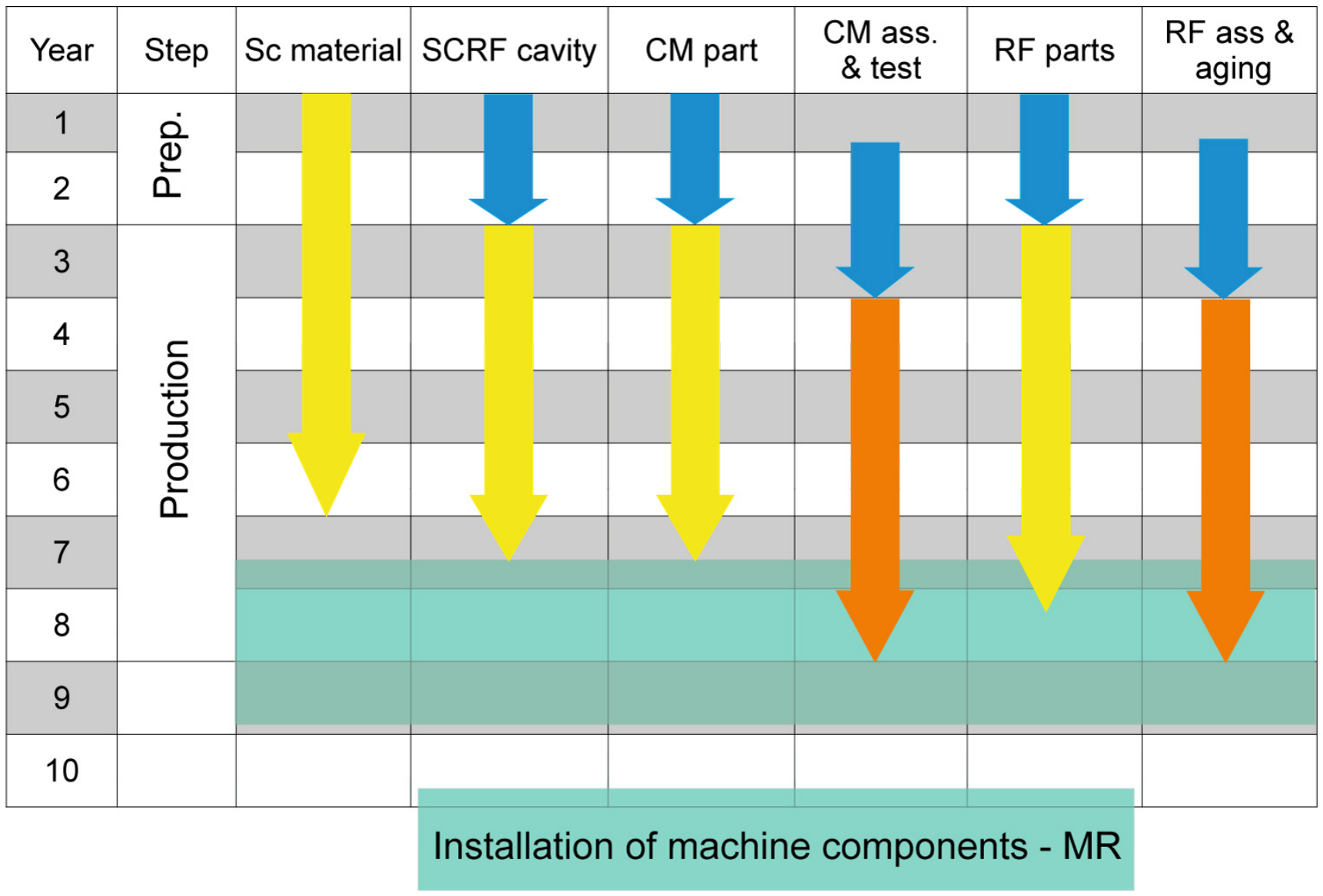

**Figure 14.6** Production plans for MR.

## 14.4 Commissioning the ILC

### 14.4.1 Early Commissioning (BDS, part of ML, DR, and PLTR)

An early commissioning draft plan has been prepared for this study. The key objectives, together with their foreseen durations, are:

- test of the $e^-$ injector system to 5 GeV and dump: 3 months;
- test of the $e^+$ source and systems to 5 GeV and dump utilising the auxiliary low current $e^-$ source to produce $e^+$: 3 months;
- hardware commissioning of injection lines and both damping rings: 3 months;
- commission both rings with beams from injectors with extraction only into first dump in the PLTR (beam still in injection/extraction tunnels): 9 months.

This commissioning exercise requires the availability of the BDS and main linac up to PM7/AH1, the PLTR, and both Damping Rings. In what follows, only the FT sites are considered; a similar study can be carried out for the MR site.

It has been established that the BDS and necessary sections of the Main Linac will become available in Y7 Q2 for the FT sites. In order to assess the feasibility of the early commissioning plan, a construction schedule of the DR and PLTR has been built.

The Damping Rings are houses in a 6 m diameter, 3259 m-long tunnel. It will be excavated using a road header at a rate of 150 m/week using 3 shifts a day. The PLTR consists in two 6-8 m diameter, 250 m long tunnels. These tunnels are to be excavated using road headers at a rate of 30 m/week using 3 shifts a day.

Table 14.6 shows the assumptions made for developing the DR and PLTR installation plan.

**Table 14.6**
Assumed progress rate for installation in the DR and PLTR tunnels.

| Progress rate | DR (m/w) | PLTR (m/w) |
|---|---|---|
| Invert and finishing | 250 | 250 |
| Survey | 120 | 120 |
| Electrics | 80 | 120 |
| Piping and ventilation | 80 | 120 |
| Cabling | 80 | 120 |
| Supports | 250 | 250 |
| Accelerator components | 50 | 100 |

The Gantt chart in Fig. 14.7 shows that the DR and PLTR should be ready for early commissioning by Y7 Q1. This means that by the time the BDS and ML become available for early commissioning, the DR and PLTR should also be ready. The commissioning exercise would then start by Y7 Q2.

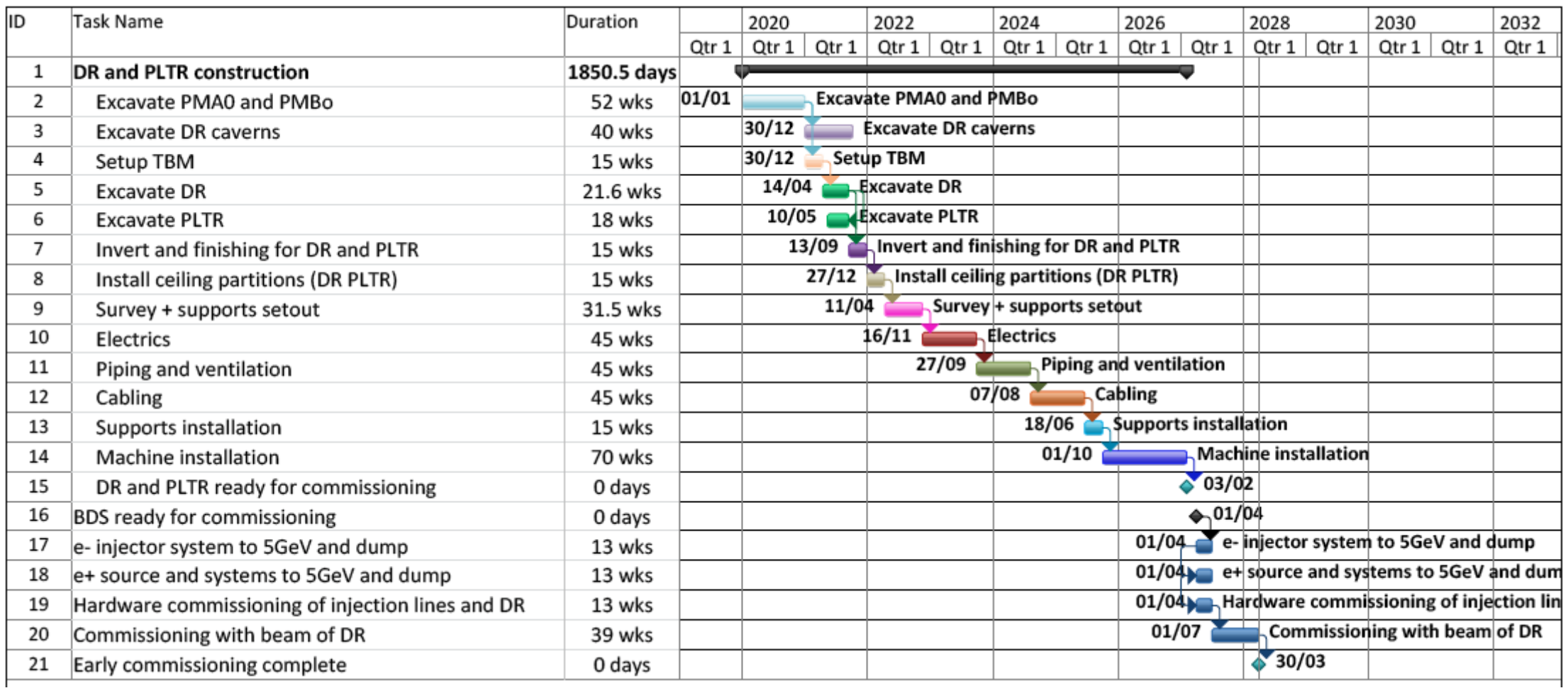

**Figure 14.7.** Draft schedule for delivery of DR and PLTR for commissioning

## 14.4.2 Global commissioning

It is challenging to assess how much time would be needed for the commissioning of the ILC. Based on experience at LHC, 6 months of pre-commissioning per sector would be necessary. In addition, 12 months would be needed to complete the final global commissioning.

Global commissioning is indicated in year 10 in Fig. 14.2 and Fig. 14.3 for flat and mountain topography sites respectively. Level-1 milestones for the commissioning phase are listed in Table 14.7.

**Table 14.7**
The fourth set of level-1 milestones.

| Milestone | Flat topography | Mountainous region |
|---|---|---|
| Civil Engineering work complete | Y4, Q4 | Y5, Q1 |
| Common Facilities installed | Y7, Q3 | Y8, Q2 |
| Accelerator ready for early commissioning (BDS and ML up to PM7/AH1) | Y7, Q2 | Y8, Q2 |
| ILC ready for full commissioning (whole accelerator available) | Y9, Q4 | Y9, Q4 |
| ILC ready for physics programme | Y10, Q4 | Y10, Q4 |

This study shows that it would be possible to build and commission the ILC in less than 10 years. This statement holds for both the FT and MR sites. The scheduling studies will continue so as to incorporate any new modifications to the designs and implications on the availability of resources.

## 14.5 Detectors

This study presents the basic structure of an ILC detector construction, underground installation and commissioning schedule. The scenario considered focuses on the ILD detector for a Flat Topography site. The applicable layout of the interaction region is shown in Fig. 14.8.

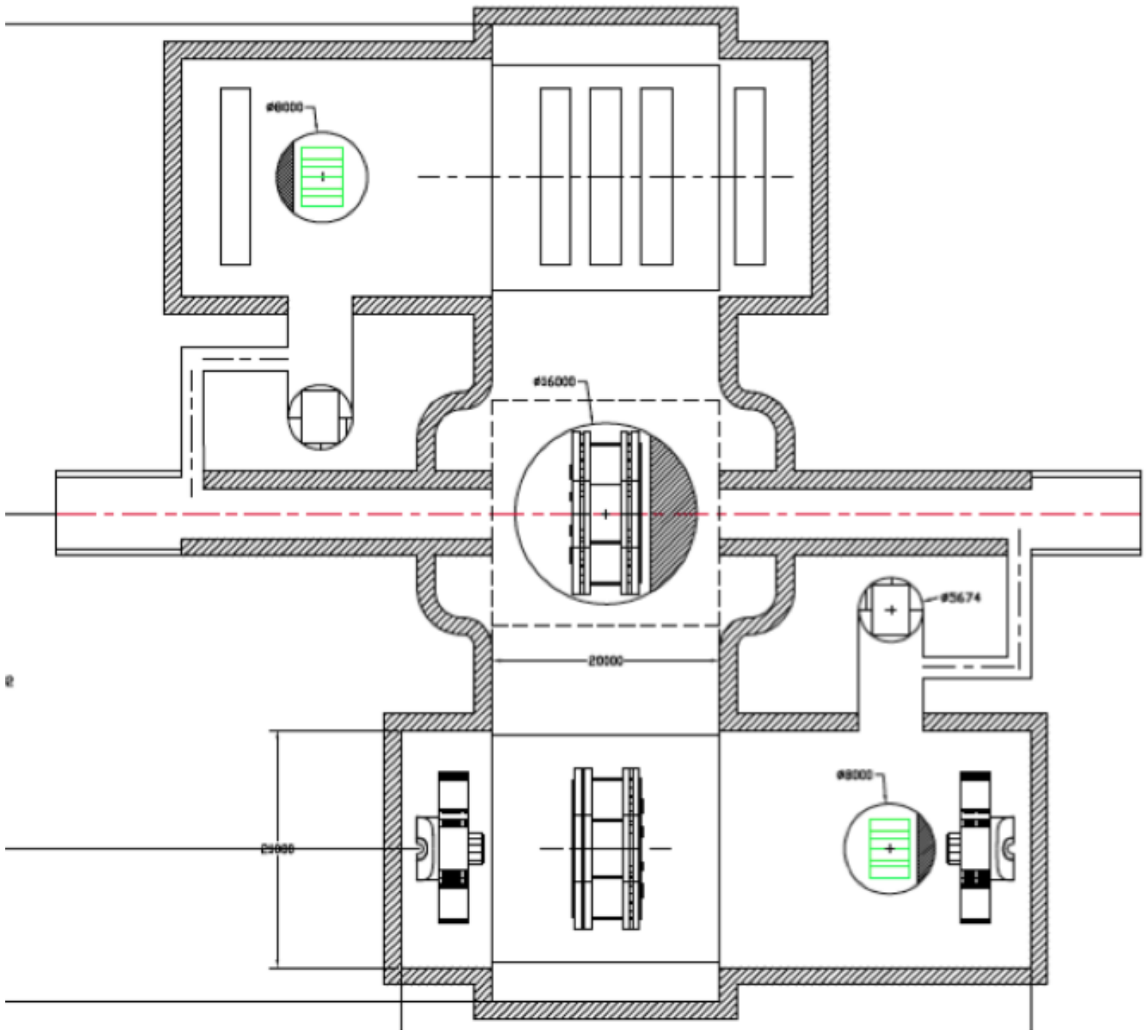

**Figure 14.8** Layout of the interaction region in a Flat Topography site

Using the CMS concept, the ILD detector is to be assembled in a surface hall before being lowered to the underground facilities. This allows work underground to proceed unaffected by the construction of the detector. In a first stage, two-thirds of the surface hall will be assembled and handed over to the detector-construction crew. At a later stage, the building will be completed to include the shaft linking the building to the underground facilities. The important milestones to extract from this scheduling study are the "Caverns ready for beneficial occupancy", "Detector ready to be lowered" and "Detector ready for commissioning with beam". The Gantt chart in Fig. 14.9 presents a preliminary schedule for the construction of the ILC interaction region.

| | Task Name | Duration |
|---|---|---|
| 1 | ILC construction started | 0 days |
| 2 | Excavate 2 shafts | 12 mons |
| 3 | Construct of 2/3 of Surface Assembly Halls | 12 mons |
| 4 | Excavate and finish Detector Caverns | 24 mons |
| 5 | Assemble and commission Detectors | 1280 days |
| 6 | Construct Surface Service Buildings | 24 mons |
| 7 | Install Caverns Infrastructure | 44 mons |
| 8 | Complete 1st Surface Assembly Hall | 12 mons |
| 9 | Complete 2nd Surface Assembly Hall | 12 mons |
| 10 | Install Crane for Heavy Lowering | 4 mons |
| 11 | Detector Caverns ready for beneficial occupancy | 0 days |
| 12 | Lower Detectors | 150 days |
| 13 | Connect detectors to their services | 11 mons |
| 14 | Commission Detectors & Push-Pull | 12 mons |
| 15 | Ready for commissioning with beam | 0 days |

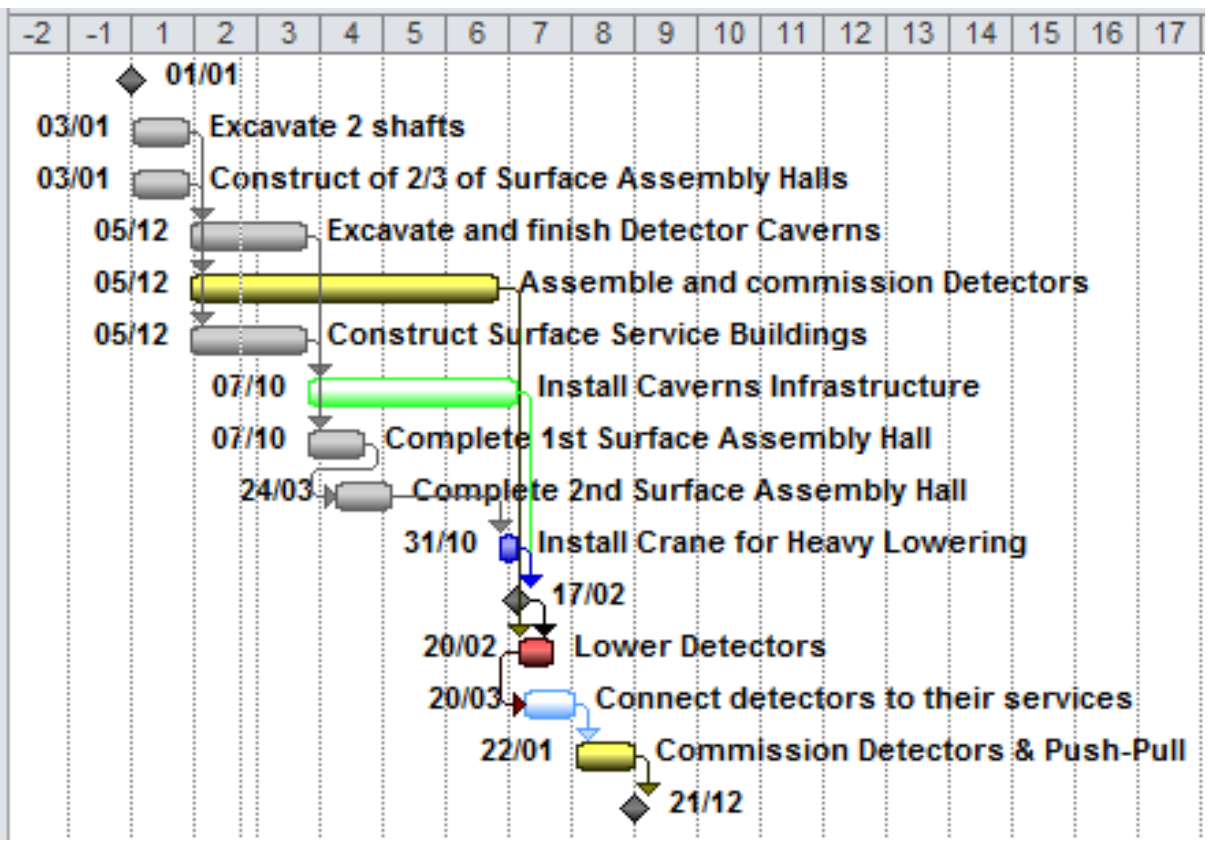

**Figure 14.9.** Flat Topography Interaction Region construction schedule

From this study, it appears that the caverns should become available for detector installation by Y7Q1.

Using this milestone, the Gantt chart in Fig. 14.10 originally produced by the ILD community has been modified. It shows the three phases of the detector activities:

- detector construction on the surface;
- lowering and installation in the underground cavern;
- commissioning without beam.

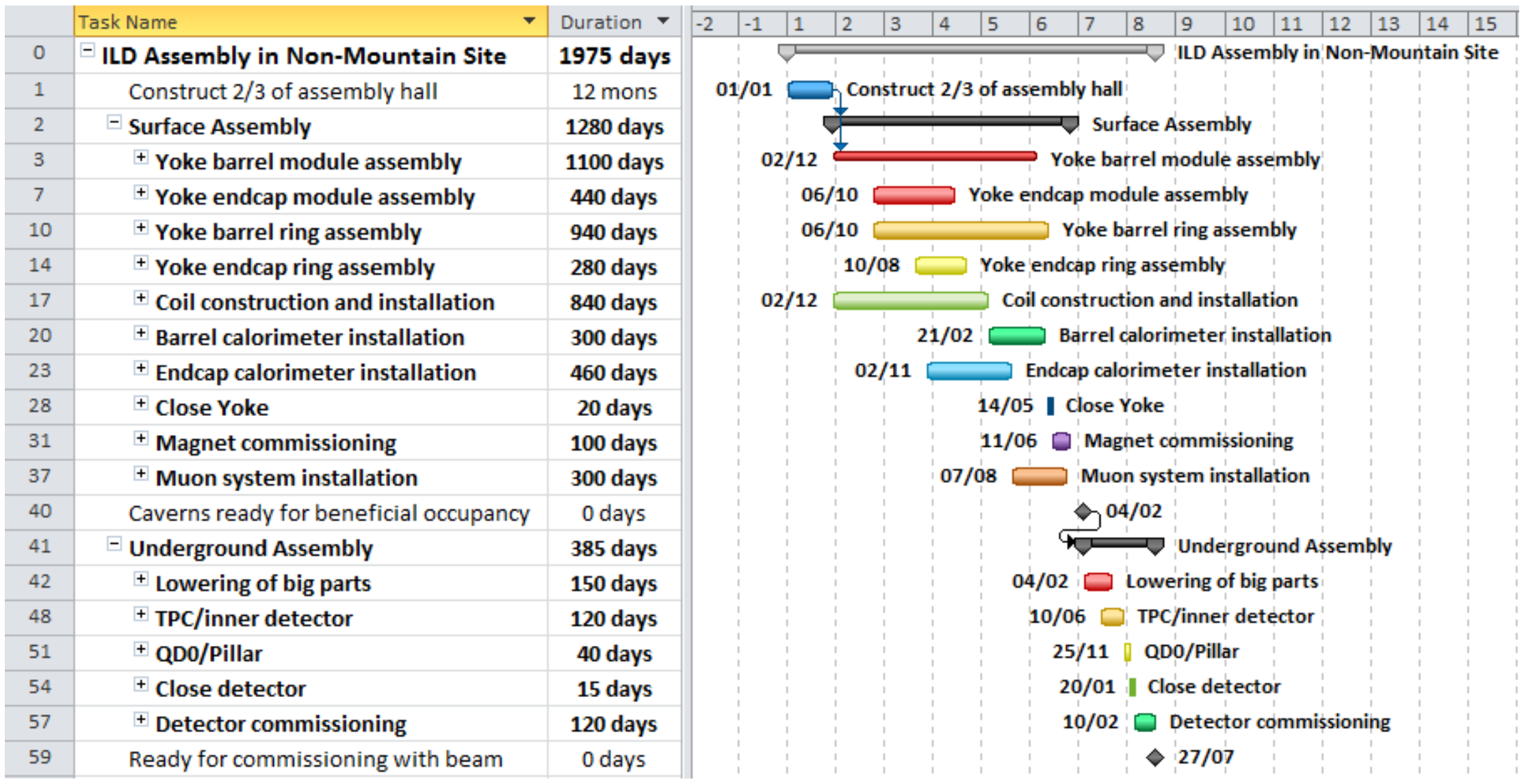

| | Task Name | Duration |
|---|---|---|
| 0 | ILD Assembly in Non-Mountain Site | 1975 days |
| 1 | Construct 2/3 of assembly hall | 12 mons |
| 2 | Surface Assembly | 1280 days |
| 3 | Yoke barrel module assembly | 1100 days |
| 7 | Yoke endcap module assembly | 440 days |
| 10 | Yoke barrel ring assembly | 940 days |
| 14 | Yoke endcap ring assembly | 280 days |
| 17 | Coil construction and installation | 840 days |
| 20 | Barrel calorimeter installation | 300 days |
| 23 | Endcap calorimeter installation | 460 days |
| 28 | Close Yoke | 20 days |
| 31 | Magnet commissioning | 100 days |
| 37 | Muon system installation | 300 days |
| 40 | Caverns ready for beneficial occupancy | 0 days |
| 41 | Underground Assembly | 385 days |
| 42 | Lowering of big parts | 150 days |
| 48 | TPC/inner detector | 120 days |
| 51 | QD0/Pillar | 40 days |
| 54 | Close detector | 15 days |
| 57 | Detector commissioning | 120 days |
| 59 | Ready for commissioning with beam | 0 days |

**Figure 14.10.** ILD construction schedule for Flat Topography sites

One conclusion of this study is that by the time the detector community is ready to start the detector-lowering phase (Y7Q1), the cavern will be available for beneficial occupancy. It also shows that by the end of year 8 the detector should be ready to be commissioned with beam. This means that the Push-Pull system should also be ready.

The commissioning of the detector together with the Push-Pull system should be seen as a separate exercise. This is reflected in Fig. 14.9 were the commissioning task lasts till Y8 Q4. The commissioning task in Fig. 14.10 only reflects the commissioning of the detector in parking position (finished by Y8Q3).

The detector readiness for beam should therefore coincide with the beginning of the accelerator final-commissioning phase. However, in order to decouple the commissioning of the accelerator from the commissioning of the detectors, it would be wise to plan the use of a temporary beam-pipe that would allow the beam to be circulated through the interaction region even if no detector can be put in beam position.

A final set of milestones can be extracted from the detector scheduling studies (see Table 14.8).

**Table 14.8**
The fifth set of level-1 milestones.

| Milestone | Flat topography | Mountainous region |
|---|---|---|
| Civil Engineering work complete | Y4, Q4 | Y5, Q1 |
| Common Facilities installed | Y7, Q3 | Y8, Q2 |
| Accelerator ready for early commissioning (BDS and ML up to PM7/AH1) | Y7, Q2 | Y8, Q2 |
| ILC ready for full commissioning (whole accelerator available) | Y9, Q4 | Y9, Q4 |
| ILC ready for beam | Y10, Q4 | Y10, Q4 |
| Caverns ready for beneficial occupancy | Y7, Q1 | |
| Detector ready to be lowered | Y7, Q1 | |
| Detector ready for commissioning with beam | Y8, Q3 | |

The detector hall for the Mountain Topography layout differs significantly from that of the Flat Topography (see Fig. 14.11). Access to the hall is via horizontal access tunnels rather than shafts, which limits the size of the components that can be brought into the detector hall. Therefore much of the detectors will need to be constructed in the detector hall similar to the approach used for ATLAS at the LHC. Although not as detailed as the surface construction schedule for the Flat Topography site, preliminary studies have indicated that the underground detector assembly is feasible within the same overall time frame.

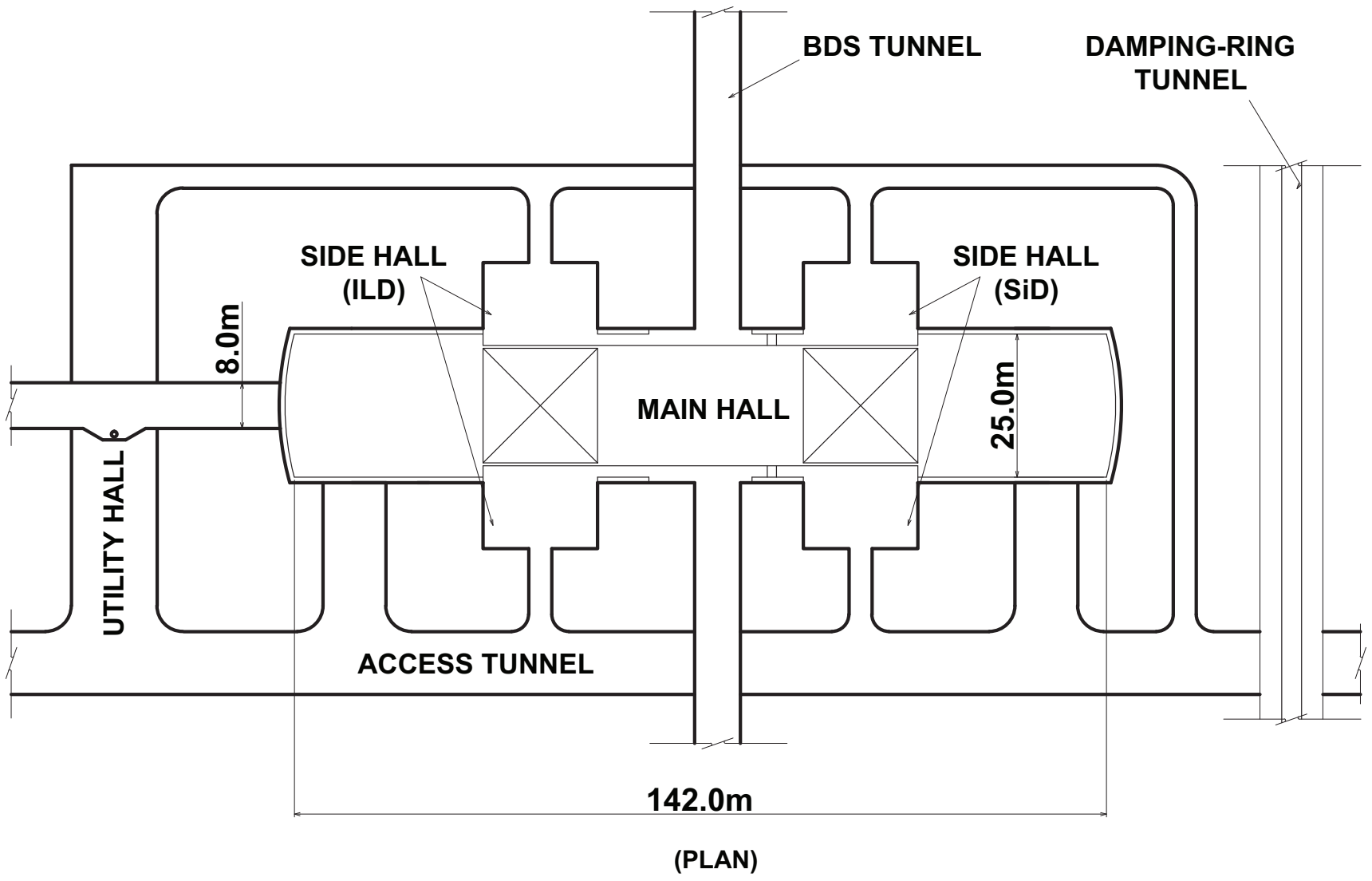

**Figure 14.11** Layout of the interaction region in a MoutainTopography site

# Chapter 15
# ILC TDR Value Estimate

## 15.1 Introduction

This chapter presents a high-level summary of the International Linear Collider Technical Design Report (ILC TDR) Value estimate. The details of the costs and the cost basis are contained in the Engineering Document Management System (EDMS) technical-design documentation files, many of which are referenced in this chapter.

Throughout the chapter, the estimate will be referred to as a "Value" estimate. This is to emphasise that its scope is limited relative to cost estimates which may be developed based on region-specific guidelines or practices. The Value estimate omits a number of items (for example, pre-construction, contingency, escalation during project execution, commissioning with beam, etc.) that would be included in some region-specific cost estimates. The precise scope of the Value estimate will be presented in detail below.

### 15.1.1 Overview

This overview provides a brief summary of the contents of each section of this chapter.

#### 15.1.1.1 Goals and scope

The ILC TDR Value estimate is for a machine of 500 GeV centre-of-mass energy, but includes some items rated for 1 TeV to enable a later energy upgrade. The scope is clearly defined by stating what is included, and what is excluded, in the estimate.

#### 15.1.1.2 Methodology

As is appropriate for a project likely to be funded mainly by in-kind contributions, the Value-estimating methodology is used. The reference currency (the "ILCU") is the United States dollar (USD) as of January, 2012. In order to eliminate regional price distortions related to exchange rates, conversions from other currencies to ILCU are based on purchasing-power-parity (PPP) indices published by the Organisation for Economic Co-operation and Development (OECD). The motivations for, and issues related to, the use of PPP indices are discussed, and the PPP indices are compared graphically with exchange rates over the past 6 years.

#### 15.1.1.3 Cost Guidelines and Learning Curves

One important general guideline is that at least two vendors are assumed for industrial procurements. Specific guidelines are defined for specialised cost elements, such as cavities, cryomodules, and conventional facilities. One important specific guideline is that the cavities and cryomodules are fabricated by industrial vendors based on build-to-print specifications. Performance is guaranteed through testing using labour and facilities provided by collaborating host laboratories. General considerations on the use of learning curves are presented in the final part of this section.

#### 15.1.1.4 Development and Format of the Estimate

Approximately 75 % of the TDR estimate has a new cost basis; the remaining 25 % is taken from the RDR. The format in which the TDR estimate will be documented in EDMS is presented.

#### 15.1.1.5 Cost Basis

For cavities and cryomodules, the cost bases selected for the TDR are described. Other sources of cost information, including those used for the International Linear Collider *Reference Design Report* [3] (RDR) estimate, are presented, and the reasons for selecting the TDR cost bases are explained. For L-band high-level RF systems, a similar discussion is provided. In this case, there are different bases for the flat- and mountainous-topography sites.

The cost bases for the major elements of the conventional facilities are presented. The civil-engineering cost bases for the three regional sites are discussed. Differences between the regional cost bases, resulting from site-specific factors, are explained. The cost bases for the conventional electrical, mechanical and safety systems for the Asian and Americas sites are also discussed. Finally, the cost bases for handling systems, and for survey and alignment are described.

Subsequently, the cost bases for the following technical or administrative areas are presented: installation, cryogenics, magnets and power supplies, vacuum, instrumentation, controls and computing infrastructure, other high-level RF systems, management and administration, and cost elements specific to some of the accelerator systems.

#### 15.1.1.6 Value and Labour Estimates

To provide a reference for comparison, the RDR estimate is re-stated in terms of the 2012 ILCU, using the inflation indices presented in Section 15.1.2. The TDR Value estimate is then presented, broken down into subsystems. The total explicit Labour estimate, broken down by subsystem, is presented. The site dependences of the Value and Labour estimates are presented and discussed.

#### 15.1.1.7 Cost Uncertainty

Uncertainty estimates have been made for each cost element in the TDR. The uncertainties are based on the design maturity of the item, the level of associated technical risk, the source and quality of the cost or Labour information, and the extent, if any, of scaling to large quantities. Tables of the cost-uncertainty parameters for the major Value elements in the TDR estimate are presented.

For each cost element, the fractional level of cost increase required to reach the 84 % confidence level was computed. This cost increase is called a "cost premium". The cost premiums were summed over groups of cost elements, to provide a conservative estimate of the overall cost premium required to reach the 84 % confidence level for that cost-element group. These summed cost premiums are presented graphically, broken down by system, and the overall cost premium for the total project is presented.

In a similar way, premiums on the Labour estimate are developed. Summed labour premiums, broken down by system, are also presented, and the overall Labour premium required to reach the 84 % confidence level on the Labour estimate is presented.

#### 15.1.1.8 Value and Labour Time Profiles

Given the schedule described in Chapter 14, and the Value and Labour estimates given in Section 15.8, profiles describing the Value and Labour resources needed as a function of time have been developed. These profiles assume a flat funding profile for the major civil and technical procurements for each accelerator system, which is a crude assumption, but one which captures the essential features of the overall project resource requirements.

#### 15.1.1.9 Value and Labour Estimates for Operations

A top-down estimate of projected operating costs is presented.

#### 15.1.1.10 Value and Labour Estimates for Upgrade and Staging Options

This section estimates the Value and Labour changes associated with the upgrade and staging options described in Chapter 12 of the *Technical Design Report*.

### 15.1.2 Inflation Indices

Many of the cost bases used for the TDR are stated in currencies as of a date different from the TDR reference date (January, 2012). For example, any estimates taken from the RDR are generally stated in 2006 or 2007 currencies. All costs used in the TDR estimate have been escalated to the TDR reference date. The inflation index used depends on the cost element type (either "civil engineering", which refers to all civil engineering cost elements, or "machinery and equipment", which refers to all other cost elements), and on the region in which the estimate was made. The regional [1] inflation indices [241–247] are shown in Fig. 15.1. After escalation, costs were re-stated in ILCU, using PPP indices for currencies other than the USD, as discussed in Section 15.4.2.

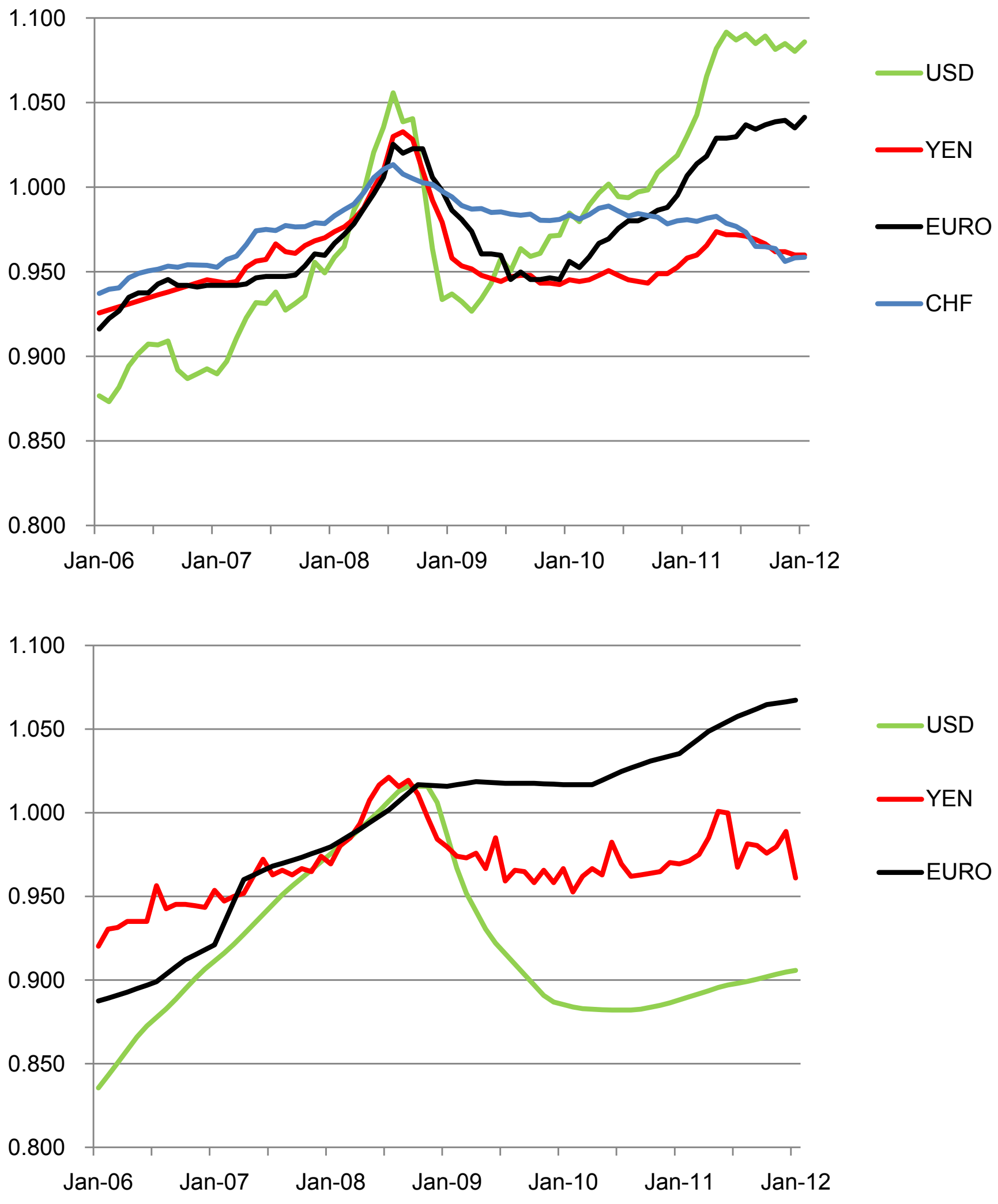

**Figure 15.1** National inflation indices. Top: machinery and equipment; Bottom: civil engineering

[1]The inflation indices for the Euro were taken to be those of Germany, the dominant manufacturing economy which uses the Euro.

## 15.2 Goals

The ILC TDR Value estimate is a comprehensive and well-documented estimate of the resources required to build the ILC, as described in the *Technical Design Report*. The project is intended to be funded by a mixture of cash and in-kind contributions from a collaboration of countries or regions around the world. The ILC TDR Value estimate should:

1. allow funding agencies in nations that are considering in-kind contributions to the ILC project to assess the nature and scope of resources needed for the project;
2. provide detailed information on cost drivers and cost trade-offs which can be used in the pre-construction phase for further cost-optimisation of the project through value engineering and R&D.

## 15.3 Scope

The ILC TDR Value estimate is made for a machine of 500 GeV centre-of-mass energy, but it includes some items rated for 1 TeV, such as the beam dumps and the Beam Delivery tunnel, to enable a later energy upgrade. The estimate does not include the cost of the detectors. They are assumed to be funded by a separate agreement between the collaborating institutes, in the way the LEP and LHC detectors were built. The estimate does include civil engineering work for the detectors, e.g. assembly buildings, underground experimental halls, shafts, etc. Table 15.1 summarises the items that are included in, or excluded from, the estimate.

The estimate assumes a 9-year construction period (see Chapter 14). The estimate for a given item covers the cost from the day the project obtains funding until that item is installed, tested, and ready for commissioning. Commissioning in one area may overlap with construction elsewhere. The construction period covered by the Value estimate ends when the last component has been installed and tested, and the machine is ready for commissioning with beam.

**Table 15.1**
Summary of the items that are included in, or excluded from, the ILC TDR Value estimate.

| Included | Excluded |
|---|---|
| Construction, installation, and hardware commissioning costs for a 500 GeV machine | Beam commissioning, operations, decommissioning |
| Tooling-up industry, final engineering designs and construction management | Engineering, design, or preparation activities that can be accomplished before construction starts, such as research & development, and prototype systems tests |
| Construction of all conventional facilities, including the tunnels, surface buildings, access shafts and other facilities | Pre-construction costs (e.g. architectural engineering, conceptual and construction drawings, component and system designs), surface land acquisition and underground easement acquisition costs |
| Construction of the detector-assembly building, underground experimental halls and detector-access shafts | Experimental detectors |
| Explicit labour, including that for management and administrative personnel. | Taxes, contingency and escalation |
| Costs for upgrading the machine to 1 TeV which would be very difficult to provide after construction of the 500 GeV machine (e.g., beam dumps, BDS length). | Additional costs due to potential overheads related to management of in-kind contributions |

## 15.4 Methodology

### 15.4.1 Definition of Value Estimating

In order to achieve the first goal stated in Section 15.2, the ILC TDR Value estimate must be structured so that it can be useful to all potential collaborators. Each of these collaborators may have different currencies, and different traditions and conventions for planning and estimating the cost of large projects. In order to divide up these in-kind contributions equitably among the collaborators, a project estimate that is independent of any particular accounting system but compatible with all of them must be developed. The "value-estimating methodology" for stating this estimate has become the standard for such international projects. It was adopted by ITER (the International Thermonuclear Experimental Reactor project) and by the LHC (Large Hadron Collider) experiments, among others.

As expressed using the "value-estimating methodology", the ILC TDR Value estimate consists of two important parts:

- Value[2] (in terms of currency units) for items procured from vendors. The Value of a component is defined as the lowest reasonable estimate of the procurement cost of an item with the required specification and in the appropriate quantity, based on production costs in a major industrial nation. It is expressed in current-year currency units (not escalated to the years in which the funds are projected to be spent) and does not include R&D, pre- or post-construction, beam commissioning, operating costs, taxes or contingency. It is effectively the barest cost estimate that would be used by any of the funding agencies. Individual regions can then add to the base Value any other items usually included in their own estimating system;
- Labour[3] (in terms of person-hours). In this context, Labour is defined as "explicit" labour, which may be provided by the collaborating laboratories and institutions, or may be purchased from industrial firms. This to be distinguished from a company's "implicit" labour associated with the industrial production of components and contained (implicitly) within the purchase price. The implicit labour is included in the Value part of the estimate.

The ILC TDR Value estimate, stated in terms of Value and Labour, is the basis on which contributions may be apportioned among the collaborators. Each participant makes an agreement with the ILC project management to provide a certain amount of Value and Labour to the project, which may be in the form of in-kind component and service contributions. The ILC TDR Value estimate documents the specific project items associated with these Value and Labour contributions. The collaborators are then responsible for providing these contracted items, independently of what they may cost as measured by national accounting systems.

### 15.4.2 Definition of the ILCU for the TDR

One of the key elements of the Value-estimating methodology is the definition of a common currency unit (the "ILCU"), and the development of a consistent and reasonable procedure [248] for converting costs in national currencies into this unit.

#### 15.4.2.1 Purchasing Power Parity

For the 2007 ILC *Reference Design Report* (RDR), the ILCU was defined to be equal to the USD as of January 1, 2007. Conversions from other regional currencies to ILCU were based on averages of currency exchange rates to the USD over the previous 5 years. Explicitly, the RDR ILCU was equal to 0.8333 Euro and 116.7 Yen.

A similar procedure for the TDR estimate is problematic, however. This is because, in general, and particularly in times of wide fluctuations in monetary supplies related to financial crises, exchange rates do not necessarily represent true comparative prices between items manufactured in different regions of the world. Exchange rates can be strongly influenced by the supply and demand for different currencies, and the supply and demand for currencies are influenced by factors such as capital flows between countries and currency speculation, rather than simply by the needs of international trade.

[2] Value is capitalised in this document when it has the specific meaning described in this paragraph.
[3] Labour is capitalised in this document when it has the specific meaning described in this paragraph.

International economists have introduced the concept of "purchasing power parity" (PPP) to deal with this issue. Compiled through extensive surveys by the research arm of the Organisation for Economic Co-operation and Development (OECD) and Eurostat (the European Union's statistical agency), PPP indices [249] are price relatives derived from the ratio of the prices in national currencies of the same good or service in different countries.

#### 15.4.2.2 Motivations for the Use of PPP Indices

There are two primary motivations for the use of PPP indices. The first motivation is related to the development of the estimate, while the second motivation is related to the use of the estimate.

First, in the development of an estimate of an item's Value, it is common to have Value estimates for the same item, from different regions, stated in different currencies. These estimates are developed in the region, and typically correspond to the prices of the items from fabricators in the region in which the estimate is made. The proper way to compare these costs, so as to arrive at the lowest reasonable cost (the Value of the item), is on the basis of PPP indices, which are specifically designed to compare prices of similar items across national boundaries, avoiding the distortions associated with the use of exchange rates.

Secondly, as discussed above, the ILC project is expected to be funded through in-kind contributions. Since the Value of each item is stated in terms of an ILCU based on PPP indices, regional collaborators, seeking to assess the local resource requirements for their in-kind contributions, can use the PPP indices to translate components of the Value estimate into an estimate of the local currency required to build those components in their region. This can be done in a way which is not dependent on volatile exchange rates, now or in the future. Note that the local currency estimate derived in this way is independent of the region and currency in which the original estimate was made.

#### 15.4.2.3 Issues with the Use of PPP Indices

There are several issues related to the use of PPP indices:

1. **Cost-element-type dependence.** Similar to inflation indices, PPP indices depend on the type of cost element (e.g. consumer goods, food, technical equipment, etc.). This dependence has been recognized by the OECD and other organizations which compile PPP indices, and separate indices have been derived and published for different types of elements. The two types of cost elements that are important for the ILC are civil engineering and technical equipment and machinery. The accounting framework for the ILC estimate, which was designed to handle different inflation indices for these two types of cost element, can also handle different PPP indices for these types.

2. **Accuracy.** Since the Value estimate uses PPP indices for conversions of estimates other than those in USD, the accuracy of the estimate depends on the accuracy of these indices. The Eurostat-OECD price surveys used to determine the PPP indices are comprehensive and well documented. The PPP indices from these surveys are used by many international organisations throughout the world. For industrialised nations, estimates [250] of the standard errors for PPP indices are in the range of 5-8 %. These uncertainties are considerably smaller than price distortions which would be introduced by the use of exchange rate and are typically also smaller than the overall uncertainties associated with the local-currency Value estimates themselves.

3. **Extrapolations.** The most recent published PPP benchmark survey was in 2008 [251], and the results from the next one, made in 2011, will not be available until 2013[4]. Following OECD recommended practice, the current PPP index can be obtained by extrapolation from the 2008 PPP index, based on the relative national[5] rates of inflation [241–247] from 2008 to the present, for the two national currencies that are related by the index. Errors introduced by this extrapolation are expected to be $\sim 5\,\%$ [248]. Once the 2011 Eurostat-OECD PPP benchmarks become available, the Value estimate could easily be updated.

4. **Regional versus global procurements.** The Value of a component, translated into the local currency in a region using PPP indices, corresponds to the price of this component if it is made and purchased in that region using the local currency. If, as a result of exchange-rate fluctuations, the region happens to have an overvalued currency relative to some other region, it is possible that the price of the component could be lower, in local currency units, if the component were purchased from the other region. In this case, a regional collaborator could choose between regional production of the component (which could benefit regional industry) or procurement from another region (which would require less local currency). But the Value estimate, translated to local currency using PPP indices, always establishes the required local currency for the component.

#### 15.4.2.4 ILCU Definition in terms of PPP Indices

For the TDR, the ILCU will be defined as equal to the USD on January 1, 2012. Conversions of estimates obtained in currencies other than USD to ILCU will be based on PPP indices (as of January 1, 2012) relating those currencies to the USD. The only exception to this rule is for the superconducting material for the cavities. There is only one supplier of RRR-niobium raw material in the world. Thus, it is appropriate to consider this cost element to be a commodity which must be purchased on the international market. In preparing the Value estimate, conversions from currencies other than USD to ILCU for this cost element have been based on exchange rates as of January, 2012. The PPP indices of four regional[6] currencies, relative to the USD, together with exchange rates, are shown in Fig. 15.2. Numerical values of the PPP indices and exchange rates for January, 2012, which are used in the Value estimate, are given in Table 15.2.

**Table 15.2.** Currency conversion factors between ILCU and national currencies (January, 2012). To convert a cost element from ILCU to the indicated currency, multiply by the factor appropriate for the type of cost element.

| Cost element type | ILCU→USD | ILCU→Euro | ILCU→Yen | ILCU→CHF |
|---|---|---|---|---|
| Civil construction (PPP) | 1 | 0.939 | 109.3 | 1.303 |
| Machinery and equipment (PPP) | 1 | 0.923 | 127.3 | 1.480 |
| Superconducting material (EX) | 1 | 0.776 | 76.9 | 0.939 |

[4] PPP indices for European Union nations are compiled annually [252] by Eurostat.

[5] The PPP and inflation indices for the Euro were taken to be those of Germany, the dominant manufacturing economy which uses the Euro.

[6] The PPP indices for the Euro were taken to be those of Germany, the dominant manufacturing economy which uses the Euro.

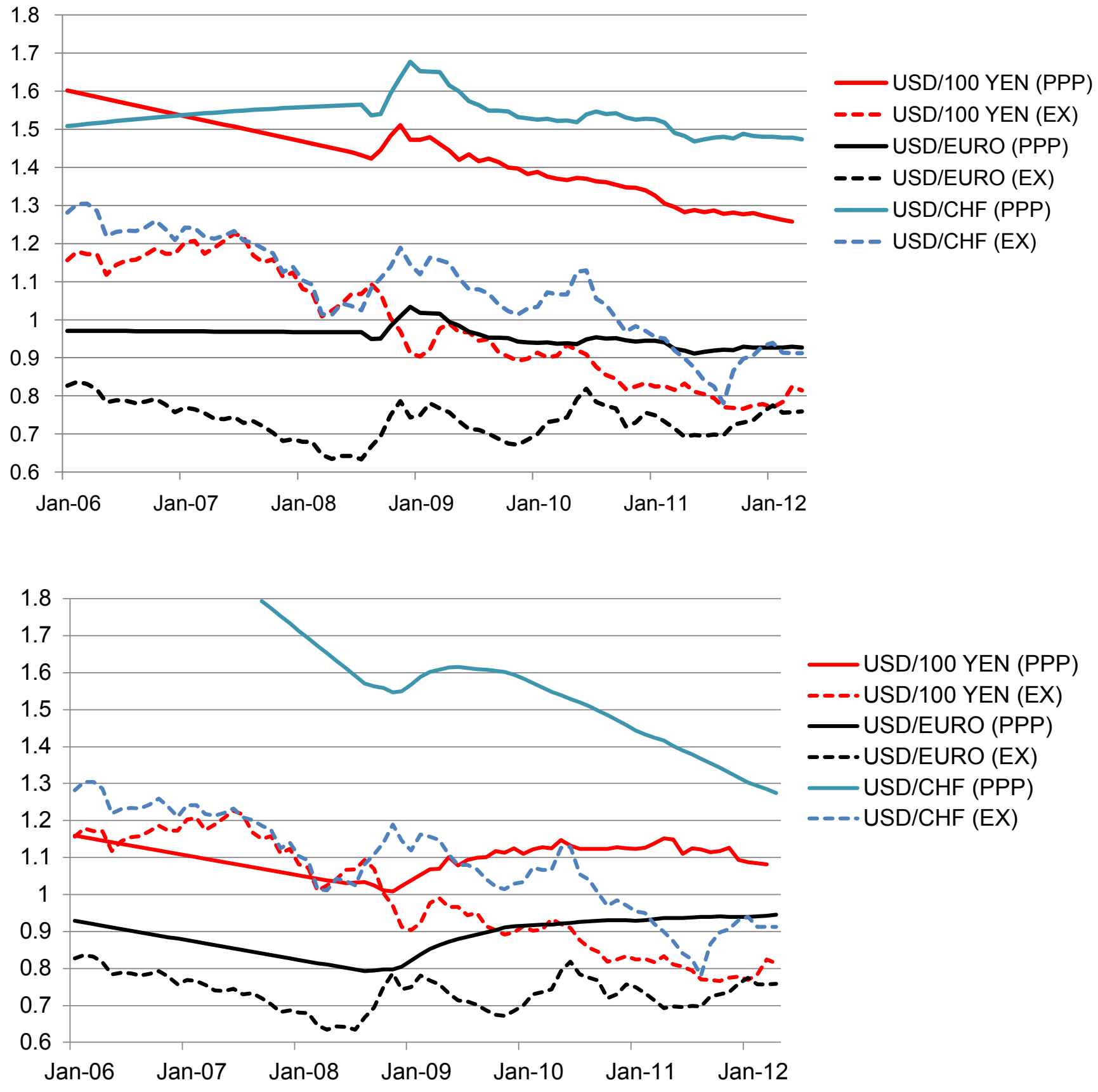

**Figure 15.2** PPP indices and exchange rates. Top: machinery and equipment. Bottom: civil engineering. Indices between 2005 and 2008 are based on a linear interpolation between the 2005 [253] and 2008 [251] survey points. Indices after 2008 are extrapolated from the 2008 survey point, based on relative inflation rates. For the case of CHF, since a civil engineering inflation index is not readily available, the CHF civil engineering PPP index was based on scaling from the Euro PPP index, using the annual Eurostat PPP indices to relate the Euro and CHF PPP indices.

## 15.5 Cost Guidelines and Learning Curves

The specific cost guidelines used in preparing the Value estimate for the TDR were similar to those used for the RDR [254, 255].

### 15.5.1 General guidelines

The key elements of the general guidelines, which apply to all cost elements, are:

- the estimates for each cost element are median estimates. In other words, the estimate corresponds to the 50 % probability point in the cumulative cost-distribution curve. Thus, if a given item were to be offered independently for bid many times, taking the lowest world-wide bid each time, half of the lowest bids would be below the median estimate, and half above. Due to limited resources, no estimate was actually obtained by taking the median of a large number of bidding attempts. The TDR estimate for a given cost element was typically obtained by choosing from the available sources (such as vendor quotes, engineering estimates, or industrial studies) that estimate which was deemed to be the most reliable representative of a median lowest-bid;
- the specifications and quantities for each component or subsystem in the estimate correspond to the design documented in the TDR, which is intended to represent the optimal balance between technical performance, reliability, acquisition cost, and operating cost (over a 10 year lifetime);
- the Value estimate for a component or subsystem is the lowest world-wide vendor cost for the item, as determined by PPP, that is practical, feasible, and reasonable, for the required specification and quantity, with a procurement time consistent with the project schedule;

- in estimating the costs, at least two vendors were assumed for all industrial procurements for which the cost model has an explicit dependence on the number of items;
- the manufacturer's implicit labour for all fabrication activities, and for Engineering, Design, Inspection and Acceptance (EDIA), quality control, quality assurance and technical testing, is included in the item's Value;
- tooling, instrumentation and infrastructure necessary for the fabrication, acceptance and testing of the item or subsystem is included in the Value, if it does not exist or is not available at collaborating laboratories or other associated institutions; transportation costs are included;
- only those spares that are installed in the accelerator complex and are required for operational or reliability margins (as specified in the TDR) are included;
- explicit Labour (labour at the ILC, collaborating laboratories or other associated institutions, or labour purchased from an industrial vendor) includes, for example, final engineering design to prepare bid packages (after construction start), contract management, sustaining engineering, vendor liaison, inspection and acceptance tests, quality assurance, installation, system integration, alignment, and initial checkout (without beam). Commissioning with beam is not included. This Labour is estimated, separately from the item's Value, in person-hours. To convert person-years to person-hours, the number of hours per year was taken to be 1,700. Four classes of explicit manpower are included: engineer, scientist, technical staff, and administrative staff.

## 15.5.2 Specific Guidelines

In addition to the general guidelines, specific guidelines were applied to cavities and cryomodules, and to conventional facilities.

### 15.5.2.1 Cavities and Cryomodules

Cavity fabrication costs are based on a build-to-print specification to industrial vendors with minimum acceptance criteria, which must include vacuum leak testing, room-temperature RF tuning, high-pressure test, etc. but without a guarantee of accelerating-gradient performance. Superconducting material will be supplied by the project to the vendors.

All of the cavities will be vertically tested. Testing and quality assurance for the cavities and quadrupoles, and high-power processing of the couplers, is the responsibility of the project and its multi-region partner institutions; the effort is included in the explicit Labour estimate. Cryomodule fabrication and assembly is also based on a build-to-print specification to industrial vendors.

The overall performance of the cavities and cryomodules will be guaranteed by the project and its partner institutions. Approximately one third of the cryomodules will be fully tested. The required testing and QA effort is included in the explicit Labour estimate.

### 15.5.2.2 Conventional Facilities

There is one common machine design. The footprints for the sites have small geology-driven differences, such as shaft and hall locations, and minor differences in tunnel lengths. Nevertheless, the costs for many aspects of conventional facilities are site-specific and there are separate estimates for each sample site (one in each region: the Americas, Europe and Asia[7]). These are driven by real considerations, e.g. different geology and landscape, availability of electrical power and cooling water, etc. The cost of surface land and underground easements, and site-dependent costs due to local regulations, are not included.

[7] There are two Asian site candidates. The Asian TDR estimate corresponds to an average of the costs for these two sites.

The site-specific costs are combined into a single average conventional facilities (CFS) Value estimate for the TDR. The reasons for any major differences between the site-specific costs are discussed in Section 15.7.3.

### 15.5.3 Learning curves

Many of the cost bases used for the ILC TDR Value estimate correspond to smaller numbers of units that those required for the ILC. To account for expected economies of scale, when the cost basis explicitly corresponds to a much smaller number of units than required for the ILC, the unit Value estimate for ILC quantities has been obtained by applying a discount based on a (Crawford) learning curve [256].

In using learning curves to estimate quantity discounts, care must be used in the choice of the learning-curve slope. Table 15.3 [256, p. 180] shows the typical range of learning-curve slopes that have been found for various types of manufacturing processes or general categories of items. The components manufactured for the ILC generally correspond to mixtures of the first five lines in this table. Components for the ILC components would thus be expected to have learning curve slopes roughly in the range of 85-95%.

**Table 15.3**
Typical learning curve slopes [256, p. 180].

| Manufacturing process or item | Range of learning curve slopes |
|---|---|
| Raw materials | 93-96% |
| Repetitive electronics manufacturing | 90-95% |
| Repetitive machining or punch-press operations | 90-95% |
| Repetitive welding operations | 90% |
| Purchased parts | 85-88% |
| Repetitive clerical operations | 75-85% |
| Construction operations | 70-90% |

The discount is a sensitive function of the slope, as shown in Fig. 15.3, which demonstrates that the discount varies by a factor of about 2 in going from a 95% slope to a 90% slope.

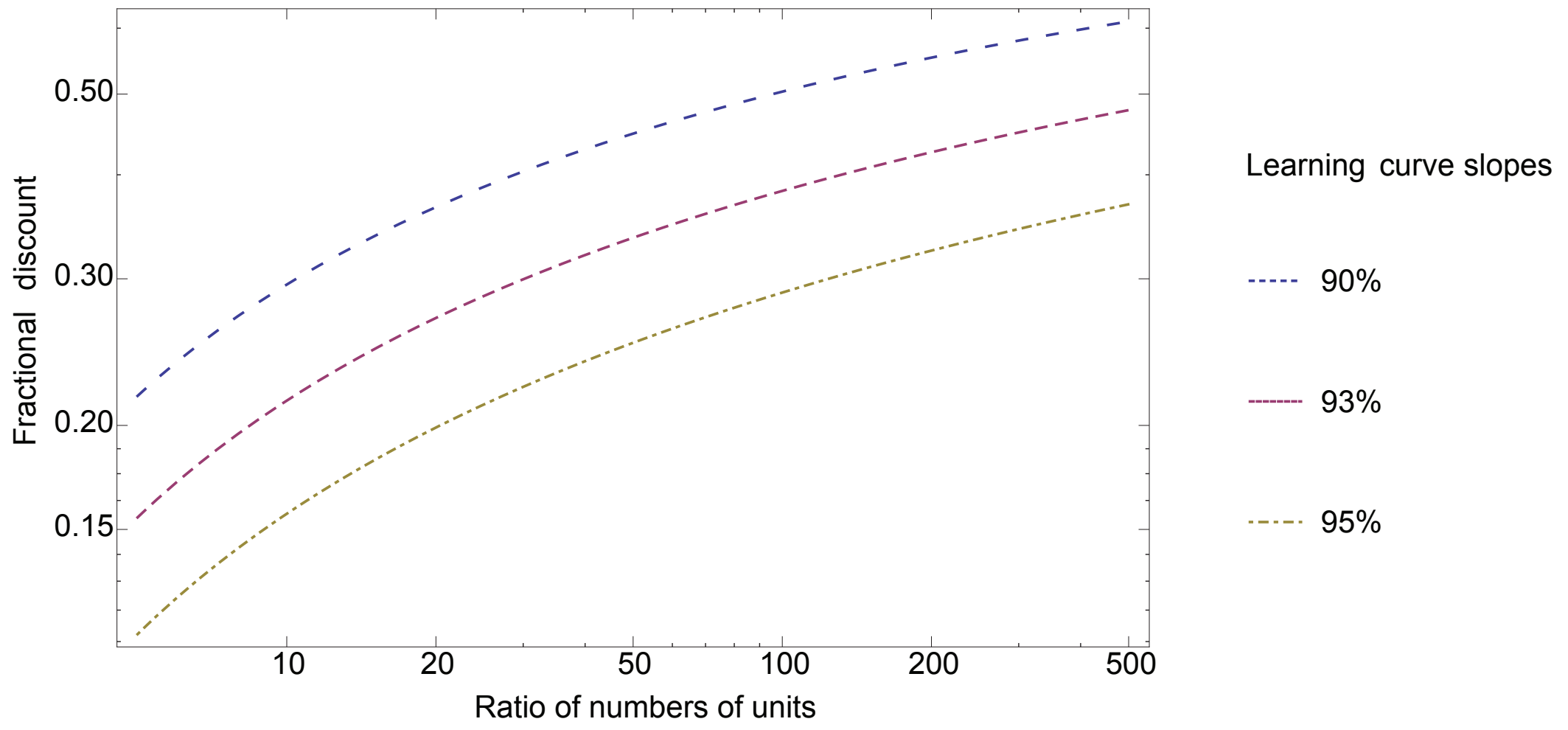

**Figure 15.3.** Discount versus the ratio of the number of units in the discounted estimate to that in the original estimate, for learning curves with 90%, 93%, and 95% slopes.

Without a specific manufacturing model for an ILC component, it is difficult to determine precisely what learning-curve slope should be used. Thus, when there is no specific manufacturing study (or vendor quote in the appropriate quantity) for a component, a conservative choice of a learning curve slope at the upper end of the expected range (95%) has been made. No saturation of the learning curve has been assumed.

For the L-band modulators, the engineering cost estimate does have a specific breakdown of labour and materials costs. Thus, for this component, a 90% learning curve for the labour cost and a 95% learning curve for the materials cost have been used.

No learning curves have been used when the component cost basis has explicitly accounted for the correct quantity of items required from a given vendor for the ILC. The most important examples of such components are the industrial studies for production of the cavity resonator and assembly of the cryomodule, and vendor quotes for the klystrons. For these cases, the extensively documented work explains precisely how the discount is obtained. The effective learning curves that can be derived from the explicit studies or the vendor quotes are in fact between 87% and 92%, illustrating the conservatism of the choice of a 95% slope.

## 15.6 Development and Format of the Estimate

### 15.6.1 Estimate Development

A complete Value estimate for the ILC was developed at the time of the *Reference Design Report* (2007). Subsequently, value engineering, design development, and component R&D have led to the evolution of a more cost-optimised machine design, and to more mature concepts for many of the components. For the TDR, new estimates for the conventional facilities, the superconducting cavities and their cryomodules, the L-band high-level RF systems, and the cryogenic facilities have been made. In addition, since the damping-ring design in the TDR is substantially changed from the RDR, major elements of the damping ring were re-estimated, as were significant portions of the positron source. Overall, the new estimates comprise about 75 % of the total cost of the project.

For the remaining approximately 25 % of the estimate, the estimated Value and explicit Labour per component unit were generally taken from those provided at the time of the RDR. Unit costs were escalated from the RDR estimate date to the TDR estimate date based on the regional escalation indices shown in Fig. 15.1, and converted to 2012 ILCU using the PPP indices shown in Fig. 15.2.

### 15.6.2 Format of the Estimate

The Value estimate has been formatted using the ILC Cost Estimating Tool (ICET). The tool is a series of Windows and Visual Basic scripts which allows the manipulation of custom Excel spreadsheets, called “Cost Estimating Modules (CEMs)”, that contain the cost data and links to the cost-basis documents in EDMS. The ICET organises the CEM cost data into a WBS structure, where each cost element includes a description, basis of estimate, quantity required, materials and services estimate, explicit labour, and an uncertainty characterisation. ICET can be used to load the cost data into a database. Subsequently, a series of reports detailing and cross-cutting the estimate can be generated from the database. There are approximately 700 cost elements in the cost database for the TDR estimate.

## 15.7 Cost Basis

The bases of estimate for each major cost-element category are discussed in the following sections.

### 15.7.1 Cavities and Cryomodules

The cavities and cryomodules represent a substantial fraction (about 1/3) of the total ILC project Value. For the RDR, the Value estimate was based on the cost studies carried out for the TESLA TDR [13] more than 10 years ago. Since the publication of the RDR, there have been two developments which allow this TDR estimate to have a much more diverse and mature cost basis:

- substantial R&D has been carried out in all three regions, so that a more extensive world-wide experience base now exists, allowing industrial cost studies for cavity fabrication and cryomodule assembly to be carried out in all regions;

- extensive experience from procurements for the European XFEL (EXFEL) is available, which provides actual costs for all the key components of the ILC cavities and cryomodules, albeit for a smaller number of units than for the ILC.

Based on these industrial studies and the EXFEL experience, and following the cost guidelines outlined above, Value and Labour estimates have been developed for the cavities and cryomodules. A brief description of the cost basis of each subcomponent is given in the following subsections.

For almost all the elements of the cavities and cryomodules, the TDR estimates are substantially higher than those of the RDR. For the total cavity and cryomodule cost, the TDR estimate is about 1.7 times higher than the escalated RDR estimate. This difference arises from the much broader R&D and procurement experience base on which the TDR cost basis rests.

The essential elements of this broader experience base are:

- procurements of cavities and cryomodules made for the EXFEL project;
- industrial studies by qualified vendors focused on fabrication and processing of the cavity resonator;
- industrial studies by qualified vendors focused on assembly of the cryomodule.

To provide a clear picture of the relationship between these elements, the most important cost drivers for the cavity and cryomodule are listed in Table 15.4, and the TDR cost basis is listed. For most of the items, the TDR estimate is based on the EXFEL procurement, with a quantity discount (on the basis of a 95 % learning curve) associated with the increase in the number of items by a factor of 10. For the cavity resonator, the TDR estimate is based on detailed industrial studies performed by qualified cavity vendors. For the magnet package, the TDR estimate is based on a direct vendor quote. For the cryomodule assembly, the TDR estimate is based on a detailed industrial study, again performed by an experienced company.

For the sum of these major cost drivers for an 8-cavity cryomodule with quadrupole, the TDR estimate is about 72% of the actual EXFEL procurement cost; that is, a reduction of 28 %. About 16 % of the overall reduction is due to quantity discounts, based on a 95 % learning curve. The remaining 12% is primarily due to the cost reductions resulting from the industrial study.

**Table 15.4**
Cavity and cryomodule cost drivers: Summary of the TDR cost basis.

| Item | TDR Cost Basis |
|---|---|
| Superconducting material | EXFEL procurement† |
| Cavity resonator | industrial study |
| Power coupler | EXFEL procurement† |
| Tuner | EXFEL procurement† |
| Helium vessel | EXFEL procurement† |
| Magnet package | vendor quote |
| Cryostat materials | EXFEL procurement† |
| Cryomodule assembly | industrial study |

† discount based on a 95 % learning curve

More details about all the elements of the TDR cavity and cryomodule estimate are presented in the following sections.

#### 15.7.1.1 Dressed and Qualified Cavity

15.7.1.1.1 Superconducting material Vendor quotes and recent procurements for the EXFEL and Fermilab Cryomodule 3 (CM3) were reviewed to establish the basis of estimate for the cavity superconducting material. The most reliable cost basis was judged to be the EXFEL procurement, since this was a true price obtained in the world-wide market for a substantial number of cavities. Thus, the TDR estimate is based on the cost of the EXFEL superconducting material. This cost includes material quality control.

About half of the high-purity superconducting material cost is for the niobium raw material; the remainder is smelting infrastructure ($\sim$ 25 %) and smelting and rolling operations labour ($\sim$ 25 %). A 95 % learning curve has been used to estimate the discount associated with the increase in the amount of superconducting material from the EXFEL quantity to that required for the ILC (assuming 2 vendors). The learning curve discount in this case is about 16 %.

To derive the total costs, a 90 % cavity production yield was assumed to determine the superconducting material required for the total number of ILC cavities (17,804).

15.7.1.1.2 Cavity resonator Cost information on fabrication and chemical processing of the cavity resonator is available from the EXFEL procurement, and from three industrial studies, one in each region.

The cost estimate was based on the industrial studies adjusted where needed to a nominal series production of both 9,000 and 18,000 cavities. The estimate assumed a 2.67-year series-production schedule, and includes the costs for an additional 3 year ramp-up period, which includes 1 year of pre-series cavity production. Ramp-down costs were also included, including a modest (10 %) infrastructure cost recovery. The unit costs for a total production of 9,000 cavities, approximately corresponding to a 2 vendor procurement for the ILC, were used.

The scope of work includes cavity fabrication and chemical processing, and welding of the cavity into the helium vessel. The total cost includes all manufacturing and labour for the total number of cavities, and infrastructure for two fabrication sites.

Where necessary, adjustments have been applied to the estimates presented in the industrial studies. An example is the late adoption in the ILC baseline an internal magnetic shield, the fabrication and integration of which was not included in the industrial studies. To account for the mounting of this shield, the cost of the additional labour was added The cost of material for the magnetic shield is accounted for separately below. It should be noted that these items have a relatively minor cost impact.

Since industrial studies were done for two different numbers of cavities, the average costs for these two cavity production numbers can be used to derive an effective learning curve slope. This slope is 87 % for the mechanical fabrication of the cavities, and 89 % for cavity chemical processing.

To compute the total costs, a 90 % overall cavity-fabrication and chemical-processing yield was assumed to determine the required number of ILC cavities to be fabricated (17,804). The cost of a second chemical processing for 20 % of the cavities (3,561) has been included.

15.7.1.1.3 Cavity qualification Materials and supplies (M&S) have been included for cavity qualification, based on an internal ILC cost study at Fermilab done in 2007 in connection with the development of the RDR.

This task is accounted for as explicit Labour. The labour estimate is also based on the Fermilab internal ILC cost study. The total labour and M&S costs cover qualification of the total number of processed cavities (21,365).

15.7.1.1.4 Power coupler Cost information on the high-power coupler is available from the procurements for Fermilab CM3 and EXFEL, and from a 2008 industrial study on cost-reduction options for an ILC cryomodule, made by a qualified experience vendor.

The TDR cost basis is derived from the EXFEL procurement, which is the most reliable in-quantity recent cost. To allow for economies of scale, a 95 % learning curve has been used to estimate the unit cost discount for ILC quantities. This discount is 16 %. The total cost covers the baseline number of couplers (16,024).

15.7.1.1.5 Coupler processing Materials and supplies costs have been included for coupler processing, based on the the same 2008 industrial study quoted above. This task is accounted for as explicit

Labour. The labour estimate is based on a scaling from EXFEL experience with coupler processing.

15.7.1.1.6 Tuner Cost information on the tuner is available from the procurements for Fermilab CM3 and EXFEL, and an industrial study conducted in 2012. The EXFEL costs are for a Saclay/DESY-style tuner, while the Fermilab CM3 and industrial study costs are for blade tuners. The costs include the mechanical parts, motor drives and piezoelectric devices.

The TDR baseline is the blade tuner, but no large quantity procurements of these tuners are available to provide a sound cost basis. Thus, the TDR cost basis is derived from the EXFEL cost, which is a reliable in-quantity recent procurement of an item similar to the TDR baseline design. To allow for economies of scale, a 95 % learning curve has been used to estimate the unit cost discount for ILC quantities. This discount is 16 %.

Post-TDR research and development will be required to develop either a shorter Saclay/EXFEL style tuner for the ILC, or a different design with an equivalent cost. The total cost covers the baseline number of tuners (16,024).

15.7.1.1.7 Helium vessel Cost information on the helium vessel is available from the procurements for Fermilab CM3 and EXFEL, and several industrial studies. The EXFEL costs are for a helium vessel matched to a Saclay/DESY style tuner, while the Fermilab CM3, and industrial study costs are for a helium vessel appropriate for a blade tuner.

The TDR baseline is the helium vessel appropriate for a blade tuner, but no large quantity procurements of these vessels are available to provide a sound cost basis. Thus, the TDR cost basis is derived from the EXFEL cost, which is a reliable in-quantity recent procurement of an item similar to the TDR baseline design. To allow for economies of scale, a 95 % learning curve has been used to estimate the unit cost discount for ILC quantities. This discount is 16 %.

Post-TDR tuner research and development will include development of a helium vessel appropriate for a blade tuner, with a cost equivalent to that of a vessel matched to the Saclay/DESY-style tuner. Since the vessel is carried with the cavity in vertical test, the total cost covers the number of fabricated cavities (17,804).

15.7.1.1.8 Cavity magnetic shield The cost basis for materials for this item is a vendor quote contained in recent industrial study. Since the shield is carried with the cavity in vertical test, the total cost covers the number of fabricated cavities (17,804).

15.7.1.1.9 Cavity shipping and handling The cost basis for this item is an internal ILC cost study done at Fermilab in 2007. The total cost covers the baseline number of processed cavities (21,365).

### 15.7.1.2 Quadrupole-Magnet Package

This item refers to the superconducting quadrupole and the correction dipole, together with their current leads and associated hardware. (The beam-position monitor is included in the Instrumentation estimate.)

Cost information on the magnet package corresponding to the RDR design is available from the procurements for Fermilab CM3 and EXFEL, qualified vendor industrial studies, and the Fermilab internal ILC cost study. Cost information on a magnet package corresponding to the conduction-cooled TDR design is available from a direct vendor quote based on fabrication of 300 or 600 units.

The vendor quote unit cost estimate for the conduction-cooled magnet (for 300 units) is used as the TDR cost basis since it corresponds to the TDR design and is a well-developed estimate. The total number of magnet packages required for the ILC is 673. The M&S costs for quadrupole qualification, and the associated explicit Labour, are also included. For these items, the cost basis is the Fermilab internal ILC cost study.

### 15.7.1.3 Cryomodule

15.7.1.3.1 Cryomodule EDIA  Final engineering design, and sustaining engineering, for the cryomodule and all of its components, at the ILC or collaborating institutions, is included as cryomodule EDIA explicit Labour. The cost basis is an engineering estimate taken from the Fermilab internal ILC cost study.

15.7.1.3.2 Cryostat material  This item refers to the materials which comprise the cryostat. Cost information for this item is available from the procurements for Fermilab CM3 and EXFEL, the Fermilab internal ILC cost study, qualified vendor industrial studies, as well as a direct vendor quote.

The TDR cost basis is derived from the EXFEL procurement, which is the most reliable in-quantity recent cost. The EXFEL procurement has been scaled up by 5 %, to account for the length difference between the EXFEL and ILC designs. To allow for economies of scale, a 95 % learning curve has been used to estimate the unit cost discount for ILC quantities. This discount is 15 %. The total cost covers the baseline number of cryomodules (1,855).

15.7.1.3.3 Cryomodule assembly  Cost information for assembly of the cryostat is available from the procurements for the EXFEL, the Fermilab internal ILC cost study and qualified vendor industrial studies, for both 1950 and 650 cryomodules.

The cost basis for the TDR is based on detailed industrial studies for 650 cavities, corresponding to a 3-vendor procurement. The particular study was chosen because of its depth, detail, and comprehensive scope. The study assumed a 2.6-year ramp-up period, followed by 3.5 years of cryomodule production. The Value estimate includes all labour and infrastructure for three assembly sites. Ramp-up and ramp-down costs were included, but no cost recovery for infrastructure was assumed. Following the EXFEL procedure, the effort includes assembly of the coupler and tuner onto the helium vessel containing the cavity and magnetic shield, which is received after cavity qualification, followed by assembly of the cavity string with quadrupole, alignment, and completion of the vacuum-vessel assembly. Labour rates (a variable across different studies) have been harmonised where applicable.

Since estimates are available for two different numbers of cryomodules, the average unit costs for these two numbers of assembled cryomodules can be used to derive an effective learning-curve slope. This slope is 89 % for the total cryomodule assembly labour.

The total cost covers the baseline number of cryomodules (1855), plus an additional 5 % assumed to need re-work (see below, Section 15.7.1.3.6), for a total of 1948.

15.7.1.3.4 Cryomodule shipping and handling  The cost basis for this item is an internal ILC cost study done at Fermilab in 2007. The total cost covers the baseline number of cryomodules, including rework (1,948).

15.7.1.3.5 Cryomodule qualification  Materials and supplies have been included for cryomodule qualification, based on the Fermilab internal ILC cost study. In addition, the cost of electrical power for RF, cryogenics, and water for the test stands has been included, based on scaling from EXFEL experience.

This task is accounted for as explicit Labour. The Labour estimate is based on a scaling from EXFEL experience[8]. The number of cryomodules to be tested is based on the assumption that, at peak production rate, 33 % of the cryomodules will be tested before installation. An additional 5 % has been added for testing during the ramp-up period.

15.7.1.3.6 Cryomodule hardware commissioning in the tunnel enclosure  Additional explicit Labour will be required to commission the cryomodules in the tunnel. This Labour covers the preparation and conditioning of untested cryomodules. To account for cryomodule failures which occur either during testing or during hardware commissioning in the tunnel, an allowance for re-working 5 % of the cryomodules has been included in the Value element for the total cryomodule assembly. The unit cost of a rework has been assumed to be the same as the unit cost for assembly of a cryomodule.

[8] It is assumed that all of the testing labour reported for the EXFEL is for cryomodule testing.

#### 15.7.1.4 Coupler Processing, and Cavity and Cryomodule Test Infrastructure

Coupler processing, cavity qualification, magnet testing and cryomodule qualification will be done in existing facilities supplied by international institutional collaborators.

The total test and processing facility costs have been estimated based on the Fermilab internal ILC cost study. The relative breakdown of the costs for each of the facilities is presented in Table 15.5.

The facilities are owned by the institutional collaborators and used by the project for the duration of the cavity and cryomodule production period, which is about 5 years. Maintenance and required upgrades to the facilities during this period are assumed to cost about 10 % per year. Thus, the cost to the project associated with this infrastructure has been taken to be equal to 50 % of the estimated total facility cost given in the Fermilab study.

**Table 15.5**
Processing and test infrastructure relative cost

| Facility | Fraction of total cost of processing and test infrastructure |
|---|---|
| Cavity vertical test | 26 % |
| Coupler processing | 27 % |
| Magnet qualification | 4 % |
| Cryomodule qualification | 43 % |

#### 15.7.1.5 Cryomodule Vacuum System

This system includes the beam-line vacuum, the insulating vacuum for the cryomodule, and the coupler vacuum. The TDR unit-cost basis for the cryomodule vacuum system was taken from the RDR.

The RDR cost was based on a bottoms-up accounting of the parts associated with the vacuum-system hardware. The prices were based on price lists and vendor quotes, with some adjustments for quantity price reductions. Associated Labour (EDIA) is included.

#### 15.7.1.6 Summary

The basis for the components of the estimates for cavities and cryomodules is summarised in Table 15.6 and Table 15.7.

### 15.7.2 L-band High-Level RF Systems

The principal components of the L-band high-level RF system are the multi-beam klystrons (MBKs), the modulators, and the RF distribution system.

The flat and mountainous topography sites described in the TDR require different designs for the high-level RF systems. These designs (called the Klystron Cluster System (KCS) for the flat sites, and the Distributed Klystron System (DKS) for the mountainous site) require different numbers of klystrons and modulators, and different RF-distribution systems. Consequently, a separate Value estimate was developed for each design.

**Table 15.6**
Value basis for cavities and cryomodules

| Sub-component | Basis Type | Basis Source (Date) | Number of units |
|---|---|---|---|
| Superconducting material | Procurement[a] | EXFEL (2011) | 17804[b] |
| Cavity fabrication | Industrial study | qualified vendor | 17804[b] |
| Cavity chemical processing | Industrial study | qualified vendor | 21365[c] |
| Cavity qualifications | Engineering estimate | FNAL (2007) | 21365[c] |
| Power coupler | Procurement[a] | EXFEL (2011) | 16024 |
| Power coupler processing | Engineering estimate | qualified vendor | 16024 |
| Tuner | Procurement[a] | EXFEL (2011) | 16024 |
| Helium vessel | Procurement[a] | EXFEL (2011) | 17804[b] |
| Magnetic shield | Vendor quote[d] | qualified vendor | 17804[b] |
| Cavity shipping and handling | Engineering estimate | FNAL (2007) | 21365[c] |
| Magnet system | Vendor quote | qualified vendor | 673 |
| Magnet qualification | Engineering estimate | FNAL (2007) | 673 |
| Cryostat materials | Procurement[e] | EXFEL (2011) | 1855 |
| Cryomodule assembly | Industrial study | qualified vendor | 1948[f] |
| Cryomodule qualification | Engineering qualification | FNAL (2007) | 711[g] |
| Cryomodule shipping and handling | Engineering estimate | FNAL (2007) | 1948[f] |
| Cryomodule vacuum | Vendor quote | INFN (2007) | 1855 |
| Cavity fabrication and chemical procesing infrastructure | Industrial study | qualified vendor | 2 |
| Cryomodule assembly infrastructure | Industrial study | qualified vendor | 3 |
| Infrastructure for coupler processing, and for cavity and cryomodule test | Engineering[h] estimate | FNAL (2007) | $\geq 3$ |

[a] discounted by 16 %, based on a 95 % learning curve
[b] assuming 90 % overall yield
[c] assuming 80 % first pass yield
[d] discounted by 6 %, based on a 95 % learning curve
[e] discounted by 15 %, based on a 95 % learning curve
[f] assumes 5 % more than the baseline, to account for re-work
[g] assumes 38.3 % of CM's are tested
[h] cost for maintenance and upgrades of existing facilities at collaborating labs is taken as 50 % of the estimated facility cost

**Table 15.7**
Explicit Labour basis for cavities and cryomodules

| Task | Basis Type | Basis Source (Date) | Number of units |
|---|---|---|---|
| Cavity qualification | Engineering estimate | FNAL internal cost study (2007) | 21365[†] |
| Coupler processing | Lab experience | EXFEL (2011) | 16024 |
| Quadrupole qualification | Engineering estimate | FNAL internal cost study (2007) | 673 |
| Cryomodule EDIA | Engineering estimate | FNAL internal cost study (2007) | 1855 |
| Cryomodule qualification | Lab experience | EXFEL (2011) | 711[‡] |
| Cryomodule hardware commissioning in the tunnel | Engineering estimate | GDE (2012) | 1855 |
| Cryomodule vacuum EDIA | Engineering estimate | RDR (estimate from INFN) (2007) | 1855 |

[†] assuming 80 % first pass yield
[‡] assuming 38.3 % of CM's are tested

#### 15.7.2.1 Klystron

15.7.2.1.1 Klystron Tube Cost information for the klystron is available from direct vendor quotes. The procurement costs for the klystrons acquired for the EXFEL project are also available.

For the TDR, the most current vendor quote has been chosen as the cost basis. The vendor gave quotes for two quantities of klystrons, one quantity lower than required for DKS, and one quantity higher than required for KCS. The two quotes allow the standard (Crawford) learning-curve parameters to be determined[9], which were used to estimate the unit costs corresponding to the number of klystrons to be procured (assuming two vendors) for the KCS and DKS configurations.

The quotes correspond to the procurement of fully processed klystrons. The full cost of labour at the vendor for the processing has been assumed, but the processing infrastructure costs given by the vendor have been discounted by 50 %. The infrastructure costs in the estimate are the costs of modulators and associated controls. However, the project will have access to modulators and controls which can be supplied to the vendor for the processing, substantially reducing the costs.

15.7.2.1.2 Klystron Accessories The vendor quotes correspond to a fully processed tube body, including the focusing magnet, mounting hardware, tube socket, and oil tank. However, there are additional costs associated with the solenoid, filament and ion-pump power supplies, the RF pre-driver, and related controls, software, and interlock systems. These additional costs are included in the TDR, with a unit cost basis taken from the RDR (after adjustment for quantity discounts).

#### 15.7.2.2 Modulator

The baseline modulator for the TDR is a Marx modulator. Cost information for this device is available from vendor quotes and from an engineering estimate carried out at SLAC. The procurement costs for the modulators acquired for the EXFEL project are also available.[10]

Based on the extensive experience at SLAC in developing the Marx modulator, the cost basis for this item is the SLAC engineering estimate. This is a bottoms-up estimate of the material and fabrication costs. This cost is intended to be that of an industrially produced item, so the estimate uses industrial labour rates (derived from 2012 RSMeans[11] contractor labour rate tables) and includes a typical profit margin (15 %). The first article unit costs have been discounted using a 95 % learning curve for materials (which are 80 % of the total first article cost), and a 90 % learning curve for labour (the remaining 20 %). The number of units is based on a two-vendor procurement of the number of modulators needed for the ILC. The average quantity discount from the first article cost is about 31 %.

#### 15.7.2.3 RF-distribution system

The RF-distribution system brings the high-level RF from the klystrons to the power couplers. For the KCS layout, the RF must be transported from the surface to the cryomodules in the tunnel, and then distributed locally to the power couplers. For the DKS layout, only the local power-distribution system is needed. For the KCS system, the components are (by cost) 68 % standard catalogue microwave components, and 32 % specialised devices (CTO's, loads, variable hybrids, and phase shifters).

The procurement costs for the RF-distribution system acquired for the EXFEL project are available. However, these costs cannot be used directly, since increased functionality[12] is needed in the TDR systems.

For both types of distribution system, the cost was estimated based on catalogue unit prices for the standard microwave components, and vendor quotes for the specialised devices. The prices were developed jointly by the engineering teams at SLAC and KEK. The system costs in quantity were then estimated based on a discount derived from a 95 % learning curve, and assuming a two-vendor procurement for each component. The average quantity discount from the single-item unit prices is about 33 %.

[9] The effective learning curve slope for the klystron fabrication is 92%; for the processing labour, it is 87 %.
[10] Note that these devices do not use the Marx modulator technology.
[11] North America's leading supplier of construction cost information
[12] The increased functionality is needed to cope with the cavity gradient spread allowed by the TDR specifications.

The TDR-DKS unit costs are larger than those of the RDR, and larger than the EXFEL procurement costs, because of the increased functionality required by the TDR design requirements. The TDR-KCS unit costs are larger than those of the TDR-DKS because of the additional microwave hardware needed to bring RF from the klystron clusters on the surface to the cryomodules in the tunnel.

#### 15.7.2.4 Supporting Infrastructure in the Tunnel

There are additional costs for the high-level RF system related to cabling, instrumentation, and electrical distribution. The unit Value estimate for these items has been taken from the RDR.[13]

#### 15.7.2.5 High-level RF Explicit Labour

Final engineering design, and sustaining engineering, for the L-band high-level RF systems, is accounted for as explicit Labour. This includes documentation and supervision of bid packages for delivery of tested klystrons, modulators and distribution assemblies; support of factory testing, factory verification of testing, and acceptance to ship; on-site final assembly area development, supervision, and management; systems engineering and documentation for production and final assembly; engineering team development, support of final integration and equipment commissioning in tunnels; and sustaining engineering and technical maintenance of tested systems during commissioning.

The estimate is taken to be that developed for the RDR, but rescaled for the TDR, based on the ratio of M&S costs between the RDR and the TDR.

#### 15.7.2.6 Summary

The basis for the components of the estimates for the high-level RF system is summarized in Table 15.8.

**Table 15.8**
Basis for high-level RF system

| Sub-component | Basis Type | Basis Source (Date) | Number of units (DKS) | Number of units (KCS) |
|---|---|---|---|---|
| Klystrons | Vendor quote* | qualified vendor | 426 | 461 |
| Klystron auxiliary items | Engineering estimate | RDR (2007) | 426 | 461 |
| Modulators | Engineering estimate† | SLAC (2012) | 426 | 461 |
| KCS RF distribution | Catalog prices and vendor quotes‡ | SLAC, KEK (2012) | 0 | 567 |
| Local RF distribution | Catalog prices and vendor quotes‡ | SLAC, KEK (2012) | 616 | 616 |
| Explicit manpower | Engineering estimate | RDR (2007) | | |

* includes processing labour and 50 % of processing infrastructure
† discounted by 30 %, based on a 90 % (labour) and 95 % (materials) learning curve
‡ discounted by about 33 %, based on a 95 % learning curve

[13] The cost per RF unit has been assumed to be the same for KCS and DKS.

## 15.7.3 Conventional Facilities (CFS)

### 15.7.3.1 Introduction

The Value estimate for the ILC CFS has been developed internationally with teams in each of the three regions (Americas, Asia and Europe). These teams have worked closely together to optimise the CFS design, based on the requirements supplied by the Accelerator and Technical Systems. The three estimates have been formatted using the same detailed WBS structure up to level 5 of the WBS. At deeper levels, there are site-specific differences.

Information was drawn from consultant engineers, historical data from other accelerator or similar projects, industry-standard cost-estimating guides, and, where applicable, the scaling of costs from similar systems. In all cases, the estimates reflect a median value for the work based on the criteria provided to date.

The Americas estimate includes a small increase to represent typical cost growth from a design estimate to the "Final Construction Cost". It accounts for Americas-specific features, such as possible claims resulting from design immaturity. However, it does not correspond to contingency. There are no explicit factors for contingency contained in any of the CFS Value estimates.

For each category of the CFS estimate, costs associated with ongoing engineering and documentation are included in the category labeled "Engineering, study work and documentation". Part of this work is done through A&E firms, and this is included in the Value estimates described below. Additional work, which may be done at the ILC laboratory or collaborating institutions, is included as person-hours in the CFS component of the explicit Labour estimate.

### 15.7.3.2 Civil Engineering

Due to the differences in geology and topography at the different sites, separate Value estimates were developed for each of the three sites. These estimates were developed using the same criteria. The drawings for each site reflected necessary site-specific differences.

Costs for activities that take place prior to the construction start are explicitly not included in the estimate. Some examples of such costs are A&E Services before the start of construction, development costs for geotechnical and environmental investigation, land-acquisition costs, and costs incurred for compliance with local governmental statutes and regulations. These costs cannot be accurately identified until a specific site selection is made.

15.7.3.2.1 Underground construction All major elements of the civil engineering, such as tunnels, shafts, caverns, halls, etc. are included in the estimate. The costs have been estimated with the help of consultants, using information from similar projects and standard civil engineering practices. All temporary facilities needed for construction work, as well as the necessary site preparation before start of the work, are included in the Value estimate.

The local geology at the European site near CERN, consisting of stable Molasse rock, permits the creation of underground facilities using standard excavation methods. Tunnel-boring machines (TBM's) will be used to excavate beam tunnels with finished inside diameters varying from 5.2 m to 8 m. Shafts will be constructed using traditional excavation methods for the dry moraines (upper 50 m) and ground freezing techniques for the wet moraines. When the Molasse rock is reached, the shafts will be further excavated using rock breakers and roadheaders. These machines are also used for cavern excavation. No drilling and blasting is required for the European site.

The Americas site uses TBM's to excavate beam tunnels with finished inside diameters of 4.5 m and 5 m. At the Americas site, due to the local hard rock (Dolomite), shaft and cavern excavation requires drilling and blasting.

All Asian underground construction is carried out using drill and blast excavation, following the New Austrian Tunneling Method (NATM).

The different excavation methods used at the three sites result in different unit costs, as discussed below.

15.7.3.2.2 Surface buildings The type, number and dimensions of the buildings include only those surface facilities required for construction, installation and operation of the project, taking into account the specifics of each of the three sample sites. For instance, for the Americas and European sites, additional infrastructure such as seminar rooms, guest houses, restaurants, administrative facilities, warehouses, etc. are assumed to be supplied by a nearby (host) laboratory, and are not included in the Value estimate. The Asian sample site does not have a nearby laboratory so that the Asian estimate does include such central campus facilities. It also includes 3 large warehouses to be used during installation.

For the areas where surface buildings are located (central campus, shaft positions), the following items have been included in the cost estimate:

- fences and gates;
- roads and car parks within fences and from fence to existing road network;
- pedestrian walkways;
- lighting for walkways and around buildings including buried electrical connections;
- all necessary drains along roads and car parks, including sumps, water treatment facilities and connections to existing mains;
- all needed water supply pipes, tanks and connections to existing water-supply networks;
- landscaping and planting of trees, bushes, seeding of grass as required;
- spoil dumps (where applicable) created close to the building areas, including landscaping.

15.7.3.2.3 Tunnel volumes and surface-building areas The total beam- and service-tunnel volumes, and the total surface-building areas, for each regional site, are shown in Table 15.9.

**Table 15.9**
Beam- and service-tunnel volumes, and surface-building areas. The Asian surface building total includes 3 warehouses, each with an area of 6,000 $m^2$.

| Region | Tunnel Volume ($m^3$) | Cavern Volume ($m^3$) | IR Hall Volume ($m^3$) | Surface Building Area ($m^2$) |
|---|---|---|---|---|
| Americas | 904,881 | 133,755 | 135,703 | 74,599 |
| Europe | 1,070,268 | 156,232 | 127,100 | |
| Asia | 2,091,630 | 330,360 | 189,381 | 109,275 |

15.7.3.2.4 Unit costs The CFS group established a set of similar definitions for underground construction unit costs. This ensured consistency across all three regions. Estimates for each unit cost were independently produced by experts and consultant engineering firms in each of the three regions. These unit costs, together with the civil-engineering design details specific to each site, were used to develop the site-specific civil-engineering costs.

*Unit cost definitions* For the European region, underground civil-engineering costs have been estimated for tunnels, shafts and caverns. Excavation is performed using TBM, rock breakers and roadheader machines. The unit prices for these elements include items shown in Table 15.10.

The unit costs for the Asian site always correspond to drill and blast excavation, and include both direct and indirect costs. The direct costs include blast excavation, transport of muck, sprayed concrete finish in the sloping access tunnels, rock bolts, inflow water treatment, a mold frame for lining, a concrete-lining finish in beam/service tunnels, floor concrete finish, survey, grouting (20 % of direct costs), and construction support. Indirect costs are 100 % of the direct costs. Site preparation costs (8 % of total costs) are contained in a separate WBS element.

**Table 15.10**
European civil-estimate unit costs. This table shows what is included in the civil engineering costs for the European estimate. The percentages given are of the "direct" costs, defined as the sum of all costs without a percentage after them.

| Item | Tunnels | Shafts | Caverns |
|---|---|---|---|
| Manpower (no overhead)† | X | X | X |
| Excavation & deposit | X | X | X |
| Outer lining incudes: | | | |
| Bolts | | X | X |
| Shotcrete with fibre | | X | X |
| Ring gap/tolerance filling | X | X | X |
| Sealing | | X | X |
| Inner lining includes: | | | |
| Hydroshield | X | X | X |
| Double-sided formwork | | X | X |
| reinforcement | X | X | X |
| Ceiling | | | X |
| Finishing (floor) | X | | |
| Special works includes | | | |
| Moraines | | X | |
| Walls ceilings (10 %) | | | X |
| Drainage & dewatering (4 %)‡ | X | X | X |
| Probes (2 %)‡ | X | X | X |
| Auxiliary measures (15 %)‡ | X | X | X |
| Installation $(25 - 35\,\%)$‡ | X | X | X |
| Overhead $(\sim 16\,\%)$‡ | X | X | X |

† Manpower includes 4 shifts working 24 h/day, 7 days/week and 320 days/yr

‡ These percentages are different between the Americas and European regions

The Americas unit costs include the direct and indirect costs for the shaft, tunnels and cavern excavation. Unit costs have been developed from both parametric scaling of excavations in the Chicago and Milwaukee areas, and bottoms-up estimates for items for which the parametric data is not available. Shaft costs include soil rock interface grouting, earth retention for the overlaying soft-ground excavation and drill and blast excavation in the hard rock. Shafts costs also include feature grouting and concrete lining. Tunnel unit costs include the excavation and mucking of the hard rock, feature grouting, invert and lining concrete. Caverns and hall unit costs include the drill-and-blast excavation, mucking of spoils, feature grouting, invert slab and shotcrete lining. Rock bolts or rock dowels will be required for permanent rock support for the shafts, tunnels, caverns and halls.

*Vertical shafts and horizontal sloping access tunnels* For the same cross-sectional area, the unit costs for the horizontal sloping tunnels used for access in the Asian site are much lower than the vertical shaft unit costs used in the other regions. The vertical shafts are generally more expensive because of more expensive surface construction support, and the need for more excavation time per unit length (since, for example, transportation or movement of the excavation machine is limited).

The vertical-shaft unit cost is higher for the Americas than in Europe because the excavation methods used for the Americas site are more complex. For the Americas site, two different means and methods are required to excavate the softer overlaying soils and the harder rock below. The soils require an earth retention system with a grout curtain installed to control water inflow, and the harder rock requires drill and blast equipment. As noted above, in the European region, no drilling and blasting is required. In addition, underground potable water and sanitary sewers are not included in the civil unit costs in the European region. Furthermore, both regions use different overheads percentages. In the Americas region, having two different construction means and methods results in higher overhead costs.

*Beam and service tunnels* Tunneling units costs used for American and European estimates are based on the use of TBM's, while the Asian estimate is based on the NATM drill-and-blast technique. For the same cross-sectional area, the NATM costs in the Asian region are lower than the TBM costs in the other regions. Part of this difference is related to labour cost differences between the regions.

The unit cost for the large "Kamaboko" tunnel in the Asian site is about 50 % greater than for the smaller $\sim 5\,\mathrm{m}$ Main Linac tunnels in the Americas and European regions. The unit cost difference increases to more than a factor of 2, however, when the cost of the shield wall is included.

*Caverns* Cavern unit costs are generally lower for the European region than for the Americas, for the same reasons as for the shaft unit prices, namely a difference in excavation methods and overheads.

The Asian cavern unit costs are generally considerably lower than those at the American site, for the same cross section, despite the fact that both use drill and blast excavation method. Part of this difference is related to labour-cost differences between the two regions.

*Halls* The IR hall unit costs are lowest for the Asian region and highest for the Americas region. Unit hall prices are lower for the European region than for the Americas, for the same reasons as for the shaft and cavern unit prices, namely a difference in excavation methods and division of overhead percentages. One of the reasons for the lower Asian unit costs is related to regional labour-cost differences.

15.7.3.2.5 Total costs In general, the distribution of total costs among tunnels, shafts, caverns and halls for the American and European sites is similar, as expected given the similarity in their designs. The beam-tunneling unit costs are similar for the two regions, as is the total cost for beam tunnels. Although the European region has fewer caverns, the volumes are larger, so that, despite the lower cavern unit costs in Europe, the overall cavern prices for both regions are similar.

The total shaft costs are significantly higher in the Americas estimate. This is due in part to the higher vertical shaft unit costs for the Americas region. In addition, there is a difference in shaft bottom definition between the two regions. The European region delineates the length of a shaft from the surface to the top of the connecting cavern, while the Americas region defines the shaft length from the surface to the bottom of the connecting cavern. This makes the Americas shafts 30% longer than the European shafts (130 m versus 100 m, respectively).

Relative to the flat topography sites, the beam tunnels are a larger fraction of the total civil engineering cost for the Asian site, because of the larger cross section of the "Kamaboko" tunnel, and consequent higher unit costs. The horizontal sloping access tunnels are a smaller fraction of the total, due to low unit costs for horizontal access tunnel excavation. Similarly, although the cavern volumes are larger for the Asian site, the cavern-excavation unit costs are much lower, leading to an overall cavern cost similar to that of the flat topography sites.

Since it is a "greenfield" site, the costs for site development are higher for the Asian site than for the other sites. The "greenfield" Asian site also contains additional surface structures associated with the central lab, office buildings, user facilities, and warehouses. However, there are fewer service buildings required for the mountainous region site, so the overall surface structure costs are comparable to those of the flat topography sites, which are located close to existing labs.

### 15.7.3.3 Conventional Electrical Systems

Conventional electrical systems include high- and low-voltage equipment and power distribution networks, emergency power sources, and communications and power network monitoring equipment.

The machine designs for the flat and mountainous sites utilise different distribution systems for high-level RF and have different layout configurations for electrical systems. Consequently, separate Value estimates for the conventional electrical systems for the Americas and Asian sites were developed by consultant engineering firms. These estimates, which were based on the RDR electrical requirements, were modified as needed and adjusted to correspond to the TDR design power requirement for each site.

The final Americas (Asian) electrical system Value estimate is based on a total nominal operational power requirement of 161 (164) MW.

The differences in the distribution of costs between the estimates for the two regions reflect the different equipment layouts (e.g. high-voltage distribution in tunnels or on the surface) and differences in system redundancies for the two types of sites. The details are documented in Chapter 11. Based on the design similarities between the flat topography sites, the Americas estimate for the conventional electrical systems was used for the European site.

#### 15.7.3.4 Conventional Mechanical Systems

Conventional mechanical systems include HVAC equipment, piped utilities (sump systems and fire suppression systems), and primary and secondary process (cooling) water systems.

The different topographies of the flat and mountainous sites result in substantially different solutions and costs for conventional mechanical systems. Again, consultant engineering firms were used to develop Value estimates for the conventional mechanical systems for both the Americas and Asian sites. These estimates, which were based on the RDR cooling requirements, were modified as needed and adjusted to correspond to the TDR design power requirement for each site.

Part of the difference in conventional mechanical system costs between the Asian and the Americas sites is due to the different main-linac high-level-RF configurations (KCS for the Americas, versus DKS for Asia). The KCS configuration places the majority of the heat loads at the surface; these are less expensive to manage than the DKS heat loads, which are in the tunnel.

Another difference is due to the different cooling approaches used at the two sites. The Americas region utilises a single water system in the main linac, which is a single-process low-conductivity water (LCW) system serving the main-linac tunnel loads. The Asian site follows the RDR approach of having two systems (both process-LCW and chilled-water systems) serving the tunnel loads.

Finally, for the KCS configuration, water-cooling systems for surface cryo plants are simplified because of the close proximity of the plants to the cooling towers. For the DKS configuration, the cryogenic plants are in the underground caverns.

Since the conventional mechanical systems design is similar for the Americas and European sites, the Americas estimate has been used for the European site. An important difference between the two regions is the ventilation scheme adopted by CERN, which is chosen to be transversal, mainly to satisfy regional safety regulations and for temperature stabilisation. A costing study has been performed by CERN for the transversal ventilation system for the Compact Linear Collider at CERN (CLIC) machine, but not for the ILC machine. It is expected that the costs found in this CLIC study are higher than would be required for the ILC, due to the higher CLIC heat loads, and the tighter temperature stabilisation requirements. However, a realistic comparison is not possible at this stage due to the different machine designs.

#### 15.7.3.5 Handling Equipment

Handling equipment estimates were made both for installed equipment such as overhead traveling cranes and elevators, and for mobile equipment such as the special vehicles used for installation in the tunnel. The mobile equipment estimates are included under Installation (Section 15.7.4). The estimate is based on the European site layout. It was used for the Asian site without modification. For the Americas site, the only change was an increase in the number of lifts in the shafts, which is required in the Americas design.

The handling-equipment Value estimates for installed equipment (overhead traveling cranes, elevators and hoists) are based on European-supplier cost information provided at the time of the RDR. The Value estimate for the rented gantry for detector lowering is extrapolated from contract costs for the rented gantry used to lower the LHC CMS experiment at CERN.

Manpower estimates for installed equipment cover the engineering activities such as finalisation of requirements, agreement of interfaces with other infrastructure groups, specification, tendering, contract management, installation organisation and supervision, commissioning and maintenance management during the installation period. The external contractor support is taken as 2 % of purchase costs. The internal manpower estimate is extrapolated from LHC experience, based on crane and elevator quantities, and then shared among the area systems in proportion to purchase costs.

#### 15.7.3.6 Safety Equipment

Safety equipment includes primarily alarm systems. The estimate was made independently for the Americas and Asian sites, by experts at Fermilab and KEK, and was based on the regional-code requirements that dictate the requirements of the safety system. No detailed studies have been performed specifically for the European ILC site. For this reason, the Americas estimate was used for the European site.

#### 15.7.3.7 Survey and Alignment

Survey and alignment covers a very broad spectrum of activities, starting from the conceptual design of the project, through the commissioning of the machines, to the end of operations. The Value estimate developed covers the work necessary until successful completion of the machine installation. It includes equipment needed for the tasks to be performed. However, it does not include equipment for a dedicated calibration facility and workshops. It does include the staff that undertake the field work, and the temporary manpower for the workshops. Full-time staff are considered to be mainly dedicated to organisational, management, quality control, and special alignment tasks.

The Value estimate is mostly based on scaling the equivalent costs of the LHC to the ILC scope. The estimate was made for the European site and was used for the Americas and Asian sites without modification.

### 15.7.4 Installation

The installation estimate developed for the RDR was revised for the TDR. Because of the differences in tunnel layout and RF-component placement between the flat and mountainous sites, two different installation estimates were prepared, one for each site. However, the cost basis and methodology for each estimate was essentially the same as that for the RDR. The major changes were in areas for which the TDR design differs substantially from that of the RDR (main linacs with the KCS HLRF configuration, and the damping rings).

The installation Value estimate is characterised almost exclusively as explicit Labour, with minimum costs for material-handling equipment. The reason for this is that part of the installation and system check-out labour at the ILC site could be contributed by the staff of the ILC laboratory, or of collaborating institutions or laboratories. The degree to which laboratory staff could contribute depends on the availability of the necessary skilled manpower and local labour regulations. Because of the size of the project, it is likely that many tasks like electrical and plumbing work will need to be outsourced to industry. Trade-offs and translations are likely between using in-house labour and external contracts. Since the details of these trade-offs are not known, the installation manpower has been accounted for entirely as explicit Labour. It is estimated that a minimum of 10 % of the installation task must be management and supervision by in-house manpower.

The Value estimate was based on scaled information from a variety of sources, including the actual manpower used for the installation of recent accelerator projects. The installation cost model used a work-breakdown structure (WBS) that listed all of the activities required for installation of the technical systems, including the management, planning, and engineering support. The WBS was broken down into two major level-of-effort categories: General Installation and Accelerator-Systems Installation. General Installation included all common activities and preparations and associated logistics on the surface. Accelerator-system Installation included all efforts required for complete installation of the components underground.

To populate the WBS, a comprehensive list of components was compiled and interfaces and boundaries with the technical systems carefully defined. The estimates for labour and equipment required to install the components came from a wide variety of sources. For conventional components, like beam pipes and magnets, the technical system leaders provided estimates, based on experience with other projects. Visits to CERN and DESY provided data on installation of cryomodules, LHC magnets and the CMS detector as well as the opportunity to observe actual installation procedures. RSMeans 2006 cost data was used in estimating total work-hours needed for installing equivalent size/weight equipment under similar conditions. Since the main linac is a major cost driver, the installation of cryomodules and RF sources was modelled in detail. For other systems where there was no appropriate experience base, the estimates were scaled from similar installation tasks based on an assessment of complexity.

The Value estimate for mobile equipment for installation was built up by identifying suitable technical solutions for cryomodule and RF installation with allowances for powering and guidance infrastructure. The number of convoys was estimated based on the time available for installation, distances, speeds and estimated times for loading and unloading. An estimate for ad-hoc solutions for load interfaces to allow installation of other equipment was added. The cost of the individual vehicles is based on European costs for similar equipment purchased by CERN. Manpower estimates for mobile-equipment engineering activities were included.

At the time of the RDR, the estimates were reviewed by experts and crosschecked. The estimates were also compared with individual estimates from other sources, and with the actual manpower used for the installation of recent accelerator projects. There was also a bottoms-up study for installation of the cryomodules for the main linac done by two separate engineering teams, with comparable results.

Scientists, engineers and administrators comprise approximately 15 % of the installation manpower; the remainder is technician manpower.

### 15.7.5 Cryogenic Systems

Due to the different topologies of the flat and mountainous sites, which necessitate a different cryogenic layout, a separate Value estimate was developed for each site. However, the same cost basis was used for each. It was assumed that there was no difference in cost due to the fact that the cryoplants must be installed in underground caverns in the mountainous site.

The cost-estimating relationship for a cryogenic plant is a non-linear parametric relation between cost and plant power. For cryogenic distribution systems, linear relations between cost and length were used. The parameters in these relations are based on experience in procurement of cryogenic plants and distribution equipment at Fermilab, CERN, DESY, and other laboratories.

For the main linac plants, the cost basis is the same as that used for the RDR. This basis was independently validated through a comparison with experience in cryogenic-plant procurements at Jefferson Lab in 2010.

For the smaller plants used by the injectors and the damping rings, the cost basis was derived from recent procurement costs of the plant at Fermilab's New Muon Laboratory.

The explicit Labour estimate is based on the experience of staffing levels for cryogenic systems at the Superconducting Super Collider (SSC).

### 15.7.6 Magnet and Magnet Power-Supply Systems

Except for the damping rings, the unit-cost basis for all magnets and their power supplies is the same as that used for the RDR. The component counts for each area system were updated based on the TDR design.

The unit Value estimates were based on conceptual designs for magnets, power systems, stands and movers, with additional assumptions about estimated costs of material and labour. Given time and resource limitations, detailed conceptual designs were developed for only a small number of the magnet styles. The majority of the estimates are engineering estimates based on existing designs with similar requirements. Standardised labour rates were determined from laboratory and industrial sources in the Americas region.

In order to determine the material costs, the weights of magnet and cable materials, primarily copper and iron, were estimated and summed, and based on world commodity prices obtained at the time of the RDR (2007). Similarly, prices were obtained at the time of the RDR for commercially available electronic components such as power supplies, FPGAs and PLCs, controllers and Ethernet interfaces. These prices were escalated to the TDR estimate date (January, 2012) based on escalation rates for manufactured items. While this procedure is generally reliable for typical manufactured items, it may have limited accuracy for items whose cost is dominated by volatile commodity prices.

At the time of the RDR, a design and a complete set of drawings was developed for a positron-source transfer-line quadrupole, and a request for quote sent to a number of magnet vendors. The vendor quotes obtained were in reasonable agreement with an internal estimate: the average agreed within a few percent of the internal estimate, with a spread of $\sim 25\,\%$.

For a few magnet systems, more detailed Value estimates were provided based on R&D prototypes already in progress at the time of the RDR (e.g. the Daresbury/Rutherford undulators); in a similar fashion, Brookhaven provided detailed Value estimates for the superconducting insertion magnets at the IR based on experience with similar magnet designs.

Estimates of Engineering, Design, Inspection and Acceptance (EDIA) labour costs were based upon reviews of recent large accelerator magnet and power-supply projects at SLAC and Fermilab, where the materials, fabrication and EDIA labour fractions are well known. The fractional distribution of EDIA among several types of labourers, which were estimated at the standardised labour rates, was assigned on the basis of project-management experience.

Because of the major changes made to the design of the damping rings, for the TDR, new magnet unit Value estimates were developed. A new distributed power-supply system was designed and its cost was estimated for the TDR. The Value estimates were produced by two of the same engineers involved with the RDR estimates. The same general description given above for the RDR applies to the new TDR damping-ring magnet and power-supply cost bases.

The engineering and fabrication experience for the CESR-c wigglers were used to provide reliable Value estimates for the ILC damping wigglers, taking proper account of the well-defined differences in specification. New estimates were made for the TDR, since the damping wiggler design was modified from that in the RDR.

Costs for the damping-ring kicker pulser were based on a commercially available pulser (a fast ionisation dynistor, or FID, device) that comes close to meeting the specifications for the damping-ring injection/extraction kickers; this cost dominates the total cost of the injection/extraction system. Other components, including the strip-line electrodes and the septa, are relatively conventional, and costs were based on similar existing devices. The unit Value estimates developed for the RDR were taken for the TDR.

The costs of the main-linac quadrupoles and correctors are included with that of the cryomodules. The costs of the quadrupole power supplies are included here.

### 15.7.7 Vacuum Systems

Except for the damping rings, the unit-cost basis for all vacuum systems is the same as that used for the RDR. The component counts for each area system were updated based on the TDR design.

The main parts of the vacuum systems were obtained from quotations from vendors and from recent large-quantity procurements. "Consumables," such as flanges, gaskets, bolts and nuts, cables, etc, were either not included or were estimated for quantity discounts of catalog items.

For the damping rings, since the TDR vacuum system is considerably different from that specified for the RDR, new vacuum-system Value estimates were developed. These include the costs of the surface treatments and antechamber designs, which are required to mitigate the electron-cloud effect in the positron ring. The Value estimate is based on estimates made for the Super-KEKB positron ring, which has a very similar vacuum system. The cryomodule vacuum system cost is included with that of the cryomodules.

### 15.7.8 Instrumentation

The unit-cost basis for all instrumentation systems is the same as that used for the RDR. The component counts for each area system were were taken to be the same as in the RDR, except for the positron source[14], for which they were updated based on the TDR design. The cost of the BPM's in the cryomodules is included under Instrumentation.

For beam monitors the Instrumentation Value estimate covers:

- all pickup stations, as part of the vacuum system;
- scintillators, PMTs, laser systems, calibration systems;
- RF systems and infrastructure for the DMC-based bunch-length monitors;
- associated motors, switches, and mechanical set up;
- signal and control cables, connectors, patch cables, etc.;
- dedicated read-out electronics (analog & digital), control units, local timing electronics, calibration electronics, local software and firmware.

Except for special cases, e.g. certain feedback systems, data-acquisition infrastructure is covered by the control-system Value estimate.

For costing purposes, instrumentation was classified into 17 different *systems*. Core cost and manpower information was estimated for each individual component of an instrumentation system and its subcomponents, including the cost reductions due to volume and/or technology advances. No spares were included. Counts of control racks required for data acquisition were generated from the above data. Labour information (in person-years) was estimated separately for Prototyping, Testing and Installation. The Installation labour was then incorporated into the Installation estimate and not included in Instrumentation.

The cost for the S-band dipole-mode structures, used for bunch-length measurements, was developed specifically for the RTML. The RDR estimate, which was used for the TDR, was based on recent experience with accelerator construction at IHEP.

[14] The redesign of the damping rings should also result in some changes to the RDR Instrumentation estimate for this system, but these changes are expected to be small and were ignored.

### 15.7.9 Dumps and Collimators

The unit-cost basis for all dumps and collimators is the same as that used for the RDR. The component counts for each system were updated based on the TDR design.

The systems that put water into direct contact with the beam dominate the Value estimate of this technical system. For the main beam dumps, the Value estimate is based on industrial studies [257, 258] by two German companies expert in nuclear reactor technology. At the time of the RDR, their estimates were examined by the staff responsible for the ISIS neutron-spallation target and adjusted, for example, to add the costs of the remote-controlled window-replacement system and air drying systems. For the aluminium-ball dumps that do not operate at high pressure, the cost of the 2006 ISIS target-cooling system was used as the basis of estimate.

Items with peripheral cooling supplied by the tunnel LCW system have only mechanical design and construction costs. Whether for collimators or solid dumps, these costs are estimated based on the production costs of similar devices in use at SLAC.

### 15.7.10 Integrated Controls and Low-Level RF (LLRF)

The scope of the Integrated Controls and LLRF system includes:

- global control system hardware and software;
- central computers for the accelerator control system;
- control-system databases;
- control-system network infrastructure;
- control-system front-end electronics and cabling;
- LLRF electronics and cabling;
- Personnel-Protection-System and Machine-Protection-System logic; and
- 5-Hz-feedback infrastructure.

The unit-cost basis for all controls systems is the same as that used for the RDR. The component counts were taken to be the same as in the RDR[15]. An inherent assumption is that the control-system hardware model can be implemented largely using COTS equipment.

Manpower estimates were developed top down, using assumptions about the level of effort required to implement a control system for ILC, and, at the time of the RDR, were compared with levels of effort from recent accelerator projects. It is assumed that the ILC control-system software framework is founded on an existing framework, rather than developing a new framework from the ground up. Assumptions were made on the level of extra effort needed to implement high-availability control-system hardware and software.

Materials and Services Value estimates were derived from a bottom-up assessment of the controls requirements from each accelerator and technical system. Costs for computing infrastructure (servers, networking, storage) were based on current commodity-computing vendor prices, with an inherent assumption that technology advances will bring commodity computing to the level of performance required for the ILC by the time of project construction. Estimates for the distribution of the RF phase reference were developed from a reference design and vendor quotes. Estimates for ATCA front-end electronics were based on technically comparable components in other electronics platforms since, at the time of the RDR, equivalent components were not yet available (or at least not in quantity) for ATCA.

[15] This is not precisely correct, especially for the positron source and the damping rings, but the error in the TDR estimate resulting from this approximation is expected to be much less than the uncertainty in the estimate.

Costs of the ILC damping-ring fast-feedback systems were taken directly from comparable systems in existing machines. Power amplifiers dominate the cost of the fast-feedback systems. Amplifiers operating in the appropriate parameter regime are available commercially, and costs for these were obtained from an experienced manufacturer.

### 15.7.11 Computing Infrastructure

Computing infrastructure costs required to operate the facility were taken from the RDR estimate. They include business computing facilities and software, a core campus network with associated software, central computing services, a computer security system, and engineering software.

The IT infrastructure estimates were based on actual costs for building and running IT infrastructure at Fermilab, assuming that an ILC laboratory requires equivalent functionality at approximately the same scale.

### 15.7.12 Other High-Level RF

This item refers to all systems generating RF power except the L-band systems. Specifically, this includes the warm sub-harmonic bunching system in the electron source, the warm high-level RF systems in the positron source, and the 650 MHz RF systems in the damping rings.

For the sources, the unit costs for the RDR were based on engineering estimates from warm RF experts at SLAC. For the TDR, the same unit-cost basis is used. The total costs were adjusted based on the TDR design requirements and component counts.

For the damping rings, the RF system is CW and operates at 650 MHz, a different frequency from the RF systems used elsewhere in the ILC. The designs of high-power RF components, such as klystrons and circulators, were scaled from commercially available 500 MHz devices. Estimates from klystron manufacturers indicated that development costs would increase the total cost by roughly the cost of one additional unit at the standard catalogue price.

The TDR estimate for this system is based on the unit Value estimates developed for the RDR. The TDR estimate was derived from these unit costs using the RF component counts associated with the TDR design.

### 15.7.13 Accelerator-Area-specific cost bases

In the following subsections, the cost basis for items specific to an accelerator system are described. For all systems, the Labour estimate includes the EDIA resources required for the system-specific Value elements (if any), together with the staff required for overall accelerator system integration during project construction and hardware commissioning. Except where noted, all estimates are those made at the time of the RDR.

#### 15.7.13.1 Electron Source

The costs for the following items were estimated specifically for the electron source: the laser systems, the polarised-electron guns, the sub-harmonic bunchers and the travelling-wave bunchers. The costs were engineering estimates made at SLAC at the time of the RDR based on experience with polarised electron sources.

#### 15.7.13.2 Positron Source

The costs for the following items were estimated specifically for the positron source: the positron-production target and its housing, the optical-matching device, the standing-wave and travelling-wave warm accelerators, the auxiliary source, and the target remote-handling system. Except for the last two items, the estimates were taken from engineering estimates made at the time of the RDR by engineers at SLAC.

The estimate for the new TDR auxiliary source was developed at ANL. This is an engineering estimate, based on experience with electron sources at that laboratory. The engineering estimate for the target remote handling is also new for the TDR, and is based on experience with remote handling of similar systems at IHEP.

#### 15.7.13.3 Damping Rings

The cost of the cavities and cryomodules for the CW 650-MHz system were estimated specifically for the damping ring. The TDR estimate is based on the unit Value estimates developed for the RDR by engineers from INFN. Manufacturing costs for the cavities and cryomodules were assumed to be the same as for commercial versions of 500 MHz systems developed at Cornell and KEKB, with increased engineering effort to account for the rescaling, or in some cases redesign, of the existing subcomponents. The TDR estimate was derived from these unit costs using the cryomodule-component counts associated with the TDR design. For the TDR, the EDIA estimate for the cavities and cryomodules was taken to be 10 % of the M&S costs.

### 15.7.14 Management and Administration

As for the RDR[16], the model for management and administration staff is based on 50 % of the actual staffing levels during the construction phase (March, 1992) of the Superconducting Super Collider (SSC), but without central computing staff, which are included in Computing Infrastructure. A detailed breakdown is given in Table 15.11.

**Table 15.11** Composition of the management model at ILC. The numbers indicate the percentage of the total management and administration manpower associated with that function.

| Unit | Percent of total | Responsibilities |
|---|---|---|
| Directorate | 10 | Director's Office, Planning, ES&H Oversight, Legal, External Affairs, Education, International Coordination, Technology Transfer |
| Management Division | 4 | Quality Assurance, ES&H |
| Laboratory Technical Services | 42 | Facilities Services, Engineering Support, Material and Logistical Services, Laboratory Fabrication Shops, Staff Services |
| Administrative Services | 32 | Personnel, Finance, Procurement, Minority Affairs |
| Project-Management Division | 11 | Management, Administrative, Project-Management Division Office |

The total management manpower is estimated by taking the FTE count for eight years, and converting to hours using 1700 hrs/yr. The eight-year duration assumes a two-year linear staffing ramp-up at the start of the nine-year construction project, followed by seven years at full staffing levels.

It is the practice in some regions to apply general and administrative overheads to purchases and labour for projects. These overheads are applied as a multiplier on the underlying Labour and Value, and cover the costs of the behind-the-scenes support personnel. In this estimate, such personnel are explicitly enumerated as labour under Directorate, Management Division, Laboratory Technical Service, and Administrative Services in Table 15.11. Therefore, the overheads are included as additional explicit Labour, rather than as a multiplier on Value and technical Labour.

[16] The total FTE count is reduced slightly from the RDR estimate due to a correction to the SSC project management office staff.

It should be noted that this model for management and administrative staff is based on a project which was centrally managed. The ILC project, with its strong in-kind-contribution character, may require additional management or administrative staffing located centrally or in collaborating regions. These additional resources, if any, are difficult to estimate without a specific in-kind model, and so have not been included in this estimate.

### 15.7.15 Summary

The cost bases for the Value and Labour estimates developed for the ILC TDR have been presented and discussed in the previous sections.

A breakdown of the Value estimate, by type of cost basis, is shown in Fig. 15.4. A breakdown of the Labour estimate, by type of cost basis, is shown in Fig. 15.5.

**Figure 15.4**
Breakdown of the ILC TDR Value estimate, by cost basis type

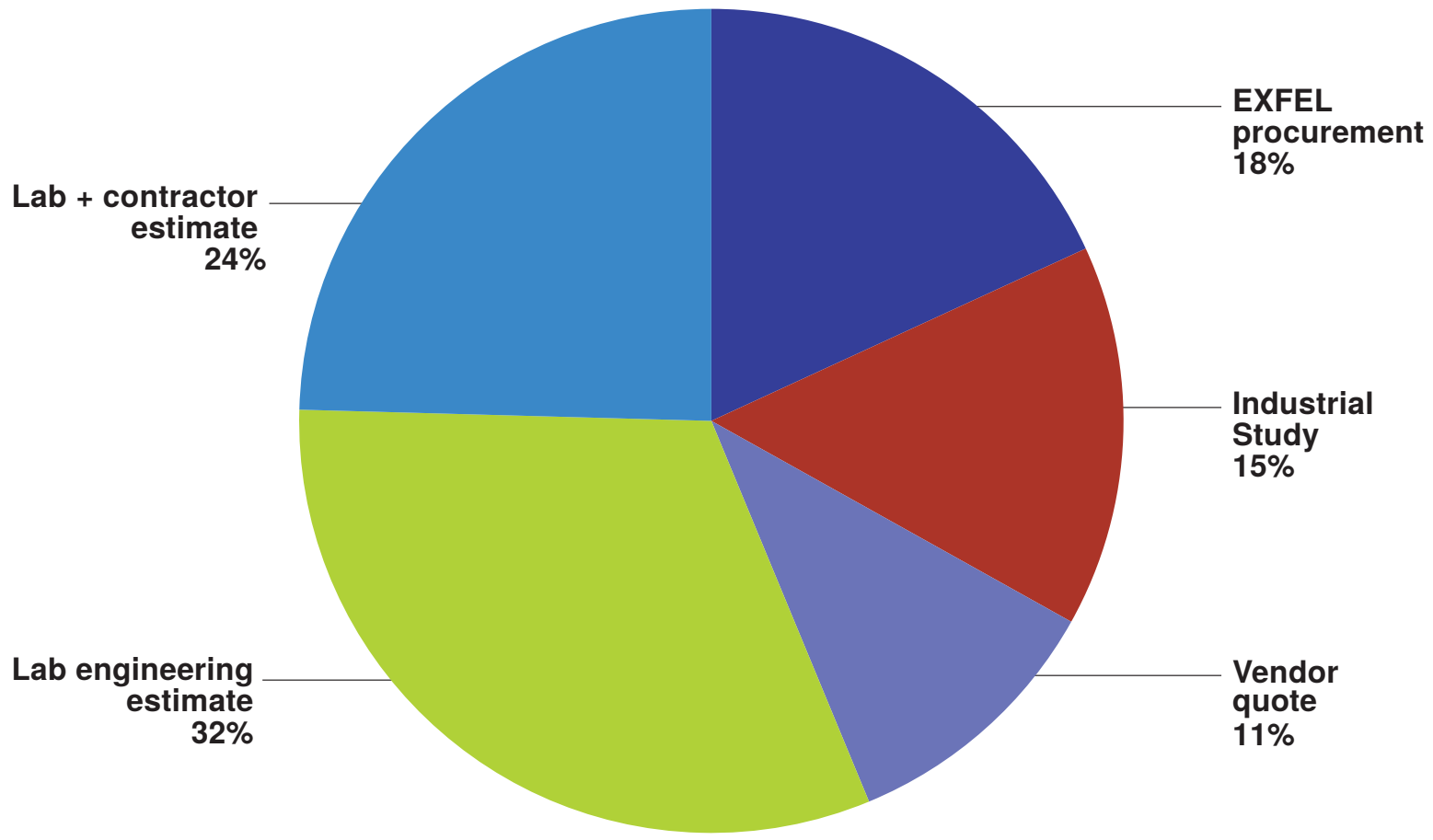

**Figure 15.5**
Breakdown of the ILC TDR Labour estimate, by cost basis type

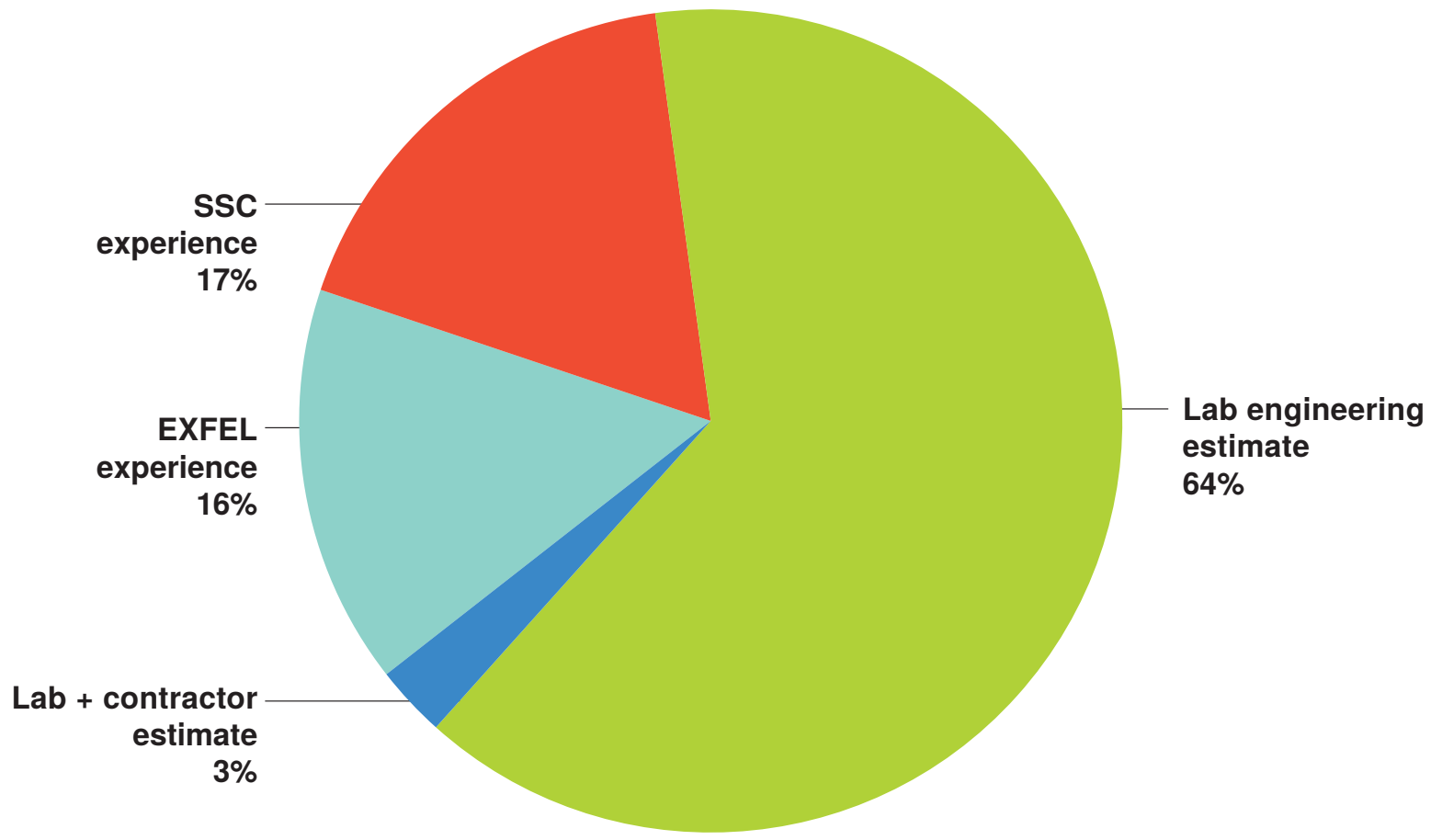

## 15.8 Value and Labour Estimates for the Construction of the ILC

### 15.8.1 Escalation and re-statement of the RDR

As discussed in Section 15.4.2.4 above, the ILCU for the TDR is defined as equal to the USD on January 1, 2012. Conversions of estimates obtained in currencies other than USD to ILCU are based on PPP indices (as of January 1, 2012) relating those currencies to the USD, except for the cavity superconducting material, for which exchange rates are used. With this definition of the ILCU for the TDR, the RDR estimate can be re-stated in these units by escalating the elements of the RDR Value estimate from their original date to 2012, based on the regional escalation indices shown in Fig. 15.1, and converting to 2012 ILCU using the PPP indices shown in Fig. 15.2.

The resulting breakdown of the escalated RDR, in units of 2012 ILCU, is shown in Fig. 15.6. The cost breakdown categories are the same as those presented in Section 15.7. The total Value for the escalated RDR is 7,266 MILCU. This would be the TDR estimate if the TDR design and cost basis were identical to that of the RDR.

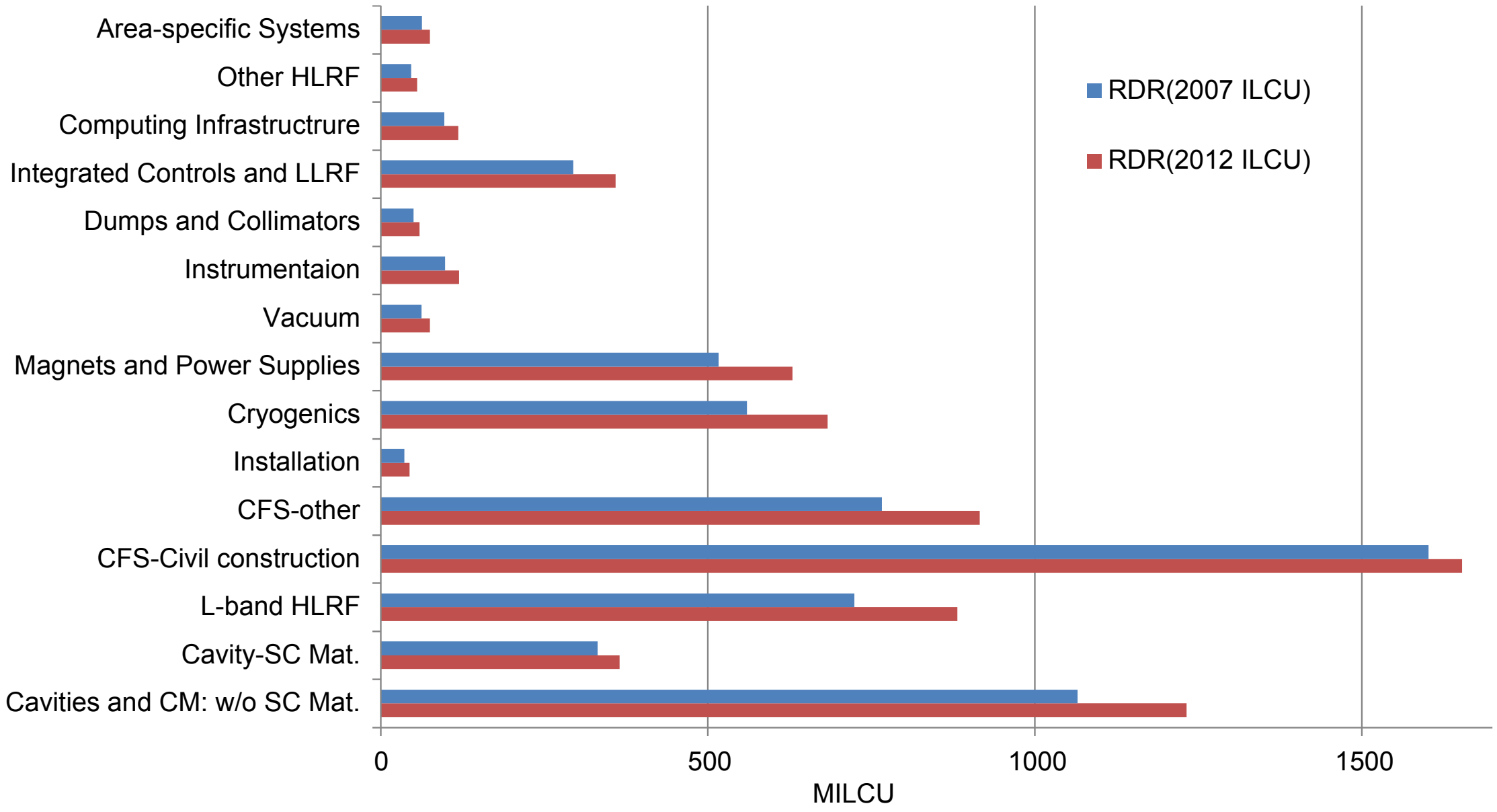

**Figure 15.6.** RDR Value escalation from 2007 to 2012. The total escalated RDR Value (red bars) is 7,266 (2012 MILCU). Also shown (blue bars) is the RDR stated in terms of an ILCU based on 2007 PPP indices. In these units, the RDR Value is 6,312 MILCU. The ratio between the RDR in 2012 ILCU (7,266) and in 2007 ILCU (6,312) is 1.15, which is the inflation rate (in USD, averaged over the project cost element types) over the period from 2007 to 2012.

### 15.8.2 Value Estimate for the TDR

The Value estimate for the cost of the ILC design as presented in this *Technical Design Report*, averaged over the three regional sites, is 7,780 MILCU. This may be compared with the escalated RDR estimate of 7,266 MILCU.

The cost optimisation of the machine design discussed in Section 15.6.1 resulted in a cost decrease of approximately 9% in the total project cost. The TDR estimate for the fabrication of cavities and cryomodules, which is based on extensive experience not available at the time of the RDR, increased relative to the RDR estimate by about 16% of the total project cost. The net overall effect (after correction for inflation) is a cost growth of approximately 7% from the RDR to the TDR.

The breakdown of the Value estimate for the TDR, in units of 2012 ILCU, is shown in Fig. 15.7. The cost breakdown categories are the same as those presented in Section 15.7. All estimates have been averaged over the three regional sites. For comparison, the escalated RDR is also shown. The superconducting RF components, including their cryogenic systems and RF-power systems, represent about 76% of the estimate for all non-CFS components.

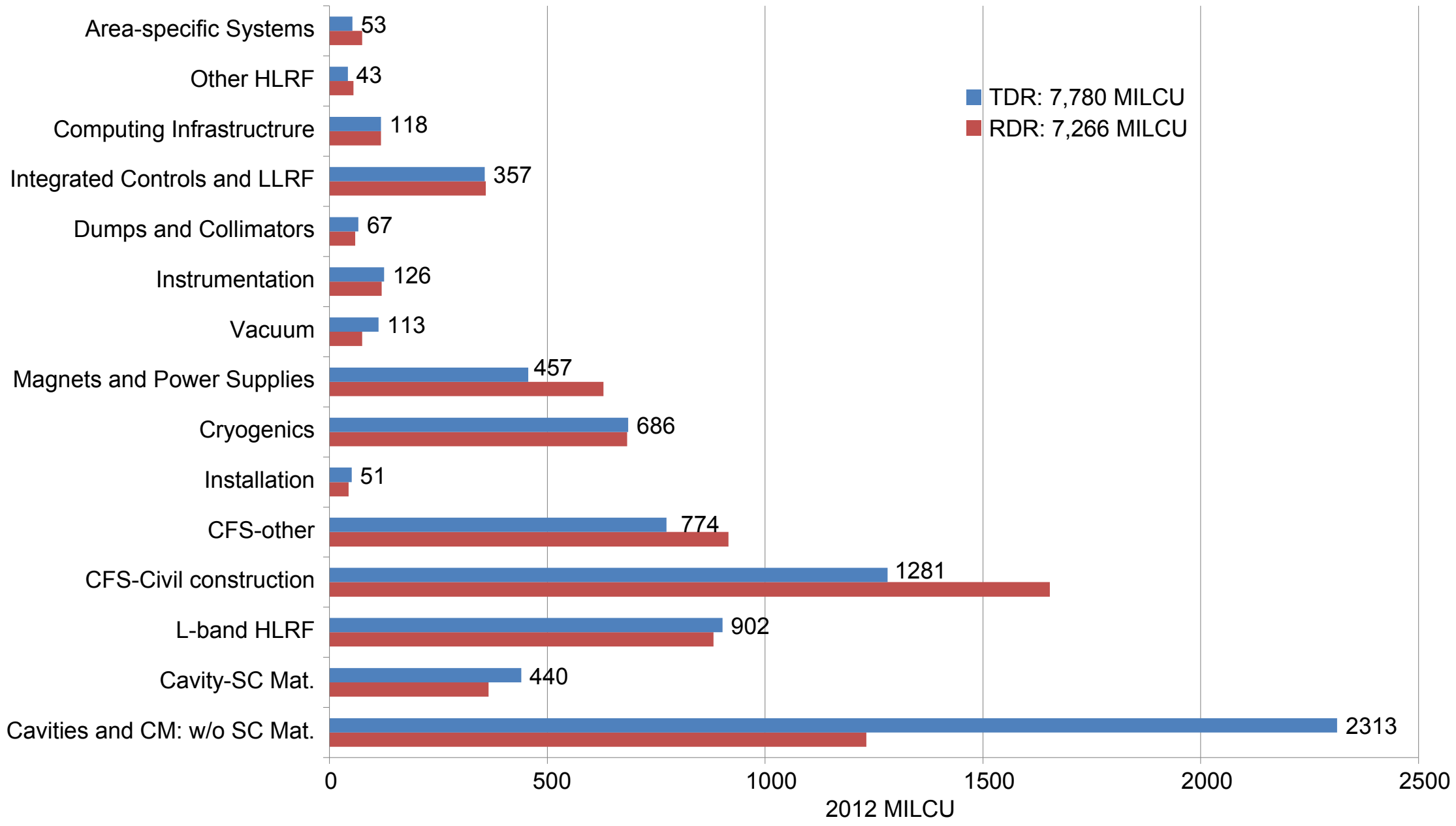

**Figure 15.7.** TDR Value estimate by technical system. Also shown for comparison is the escalated RDR. The numbers give the TDR estimate for each system in MILCU.

The Value estimates broken down by Area (Accelerator) System are shown separately for both the conventional facilities and the components in Fig. 15.8. The system labeled "Common" refers to infrastructure elements such as computing infrastructure, high-voltage transmission lines and main substation, common control system, general installation equipment, site-wide alignment monuments, temporary construction utilities, soil borings and site characterisation, safety systems and communications.

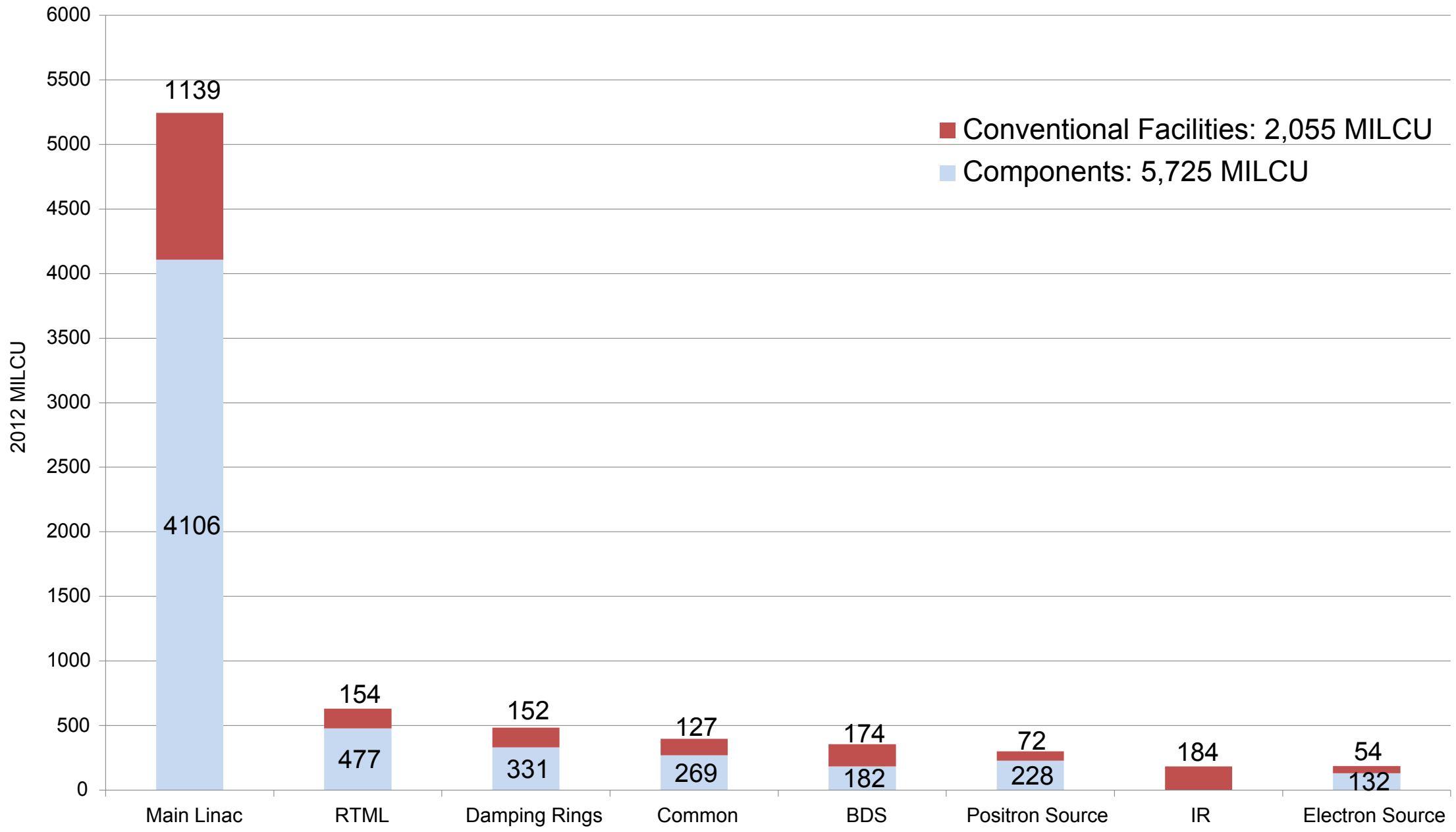

**Figure 15.8.** Distribution of the ILC value estimate by system and common infrastructure, in ILC Units. The numbers give the TDR estimate for each system in MILCU.

The component value estimates for each of the Accelerator Systems include their respective RF

sources and cryomodules, cryogenics, magnets and power supplies, vacuum system, beam stops and collimators, controls, low-level RF, instrumentation, installation, etc. The main linac comprises about 67% of the total project Value.

### 15.8.3 Explicit Labour Estimate for the TDR

The explicit Labour for the technical systems, and specific specialty items for Electron Source, Positron Source, Damping Rings, and Ring to Main Linac, includes the scientific, engineering, and technical staff needed to plan, execute, and manage those elements including specification, design, procurement oversight, vendor liaison, quality assurance, acceptance testing, integration, installation, and preliminary check-out of the installed systems.

The Labour estimate for the ILC design as presented in this Technical Design Report, averaged over the three regional sites, is 22,613 thousand person-hrs. This may be compared with the RDR estimate: 24,427 thousand person-hrs. The overall reduction of about 7% results partially from the cost optimization of the machine design discussed in Section 15.6.1, and partially from re-estimates of management and system-integration manpower.

The breakdown of the Labour estimate for the TDR, in units of thousand person-hrs, is shown in Fig. 15.9. The cost breakdown categories are the same as those presented in Section 15.7. All estimates have been averaged over the three regional sites. For comparison, the RDR is shown also. Installation is the largest fraction of explicit Labour, about 24%. Laboratory management and administration is the second largest fraction at about 18%, followed by L-band cavities and cryomodules at 16%.

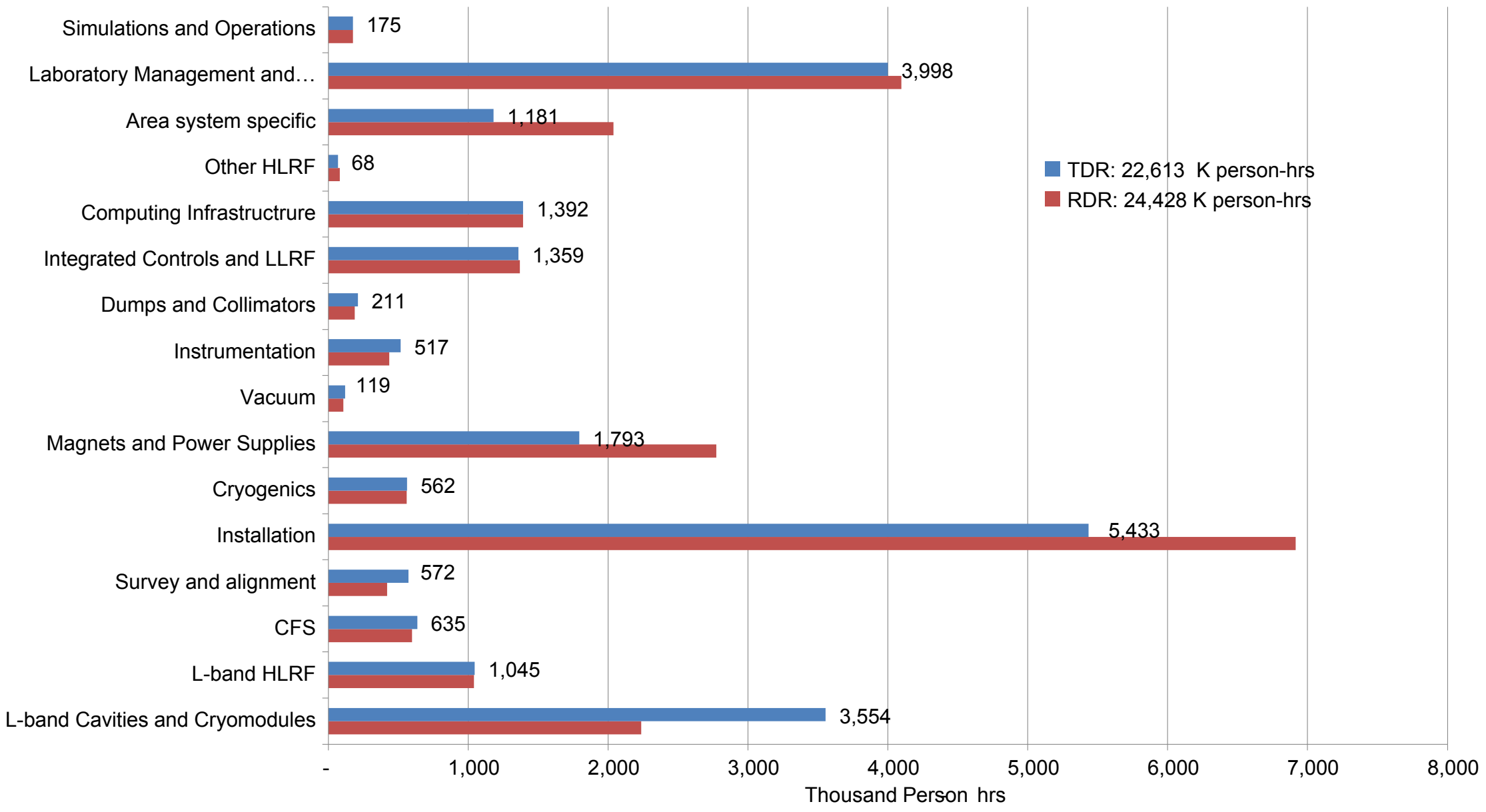

**Figure 15.9.** Explicit Labour, which may be supplied by collaborating laboratories or institutions, listed by technical system, and some Accelerator-specific systems. The numbers give the TDR estimate for each system in thousand person-hours. Also shown for comparison is the RDR.

The Labour estimates broken down by Area (Accelerator) System are shown separately for both the installation and all other labour elements in Fig. 15.10. The system labeled "Common" refers to computing infrastructure labour, laboratory management and administration, simulation and operations labour, and global elements of CFS, installation, and controls labour.

The component Labour estimates for each of the Accelerator Systems include system installation labour, EDIA for all accelerator components, and system integration staff. The main linac comprises about 37% of the total project Labour, followed by "Common" at 31%.

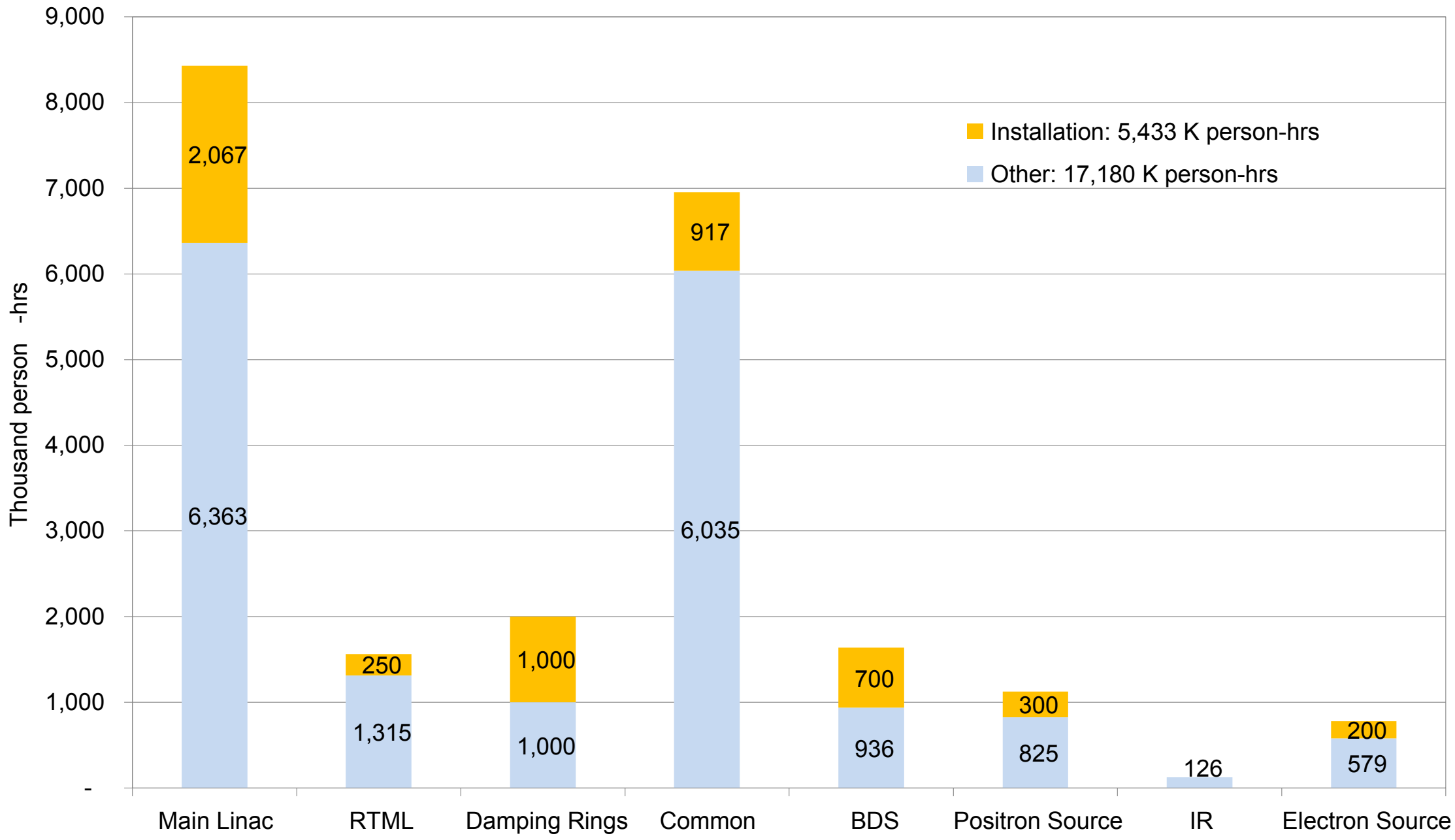

**Figure 15.10.** Distribution of the ILC Labour estimate by accelerator system. The numbers give the TDR estimate for each system in thousand person-hrs.

### 15.8.4 Site Dependence of the Value and Labour Estimates

The Value estimate may be broken down into two parts: the Value for site-specific costs, and the Value for shared parts. In one possible model for the division of responsibilities among the collaborating regions, the host region is expected to provide the site-specific parts, because of the size, complexity, and specific nature of these elements. The site-specific elements include all the civil engineering (tunnels, shafts, underground halls and caverns, surface buildings, and site development work); the primary high-voltage electrical-power equipment, main substations, medium-voltage distribution, and transmission lines; and the primary water-cooling towers, primary pumping stations, and piping. Responsibilities for the other parts of the conventional facilities: low-voltage electrical power distribution, emergency power, communications, HVAC, plumbing, fire suppression, secondary water-cooling systems, elevators, cranes, hoists, safety systems, and survey and alignment, could be shared between the host and non-host regions. All other technical components for the machine could, of course, also be shared between the host and non-host regions.

The Value estimates corresponding to this division of costs, for each regional site, are summarised in Table 15.12. The shared costs are higher for the flat topography sites because they require the more expensive KCS high-level RF-system configuration. The rms spread in the total costs among the three regional sites is 147 MILCU (1.9%).

**Table 15.12**
Possible division of Value for the 3 sample sites (2012 MILCU).

| Region | Site-Specific | Shared | Total |
|---|---|---|---|
| Asia | 1,756 | 6,226 | 7,982 |
| Americas | 1,413 | 6,310 | 7,723 |
| Europe | 1,330 | 6,304 | 7,634 |
| Average | 1,499 | 6,281 | 7,780 |

Similarly, the Labour estimate may be broken down into two parts: the Labour for site-specific parts, and the Labour for shared parts. The Labour for site-specific costs is the EDIA associated with the site-specific Value elements, together with Laboratory management and administration.

The Labour estimates corresponding to this division, for each regional site, are summarized in

Table 15.13. The rms spread in the total Labour among the three regional sites 1%.

**Table 15.13**
Possible division of Labour for the 3 sample sites (thousand person-hrs).

| Region | Site-Specific | Shared | Total |
|---|---|---|---|
| Asia | 4,536 | 18,356 | 22,892 |
| Americas | 4,272 | 18,096 | 22,368 |
| Europe | 4,496 | 18,084 | 22,580 |
| Average | 4,435 | 18,178 | 22,613 |

## 15.9 Cost Uncertainties, Confidence Levels and Cost Premiums

In this section, estimates of the uncertainties in the ILC TDR Value and Labour estimates are given. It is important to understand that these uncertainty estimates do not in general correspond to what is often referred to as contingency. Contingency is a broader term, and includes not only cost uncertainties but also, for example, allowances for missing items.

Cost risk is due to uncertainties or errors in the cost basis (e.g, procurement of a similar item, quantity discount from a single unit price, engineering estimate, etc.) on which the cost of a specific item is based. Technical risk is related to failure of a specific item to achieve the design performance, requiring a redesign which may result in schedule delays and increase the cost. Schedule risk is related to failure to supply a specific item on schedule, requiring delays which may increase the cost (typically by introducing inefficiencies and additional manpower requirements). Market risk is related to deviations in procurement costs from the estimate, due to changes in economic market conditions between when the estimate was made, and when the procurement is made.

The cost uncertainties estimated for the TDR only express the cost risk. They do not cover cost increases due to technical, schedule, or market risk, or to items that have been inadvertently omitted from the estimate. They also do not include allowances related to the potential cost and schedule risks associated with projects having large in-kind contribution components from several different regions of the world.

### 15.9.1 General Methodology

#### 15.9.1.1 Confidence Level for the TDR Estimate

Cost estimation always involves some degree of uncertainty, which can be characterised by the width of the differential cost-distribution function. The cost of each element in the ILC TDR cost estimate corresponds to the median of the distribution: that is, it corresponds to the 50% probability point on the cumulative cost-distribution function. For simplicity, all cost distribution functions used in the ILC TDR estimate were taken to be symmetric Gaussian distributions. For such distributions, the median and the mean are identical.

The confidence level of an estimate is the probability that the actual cost of the item will be less than the estimate. Thus, the confidence level of each cost element in the ILC TDR cost estimate is 50%.

#### 15.9.1.2 Cost-Element Uncertainty Characterisation

15.9.1.2.1 Development of Cost Uncertainties during the RDR.

*Description of the uncertainties* The uncertainty associated with each cost element depends on the nature, quality and maturity of the basis of estimate for that element. During the development of the RDR estimate, for each cost element, the estimator was required to evaluate the uncertainty. General guidelines were established to associate cost uncertainty with quality of the cost basis. These guidelines included the following elements:

- the maturity of the item's design (conceptual, preliminary, or detailed);

- the level of technical risk involved in the design and manufacture of the item;
- the impact of delays in this item on the project schedule (critical-path impact, non-critical-path impact, no schedule impact on any other item);
- the source of the cost information (engineering estimate based on minimal experience, engineering estimate based on extensive experience, vendor quote, industrial study, catalogue price);
- the extent, if any, of cost scaling to large quantities.

Using these guidelines, or by other means, the estimator identified the shape of the differential cost-distribution function for the cost element, choosing from three possible shapes: rectangular, Gaussian, or triangular. The estimator then characterized the upper($\sigma_U$) and lower($\sigma_L$) root-mean-square (rms) widths of the curve. The widths of the curves are measured from the mode (the cost corresponding to the maximum of the cost-distribution curve).

*Cost premiums at the 84% confidence level* From the differential cost-distribution function and the median estimate, the cost increase required to achieve a higher confidence level (84%) was computed for each element. In this section, this cost increase is called the "cost premium" $(P)$. The premium is the cost that must be added to the median estimate $(M)$ to obtain a "high-confidence estimate". Assuming that the cost distribution curve properly describes all sources of cost uncertainty for this element, the chance of the "high-confidence estimate" $(M + P)$ being exceeded during project execution is 16%.

15.9.1.2.2 Treatment of cost uncertainties for the TDR. For the TDR, a simpler approach to the description of cost uncertainties has been adopted. This is based on the fact that the cost premium is the only information required from the cost distribution for the TDR.

For the TDR, only symmetric Gaussian distributions are used for all cost elements. To describe the uncertainty for such distributions, only one parameter is required: the rms width $\sigma$. Moreover, for such a distribution, the cost premium $P$ at the 84% confidence level is simply equal to $\sigma$. A $\sigma$ is determined for all cost elements that have a new estimate developed for the TDR, using guidelines similar to those developed for the RDR. For all other cost elements, $\sigma$ is set equal to the cost premium computed from the original cost distribution given for the RDR. This procedure ensures that, for cost elements taken from the RDR, the cost premium in the TDR estimate is the same as that specified for the RDR estimate. The shapes of the distributions may be different in the TDR, since the normal distribution is used for all cost elements. However, since the cost premium is the only information required from the cost distribution for the TDR, the change in shape is of no consequence, as long as the premium is the same.

### 15.9.2 Median Estimates and Cost Premiums for Groups of Cost Elements

For any group of cost elements, the median estimates for all the cost elements in the group are summed to give the median estimate for the total cost of the group. This also applies to the total project cost. Thus, the confidence level of the total project cost stated in the ILC TDR cost estimate is taken to be 50%.

Similarly, for any group of cost elements, the cost premiums were summed over all the cost elements to approximate the cost premium on the total group cost. If the cost elements were completely uncorrelated, taking a summation in quadrature of the cost premiums would be approximately correct. However, the cost elements are correlated to some degree. In such a case, the use of a linear sum provides a relatively conservative estimate of the cost premium on the total group cost. For the group of all the cost elements, corresponding to the total project cost, the TDR estimate plus the total cost premium represents a "high-confidence estimate" with a confidence level of 84%.

### 15.9.3 Methodology for Assigning Cost Premiums for TDR Cost Elements

For cost elements whose estimates were newly developed for the TDR, a general methodology was developed to make an estimate of $\sigma/M$ for a symmetric Gaussian (i.e. the relative cost premium). In this methodology, a "basic" (relative) premium was assigned, depending on the nature of the cost basis used for the cost element. To account for the uncertainty in the quantity discount used in the TDR estimate, an additional premium was added, equal to half of the quantity discount. Finally, to account for any additional sources of cost risk specific to a particular cost element, a "special" premium was added, if necessary. The total premium used for the cost element is the linear sum of the basic premium, half the quantity discount, and the special premium, if any.

### 15.9.4 Overall Cost Premiums for the TDR Value Estimate

Figure 15.11 shows the relative cost premiums for the TDR, broken down by the cost categories described in Section 15.7. For cavities and cryomodules, L-band high-level RF, conventional facilities, and installation, the cost premiums were developed using the methodology described in Section 15.9.3. For all other technical systems, the cost premiums are those developed during the RDR. The overall relative Value premium for the total ILC TDR Value estimate is 26%.

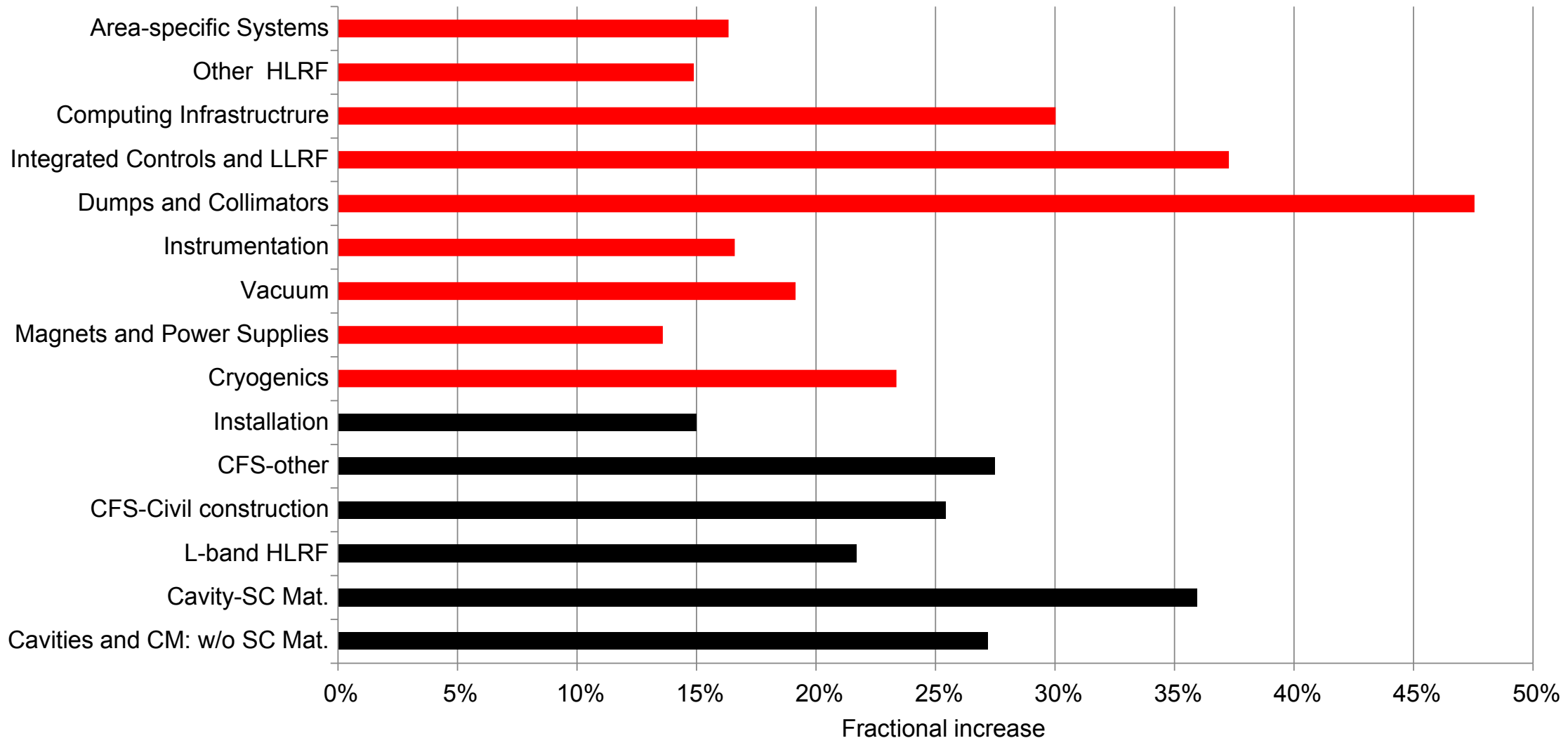

**Figure 15.11.** Relative Value premiums, broken down by subsystem, for the TDR.The red bars are taken from the RDR; the black bars correspond to premiums developed for the TDR.

### 15.9.5 Overall Cost Premiums for the TDR Labour Estimate

New estimates were made for installation labour, for cavity and cryomodule test and commissioning labour, for coupler processing labour, for conventional facilities labour (including survey and alignment), for management and administration labour, and for accelerator system-integration labour. The premiums for these cost elements were developed using the methodology described in Section 15.9.3. No special premiums were applied. For all other technical systems, the labour premiums are those developed during the RDR.

Figure 15.12 shows the relative Labour premiums for the TDR, broken down by the categories described in Section 15.7. The overall relative Labour premium for the total ILC TDR Labour estimate is 24%.

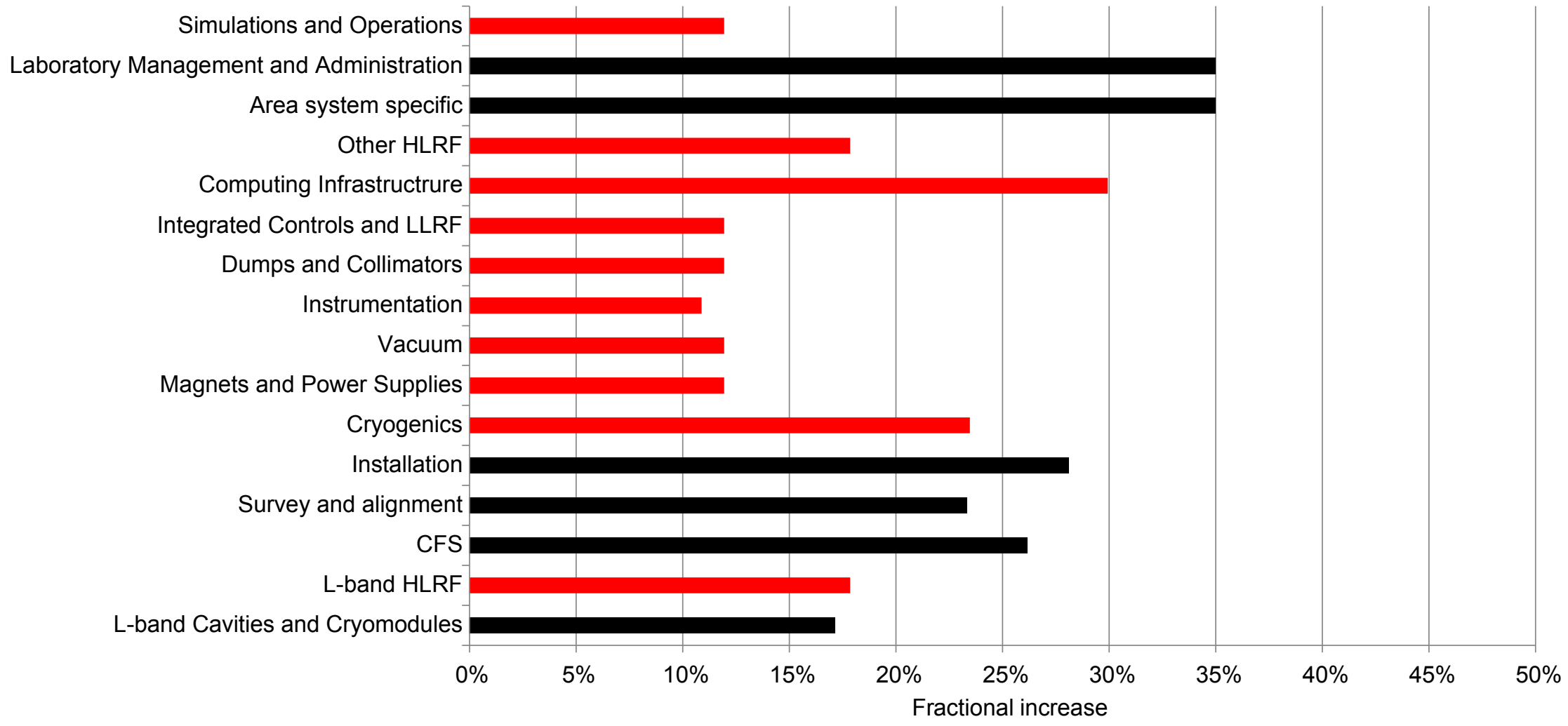

**Figure 15.12.** Relative Labour premiums, broken down by subsystem, for the TDR. The red bars are taken from the RDR; the black bars correspond to premiums developed for the TDR.

## 15.10 Value and Labour Time Profiles

Given the schedule described in Chapter 14, and the Value and Labour estimates given in Section 15.8, profiles describing the Value and Labour resources needed as a function of time can be developed. These profiles assume a flat funding profile for the major civil and technical procurements for each accelerator system, which is a crude assumption, but one which captures the essential features of the overall project-resource requirements.

The Value profile is shown in Fig. 15.13, broken down by technical system. The profile shows the front-loading of the civil construction effort, and the overall roughly six-year period for ramp-up and production of the cavities and cryomodules, which completes in year 7. The peak Value requirement is about 1,200 MILCU in years 4 and 5.

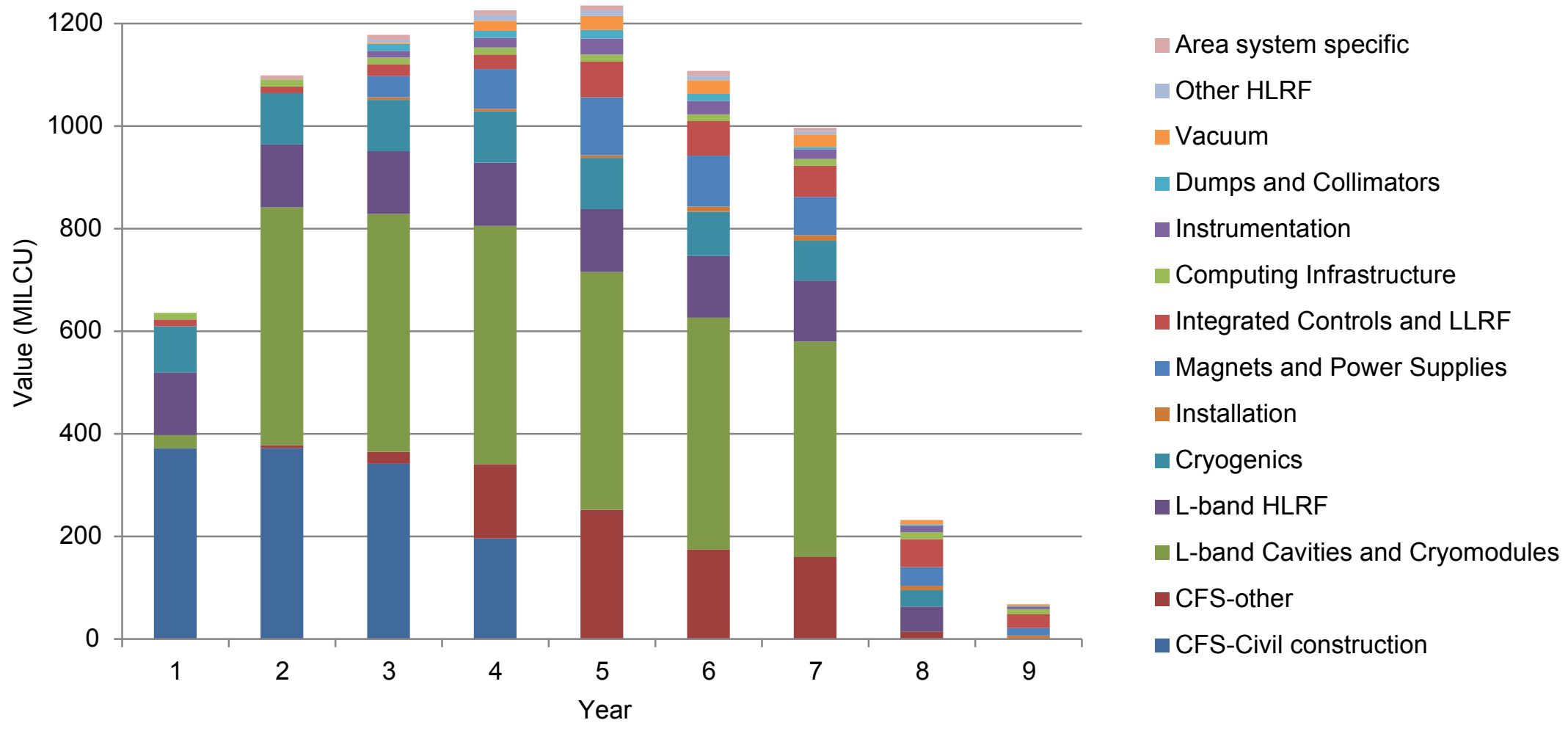

**Figure 15.13.** Value profile vs. project year. broken down by technical system.

The Labour profile (excluding installation) is shown in Fig. 15.14, broken down by technical system, while the installation Labour is shown in Fig. 15.15. The Labour profiles are in FTE, and assume 1700 person-hrs per year[17]. The back-loading of the installation profile is evident. The peak manpower requirements are about 1600 FTE in year 5 for all tasks except installation, and about

[17] 2000 person-hrs per year for installation.

950 FTE in year 7 for installation.

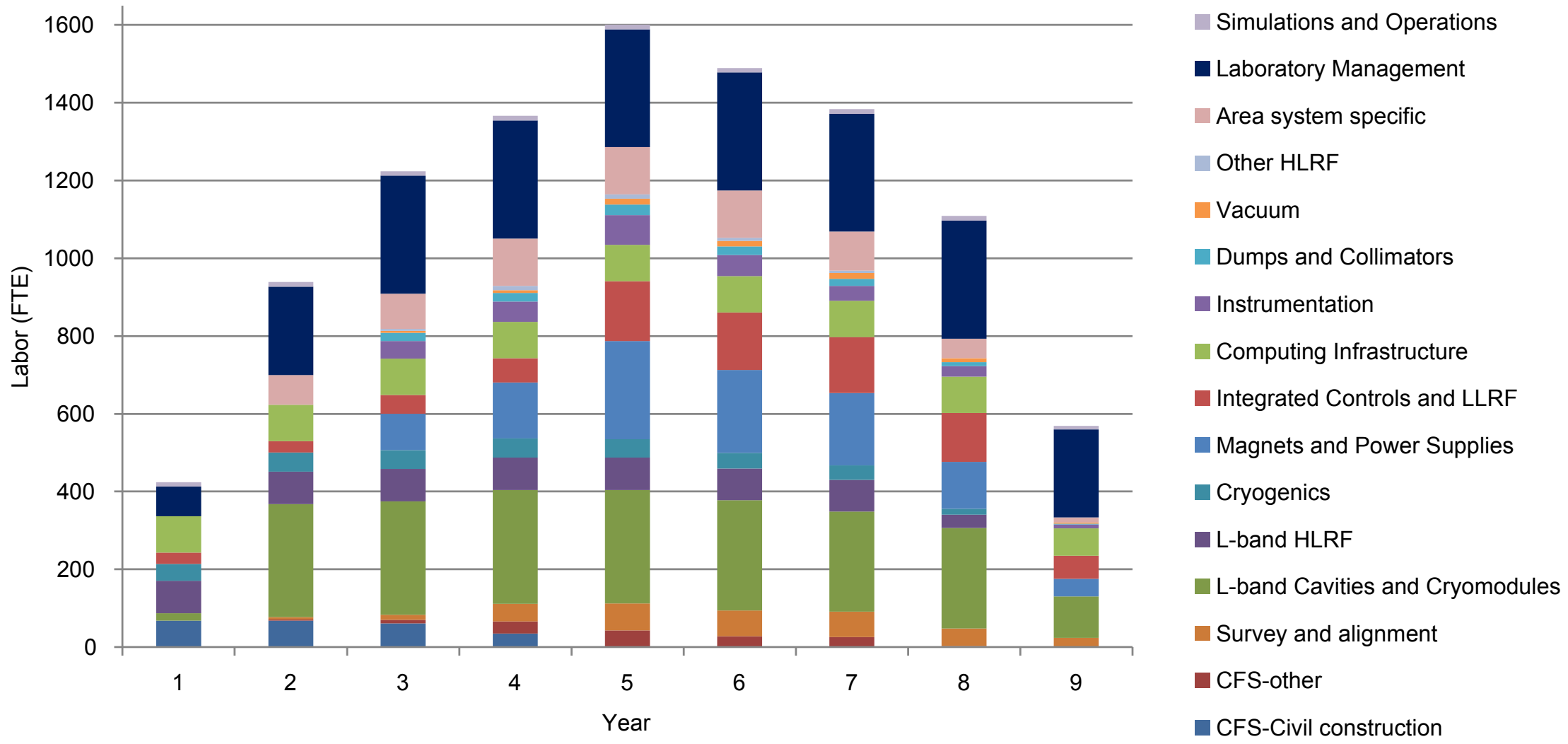

**Figure 15.14.** Profile for explicit Labour (excluding installation) vs. project year, broken down by technical system.

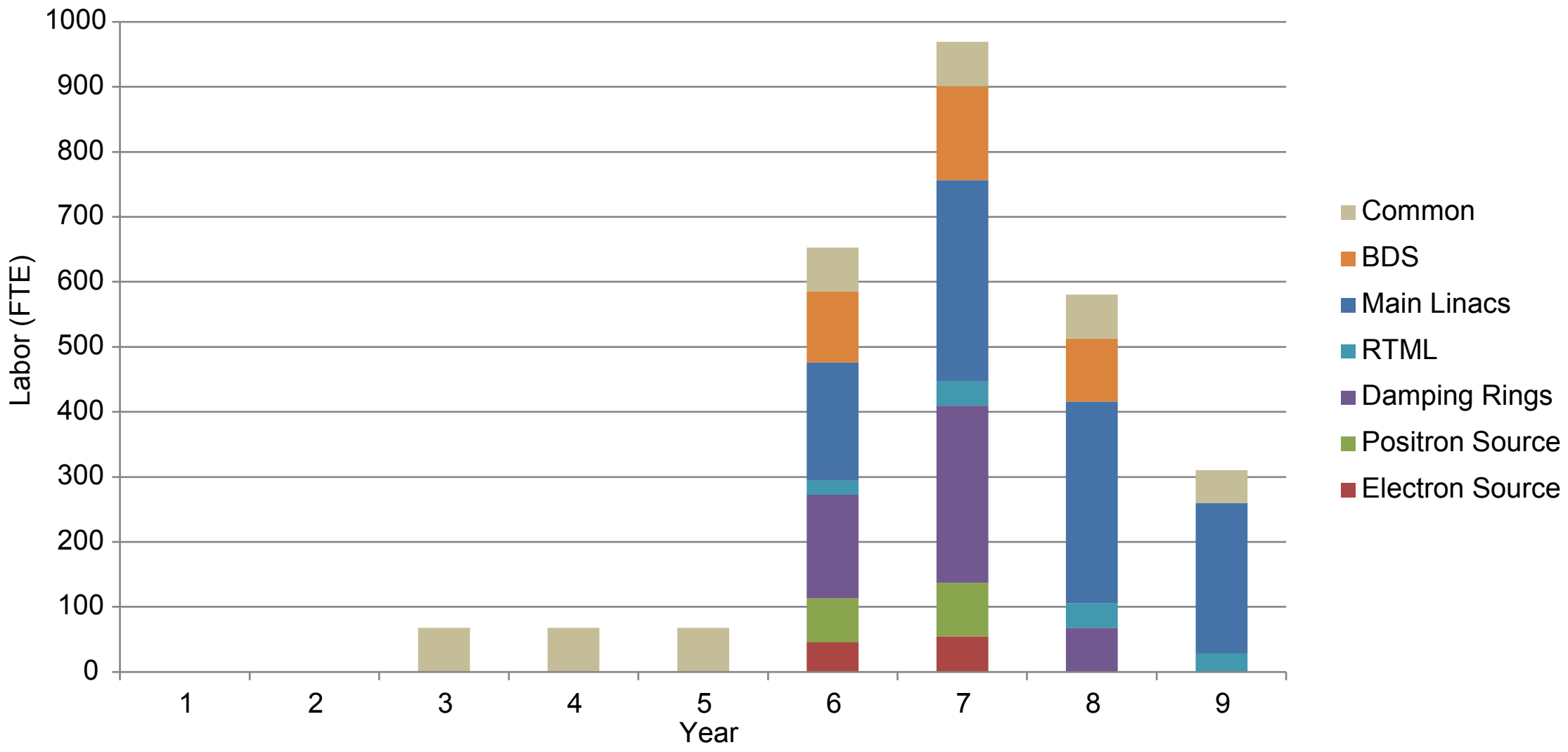

**Figure 15.15.** Profile for explicit installation Labour vs. project year

## 15.11 Value and Labour Estimates for Operations

Operating costs are not included in the estimate for the construction project, but a very preliminary estimate is given in this section. It is also to be noted that spare components (those stored in warehouses and not the installed redundant components), although fabricated along with the installed components, are assumed to be financed through operating funds, and are not considered part of the construction project.

Major factors in the Value estimate for operations include electrical power, maintenance and repairs, helium and nitrogen consumables, and components that have a limited life expectancy and need continuous replacement or refurbishment, like klystrons. The electric power costs and the cost for material and supplies during operation are estimated to lie in a range of 280 to 510 MILCU per year. The Value estimate for operations is taken to be the center of this range: 390 MILCU per year. The cost premium for this estimate is taken to be the standard deviation of the upper or lower estimate from the center, which gives a premium of about 40%.

The Labour estimate, corresponding to the continuing operations and administrative staff, is expected to be comparable to that at existing facilities (not including support of the scientific program). On this basis, the operations staff is estimated to be between 700 and 1000 FTE. Using the center of this range to set the Labour estimate, and the standard deviation of the upper or lower estimate from the center to set the premium, the Labour estimate for operations is 850 FTE, with a premium of about 25%.

Operating costs are anticipated to gradually increase over the fourth through ninth years of construction from zero up to the full level of long-term operations at the end of the 9 year construction phase.

## 15.12 Value and Labour Estimates for Upgrade and Staging Options

This section estimates the Value and Labour changes associated with the upgrade and staging options described in Chapter 12 of this *Technical Design Report*.

### 15.12.1 Value and Labour Estimate for Luminosity Upgrade

In Section 12.3, a luminosity upgrade for the 500 GeV baseline machine is discussed. The luminosity upgrade is accomplished by doubling the number of bunches, resulting in a doubling of the average beam power. Additional RF power sources (klystrons and modulators) are added to the main linacs, and an additional positron damping ring is installed in the damping ring enclosure.

#### 15.12.1.1 Technical system scope, Value and Labour changes

15.12.1.1.1 L-band high-level RF systems In the main linacs, additional klystrons and modulators are added, and microwave power distribution systems must be modified. The technical scope changes differ for the flat and mountainous topography sites, due to their different RF power source configurations.

For the KCS configuration at the flat topography sites, an additional 10 klystrons and modulators are added at each of the 22 klystron clusters. The total number of klystrons and modulators thus increases by 220. In addition, some additional microwave hardware is required in the KCS power distribution system.

For the DKS configuration at the mountainous topography sites, the local PDS systems are reconfigured so that 26, rather than 39, cavities are driven by each klystron. This increases the required number of klystrons and modulators in the main linacs from the baseline number (378) to $378 \times 39/26 = 567$, which corresponds to an increase of 189 klystrons and modulators.

In addition to these changes in the main linacs, 3 more klystrons and modulators must be added to the 5 GeV booster in the positron source.

The Value estimates for the additional klystrons, modulators and associated microwave hardware have been made assuming the components are procured in the required numbers from 2 vendors. The unit costs have been adjusted for the number of procured components, using the same cost estimating relationship that was used for the baseline estimates. The Labour estimates were based on simple scaling with Value from the baseline Labour estimates.

15.12.1.1.2 Conventional facilities Conventional facilities support systems (electrical and mechanical) must be upgraded to handle the increased beam power in all accelerator systems, and the increased number of RF power sources in the main linacs. The Value estimates for the cost of these upgraded support systems are based on information developed for the Americas region conventional systems baseline estimate. The Labour estimates were based on simple scaling with Value from the baseline Labour estimates.

15.12.1.1.3 Damping rings Since the number of bunches is doubled, the bunch spacing in the damping rings is halved. To maintain the same bunch spacing for positrons[18], an additional positron damping ring is added, in the same tunnel as the baseline positron and electron rings.

Low-energy operation at 10 Hz with twice the beam current in the electron ring requires the addition of more RF cavities. The available space in the lattice permits 4 more cavities to be added.

The Value and Labour estimates for the new positron damping ring are based on simple scaling from the baseline damping rings estimate. The small additional cost of 4 more RF cavities in the electron damping ring has been included.

15.12.1.1.4 Common In addition to the items noted above, changes in the scope of Common elements of the associated technical systems (conventional facilities, installation, and control systems) are also included. This is based on a simple scaling with changes in the associated technical system Value or Labour. No changes in laboratory-wide Common Value or Labour elements, (i.e. computing infrastructure, laboratory management and administration, and simulation and operations) are included, as it is not clear how these elements would scale with the technical scope changes.

#### 15.12.1.2 Summary of Value and Labour changes

The total Value change associated with the luminosity upgrade is 483 MILCU. This is about 6% of the 500 GeV baseline Value estimate. The total Labour change associated with the luminosity upgrade is 1,537 thousand person-hrs. This is about 7% of the 500 GeV baseline Labour estimate.

### 15.12.2 Value Estimate and Labour Estimate for 1 TeV Energy Upgrade

In Section 12.4, the upgrade of the baseline machine to 1 TeV center-of-mass is discussed. The beam current needed for the 1 TeV machine requires the luminosity upgrade discussed in Section 12.3. Consequently, in evaluating the Value and Labour changes for the 1 TeV upgrade, the technical scope, and corresponding Value and Labour estimates, have been taken to be that of the baseline with the luminosity upgrade.

The beam energy upgrade is accomplished by extending the main SCRF linacs to provide the additional 250 GeV beam energy. The main linac tunnels are lengthened to accommodate the additional SCRF hardware, new RTML turn-arounds and bunch compressor systems are constructed at the new low-energy ends of the main linacs, and the long 5 GeV transfer line is extended. The positron-production undulator is replaced with one suitable for 500 GeV beam energy, and additional dipoles are added in the BDS to provide the required higher integrated field strength.

Three possible scenarios for the re-configuration of the main linacs are presented:

- **Scenario A:** The linac extension is accomplished using the baseline SCRF technology, i.e. cavities with an average gradient of 31.5 MV/m.
- **Scenario B:** The linac extension is accomplished using improved SCRF technology, i.e. cavities with an average gradient of 45 MV/m.
- **Scenario C:** The entire main linac is removed and replaced with improved SCRF technology, i.e. cavities with an average gradient of 45 MV/m, with a length sufficient to provide a beam energy of 500 GeV.

[18]The minimum bunch spacing is determined by the electron cloud effect in the positron ring.

#### 15.12.2.1 Technical system scope, Value and Labour changes

15.12.2.1.1 Main linacs The changes in the main linacs and associated civil construction depend on the scenario, as detailed in the following bullets:

- **Scenario A:** An additional 260 GeV of main linac is added. Since the baseline linac energy is 235 GeV, the addition is essentially the baseline scaled up in energy by 260/235=1.106. The gradient for the new linac is the same as for the baseline, so the additional tunnel and linac have a length scaled up by the same factor as the energy. The additional cryogenics load is 70% of the baseline.
- **Scenario B:** Again, an additional 260 GeV of main linac is added. The addition is essentially the baseline scaled up in energy by 1.106. However, since the average gradient for the new linac is 45 MV/m, the additional tunnel and linac have a length scaled up by $1.106 \times 31.5/45=0.774$. The additional cryogenics load is 60% of the baseline.
- **Scenario C:** In this case, as far as power sources, cryogenics, and conventional facilities support are concerned, an additional 260 GeV of main linac is again added. However, the cavities and cryomodules for the entire linac are replaced: this corresponds to adding 485 GeV of linac. The associated linac length scale factor relative to the baseline is $485/235 \times 31.5/45=1.445$. However, the scale factor for additional linac tunnel is 0.445. The additional cryogenics load is 50% of the baseline.

In estimating the Value changes for the cavities and cryomodules based on improved SCRF technology, the cost per unit length is taken to be the same as for the baseline SCRF technology. Note that the system Value or Labour is simply scaled with the change in associated technical scope, assuming that component unit costs do not change. In fact, the change in the numbers of components would result in unit cost changes, but this effect is neglected.

15.12.2.1.2 Other accelerator systems In addition to the linac changes, for all scenarios, the baseline RTML is essentially duplicated, and installed at the low energy end of the 1 TeV machine. Consequently, the Value and Labour for the baseline RTML is added, for all scenarios[19]. The Value of the baseline undulator is also added, to approximate the cost of the required new undulator. Finally, the Value of 10% of the baseline BDS magnets and power supplies is included, to approximate the costs of the new BDS components required for 1 TeV.

15.12.2.1.3 Common In addition to changes in the linacs and RTML, changes in the scope of Common elements of the associated technical systems (CFS, installation, and control systems) are also included. This based on a simple scaling with changes in the associated technical system Value or Labour. No changes in laboratory-wide Common Value or Labour elements are included, (i.e. computing infrastructure, laboratory management and administration, and simulation and operations), as it is not clear how these elements would scale with the technical scope changes.

#### 15.12.2.2 Summary of Value and Labour changes

The total Value changes associated with scenario A, B and C are 6,706, 5,489 and 7,082 MILCU, respectively. These increases correspond to 81%, 66%, and 86%, respectively, of the 500 GeV Value estimate for the baseline with luminosity upgrade. The total Labour changes associated with scenario A, B and C are 11,988, 9,416 and 14,256 thousand person-hrs, respectively. These increases correspond to 50%, 42%, and 59%, respectively, of the 500 GeV baseline Labour estimate with luminosity upgrade.

[19] This is not quite correct, since some of the baseline RTML Value and Labour is associated with the beamlines from the damping rings to the long 5 GeV transfer line. The RTML contribution to the 1 TeV upgrade is thus slightly overestimated.

### 15.12.3 Value and Labour Estimates for a Light Higgs Factory as a First-Stage Option

In Section 12.5, a 250 GeV center-of-mass machine is discussed, which could be implemented as the first stage of a route to the baseline 500 GeV ILC. The first stage machine would require the installation of approximately half of the baseline linacs. Two possible scenarios are presented:

1. Only the tunnel and support shafts (access ways) required for the 250 GeV machine are constructed, and the linacs are installed in this tunnel.
2. The complete tunnel and support shafts (access ways) for the 500 GeV machine are constructed as part of the first stage, and the linacs are installed in the first half of each tunnel, followed by a beam transfer line to the central region.

The first scenario is conceptually the same as that proposed for the 1 TeV upgrade, although half the scale. It is likely to represent the minimum cost for the initial phase machine. The second scenario requires greater investment for the initial phase (for the civil construction), but increasing the centre-of-mass energy then becomes relatively straightforward, and opens up the possibility for a more adiabatic approach to increasing the energy.

#### 15.12.3.1 Technical system scope, Value and Labour changes

For the first stage machine described in Section 12.5 of the TDR, the electron and positron sources, the damping rings, and the beam delivery systems, are identical to those of the baseline machine. Both scenarios require a 150 GeV electron linac operating at 10 Hz[20] (for positron production) and a 125 GeV positron linac. Scenario 1 also requires an RTML with a transfer line only about half the length of the baseline system, while scenario 2 requires essentially the baseline RTML.

<u>15.12.3.1.1 Scenario 1</u> The technical scope reductions in this scenario are the removal of 100 GeV of electron linac (together with its tunnel), 125 GeV of positron linac (and tunnel), and the corresponding lengths of the RTML long transfer lines. The ratio of the electron linac removed to the total baseline linac is 100/470 = 0.212. The ratio of the positron linac removed to the total baseline linac is 125/470 = 0.266. The fraction of the RTML long transfer line removed is the sum: 0.212+0.266 = 0.479.

In evaluating the Value and Labour changes associated with these technical scope reductions, the system Value or Labour are simply scaled with the change in associated technical scope, assuming that component unit costs do not change. In fact, the change in the numbers of components would result in unit cost changes, but this effect is neglected.

In addition to changes in the linacs and RTML, changes in the scope of Common elements of the associated technical systems (CFS, installation, and control systems) are also included. This based on a simple scaling with changes in the associated technical system Value or Labour. No changes in laboratory-wide Common Value or Labour elements, (i.e. computing infrastructure, laboratory management and administration, and simulation and operations) are included, as it is not clear how these elements would scale with the technical scope changes.

The total Value change associated with scenario 1 is -2,425 MILCU. This is about 31% of the 500 GeV baseline Value estimate. The total Labour change associated with scenario 1 is -4,583 thousand person-hrs. This is about 20% of the 500 GeV baseline Labour estimate.

<u>15.12.3.1.2 Scenario 2</u> Relative to scenario 1, this scenario simply adds back the tunnels for 100 GeV of electron linac and 125 GeV of positron linac, and the corresponding lengths of the RTML long transfer lines. In addition, a beamline is required to transport the 125 or 150 GeV beam from the end of the linac to the entry to the BDS. As a crude approximation, the cost of this beamline is taken to be the same as that of an equal length of 5 GeV RTML beamline.

[20] The need for the 10-Hz mode could be removed by increasing the length of the superconducting helical undulator from the baseline length of 147 m to approximately 250 m. The electron linac would now only require an additional 3.5 GeV beyond 125 GeV to drive the undulator, and only needs to run at 5 Hz.

The net Value change associated with scenario 2, relative to the baseline, is -1,934 MILCU. This is about 25% of the 500 GeV baseline Value estimate. The net Labour change associated with scenario 2, relative to the baseline, is -3,563 thousand person-hrs. This is about 16% of the 500 GeV baseline Labour estimate.

### 15.12.4 Summary

Relative to the baseline, the Value and Labour estimates for the options discussed in Chapter 12 of the TDR are plotted in Fig. 15.16 and Fig. 15.17.

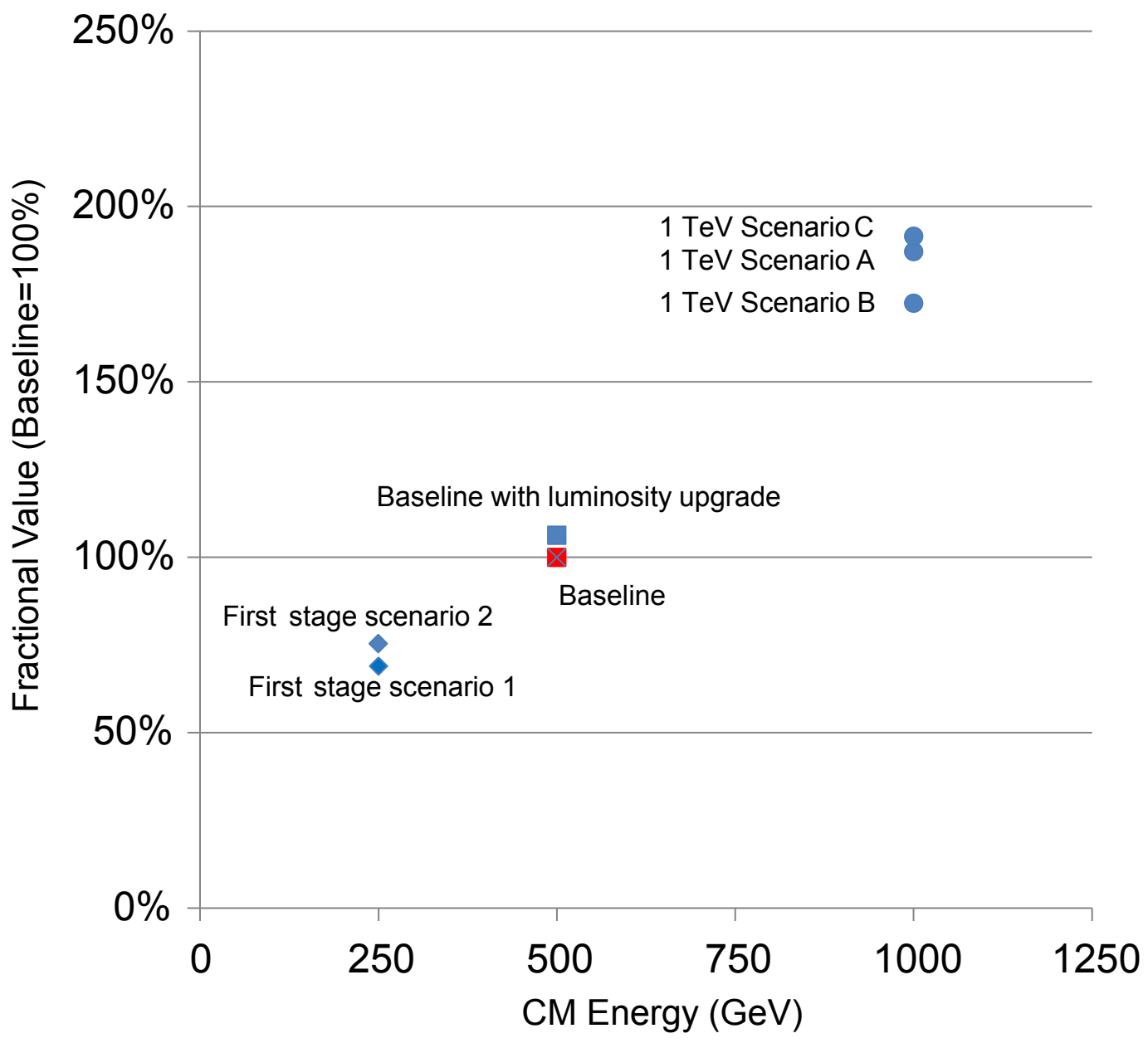

**Figure 15.16**
Relative Value estimates for upgrade and staging options.

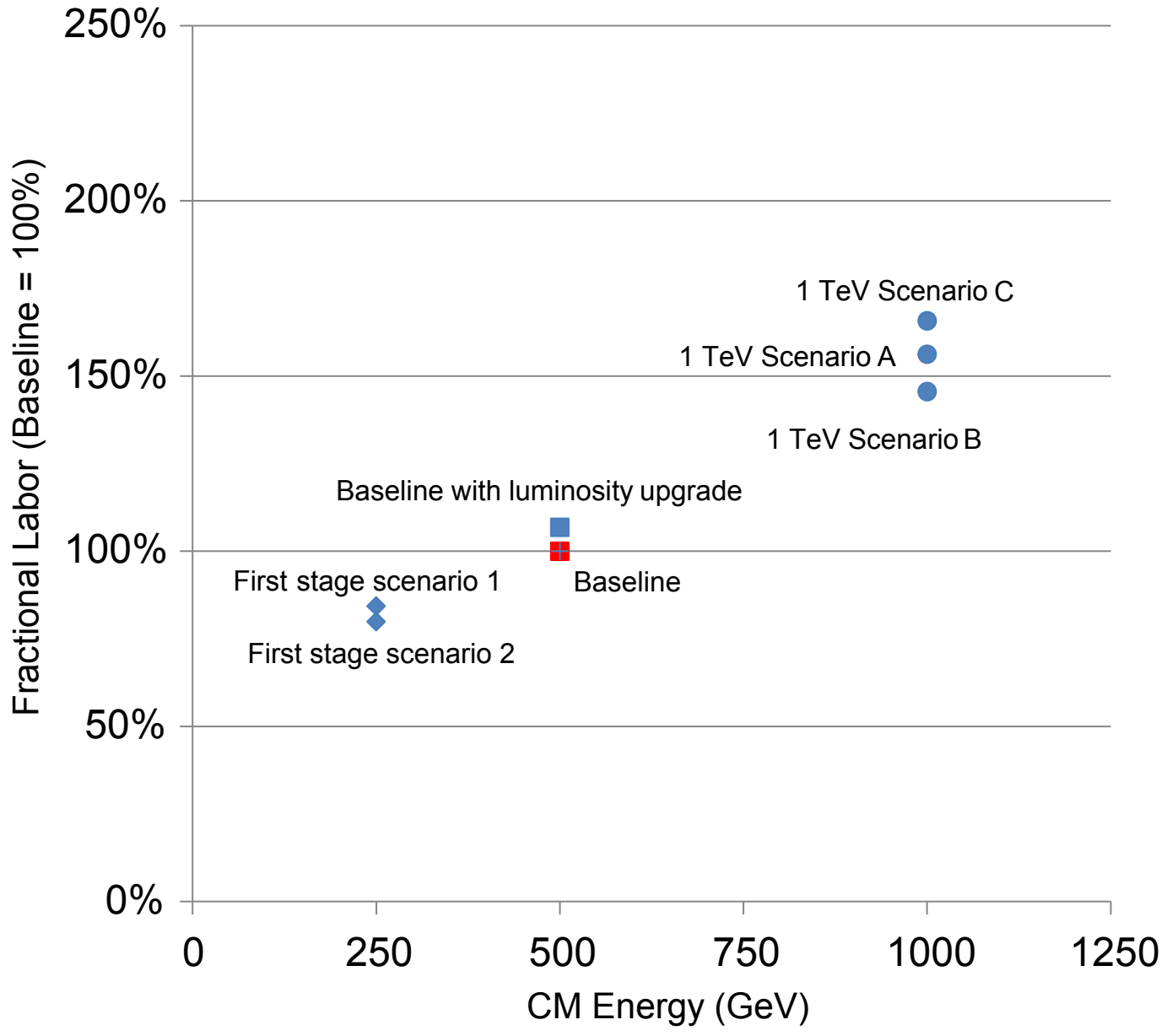

**Figure 15.17**
Relative Labour estimates for upgrade and staging options.

# Appendix A
# Evolution of the ILC design in the Technical Design Phase

## A.1 The goals of the Technical Design Phase

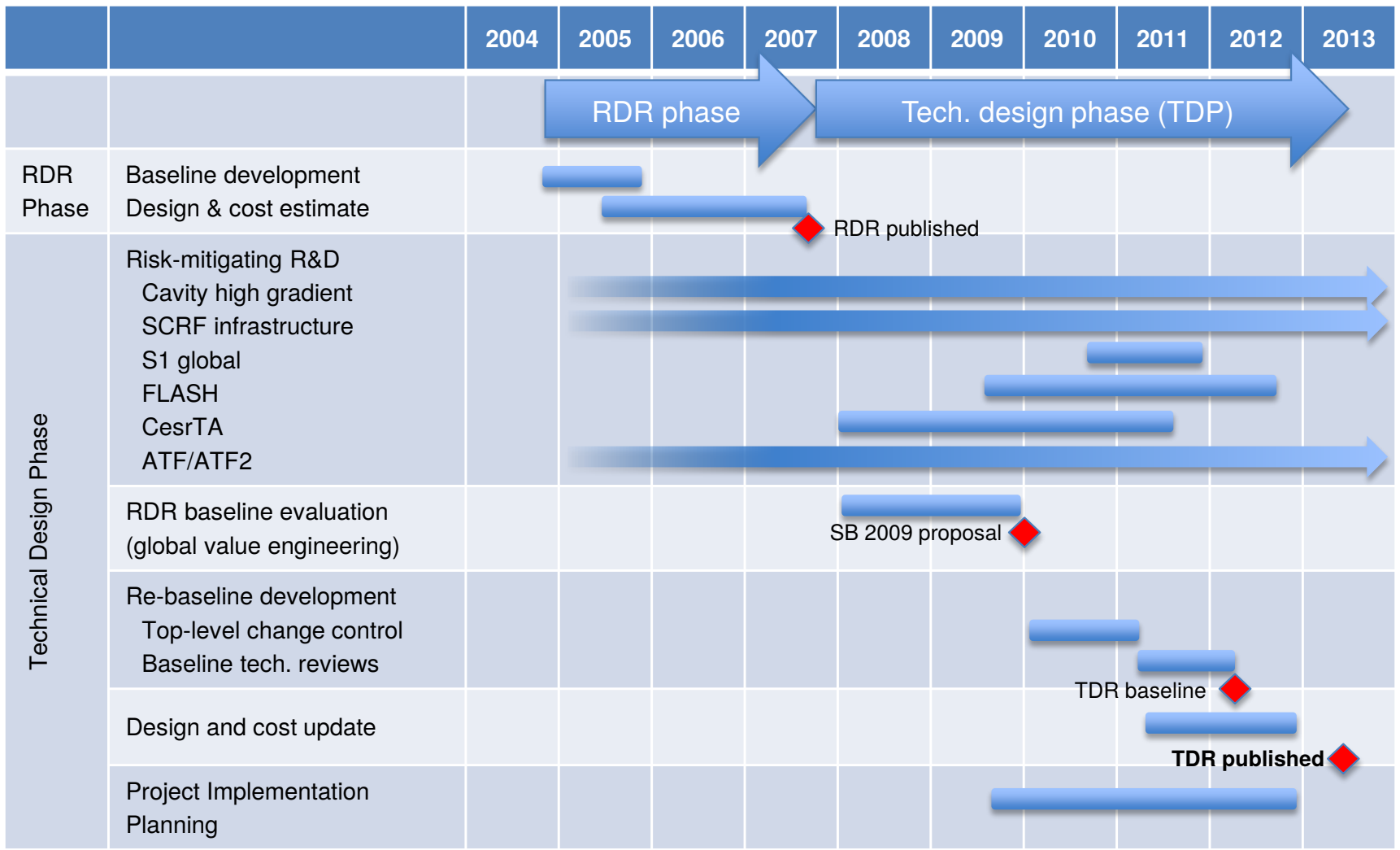

**Figure A.1** Path to the ILC Technical Design Report, indicating the two distinct project phases of the Global Design Effort: the RDR phase, which focused on design and cost-estimate work for the GDE first major deliverable, the 2007 Reference Design Report; and the subsequent Technical Design Phase, which focused on risk-mitigating R&D and worldwide development of SCRF technology, and a re-evaluation of the RDR baseline and updated cost estimate.

The Technical Design (TD) phase of the ILC Global Design Effort (GDE) began after the publication of the ILC Reference Design Report (RDR) in 2007 [3]. The main objectives have been mitigation of the remaining identified high-risk issues associated with the RDR baseline design, and to further refine that design with a strong emphasis on cost optimisation. The primary GDE deliverables summarised in the TDR are:

- an updated technical description of the ILC Technical Design in sufficient detail to justify the associated VALUE estimate;
- results from critical R&D programmes and test facilities, which either demonstrate or support the choice of key parameters in the machine design;
- one or more models for a Project Implementation Planning (PIP), including scenarios for globally distributed mass-production of high-technology components as "in-kind" contributions;
- an updated and robust VALUE estimate and construction schedule consistent with the scope of the machine.

Figure A.1 shows the GDE's top-level phases, while Fig. A.2 shows how the R&D programmes together with the Accelerator Design and Integration (AD&I) activities factor into the TDR and also the PIP. The five themes identified (risk-mitigating R&D, of which SCRF R&D is a special case, AD&I cost and schedule and finally risk assessment) form an integrated approach to producing a mature and relative low-risk design for the ILC.

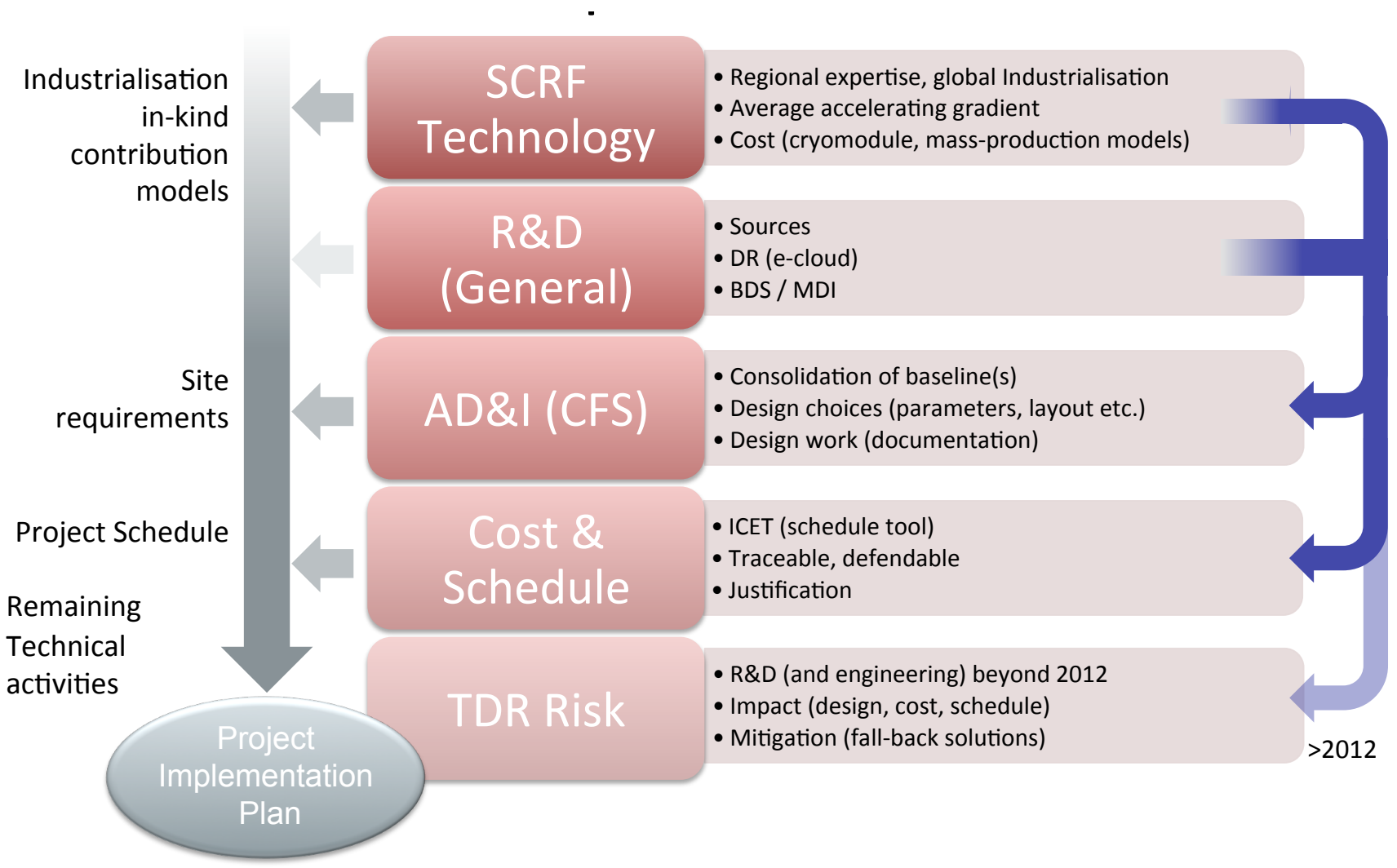

**Figure A.2**
The primary themes of the GDE's Technical Design Phase, and how they relate to the key deliverables of the Technical Design Report.

To coordinate the TD-Phase plans, the GDE implemented a monolithic project management structure shown in Fig. A.3. The project was divided into three main Technical Areas, each representing about one third of the total project cost: SCRF Technology; Conventional Facilities and Siting (CFS), together with global systems; and finally Accelerator Systems, which effectively covered the accelerator design of the sources, damping rings and beam-delivery system. Each Technical Area was managed by one of three project managers, who formed a central management team. Under each project manager, a number of Technical Area Groups were identified. The Technical Area project managers and Technical Area group leaders – together with integration and documentation technical support, formed the central ILC design group for the TD phase.

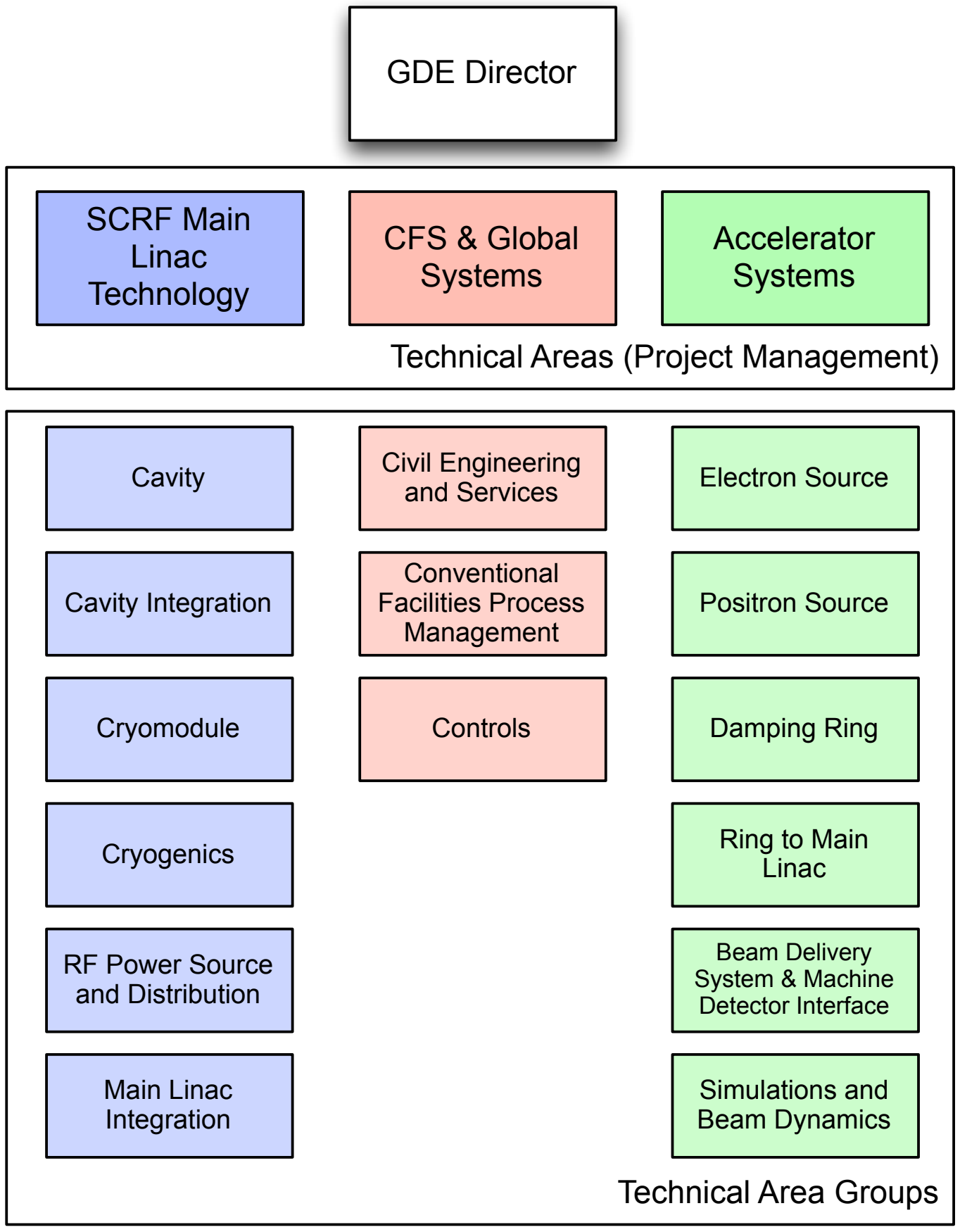

**Figure A.3**
The GDE project structure for the Technical Design phase.

## A.2 Approach to cost constraint and re-baselining the ILC

The 2007 RDR published a value estimate of 6.7 Billion ILCU together with an estimated 14,200 person-years of institutional labour, for the construction costs of the machine. As part of the overall project risk mitigation, the need to constrain the total construction cost was mandated early on in the technical design phase. This resulted in an approach to the risk-mitigation R&D and in particular the AD&I activities which maintained a strong emphasis on cost impact. For the RDR, the cost of the SCRF technology included in the RDR was based entirely on the European estimates developed for the TESLA project and subsequently updated for the European XFEL, and assumed a single-vendor model. An important goal of the TD phase was to bring the SCRF capabilities of the Americas and Asian region, including industry, and to reconsider the single-vendor model in the light of possible global distribution of the manufacturing, as well as risk reduction. Similarly, the CFS costs – in particular for civil engineering – were consider a risk item as the constraints and issues arising from site-dependent designs became more apparent as they were developed during the TD phase.

With these potential cost risks in mind, a complete high-level cost-driven review of the RDR machine layout and design was undertaken early in the TD phase, in order to reduce the RDR cost and provide margin to hedge any component-level unit cost from the TD phase R&D programmes. The approach adopted to re-baselining was based on an assumption that the RDR design – although sound – was conservative in many of its design decisions, relatively immature from a detailed engineering standpoint, and was "performance-driven" as opposed to cost optimised. Conventional Facilities and Siting (CFS) was identified early on as a strong focus for design optimisation; in particular the reduction of underground civil construction, achieved by a critical re-evaluation of the criteria driven by the accelerator design assumptions. On analysis of the RDR costs drivers, it quickly became apparent that no major cost savings (i.e. tens of percent) where achievable without a change in project scope. Value engineering was expecting to provide savings on the order of $\sim 10\,\%$ total project cost, by consolidating many detailed design elements at the $< 1\,\%$ level. The engineering resources required for such detailed design work were not available to the GDE during the TD phase, and value engineering is now considered part of the post-TDR work, likely as part of a pre-construction project. With this in mind, a strategic decision was made to focus the limited design resources available on relatively high-level layout and design modifications, each of which could provide 1—2% cost savings (based on the RDR costs). The RDR design review or "global value engineering" as it later became known, was based on the following premises:

- **overall cost reduction** – Any opportunities for cost reduction should be taken, in so far as they do not unacceptably impact performance or increase technical risk;
- **improved cost balancing** – Cost margins created as part of the cost-reduction exercise can be made available for other subsystems which incur increased (estimated) construction costs.
- **improved understanding of system functionality** – Understanding how a given system's requirements and functionality impact cost forced a careful analysis of the system's strengths and vulnerabilities; this has a critical value on its own beyond cost-reduction;
- **more complete and robust design** – Revisiting many of the design and implementation details that were not completely covered during the RDR design phase.

The analysis and subsequent review resulted in six major design modifications reflecting an approximate 10 % reduction in the 2007 RDR cost estimate. The final proposed modifications were captured in the "straw-man baseline" SB2009 proposal report [209], submitted by the project management to the GDE Director. To achieve the global consensus of all stake-holders required, a formal process known as Top-Level Change Control was initiated, which was developed over a twelve-month period. A second phase of lower-level change control followed, consolidating more detailed design decisions. Each phase of the design and evaluation process culminated in a focus workshop where a particular subset of the proposed design modifications underwent a final management-level review before a consensus decision was made. For the initial TLCC, a series of four Baseline Assessment Workshops (BAW) were held, each of which resulted in a written proposal to the GDE Director. These workshops focused on the primary high-level concepts outlined in the SB2009 proposal [209]. The second phase was a more comprehensive and detailed review of the entire machine layout, in order to consolidated and document the results of the TLCC decisions, as well as the many lower-level technical decisions that still required resolution. This phase was also accomplished by a series of focus workshops (Baseline Technical Reviews, BTR). Table A.1 summarises the workshops and their focus. The process successfully established the updated baseline for the TDR which is presented in Part II: *ILC Baseline Design*

**Table A.1.** The ILC baseline re-evaluation process during the Technical Design Phase

| | | | | |
|---|---|---|---|---|
| **Top Level Change Control: Baseline Assessment Workshops (BAW)** | | | | |
| BAW 1 | 7-8.10.2009 | KEK | Choice of average accelerating gradient, including margins for installation and operation.<br>RF power overhead for support of ±20% spread in cavity gradients, including design of power distribution system and impact on low-level RF control. | [259] |
| BAW 2 | 9-10.10.2009 | KEK | Removal of Main Linac service tunnel (single-tunnel solutions)<br>RF power generation and distribution for single tunnel solutions (Klystron Cluster scheme and Distributed RF Source scheme). | [260] |
| BAW 3 | 18-19.01.2010 | SLAC | Relocation of the undulator-based positron source to the exit of the main electron linac (nominal 250 GeV beam energy), including integration into central region. Considerations for low centre-of-mass energy running (10 Hz operation mode). | [261] |
| BAW 4 | 20-21.01.2011 | SLAC | Reduction of the number of bunches per pulse by 50 % (reduced beam power). Associated reduction of the damping ring circumference and main linac klystron and modulator count. Luminosity recoverted by stronger beam-beam interaction at the interaction point (stronger focusing in the final-focus system). | [262] |
| **Baseline Technical Reviews (BTR)** | | | | |
| BTR 1 | 6-8.07.2011 | INFN Frascati | Damping rings | [263] |
| BTR 2 | 24-27.10.2011 | DESY | Electron source<br>Positron source<br>Ring to main linac (bunch compressor)<br>Beam-delivery system and machine-detector interface | [264] |
| BTR 3 | 19-20.01.2012 | KEK | Superconducting RF technology<br>Main-linac layout | [265] |
| BTR 4 | 20-23.03.2012 | CERN | Conventional facilities and siting:<br>▪ civil construction<br>▪ mechanical and electrical systems<br>▪ site variant designs<br>▪ schedule, installation and alignment<br>▪ detector hall | [266] |

## A.3 Proposed top-level design modifications and their impact

The global value engineering process briefly outlined above culminated in six top-level modifications to the published 2007 Reference Design.

1. A Main Linac length consistent with an average accelerating gradient of 31.5 MV/m and maximum operational beam energy of 250 GeV, together with a RF distribution scheme which optimally supports a spread ≤20% of individual cavity gradients. This differs from the RDR assumption that all cavities operated at 31.5 MV/m. The inclusion of the operational gradient spread, allowing acceptance of cavities achieving as low as 28 MV/m in the vertical test, increases the effective yield seen in mass production and thus produces a cost benefit. It is assumed that the average 35 MV/m (vertical test) is maintained by cavities achieving ≥42 MV/m (vertical test), which has been demonstrated. Operation with a spread in cavity gradients requires a more complex RF distribution system, and places higher demands on the low-level RF control systems, as well as requiring an additional RF power overhead of approximately 6%, all of which adds cost. However the net cost benefit is considered to be positive.

2. The RDR main linac adopted a two-tunnel solution, where one tunnel housed the accelerator (beam tunnel), while the second service tunnel housed the klystrons, modulators and other support equipment. This solution was arrived at by initial considerations of life-safety egress requirements, as well as machine operational availability. In order to reduce significantly the scope and cost of the underground construction work, a single-tunnel solution was further evaluated, and was subsequently shown to be feasible both from the perspective of life safety and availability. The evaluation process highlighted the need for site specific rather than generic solutions, and resulted in two different approaches to the RF power distribution:

   a) A Klystron Cluster Scheme (KCS), which places 10 MW multi-beam klystrons (MBK) and modulators on the surface in "clusters" every two kilometres. The RF power from a cluster is combined into a single over-moded waveguide and transported as microwave power from the surface building into the tunnel, where it is then incrementally tapped-off to feed units of three cryomodules (26 SCRF cavities). This novel solution has many attractive features, but the cost savings are partially offset by the need for additional shafts and surface buildings, as well as additional klystrons to compensate the higher RF losses in the long waveguides. Significant R&D on the distribution system is also still required.

   b) Distributed RF Source scheme (DRFS), which installs many small 850 MW modulated anode klystrons and modulators in the single beam tunnel, in a high-availability configuration, with each klystron driving 4 cavities. This solution does not require the surface buildings needed by KCS and was considered more cost-effective for mountainous topographies, such as the proposed Japanese sites. This solution was later dropped in favour of the more cost-effective and established 2007 Reference Design concept using distributed 10 MW MBKs, after more detail considerations of tunnel-construction methods in mountainous geology showed that a single wide tunnel was both cost effective and provided the same functionality as the original twin-tunnel solution.

3. undulator-based positron source was relocated from the nominal (and fixed) 150 GeV point in the main electron linac to its exit (nominal 250 GeV). This effectively consolidated all the source infrastructure in the central region of the accelerator, as well as removing the need for a long transfer line from the source to the damping rings. Low centre-of-mass energy operation ($< 300\,\mathrm{GeV}$) now requires a second electron pulse to generate positrons in a 10 Hz operation mode, which also has implications for the damping rings (half the damping time).

4. A lower beam-power parameter set with the number of bunches per pulse reduced by a factor of two ($n_b = 1312$), as compared to the nominal RDR parameter set ($n_b = 2625$). The luminosity is approximately restored by a stronger beam-beam interaction, at the cost of tighter tolerances on the beam collision. The reduced beam power (beam current) allows significant cost savings by reducing the required number of klystrons and modulators by about 33%, as well as halving the circumference of the damping rings to 3.2 km. The possibility of restoring the full 2007 RDR parameters has been maintained in the baseline design as a potential future luminosity upgrade. In particular, the damping-ring tunnel can accommodate installation of a third damping ring (second positron ring) if the higher current in the single ring is limited by electron-cloud effects.
5. The new design of damping rings can provide a 6 mm bunch length as opposed to the 9 mm length reported in the RDR. This opened up the possibility to consider a single-stage bunch compressor with a compression ratio of 20, as compared to the RDR two-stage solution. Although a cheaper solution, during the formal change-control review process, the small savings were not considered substantial enough to merit the loss of tuning range and margin of the bunch length implied by the single-stage design. Consequently the TDR remains with the two-stage concept.
6. Further integration of the positron and electron sources into a common central-region beam tunnel, together with the Beam-Delivery System, resulting in an overall simplification of civil construction in the central region.

The result of the re-baselining has produced a machine design that is both more robust, generally lower risk and more cost effective than the 2007 Reference Design. The process by which the new baseline was established followed the GDE mandate to provide a global-consensus-driven design which included all stakeholders. In particular, items 3, 4 and 5 above had potential physics-scope impact, requiring studies by the physics and detector groups.

The complete design of the ILC encompasses a mechanical and geometric description of the planned facility, a description of its function suitable for simulations, a cost estimate and an implementation plan. The aim of Design Integration is to ensure that this overall design is complete, correct, and self-consistent. During the design-integration process, the separate design results from the various accelerator systems and the technical groups are brought together. During the Technical Design Phase II, the design integration focussed on the lattice as a central description of the overall accelerator layout. First, the individual lattices of the accelerator systems were fit together with the help of treaty points that had been negotiated and agreed upon by the lattice designers and integration team. Then, using simple 3D visualisations of the lattices, the lattice geometry was optimised in order to avoid collisions between beamlines, to ensure there was sufficient space for installation of the components, and to assess whether or not it would be possible to reduce tunnel cross sections by a suitable alignment of the beamlines.

In addition to this horizontal integration work across accelerator systems, the design was integrated vertically between different technical areas. The geometrically-integrated lattice was translated into coordinate sets that were communicated to the CFS group, who based the final tunnel layout on the lattice geometry. This ensures consistency between the accelerator and tunnel geometry as well as correctness and completeness (for instance with respect to space requirements of the various dump locations). Combining the 3D visualisation of the beamlines with a 3D tunnel design facilitates further planning and optimisation with regard to installation, accessibility and egress and life safety. Figure A.4 shows a particularly complex region around the branch off of the transfer tunnel, where the beamline geometry was substantially altered in the integration process after the inspection of the 3D model of an earlier design. By sharing a common vision of the machine through 3D modelling, the involved parties can evaluate the design at an early stage and agree on necessary modifications, which may affect the tunnel layout, the lattice geometry, or both.

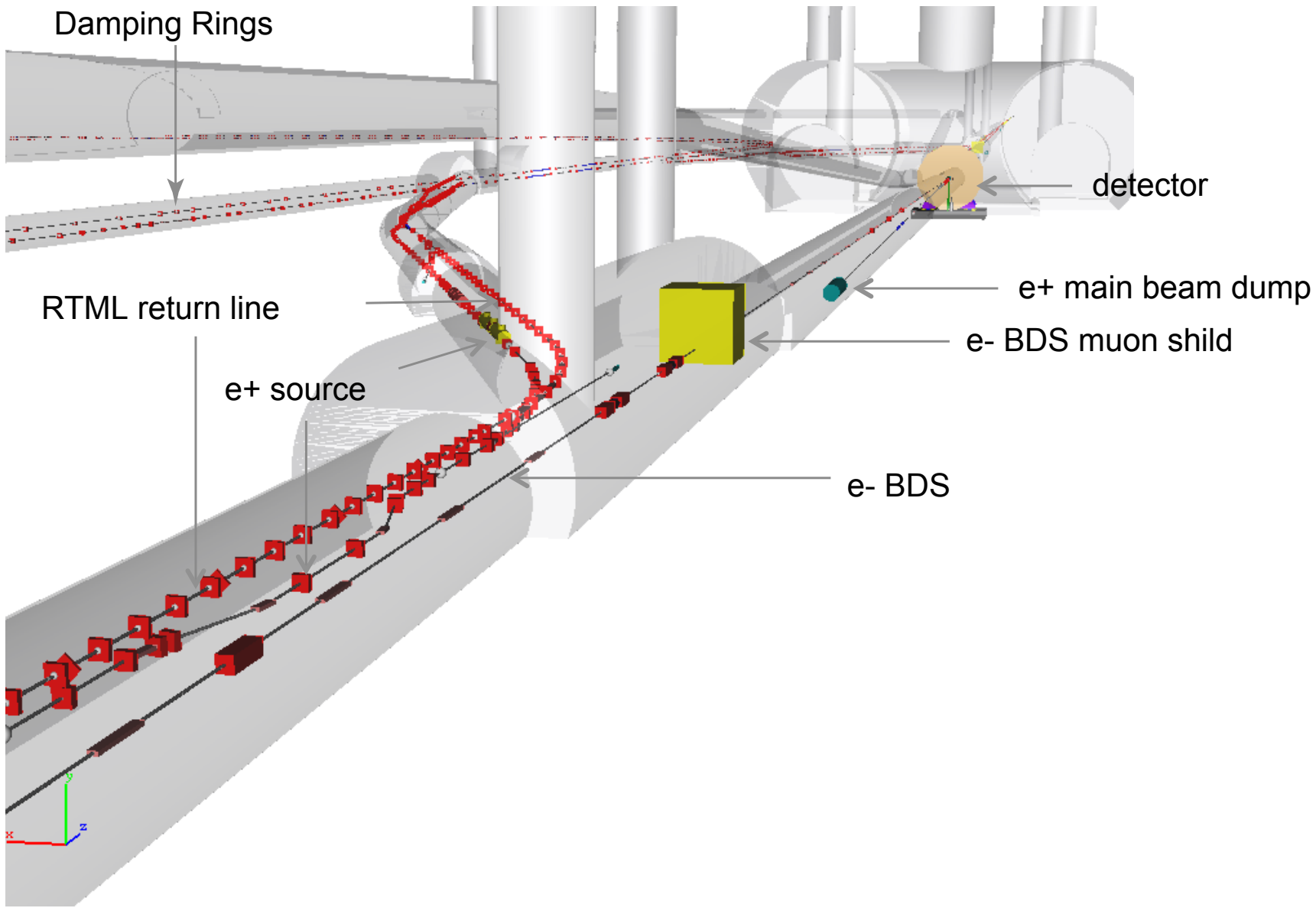

**Figure A.4** Example for design integration: the region where the transfer-tunnel branches off from the electron main tunnel towards the damping rings is shown. The European tunnel is shown together with a visualisation of the electron RTML and BDS and the positron-source beamlines. The transfer-tunnel geometry was changed in the central-region integration process in order to avoid the region around the positron main dump and the electron BDS muon shield.

More detailed designs of individual components are also incorporated into the lattice as they become available.

In particular, dimensions of complex components such as cryomodules or space-requirements of special devices such as dumps, targets, or instrumentation can be incorporated into the lattice. By structuring the lattice such that it represents the physical dimensions of the components, and by capturing the correspondence between the components and their lattice representation, it is possible to extract accurate and up-to-date component counts from the lattice, which can be provided to the installation-planning team or to the cost estimate. The availability of automated procedures for the extraction of this information makes it possible to track the effect of design changes efficiently and propagate their consequences. Thus, a process has been introduced that allows a real-time view on the various facets of the overall design to be kept while the design evolves.

In summary, design integration is an inherently important task that is essential for a coherent execution of the project. A central design office that collects and provides design information in a uniform manner under quality control and develops, establishes, and coordinates the integration process is instrumental for a successful and efficient implementation of design integration.

# List of Signatories

The following list of signatories represents a comprehensive list of those people who have contributed to the R&D and design work, for both the accelerator and the detectors, which is summarised in this report. The list also includes those people who wish to indicate their support for the next phases of the worldwide ILC effort.

It should be noted that inclusion in this list does not indicate any formal commitment by the signatories. It does not indicate commitment to the specific detector designs presented, nor exclusive support for ILC over other collider programs.

A. Abada[171], T. Abe[24], T. Abe[236], J. M. Abernathy[379], H. Abramowicz[265], A. Abusleme[231], S. Aderhold[47], O. Adeyemi[333], E. Adli[357,251], C. Adloff[164], C. Adolphsen[251], K. Afanaciev[209], M. Aguilar[31], S. Ahmad[93], A. Ahmed[382], H. Aihara[375], R. Ainsworth[237,139], S. Airi[154], M. Aizatskyi[208], T. Akagi[73], M. Akemoto[71], A. Akeroyd[367], J. Alabau-Gonzalvo[108], C. Albajar[46], J. E. Albert[379], C. Albertus[281], J. Alcaraz Maestre[31], D. Alesini[174], B. Alessandro[128], G. Alexander[265], J. P. Alexander[43], A. Alhaidari[243], N. Alipour Tehrani[33], B. C. Allanach[323], O. Alonso[311], J. Alwall[210], J. W. Amann[251], Y. Amhis[167], M. S. Amjad[167], B. Ananthanarayan[83], A. Andreazza[386,122], N. Andreev[58], L. Andricek[186], M. Anduze[172], D. Angal-Kalinin[258], N. Anh Ky[106,394], K. A. Aniol[18], K. I. Aoki[148], M. Aoki[148], H. Aoyagi[137], S. Aoyama[250], S. J. Aplin[47], R. B. Appleby[343,40], J. Arafune[96], Y. Arai[71], S. Araki[71], L. Arazi[404], A. Arbey[50], D. Ariza[47], T. Arkan[58], N. D. Arnold[7], D. Arogancia[194], F. Arteche[113], A. Aryshev[71], S. Asai[375], T. Asaka[137], T. Asaka[212], E. Asakawa[220], M. Asano[333], F. B. Asiri[58], D. Asner[225], M. Asorey[286], D. Attié[21], J. E. Augustin[169], D. B. Augustine[58], C. S. Aulakh[226], E. Avetisyan[47], V. Ayvazyan[47], N. Azaryan[142], F. Azfar[358], T. Azuma[246], O. Bachynska[47], H. Baer[355], J. Bagger[141], A. Baghdasaryan[407], S. Bai[102], Y. Bai[384], I. Bailey[40,176], V. Balagura[172,107], R. D. Ball[329], C. Baltay[405], K. Bamba[198], P. S. Bambade[167], Y. Ban[9], E. Banas[268], H. Band[384], K. Bane[251], M. Barbi[362], V. Barger[384], B. Barish[17,65], T. Barklow[251], R. J. Barlow[392], M. Barone[58,65], I. Bars[368], S. Barsuk[167], P. Bartalini[210], R. Bartoldus[251], R. Bates[332], M. Battaglia[322,33], J. Baudot[94], M. Baylac[170], P. Bechtle[295], U. Becker[185,33], M. Beckmann[47], F. Bedeschi[126], C. F. Bedoya[31], S. Behari[58], O. Behnke[47], T. Behnke[47], G. Belanger[165], S. Belforte[129], I. Belikov[94], K. Belkadhi[172], A. Bellerive[19], C. Belver Aguilar[108], A. Belyaev[367,259], D. Benchekroun[184], M. Beneke[57], M. Benoit[33], A. Benot-Morell[33,108], S. Bentvelsen[213], L. Benucci[331], J. Berenguer[31], T. Bergauer[224], S. Berge[138], E. Berger[7], J. Berger[251], C. M. U. Berggren[47], Z. Bern[318], J. Bernabeu[108], N. Bernal[295], G. Bernardi[169], W. Bernreuther[235], M. Bertucci[119], M. Besancon[21], M. Bessner[47], A. Besson[94,306], D. R. Bett[358,140], A. J. Bevan[234], A. Bhardwaj[326], A. Bharucha[333], G. Bhattacharyya[241], B. Bhattacherjee[150], B. Bhuyan[86], M. E. Biagini[174], L. Bian[102], F. Bianchi[128], O. Biebel[182], T. R. Bieler[190], C. Biino[128], B. Bilki[7,337], S. S. Biswal[221], V. Blackmore[358,140], J. J. Blaising[164], N. Blaskovic Kraljevic[358,140], G. Blazey[217], I. Bloch[48], J. Bluemlein[48], B. Bobchenko[107], T. Boccali[126], J. R. Bogart[251], V. Boisvert[237], M. Bonesini[121], R. Boni[174], J. Bonnard[168], G. Bonneaud[169], S. T. Boogert[237,139], L. Boon[233,7],

G. Boorman[237,139], E. Boos[180], M. Boronat[108], K. Borras[47], L. Bortko[48], F. Borzumati[272], M. Bosman[294], A. Bosotti[119], F. J. Botella[108], S. Bou Habib[401], P. Boucaud[171], J. Boudagov[142], G. Boudoul[89], V. Boudry[172], D. Boumediene[168], C. Bourgeois[167], A. Boveia[52], A. Brachmann[251], J. Bracinik[315], J. Branlard[47,58], B. Brau[346], J. E. Brau[356], R. Breedon[320], M. Breidenbach[251], A. Breskin[404], S. Bressler[404], V. Breton[168], H. Breuker[33], C. Brezina[295], C. Briegel[58], J. C. Brient[172], T. M. Bristow[329], D. Britton[332], I. C. Brock[295], S. J. Brodsky[251], F. Broggi[119], G. Brooijmans[42], J. Brooke[316], E. Brost[356], T. E. Browder[334], E. Brücken[69], G. Buchalla[182], P. Buchholz[301], W. Buchmuller[47], P. Bueno[111], V. Buescher[138], K. Buesser[47], E. Bulyak[208], D. L. Burke[251], C. Burkhart[251], P. N. Burrows[358,140], G. Burt[40], E. Busato[168], L. Butkowski[47], S. Cabrera[108], E. Cabruja[32], M. Caccia[389,122], Y. Cai[251], S. S. Caiazza[47,333], O. Cakir[6], P. Calabria[89], C. Calancha[71], G. Calderini[169], A. Calderon Tazon[110], S. Callier[93], L. Calligaris[47], D. Calvet[168], E. Calvo Alamillo[31], A. Campbell[47], G. I. E. Cancelo[58], J. Cao[102], L. Caponetto[89], R. Carcagno[58], M. Cardaci[201], C. Carloganu[168], S. Caron[97,213], C. A. Carrillo Montoya[123], K. Carvalho Akiba[291], J. Carwardine[7], R. Casanova Mohr[311], M. V. Castillo Gimenez[108], N. Castro[175], A. Cattai[33], M. Cavalli-Sforza[294], D. G. Cerdeno[111], L. Cerrito[234], G. Chachamis[108], M. Chadeeva[107], J. S. Chai[260], D. Chakraborty[217], M. Champion[254], C. P. Chang[201], A. Chao[251], Y. Chao[210], J. Charles[28], M. Charles[358], B. E. Chase[58], U. Chattopadhyay[81], J. Chauveau[169], M. Chefdeville[164], R. Chehab[89], A. Chen[201], C. H. Chen[203], J. Chen[102], J. W. Chen[210], K. F. Chen[210], M. Chen[330,102], S. Chen[199], Y. Chen[1], Y. Chen[102], J. Cheng[102], T. P. Cheng[351], B. Cheon[66], M. Chera[47], Y. Chi[102], P. Chiappetta[28], M. Chiba[276], T. Chikamatsu[193], I. H. Chiu[210,210], G. C. Cho[220], V. Chobanova[186], J. B. Choi[36,36], K. Choi[157], S. Y. Choi[37], W. Choi[375,248], Y. I. Choi[260], S. Choroba[47], D. Choudhury[326], D. Chowdhury[83], G. Christian[358,140], M. Church[58], J. Chyla[105], W. Cichalewski[263,47], R. Cimino[174], D. Cinca[337], J. Clark[58], J. Clarke[258,40], G. Claus[94], E. Clement[316,259], C. Clerc[172], J. Cline[187], C. Coca[206], T. Cohen[251], P. Colas[21], A. Colijn[213], N. Colino[31], C. Collard[94], C. Colledani[94], N. Collomb[258], J. Collot[170], C. Combaret[89], B. Constance[33], C. A. Cooper[58], W. E. Cooper[58], G. Corcella[174], E. Cormier[29], R. Cornat[172], P. Cornebise[167], F. Cornet[281], G. Corrado[123], F. Corriveau[187], J. Cortes[286], E. Cortina Gil[303], S. Costa[308], F. Couchot[167], F. Couderc[21], L. Cousin[94], R. Cowan[185], W. Craddock[251], A. C. Crawford[58], J. A. Crittenden[43], J. Cuevas[283], D. Cuisy[167], F. Cullinan[237], B. Cure[33], E. Currás Rivera[110], D. Cussans[316], J. Cvach[105], M. Czakon[235], K. Czuba[401], H. Czyz[365], J. D'Hondt[399], W. Da Silva[169], O. Dadoun[167], M. Dahiya[327], J. Dai[102], C. Dallapiccola[346], C. Damerell[259], M. Danilov[107], D. Dannheim[33], N. Dascenzo[47,238], S. Dasu[384], A. K. Datta[67], T. S. Datta[115], P. Dauncey[80], T. Davenne[259], J. David[169], M. Davier[167], W. De Boer[90], S. De Cecco[169], S. De Curtis[120], N. De Groot[97,213], P. De Jong[213], S. De Jong[97,213], C. De La Taille[93], G. De Lentdecker[307], S. De Santis[177], J. B. De Vivie De Regie[167], A. Deandrea[89], P. P. Dechant[49], D. Decotigny[172], K. Dehmelt[257], J. P. Delahaye[251,33], N. Delerue[167], O. Delferriere[21], F. Deliot[21], G. Della Ricca[388], P. A. Delsart[170], M. Demarteau[7], D. Demin[142], R. Dermisek[87], F. Derue[169], K. Desch[295], S. Descotes-Genon[171], A. Deshpande[252], A. Dexter[40], A. Dey[81], S. Dhawan[405], N. Dhingra[226], V. Di Benedetto[58,123], B. Di Girolamo[33], M. A. Diaz[231], A. Dieguez[311], M. Diehl[47], R. Diener[47], S. Dildick[331], M. O. Dima[206], P. Dinaucourt[167,93], M. S. Dixit[19], T. Dixit[252], L. Dixon[251], A. Djouadi[171], S. Doebert[33], M. Dohlus[47], Z. Dolezal[34], H. Dong[102], L. Dong[102], A. Dorokhov[94], A. Dosil[112], A. Dovbnya[208], T. Doyle[332], G. Doziere[94], M. Dragicevic[224], A. Drago[174], A. J. Dragt[345], Z. Drasal[34], I. Dremin[179], V. Drugakov[209], J. Duarte Campderros[110], F. Duarte Ramos[33], A. Dubey[272], A. Dudarev[142], E. Dudas[171,171], L. Dudko[180], C. Duerig[47], G. Dugan[43], W. Dulinski[94], F. Dulucq[93], L. Dumitru[206], P. J. Dunne[80], A. Duperrin[27], M. Düren[147], D. Dzahini[170], H. Eberl[224], G. Eckerlin[47], P. Eckert[297], N. R. Eddy[58],

W. Ehrenfeld[295], G. Eigen[314], S. Eisenhardt[329], L. Eklund[332], L. Elementi[58], U. Ellwanger[171], E. Elsen[47], I. Emeliantchik[209], L. Emery[7], K. Enami[71], K. Endo[71], M. Endo[375], J. Engels[47], C. Englert[49], S. Eno[345], A. Enomoto[71], S. Enomoto[197], F. Eozenou[21], R. Erbacher[320], G. Eremeev[269], J. Erler[287], R. Escribano[294], D. Esperante Pereira[108], D. Espriu[311], E. Etzion[265,33], S. Eucker[47], A. Evdokimov[336,107], E. Ezura[71], B. Faatz[47], G. Faisel[201], L. Fano[125], A. Faraggi[340], A. Fasso[251], A. Faus-Golfe[108], L. Favart[307], N. Feege[257], J. L. Feng[321], T. Ferber[47], J. Ferguson[33], J. Fernández[283], P. Fernández Martínez[108], E. Fernandez[294,293], M. Fernandez Garcia[110], J. L. Fernandez-Hernando[130], P. Fernandez-Martinez[32], J. Fernandez-Melgarejo[33], A. Ferrer[108], F. Ferri[21], S. Fichet[117], T. Fifield[347], L. Filkov[179], F. Filthaut[97,213], A. Finch[176], H. E. Fisk[58], T. Fiutowski[2], H. Flaecher[316], J. W. Flanagan[71], I. Fleck[301], M. Fleischer[47], C. Fleta[32], J. Fleury[93], D. Flores[32], M. Foley[58], M. Fontannaz[171], K. Foraz[33], N. Fornengo[128], F. Forti[126,391], B. Foster[47,140], M. C. Fouz[31], P. H. Frampton[353], K. Francis[7], S. Frank[224], A. Freitas[361], A. Frey[64], R. Frey[356], M. Friedl[224], C. Friedrich[48], M. Frigerio[163], T. Frisson[167], M. Frotin[172], R. Frühwirth[224], R. Fuchi[378], E. Fuchs[47], K. Fujii[71], J. Fujimoto[71], H. Fuke[134], B. Fuks[94,33], M. Fukuda[71], S. Fukuda[71], K. Furukawa[71], T. Furuya[71], T. Fusayasu[195], J. Fuster[108], N. Fuster[108], Y. Fuwa[95,159], A. Gaddi[33], K. Gadow[47], F. Gaede[47], R. Gaglione[164], S. Galeotti[126], C. Gallagher[356], A. A. Gallas Torreira[112], L. Gallin-Martel[170], A. Gallo[174], D. Gamba[140,33], D. Gamba[128], J. Gao[102], Y. Gao[24], P. H. Garbincius[58], F. Garcia[69], C. Garcia Canal[288], J. E. Garcia Navarro[108], P. Garcia-Abia[31], J. J. Garcia-Garrigos[108], L. Garcia-Tabares[31], C. García[108], J. V. García Esteve[286], I. García García[108], S. K. Garg[408], L. Garrido[311], E. Garutti[333], T. Garvey[167,261], M. Gastal[33], F. Gastaldi[172], C. Gatto[58,123], N. Gaur[326], D. Gavela Pérez[31], P. Gay[168], M. B. Gay Ducati[109], L. Ge[251], R. Ge[102], A. Geiser[47], A. Gektin[100], A. Gellrich[47], M. H. Genest[170], R. L. Geng[269], S. Gentile[387,127], A. Gerbershagen[33,358], R. Gerig[7], S. German[111], H. Gerwig[33], S. Ghazaryan[47], P. Ghislain[169], D. K. Ghosh[81], S. Ghosh[115], S. Giagu[387,127], L. Gibbons[43], S. Gibson[139,33], V. Gilewsky[143], A. Gillespie[328], F. Gilman[20], B. Gimeno Martínez[108], D. M. Gingrich[312,279], C. M. Ginsburg[58], D. Girard[164], J. Giraud[170], G. F. Giudice[33], L. Gladilin[180], P. Gladkikh[208], C. J. Glasman[343,40], R. Glattauer[224], N. Glover[49], J. Gluza[365], K. Gnidzinska[263], R. Godbole[83], S. Godfrey[19], F. Goertz[54], M. Goffe[94], N. Gogitidze[179,47], J. Goldstein[316], B. Golob[144,341], G. Gomez[110], V. Goncalves[290], R. J. Gonsalves[256], I. González[283], S. González De La Hoz[108], F. J. González Sánchez[110], G. Gonzalez Parra[294], S. Gopalakrishna[103], I. Gorelov[352], D. Goswami[86], S. Goswami[229], T. Goto[71], K. Gotow[398], P. Göttlicher[47], M. Götze[385], A. Goudelis[165], P. Goudket[258], S. Gowdy[33], O. A. Grachov[309], N. A. Graf[251], M. Graham[251], A. Gramolin[15], R. Granier De Cassagnac[172], P. Grannis[257], P. Gras[21], M. Grecki[47], T. Greenshaw[339], D. Greenwood[181], C. Grefe[33], M. Grefe[111], I. M. Gregor[47], D. Grellscheid[49], G. Grenier[89], M. Grimes[316], C. Grimm[58], O. Grimm[53], B. Grinyov[100], B. Gripaios[323], K. Grizzard[141], A. Grohsjean[47], C. Grojean[294,33], J. Gronberg[178], D. Grondin[170], S. Groote[370], P. Gros[240], M. Grunewald[310], B. Grzadkowski[382], J. Gu[102], M. Guchait[262], S. Guiducci[174], E. Guliyev[172], J. Gunion[320], C. Günter[47], C. Gwilliam[339], N. Haba[75], H. Haber[322], M. Hachimine[197], Y. Haddad[172], L. Hagge[47], M. Hagihara[378], K. Hagiwara[71,158], J. Haley[216], G. Haller[251], J. Haller[333], K. Hamaguchi[375], R. Hamatsu[276], G. Hamel De Monchenault[21], L. L. Hammond[58], P. Hamnett[47], L. Han[364], T. Han[361], K. Hanagaki[223], J. D. Hansen[211], K. Hansen[47], P. H. Hansen[211], X. Q. Hao[70], K. Hara[71], K. Hara[378], T. Hara[71], D. Harada[83], K. Harada[161], K. Harder[259], T. Harion[297], R. V. Harlander[385], E. Harms[58], M. Harrison[13], O. Hartbrich[47,385], A. Hartin[47], T. Hartmann[300], J. Harz[47], S. Hasegawa[197], T. Hasegawa[71], Y. Hasegawa[248], M. Hashimoto[38], T. Hashimoto[62], C. Hast[251], S. Hatakeyama[135], J. M. Hauptman[118], M. Hauschild[33], M. Havranek[105], C. Hawkes[315], T. Hayakawa[197], H. Hayano[71], K. Hayasaka[198], M. Hazumi[71,253], H. J. He[24], C. Hearty[317,104], H. F. Heath[316],

T. Hebbeker[235], M. Heck[90], V. Hedberg[183], D. Hedin[217], S. M. Heindl[90], S. Heinemeyer[110], I. Heinze[47], A. Hektor[205], S. Henrot-Versille[167], O. Hensler[47], A. Heo[23], J. Herbert[258], G. Herdoiza[138], B. Hermberg[47], J. J. Hernández-Rey[108], M. J. Herrero[111], B. Herrmann[165], A. Hervé[384], J. Hewett[251], S. Hidalgo[32], B. Hidding[333,318], N. Higashi[375], N. Higashi[71], T. Higo[71], E. Higón Rodríguez[108], T. Higuchi[150], M. Hildreth[354], C. T. Hill[58], S. Hillert[295], S. Hillier[315], T. Himel[251], A. Himmi[94], S. Himori[272], Z. Hioki[374], B. Hippolyte[94], T. Hiraki[159], K. Hirano[136], S. Hirano[197], K. Hirata[71], T. Hirose[276], M. Hirsch[108], J. Hisano[197], P. M. Ho[210], A. Hoang[302], A. Hocker[58], A. Hoecker[33], M. Hoeferkamp[352], M. Hoffmann[47], W. Hollik[186], K. Homma[72], Y. Homma[154], S. Honda[378], T. Honda[71], Y. Honda[71], N. T. Hong Van[106], K. Honkavaara[47], T. Honma[71], T. Hori[236], T. Horiguchi[272], Y. Horii[197], A. Horio[196], R. Hosaka[377], Y. Hoshi[271], H. Hoshino[197], K. Hosoyama[71], J. Y. Hostachy[170], G. W. Hou[210], M. Hou[102], A. Hoummada[184], M. S. Hronek[58], T. Hu[102], C. Hu-Guo[94], M. Huang[24], T. Huang[102], E. Huedem[58], F. Hügging[295], J. L. Hugon[330], C. Hugonie[173], K. Huitu[335], P. Q. Hung[380], C. Hunt[80], U. Husemann[90], G. Hussain[24], D. Hutchcroft[339], Y. Hyakutake[79], J. C. Ianigro[89], L. E. Ibanez[111], M. Ibe[96], M. Idzik[2], H. Igarashi[78], Y. Igarashi[71], K. Igi[236], A. Ignatenko[209], O. Igonkina[213], T. Iijima[198,197], M. Iinuma[73], Y. Iiyama[20], H. Ikeda[134], K. Ikeda[71], K. Ikematsu[301], J. I. Illana[281], V. A. Ilyin[207,180], A. Imhof[333], T. Inagaki[236], T. Inagaki[197], K. Inami[197], S. Inayoshi[248], K. Inoue[161], A. Irles[108], S. Isagawa[71], N. Ishibashi[378], A. Ishida[375], K. Ishida[212], N. Ishihara[71], S. Ishihara[77], K. Ishii[71], A. Ishikawa[272], K. Ishikawa[375], K. I. Ishikawa[72], K. Ishikawa[75], T. Ishikawa[71], M. Ishitsuka[275], K. Ishiwata[17], G. Isidori[174], A. Ismail[251], S. Iso[71], T. Isogai[197], C. Issever[358], K. Itagaki[272], T. Itahashi[223], A. Ito[275], F. Ito[378], S. Ito[272], R. Itoh[71], E. Itou[71], M. I. Ivanyan[26], G. Iwai[71], S. Iwamoto[375], T. Iwamoto[116], H. Iwasaki[71], M. Iwasaki[71], Y. Iwashita[95], S. Iwata[71], S. Iwata[276], T. Izubuchi[13,236], Y. Izumiya[272], S. Jablonski[401], F. Jackson[258], J. A. Jacob[316], M. Jacquet[167], P. Jain[40], P. Jaiswal[59], W. Jalmuzna[263], E. Janas[401], R. Jaramillo Echeverría[110], J. Jaros[251], D. Jatkar[67], D. Jeans[375], R. Jedziniak[58], M. J. Jenkins[176,40], K. Jensch[47], C. P. Jessop[354], T. Jezynski[47], M. Jimbo[35], S. Jin[102], O. Jinnouchi[275], M. D. Joergensen[211], A. S. Johnson[251], S. Jolly[309], D. T. Jones[340], J. Jones[258,40], R. M. Jones[343,40], L. Jönsson[183], N. Joshi[237], C. K. Jung[257,150], N. Juntong[343,40], A. Juste[88,294], W. Kaabi[167], M. Kadastik[205], M. Kado[167,33], K. Kadota[197], E. Kajfasz[27], R. Kajikawa[197], Y. Kajiura[197], M. Kakizaki[377], E. Kako[71], H. Kakuhata[377], H. Kakuno[276], A. Kalinin[258], J. Kalinowski[382], G. E. Kalmus[259], K. Kamada[47], J. Kaminski[295], T. Kamitani[71], Y. Kamiya[116], Y. Kamiya[71], R. Kammering[47], T. Kamon[266], J. I. Kamoshita[60], T. Kanai[275], S. Kananov[265], K. Kanaya[378], M. Kaneda[33], T. Kaneko[71,253], S. Kanemura[377], K. Kaneta[75], W. Kang[102], D. Kanjilal[115], K. Kannike[205], F. Kapusta[169], D. Kar[332], P. Karataev[237,139], P. E. Karchin[403], D. Karlen[379,279], S. Karstensen[47], Y. Karyotakis[164], M. Kasemann[47], V. S. Kashikhin[58], S. Kashiwagi[273], A. Kataev[98], V. Katalev[48], Y. Kataoka[116], N. Katayama[150], R. Katayama[375], E. Kato[272], K. Kato[155], S. Kato[71], Y. Kato[153], T. Katoh[71], A. Kaukher[47], S. Kawabata[71], S. I. Kawada[73], K. Kawagoe[161], M. Kawai[71], T. Kawamoto[116], H. Kawamura[71], M. Kawamura[71], Y. Kawamura[248], S. Kawasaki[71], T. Kawasaki[212], H. Kay[47], S. Kazama[375], L. Keegan[111], J. Kehayias[150], L. Keller[251], M. A. Kemp[251], J. J. Kempster[237], C. Kenney[251], I. Kenyon[315], R. Kephart[58], J. Kerby[7], K. Kershaw[33], J. Kersten[333], K. Kezzar[151], V. G. Khachatryan[26], M. A. Khan[23], S. A. Khan[242], Y. Khoulaki[184], V. Khoze[49], H. Kichimi[71], R. Kieffer[33], C. Kiesling[186], M. Kikuchi[377], Y. Kikuta[71], M. Killenberg[47], C. S. Kim[408], D. W. Kim[63], D. Kim[23], E. J. Kim[36,36], E. S. Kim[23], G. Kim[23], H. S. Kim[23], H. D. Kim[245], J. Kim[63], S. H. Kim[378], S. K. Kim[245], S. G. Kim[87], Y. I. Kim[358,140], Y. Kimura[71], R. E. Kirby[251], F. Kircher[21], Y. Kishimoto[96], L. Kisslinger[20], T. Kitahara[375], R. Kitano[272], Y. Kiyo[146,71], C. Kleinwort[47], W. Klempt[33], P. M. Kluit[213], V. Klyukhin[180,33], M. Knecht[28], J. L. Kneur[163], B. A. Kniehl[333], K. Ko[251], P. Ko[158], D. Kobayashi[275], M. Kobayashi[71], N. Kobayashi[71],

A. Mehta[339], B. Mele[127], R. E. Meller[43], I. A. Melzer-Pellmann[47], L. Men[102], G. Mendiratta[83], Z. Meng[316], M. H. Merk[213,400], M. Merkin[180], A. Merlos[32], L. Merminga[279], A. B. Meyer[47], A. Meyer[235], N. Meyners[47], Z. Mi[102], P. Michelato[119], S. Michizono[71], S. Mihara[71], A. Mikhailichenko[43], D. J. Miller[309], C. Milstene[403], Y. Mimura[210], D. Minic[398], L. Mirabito[89], S. Mishima[387], T. Misumi[13], W. A. Mitaroff[224], T. Mitsuhashi[71], S. Mitsuru[71], K. Miuchi[154], K. Miyabayashi[200], A. Miyamoto[71], H. Miyata[212], Y. Miyazaki[161], T. Miyoshi[71], R. Mizuk[107], K. Mizuno[3], U. Mjörnmark[183], J. Mnich[47], G. Moeller[47], W. D. Moeller[47], K. Moenig[48], K. C. Moffeit[251], P. Mohanmurthy[269], G. Mohanty[262], L. Monaco[119], S. Mondal[81], C. Monini[170], H. Monjushiro[71], G. Montagna[359,124], S. Monteil[168], G. Montoro[298], I. Montvay[47], F. Moortgat[53], G. Moortgat-Pick[333,47], P. Mora De Freitas[172], C. Mora Herrera[30], G. Moreau[171], F. Morel[94], A. Morelos-Pineda[280], M. Moreno Llacer[108], S. Moretti[367,259], V. Morgunov[47,107], T. Mori[71], T. Mori[272], T. Mori[116], Y. Morita[71], S. Moriyama[96,150], L. Moroni[121], Y. Morozumi[71], H. G. Moser[186], A. Moszczynski[268], K. Motohashi[275], T. Moulik[249], G. Moultaka[163], D. Moya Martin[110], S. K. Mtingwa[215], G. S. Muanza[27], M. Mühlleitner[91], A. Mukherjee[85], S. Mukhopadhyay[241], M. Mulders[33], D. Muller[251], F. Müller[47], T. Müller[90], V. S. Mummidi[83], A. Münnich[47], C. Munoz[111], F. J. Muñoz Sánchez[110], H. Murayama[375,319], R. Murphy[7], G. Musat[172], A. Mussgiller[47], R. Muto[71], T. Nabeshima[377], K. Nagai[71], K. Nagai[378], S. Nagaitsev[58], T. Nagamine[272], K. Nagano[71], K. Nagao[71], Y. Nagashima[223], S. C. Nahn[185], P. Naik[316], D. Naito[159], T. Naito[71], H. Nakai[71], K. Nakai[71,223], Y. Nakai[161], Y. Nakajima[177], E. Nakamura[71], H. Nakamura[71], I. Nakamura[71], S. Nakamura[159], S. Nakamura[71], T. Nakamura[116], K. Nakanishi[71], E. Nakano[222], H. Nakano[212], H. Nakano[272], Y. Namito[71], W. Namkung[230], H. Nanjo[159], C. D. Nantista[251], O. Napoly[21], Y. Nara[4], T. Narazaki[272], S. Narita[131], U. Nauenberg[324], T. Naumann[48], S. Naumann-Emme[47], J. Navarro[108], A. Navitski[47], H. Neal[251], K. Negishi[272], K. Neichi[270], C. A. Nelson[255], T. K. Nelson[251], S. Nemecek[105], M. Neubert[138], R. Neuhaus[218], L. J. Nevay[237], D. M. Newbold[316,259], O. Nezhevenko[58], F. Nguyen[175], M. Nguyen[172], M. N. Nguyen[251], T. T. Nguyen[106], R. B. Nickerson[358], O. Nicrosini[124], C. Niebuhr[47], J. Niehoff[58], M. Niemack[43], U. Nierste[92], H. Niinomi[159], I. Nikolic[169], H. P. Nilles[295], S. Nishida[71], H. Nishiguchi[71], K. Nishiwaki[67], O. Nitoh[277], L. Niu[24], R. Noble[251], M. Noji[71], M. Nojiri[71,150], S. Nojiri[197,198], D. Nölle[47], A. Nomerotski[358], M. Nomura[135], T. Nomura[201], Y. Nomura[319,177], C. Nonaka[198], J. Noonan[7], E. Norbeck[337], Y. Nosochkov[251], D. Notz[47], O. Novgorodova[48,12], A. Novokhatski[251], J. A. Nowak[349], M. Nozaki[71], K. Ocalan[192], J. Ocariz[169], S. Oda[161], A. Ogata[71], T. Ogawa[248], T. Ogura[248], A. Oh[343], S. K. Oh[156], Y. Oh[23], K. Ohkuma[61], T. Ohl[145], Y. Ohnishi[71], K. Ohta[189], M. Ohta[71,253], S. Ohta[71,253], N. Ohuchi[71], K. Oishi[161], R. Okada[71], Y. Okada[71,253], T. Okamura[71], H. Okawa[13], T. Okugi[71], T. Okui[59], K. I. Okumura[161], Y. Okumura[52,58], L. Okun[107], H. Okuno[236], C. Oleari[390], C. Oliver[31], B. Olivier[186], S. L. Olsen[245], M. Omet[253,71], T. Omori[71], Y. Onel[337], H. Ono[214], Y. Ono[272], D. Onoprienko[251], Y. Onuki[116,150], P. Onyisi[371], T. Oogoe[71], Y. Ookouchi[159], W. Ootani[116], M. Oreglia[52], M. Oriunno[251], M. C. Orlandea[206], J. Orloff[168], M. Oroku[375,71], R. S. Orr[376], J. Osborne[33], A. Oskarsson[183], P. Osland[314], A. Osorio Oliveros[282], L. Österman[183], H. Otono[223], M. Owen[343], Y. Oyama[71], A. Oyanguren[108], K. Ozawa[71,375], J. P. Ozelis[58,191], D. Ozerov[47], G. Pásztor[304,99], H. Padamsee[43], C. Padilla[294], C. Pagani[119,386], R. Page[316], R. Pain[169], S. Paktinat Mehdiabadi[101], A. D. Palczewski[269], S. Palestini[33], F. Palla[126], M. Palmer[58], F. Palomo Pinto[285], W. Pan[102], G. Pancheri[174], M. Pandurovic[396], O. Panella[125], A. Pankov[227], Y. Papaphilippou[33], R. Paparella[119], A. Paramonov[7], E. K. Park[75], H. Park[23], S. I. Park[23], S. Park[260], S. Park[373], W. Park[23], A. Parker[323], B. Parker[13], C. Parkes[343], V. Parma[33], Z. Parsa[13], R. Partridge[251], S. Pastor[108], E. Paterson[251], M. Patra[83], J. R. Patterson[43], M. Paulini[20], N. Paver[129], S. Pavy-Bernard[167],

B. Pawlik[268], A. Pérez Vega-Leal[285], B. Pearson[355], J. S. Pedersen[211], A. Pedicini[397],
S. Pedraza López[108], G. Pei[102], S. Pei[102], G. Pellegrini[32], A. Pellegrino[213], S. Penaranda[286],
H. Peng[364], X. Peng[102], M. Perelstein[43], E. Perez[112], M. A. Perez-Garcia[284,114],
M. Perez-Victoria[281], S. Peris[293], D. Perret-Gallix[164], H. Perrey[47], T. M. Perry[384], M. E. Peskin[251],
P. Petagna[33], R. Y. Peters[64,47], T. C. Petersen[211], D. P. Peterson[43], T. Peterson[58],
E. Petrakou[210], A. A. Petrov[403], A. Petrukhin[89,107], S. Pfeiffer[47], H. Pham[94], K. H. Phan[395,71],
N. Phinney[251], F. Piccinini[124], A. Pich[108], R. Pichai[85], J. Piedra[283], J. Piekarski[401], A. Pierce[348],
P. Pierini[119], N. Pietsch[333,47], A. Pineda[293], J. Pinfold[312,152], A. Piotrowski[263], Y. Pischalnikov[58],
R. Pittau[281], M. Pivi[251], W. Placzek[132], T. Plehn[296], M. A. Pleier[13], M. Poelker[269], L. Poggioli[167],
I. Pogorelsky[13], V. Poireau[164], M. E. Pol[30], I. Polak[105], F. Polci[169], M. Polikarpov[107],
T. Poll[316,259], M. W. Poole[258,40], W. Porod[145], F. C. Porter[17], S. Porto[333], J. Portolés[108],
R. Pöschl[167], S. Poss[33], C. T. Potter[356], P. Poulose[86], K. T. Pozniak[401], V. Prahl[47], R. Prepost[384],
C. Prescott[251], D. Price[87], T. Price[315], P. S. Prieto[58], D. Protopopescu[332], D. Przyborowski[2],
K. Przygoda[263], H. Przysiezniak[164], F. Ptochos[325], J. Puerta-Pelayo[31], C. Pulvermacher[90],
M. Purohit[366], Q. Qin[102], F. Qiu[102], H. Qu[102], A. Quadt[64], G. Quast[90], D. Quirion[32], M. Quiros[88],
J. Rademacker[316], R. Rahmat[350], S. Rai[67], M. Raidal[205], S. Rakshit[84], M. Ramilli[333], F. Rarbi[170],
P. Ratoff[176], T. Raubenheimer[251], M. Rauch[91], L. Raux[93], G. Raven[400,213], P. Razis[325], V. Re[124],
S. Redford[33], C. E. Reece[269], I. Reichel[177], A. Reichold[358,140], P. Reimer[105], M. Reinecke[47],
A. Rekalo[100], J. Repond[7], J. Resta-Lopez[108], J. Reuter[47], J. T. Rhee[156], P. M. Ribeiro Cipriano[47],
A. Ribon[33], G. Ricciardi[292,123], F. Richard[167], E. Richter-Was[132], G. Riddone[33], S. Riemann[48],
T. Riemann[48], M. Rijssenbeek[257], K. Riles[348], C. Rimbault[167], R. Rimmer[269], S. D. Rindani[229],
A. Ringwald[47], L. Rinolfi[33], I. Ripp-Baudot[94], I. Riu[294], T. G. Rizzo[251], P. Robbe[167],
J. Roberts[140,33], A. Robson[332], G. Rodrigo[108], P. Rodriguez[251], P. Rodriguez Perez[112],
K. Rolbiecki[111], P. Roloff[33], R. S. Romaniuk[401], E. Romero Adam[108], A. Ronzhin[58], L. Roos[169],
E. Ros[108], A. Rosca[47], C. Rosemann[47], J. Rosiek[382], M. C. Ross[251], R. Rossmanith[90], S. Roth[235],
J. Rouëné[167], A. Rowe[58], P. Rowson[251], A. Roy[115], L. Royer[168], P. Royole-Degieux[93], C. Royon[21],
A. Rozanov[27], M. Ruan[172], D. L. Rubin[43], I. Rubinskiy[47], R. Rückl[145], R. Ruiz[361],
R. Ruiz De Austri[108], P. Ruiz Valls[108], P. Ruiz-Femenía[108], A. Ruiz-Jimeno[110], R. Ruland[251],
V. Rusinov[107], J. J. Russell[251], I. Rutkowski[401], V. Rybnikov[47], A. Ryd[43], A. Sabio Vera[111],
B. Sabirov[142], J. J. Saborido Silva[112], H. F. W. Sadrozinski[322], T. Saeki[71], B. Safarzadeh[101],
P. Saha[305], H. Sahoo[7], A. Sailer[33], N. Saito[71], T. Saito[272], T. Sakaguchi[13], H. Sakai[71], K. Sakai[71],
K. Sakaue[402], K. Sakurai[47], R. Salerno[172], J. Salfeld-Nebgen[47], J. Salt[108], L. Sanchez[31],
M. A. Sanchis Lozano[108], J. Sandweiss[405], A. Santa[377], A. Santagata[286], A. Santamaria[108],
P. Santorelli[292], T. Sanuki[272], A. A. Sapronov[142], M. Sasaki[96], H. Sato[248], N. Sato[71], Y. Sato[272],
M. Satoh[71], E. Sauvan[164], V. Saveliev[238,47], A. Savoy-Navarro[166,126], M. Sawabe[71], R. Sawada[116],
H. Sawamura[402], L. Sawyer[181], O. Schäfer[300,47], R. Schäfer[47], J. Schaffran[47], T. Schalk[322,251],
R. D. Schamberger[257], J. Scheirich[34], G. Schierholz[47], F. P. Schilling[90], F. Schirra[89],
F. Schlander[47], H. Schlarb[47], D. Schlatter[33], P. Schleper[333], J. L. Schlereth[7], R. D. Schlueter[177],
C. Schmidt[47], U. Schneekloth[47], S. Schnetzer[239], T. Schoerner-Sadenius[47], M. Schram[225],
H. J. Schreiber[48], S. Schreiber[47], K. P. Schüler[47], D. Schulte[33], H. C. Schultz-Coulon[297],
M. Schumacher[5], S. Schumann[64], B. A. Schumm[322], M. H. Schune[167], S. Schuwalow[333,48],
C. Schwanda[224], C. Schwanenberger[343], F. Schwartzkopff[295], D. J. Scott[258,58], F. Sefkow[47,33],
A. Segui[286], N. Seguin-Moreau[93], S. Seidel[352], Y. Seiya[222], J. Sekaric[338], K. Seki[197], S. Sekmen[33],
S. Seletskiy[13], S. Sen[337], E. Senaha[158], K. Senyo[406], S. Senyukov[94], I. Serenkova[227],
D. A. Sergatskov[58], H. Sert[47,333], D. Sertore[119], A. Seryi[358,140], O. Seto[74], R. Settles[186],
P. Sha[102], S. Shahid[301], A. Sharma[33], G. Shelkov[142], W. Shen[297], J. C. Sheppard[251], M. Sher[41],
C. Shi[102], H. Shi[102], T. Shidara[71], W. Shields[237,139], M. Shimada[71], H. Shimizu[71], Y. Shimizu[272],

M. Shimojima[195], S. Shimojima[276], T. Shindou[155], N. Shinoda[272], Y. Shinzaki[272], M. Shioden[78,71], I. Shipsey[233], S. Shirabe[161], M. J. Shirakata[71], T. Shirakata[71], G. Shirkov[142], T. Shishido[71], T. Shishido[71], J. G. Shiu[210], R. Shivpuri[326], R. Shrock[257], T. Shuji[71], N. Shumeiko[209], B. Shuve[228,188], P. Sicho[105], A. M. Siddiqui[133], P. Sievers[33], D. Sikora[401], D. A. Sil[86], F. Simon[186], N. B. Sinev[356], W. Singer[47], X. Singer[47], B. K. Singh[10], R. K. Singh[82], N. Sinha[103], R. Sinha[103], K. Sinram[47], T. Sinthuprasith[14], P. Skubic[355], R. Sliwa[167], I. Smiljanic[396], J. R. Smith[373,7], J. C. Smith[251,43], S. R. Smith[251], J. Smolík[105,45], J. Snuverink[237,139], B. Sobloher[47], J. Sola[311], C. Soldner[186,57], S. Soldner-Rembold[343], D. Son[23], H. S. Song[260], N. Sonmez[51], A. Sopczak[44], D. E. Soper[356], P. Spagnolo[126], S. Spannagel[47], M. Spannowsky[49], A. Sparkes[329], C. M. Spencer[251], H. Spiesberger[138], M. Spira[261], M. Stahlhofen[47], M. Stanescu-Bellu[48], M. Stanitzki[47], S. Stapnes[33], P. Starovoitov[209], F. Staufenbiel[48], L. Steder[47], M. Steder[47], A. Steen[89], G. Steinbrueck[333], M. Steinhauser[92], F. Stephan[48], W. Stephen[237], S. Stevenson[358], I. Stewart[185], D. Stöckinger[264], H. Stoeck[369], M. Strauss[355], S. Striganov[58], D. M. Strom[356], R. Stromhagen[47], J. Strube[33], A. Strumia[205], G. Stupakov[251], N. Styles[47], D. Su[251], F. Su[102], S. Su[313], J. Suarez Gonzalez[209], Y. Sudo[161], T. Suehara[116], F. Suekane[274], Y. Suetsugu[71], R. Sugahara[71], A. Sugamoto[220], H. Sugawara[71], Y. Sugimoto[71], A. Sugiyama[240], H. Sugiyama[377], M. K. Sullivan[251], Y. Sumino[272], T. Sumiyoshi[276,71], H. Sun[102], M. Sun[20], X. Sun[170], Y. Sun[102], Y. Susaki[197], T. Suwada[71], A. Suzuki[71], S. Suzuki[240], Y. Suzuki[71], Y. Suzuki[73], Z. Suzuki[272], K. Swientek[2], C. Swinson[13], Z. M. Szalata[251], B. Szczezepanski[47], M. Szelezniak[94], J. Szewinski[202], A. Sznajder[289], L. Szymanowski[202], H. Tabassam[329], K. Tackmann[47], M. Taira[71], H. Tajima[197,251], F. Takahashi[272], R. Takahashi[71], R. Takahashi[75], T. Takahashi[73], Y. Takahashi[197], K. Takata[71], F. Takayama[160], Y. Takayasu[75], H. Takeda[154], S. Takeda[71], T. Takeshita[248], A. Taketani[236], Y. Takeuchi[378], T. Takimi[262], Y. Takubo[71], Y. Tamashevich[47], M. Tamsett[181], M. Tanabashi[198,197], T. Tanabe[116], G. Tanaka[161], M. M. Tanaka[71], M. Tanaka[223], R. Tanaka[73], H. Taniuchi[377], S. Tapprogge[138], E. Tarkovsky[107], M. A. Tartaglia[58], X. R. Tata[334], T. Tauchi[71], M. Tawada[71], G. Taylor[347], A. M. Teixeira[168], V. I. Telnov[15,219], P. Tenenbaum[251], E. Teodorescu[206], S. Terada[71], Y. Teramoto[222], H. Terao[200], A. Terashima[71], S. Terui[71], N. Terunuma[71], M. Terwort[47], M. Tesar[186], F. Teubert[33], T. Teubner[340], R. Teuscher[376], T. Theveneaux-Pelzer[168], D. Thienpont[93,172], J. Thom-Levy[43], M. Thomson[323], J. Tian[71], X. Tian[366], M. Tigner[43], J. Timmermans[213], V. Tisserand[164], M. Titov[21], S. Tjampens[164], K. Tobe[197], K. Tobioka[150,319], K. Toda[278], M. Toda[71], N. Toge[71], J. Tojo[161], K. Tokushuku[71], T. Toma[49], R. Tomas[33], T. Tomita[161], A. Tomiya[223], M. Tomoto[197,198], K. Toms[352], M. Tonini[47], F. Toral[31], E. Torrence[356], E. Torrente-Lujan[33], N. Toumbas[325], C. Touramanis[339], F. Toyoda[161], K. Toyomura[71], G. Trahern[55], T. H. Tran[172], W. Treberspurg[224], J. Trenado[311], M. Trimpl[58], S. Trincaz-Duvoid[169], M. Tripathi[320], W. Trischuk[376], M. Trodden[360], G. Trubnikov[142], H. C. Tsai[39], J. F. Tsai[210], K. H. Tsao[336], R. Tschirhart[58], E. Tsedenbaljir[210], S. Y. Tseng[201], T. Tsuboyama[71], A. Tsuchiya[250], K. Tsuchiya[71], T. Tsukamoto[71], K. Tsumura[197], S. Tsuno[71], T. Tsurugai[189], T. Tsuyuki[96], B. Tuchming[21], P. V. Tyagi[68,254], I. Tyapkin[142], M. Tytgat[331], K. Uchida[295], F. Uchiyama[71], Y. Uchiyama[116], S. Uehara[71], H. Ueno[161], K. Ueno[71], K. Ueno[71], K. Ueshima[274], Y. Uesugi[73], N. Ujiie[71], F. Ukegawa[378], N. Ukita[378], M. Ullán[32], H. Umeeda[72], K. Umemori[71], Y. Unno[66], S. Uozumi[23], J. Urakawa[71], A. M. Uranga[111], J. Urresti[32], A. Ushakov[333], I. Ushiki[272], Y. Ushiroda[71], A. V[83], P. Vázquez Regueiro[112], L. Vacavant[27], G. Valencia[118], L. Valery[168], I. Valin[94], J. W. Valle[108], C. Vallee[27], N. Van Bakel[213], H. Van Der Graaf[213], N. Van Der Kolk[167], E. Van Der Kraaij[33], B. Van Doren[338], B. Van Eijk[213], R. Van Kooten[87], W. T. Van Oers[279], D. Vanegas[108], P. Vanhoefer[186], P. Vankov[47], P. Varghese[58], A. Variola[167], R. Varma[85], G. Varner[334], G. Vasileiadis[162], A. Vauth[47], J. Velthuis[316], S. K. Vempati[83], V. Vento[108], M. Venturini[177], M. Verderi[172], P. Verdier[89], A. Verdugo[31],

A. Vicente[171], J. Vidal-Perona[108], H. L. R. Videau[172], I. Vila[110], X. Vilasis-Cardona[299], E. Vilella[311], A. Villamor[32], E. G. Villani[259], J. A. Villar[286], M. A. Villarejo Bermúdez[108], D. Vincent[169], P. Vincent[169], J. M. Virey[28], A. Vivoli[58], V. Vogel[47], R. Volkenborn[47], O. Volynets[47], F. Von Der Pahlen[110], E. Von Toerne[295], B. Vormwald[47], A. Voronin[180], M. Vos[108], J. H. Vossebeld[339], G. Vouters[164], Y. Voutsinas[94,47], V. Vrba[105,45], M. Vysotsky[107], D. Wackeroth[256], A. Wagner[47], C. E. Wagner[7,52], R. Wagner[7], S. R. Wagner[324], W. Wagner[385], J. Wagner-Kuhr[90], A. P. Waite[251], M. Wakayama[197], Y. Wakimoto[276], R. Walczak[358,140], R. Waldi[300], D. G. E. Walker[251], N. J. Walker[47], M. Walla[47], C. J. Wallace[49], S. Wallon[171,393], D. Walsh[328], S. Walston[178], W. A. T. Wan Abdullah[342], D. Wang[102], G. Wang[102], J. Wang[251], L. Wang[251], L. Wang[52], M. H. Wang[251], M. Z. Wang[210], Q. Wang[102], Y. Wang[102], Z. Wang[24], R. Wanke[138], C. Wanotayaroj[356], B. Ward[8], D. Ward[323], B. Warmbein[47], M. Washio[402], K. Watanabe[71], M. Watanabe[212], N. Watanabe[71], T. Watanabe[155], Y. Watanabe[71], S. Watanuki[272], Y. Watase[71], N. K. Watson[315], G. Watts[383], M. M. Weber[90], H. C. Weddig[47], H. Weerts[7], A. W. Weidemann[251], G. Weiglein[47], A. Weiler[47], S. Weinzierl[138], H. Weise[47], A. Welker[138], N. Welle[47], J. D. Wells[33,348], M. Wendt[58,33], M. Wenskat[47], H. Wenzel[58], N. Wermes[295], U. Werthenbach[301], W. Wester[58], L. Weuste[186,57], A. White[373], G. White[251], K. H. Wichmann[47], M. Wielers[183], R. Wielgos[58], W. Wierba[202], T. Wilksen[47], S. Willocq[346], F. F. Wilson[259], G. W. Wilson[338], P. B. Wilson[251], M. Wing[309], M. Winter[94], K. Wittenburg[47], P. Wittich[43], M. Wobisch[181], A. Wolski[339,40], M. D. Woodley[251], M. B. Woods[251], M. Worek[385], S. Worm[33,259], G. Wormser[167], D. Wright[178], Z. Wu[251], C. E. Wulz[224], S. Xella[211], G. Xia[40,343], L. Xia[7], A. Xiao[7], L. Xiao[251], M. Xiao[102], Q. Xiao[102], J. Xie[7], C. Xu[102], F. Xu[210], G. Xu[102], K. Yagyu[201], U. A. Yajnik[85], V. Yakimenko[251], S. Yamada[71,116], S. Yamada[71], Y. Yamada[272], Y. Yamada[402], A. Yamaguchi[274], D. Yamaguchi[275], M. Yamaguchi[272], S. Yamaguchi[272], Y. Yamaguchi[375], Y. Yamaguchi[75], A. Yamamoto[71,375], H. Yamamoto[272], K. Yamamoto[222], K. Yamamoto[118], M. Yamamoto[71], N. Yamamoto[197], N. Yamamoto[71], Y. Yamamoto[71], Y. Yamamoto[375], T. Yamamura[375], T. Yamanaka[116], S. Yamashita[116], T. Yamashita[3], Y. Yamashita[214], K. Yamauchi[197], M. Yamauchi[71], T. Yamazaki[375], Y. Yamazaki[154], J. Yan[375,71], W. Yan[364], C. Yanagisawa[257,11], H. Yang[247], J. Yang[56], U. K. Yang[245,343], Z. Yang[24], W. Yao[177], S. Yashiro[71], F. Yasuda[375], O. Yasuda[276], I. Yavin[188,228], E. Yazgan[331], H. Yokoya[377], K. Yokoya[71], H. Yokoyama[375], S. Yokoyama[275], R. Yonamine[71], H. Yoneyama[240], M. Yoshida[71], T. Yoshida[62], K. Yoshihara[116,33], S. Yoshihara[116,33], M. Yoshioka[71,272], T. Yoshioka[161], H. Yoshitama[73], C. C. Young[251], H. B. Yu[348], J. Yu[373], C. Z. Yuan[102], F. Yuasa[71], J. Yue[102], A. Zabi[172], W. Zabolotny[401], J. Zacek[34], I. Zagorodnov[47], J. Zalesak[105,58], A. F. Zarnecki[381], L. Zawiejski[268], M. Zeinali[101], C. Zeitnitz[385], L. Zembala[401], K. Zenker[47], D. Zeppenfeld[91], D. Zerwas[167], P. Zerwas[47], M. Zeyrek[192], A. Zghiche[164], J. Zhai[102], C. Zhang[102], J. Zhang[102], J. Zhang[7], Y. Zhang[24,33], Z. Zhang[167], F. Zhao[102], F. Zhao[102], T. Zhao[102], Y. Zhao[251], H. Zheng[102], Z. Zhengguo[364], L. Zhong[24], F. Zhou[251], X. Zhou[364,102], Z. Zhou[102], R. Y. Zhu[17], X. Zhu[24], X. Zhu[102], M. Zimmer[47], F. Zomer[167], T. Zoufal[47], R. Zwicky[329] 1 Academia Sinica - 128 Sec. 2, Institute of Physics, Academia Rd., Nankang, Taipei 11529, Taiwan, R.O.C.
2 AGH University of Science and Technology, Akademia Gorniczo-Hutnicza im. Stanislawa Staszica w Krakowie, Al. Mickiewicza 30 PL-30-059 Cracow, Poland
3 Aichi Medical University, Nagakute, Aichi, 480-1195, Japan
4 Akita International University, Yuwa, Akita City, 010-1292, Japan
5 Albert-Ludwigs Universität Freiburg, Physikalisches Institut, Hermann-Herder Str. 3, D-79104 Freiburg, Germany
6 Ankara Üniversitesi Fen Fakültesi, Fizik Bölümü, Dögol Caddesi, 06100 Tandoĝan Ankara, Turkey
7 Argonne National Laboratory (ANL), 9700 S. Cass Avenue, Argonne, IL 60439, USA
8 Baylor University, Department of Physics, 101 Bagby Avenue, Waco, TX 76706, USA
9 Beijing University, Department of Physics, Beijing, China 100871
10 Benares Hindu University, Benares, Varanasi 221005, India
11 Borough of Manhattan Community College, The City University of New York, Department of Science, 199 Chambers Street, New York, NY 10007, USA
12 Brandenburg University of Technology, Postfach 101344, D-03013 Cottbus, Germany
13 Brookhaven National Laboratory (BNL), P.O.Box 5000, Upton, NY 11973-5000, USA
14 Brown University, Department of Physics, Box 1843, Providence, RI 02912, USA
15 Budker Institute for Nuclear Physics (BINP), 630090 Novosibirsk, Russia
16 Calcutta University, Department of Physics, 92 A.P.C. Road, Kolkata 700009, India
17 California Institute of Technology, Physics, Mathematics and Astronomy (PMA), 1200 East California Blvd, Pasadena, CA 91125, USA
18 California State University, Los Angeles, Dept. of Physics and Astronomy, 5151 State University Dr., Los Angeles, CA 90032, USA
19 Carleton University, Department of Physics, 1125 Colonel By Drive, Ottawa, Ontario, Canada K1S 5B6
20 Carnegie Mellon University, Department of Physics, Wean Hall 7235, Pittsburgh, PA 15213, USA
21 CEA Saclay, IRFU, F-91191 Gif-sur-Yvette, France
22 CEA Saclay, Service de Physique Théorique, CEA/DSM/SPhT, F-91191 Gif-sur-Yvette Cedex, France
23 Center for High Energy Physics (CHEP) / Kyungpook National University, 1370 Sankyuk-dong, Buk-gu, Daegu 702-701, Republic of Korea
24 Center for High Energy Physics (TUHEP), Tsinghua University, Beijing, China 100084
25 Center For Quantum Spacetime (CQUeST), Sogang University, 35 Baekbeom-ro, Mapo-gu, Seoul 121-742, Republic of Korea
26 Center for the Advancement of Natural Discoveries using Light Emission (CANDLE), Acharyan 31, 0040, Yerevan, Armenia
27 Centre de Physique des Particules de Marseille (CPPM), Aix-Marseille Université, CNRS/IN2P3, 163, Avenue de Luminy, Case 902, 13288 Marseille Cedex 09, France
28 Centre de Physique Theorique, CNRS - Luminy, Universiti d"Aix - Marseille II, Campus of Luminy, Case 907, 13288 Marseille Cedex 9, France
29 Centre Lasers Intenses et Applications (CELIA), Université Bordeaux 1 - CNRS - CEA, 351 Cours de la Libération, 33405 Talence Cedex, France
30 Centro Brasileiro de Pesquisas Físicas (CBPF), Rua Dr. Xavier Sigaud, n.150 22290-180, Urca - Rio de Janeiro, RJ, Brazil
31 Centro de Investigaciones Energéticas, Medioambientales y Tecnológicas, CIEMAT, Avenida Complutense 22, E-28040 Madrid, Spain
32 Centro Nacional de Microelectrónica (CNM), Instituto de Microelectrónica de Barcelona (IMB), Campus UAB, 08193 Cerdanyola del Vallès (Bellaterra), Barcelona, Spain
33 CERN, CH-1211 Genève 23, Switzerland
34 Charles University, Institute of Particle & Nuclear Physics, Faculty of Mathematics and Physics, V Holesovickach 2, CZ-18000 Prague 8, Czech Republic
35 Chiba University of Commerce, 1-3-1 Konodai, Ichikawa-shi, Chiba, 272-8512, Japan
36 Chonbuk National University, Division of Science Education, Jeonju 561-756, Republic of Korea
37 Chonbuk National University, Physics Department, Jeonju 561-756, Republic of Korea
38 Chubu University, 1200 Matsumoto-cho, Kasugai-shi, Aichi, 487-8501, Japan
39 Chung Yuan Christian University, Department of Physics, 200 Chung Pei Rd., Chung Li 32023 Taiwan, R.O.C
40 Cockcroft Institute, Daresbury, Warrington WA4 4AD, UK
41 College of William and Mary, Department of Physics, Williamsburg, VA, 23187, USA
42 Columbia University, Department of Physics, New York, NY 10027-6902, USA
43 Cornell University, Laboratory for Elementary-Particle Physics (LEPP), Ithaca, NY 14853, USA
44 Czech Technical University in Prague, Institute of Experimental and Applied Physics (IEAP), Horska 3a/22, 12800 Prague 2, Czech Republic
45 Czech Technical University, Faculty of Nuclear Science and Physical Engineering, Brehova 7, CZ-11519 Prague 1, Czech Republic
46 Departamento de Física Teórica, Facultad de Ciencias, Módulo 15 (antiguo C-XI) y Módulo 8, Universidad Autónoma de Madrid, Campus de Cantoblanco, 28049 Madrid, Spain
47 Deutsches Elektronen-Synchrotron DESY, A Research Centre of the Helmholtz Association, Notkestrasse 85, 22607 Hamburg, Germany (Hamburg site)
48 Deutsches Elektronen-Synchrotron DESY, A Research Centre of the Helmholtz Association, Platanenallee 6, 15738 Zeuthen, Germany (Zeuthen site)
49 Durham University, Department of Physics, Ogen Center for Fundamental Physics, South Rd., Durham DH1 3LE, UK
50 École Normale Supérieure de Lyon, 46 allée d'Italie, 69364 Lyon Cedex 07, France
51 Ege University, Department of Physics, Faculty of Science, 35100 Izmir, Turkey

52 Enrico Fermi Institute, University of Chicago, 5640 S. Ellis Avenue, RI-183, Chicago, IL 60637, USA
53 ETH Zürich, Institute for Particle Physics (IPP), Schafmattstrasse 20, CH-8093 Zürich, Switzerland
54 ETH Zürich, Institute for Theoretical Physics (ITP), Wolfgang-Pauli-Str. 27, Zürich, Switzerland
55 European Spallation Source ESS AB, Box 176, 221 00 Lund, Sweden
56 Ewha Womans University, 11-1 Daehyun-Dong, Seodaemun-Gu, Seoul, 120-750, Republic of Korea
57 Excellence Cluster Universe, Technische Universität München, Boltzmannstr. 2, 85748 Garching, Germany
58 Fermi National Accelerator Laboratory (FNAL), P.O.Box 500, Batavia, IL 60510-0500, USA
59 Florida State University, Department of Physics, 77 Chieftan Way, Tallahassee, FL 32306-4350, USA
60 Fujita Gakuen Health University, Department of Physics, Toyoake, Aichi 470-1192, Japan
61 Fukui University of Technology, 3-6-1 Gakuen, Fukui-shi, Fukui 910-8505, Japan
62 Fukui University, Department of Physics, 3-9-1 Bunkyo, Fukui-shi, Fukui 910-8507, Japan
63 Gangneung-Wonju National University, 210-702 Gangneung Daehangno, Gangneung City, Gangwon Province, Republic of Korea
64 Georg-August-Universität Göttingen, II. Physikalisches Institut, Friedrich-Hund-Platz 1, 37077 Göttingen, Germany
65 Global Design Effort
66 Hanyang University, Department of Physics, Seoul 133-791, Republic of Korea
67 Harish-Chandra Research Institute, Chhatnag Road, Jhusi, Allahabad 211019, India
68 Helmholtz-Zentrum Berlin für Materialien und Energie (HZB), Wilhelm-Conrad-Röntgen Campus, BESSY II, Albert-Einstein-Str. 15, 12489 Berlin, Germany
69 Helsinki Institute of Physics (HIP), P.O. Box 64, FIN-00014 University of Helsinki, Finland
70 Henan Normal University, College of Physics and Information Engineering, Xinxiang, China 453007
71 High Energy Accelerator Research Organization, KEK, 1-1 Oho, Tsukuba, Ibaraki 305-0801, Japan
72 Hiroshima University, Department of Physics, 1-3-1 Kagamiyama, Higashi-Hiroshima, Hiroshima 739-8526, Japan
73 Hiroshima University, Graduate School of Advanced Sciences of Matter, 1-3-1 Kagamiyama, Higashi-Hiroshima, Hiroshima 739-8530, Japan
74 Hokkai-Gakuen University, 4-1-40 Asahimachi, Toyohira-ku, Sapporo 062-8605, Japan
75 Hokkaido University, Department of Physics, Faculty of Science, Kita, Kita-ku, Sapporo-shi, Hokkaido 060-0810, Japan
76 Humboldt Universität zu Berlin, Fachbereich Physik, Institut für Elementarteilchenphysik, Newtonstr. 15, D-12489 Berlin, Germany
77 Hyogo University of Teacher Education, 942-1 Shimokume, Kato-city, Hyogo 673-1494, Japan
78 Ibaraki National College of Technology, 866 Nakane, Hitachinaka, Ibaraki 312-8508, Japan
79 Ibaraki University, College of Technology, Department of Physics, Nakanarusawa 4-12-1, Hitachi, Ibaraki 316-8511, Japan
80 Imperial College, Blackett Laboratory, Department of Physics, Prince Consort Road, London, SW7 2BW, UK
81 Indian Association for the Cultivation of Science, Department of Theoretical Physics and Centre for Theoretical Sciences, Kolkata 700032, India
82 Indian Institute of Science Education and Research (IISER) Kolkata, Department of Physical Sciences, Mohanpur Campus, PO Krishi Viswavidyalaya, Mohanpur 741252, Nadia, West Bengal, India
83 Indian Institute of Science, Centre for High Energy Physics, Bangalore 560012, Karnataka, India
84 Indian Institute of Technology Indore, IET Campus, M-Block, Institute of Engineering and Technology (IET), Devi Ahilya Vishwavidyalaya Campus, Khandwa Road, Indore - 452017, Madhya Pradesh, India
85 Indian Institute of Technology, Bombay, Powai, Mumbai 400076, India
86 Indian Institute of Technology, Guwahati, Guwahati, Assam 781039, India
87 Indiana University, Department of Physics, Swain Hall West 117, 727 E. 3rd St., Bloomington, IN 47405-7105, USA
88 Institucio Catalana de Recerca i Estudis, ICREA, Passeig Lluis Companys, 23, Barcelona 08010, Spain
89 Institut de Physique Nucléaire de Lyon (IPNL), Domaine scientifique de la Doua, Bâtiment Paul Dirac 4, rue Enrico Fermi, 69622 Villeurbanne, Cedex, France
90 Institut für Experimentelle Kernphysik, KIT,Universität Karlsruhe (TH), Wolfgang-Gaede-Str. 1, Postfach 6980, 76128 Karlsruhe, Germany
91 Institut für Theoretische Physik (ITP), Karlsruher Institut für Technologie (KIT), Fakultät für Physik, Postfach 6980, 76049 Karlsruhe, Germany
92 Institut für Theoretische Teilchenphysik, Campus Süd, Karlsruher Institut für Technologie (KIT), 76128 Karlsruhe, Germany
93 Institut National de Physique Nucleaire et de Physique des Particules, 3, Rue Michel- Ange, 75794 Paris Cedex 16, France
94 Institut Pluridisciplinaire Hubert Curien, 23 Rue du Loess - BP28, 67037 Strasbourg Cedex 2, France
95 Institute for Chemical Research, Kyoto University, Gokasho, Uji, Kyoto 611-0011, Japan
96 Institute for Cosmic Ray Research, University of Tokyo, 5-1-5 Kashiwa-no-Ha, Kashiwa, Chiba 277-8582, Japan
97 Institute for Mathematics, Astrophysics and Particle Physics (IMAPP), P.O. Box 9010, 6500 GL Nijmegen, Netherlands
98 Institute for Nuclear Research, Russian Academy of Sciences (INR RAS), 60-th October Anniversary Prospect 7a, 117312, Moscow, Russia
99 Institute for Particle and Nuclear Physics, Wigner Research Centre for Physics, Hungarian Academy of Sciences, P.O. Box 49, 1525 Budapest, Hungary
100 Institute for Scintillation Materials (ISMA), 60 Lenina Ave, 61001, Kharkiv, Ukraine
101 Institute for studies in fundamental sciences (IPM), Niavaran Square, P.O. Box 19395-5746, Tehran, Iran
102 Institute of High Energy Physics - IHEP, Chinese Academy of Sciences, P.O. Box 918, Beijing, China 100049
103 Institute of Mathematical Sciences, Taramani, C.I.T. Campus, Chennai 600113, India
104 Institute of Particle Physics, Canada

105 Institute of Physics, ASCR, Academy of Science of the Czech Republic, Division of Elementary Particle Physics, Na Slovance 2, CZ-18221 Prague 8, Czech Republic
106 Institute of Physics, Vietnam Academy of Science and Technology (VAST), 10 Dao-Tan, Ba-Dinh, Hanoi 10000, Vietnam
107 Institute of Theoretical and Experimetal Physics, B. Cheremushkinskawa, 25, RU-117259, Moscow, Russia
108 Instituto de Fisica Corpuscular (IFIC), Centro Mixto CSIC-UVEG, Edificio Investigacion Paterna, Apartado 22085, 46071 Valencia, Spain
109 Instituto de Física da Universidade Federal do Rio Grande do Sul (UFRGS), Av. Bento Gonçalves 9500, Caixa Postal 15051, CEP 91501-970, Porto Alegre, RS, Brazil
110 Instituto de Fisica de Cantabria, (IFCA, CSIC-UC), Facultad de Ciencias, Avda. Los Castros s/n, 39005 Santander, Spain
111 Instituto de Física Teórica UAM/CSIC, C/ Nicolás Cabrera 13-15, Universidad Autónoma de Madrid, Cantoblanco, 28049 Madrid, Spain
112 Instituto Galego de Fisica de Altas Enerxias (IGFAE,USC) Facultad de Fisica, Campus Sur E-15782 Santiago de Compostela, Spain
113 Instituto Tecnológico de Aragón (ITA), C/ María de Luna 7-8, 50018 Zaragoza, Spain
114 Instituto Universitario de Física Fundamental y Matemáticas de la Universidad de Salamanca (IUFFyM), Casas del Parque, 37008 Salamanca, Spain
115 Inter-University Accelerator Centre, Aruna Asaf Ali Marg, Post Box 10502, New Delhi 110067, India
116 International Center for Elementary Particle Physics, University of Tokyo, Hongo 7-3-1, Bunkyo District, Tokyo 113-0033, Japan
117 International Institute of Physics, Federal University of Rio Grande do Norte, Av. Odilon Gomes de Lima, 1722 - Capim Macio - 59078-400 - Natal-RN, Brazil
118 Iowa State University, Department of Physics, High Energy Physics Group, Ames, IA 50011, USA
119 Istituto Nazionale di Fisica Nucleare (INFN), Laboratorio LASA, Via Fratelli Cervi 201, 20090 Segrate, Italy
120 Istituto Nazionale di Fisica Nucleare (INFN), Sezione di Firenze, Via G. Sansone 1, I-50019 Sesto Fiorentino (Firenze), Italy
121 Istituto Nazionale di Fisica Nucleare (INFN), Sezione di Milano Bicocca, Piazza della Scienza 3, I-20126 Milano, Italy
122 Istituto Nazionale di Fisica Nucleare (INFN), Sezione di Milano, Via Celoria 16, I-20133 Milano, Italy
123 Istituto Nazionale di Fisica Nucleare (INFN), Sezione di Napoli, Complesso Universitá di Monte Sant'Angelo,via, I-80126 Naples, Italy
124 Istituto Nazionale di Fisica Nucleare (INFN), Sezione di Pavia, Via Bassi 6, I-27100 Pavia, Italy
125 Istituto Nazionale di Fisica Nucleare (INFN), Sezione di Perugia, Via A. Pascoli, 06123 Perugia, Italy
126 Istituto Nazionale di Fisica Nucleare (INFN), Sezione di Pisa, Edificio C - Polo Fibonacci Largo B. Pontecorvo, 3, I-56127 Pisa, Italy
127 Istituto Nazionale di Fisica Nucleare (INFN), Sezione di Roma, c/o Dipartimento di Fisica - Università degli Studi di Roma "La Sapienza", P.le Aldo Moro 2, I-00185 Roma, Italy
128 Istituto Nazionale di Fisica Nucleare (INFN), Sezione di Torino, c/o Universitá di Torino, facoltá di Fisica, via P Giuria 1, 10125 Torino, Italy
129 Istituto Nazionale di Fisica Nucleare (INFN), Sezione di Trieste, Padriciano 99, I-34012 Trieste (Padriciano), Italy
130 ITER Organization, Route de Vinon-sur-Verdon, 13115 St. Paul-lez-Durance, France
131 Iwate University, 4-3-5 Ueda, Morioka, Iwate, 020-8551, Japan
132 Jagiellonian University, Institute of Physics, Ul. Reymonta 4, PL-30-059 Cracow, Poland
133 Jamia Millia Islamia, Department of Physics, Jamia Nagar, New Delhi 110025, India
134 Japan Aerospace Exploration Agency, Sagamihara Campus, 3-1-1 Yoshinodai, Sagamihara, Kanagawa 220-8510 , Japan
135 Japan Atomic Energy Agency, 4-49 Muramatsu, Tokai-mura, Naka-gun, Ibaraki 319-1184, Japan
136 Japan Atomic Energy Agency, Tokai Research and Development Center, 2-4 Shirane Shirakata, Tokai-mura, Naka-gun, Ibaraki 319-1195, Japan
137 Japan Synchrotron Radiation Research Institute (JASRI), 1-1-1, Kouto, Sayo-cho, Sayo-gun, Hyogo 679-5198, Japan
138 Johannes Gutenberg Universität Mainz, Institut für Physik, 55099 Mainz, Germany
139 John Adams Institute for Accelerator Science at Royal Holloway University of London, Egham Hill, Egham, Surrey TW20 0EX, UK
140 John Adams Institute for Accelerator Science at University of Oxford, Denys Wilkinson Building, Keble Road, Oxford OX1 3RH, UK
141 Johns Hopkins University - Henry A. Rowland Department of Physics & Astronomy 3701 San Martin Drive, Baltimore, Maryland (MD) 21218, USA
142 Joint Institute for Nuclear Research (JINR), Joliot-Curie 6, 141980, Dubna, Moscow Region, Russia
143 Joint Institute for Power and Nuclear Research "Sosny" at National Academy of Sciences of Belarus, 99 Academician A.K.Krasin Str., Minsk BY-220109, Belarus
144 Jozef Stefan Institute, Jamova cesta 39, 1000 Ljubljana, Slovenia
145 Julius-Maximilians-Universität Würzburg, Fakultät für Physik und Astronomie, Am Hubland, 97074 Würzburg, Germany
146 Juntendo University, School of Medicine, Dept. of Physics, Hiraga-gakuendai 1-1, Inzai-shi, Chiba 270-1695, Japan
147 Justus-Liebig-Universität Gießen, II. Physikalisches Institut, Heinrich-Buff-Ring 16, 35392 Gießen, Germany
148 Kanazawa University, Institute for Theoretical Physics (KITP), School of Mathematics and Physics, College of Science and Engineering, Kakuma-machi, Kanazawa city, Ishikawa 920-1192, Japan
149 Kansas State University, Department of Physics, 116 Cardwell Hall, Manhattan, KS 66506, USA
150 Kavli Institute for the Physics and Mathematics of the Universe (Kavli IPMU), University of Tokyo, 5-1-5 Kashiwanoha, Kashiwa, 277-8583, Japan
151 King Saud University (KSU), Dept. of Physics, P.O. Box 2454, Riyadh 11451, Saudi Arabia
152 King's College London - Department of physics, Strand, London WC2R 2LS, London, UK
153 Kinki University, Department of Physics, 3-4-1 Kowakae, Higashi-Osaka, Osaka 577-8502, Japan
154 Kobe University, Department of Physics, 1-1 Rokkodai-cho, Nada-ku, Kobe, Hyogo 657-8501, Japan

155 Kogakuin University, Department of Physics, Shinjuku Campus, 1-24-2 Nishi-Shinjuku, Shinjuku-ku, Tokyo 163-8677, Japan

156 Konkuk University, 93-1 Mojin-dong, Kwanglin-gu, Seoul 143-701, Republic of Korea

157 Korea Advanced Institute of Science & Technology, Department of Physics, 373-1 Kusong-dong, Yusong-gu, Taejon 305-701, Republic of Korea

158 Korea Institute for Advanced Study (KIAS), School of Physics, 207-43 Cheongryangri-dong, Dongdaemun-gu, Seoul 130-012, Republic of Korea

159 Kyoto University, Department of Physics, Kitashirakawa-Oiwakecho, Sakyo-ku, Kyoto 606-8502, Japan

160 Kyoto University, Yukawa Institute for Theoretical Physics, Kitashirakawa-Oiwakecho, Sakyo-Ku, Kyoto 606-8502, Japan

161 Kyushu University, Department of Physics, 6-10-1 Hakozaki, Higashi-ku, Fukuoka 812-8581, Japan

162 L.P.T.A., UMR 5207 CNRS-UM2, Université Montpellier II, Case Courrier 070, Bât. 13, place Eugène Bataillon, 34095 Montpellier Cedex 5, France

163 Laboratoire Charles Coulomb UMR 5221 CNRS-UM2, Université Montpellier 2, Place Eugène Bataillon - CC069, 34095 Montpellier Cedex 5, France

164 Laboratoire d'Annecy-le-Vieux de Physique des Particules (LAPP) , Université de Savoie, CNRS/IN2P3, 9 Chemin de Bellevue, BP 110, F-74941 Annecy-Le-Vieux Cedex, France

165 Laboratoire d'Annecy-le-Vieux de Physique Theorique (LAPTH), Chemin de Bellevue, BP 110, F-74941 Annecy-le-Vieux Cedex, France

166 Laboratoire d'AstroParticules et Cosmologie (APC), Université Paris Diderot-Paris 7 - CNRS/IN2P3, Bâtiment Condorcet, Case 7020, 75205 Paris Cedex 13, France

167 Laboratoire de l'Accélérateur Linéaire (LAL), Université Paris-Sud 11, Bâtiment 200, 91898 Orsay, France

168 Laboratoire de Physique Corpusculaire de Clermont-Ferrand (LPC), Université Blaise Pascal, I.N.2.P.3./C.N.R.S., 24 avenue des Landais, 63177 Aubière Cedex, France

169 Laboratoire de Physique Nucléaire et des Hautes Energies (LPNHE), UPMC, UPD, IN2P3/CNRS, 4 Place Jussieu, 75005, Paris Cedex 05, France

170 Laboratoire de Physique Subatomique et de Cosmologie (LPSC), Université Joseph Fourier (Grenoble 1), CNRS/IN2P3, Institut Polytechnique de Grenoble, 53 rue des Martyrs, F-38026 Grenoble Cedex, France

171 Laboratoire de Physique Theorique, Université de Paris-Sud XI, Batiment 210, F-91405 Orsay Cedex, France

172 Laboratoire Leprince-Ringuet (LLR), École polytechnique – CNRS/IN2P3, Route de Saclay, F-91128 Palaiseau Cedex, France

173 Laboratoire Univers et Particules de Montpellier (LUPM) - UMR5299, Université de Montpellier II, Place Eugène Bataillon - Case courrier 72, 34095 Montpellier Cedex 05, France

174 Laboratori Nazionali di Frascati, via E. Fermi, 40, C.P. 13, I-00044 Frascati, Italy

175 Laboratório de Instrumentação e Física Experimental de Partículas (LIP LISBOA), Av. Elias Garcia 14 - 1°, 1000-149 Lisbon, Portugal

176 Lancaster University, Physics Department, Lancaster LA1 4YB, UK

177 Lawrence Berkeley National Laboratory (LBNL), 1 Cyclotron Rd, Berkeley, CA 94720, USA

178 Lawrence Livermore National Laboratory (LLNL), Livermore, CA 94551, USA

179 Lebedev Physical Institute, Leninsky Prospect 53, RU-117924 Moscow, Russia

180 Lomonosov Moscow State University, Skobeltsyn Institute of Nuclear Physics (MSU SINP), 1(2), Leninskie gory, GSP-1, Moscow 119991, Russia

181 Louisiana Tech University, Department of Physics, Ruston, LA 71272, USA

182 Ludwig-Maximilians-Universität München, Fakultät für Physik, Am Coulombwall 1, D - 85748 Garching, Germany

183 Lunds Universitet, Fysiska Institutionen, Avdelningen för Experimentell Högenergifysik, Box 118, 221 00 Lund, Sweden

184 L'Université Hassan II, Aïn Chock, "Réseau Universitaire de Physique des Hautes Energies" (RUPHE), Département de Physique, Faculté des Sciences Aïn Chock, B.P 5366 Maarif, Casablanca 20100, Morocco

185 Massachusetts Institute of Technology (MIT), Laboratory for Nuclear Science, 77 Massachusetts Avenue, Cambridge, MA 02139, USA

186 Max-Planck-Institut für Physik (Werner-Heisenberg-Institut), Föhringer Ring 6, 80805 München, Germany

187 McGill University, Department of Physics, Ernest Rutherford Physics Bldg., 3600 University Street, Montreal, Quebec, H3A 2T8 Canada

188 McMaster University, Department of Physics & Astronomy, 1280 Main Street West, Hamilton, ON, L8S 4M1, Canada

189 Meiji Gakuin University, Department of Physics, 2-37 Shirokanedai 1-chome, Minato-ku, Tokyo 244-8539, Japan

190 Michigan State University, Department of Chemical Engineering & Materials Science, 2527 Engineering Building East Lansing, MI 48824-1226, USA

191 Michigan State University, Department of Physics and Astronomy, East Lansing, MI 48824, USA

192 Middle East Technical University, Department of Physics, TR-06531 Ankara, Turkey

193 Miyagi Gakuin Women's University, Faculty of Liberal Arts, 9-1-1 Sakuragaoka, Aoba District, Sendai, Miyagi 981-8557, Japan

194 MSU-Iligan Institute of Technology, Department of Physics, Andres Bonifacio Avenue, 9200 Iligan City, Phillipines

195 Nagasaki Institute of Applied Science, 536 Abamachi, Nagasaki-Shi, Nagasaki 851-0193, Japan

196 Nagoya University, Department of Materials Science and Engineering, Furo-cho, Chikusa-ku, Nagoya, 464-8603, Japan

197 Nagoya University, Department of Physics, School of Science, Furo-cho, Chikusa-ku, Nagoya, Aichi 464-8602, Japan

198 Nagoya University, Kobayashi-Maskawa Institute for the Origin of Particles and the Universe (KMI), Furo-cho, Chikusa-ku, Nagoya Aichi 464-8602, Japan

199 Nanjing University, Department of Physics, Nanjing, China 210093

200 Nara Women's University, High Energy Physics Group, Kitauoya-Nishimachi, Nara 630-8506, Japan

201 National Central University, High Energy Group, Department of Physics, Chung-li, Taiwan 32001, R.O.C

202 National Centre of Nuclear Research (NCBJ), ul. Andrzeja Soltana 7, 05-400 Otwock-Swierk, Poland

203 National Cheng Kung University, Physics Department, 1 Ta-Hsueh Road, Tainan, Taiwan 70101, R.O.C

204 National Chiao-Tung University, Institute of Physics, 1001 Ta Hsueh Rd, Hsinchu, Taiwan 300, R.O.C.

205 National Institute of Chemical Physics and Biophysics (NICPB), Ravala pst 10, 10143 Tallinn, Estonia

206 National Institute of Physics and Nuclear Engineering “Horia Hulubei” (IFIN-HH), Str. Reactorului no.30, P.O. Box MG-6, R-76900 Bucharest - Magurele, Romania

207 National Research Centre “Kurchatov Institute”, 1 Akademika Kurchatova pl., Moscow, 123182, Russia

208 National Science Center - Kharkov Institute of Physics and Technology (NSC KIPT), Akademicheskaya St. 1, Kharkov, 61108, Ukraine

209 National Scientific & Educational Centre of Particle & High Energy Physics (NCPHEP), Belarusian State University, M.Bogdanovich street 153, 220040 Minsk, Belarus

210 National Taiwan University, Physics Department, Taipei, Taiwan 106, R.O.C

211 Niels Bohr Institute (NBI), University of Copenhagen, Blegdamsvej 17, DK-2100 Copenhagen, Denmark

212 Niigata University, Department of Physics, Ikarashi, Niigata 950-218, Japan

213 Nikhef, National Institute for Subatomic Physics, P.O. Box 41882, 1009 DB Amsterdam, Netherlands

214 Nippon Dental University School of Life Dentistry at Niigata, 1-8 Hamaura-cho, Chuo-ku, Niigata 951-1500, Japan

215 North Carolina A&T State University, 1601 E. Market Street, Greensboro, NC 27411, USA

216 Northeastern University, Physics Department, 360 Huntington Ave, 111 Dana Research Center, Boston, MA 02115, USA

217 Northern Illinois University, Department of Physics, DeKalb, Illinois 60115-2825, USA

218 Northwestern University, Department of Physics and Astronomy, 2145 Sheridan Road., Evanston, IL 60208, USA

219 Novosibirsk State University (NGU), Department of Physics, Pirogov st. 2, 630090 Novosibirsk, Russia

220 Ochanomizu University, Department of Physics, Faculty of Science, 1-1 Otsuka 2, Bunkyo-ku, Tokyo 112-8610, Japan

221 Orissa University of Agriculture & Technology, Bhubaneswar 751003, Orissa, India

222 Osaka City University, Department of Physics, Faculty of Science, 3-3-138 Sugimoto, Sumiyoshi-ku, Osaka 558-8585, Japan

223 Osaka University, Department of Physics, 1-1 Machikaneyama, Toyonaka, Osaka 560-0043, Japan

224 Österreichische Akademie der Wissenschaften, Institut für Hochenergiephysik, Nikolsdorfergasse 18, A-1050 Vienna, Austria

225 Pacific Northwest National Laboratory, (PNNL), PO Box 999, Richland, WA 99352, USA

226 Panjab University, Chandigarh 160014, India

227 Pavel Sukhoi Gomel State Technical University, ICTP Affiliated Centre & Laboratory for Physical Studies, October Avenue, 48, 246746, Gomel, Belarus

228 Perimeter Institute for Theoretical Physics, 31 Caroline Street North, Waterloo, Ontario N2L 2Y5, Canada

229 Physical Research Laboratory, Navrangpura, Ahmedabad 380 009, Gujarat, India

230 Pohang Accelerator Laboratory (PAL), San-31 Hyoja-dong, Nam-gu, Pohang, Gyeongbuk 790-784, Republic of Korea

231 Pontificia Universidad Católica de Chile, Avda. Libertador Bernardo OHiggins 340, Santiago, Chile

232 Princeton University, Department of Physics, P.O. Box 708, Princeton, NJ 08542-0708, USA

233 Purdue University, Department of Physics, West Lafayette, IN 47907, USA

234 Queen Mary, University of London, Mile End Road, London, E1 4NS, United Kingdom

235 Rheinisch-Westfälische Technische Hochschule (RWTH), Physikalisches Institut, Physikzentrum, Otto-Blumenthal-Straße, 52056 Aachen

236 RIKEN, 2-1 Hirosawa, Wako, Saitama 351-0198, Japan

237 Royal Holloway, University of London (RHUL), Department of Physics, Egham, Surrey TW20 0EX, UK

238 Russian Academy of Science, Keldysh Institute of Applied Mathematics, Muiskaya pl. 4, 125047 Moscow, Russia

239 Rutgers, The State University of New Jersey, Department of Physics & Astronomy, 136 Frelinghuysen Rd, Piscataway, NJ 08854, USA

240 Saga University, Department of Physics, 1 Honjo-machi, Saga-shi, Saga 840-8502, Japan

241 Saha Institute of Nuclear Physics, 1/AF Bidhan Nagar, Kolkata 700064, India

242 Salalah College of Technology (SCOT), Engineering Department, Post Box No. 608, Postal Code 211, Salalah, Sultanate of Oman

243 Saudi Center for Theoretical Physics, King Fahd University of Petroleum and Minerals (KFUPM), Dhahran 31261, Saudi Arabia

244 Seikei University, Faculty of Science and Technology, 3-3-1 Kichijoji-Kitamachi, Musashino-shi, Tokyo 180-8633, Japan

245 Seoul National University, San 56-1, Shinrim-dong, Kwanak-gu, Seoul 151-742, Republic of Korea

246 Setsunan University, Institute for Fundamental Sciences, 17-8 Ikeda Nakamachi, Neyagawa, Osaka, 572-8508, Japan

247 Shanghai Jiao Tong University, Department of Physics, 800 Dongchuan Road, Shanghai, China 200240

248 Shinshu University, 3-1-1, Asahi, Matsumoto, Nagano 390-8621, Japan

249 Shiv Nadar University, Village Chithera, Tehsil Dadri, District Gautam Budh Nagar, 203207 Uttar Pradesh, India

250 Shizuoka University, Department of Physics, 836 Ohya, Suruga-ku, Shizuoka 422-8529, Japan

251 SLAC National Accelerator Laboratory, 2575 Sand Hill Road, Menlo Park, CA 94025, USA

252 Society for Applied Microwave Electronics Engineering and Research (SAMEER), I.I.T. Campus, Powai, Post Box 8448, Mumbai 400076, India

253 Sokendai, The Graduate University for Advanced Studies, Shonan Village, Hayama, Kanagawa 240-0193, Japan

254 Spallation Neutron Source (SNS), Oak Ridge National Laboratory (ORNL), P.O. Box 2008 MS-6477, Oak Ridge, TN 37831-6477, USA

255 State University of New York at Binghamton, Department of Physics, PO Box 6016, Binghamton, NY 13902, USA

256 State University of New York at Buffalo, Department of Physics & Astronomy, 239 Franczak Hall, Buffalo, NY 14260, USA

257 State University of New York at Stony Brook, Department of Physics and Astronomy, Stony Brook, NY 11794-3800, USA

258 STFC Daresbury Laboratory, Daresbury, Warrington, Cheshire WA4 4AD, UK

259 STFC Rutherford Appleton Laboratory, Chilton, Didcot, Oxon OX11 0QX, UK

260 Sungkyunkwan University (SKKU), Natural Science Campus 300, Physics Research Division, Chunchun-dong, Jangan-gu, Suwon, Kyunggi-do 440-746, Republic of Korea

261 Swiss Light Source (SLS), Paul Scherrer Institut (PSI), PSI West, CH-5232 Villigen PSI, Switzerland

262 Tata Institute of Fundamental Research, School of Natural Sciences, Homi Bhabha Rd., Mumbai 400005, India

263 Technical University of Lodz, Department of Microelectronics and Computer Science, al. Politechniki 11, 90-924 Lodz, Poland

264 Technische Universität Dresden, Institut für Kern- und Teilchenphysik, D-01069 Dresden, Germany

265 Tel-Aviv University, School of Physics and Astronomy, Ramat Aviv, Tel Aviv 69978, Israel

266 Texas A&M University, Physics Department, College Station, 77843-4242 TX, USA

267 Texas Tech University, Department of Physics, Campus Box 41051, Lubbock, TX 79409-1051, USA

268 The Henryk Niewodniczanski Institute of Nuclear Physics, Polish Academy of Sciences (IFJ PAN), ul. Radzikowskiego 152, PL-31342 Cracow, Poland

269 Thomas Jefferson National Accelerator Facility (TJNAF), 12000 Jefferson Avenue, Newport News, VA 23606, USA

270 Tohoku Gakuin University, Department of Business Administration, 1-3-1 Tsuchitoi, Aoba-ku Sendai, Miyagi 980-8511, Japan

271 Tohoku Gakuin University, Faculty of Technology, 1-13-1 Chuo, Tagajo, Miyagi 985-8537, Japan

272 Tohoku University, Department of Physics, Aoba District, Sendai, Miyagi 980-8578, Japan

273 Tohoku University, Research Center for Electron Photon Science, Taihaku District, Sendai, Miyagi 982-0826, Japan

274 Tohoku University, Research Center for Neutrino Science, Aoba District, Sendai, Miyagi 980-8578, Japan

275 Tokyo Institute of Technology, Department of Physics, 2-12-1 O-Okayama, Meguro, Tokyo 152-8551, Japan

276 Tokyo Metropolitan University, Faculty of Science and Engineering, Department of Physics, 1-1 Minami-Osawa, Hachioji-shi, Tokyo 192-0397, Japan

277 Tokyo University of Agriculture Technology, Department of Applied Physics, Naka-machi, Koganei, Tokyo 183-8488, Japan

278 Toyama Prefectural University, Department of Mathematical Physics, 5180 Kurokawa Imizu-shi, Toyama, 939-0398, Japan

279 TRIUMF, 4004 Wesbrook Mall, Vancouver, BC V6T 2A3, Canada

280 Universidad Autónoma de San Luis Potosí, Alvaro Obregon 64, Col. Centro, San Luis Potosí, S.L.P. 78000, México

281 Universidad de Granada, Departamento de Física Teórica y del Cosmos, Campus de Fuentenueva, E-18071 Granada, Spain

282 Universidad de los Andes, Faculty of Science, Department of Physics, Carrera 1 18A-10, Bloque Ip. Bogotá, Colombia

283 Universidad de Oviedo, Departamento de Física, Campus de Llamaquique. C/ Calvo Sotelo, s/n 33005 Oviedo, Spain

284 Universidad de Salamanca, Departamento de Física Fundamental, Plaza de la Merced, s/n., 37008 Salamanca, Spain

285 Universidad de Sevilla, Escuela Técnica Superior de Ingeniería, Departamento Ingeniería Electrónica, Camino de los Descubrimientos s/n, 41092 Sevilla, Spain

286 Universidad de Zaragoza - Departamento de Física Teórica, Pedro Cerbuna 12, E-50009 Zaragoza, Spain

287 Universidad Nacional Autónoma de México, Instituto de Física, Circuito de la Investigación Cientifica s/n, Ciudad Universitaria, CP 04510 México D.F., Mexico

288 Universidad Nacional de La Plata, Departamento de Física, Facultad de Ciencias Exactas, C.C. N 67, 1900 La Plata, Argentina

289 Universidade do Estado do Rio de Janeiro (UERJ), Rio de Janeiro, RJ - Brasil 20550-900, Brazil

290 Universidade Federal de Pelotas, Instituto de Física e Matemática, Campus Universitário, Caixa Postal 354, 96010-900 Pelotas, RS, Brazil

291 Universidade Federal do Rio de Janeiro (UFRJ), Instituto de Física, Av. Athos da Silveira Ramos 149, Centro de Tecnologia - Bloco A, Cidade Universitária, Ilha do Fundão, Rio de Janeiro, RJ, Brazil

292 Universitá degli Studi di Napoli "Federico II", Dipartimento di Fisica, Via Cintia, 80126 Napoli, Italy

293 Universitat Autònoma de Barcelona, Departament de Física, Edifici C, 08193 Bellaterra, Barcelona, Spain

294 Universitat Autònoma de Barcelona, Institut de Fisica d'Altes Energies (IFAE), Campus UAB, Edifici Cn, E-08193 Bellaterra, Barcelona, Spain

295 Universität Bonn, Physikalisches Institut, Nußallee 12, 53115 Bonn, Germany

296 Universität Heidelberg, Institut für Theoretische Physik, Philosophenweg 16, 69120 Heidelberg, Germany

297 Universität Heidelberg, Kirchhoff-Institut für Physik, Im Neuenheimer Feld 227, 69120 Heidelberg, Germany

298 Universitat Politècnica de Catalunya, Institut de Tècniques Energètiques, Campus Diagonal Sud, Edifici PC (Pavelló C). Av. Diagonal, 647 08028 Barcelona, Spain

299 Universitat Ramon Llull, La Salle, C/ Quatre Camins 2, 08022 Barcelona, Spain

300 Universität Rostock, 18051 Rostock, Germany

301 Universität Siegen, Naturwissenschaftlich-Technische Fakultät, Department Physik, Emmy Noether Campus, Walter-Flex-Str.3, 57068 Siegen, Germany

302 Universität Wien - Theoretische Physik Boltzmanngasse 5, A-1090 Vienna, Austria

303 Université catholique de Louvain, Centre for Cosmology, Particle Physics and Phenomenology (CP3), Institute of Mathematics and Physics, 2 Chemin du Cyclotron, 1348 Louvain-la-Neuve, Belgium

304 Université de Genève, Section de Physique, 24, quai E. Ansermet, 1211 Genève 4, Switzerland

305 Université de Montréal, Département de Physique, Groupe de Physique des Particules, C.P. 6128, Succ. Centre-ville, Montréal, Qc H3C 3J7, Canada

306 Université de Strasbourg, UFR de Sciences Physiques, 3-5 Rue de l'Université, F-67084 Strasbourg Cedex, France

307 Université Libre de Bruxelles, Boulevard du Triomphe, 1050 Bruxelles, Belgium

308 Universittà di Catania, Dipartimento di Fisica e Astronomia, Via Santa Sofia 64, 95123 Catania, Italy

309 University College London (UCL), High Energy Physics Group, Physics and Astronomy Department, Gower Street, London WC1E 6BT, UK

310 University College, National University of Ireland (Dublin), Department of Experimental Physics, Science Buildings, Belfield, Dublin 4, Ireland
311 University de Barcelona, Facultat de Física, Av. Diagonal, 647, Barcelona 08028, Spain
312 University of Alberta - Faculty of Science, Department of Physics, 4-181 CCIS, Edmonton AB T6G 2E1, Canada
313 University of Arizona, Department of Physics, 1118 E. Fourth Street, PO Box 210081, Tucson, AZ 85721, USA
314 University of Bergen, Institute of Physics, Allegaten 55, N-5007 Bergen, Norway
315 University of Birmingham, School of Physics and Astronomy, Particle Physics Group, Edgbaston, Birmingham B15 2TT, UK
316 University of Bristol, H. H. Wills Physics Lab, Tyndall Ave., Bristol BS8 1TL, UK
317 University of British Columbia, Department of Physics and Astronomy, 6224 Agricultural Rd., Vancouver, BC V6T 1Z1, Canada
318 University of California (UCLA), Los Angleles, CA 90095, US
319 University of California Berkeley, Department of Physics, 366 Le Conte Hall, #7300, Berkeley, CA 94720, USA
320 University of California Davis, Department of Physics, One Shields Avenue, Davis, CA 95616-8677, USA
321 University of California Irvine, Department of Physics and Astronomy, High Energy Group, 4129 Frederick Reines Hall, Irvine, CA 92697-4575 USA
322 University of California Santa Cruz, Institute for Particle Physics, 1156 High Street, Santa Cruz, CA 95064, USA
323 University of Cambridge, Cavendish Laboratory, J J Thomson Avenue, Cambridge CB3 0HE, UK
324 University of Colorado at Boulder, Department of Physics, 390 UCB, University of Colorado, Boulder, CO 80309-0390, USA
325 University of Cyprus, Department of Physics, P.O.Box 20537, 1678 Nicosia, Cyprus
326 University of Delhi, Department of Physics and Astrophysics, Delhi 110007, India
327 University of Delhi, S.G.T.B. Khalsa College, Delhi 110007, India
328 University of Dundee, Department of Physics, Nethergate, Dundee, DD1 4HN, Scotland, UK
329 University of Edinburgh, School of Physics, James Clerk Maxwell Building, The King's Buildings, Mayfield Road, Edinburgh EH9 3JZ, UK
330 University of Florida, Department of Physics, Gainesville, FL 32611, USA
331 University of Ghent, Department of Subatomic and Radiation Physics, Proeftuinstraat 86, 9000 Gent, Belgium
332 University of Glasgow, SUPA, School of Physics & Astronomy, University Avenue, Glasgow G12 8QQ, Scotland, UK
333 University of Hamburg, Physics Department, Luruper Chaussee 149, 22761 Hamburg, Germany
334 University of Hawaii, Department of Physics and Astronomy, HEP, 2505 Correa Rd., WAT 232, Honolulu, HI 96822-2219, USA
335 University of Helsinki, Department of Physical Sciences, P.O. Box 64 (Vaino Auerin katu 11), FIN-00014, Helsinki, Finland
336 University of Illinois at Chicago, Department Of Physics, 845 W Taylor St., Chicago IL 60607, USA
337 University of Iowa, Department of Physics and Astronomy, 203 Van Allen Hall, Iowa City, IA 52242-1479, USA
338 University of Kansas, Department of Physics and Astronomy, Malott Hall, 1251 Wescoe Hall Drive, Room 1082, Lawrence, KS 66045-7582, USA
339 University of Liverpool, Department of Physics, Oliver Lodge Lab, Oxford St., Liverpool L69 7ZE, UK
340 University of Liverpool, Division of Theoretical Physics, Department of Mathematical Sciences, Chadwick Building, Liverpool L69 3BX, UK
341 University of Ljubljana, Faculty of Mathematics and Physics, Jadranska ulica 19, 1000 Ljubljana, Slovenia
342 University of Malaya, Faculty of Science, Department of Physics, 50603 Kuala Lumpur, Malaysia
343 University of Manchester, School of Physics and Astronomy, Schuster Lab, Manchester M13 9PL, UK
344 University of Maribor, Faculty of Chemistry and Chemical Engineering (FKKT), Smetanova ulica 17, 2000 Maribor, Slovenia
345 University of Maryland, Department of Physics and Astronomy, Physics Building (Bldg. 082), College Park, MD 20742, USA
346 University of Massachusetts - Amherst, Department of Physics, 1126 Lederle Graduate Research Tower (LGRT), Amherst, MA 01003-9337, USA
347 University of Melbourne, School of Physics, Victoria 3010, Australia
348 University of Michigan, Department of Physics, 500 E. University Ave., Ann Arbor, MI 48109-1120, USA
349 University of Minnesota, 148 Tate Laboratory Of Physics, 116 Church St. S.E., Minneapolis, MN 55455, USA
350 University of Mississippi, Department of Physics and Astronomy, 108 Lewis Hall, PO Box 1848, Oxford, Mississippi 38677-1848, USA
351 University of Missouri – St. Louis, Department of Physics and Astronomy, 503 Benton Hall One University Blvd., St. Louis Mo 63121, USA
352 University of New Mexico, New Mexico Center for Particle Physics, Department of Physics and Astronomy, 800 Yale Boulevard N.E., Albuquerque, NM 87131, USA
353 University of North Carolina at Chapel Hill, Department of Physics and Astronomy, Phillips Hall, CB #3255, 120 E. Cameron Ave., Chapel Hill, NC 27599-3255, USA
354 University of Notre Dame, Department of Physics, 225 Nieuwland Science Hall, Notre Dame, IN 46556, USA
355 University of Oklahoma, Department of Physics and Astronomy, Norman, OK 73071, USA
356 University of Oregon, Department of Physics, 1371 E. 13th Ave., Eugene, OR 97403, USA
357 University of Oslo, Department of Physics, P.O box 1048, Blindern, 0316 Oslo, Norway
358 University of Oxford, Particle Physics Department, Denys Wilkinson Bldg., Keble Road, Oxford OX1 3RH England, UK
359 University of Pavia, Department of Physics, via Bassi 6, I-27100 Pavia, Italy
360 University of Pennsylvania, Department of Physics and Astronomy, 209 South 33rd Street, Philadelphia, PA 19104-6396, USA
361 University of Pittsburgh, Department of Physics and Astronomy, 100 Allen Hall, 3941 O'Hara St, Pittsburgh PA 15260, USA
362 University of Regina, Department of Physics, Regina, Saskatchewan, S4S 0A2 Canada

363 University of Rochester, Department of Physics and Astronomy, Bausch & Lomb Hall, P.O. Box 270171, 600 Wilson Boulevard, Rochester, NY 14627-0171 USA
364 University of Science and Technology of China, Department of Modern Physics (DMP), Jin Zhai Road 96, Hefei, China 230026
365 University of Silesia, Institute of Physics, Ul. Uniwersytecka 4, PL-40007 Katowice, Poland
366 University of South Carolina, Department of Physics and Astronomy, 712 Main Street, Columbia, SC 29208, USA
367 University of Southampton, School of Physics and Astronomy, Highfield, Southampton S017 1BJ, England, UK
368 University of Southern California, Department of Physics & Astronomy, 3620 McClintock Ave., SGM 408, Los Angeles, CA 90089-0484, USA
369 University of Sydney, Falkiner High Energy Physics Group, School of Physics, A28, Sydney, NSW 2006, Australia
370 University of Tartu, Institute of Physics, Riia 142, 51014 Tartu, Estonia
371 University of Texas at Austin, Department of Physics, 1 University Station C1600, Austin, Texas 78712, USA
372 University of Texas at Dallas, Department of Physics, 800 West Campbell Road, Richardson, Texas 75080, USA
373 University of Texas, Center for Accelerator Science and Technology, Arlington, TX 76019, USA
374 University of Tokushima, Institute of Theoretical Physics, Tokushima-shi 770-8502, Japan
375 University of Tokyo, Department of Physics, 7-3-1 Hongo, Bunkyo District, Tokyo 113-0033, Japan
376 University of Toronto, Department of Physics, 60 St. George St., Toronto M5S 1A7, Ontario, Canada
377 University of Toyama, Department of Physics, 3190 Gofuku, Toyama 930-8555, Japan
378 University of Tsukuba, Faculty of Pure and Applied Sciences, 1-1-1 Ten'nodai, Tsukuba, Ibaraki 305-8571, Japan
379 University of Victoria, Department of Physics and Astronomy, P.O.Box 3055 Stn Csc, Victoria, BC V8W 3P6, Canada
380 University of Virginia, Department of Physics, 382 McCormick Rd., PO Box 400714, Charlottesville, VA
381 University of Warsaw, Institute of Experimental Physics, Ul. Hoza 69, PL-00 681 Warsaw, Poland
382 University of Warsaw, Institute of Theoretical Physics, Ul. Hoza 69, PL-00 681 Warsaw, Poland
383 University of Washington, Department of Physics, PO Box 351560, Seattle, WA 98195-1560, USA
384 University of Wisconsin, Physics Department, Madison, WI 53706-1390, USA
385 University of Wuppertal, Gaußstraße 20, D-42119 Wuppertal, Germany
386 Università degli Studi di Milano, Dipartimento di Fisica, Via Celoria 16, 20133 Milano, Italy
387 Università degli Studi di Roma "La Sapienza", Dipartimento di Fisica, Piazzale Aldo Moro 5, 00185 Roma, Italy
388 Università degli Studi di Trieste, Dipartimento di Fisica, via A. Valerio 2, I-34127 Trieste, Italy
389 Università dell'Insubria in Como, Dipartimento di Scienze CC.FF.MM., via Vallegio 11, I-22100 Como, Italy
390 Università di Milano-Bicocca, Dipartimento di Fisica"G. Occhialin", Piazza della Scienza 3, 20126 Milano, Italy
391 Università di Pisa, Departimento di Fisica "Enrico Fermi", Largo Bruno Pontecorvo 3, I-56127 Pisa, Italy
392 Universiy of Huddersfield, International Institute for Accelerator Applications, Queensgate Campus, Huddersfield HD1 3DH, UK
393 UPMC Univ. Paris 06, Faculté de Physique (UFR 925), 4 Place Jussieu, 75252 Paris Cedex 05, France
394 Vietnam National University, Laboratory of High Energy Physics and Cosmology, Faculty of Physics, College of Science, 334 Nguyen Trai, Hanoi, Vietnam
395 Vietnam National University, University of Natural Sciences, 227 Nguyen Van Cu street, District 5, Ho Chi Minh City, Vietnam
396 VINCA Institute of Nuclear Sciences, Laboratory of Physics, PO Box 522, YU-11001 Belgrade, Serbia
397 Virginia Commonwealth University, Department of Physics, P.O. Box 842000, 701 W. Grace St.,Richmond, VA. 23284-2000, USA
398 Virginia Polytechnic Institute and State University, Physics Department, Blacksburg, VA 2406, USA
399 Vrije Universiteit Brussel, Pleinlaan 2, 1050 Brussels, Belgium
400 Vrije Universiteit, Department of Physics, Faculty of Sciences, De Boelelaan 1081, 1081 HV Amsterdam, Netherlands
401 Warsaw University of Technology, The Faculty of Electronics and Information Technology, ul. Nowowiejska 15-19, 00-665 Warsaw, Poland
402 Waseda University, Advanced Research Institute for Science and Engineering, Shinjuku, Tokyo 169-8555, Japan
403 Wayne State University, Department of Physics, Detroit, MI 48202, USA
404 Weizmann Institute of Science, Department of Particle Physics, P.O. Box 26, Rehovot 76100, Israel
405 Yale University, Department of Physics, New Haven, CT 06520, USA
406 Yamagata University, 1-4-12 Kojirakawa-cho, Yamagata-shi, Yamagata, 990-8560, Japan
407 Yerevan Physics Institute, 2 Alikhanyan Brothers St., Yerevan 375036, Armenia
408 Yonsei University, Department of Physics, 134 Sinchon-dong, Sudaemoon-gu, Seoul 120-749, Republic of Korea

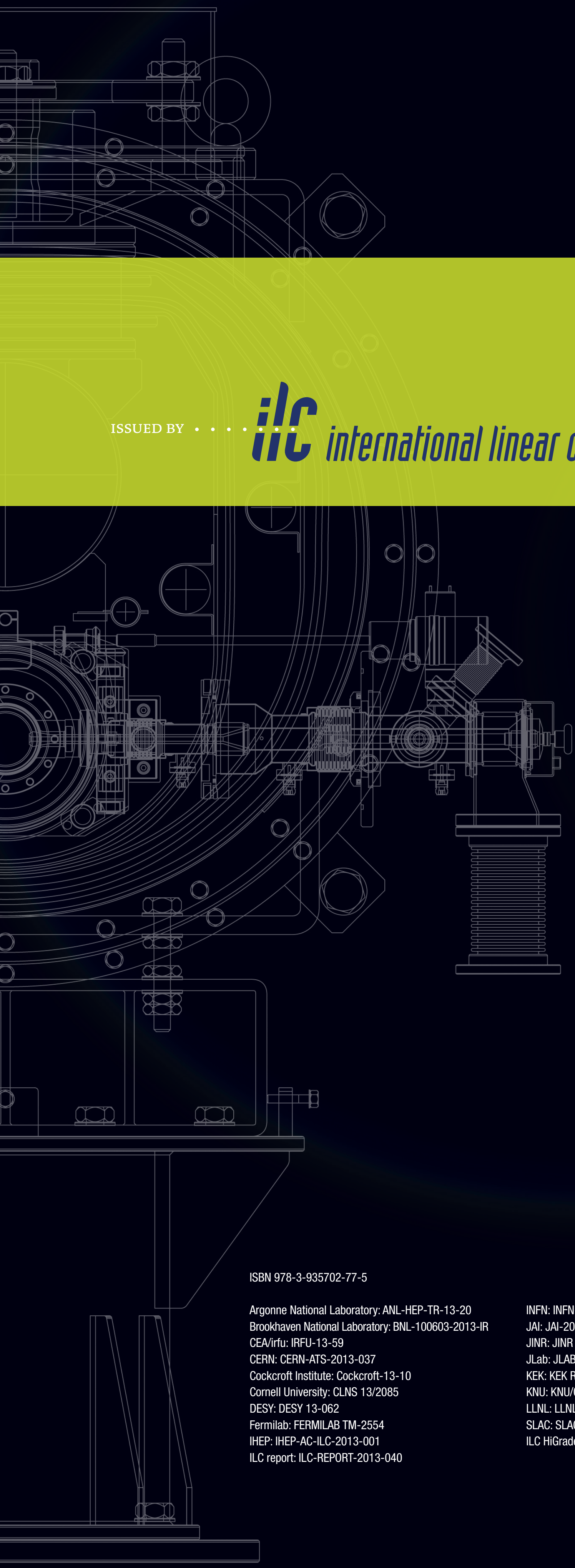

ISSUED BY international linear collider

WWW.LINEARCOLLIDER.ORG